\documentclass[envcountchap,graybox,vecphys]{svmono}

\usepackage{type1cm}
\usepackage{newtxtext}
\usepackage{newtxmath}

\usepackage{cite}			 % for compressed lists of citations
\usepackage[ddmmyyyy]{datetime}% custom formatting of date and time
\usepackage[bottom]{footmisc}% places footnotes at page bottom

\usepackage{amsfonts}   % AMS font package
\usepackage{bbold}		% blackboard bold font
\usepackage{cancel}     % for cancellations within equations
\usepackage[type={CC},modifier={by-nc-nd},version={4.0}]{doclicense}% CC license logo
\usepackage{graphicx}   % standard LaTeX graphics tool
\usepackage[pdftex,colorlinks,linkcolor=purple,citecolor=red,urlcolor=blue]{hyperref}
\usepackage{mathrsfs}	% math rsfs font
\usepackage{mathtools}	% extension of the amsmath package
\usepackage{wasysym}	% astronomical symbols

\allowdisplaybreaks
\newcommand{\probsec}{Exercise Problems}% name for the problem section

\let\a=\alpha
\let\b=\beta
\let\g=\gamma
\let\d=\delta
\let\eps=\epsilon
\let\k=\kappa
\let\l=\lambda
\let\m=\mu
\let\n=\nu
\let\p=\phi
\let\s=\sigma
\let\S=\Sigma
\let\t=\theta
\let\x=\xi
\let\o=\omega
\let\O=\Omega

\let\ve=\varepsilon% for Levi-Civita tensor
\let\vp=\varphi
\let\vr=\varrho

\newcommand{\C}{\mathbb{C}}% complex numbers
\newcommand{\R}{\mathbb{R}}% real numbers
\newcommand{\Z}{\mathbb{Z}}% integer numbers

\newcommand{\un}{\mathbb{1}}% unit matrix symbol
\newcommand{\df}{F}% infinitesimal transformation of fields
\newcommand{\DF}{\mathbb{F}}% finite transformation of fields
\newcommand{\Ha}{\mathscr{H}}% Hamiltonian density
\newcommand{\La}{\mathscr{L}}% Lagrangian density
\newcommand{\Ta}{\mathscr{T}}% kinetic energy density
\newcommand{\Va}{\mathscr{V}}% potential energy density
\newcommand{\Fa}{\mathscr{F}}% functional density
\newcommand{\Ga}{\mathscr{G}}% alternative functional density
\newcommand{\bigO}{\mathcal{O}}% the big O

\newcommand{\abs}[1]{\left\lvert{#1}\right\rvert}% absolute value
\newcommand{\gr}[1]{\mathrm{#1}}% font for specific group names
\newcommand{\Od}[2]{\D{#1}/\D{#2}}% ordinary derivative in text
\newcommand{\OD}[2]{\frac{\D{#1}}{\D{#2}}}% ordinary derivative display
\newcommand{\Pd}[2]{\de{#1}/\de{#2}}% partial derivative in text
\newcommand{\PD}[2]{\frac{\de{#1}}{\de{#2}}}% partial derivative display
\newcommand{\skal}[2]{\vec{#1}\cdot\vec{#2}}% scalar product of two vectors
\newcommand{\vekt}[2]{\vec{#1}\times\vec{#2}}% vector product of two vectors
\newcommand{\at}[2]{\left.{#1}\right\rvert_{#2}}% evaluation of #1 at the point #2
\newcommand{\scal}[2]{(#1,#2)}% abstract scalar product on a vector space
\newcommand{\he}[1]{{#1}^\dagger}% Hermitian conjugate

\let\de=\partial% partial derivative
\newcommand{\grad}{\vec\nabla}% gradient operator
\newcommand{\divg}{\vec\nabla\cdot}% divergence operator
\newcommand{\rot}{\vec\nabla\times}% curl operator
\DeclareMathOperator{\ifeq}{\overset{?}{=}}% equality sign with a question mark on top
\DeclareMathOperator{\tr}{tr}% trace
\DeclareMathOperator{\sgn}{sgn}% sign of a permutation
\DeclareMathOperator{\diag}{diag}% diagonal matrix

\newcounter{exno}[chapter]
\renewcommand{\theexno}{\thechapter.\arabic{exno}}
\newcommand{\refex}[1]{Example~\ref{#1}}% reference to an example
\newcommand{\refpr}[1]{Problem~\ref{#1}}% reference to an exercise problem
\newenvironment{illustration}{\refstepcounter{exno}\begin{tips}{Example~\theexno}}{\end{tips}}
\newenvironment{watchout}{\begin{svgraybox}}{\end{svgraybox}}

\begin{document}

\author{Tom\'a\v{s} Brauner}
\title{\hbox to0pt{Analytical Mechanics and Field Theory}}
\subtitle{Version 1.0 (\today)}
\maketitle

%%%%%%%%%%%%%%%%%%%%%%%%%%%%%%%%%%%%%%%%%%%%%%%%%%%%%%%%%%%%
%%%%%%%%%%%%%%%%%%%%%% Frontmatter %%%%%%%%%%%%%%%%%%%%%%%%%
%%%%%%%%%%%%%%%%%%%%%%%%%%%%%%%%%%%%%%%%%%%%%%%%%%%%%%%%%%%%

\frontmatter
\preface

%%%%%%%%%%%%%%%%%%%%%%%%%%%%%%%%%%%%%%%%%%%%%%%%%%%%%%%%%%%%

This text contains lecture notes accompanying the course \emph{FYS500 Analytical Mechanics and Field Theory} at the University of Stavanger. The presentation follows closely the lectures with occasional minor additions of extra material. A set of exercise problems is suggested at the end of each \chaptername. A sketch of solutions to all the problems can be found in \appendixname~\ref{app:solutions}. It is however strongly recommended that you first try to solve the problems on your own. In case you cannot fully solve a problem without help, you should use the provided solution as a guide. You might still have to fill in some of the more straightforward intermediate steps. Parts of the course require some amount of background in linear algebra. \appendixname~\ref{app:linalg} gives a brief introduction to the subject in an extent that is more than sufficient for the purposes of the course.

Let me conclude the preface with some remarks on available literature. Mechanics is a time-honored subject on which countless books have been published. While it is the very purpose of these lecture notes to provide a self-contained reading material for our course, you might want to look for additional resources, either for an alternative presentation of the material, or for further details. I would like to help you with that by offering a collection of references that I find particularly useful. The list combines relatively new books that are likely to be available on the market with a narrow selection of venerable classics.

\runinhead{Undergraduate-Level Books} You have all taken a course on mechanics before, and I will therefore not bother you with recommendations on books that are of introductory nature. However, there are some books that start nearly from scratch and still cover to some extent the Lagrangian and Hamiltonian formulations of mechanics, which constitute the core of the course. Among these, Morin's book~\cite{Morin2008} stands out for its accessibility and wealth of exercise problems to ponder on. Also, the book by Hamill~\cite{Hamill2022} might be useful. Finally, if you prefer a concise style, try Landau and Lifshitz~\cite{Landau1976}. This classic covers much of the same ground as other books on just over 160 pages. From books at an intermediate level, I would certainly recommend Taylor~\cite{Taylor2005}, which also contains many interesting exercise problems.

\runinhead{Graduate-Level Books} The book by Goldstein was first published in 1950 and is now available in an updated third edition with two co-authors~\cite{Goldstein2013a}. Our course covers much of this book, skipping some of the more technical passages and adding bits of extra material here and there. The book is therefore certainly a very useful complement to the course. One drawback is that the exercise problems are unfortunately not rated by difficulty, and some of them can be quite intimidating. The book by Hand and Finch~\cite{Hand1998} is a good alternative to Goldstein, being less wordy and offering some original insights not found elsewhere. Finally, for a more modern exposition, try Lemos~\cite{Lemos2018}. This is essentially the same content as Goldstein including the order of the chapters, but uses a more mature mathematical language and makes occasional detours into more advanced topics.

\runinhead{Mathematically Oriented Books} For those who are more mathematically inclined, let me add a few more references; note that these are only meant as further reading for enthusiasts. Classical mechanics as a research area has developed far beyond the initial discoveries of Euler, Hamilton, Lagrange and others that one usually meets in university courses. There is quite some literature that attempts to communicate the more recent advances in the field, but this is largely inaccessible to students without rigorous mathematical training, with few exceptions. Let me begin with the remarkable book by Spivak~\cite{Spivak2010}. This starts literally from scratch, rethinking the basics of Newton's laws, but ending up at a fairly advanced level. A bit idiosyncratic for a physicist, the book does contain much unique insight. The book by Jos\'e and Saletan~\cite{Jose1998a} offers a reasonable compromise between accessibility and depth. A somewhat older alternative is the classic by Arnold~\cite{Arnold1989a}, one of the pioneers of the modern development of classical mechanics. Written in a characteristically lucid style, the book abounds in deep connections between mechanics and differential geometry and topology.

\runinhead{Books on Classical Field Theory} Most of the above-reviewed books focus of the classical mechanics of systems with a finite number of degrees of freedom. Yet, classical field theory is equally important, both due to its applications to macroscopic matter and as a springboard towards quantum field theory, which plays a key role in modern high-energy and condensed-matter theory. Sadly, very few books give classical field theory proper credit. Usually, it is either relegated to the last chapter(s) of the volume, or left out altogether. The few books dedicated wholly to classical field theory mostly focus on its important applications such as electrodynamics and general relativity. The only relatively accessible book that treats classical field theory as a subject worth attention in its own right seems to be that by Soper~\cite{Soper2008}. This has heavily influenced the style of \chaptername s~\ref{chap:elasticity} and~\ref{chap:fluid} of the present lecture notes. Other than that, if you are looking for more information on classical field theory, you will have to resort to specialized monographs on the individual topics covered here.

%%%%%%%%%%%%%%%%%%%%%%%%%%%%%%%%%%%%%%%%%%%%%%%%%%%%%%%%%%%%

\vspace{\baselineskip}
\begin{flushright}\noindent
Stavanger, Norway\hfill Tom\'a\v{s} Brauner\\
\monthname\space\the\year\hfill{\phantom{Brauner}}\\
\end{flushright}

\tableofcontents
\extrachap{Notation and Conventions}

%%%%%%%%%%%%%%%%%%%%%%%%%%%%%%%%%%%%%%%%%%%%%%%%%%%%%%%%%%%%

\section*{List of Acronyms}

\begin{description}[EoM]
\item[CMB]{Cosmic microwave background}
\item[CoM]{Center of mass}
\item[EL]{Euler--Lagrange (equation)}
\item[EM]{Energy--momentum (tensor)}
\item[EoM]{Equation of motion}
\item[IRF]{Inertial reference frame}
\item[KG]{Klein--Gordon (equation, field, theory)}
\item[LRL]{Laplace--Runge--Lenz (vector)}
\item[ODE]{Ordinary differential equation}
\item[PDE]{Partial differential equation}
\end{description}

%%%%%%%%%%%%%%%%%%%%%%%%%%%%%%%%%%%%%%%%%%%%%%%%%%%%%%%%%%%%

\section*{Mathematical Conventions}

\runinhead{Space and Spacetime Indices} Spatial vectors are indicated in boldface, such as $\vec r,\vec v,\vec E,\vec B$. Components of space vectors in a Cartesian coordinate system are distinguished using lowercase Latin indices $i,j,\dotsc$. It is common to use superscripts for components of vectors and we do so where it improves visual clarity. Where there is no danger of confusion, subscripts may be used instead. Thus, for instance, both $E^i$ and $E_i$ may refer to the Cartesian components of an electric field $\vec E$.

The symbol used specifically for the position vector in space depends somewhat on the context. In mechanics where position is a dependent variable, the position vector is denoted mostly as $\vec r$. In field theory, on the other hand, position belongs to the independent variables. Here it is more common to use the symbol $\vec x$.

Spacetime indices are indicated using lowercase Greek indices $\m,\n,\dotsc$. Components of spacetime vectors are always distinguished using superscripts, such as $x^\m$ for the coordinate four-vector. There is no widely used font reserved for four-vectors. In these lecture notes, four-vectors (or other spacetime tensors) are denoted using regular italic letters. For instance, $x$ without an index stands for the entire coordinate four-vector rather than its components.

\runinhead{Einstein Summation Convention} Whenever a pair of identical indices appears in a product, summation over all possible values of the index is automatically implied. In case of summation over spacetime indices, one of the indices in the pair must be a subscript and the other a superscript. For example, $u^iv^i,u^iv_i,u_iv^i,u_iv_i$ are all permissible expressions for the scalar product of two vectors, $\skal uv$. An example of an inappropriate use of the summation convention would be $u_iv_iw_i$, where the same index is repeated three times. In such cases, summation must be indicated explicitly.

\runinhead{Standard Space(time) Tensors} The \emph{Kronecker delta} is defined by
\begin{equation*}
\d_{ij}=1\quad\text{for }i=j\;,\qquad
\d_{ij}=0\quad\text{for }i\neq j\;.
\end{equation*}
The alternative forms $\d^{ij}$ and $\d^i_j$ are also common. When used with spatial indices, the Kronecker delta encodes the standard Euclidean metric. Thus, the scalar product of two vectors can be written, among others, as
\begin{equation*}
\skal uv=\d_{ij}u^iv^j=\d^{ij}u_iv_j\;.
\end{equation*}
In Minkowski spacetime, the \emph{Minkowski metric} plays the same role,
\begin{equation*}
g_{\m\n}=-1\quad\text{for }\m=\n=0\;,\quad
g_{\m\n}=+1\quad\text{for }\m=\n>0\;,\quad
g_{\m\n}=0\quad\text{for }\m\neq\n\;.
\end{equation*}
This allows to write the Minkowski dot product of four-vectors $p,q$ as
\begin{equation*}
p\cdot q=g_{\m\n}p^\m q^\n=-p^0q^0+\skal pq\;,
\end{equation*}
where $\vec p,\vec q$ are the spatial parts of the four-vectors. The Minkowski metric $g_{\m\n}$ and its matrix inverse $g^{\m\n}$ can also be used to raise or lower indices of spacetime tensors. When doing so, the order of indices in a tensor must always be maintained, as in
\begin{equation*}
p_\m=g_{\m\n}p^\n\;,\qquad
T^\m_{\phantom\m\n}=g_{\n\l}T^{\m\l}\;,\qquad
F^{\m\n}=g^{\m\a}g^{\n\b}F_{\a\b}\;.
\end{equation*}
This convention allows one to express the Minkowski dot product of four-vectors alternatively as
\begin{equation*}
p\cdot q=p^\m q_\m=p_\m q^\m=g^{\m\n}p_\m q_\n\;.
\end{equation*}
Note that expressions such as $p_\m q_\m$ are ill-defined and should be avoided.

%%%%%%%%%%%%%%%%%%%%%%%%%%%%%%%%%%%%%%%%%%%%%%%%%%%%%%%%%%%%
%%%%%%%%%%%%%%%%%%%%%% Mainmatter %%%%%%%%%%%%%%%%%%%%%%%%%%
%%%%%%%%%%%%%%%%%%%%%%%%%%%%%%%%%%%%%%%%%%%%%%%%%%%%%%%%%%%%

\mainmatter
\chapter{Mathematical Introduction}
\label{chap:mathintro}

\keywords{Functional, Gateaux derivative, functional derivative, Euler--Lagrange equation, cyclic coordinate, first integral, Lagrange multiplier.}

%%%%%%%%%%%%%%%%%%%%%%%%%%%%%%%%%%%%%%%%%%%%%%%%%%%%%%%%%%%%

\noindent Our course largely revolves around two advanced formulations of mechanics, discovered by Lagrange and Hamilton. These are, unlike the older Newtonian mechanics, not based on the concept of force. As we will see later, this offers numerous technical advantages. However, the greatest virtue of the Lagrange and Hamilton formalisms is that they are readily applied to other branches of physics than mechanics. Indeed, they constitute a universal framework in which \emph{all} currently known fundamental laws of nature are formulated. In order to be able to fully benefit from the versatility of the Lagrange and Hamilton formalisms, we have to start by developing an appropriate mathematical language. This will, apart from making us ready to deal with physics later, also allow us to address interesting problems of purely mathematical nature. In case you are looking for further reading supplementing this \chaptername, I warmly recommend Chap.~1 of~\cite{Stone2009a}.

%%%%%%%%%%%%%%%%%%%%%%%%%%%%%%%%%%%%%%%%%%%%%%%%%%%%%%%%%%%%

\section{Basics of Variational Calculus}

You are already familiar with calculus of functions of one or several variables. What we need to do here is promote some of the structure of calculus to mathematical objects that take one or more functions as their argument(s). Such objects are called \emph{functionals}, and the art of manipulating functionals is known as \emph{variational calculus}.\footnote{Not to be confused with ``functional calculus'' and ``functional analysis,'' which are quite different branches of mathematics.} In order to underline the basic ideas without drowning them in complicated notation, we start with the simplest case of functionals that depend on a single function of a single real variable.

%%%%%%%%%%%%%%%%%%%%%%%%%%%%%%%%%%%%%%%%%%%%%%%%%%%%%%%%%%%%

\subsection{Single Function of One Variable}

For the time being, we shall mean by a functional any map that assigns to a function of a real variable a single number. While this may sound a bit formal, you actually already know numerous examples of functionals.

\begin{illustration}%
\label{ex00:deffunctional}%
Take any real function, $f:\R\to\R$, and define the functional $F_1$ by
\begin{equation}
F_1[f]\equiv f(0)\;.
\label{ch00:deltadef}
\end{equation}
Mind the notation: the arguments of functions are indicated using parentheses, whereas those of functionals by brackets. Also, I will always use capital letters to denote functionals. Of course, there is nothing special about evaluating the function $f(x)$ at $x=0$. We could have as well defined the functional~\eqref{ch00:deltadef} by the value of $f$ at any other (fixed) point. Here are some more examples of functionals,
\begin{equation}
F_2[f]\equiv-f'(0)\;,\qquad
F_3[f]\equiv\int_{-\infty}^{+\infty}[f'(x)]^2\exp(-x^2)\,\D x\;.
\end{equation}
\end{illustration}

As you see, functionals are nothing to be afraid of. In practice, they are mostly built using already familiar operations on functions. The novelty is the emphasis on the function as the variable. From calculus, we are already used to the fact that functions always have a domain, whether explicit or implicit, on which they are defined. The same applies to functionals. Just think of what conditions the argument $f$ must satisfy so that the three functionals in~\refex{ex00:deffunctional} are well-defined.

For reasons that will become clear later, we will from now on only deal with \emph{local functionals}, which can be expressed~as
\begin{equation}
F[f]\equiv\int_a^b\Fa(f(x),f'(x),f''(x),\dotsc,x)\,\D x\;.
\label{ch00:localfunctional}
\end{equation}
Here $\Fa$ is the \emph{density} of the functional, which is an ordinary function of several variables. When evaluating $F[f]$, these variables are substituted with the values of $f$ and a finite number of its derivatives at the same point $x\in\R$. The last argument of $\Fa(f(x),f'(x),f''(x),\dotsc,x)$ indicates that it may also depend directly (explicitly) on $x$ itself. I will often simplify the notation and write the integrand in~\eqref{ch00:localfunctional} simply as $\Fa(f,f',f'',\dotsc,x)$.

\begin{illustration}%
You are also already familiar with an example of a local functional. Just think of the mass $m$ of a body whose density $\vr(\vec x)$ depends on the position $\vec x$. This is given by a simple integration of the density without any derivatives, $m=\int\vr(\vec x)\,\D V$, where $\D V$ is an infinitesimal volume element. Thus, the mass of the body is a functional of its density distribution. You will now surely be able to recall further examples of local functionals that you have already met in physics. The functional $F_3$ introduced in~\refex{ex00:deffunctional} is another example of a local functional.

Notably, the functionals $F_1$ and $F_2$ from~\refex{ex00:deffunctional} are not of the local type. There is simply no way to get $f(0)$ by integrating some combination of $f(x)$ and its derivatives without further assumptions. However, since representing a functional by its density as in~\eqref{ch00:localfunctional} is practically very convenient, it is common to introduce a fictitious ``function'' $\d(x)$ of a single variable such that
\begin{equation}
\int_a^bf(x)\d(x)\,\D x=f(0)=F_1[f]
\end{equation}
for any $a,b$ such that $a<0<b$. As you see, the infamous \emph{Dirac $\d$-function} $\d(x)$ is in fact a functional in disguise. It should not surprise you that the functional $F_2[f]$ can similarly be represented by the derivative of the $\d$-function, $\d'(x)$. Make sure you understand the origin of the minus sign in $F_2[f]=-f'(0)$ though!
\end{illustration}

\begin{watchout}%
The set of all local functionals of a function $f$ of a single real variable possesses a natural linear-algebraic structure. Namely, for any two functionals $F_1,F_2$ with densities $\Fa_1,\Fa_2$ and any two real numbers $c_1,c_2$, the functional $c_1F_1+c_2F_2$ can be defined by integrating the density $c_1\Fa_1+c_2\Fa_2$. In other words, we can add functionals and multiply them by a scalar. What we cannot do straightforwardly is taking a product of functionals. There is no easy way to represent the functional $F_1[f]F_2[f]$ in terms of a density. Vice versa, there is no easy way to relate the result of integrating the density $\Fa_1(f,f',\dotsc,x)\Fa_2(f,f',\dotsc,x)$ to the values of $F_1[f]$ and $F_2[f]$. Since it is not trivial to define even basic algebraic operations on functionals, it is hardly surprising that generalizing calculus to functionals is tricky as well. What we will be able to do is introduce the concept of a derivative of functionals. This is all we will need for the purposes of the present course. It is also possible to define integration of functionals, that is however only relevant for quantum mechanics and quantum field theory, and to some extent for statistical physics.
\end{watchout}

To see how to define a derivative of a functional, recall the concept of gradient of a function of several variables $x_1,x_2,\dotsc$. This is formally a vector with components given by the individual partial derivatives of the function,
\begin{equation}
\grad f\equiv\left(\PD f{x_1},\PD f{x_2},\dotsc\right)\;.
\end{equation}
The partial derivatives themselves are defined by a limit,
\begin{equation}
\PD f{x_i}\equiv\lim_{\eps\to0}\frac{f(x_1,\dotsc,x_{i-1},x_i+\eps,x_{i+1},\dotsc)-f(x_1,\dotsc,x_{i-1},x_i,x_{i+1},\dotsc)}{\eps}\;.
\end{equation}
If you think of $x_1,x_2,\dotsc$ as Cartesian coordinates in Euclidean space, the partial derivatives seem to be rather arbitrary, since they depend on the choice of orientation of the Cartesian coordinate frame. However, the gradient is a well-defined geometric object. This can be used to define a scalar derivative that is independent of the choice of coordinates. To that end, we pick an arbitrary constant vector $\vec v$ and set
\begin{equation}
\nabla_{\vec v}f(\vec x)\equiv\lim_{\eps\to0}\frac{f(\vec x+\eps\vec v)-f(\vec x)}{\eps}=\skal v\nabla f(\vec x)\;.
\label{ch00:directionalderivative}
\end{equation}
This is the \emph{directional derivative} of $f$ along the vector $\vec v$. It measures the rate of change of the function $f$ at point $\vec x$ in the ``direction'' specified by the vector $\vec v$. Note that the directional derivative depends on both the direction and the magnitude of $\vec v$. Being obviously linear in $\vec v$, $\nabla_{\vec v}f$ doubles if we stretch the vector to twice its length without changing its direction.

Following the analogy, we now choose a fixed function $\eta:\R\to\R$ and define, for a functional $F$ of a function of a single variable,
\begin{equation}
\boxed{D_\eta F[f]\equiv\lim_{\eps\to0}\frac{F[f+\eps\eta]-F[f]}{\eps}\;.}
\label{ch00:Gateauxder}
\end{equation}
This so-called \emph{Gateaux derivative} has the meaning of a ``directional derivative'' of the functional $F$ at the ``point'' $f$ in the ``direction'' specified by the auxiliary function~$\eta$.

\begin{illustration}%
Recall the functional $F_3$ introduced in~\refex{ex00:deffunctional}. Using this in the definition~\eqref{ch00:Gateauxder}, we find
\begin{equation}
\begin{split}
F_3[f+\eps\eta]-F_3[f]&=\int_{-\infty}^{+\infty}\bigl\{[f'(x)+\eps\eta'(x)]^2-[f'(x)]^2\bigr\}\exp(-x^2)\,\D x\\
&=\int_{-\infty}^{+\infty}\bigl\{2\eps\eta'(x)f'(x)+\eps^2[\eta'(x)]^2\bigr\}\exp(-x^2)\,\D x\;.
\end{split}
\end{equation}
This implies at once that
\begin{equation}
D_\eta F_3[f]=2\int_{-\infty}^{+\infty}\eta'(x)f'(x)\exp(-x^2)\,\D x\;.
\end{equation}
Suppose now that both $f(x)$ and $\eta(x)$ and their first derivatives are bounded functions, or at least do not grow too fast for large $x$. We can then integrate by parts to remove the derivative from $\eta'(x)$ without generating a boundary term. This leads to
\begin{equation}
D_\eta F_3[f]=-2\int_{-\infty}^{+\infty}\eta(x)\OD{}x[f'(x)\exp(-x^2)]\,\D x\;.
\label{ch00:GateauxF3}
\end{equation}
\end{illustration}

Let us pause and consider what the calculation in the above example tells us about the Gateaux derivative of an arbitrary local functional. First, $D_\eta F$ is linear in $\eta$. This is because upon a series expansion, higher powers of $\eta$ always come with higher powers of $\eps$, which drop out in the limit $\eps\to0$. We only need to expand $F[f+\eps\eta]$ to first order in $\eps$ and $\eta$. Second, whatever derivatives on $\eta$ might appear in the process can always be removed by partial integration. This step may require additional assumptions on the behavior of $f(x)$ and $\eta(x)$ at the boundary of the integration domain. Provided no boundary terms are generated by the partial integration, the Gateaux derivative can eventually be written as
\begin{equation}
\boxed{D_\eta F[f]\equiv\int\eta(x)\frac{\udelta F}{\udelta f(x)}\,\D x\;.}
\label{ch00:functionalder}
\end{equation}
This defines implicitly the \emph{functional derivative} $\udelta F/\udelta f$ of the functional $F$ at the ``point'' $f$. The functional derivative is an ordinary function of $x$; it is fixed uniquely by the requirement that~\eqref{ch00:functionalder} holds for any choice of $\eta(x)$ that guarantees the absence of boundary terms. Note the similarity of~\eqref{ch00:functionalder} to~\eqref{ch00:directionalderivative}: while the Gateaux derivative generalizes the concept of a directional derivative from functions to functionals, the functional derivative similarly generalizes the gradient of a function.

In~\eqref{ch00:functionalder}, I have dropped the integration limits. The definition applies equally to local functionals~\eqref{ch00:localfunctional} regardless of the choice of $a,b$. The only requirement is that whatever integration by parts is needed to bring the Gateaux derivative $D_\eta F[f]$ to the form~\eqref{ch00:functionalder}, it does not give rise to any boundary terms. For functionals defined by integration over the whole real axis, this is ensured by assuming that both $f(x)$ and $\eta(x)$ and their first derivatives decrease sufficiently fast for $x\to\pm\infty$. In case of integration over a finite interval, one might instead have to impose a sufficiently strong boundary condition on $\eta$ at $x=a,b$. Typically, this entails the vanishing of $\eta$ and possibly some of its derivatives at the boundary.

\begin{illustration}%
\label{ex00:Fg}%
Getting back to the functional $F_3$ from~\refex{ex00:deffunctional}, it follows at once from~\eqref{ch00:GateauxF3}~that
\begin{equation}
\frac{\udelta F_3}{\udelta f(x)}=-2\OD{}x[f'(x)\exp(-x^2)]=-2[f''(x)-2xf'(x)]\exp(-x^2)\;.
\end{equation}
For a little extra exercise, suppose that we replace the exponential $\exp(-x^2)$ in the definition of $F_3$ with an arbitrary fixed function $g$ and define $F_g[f]\equiv\int[f'(x)]^2g(x)\,\D x$. It is easy to repeat the steps leading to the Gateaux derivative~\eqref{ch00:GateauxF3} and see that in this more general case,
\begin{equation}
\frac{\udelta F_g}{\udelta f(x)}=-2\OD{}x[f'(x)g(x)]=-2[f''(x)g(x)+f'(x)g'(x)]\;.
\label{ch00:dFgdf}
\end{equation}
Below, we will develop a simple way to compute the functional derivative directly from the functional density $\Fa$ without having to first expand in powers of $\eps$ and then integrate by parts every time.
\end{illustration}

%%%%%%%%%%%%%%%%%%%%%%%%%%%%%%%%%%%%%%%%%%%%%%%%%%%%%%%%%%%%

\subsection{Multiple Functions of Several Variables}

Having gained a bit of experience, let us delve into the general case of local functionals of a set of functions, each of which itself depends on the same set of independent variables. I will denote the latter as $x_i$ without specifying the range for the index $i$; none of the results below depend on it. Likewise, I will use the notation $f_A$ for the functions that constitute the arguments of a functional. The one simplifying assumption I will make is that the functional density $\Fa$ only depends on the functions $f_A$ and their first partial derivatives; this will save us some manipulations. Altogether, we therefore consider local functionals of the type
\begin{equation}
F[f]\equiv\int_\O\D^nx\,\Fa(f,\de f,x)\;,
\label{ch00:localfunctionalgeneral}
\end{equation}
where $n$ is the number of variables $x_i$ that the functions $f_A$ depend on. The integration domain $\O$ (often suppressed in the following) can be the whole space $\R^n$ or a finite domain therein. Note the shorthand notation, where ``$f$'' as an argument of a functional or its density stands for the collection of all the functions $f_A$. The same applies to ``$x$'' as the collection of all the independent variables $x_i$, and to ``$\de f$'' as the collection of all the first partial derivatives $\de_if_A\equiv\Pd{f_A}{x_i}$ of all the functions.

Just like a function can be differentiated with respect to any of its variables, so can we now take the Gateaux or functional derivative of $F$ with respect to any of the functions $f_A$. To that end, we introduce a set of auxiliary functions $\eta_A$, one for each $f_A$. The Gateaux derivative is then defined by a straightforward generalization of~\eqref{ch00:Gateauxder}. We first take the finite difference
\begin{align}
F[f+\eps\eta]-F[f]&=\int\D^nx\,[\Fa(f+\eps\eta,\de f+\eps\de\eta,x)-\Fa(f,\de f,x)]\\
\notag
&=\eps\sum_A\int\D^nx\,\biggl[\eta_A\PD{\Fa}{f_A}+\sum_i(\de_i\eta_A)\PD{\Fa}{(\de_if_A)}\biggr]+\bigO(\eps^2)\;.
\end{align}
Next, we integrate by parts to remove the derivatives from the $\eta_A$s in the second~term under the integral. In order that this does not generate any boundary terms, we assume that all the $\eta_A$s vanish on the boundary of the integration domain. Finally, dividing by $\eps$ and subsequently taking the limit $\eps\to0$, we obtain the Gateaux derivative
\begin{equation}
D_\eta F[f]=\sum_A\int\D^nx\,\eta_A\biggl[\PD{\Fa}{f_A}-\sum_i\de_i\PD{\Fa}{(\de_if_A)}\biggr]\;,
\end{equation}
with the shorthand notation $\de_i\equiv\Pd{}{x_i}$. This leads in turn immediately to a basic formula for the functional derivative,
\begin{equation}
\boxed{\frac{\udelta F}{\udelta f_A}=\PD{\Fa}{f_A}-\sum_i\de_i\PD{\Fa}{(\de_if_A)}\;.}
\label{ch00:functionaldergeneral}
\end{equation}

\begin{watchout}%
The notation used in the derivation as well as its final result~\eqref{ch00:functionaldergeneral} is, while commonplace, a perpetual source of confusion. How do we take a derivative of $\Fa$ with respect to the partial derivative $\de_if_A$, and subsequently differentiate that with respect to $x_i$? This is a travesty that increases the blood pressure of many a mathematician. Here is how you should think about it. Imagine that $\Fa$ depends (apart from $x_i$) on two independent sets of variables, $f_A$ and $g_{iA}$. Compute the standard partial derivatives $\Pd{\Fa}{f_A}$ and $\Pd{\Fa}{g_{iA}}$. Having done this, recall that $f_A$ are in fact supposed to be functions of $x_i$. With this in mind, replace $f_A$ with $f_A(x)$ and $g_{iA}$ with $\de_if_A(x)$ everywhere. Finally, differentiate your ``$\Pd{\Fa}{(\de_if_A)}$'' as an ordinary function of coordinates with respect to $x_i$. This procedure can be formalized by using function composition. However, I do believe you can understand it by going carefully through the derivation of~\eqref{ch00:functionaldergeneral} and making sure you are comfortable with all the steps, including the mathematical meaning of all the intermediate expressions.
\end{watchout}

Just for the record, it is not a big problem to repeat all the steps leading to~\eqref{ch00:functionaldergeneral} without the assumption that the functional density $\Fa$ only depends on the functions $f_A$ and their first derivatives. For a general local functional of the type
\begin{equation}
F[f]\equiv\int_\O\D^nx\,\Fa(f,\de f,\de\de f,\dotsc,x)\;,
\label{ch00:deffunctionalhigherder}
\end{equation}
the functional derivative of $F$ with respect to $f_A$ can be computed from the functional density $\Fa$ via
\begin{equation}
\boxed{\begin{aligned}
\frac{\udelta F}{\udelta f_A}&=\PD{\Fa}{f_A}-\sum_i\de_i\PD{\Fa}{(\de_if_A)}+\sum_{i\leq j}\de_i\de_j\PD{\Fa}{(\de_i\de_jf_A)}-\dotsb\\
&=\sum_{k=0}^\infty(-1)^k\smashoperator{\sum_{i_1\leq\dotsb\leq i_k}}\de_{i_1}\dotsb\de_{i_k}\PD{\Fa}{(\de_{i_1}\dotsb\de_{i_k}f_A)}\;.
\end{aligned}}
\label{ch00:functionalderhigher}
\end{equation}
In the compact notation of the last expression, $\de_{i_1}\dotsb\de_{i_k}$ with $k=0$ evaluates to no derivative at all. Likewise, $\sum_{i_1\leq\dotsb\leq i_k}$ with $k=0$ implies no summation whatsoever.

\begin{illustration}%
\label{ex00:Fg2}%
Let us go through one example very slowly. For the functional $F_g$ defined in~\refex{ex00:Fg}, the functional density is $\Fa_g(f(x),f'(x),x)=[f'(x)]^2g(x)$. This gives
\begin{equation}
\PD{\Fa_g}{f}=0\;,\qquad
\PD{\Fa_g}{f'}=2f'g\;.
\end{equation}
Inserting this in~\eqref{ch00:functionaldergeneral}, we find
\begin{equation}
\frac{\udelta F_g}{\udelta f(x)}=-\OD{}x[2f'(x)g(x)]=-2[f''(x)g(x)+f'(x)g'(x)]\;,
\end{equation}
in accord with our previous result~\eqref{ch00:dFgdf}.
\end{illustration}

Getting back to functionals whose density only depends on the functions $f_A$ and their first derivatives, we are now ready to appreciate the broad importance of variational calculus. Namely, great many problems in mathematics and physics can be formulated as special cases of the following basic variational problem.

\begin{figure}[t]
\sidecaption[t]
\includegraphics[width=2.9in]{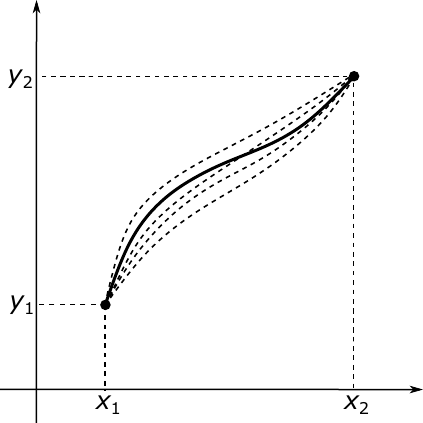}
\caption{Illustration of the space on which the minimum, maximum or stationary points of the functional~\eqref{ch00:localfunctionalgeneral} are to be found, in case of a single function $f$ of a single variable $x$. The allowed functions are subject to the boundary condition $f(x_{1,2})=y_{1,2}$ at the endpoints $x_1,x_2$ of the integration range. The dashed curves indicate some of the possible functions. The solid curve represents the putative stationary state, presumed to be uniquely determined by the imposed boundary condition.}
\label{fig00:variation}
\end{figure}

\begin{watchout}%
Find the minimum, maximum or stationary points of the local  functional~\eqref{ch00:localfunctionalgeneral},
\begin{equation}
F[f]\equiv\int_\O\D^nx\,\Fa(f,\de f,x)\;,
\end{equation}
on the space of all functions $f_A$ with prescribed fixed values on the boundary $\de\O$ of the integration domain. (See Fig.~\ref{fig00:variation} for a visualization of the space of functions in question.)
\end{watchout}

We know from calculus how to deal with such optimization problems: we have to look for points where the derivative(s) of a function vanish(es). The same applies to variational calculus. For the set of functions $f_A$ to constitute a stationary point of a functional $F$ (let alone its minimum or maximum), the Gateaux derivative of $F$ at $f_A$ must vanish in all ``allowed'' directions $\eta_A$. What do we mean by allowed? Since the boundary values of the functions are fixed, the variations of $f_A$ are restricted to $f_A\to f_A+\eps\eta_A$ with $\eta_A$ that vanish on $\de\O$. But this is exactly what we need to make sure that the boundary terms arising from partial integration vanish. Therefore, the condition that $f_A$ is a stationary point of the functional $F$ is equivalent to the vanishing of the functional derivative, $\udelta F/\udelta f_A=0$. By means of~\eqref{ch00:functionaldergeneral}, our basic variational problem thus reduces to the problem of finding the solution(s) of the \emph{Euler--Lagrange} (EL) \emph{equation},
\begin{equation}
\boxed{\PD{\Fa}{f_A}-\sum_i\de_i\PD{\Fa}{(\de_if_A)}=0\;,}
\label{ch00:EulerLagrange}
\end{equation}
with the appropriate fixed boundary condition on $\de\O$.

We have managed to do something that mathematicians like: reduce a new problem to a simpler problem that we know how to deal with. At this point, you should refresh what you learned previously about differential equations. You might recall that there is an order-of-magnitude difference between dealing with \emph{ordinary differential equations} (ODEs) and \emph{partial differential equations} (PDEs). As we will see, this is, mathematically speaking, the main difference between classical mechanics and classical field theory. Let us summarize:
\begin{itemize}
\item For functions of a single variable, \eqref{ch00:EulerLagrange} is a set of second-order ODEs, one for each function $f_A$. This is the case of mechanics, where time plays the role of the single independent variable $x$.
\item For functions of several variables, \eqref{ch00:EulerLagrange} constitutes a set of second-order PDEs, one for each function $f_A$. This is the case of field theory, where $x_i$ correspond to the spacetime coordinates and $f_A$ to the fields.
\end{itemize}
The EL equation~\eqref{ch00:EulerLagrange} is all we will need, without exception, in our course. However, in case you are curious and wonder what happens for functionals~\eqref{ch00:deffunctionalhigherder} depending on higher derivatives, it should not surprise you that the corresponding generalized EL equation (sometimes called \emph{Euler--Poisson equation}) is obtained by setting~\eqref{ch00:functionalderhigher} to zero. One just has to be a bit careful about the boundary condition, which may now require not just $\eta_A$, but also some of their derivatives, to vanish on $\de\O$. For functionals depending on up to the $k$-th derivatives of $f_A$, the Euler--Poisson equation will generally be an ordinary or partial differential equation of order $2k$.

%%%%%%%%%%%%%%%%%%%%%%%%%%%%%%%%%%%%%%%%%%%%%%%%%%%%%%%%%%%%

\section{First Integrals}
\label{sec:firstintegrals}

The one type of differential equation that is generally easy to deal with are first-order ODEs. However, even for the basic type of local functional~\eqref{ch00:localfunctionalgeneral}, the EL equation is of second order. Already for second order ODEs, only some special cases are known to be solvable in a closed form. Any trick that might help therefore counts. Here I will give a first sketch of a method that, under certain simplifying assumptions, allows one to reduce the order of the EL equation(s). In \chaptername~\ref{chap:symmetries}, we will develop this into a powerful scheme that will reveal the origin of conservation laws in physics. From now on, until we start talking about field theory in \chaptername~\ref{chap:LagHamcont}, we will only consider the case of a single independent variable $x$ and functional densities of the type $\Fa(f,f',x)$. In this special case, the EL equation~\eqref{ch00:EulerLagrange} takes the form
\begin{equation}
\PD{\Fa}{f_A}-\OD{}x\PD{\Fa}{f'_A}=0\;.
\label{ch00:EulerLagrange1d}
\end{equation}

Suppose that the functional density $\Fa$ depends on the derivatives $f'_A(x)$ and on the independent variable $x$, but not directly on $f_A(x)$ themselves, or on a subset thereof. Mathematically, this means that $\Pd{\Fa}{f_A}=0$ for some of the $f_A$s. Such functions are called \emph{cyclic coordinates}.\footnote{This terminology will only make a bit of sense when we actually start talking about mechanics, where $f_A$ are referred to as \emph{generalized coordinates}.} For any such a cyclic coordinate, the EL equation~\eqref{ch00:EulerLagrange1d} reduces to
\begin{equation}
\OD{}x\PD{\Fa}{f_A'}=0\;.
\end{equation}
We know how to solve this! It means that $\Pd{\Fa}{f_A'}$ must be constant. In accord with the fact that we have done a trivial integration of the EL equation, the constant is usually called \emph{first integral}. I will denote it as $I_{f_A}$. Then, for any cyclic coordinate $f_A$, the EL equation as a second-order ODE reduces to the first-order ODE
\begin{equation}
\boxed{I_{f_A}=\PD{\Fa}{f_A'}\;.}
\label{ch00:firstintcyclic}
\end{equation}
This is of great help, since we know how to deal with first-order ODEs, remember?

\begin{illustration}%
Recall once again the functional $F_g$ discussed in~\refex{ex00:Fg} and~\refex{ex00:Fg2}. Its functional density, $\Fa_g(f(x),f'(x),x)=[f'(x)]^2g(x)$, does not actually depend on $f(x)$, only on its derivative. Hence, $f$ is a cyclic coordinate. It follows that the EL equation arising from setting $\udelta F_g/\udelta f=0$ reduces to $2f'(x)g(x)=I_f$ with some constant $I_f$. For given function $g(x)$, this is readily integrated as
\begin{equation}
f(x)=\frac{I_f}2\int\frac{\D x}{g(x)}\;.
\end{equation}
The indefinite integral on the right-hand side involves an integration constant. Together with $I_f$, this makes the solution of the EL equation depend on two parameters, as appropriate for a second-order ODE. The parameters can be fixed using the boundary condition.
\end{illustration}

A similar simplification can be achieved when the functional density $\Fa$ does not depend explicitly on $x$, although it may depend on both $f_A$ and $f'_A$. In this case, there is another first integral, that is another function of $f_A$ and $f_A'$ that is constant as a consequence of the EL equation~\eqref{ch00:EulerLagrange1d},
\begin{equation}
\boxed{I_x\equiv\Fa-\sum_Af_A'\PD{\Fa}{f_A'}\;.}
\label{ch00:firstintenergy}
\end{equation}
This relation is sometimes called \emph{Beltrami identity}. It is easy to prove that $I_x$ as defined by~\eqref{ch00:firstintenergy} is indeed constant. All we need is the chain rule. However, keep in mind that we first treat $\Fa$ as a function of $f_A,f_A'$ and then replace the latter with $f_A(x),f_A'(x)$, which makes everything a function of $x$,
\begin{equation}
\begin{split}
\OD{I_x}x&=\sum_A\biggl(\PD{\Fa}{f_A}f_A'+\cancel{\PD{\Fa}{f_A'}f_A''}\biggr)-\sum_A\left(\cancel{f_A''\PD{\Fa}{f_A'}}+f_A'\OD{}x\PD{\Fa}{f_A'}\right)\\
&=\sum_Af_A'\left(\PD{\Fa}{f_A}-\OD{}x\PD{\Fa}{f_A'}\right)=0\;.
\end{split}
\end{equation}
Can you spot where we used the EL equation~\eqref{ch00:EulerLagrange1d} and where the assumption that $\Fa$ does not depend explicitly on $x$?

%%%%%%%%%%%%%%%%%%%%%%%%%%%%%%%%%%%%%%%%%%%%%%%%%%%%%%%%%%%%

\section{Applications}

It is now time for some illustrative examples. These will be mostly mathematical in nature. Some more physical applications of variational calculus are postponed to the exercise problems.

%%%%%%%%%%%%%%%%%%%%%%%%%%%%%%%%%%%%%%%%%%%%%%%%%%%%%%%%%%%%

\subsection{Path of Shortest Distance}
\label{subsec:shortestdistance}

\begin{figure}[t]
\sidecaption[t]
\includegraphics[width=2.9in]{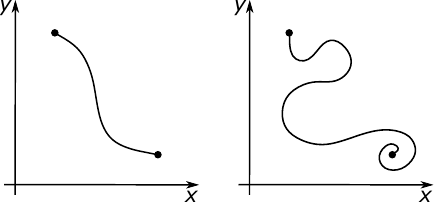}
\caption{Two qualitatively different types of paths connecting the same two given points in the Euclidean plane. Left panel: a path that can be represented by a function of a single variable, $y=f(x)$. Right panel: a path that requires parameterization by a variable $t$ (``time'') that cannot be identified with either~$x$~or~$y$.}
\label{fig00:shortestpath}
\end{figure}

Probably the simplest example that leads to a variational problem is concerned with finding the shortest path in a plane that connects two given points. (You know the answer, right?) The first thing we have to do is to give the colloquially defined problem a precise mathematical formulation. We introduce Cartesian coordinates in the plane, and assume that the path connecting our two points can be viewed as the graph of a function, $y=f(x)$; see the left panel of Fig.~\ref{fig00:shortestpath}. The length of an infinitesimal segment of the path can then be expressed as
\begin{equation}
\D s=\sqrt{\D x^2+\D y^2}=\D x\sqrt{1+\left(\OD yx\right)^2}=\D x\sqrt{1+[f'(x)]^2}\;.
\label{ch00:lineelement}
\end{equation}
The total length of the path is thus defined as a functional with $f$ as the sole argument,
\begin{equation}
L[f]=\int_a^b\D x\,\sqrt{1+[f'(x)]^2}\;,
\end{equation}
where $a,b$ are the $x$-coordinates of the fixed endpoints of the path. Finding the path that minimizes this functional matches precisely our basic variational problem. The density of the functional $L[f]$ is $\La(f,f',x)=\sqrt{1+(f')^2}$. This depends neither on $f$ nor on $x$. We therefore have the luxury of choosing between~\eqref{ch00:firstintcyclic} and~\eqref{ch00:firstintenergy} to find a first integral of the corresponding EL equation. In either case, the resulting first-order differential equation is equivalent to the requirement that $f'(x)$ be constant. There is only one function with constant derivative that satisfies the given boundary condition, i.e.~passes through the two prescribed points. The solution is therefore unique and it is, as expected, a straight line.

So far so good. However, we made a simplifying assumption at the initial stage that was not quite innocuous. Namely, not all (even smooth) paths connecting two points in a plane correspond to the graph of a function; see the right panel of Fig.~\ref{fig00:shortestpath} for an example. It is then more appropriate to introduce a parameterization of the path $(x(t),y(t))$ in terms of a new variable $t$. If you wish, you can think of $t$ as the ``time'' that parameterizes motion along the path. To be able to impose a fixed boundary condition, we assume that for any path connecting the same endpoints, the parameter runs from $t=0$ to $t=1$ as the path is traversed. Even with this restriction, there are still infinitely many allowed parameterizations of the same geometric path. (It might help to think of the analogy with physical motion to understand this.)

With the basic formalities out of the way, we now express the line element~\eqref{ch00:lineelement}~as
\begin{equation}
\D s=\D t\sqrt{[\dot x(t)]^2+[\dot y(t)]^2}\;,
\end{equation}
where the dot, as common in physics, indicates a derivative with respect to $t$. This leads to a formally quite different variational problem, based on the functional
\begin{equation}
L[x,y]=\int_0^1\D t\,\sqrt{[\dot x(t)]^2+[\dot y(t)]^2}\;.
\end{equation}
While we still have a single independent variable $t$, our length functional now depends on two arguments, $x(t)$ and $y(t)$. There will accordingly be two independent EL equations. Luckily, we do not have to deal with them directly since the functional density again does not depend on $x(t),y(t)$, or $t$ itself. Interestingly, the first integral implied by~\eqref{ch00:firstintenergy} is identically zero in this case. However, we still get two nontrivial first integrals from the fact that $x,y$ are cyclic. Using~\eqref{ch00:firstintcyclic} we thus find that both
\begin{equation}
I_x=\frac{\dot x}{\sqrt{\dot x^2+\dot y^2}}\quad\text{and}\quad
I_y=\frac{\dot y}{\sqrt{\dot x^2+\dot y^2}}
\label{ch00:shortestpathxyt}
\end{equation}
are constant. This of course implies that $\Od yx=\dot y/\dot x=I_y/I_x$ is also constant, which again leads to a straight line as the shortest path connecting the two given points.

Let us be a little more careful though. We see from~\eqref{ch00:shortestpathxyt} that $\sqrt{I_x^2+I_y^2}=1$. The two first integrals are therefore not mutually independent, but rather can be related to each other by $I_x=\cos\vp$ and $I_y=\sin\vp$, where $\vp$ is a constant angle determining the slope of the shortest path. It follows that
\begin{equation}
\dot x(t)=v(t)\cos\vp\;,\qquad
\dot y(t)=v(t)\sin\vp
\end{equation}
with some positive function $v(t)$ (``speed'') automatically satisfies the reduced EL equations~\eqref{ch00:shortestpathxyt}. The function $v(t)$ itself is only constrained by the boundary condition, which requires that $\int_0^1\D t\,v(t)$ is the actual length of the shortest path, i.e.~the distance of the endpoints. 

\begin{watchout}%
Let us ponder on what we have learned by this example. We found that the same geometric (or physical, to that matter) question may be formalized by multiple variational problems with vastly different solutions. Concretely, we saw how the length of a path in the plane can be cast as a functional of a single function, $f:\R\to\R$, or of two functions, $(x,y):\R\to\R^2$. Depending on the concrete formulation, the solution of the variational problem may but need not be unique. In fact, it is easy to understand why our second approach gave formally infinitely many solutions. This is traced back to the fact that we had to invent the parameter $t$ that does not feature in the original, geometric formulation of the shortest-path problem. The benefit of doing so is that our second solution has an immediate generalization to Euclidean spaces of arbitrary dimension. In fact, it can also be used to find the shortest path of two points lying on any curved surface. We will get back to this in \chaptername~\ref{chap:geometryclassmech}.
\end{watchout}

%%%%%%%%%%%%%%%%%%%%%%%%%%%%%%%%%%%%%%%%%%%%%%%%%%%%%%%%%%%%

\subsection{The Brachistochrone Problem}

\begin{figure}[t]
\sidecaption[t]
\includegraphics[width=2.9in]{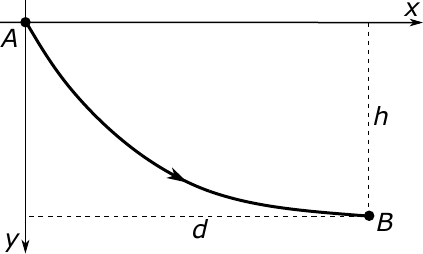}
\caption{Illustration of the brachistochrone problem. A point mass is released from point $A$ and constrained to move in a uniform gravitational field along the curve in bold, eventually reaching point $B$. We are to find the shape of the curve that minimizes the time it takes the mass to reach $B$ from $A$.}
\label{fig00:brachistochrone}
\end{figure}

As the next illustration, we consider the so-called brachistochrone problem, which played a central role in the historical development of variational calculus. The problem is as follows. We are given two points $A,B$ in a vertical plane, with horizontal distance $d$ and vertical distance $h$ (see Fig.~\ref{fig00:brachistochrone}). The whole system is subject to a uniform gravitational field $g$, oriented as usual vertically downwards. We take a point mass $m$ and release it from $A$, initially at rest. The mass is constrained to move without friction along a curve that ends at $B$. What is the shape of the curve the minimizes the time it takes the mass to reach the point $B$?

To formalize the problem, we introduce Cartesian coordinates as shown in Fig.~\ref{fig00:brachistochrone}. Energy conservation requires that the speed $v$ of the mass is related to its vertical position $y$ by $(1/2)mv^2=mgy$, hence $v=\sqrt{2gy}$. The time required to traverse a segment of the curve of length $\D s$ is then $\D t=\D s/\sqrt{2gy}$. It remains to represent the curve, as in the previous problem, by a single function, $y=f(x)$. The total time of travel between $A$ and $B$ is then expressed with the help of~\eqref{ch00:lineelement} by the functional
\begin{equation}
T[f]=\int_0^d\D x\,\sqrt{\frac{1+[f'(x)]^2}{2gf(x)}}\;.
\end{equation}
Combined with the fixed boundary condition, $f(0)=0$ and $f(d)=h$, this matches our basic variational problem, the corresponding functional density being
\begin{equation}
\Ta(f,f',x)=\sqrt{\frac{1+(f')^2}{2gf}}\;.
\label{ch00:brachistochroneTa}
\end{equation}
Note that the physics (presence of gravity) only enters the functional $\Ta[f]$ through the overall factor $1/\sqrt{2g}$. The problem is therefore purely geometric, and we expect to find a unique solution that only depends on the geometric parameters $d$ and $h$.

Now observe that the functional density~\eqref{ch00:brachistochroneTa} does not depend explicitly on $x$. We can therefore use~\eqref{ch00:firstintenergy} to deduce a first integral of the EL equation,
\begin{equation}
I_x=\frac1{\sqrt{2gf}}\frac{1}{\sqrt{1+(f')^2}}\;.
\end{equation}
With the shorthand notation $c\equiv1/(2gI_x^2)$, we infer that $f[1+(f')^2]=c$ is an $x$-independent constant. This first-order ODE can in principle be solved directly. However, we can save computational effort by using a bit of physics insight. Imagine that instead of the function $y=f(x)$, we represent the trajectory of the point mass in the parametric form as $(x(t),y(t))$, where $t$ is the physical time. The relation $v=\sqrt{2gy}$ that we obtained from energy conservation can then be expressed as
\begin{equation}
\dot x^2+\dot y^2=2gy\;.
\label{ch00:brachistochronespeed}
\end{equation}
The previously derived condition $f[1+(f')^2]=c$ is likewise equivalent to $y(\dot x^2+\dot y^2)=c\dot x^2$. In combination with~\eqref{ch00:brachistochronespeed}, we thus get $2gy^2=c\dot x^2$. Upon inserting this back into~\eqref{ch00:brachistochronespeed} and taking another time derivative, we eventually arrive at a second-order ODE for $y(t)$ alone,
\begin{equation}
\ddot y+\frac{2g}cy=g\;.
\end{equation}

This is mathematically equivalent to the equation of motion of a mass suspended on a linear spring in the gravitational field. It describes harmonic motion with angular frequency $\o=\sqrt{2g/c}$. There is a unique solution that satisfies the appropriate initial condition $x(0)=y(0)=\dot x(0)=\dot y(0)$,
\begin{equation}
x(t)=\frac c2(\o t-\sin\o t)\;,\qquad
y(t)=\frac c2(1-\cos\o t)\;.
\label{ch00:cycloid}
\end{equation}
The price for the mathematical elegance is that our solution is only implicit. The total time $T$ of travel between $A$ and $B$ and the integration constant $c$ are determined by the boundary condition at $B$, namely $x(T)=d$ and $y(T)=h$. The geometric curve defined by~\eqref{ch00:cycloid} is called \emph{cycloid}.

%%%%%%%%%%%%%%%%%%%%%%%%%%%%%%%%%%%%%%%%%%%%%%%%%%%%%%%%%%%%

\section{Constrained Optimization}
\label{sec:constrainedoptimization}

There is a large class of optimization problems that involve a constraint on the domain in which a function or a functional is to be minimized or maximized. In ordinary calculus, such problems can be dealt with using the technique of \emph{Lagrange multipliers}. It is advisable that before proceeding, you review what you learned about Lagrange multipliers before; use your favorite textbook or even just the corresponding \href{https://en.wikipedia.org/wiki/Lagrange_multiplier}{Wikipedia page}. Luckily, the method extends straightforwardly to variational calculus. Instead of working out the general theory, we can therefore focus on a couple of illustrative examples. The one novel feature of constrained variational problems is that we have to distinguish two different ways how the constraint(s) can be implemented.

%%%%%%%%%%%%%%%%%%%%%%%%%%%%%%%%%%%%%%%%%%%%%%%%%%%%%%%%%%%%

\subsection{Global Constraints}

Suppose we want to find the minimum, maximum or stationary points of a functional $F$ on a space of functions $f_A$ constrained by the conditions $G_i[f]=0$, where $G_i$ are themselves some functionals. This problem is solved by modifying the functional~$F$~to
\begin{equation}
F_\l[f]\equiv F[f]-\sum_i\l_iG_i[f]\;,
\end{equation}
where $\l_i$ is a set of Lagrange multipliers, one for each constraint. One then searches for unconstrained stationary points of $F_\l$ by solving the EL equation $\udelta F_\l/\udelta f_A=0$.

\begin{figure}[t]
\sidecaption[t]
\includegraphics[width=2.9in]{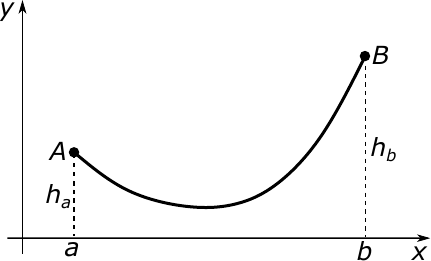}
\caption{Illustration of the catenary problem. A uniform chain (cable, rope) of length $\ell$ is suspended at points $A,B$. We are to find the shape the chain assumes under the influence of a uniform gravitational field oriented along the negative $y$-semiaxis.}
\label{fig00:catenary}
\end{figure}

For an illustration of how this works, let us look at the \emph{catenary problem}, which is to find the shape of a freely hanging chain suspended at its endpoints (see Fig.~\ref{fig00:catenary}). The main idea is that the shape we are looking for should minimize the potential energy of the chain in the external gravitational field. To formalize the problem, we will describe the vertical position of a point on the chain by a function of the horizontal coordinate, $y=f(x)$. The potential energy of an infinitesimal segment of the chain of length $\D s$ is then
\begin{equation}
\D E=\vr g f(x)\D s=\vr gf(x)\sqrt{1+[f'(x)]^2}\,\D x\;,
\end{equation}
where we used~\eqref{ch00:lineelement} and denoted by $\vr$ and $g$ the linear mass density of the chain and the gravitational acceleration. We would therefore like to minimize the functional
\begin{equation}
E[f]\equiv\vr g\int_a^b\D x\,f(x)\sqrt{1+[f'(x)]^2}\;.
\end{equation}
However, not every function $f(x)$ represents a possible configuration of the chain, since the total length $\ell$ of the chain is fixed. This leads to a single constraint functional
\begin{equation}
L[f]\equiv-\ell+\int_a^b\D x\,\sqrt{1+[f'(x)]^2}\;.
\end{equation}

The rest of the analysis proceeds as usual. We define the modified functional $E_\l[f]\equiv E[f]-\l L[f]$ using a single Lagrange multiplier $\l$. Since the corresponding functional density does not depend explicitly on $x$, we can make use of the first integral~\eqref{ch00:firstintenergy},
\begin{equation}
I_x=\frac{\vr gf-\l}{\sqrt{1+(f')^2}}+\l\ell\;.
\end{equation}
This first-order ODE is solved by
\begin{equation}
f(x)=\frac1{\vr g}\biggl[\l+\tilde I_x\cosh\frac{\vr g(x-x_0)}{\tilde I_x}\biggr]\;,
\end{equation}
where $\tilde I_x\equiv I_x-\l\ell$ and $x_0$ is an integration constant interpreted as the horizontal position of the lowest-lying point on the chain. The optimal shape of the chain is thus determined up to the values of the constants $I_x,\l,x_0$. The latter are fixed implicitly by the constraint $L[f]=0$ and the boundary conditions $f(a)=h_a$ and $f(b)=h_b$.

%%%%%%%%%%%%%%%%%%%%%%%%%%%%%%%%%%%%%%%%%%%%%%%%%%%%%%%%%%%%

\subsection{Local Constraints}

The constraints may also be imposed directly, pointwise, on the values of the functions $f_A$ and of their derivatives. Such constraints can be taken care of at the level of the functional density. Suppose the constraints can be expressed as a set of conditions
\begin{equation}
\Ga_i(f,\de f,\dotsc,x)=0\;.
\label{ch00:localconstraint}
\end{equation}
We then modify the original functional $F$ to
\begin{equation}
F_\l[f,\l]\equiv\int\D^nx\,\biggl[\Fa(f,\de f,\dotsc,x)-\sum_i\l_i(x)\Ga_i(f,\de f,\dotsc,x)\biggr]\;.
\end{equation}
Note that in this case, the Lagrange multipliers $\l_i(x)$ are functions of the coordinates rather than constants, and become genuine arguments of the modified functional. By construction, the EL equations for $\l_i$ take us back to the constraints~\eqref{ch00:localconstraint}.

As an example, consider the problem of finding the shortest path connecting two given points on the surface of a cylinder in $\R^3$. The cylinder is defined in Cartesian coordinates by $x^2+y^2=R^2$, $R$ being its radius. We know how to deal with this kind of problem from Sect.~\ref{subsec:shortestdistance}. We introduce an arbitrary parameterization $(x(t),y(t),z(t))$ of the path, only subject to the restriction that the fixed endpoints correspond to $t=0$ and $t=1$. Adding a Lagrange multiplier $\l(t)$ for the constraint $\Ga\equiv x^2+y^2-R^2$, we arrive at the modified functional
\begin{equation}
F_\l[x,y,z,\l]\equiv\int_0^1\D t\Bigl\{\sqrt{[\dot x(t)]^2+[\dot y(t)]^2+[\dot z(t)]^2}-\l(t)\bigl[x(t)^2+y(t)^2-R^2\bigr]\Bigr\}\;.
\end{equation}
We see that $z$ is a cyclic coordinate. The corresponding first integral~\eqref{ch00:firstintcyclic} reads
\begin{equation}
I_z=\frac{\dot z}{\sqrt{\dot x^2+\dot y^2+\dot z^2}}\;.
\label{ch00:cylinderfirstintegral}
\end{equation}
Our constraint $\Ga=0$ is, in fact, so simple that we can resolve it explicitly by introducing the polar angle $\vp$ through $(x,y)=(R\cos\vp,R\sin\vp)$. Combined with our first integral, this implies
\begin{equation}
\OD z\vp=\frac{I_zR}{\sqrt{1-I_z^2}}\;.
\end{equation}
We have managed to eliminate the arbitrary parameter $t$, and arrived at a purely geometric constraint on the path. The result, stating that $\Od z\vp$ must be constant, is easy to interpret. Upon ``unwrapping'' the cylinder into a plane, the solution corresponds to a straight path connecting the given points. This is hardly surprising.

How about the Lagrange multiplier then? We still have not used the EL equations for $x(t)$ and $y(t)$. These read
\begin{equation}
2\l x+\OD{}t\frac{\dot x}{\sqrt{\dot x^2+\dot y^2+\dot z^2}}=2\l y+\OD{}t\frac{\dot y}{\sqrt{\dot x^2+\dot y^2+\dot z^2}}=0\;.
\end{equation}
Upon eliminating the ugly square root in the denominator in favor of the first integral~\eqref{ch00:cylinderfirstintegral} and using again the parameterization of $(x,y)$ by the polar angle $\vp$, both of these EL equations boil down to the same condition,
\begin{equation}
\l(t)=\frac{\sqrt{1-I_z^2}}{2R}\dot\vp(t)\;.
\label{ch00:localLagrange}
\end{equation}
There is no unique solution for $\vp(t)$, or for $\l(t)$ to that matter. This reflects the freedom to choose a parameterization of the geometric path on the cylinder. All that we can say is that the functions $\vp(t)$ and $\l(t)$ are related by~\eqref{ch00:localLagrange}.

%%%%%%%%%%%%%%%%%%%%%%%%%%%%%%%%%%%%%%%%%%%%%%%%%%%%%%%%%%%%

\section*{\probsec}
\addcontentsline{toc}{section}{\probsec}

\begin{prob}
\label{pr00:Lagmech}
The trajectory $\vec r(t)$ of a particle in $n$-dimensional Euclidean space can be viewed as a map $\vec r:\R\to\R^n$. Suppose that we know the position of the particle at two instants of time, $t_1$ and $t_2$. Construct the functional
\begin{equation}
S_\mathrm{L}[\vec r]\equiv\int_{t_1}^{t_2}\D t\,\biggl\{\frac12m[\dot{\vec r}(t)]^2-V(\vec r(t))\biggr\}\;,
\end{equation}
where $m$ is the mass of the particle and $V(\vec r)$ its potential energy in an external conservative field. Find the EL equation, describing stationary states of the functional $S_\mathrm{L}$ on the space of trajectories with a fixed boundary condition. If your result looks familiar, you have just discovered the Lagrange formulation of mechanics!
\end{prob}

\begin{prob}
\label{pr00:Hammech}
Following up on~\refpr{pr00:Lagmech}, consider the functional
\begin{equation}
S_\mathrm{H}[\vec r,\vec p]\equiv\int_{t_1}^{t_2}\D t\,\biggl\{\vec p(t)\cdot\dot{\vec r}(t)-\frac{[\vec p(t)]^2}{2m}-V(\vec r(t))\biggr\}\;,
\end{equation}
which depends on two different functions of time, $\vec r(t)$ and $\vec p(t)$. What boundary condition at $t_1,t_2$ do you need to assume for $\vec p(t)$ so that you do not have to worry about boundary terms? Derive the EL equations for $\vec r(t)$ and $\vec p(t)$, and use them to find a physical interpretation for the latter. Now you have also discovered the Hamilton formulation of mechanics!
\end{prob}

\begin{prob}
\label{pr00:SGkink}
Find a function $\p(x)$ that minimizes the functional
\begin{equation}
F[\p]\equiv\int_{-\infty}^{+\infty}\D x\,\biggl\{\frac12[\p'(x)]^2+[1-\cos\p(x)]\biggr\}
\end{equation}
given the boundary condition $\p(-\infty)=0$ and $\p(+\infty)=2\pi$. How do you know that your solution is an actual minimum and not a mere stationary point of the functional? This may well be the first time you have met a soliton; this concrete example is known as the \emph{sine-Gordon kink}.
\end{prob}

\begin{prob}
\label{pr00:resistance}
The relation between the current density $\vec J$ and electric field $\vec E$ in a conducting medium is given by Ohm's law, $\vec J=\s\vec E$. In case the medium is inhomogeneous, the conductivity $\s$ may depend on position. Suppose that the current distribution in the medium is steady (time-independent). This can be achieved by connecting the medium to a source of constant current or voltage. The amount of electric energy dissipated in a unit volume of the medium per unit of time equals $\skal JE$. Minimize the total power dissipated in the medium under the constraint that the electric field is conservative everywhere. Interpret your result in terms of Kirchhoff's laws for electric circuits. Hint: there are (at least) two different routes you can take. On the one hand, you can implement the constraint $\rot\vec E=\vec0$ explicitly using a vector of Lagrange multipliers. On the other hand, you can resolve the constraint simply by trading $\vec E$ for its scalar potential.
\end{prob}

\begin{prob}
\label{pr00:isoperimetric}
You are given a smooth, simple (non-self-intersecting) closed curve $\Gamma$ in the Euclidean plane. Denote as $\O$ the domain bounded by the curve. Assuming that the length of $\Gamma$ is fixed, what shape should it take so that the area of $\O$ is maximum possible? If it helps, you may assume that the domain $\O$ is convex. (Can you guess why?) Hint: also here, there are multiple ways to the correct answer. You can place the origin somewhere inside $\O$ and use polar coordinates $r,\vp$, representing the curve $\Gamma$ by a function $r=f(\vp)$. You can also use Cartesian coordinates and encode the curve parameterically as $(x(t),y(t))$. If you are wondering how to express the area of $\O$ in such a parameterization, think of the line integral of the vector field $(-y,0)$ or of $(0,x)$, and recall the Stokes (Green's) theorem.
\end{prob}

\begin{prob}
\label{pr00:snell}
The optical properties of insulators are characterized by the refractive index~$n$, which determines the phase velocity of light propagation in terms of the speed of light $c$ in vacuum through $c/n$. Suppose that an optical medium is inhomogeneous so that the refraction index $n$ depends on position. Consider a path connecting two fixed points $A$ and $B$, and compute the time it would take a light beam to propagate from $A$ to $B$ along this path. According to \emph{Fermat's principle}, the actual trajectory along which light propagates is a stationary point of the time as a functional of the path. Use this to derive a differential equation that governs the geometric shape of the trajectory of light. You may want to express the equation in terms of the unit tangent vector to the trajectory. Show that in the special case that the refraction index $n$ depends only on a single Cartesian coordinate, your result reproduces the well-known Snell's law of refraction.
\end{prob}
\chapter{Lagrangian Mechanics}
\label{chap:Lagmechanics}

\keywords{Generalized coordinates and velocities, configuration space, Lagrangian, action, Hamilton's principle, Lagrange equations, constants (integrals) of motion, conservation laws.}

%%%%%%%%%%%%%%%%%%%%%%%%%%%%%%%%%%%%%%%%%%%%%%%%%%%%%%%%%%%%

\noindent Having covered the basics of variational calculus, we now have the tools we need to introduce a reformulation of Newtonian mechanics due to Lagrange. This is sometimes done by means of a mathematical derivation, utilizing the relation between force and work; see e.g.~Chap.~1 of~\cite{Goldstein2013a} for details. Here I will take a shortcut: I will simply postulate the basic principles of Lagrangian mechanics. Subsequently, we will verify that this reproduces the older laws of Newton. We will thus have more time left to practice the application of Lagrangian mechanics to concrete problems. However, the main reason for not serving you a derivation of Lagrangian mechanics from Newton's laws is that it would be conceptually misleading. Namely, Lagrangian mechanics has a much broader scope. As we will see later, it can accommodate equally easily relativistic mechanics but also nonmechanical systems such as electromagnetism. Indeed, all currently known fundamental laws of nature have a Lagrangian formulation. Let us therefore get started.

%%%%%%%%%%%%%%%%%%%%%%%%%%%%%%%%%%%%%%%%%%%%%%%%%%%%%%%%%%%%

\section{Basic Setup}

The Lagrangian formulation of mechanics is based on a variational principle for certain functional that depends on the trajectory of the system as a variable. Solving the corresponding variational problem gives the actual, physical trajectory. This is a completely different approach than Newton's! Newtonian mechanics is ``local'' in that the instantaneous acceleration of an object is determined by forces acting on it at the given point in space and time. On the other hand, the Lagrangian approach is ``nonlocal'' and requires guessing the entire trajectory of the system. Remarkably, this eventually leads to the same differential \emph{equation of motion} (EoM).

%%%%%%%%%%%%%%%%%%%%%%%%%%%%%%%%%%%%%%%%%%%%%%%%%%%%%%%%%%%%

\subsection{Hamilton's Variational Principle}

In the first step towards the Lagrange formulation of mechanics, we have to deal with kinematics. This refers generally to the description of a state of a system in motion in terms of certain set of independent variables. (If this sounds too abstract, think of the position and velocity of a particle.) The kinematical data consist of two parts. Suppose first that we take a snapshot of the system; this captures its position but misses all information about its motion. The snapshot gives us the instantaneous \emph{configuration} of the system. The set of all possible configurations forms the \emph{configuration space}. We assume that the configuration space can be parameterized uniquely by a set of independent variables $q_i$, called \emph{generalized coordinates}.

\begin{illustration}%
\label{ex01:kinematics}%
Take a point particle in $\R^3$. The Cartesian coordinates $x,y,z$ defining its position form a suitable set of generalized coordinates. You can of course use any other,~curvilinear coordinates you wish, as long as they are well-defined. Thus, for instance, we can replace Cartesian coordinates with the standard spherical coordinates $r,\t,\vp$, provided the particle cannot reach the $z$-axis where the azimuthal angle $\vp$ is ill-defined. In any case, the configuration space of a particle in $\R^3$ is three-dimensional, and can be identified with $\R^3$ itself.

Let us now take the level up a notch and consider two particles in $\R^3$. To describe their configuration uniquely, we need two sets of three coordinates each. These can be collected in the position vectors $\vec r_1,\vec r_2$ of the two particles. Thus, the two-particle system will evolve in a six-dimensional configuration space. You probably see the pattern now. For a system of $N$ particles in an $n$-dimensional Euclidean space, we need altogether $N\times n$ generalized coordinates. The configuration space of the $N$-particle system can therefore be identified with $\R^{N\times n}$. Remember that the choice of generalized coordinates is arbitrary as long as they are well-defined and mutually independent. It is not necessary that they can be meaningfully split into $N$ sets of $n$ variables, one for each of the particles. We could for instance use the hyperspherical coordinates in $\R^{N\times n}$. The ability to use any generalized coordinates in the configuration space is one of the strengths of the Lagrangian formalism.
\end{illustration}

The second part of kinematical data is needed to describe the change of configuration in time, i.e.~motion. Any choice of functions $q_i(t)$ of time defines a possible \emph{trajectory} of the system. Mathematically, the trajectory is thus a curve in the configuration space, parameterized by time. The description of the instantaneous kinematical state of the system is completed by adding to $q_i$ the \emph{generalized velocities} $\dot q_i$.

Next, we would like to know which of the possible trajectories $q_i(t)$ corresponds to the actual motion of the system. To that end, we need the \emph{Lagrange function} (or simply \emph{Lagrangian}) $L(q,\dot q,t)$. This is a function of the generalized coordinates and velocities, treated as independent variables, and may also depend explicitly on time. From the Lagrangian, we construct the \emph{action}
\begin{equation}
\boxed{S[q]\equiv\int_{t_1}^{t_2}\D t\,L(q(t),\dot q(t),t)\;.}
\label{ch01:action}
\end{equation}
The action is a functional of the trajectory $q_i(t)$ over the time interval $t_1\leq t\leq t_2$. In terminology of~\chaptername~\ref{chap:mathintro}, $S[q]$ is a local functional of a set of functions $q_i$, each of which depends on a single variable $t$ (time). The Lagrangian $L$ is the corresponding functional density. I tacitly assumed that the Lagrangian does not depend on higher than first time derivatives of $q_i(t)$. The ensuing \emph{Euler--Lagrange} (EL) \emph{equation} will therefore be of at most second order. Hopefully you are now starting to see how variational calculus is ideally suited for application to mechanics.

It remains to postulate the fundamental dynamical principle of Lagrangian mechanics. Somewhat confusingly, this variational principle is not named after Lagrange, but after Hamilton. Here it is. The actual, physical trajectory of the system that passes through the point $q_{i1}$ at time $t_1$ and through $q_{i2}$ at time $t_2$ is a stationary point of the action $S[q]$ in the space of trajectories with a fixed boundary condition,
\begin{equation}
q_i(t_1)=q_{i1}\;,\qquad
q_i(t_2)=q_{i2}\;.
\end{equation}
It follows that the dynamics of the system is governed by the EL equation
\begin{equation}
\boxed{\frac{\udelta S}{\udelta q_i(t)}=0\;.}
\label{ch01:ELeqgeneral}
\end{equation}

\begin{watchout}%
This is all very elegant, but why does it actually work? The deeper reason why the laws of classical mechanics can be cast in the extremely compact form~\eqref{ch01:ELeqgeneral} is only revealed by quantum mechanics. The latter has a beautiful formulation wherein the probability amplitude for a system to transition from configuration $q_{i1}$ at time $t_1$ to configuration $q_{i2}$ at time $t_2$ is given by a \emph{functional integral},
\begin{equation}
\int\text{``$\D q$''}\,\exp\biggl(\frac\I\hbar S[q]\biggr)\;.
\end{equation}
The integration is done over the infinite-dimensional space of trajectories~with a fixed boundary condition. Importantly, every possible classical trajectory contributes a complex phase to the probability amplitude. For macroscopic systems, the action $S[q]$ is larger than the Planck constant $\hbar$ by many orders of magnitude. Hence, even a slight variation in the trajectory leads to huge oscillations of the complex phase, and the contributions of most trajectories cancel each other. The exception to the rule is the trajectory for which $S[q]$ does not change appreciably as the trajectory is slightly varied. But this is exactly the definition of a stationary point of $S[q]$! We conclude that in the classical limit where $\hbar\ll S[q]$, the quantum-mechanical probability amplitude is dominated by the stationary point(s) of the action.
\end{watchout}

In order to have a complete, self-contained variational framework for mechanics, we still need to specify the Lagrange function. This is the tricky bit: there is no general algorithm for finding the Lagrangian. However, things are not that bad. The Lagrangian is known, among others, for all mechanical systems consisting of point particles interacting via conservative forces, possibly moving in a background conservative field. This will take us pretty far. For such systems, the rule of thumb is that the Lagrangian equals the kinetic energy of the system minus its potential energy, where both the interactions between the particles and their interaction with the external field are taken into account. In the special case of a single particle in an external field, you already verified that this leads to the correct EoM in~\refpr{pr00:Lagmech}.

In less fortunate cases where the Lagrangian is not known, one has to resort to various forms of guesswork. Alternatively, it is possible to simply construct the most general Lagrangian that is compatible with all the symmetries of the system. This more advanced approach, which we will not follow here, lies at the heart of the modern \emph{effective field theory} program.

%%%%%%%%%%%%%%%%%%%%%%%%%%%%%%%%%%%%%%%%%%%%%%%%%%%%%%%%%%%%

\subsection{Lagrange Equations of Motion}
\label{subsec:LagrangeEoM}

It is easy to turn the above basic dynamical principle into an algorithmic workflow that will allow us to deal, at least in principle, with nearly any mechanical problem. For your convenience, I will spell it out step by step, and then append some comments.
\begin{enumerate}
\item[(1)] Identify the configuration space and choose suitable generalized coordinates $q_i$. This is a kind of art in its own right. Any choice of coordinates is in principle good, but a judicious choice can save you a lot of effort in the subsequent steps.
\item[(2)] Construct the Lagrangian $L(q,\dot q,t)$. Remember that for (nonrelativistic) mechanical systems involving only conservative forces, the Lagrangian is given by the difference of kinetic and potential energy. Moreover, the kinetic energy is typically a bilinear function of the generalized velocities. The entire Lagrangian then acquires the simple form
\begin{equation}
L(q,\dot q)=\frac12m_{ij}(q)\dot q_i\dot q_j-V(q)\;,
\end{equation}
where $m_{ij}(q)$ is a positive-definite quadratic form, depending on the choice of coordinates, and $V(q)$ is the potential energy.
\item[(3)] Construct the set of equations of motion, one for each $q_i$, called \emph{Lagrange equations}. These are a special case of~\eqref{ch00:EulerLagrange}, usually written as
\begin{equation}
\boxed{\OD{}t\PD L{\dot q_i}=\PD L{q_i}\;.}
\label{ch01:Lagrangeeq}
\end{equation}
\item[(4)] Solve the Lagrange equations.
\end{enumerate}

It is important to keep in mind that despite the impression, the Lagrangian for a given system is not unique. Suppose that we trade $L(q,\dot q,t)$ for another Lagrangian $L'(q,\dot q,t)$ that differs from $L$ only by a total time derivative of some function $f(q,t)$ of the generalized coordinates and time. What this means precisely is that once $q_i(t)$ are treated as functions of time and $\dot q_i(t)$ as their time derivatives, the two Lagrangians are related by
\begin{equation}
L'(q(t),\dot q(t), t)\ifeq L(q(t),\dot q(t),t)+\OD{f(q(t),t)}{t}\;.
\end{equation}
The claim now is that the two Lagrangians have identical Lagrange equations. This is easy to understand from the relation between the corresponding actions,
\begin{equation}
S'[q]=S[q]+f(q_2,t_2)-f(q_1,t_1)\;.
\end{equation}
With fixed boundary conditions, the two actions differ by a mere constant and thus have the same stationary points. This simple feature is surprisingly useful.

Finally, let us formalize the observation that any choice of generalized coordinates is equally good. Suppose that we decide to trade the generalized coordinates $q_i$ for another set of coordinates $\tilde q_i$, defined implicitly by
\begin{equation}
q_i=q_i(\tilde q,t)\;.
\label{ch01:pointtransfo}
\end{equation}
Such a change of variables is called \emph{point transformation} and as indicated, it may be time-dependent. Using the chain rule, we find that the corresponding generalized velocities are related by
\begin{equation}
\dot q_i=\PD{q_i}t+\PD{q_i}{\tilde q_j}\dot{\tilde q}_j\;.
\label{ch01:pointtransfovel}
\end{equation}
Simply inserting everything into the Lagrangian $L_q(q,\dot q,t)$, we get a new Lagrangian in terms of $\tilde q_i$, $\dot{\tilde q}_i$ and $t$,
\begin{equation}
L_{\tilde q}(\tilde q,\dot{\tilde q},t)\equiv L_q(q(\tilde q,t),\Pd{q(\tilde q,t)}t+\dot{\tilde q}_j\Pd{q(\tilde q,t)}{\tilde q_j},t)\;.
\end{equation}
By construction, the Lagrangians $L_q$ and $L_{\tilde q}$ are equal to each other when evaluated on trajectories $q_i(t)$ and $\tilde q_i(t)$ connected by~\eqref{ch01:pointtransfo}. Hence also the corresponding actions $S_q[q]$ and $S_{\tilde q}[\tilde q]$ are equal. Furthermore, the form of the transformation~\eqref{ch01:pointtransfo} guarantees that a fixed boundary condition for $q_i(t_{1,2})$ translates into a fixed boundary condition for $\tilde q_i(t_{1,2})$. As a consequence, the stationary points of the actions $S_q$ and $S_{\tilde q}$ are in a one-to-one correspondence and are also related by the point transformation~\eqref{ch01:pointtransfo}. In more human terms, if you change variables at the level of the Lagrangian, the Hamilton variational principle will automatically give you the correct EoM in the new generalized coordinates.

%%%%%%%%%%%%%%%%%%%%%%%%%%%%%%%%%%%%%%%%%%%%%%%%%%%%%%%%%%%%

\section{Examples}
\label{sec:Lagexamples}

So far, I have intentionally omitted detailed examples in order to keep the exposition of Lagrangian mechanics compact. If you look back, you will hopefully agree that the basic theory is fairly simple. In order to help you master it actively, I will now work out several examples on which I will illustrate various mentioned features such as the construction of the Lagrangian in given coordinates, and the freedom to choose and change generalized coordinates. It may be a useful exercise to identify explicitly the steps (1)--(4) of our basic workflow in each of the examples.

%%%%%%%%%%%%%%%%%%%%%%%%%%%%%%%%%%%%%%%%%%%%%%%%%%%%%%%%%%%%

\subsection{Particle in a Conservative Field}
\label{subsec:singleparticle}

Let us start slowly. Consider a single point particle of mass $m$ moving along a straight line under the effect of an external force. The notion of a conservative force is trivial in one dimension, the only real assumption being that the force is time-independent. The configuration space is $\R$ and we thus need a single generalized coordinate. An obvious choice is the Cartesian coordinate, $x$. Denoting the potential energy of the particle in the external force field as $V(x)$, the Lagrangian thus reads
\begin{equation}
L(x,\dot x)=\frac12m\dot x^2-V(x)\;.
\label{ch01:Lparticle1d}
\end{equation}
The corresponding Lagrange equation~\eqref{ch01:Lagrangeeq} is $m\ddot x=-V'(x)$, where the prime indicates a derivative with respect to $x$. Recognizing the right-hand side as the force acting on the particle, we have recovered the second law of Newton.

So far so good, let us move on to the next level. Suppose the particle lives in an $n$-dimensional Euclidean space. The configuration space is thus $\R^n$. We can now use the Cartesian coordinates $x_1,\dotsc,x_n$ and collect them in the position vector $\vec r$. The Lagrangian~\eqref{ch01:Lparticle1d} is obviously generalized to
\begin{equation}
L(\vec r,\dot{\vec r})=\frac12m\dot{\vec r}^2-V(\vec r)\;,
\label{ch01:Lparticlend}
\end{equation}
where $V(\vec r)$ is the potential energy of the particle in the external conservative field. The corresponding EoM is $m\ddot{\vec r}=-\grad V(\vec r)$, which again recovers the prediction of Newton's second law.

In order not to get too attached to Cartesian coordinates, let us try something else. For some systems, especially those with forces depending on distance but not on direction, spherical coordinates are more natural. To keep things simple, let us restrict ourselves to two dimensions. We trade the Cartesian coordinates $\vec r=(x,y)$ in $\R^2$ for the polar coordinates $r,\vp$ by the point transformation $x=r\cos\vp$, $y=r\sin\vp$. Using~\eqref{ch01:pointtransfovel}, we find that the generalized velocities are to be recalculated as
\begin{equation}
\dot x=\dot r\cos\vp-r\dot\vp\sin\vp\;,\qquad
\dot y=\dot r\sin\vp+r\dot\vp\cos\vp\;.
\end{equation}
Inserting this into the Lagrangian~\eqref{ch01:Lparticlend}, we get
\begin{equation}
L(r,\vp,\dot r,\dot\vp)=\frac12m(\dot r^2+r^2\dot\vp^2)-V(r,\vp)\;,
\label{ch01:Lparticlepolar}
\end{equation}
where $V(r,\vp)$ is the same potential energy, expressed in terms of the polar coordinates. It is instructive to write down explicitly the corresponding Lagrange equations,
\begin{equation}
m(\ddot r-r\dot\vp^2)=-\PD Vr\;,\qquad
\OD{}t(mr^2\dot\vp)=-\PD V\vp\;.
\label{ch01:LparticlepolarEoM}
\end{equation}
Should the external force only depend on distance but not on direction, the potential energy must be independent of $\vp$. The second EoM then takes the form of a conservation law $\Od{(mr^2\dot\vp)}t=0$. The corresponding first integral, $mr^2\dot\vp$, is the angular momentum with respect to the origin. We will get back to this example in~\chaptername~\ref{chap:centralfields} where we will analyze motion in central fields in great detail.

\begin{watchout}%
While reproducing successfully the predictions of Newton's second law, we have completely ignored the important fact that the latter is only supposed to hold in inertial reference frames. What is going on here? Is the Lagrangian formalism also valid only in inertial frames? Definitely not! How do we distinguish between inertial and noninertial frames then? To answer this question, recall that a transformation from one reference frame to another amounts to a specific change of coordinates. In the Lagrangian formalism, this can be reproduced by a time-dependent point transformation of the type~\eqref{ch01:pointtransfo}. What we can do is thus identify~\eqref{ch01:Lparticlend} with a Lagrangian in some inertial reference frame, and then change frame by a point transformation at the level~of~the~Lagrangian.
\end{watchout}

For a simple illustration, let us stay in two dimensions and consider a reference frame that rotates around the origin in an arbitrary manner. We keep the polar coordinates $r,\vp$ in a fixed inertial reference frame, and denote the corresponding coordinates in the rotating frame as $R,\Phi$. Since rotations preserve distance, we have obviously $r=R$. The polar angles are related by
\begin{equation}
\vp(t)=\Phi(t)+\a(t)\;,
\end{equation}
where the angle $\a(t)$ defines the orientation of the rotating frame with respect to the inertial frame as a function of time. The Lagrangian in the rotating frame is obtained from~\eqref{ch01:Lparticlepolar} by a simple substitution,
\begin{equation}
L(R,\Phi,\dot R,\dot\Phi)=\frac12m[\dot R^2+R^2(\dot\Phi+\dot\a)^2]-V(R,\Phi+\a)\;.
\end{equation}
Accordingly, the Lagrange equations~\eqref{ch01:LparticlepolarEoM} become
\begin{equation}
m\ddot R-mR(\dot\Phi+\dot\a)^2=-\PD VR\;,\qquad
\OD{}t[mR^2(\dot\Phi+\dot\a)]=-\PD V\Phi\;.
\label{ch01:LparticlepolarEoM2}
\end{equation}
By comparing~\eqref{ch01:LparticlepolarEoM2} with~\eqref{ch01:LparticlepolarEoM}, we find three new types of terms in the Lagrange equations, corresponding to the three standard inertial forces in a rotating reference frame. The EoM for $R$ contains a new term $-2mR\dot\Phi\dot\a$, arising from the \emph{Coriolis force}, and another term, $-mR\dot\a^2$, representing the \emph{centrifugal force}. In case the rotation of the noninertial frame is nonuniform, there is a new term in the EoM for $\Phi$, namely $mR^2\ddot\a$. This arises from the so-called \emph{Euler force}. Finally, the EoM for $\Phi$ contains another new term, $2mR\dot R\dot\a$, which also arises from the Coriolis force. Unlike the centrifugal force that is always radial and Euler force that is always azimuthal, the Coriolis force generally has both a radial and an azimuthal~component.

It is of course also possible to derive all the three types of inertial forces in a rotating reference fame directly by a transformation of the acceleration vector. This approach is however more elaborate than the one outlined above. In case of interest, you will find more details including a thorough discussion of the physics in rotating reference frames in Chap.~10 of~\cite{Morin2008}.

%%%%%%%%%%%%%%%%%%%%%%%%%%%%%%%%%%%%%%%%%%%%%%%%%%%%%%%%%%%%

\subsection{Two-Body Problem}
\label{subsec:2body}

\begin{figure}[t]
\sidecaption[t]
\includegraphics[width=2.0in]{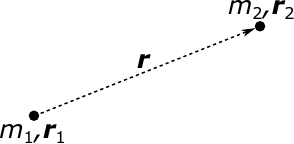}% This figure does not include so many details to warrant a full width of 2.9 in!
\caption{Illustration of variables used to describe the configuration of a system of two point particles of masses $m_1,m_2$. The relative position $\vec r$ is defined by $\vec r\equiv\vec r_2-\vec r_1$.}
\label{fig01:twobody}
\end{figure}

Let us now move on to systems consisting of two point particles. As follows from \refex{ex01:kinematics}, the configuration space of a system of two point particles in $\R^n$ is $\R^{2n}$. The simplest choice of generalized coordinates amounts to two sets of Cartesian coordinates, collected in the position vectors $\vec r_1,\vec r_2$. See Fig.~\ref{fig01:twobody} for a visualization of the configuration variables.

Suppose the particles interact by a pair force that only depends on the relative position, $\vec r\equiv\vec r_2-\vec r_1$, of the particles. Assuming that the force is conservative, the Lagrangian of the system becomes
\begin{equation}
L(\vec r_1,\vec r_2,\dot{\vec r}_1,\dot{\vec r}_2)=\frac12m_1\dot{\vec r}_1^2+\frac12m_2\dot{\vec r}_2^2-V(\vec r_2-\vec r_1)\;,
\label{ch01:Lag2body}
\end{equation}
where $m_1,m_2$ are the masses of the particles. The associated Lagrange equations are
\begin{equation}
m_1\ddot{\vec r}_1=\grad V(\vec r_2-\vec r_1)\;,\qquad
m_2\ddot{\vec r}_2=-\grad V(\vec r_2-\vec r_1)\;,
\label{ch01:EoM2body}
\end{equation}
where the gradient of the potential energy is taken with respect to its sole variable $\vec r_2-\vec r_1$. The forces acting on the two particles are opposite to each other: we have recovered Newton's third law. In hindsight, this is a consequence of the assumption that the interaction between the particles only depends on the their instantaneous relative positions.

The Lagrange equations~\eqref{ch01:EoM2body} form a set of $2n$ coupled second-order \emph{ordinary differential equations} (ODEs). This looks hopeless even in two dimensions. Fortunately, the equations can always be split into two independent sets, each of which is much easier to solve. To that end, we use the freedom to choose generalized coordinates at will. Instead of $\vec r_1,\vec r_2$, it turns out to be more convenient to use the relative position $\vec r$ and the position of the \emph{center of mass} (CoM) $\vec R$, defined jointly~by
\begin{equation}
\vec R\equiv\frac{m_1\vec r_1+m_2\vec r_2}{m_1+m_2}\;,\qquad
\vec r\equiv\vec r_2-\vec r_1\;.
\label{ch01:2bodyrelativecoords}
\end{equation}
It easy to check that in these variables, the Lagrangian~\eqref{ch01:Lag2body} takes the form
\begin{equation}
\boxed{L(\vec R,\vec r,\dot{\vec R},\dot{\vec r})=\underbrace{\frac12M\dot{\vec R}^2}_{\text{CoM motion}}+\underbrace{\frac12\m\dot{\vec r}^2-V(\vec r)}_{\text{relative motion}}\;,}
\label{ch01:2bodyLseparated}
\end{equation}
with mass parameters
\begin{equation}
M\equiv m_1+m_2\;,\qquad
\m\equiv\frac{m_1m_2}{m_1+m_2}
\end{equation}
giving the total mass of the system and the so-called \emph{reduced mass}, $\m$. As indicated by the braces, the Lagrangian has separated into two independent parts, one that only depends on the CoM position and another that only depends on the relative position. Accordingly, we end up with two separate equations of motion. The EoM for $\vec R$ is trivial and expresses the conservation of total momentum, $M\dot{\vec R}$. The EoM for $\vec r$ corresponds to an effective motion of a single particle of mass $\m$ with potential energy $V(\vec r)$. This separation of CoM and relative motion in the two-particle system plays a key role in the analysis of planetary motion that we will develop in~\chaptername~\ref{chap:centralfields}.

Before moving on to another example, I will briefly sketch what one can learn about the dynamics of a system of $N$ point particles with arbitrarily high $N$. Incidentally, the CoM and relative coordinates have an elegant generalization called \emph{Jacobi coordinates}; see the relevant \href{https://en.wikipedia.org/wiki/Jacobi_coordinates}{Wikipedia page} for more information. However, I will follow a simpler route that, while less powerful, will still allow us to gain some insight in the physics. We initially use the position vector $\vec r_i$ as the set of generalized coordinates for the $i$-th of the $N$ particles. Suppose that each pair of particles $i,j$ interacts via a force giving rise to a potential energy that only depends on the relative position, $V_{ij}(\vec r_i-\vec r_j)$. In addition, we allow for an external field such as gravity that acts on each of the particles separately. Denote the potential energy of the $i$-th particle in the external field as $V_i(\vec r_i)$. The total Lagrangian then collects all the contributions from kinetic and potential energy,
\begin{equation}
L=\sum_{i=1}^N\frac12m_i\dot{\vec r}_i^2-\sum_{i=1}^NV_i(\vec r_i)-\frac12\sum_{\substack{i,j=1\\ i\neq j}}^NV_{ij}(\vec r_i-\vec r_j)\;.
\label{ch01:NparticleLag}
\end{equation}
The notation is such that $V_{ij}(\vec r_i-\vec r_j)=V_{ji}(\vec r_j-\vec r_i)$, and the factor of $1/2$ in the last term compensates for the double-counting of particle pairs in the sum over $i,j$. The Lagrange EoM for the $i$-th particle reads
\begin{equation}
m_i\ddot{\vec r}_i=-\grad V_i(\vec r_i)-\sum_{\substack{j=1\\ j\neq i}}^N\grad V_{ij}(\vec r_i-\vec r_j)\;,
\label{ch01:NparticleEoM}
\end{equation}
where the gradient of the potentials in the last term is taken with respect to the argument $\vec r_i-\vec r_j$.

The CoM of the $N$-particle system can be defined analogously to the two-particle problem as $\vec R\equiv(m_1\vec r_1+\dotsb+ m_N\vec r_N)/M$, where $M\equiv m_1+\dotsb+m_N$ is the total mass. By adding up the individual Lagrange equations for $\vec r_i$, we find that the contributions of the pair interactions cancel each other; make sure you understand this! What is left is equivalent to
\begin{equation}
M\ddot{\vec R}=-\sum_{i=1}^N\grad V_i(\vec r_i)\;.
\label{ch01:NparticleP}
\end{equation}
This shows that the motion of the CoM does not depend on the pair interactions between the particles. This is a generalization of the law of momentum conservation to systems subject to external forces.

Another important dynamical quantity is the angular momentum. For a single particle in $\R^3$, this is defined by $\vec r_i\times\vec p_i=m_i\vec r_i\times\dot{\vec r}_i$ where $\vec p_i\equiv m_i\dot{\vec r}_i$ is the momentum. To find the total angular momentum, it is convenient to introduce position variables $\vec r_i'$ relative to the CoM, $\vec r_i'\equiv\vec r_i-\vec R$. By the definition of CoM, these satisfy the constraint $m_1\vec r_1'+\dotsb+m_N\vec r_N'=\vec0$. A simple manipulation then shows that the total angular momentum of the system can be cast as
\begin{equation}
\vec J\equiv\sum_{i=1}^Nm_i\vec r_i\times\dot{\vec r}_i=M\vec R\times\dot{\vec R}+\sum_{i=1}^Nm_i\vec r_i'\times\dot{\vec r}_i'\;.
\label{ch01:NparticleJ}
\end{equation}
The angular momentum splits into two parts. The first part only depends on the motion of the CoM, and is usually called \emph{orbital}. On the other hand, the second part of $\vec J$, often called \emph{spin}, only depends on the motion of the particles with respect to the CoM. Without going into details that would lead us astray, let me add that the total angular momentum is conserved in the absence of external forces, provided the pair interactions only depend on distance and not on direction.

\begin{illustration}%
Understanding the dynamics of systems of an arbitrary number of particles is very important for statistical physics. We know that ordinary matter consists of microscopic constituents: atoms and molecules. What makes us so sure that ignoring the microscopic structure, involving potentially very large molecular forces, does not affect the validity of Newton's laws at the macroscopic level? Regarding translational motion, this is easy to understand. According to~\eqref{ch01:NparticleP}, internal forces do not affect overall momentum balance of a chunk of matter. Rotational motion is more tricky due to the presence of the spin term in~\eqref{ch01:NparticleJ}. This is usually negligible compared to the orbital angular momentum, but there are notable exceptions. Take a piece of ferromagnet and expose it to an external magnetic field. The field forces individual spins inside the material to align with each other, leading to a change of internal angular momentum that may be macroscopically measurable. Since the total angular momentum should be conserved, this will in turn affect the macroscopic kinematics of the ferromagnet. In short, you can make a ferromagnet macroscopically rotate by magnetizing it. This is called \href{https://en.wikipedia.org/wiki/Einstein%E2%80%93de_Haas_effect}{Einstein--de Haas effect}.
\end{illustration}

%%%%%%%%%%%%%%%%%%%%%%%%%%%%%%%%%%%%%%%%%%%%%%%%%%%%%%%%%%%%

\subsection{Block Sliding on an Inclined Plane}

\begin{figure}[t]
\sidecaption[t]
\includegraphics[width=2.9in]{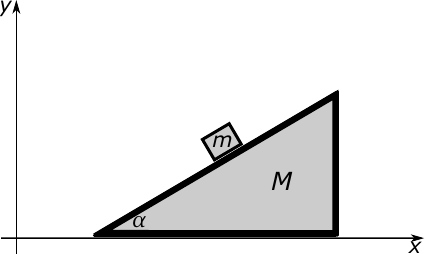}
\caption{Illustration of the setup where a block of mass $m$ slides without friction down an inclined plane of mass $M$, which itself moves horizontally without friction.}
\label{fig01:slidingblock}
\end{figure}

In all the previous examples, there was always a natural initial choice of generalized coordinates, indicating the position of one or multiple particles. It is therefore good to show at least one example where the existence of such a natural choice of coordinates may not be obvious. We will analyze the motion of the system shown in Fig.~\ref{fig01:slidingblock}, where a rectangular block of mass $m$ slides without friction along an inclined plane of total mass $M$. The plane itself is free to move along the horizontal direction. There are no external forces except of gravity, characterized by the constant acceleration $g$ in the vertical downwards direction.

In this case, it is at first not even obvious how many independent generalized coordinates there are. Let us therefore be careful. Suppose the position vector of an arbitrary but fixed point on the block that is in contact with the plane is $\vec r_m\equiv(x_m,y_m)$. Also, let us denote as $\vec r_M\equiv(x_M,y_M)$ the position vector of the lower left corner of the plane. The four variables $x_m,y_m,x_M,y_M$ are certainly sufficient to determine the configuration of the system uniquely, but they are not all independent. For one thing, the plane only moves in the horizontal direction so that $y_M$ remains constant at all times. It is therefore not a generalized coordinate of the system at all. We can (and will) even set it to zero by a suitable choice of origin of the Cartesian coordinate system as indicated in Fig.~\ref{fig01:slidingblock}. Moreover, the remaining variables $x_m,y_m,x_M$ cannot be independent because the block must always be touching the plane. This is ensured by requiring that $(y_m-y_M)/(x_m-x_M)=\tan\a$, where $\a$ is the angle of inclination of the plane. This leaves us with two independent generalized coordinates, for which there is no unique natural choice. For the sake of illustration, I will solve the problem following two different routes.

As our first choice, let us treat the coordinates $x_m,y_m$ of the block as the two independent variables. Setting $y_M=0$, it follows that $x_M=x_m-y_m\cot\a$. The Lagrangian is as usual given by the difference of kinetic and potential energy, and it takes little effort to express it in terms of our chosen generalized coordinates,
\begin{equation}
\begin{split}
L&=\frac12M\dot{\vec r}_M^2+\frac12m\dot{\vec r}_m^2-Mgy_M-mgy_m\\
&=\frac12M(\dot x_m-\dot y_m\cot\a)^2+\frac12m(\dot x_m^2+\dot y_m^2)-mgy_m\\
&=\frac12(M+m)\dot x_m^2+\frac12(M\cot^2\a+m)\dot y_m^2-M\dot x_m\dot y_m\cot\a-mgy_m\;.
\label{ch01:slidingblockL1}
\end{split}
\end{equation}
The EoM for $x_m$ does not involve the gravitational field. It allows us to eliminate one of the accelerations $\ddot x_m,\ddot y_m$ in terms of the other, for instance $\ddot y_m=(1+m/M)\ddot x_m\tan\a$. The EoM for $y_m$ reads
\begin{equation}
(M\cot^2\a+m)\ddot y_m-M\ddot x_m\cot\a=-mg\;.
\end{equation}
Upon inserting our solution for $\ddot y_m$ in terms of $\ddot x_m$, we get a single linear algebraic equation for $\ddot x_m$. You should be able to massage this until you get the final result,
\begin{equation}
\ddot x_m=-\frac{Mg\sin\a\cos\a}{M+m\sin^2\a}\;,\qquad
\ddot y_m=-\frac{(M+m)g\sin^2\a}{M+m\sin^2\a}\;.
\label{ch01:slidingblockres1}
\end{equation}
Not surprisingly, we find motion with constant acceleration.

\begin{watchout}%
Whenever the solution of a problem gives you a formula with multiple parameters, it is your sacred duty to make sure that you understand what the result says for special choices of the parameters that simplify the problem. This is a very useful strategy to check the consistency of your result. For our sliding block problem, here are some special cases that you might want to think about:
\begin{itemize}
\item The limit $\a\to0$.
\item The limit $\a\to\pi/2$.
\item The limit $m/M\to0$.
\item The limit $m/M\to\infty$.
\end{itemize}
\end{watchout}

We now solve the same problem with a different choice of generalized coordinates. It might seem natural to take as one of the variables the distance $s$ of the block from the tip of the inclined plane. The other variable can be chosen for instance as the horizontal position $x_M$ of the plane. In terms of $s,x_M$, the position vectors of the plane and the block are (setting again $y_M=0$)
\begin{equation}
\vec r_M=(x_M,0)\;,\qquad
\vec r_m=(x_M+s\cos\a,s\sin\a)\;.
\end{equation}
Expressing the Lagrangian in terms of $s,x_M$ gives
\begin{equation}
L=\frac12(M+m)\dot x_M^2+\frac12m\dot s^2+m\dot s\dot x_M\cos\a-mgs\sin\a\;.
\label{ch01:slidingblockL2}
\end{equation}
This looks simpler than~\eqref{ch01:slidingblockL1}, and the solution of the corresponding Lagrange equations now indeed requires fewer manipulations than before. The EoM for $x_M$ can be used to eliminate $\ddot x_M$ in terms of $\ddot s$, giving $\ddot x_M=-[m/(M+m)]\ddot s\cos\a$. The EoM for $s$ is then easy to solve, the final result being
\begin{equation}
\ddot s=-\frac{(M+m)g\sin\a}{M+m\sin^2\a}\;,\qquad
\ddot x_M=\frac{mg\sin\a\cos\a}{M+m\sin^2\a}\;.
\end{equation}
I recommend you as an exercise to check that this is equivalent to~\eqref{ch01:slidingblockres1}. Also, go back to the above-listed special limits of the parameters of the problem and check that the result agrees with your expectations.

%%%%%%%%%%%%%%%%%%%%%%%%%%%%%%%%%%%%%%%%%%%%%%%%%%%%%%%%%%%%

\subsection{Simple Pendulum}
\label{subsec:pendulum}

\begin{figure}[t]
\sidecaption[t]
\includegraphics[width=2.0in]{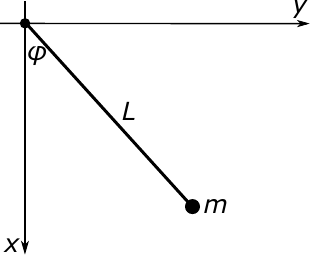}% This figure does not include so many details to warrant a full width of 2.9 in!
\caption{Geometry of a simple pendulum of mass $m$ and length $L$, oscillating in a vertical plane. The angle $\vp$ measures the deviation of the pendulum from its lowest-energy, vertical position.}
\label{fig01:pendulum}
\end{figure}

Another example that we have much to learn from is a simple physical system that you can easily build at home: a pendulum. A point mass $m$ is suspended on a string of length $L$ and allowed to oscillate in a single vertical plane under the effect of a uniform gravitational field. See Fig.~\ref{fig01:pendulum} for a visualization. We make the usual assumptions that the size of the massive object can be neglected, and so can be the mass of the string.

The solution of this problem is extremely simple provided we choose as the single generalized coordinate the angular deviation $\vp$ of the pendulum from the position of lowest potential energy. In terms of this angle, the speed of the pendulum is $L\dot\vp$. The potential energy is $-mgx=-mgL\cos\vp$. (Note the unusual choice of the coordinate axes, made in order to maintain the standard conventions for $\vp$ as the polar angle.) Altogether, the Lagrangian of the pendulum is
\begin{equation}
L=\frac12mL^2\dot\vp^2+mgL\cos\vp\;.
\label{ch01:pendulumLag}
\end{equation}
Deriving the corresponding EoM from~\eqref{ch01:Lagrangeeq} is a piece of cake,
\begin{equation}
\ddot\vp+\frac gL\sin\vp=0\;.
\label{ch01:pendulumEoM}
\end{equation}
For small-angle oscillations, we can approximate $\sin\vp\approx\vp$. The resulting linear ODE, $\ddot\vp+(g/L)\vp\approx0$, describes harmonic oscillations with angular frequency $\o=\sqrt{g/L}$. The full EoM~\eqref{ch01:pendulumEoM} can, in fact, be solved exactly in terms of the \href{https://en.wikipedia.org/wiki/Jacobi_elliptic_functions}{Jacobi elliptic functions}.

I hope you agree that the derivation of the pendulum EoM in the Lagrangian formalism is very simple. No need to draw force diagrams, consider the direction of the acceleration, or even mention the force on the mass due to the string. All we need to know about the action of the string is that it makes sure the mass moves along an arc of fixed radius $L$. However, in order to truly emphasize the power of the Lagrangian formalism, I will now tweak the pendulum problem, making it thus more nontrivial but also more interesting.

\begin{figure}[t]
\sidecaption[t]
\includegraphics[width=2.0in]{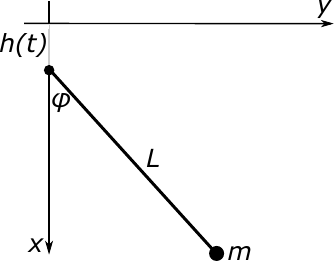}% This figure does not include so many details to warrant a full width of 2.9 in!
\caption{Visualization of a driven pendulum whose pivot point moves vertically in a controlled manner. The position of the pivot point is specified by a given function of time, $h(t)$.}
\label{fig01:pendulumdriven}
\end{figure}

Suppose the string is not attached to a fixed point, located in Fig.~\ref{fig01:pendulum} at the origin of the coordinate system. Instead, assume that the pivot point moves along the vertical axis in an a priori known manner. You can imagine, for instance, that you hold the string in your hand and, as the pendulum oscillates, wiggle it up and down. For Newtonian mechanics, this more general problem would already be fairly tricky. Yet Lagrange proceeds as before just with a minor change. Denoting the instantaneous vertical position of the pivot point as $h(t)$, the position vector of the mass $m$ is related to the angle $\vp$ by $\vec r=(h+L\cos\vp,L\sin\vp)$; see Fig.~\ref{fig01:pendulumdriven}. The potential energy is, as before, $-mgx$. The kinetic energy can be derived from the position vector as $(1/2)m\dot{\vec r}^2$. Expressing everything in terms of the sole generalized coordinate $\vp$, we find after a short manipulation that the Lagrangian becomes
\begin{equation}
L=\frac12m(L^2\dot\vp^2-2L\dot h\dot\vp\sin\vp+\xcancel{\dot h^2})+mg(\xcancel{h}+L\cos\vp)\;.
\label{ch01:pendulumLagdriven}
\end{equation}
The two terms crossed out are independent of the dynamical variable $\vp$, and thus contribute a mere constant to action. We can therefore ignore them as far as the derivation of the EoM is concerned. Recalling~\eqref{ch01:Lagrangeeq}, the modified EoM turns out~to~be
\begin{equation}
\ddot\vp+\frac{g-\ddot h}{L}\sin\vp=0\;.
\label{ch01:pendulumEoMdriven}
\end{equation}
This has a very simple interpretation, whereby acceleration of the pivot point effectively modifies the gravitational field that acts on the pendulum. A uniform motion of the pivot point, on the other hand, has no effect on the pendulum in accord with the Galilei principle of relativity.

Let us try to make this intuitive explanation of the modified EoM~\eqref{ch01:pendulumEoMdriven} a bit more precise. Following in spirit our discussion of inertial forces in a rotating frame in Sect.~\ref{subsec:singleparticle}, let us switch to the noninertial reference frame where the pivot point is at rest. Denoting the position vector of the mass $m$ in this frame as $\vec R\equiv(X,Y)=(L\cos\vp,L\sin\vp)$, the Cartesian coordinates in the two frames are related by $x=X+h$ and $y=Y$. This is mathematically exactly what we did previously, and the Lagrangian~\eqref{ch01:pendulumLagdriven} is recovered by inserting everything into~\eqref{ch01:pendulumLag} and rewriting the result in terms of $\vp$.

There is yet another, though more formal, way to see that the motion of the pivot point leads to a mere modification of the gravitational acceleration. Remember that adding to the Lagrangian a term that is a total time derivative does not change the EoM; cf.~Sect.~\ref{subsec:LagrangeEoM}. We can use this freedom to add to the Lagrangian~\eqref{ch01:pendulumLagdriven} the term $-\Od{(mL\dot h\cos\vp)}{t}$, upon which it becomes
\begin{equation}
L=\frac12mL^2\dot\vp^2+m(g-\ddot h)L\cos\vp\;.
\end{equation}
This is nothing but the original Lagrangian~\eqref{ch01:pendulumLag} for a simple pendulum with effective gravitational acceleration $g-\ddot h$, thus verifying the EoM~\eqref{ch01:pendulumEoMdriven}.

Before concluding the discussion of various applications of the Lagrangian formalism, I will use the simple pendulum in Fig.~\ref{fig01:pendulum} to illustrate yet another general technique. Suppose we were not so smart to discover right away that polar coordinates are ideally suited for the pendulum system. Could we use the standard Cartesian coordinates instead? The problem is that the two coordinates $x,y$ cannot be varied independently as long as the pendulum is forced by the string to only move along an arc of radius $L$. This is a typical example of a constrained system, which we can deal with using the approach developed in Sect.~\ref{sec:constrainedoptimization}. In our case, the relevant constraint is $x^2+y^2=L^2$. Otherwise, everything else works as before. The Lagrangian is still given by the difference of the kinetic energy, $(1/2)m(\dot x^2+\dot y^2)$, and the potential energy, $-mgx$, with an extra term taking into account the constraint, 
\begin{equation}
L=\frac12m(\dot x^2+\dot y^2)+mgx-\frac\l2(x^2+y^2-L^2)\;.
\end{equation}
The constraint is local, i.e.~must be satisfied at all times. The Lagrange multiplier $\l$ is therefore an unknown function of time. The extra factor of $1/2$ in the last term is just for convenience. The Lagrange equations implied by our Lagrangian are
\begin{equation}
m\ddot x=mg-\l x\;,\qquad
m\ddot y=-\l y\;,
\end{equation}
plus the constraint $x^2+y^2=L^2$ itself. This shows that $\l L$ has the physical interpretation of the force by which the string pulls on the mass $m$. It is not easy to solve the above system of coupled ODEs by brute force. The best thing to do is to resolve the constraint by introducing the angle $\vp$ via $(x,y)=(L\cos\vp,L\sin\vp)$. This turns the equations for $x$ and $y$ to
\begin{equation}
\begin{split}
-mL\ddot\vp\sin\vp-mL\dot\vp^2\cos\vp-mg+\l L\cos\vp&=0\;,\\
mL\ddot\vp\cos\vp-mL\dot\vp^2\sin\vp+\l L\sin\vp&=0\;.
\end{split}
\end{equation}
Multiplying the first of these by $-\sin\vp$, the second by $\cos\vp$, and adding them up eliminates the Lagrange multiplier, leaving us with $\ddot\vp+(g/L)\sin\vp=0$.

\begin{watchout}%
We have managed to reproduce the standard pendulum equation~\eqref{ch01:pendulumEoM}. However, the path to it was not quite so straightforward. Clearly, it is much more convenient to resolve the constraint on the motion of the mass $m$ directly at the level of the Lagrangian. However, this may not be always possible. In such cases, it is reassuring to have the Lagrange multiplier method at hand to fall back on. To be more explicit, suppose we can characterize the configuration of a system in terms of $n$ generalized coordinates $q_1,\dotsc,q_n$ that are not mutually independent but rather are subject to a set of $N<n$ constraints,
\begin{equation}
f_i(q_1,\dotsc,q_n,t)=0\;,\quad\text{where }i=1,\dotsc,N\;.
\end{equation}
The functions $f_i$ may depend explicitly on time but not on the generalized velocities $\dot q_1,\dotsc,\dot q_n$. Such constraints are called \emph{holonomic}. If possible, it is always advisable to resolve the constraints explicitly, that is to change the generalized coordinates to $\tilde q_1,\dotsc,\tilde q_n$ such that $\tilde q_1,\dotsc,\tilde q_N$ are fixed by the constraints. One can then construct the Lagrangian using $\tilde q_{N+1},\dotsc,\tilde q_n$ as the genuinely independent dynamical variables. In the pendulum case, this step amounts to switching from Cartesian coordinates $x,y$ to polar coordinates $r,\vp$ and using the fact that the radial variable $r$ is fixed by the constraint. Should resolving the constraints explicitly not be a viable option, one can proceed by introducing a Lagrange multiplier for each of the (unresolved) constraints.

Occasionally, one comes across physical systems with constraints that do depend on generalized velocities. Such constraints are called \emph{nonholonomic}, and are an enigma that has haunted physicists for many years. In spite of all the effort, there does not seem to be a fully general algorithm how to deal with nonholonomic constraints. Those interested will find a relatively careful discussion in~\cite{Flannery2005}.
\end{watchout}

%%%%%%%%%%%%%%%%%%%%%%%%%%%%%%%%%%%%%%%%%%%%%%%%%%%%%%%%%%%%

\section{Conservation laws}
\label{sec01:conservationlaws}

So far, we have focused mostly on the use of the Lagrangian formalism to derive the EoM of a system. The next step is to solve the EoM, which is an exercise in the theory of differential equations that requires no further insight from variational calculus, and often very little insight from physics. For a system with $n$ independent generalized coordinates, the Lagrange equations constitute a set of $n$ second-order ODEs, whose unique solution requires $2n$ integration constants. The latter are usually supplied by giving initial or boundary conditions. Some concrete examples including explicit solution of the EoM are postponed to the exercise problems.

Before wrapping up the exposition of the basic theory, let me stress that a brute-force solution of the Lagrange equation(s) is often not the best way to attack a mechanical problem. Indeed, we already know from Sect.~\ref{sec:firstintegrals} that for some variational problems, the EL equations can be reduced without doing any actual integration. There is nothing new to add to the story here; I will just translate what we learned previously to a terminology more common in physics.

In the context of mechanics, a quantity composed of the generalized coordinates and velocities, possibly depending explicitly on time, that is constant (time-independent) when evaluated on any solution of the EoM is called a \emph{constant (integral) of motion}. We also say that the quantity is \emph{conserved}. This is what we called first integral in Sect.~\ref{sec:firstintegrals}. Therein, we discovered two general classes of first integrals, so let us see what they tell us about physics. Suppose the Lagrangian $L(q,\dot q,t)$ does not depend on a generalized coordinate, $q_i$. Such $q_i$ is called a \emph{cyclic coordinate}, and we know from~\eqref{ch00:firstintcyclic} that
\begin{equation}
\boxed{p_i\equiv\PD{L}{\dot q_i}}
\label{ch01:conjugatemomentum}
\end{equation}
is a constant of motion. In physics terminology, $p_i$ is the \emph{conjugate momentum} to $q_i$.

\begin{illustration}%
The Lagrangian~\eqref{ch01:2bodyLseparated} of the two-body system that involves a pair interaction but no external fields, does not depend explicitly on the CoM coordinate $\vec R$. As a consequence, the corresponding conjugate momentum, $\vec P\equiv\Pd{L}{\dot{\vec R}}=M\dot{\vec R}$, is a constant of motion, i.e.~is conserved. This is nothing but the total momentum of the system.

For another example, have a look at the Lagrangian~\eqref{ch01:Lparticlepolar} of a particle in two dimensions, expressed in polar coordinates. Assuming that the potential energy does not depend on the angle $\vp$ but only on the radial variable $r$, $\vp$ is a cyclic coordinate. In such cases, $p_\vp\equiv\Pd{L}{\dot\vp}=mr^2\dot\vp$ is a constant of motion. This expresses the angular momentum of the particle.
\end{illustration}

\begin{watchout}%
The concept of conjugate momentum is clearly extremely useful in that it allows one to reduce the effort necessary to solve the EoM of the system. However, it is conceptually somewhat unsatisfactory that the existence of a constant of motion should depend on our ability to make such a choice of generalized coordinates that one or more of them drop out of the Lagrangian. This should not be necessary. For instance, it is known that for a particle in $\R^3$ moving under the effect of a force that only depends on distance but not on direction, the whole vector $\vec J$ of angular momentum is conserved. Yet, there is no choice of generalized coordinates such that they would all be cyclic, which would make the conservation of all components of $\vec J$ manifest simultaneously. In \chaptername~\ref{chap:symmetries}, we will develop a general theory of conservation laws that will bypass this problem and allow us to find the constants of motion of a given system regardless of the particular choice of generalized coordinates.
\end{watchout}

Another generic first integral appears in variational problems where the functional density does not depend explicitly on the sole independent variable. In physics terms, this corresponds to Lagrangians that do not depend explicitly on time. Then, according to~\eqref{ch00:firstintenergy},
\begin{equation}
\boxed{H\equiv\sum_i\dot q_i\PD{L}{\dot q_i}-L}
\label{ch01:Hamiltonian}
\end{equation}
is a constant of motion. This is the \emph{Hamilton function} (or simply \emph{Hamiltonian}) of the system. To understand its physical meaning, let us have a look at an example.

\begin{illustration}%
Consider a system of $N$ particles in $\R^n$ with masses $m_i$, defined by the Lagrangian
\begin{equation}
L=\sum_{i=1}^N\frac12m_i\dot{\vec r}_i^2-V(\vec r_1,\dotsc,\vec r_N)\;.
\label{ch01:LagrangianNparticle}
\end{equation}
The potential energy is very general: it involves the possibility of interaction with an external field as well as mutual interactions between the particles. All that matters is that it does not depend explicitly on time. The Hamilton function corresponding to the Lagrangian is
\begin{equation}
H=\sum_{i=1}^N\frac12m_i\dot{\vec r}_i^2+V(\vec r_1,\dotsc,\vec r_N)\;.
\end{equation}
\end{illustration}

The Hamiltonian of the above example clearly has the interpretation as the total energy of the system. In fact, throughout these entire lecture notes, I will treat~\eqref{ch01:Hamiltonian} as the \emph{definition} of energy for systems whose Lagrangian does not depend explicitly on time. I encourage you to check that in all the examples we have discussed so far, the Hamiltonian is given by the sum of the kinetic and potential energy of the system. 

%%%%%%%%%%%%%%%%%%%%%%%%%%%%%%%%%%%%%%%%%%%%%%%%%%%%%%%%%%%%

\section*{\probsec}
\addcontentsline{toc}{section}{\probsec}

\begin{prob}
\label{pr01:atwood}
Construct a Lagrangian for the \href{https://en.wikipedia.org/wiki/Atwood_machine}{Atwood machine}. Use it to derive the equation of motion. Check that your result agrees with one obtained from Newton's laws.
\end{prob}

\begin{prob}
\label{pr01:sphericalpendulum}
A \emph{spherical pendulum} is a generalization of the simple pendulum analyzed in Sect.~\ref{subsec:pendulum}, where the point mass $m$ is allowed to move in both horizontal directions. Find suitable generalized coordinates and construct the Lagrangian in this more general case. Derive the Lagrange equations and find a solution that amounts to uniform circular motion in a horizontal plane.
\end{prob}

\begin{figure}[t]
\sidecaption[t]
\includegraphics[width=2.9in]{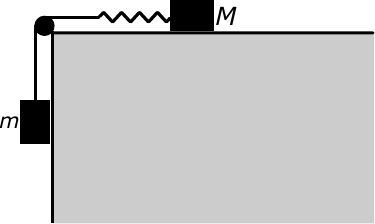}
\caption{Illustration for~\refpr{pr01:pulley}. The blocks with masses $m$ and $M$ move without friction, the former vertically and the latter horizontally. The block $m$ is suspended on a string, which passes over a pulley, its other end being connected to a spring, attached to block $M$.}
\label{fig01:pulley}
\end{figure}

\begin{prob}
\label{pr01:pulley}
Two blocks of masses $m,M$ are connected by a string and a spring as shown in Fig.~\ref{fig01:pulley}. The block $M$ slides without friction on a horizontal plane, whereas the block $m$ is suspended on the string and can only move vertically, also without friction. The other end of the string is, upon passing over a pulley, connected to the spring with spring constant $k$, which is itself attached to block $M$. Identify a set of suitable generalized coordinates parameterizing the configuration of the system. Construct the Lagrangian and derive the corresponding Lagrange equations. Use the equations to describe qualitatively the motion of the system.
\end{prob}

\begin{prob}
\label{pr01:exam2021}
A (nonrelativistic) particle of mass $m$ moves in one dimension in the potential
\begin{equation}
V(x)=-c(x^2-a^2)^2\;,
\end{equation}
where $a$ and $c$ are positive constants. Find a solution of the equation of motion that starts at $x=-a$ at time $t=-\infty$ and arrives at $x=+a$ at time $t=+\infty$.
\end{prob}

\begin{prob}
\label{pr01:freefall}
A point particle of mass $m$ is allowed to move only in the vertical direction, labeled by the Cartesian coordinate $z$. Write down the Lagrangian for the particle, assuming that the only force acting on it is that of gravity with constant acceleration $g$. Check that your Lagrangian gives the expected equation of motion. Now introduce a new coordinate $Z$ by $z\equiv Z-(1/2)gt^2$. Rewrite the Lagrangian in terms of $Z,\dot Z$ and derive the corresponding equation of motion. What is the physical interpretation of your result?
\end{prob}

\begin{prob}
\label{pr01:particleinEMfield}
The Lagrangian of a particle of mass $m$ and electric charge $q$ is given by
\begin{equation}
L(\vec r,\dot{\vec r},t)=\frac12m\dot{\vec r}^2-q\phi(\vec r,t)+q\dot{\vec r}\cdot\vec A(\vec r,t)\;,
\label{ch01:LagEMfield}
\end{equation}
where $\phi,\vec A$ are the scalar and vector potential of an electromagnetic field that the particle is exposed to. Show that this Lagrangian correctly reproduces the equation of motion of the particle in the electromagnetic field in terms of the Lorentz force, $\vec F=q(\vec E+\dot{\vec r}\times\vec B)$. Note that the Lagrangian cannot be viewed as a difference of the kinetic and potential energies of the particle. What is the corresponding Hamiltonian?
\end{prob}

\begin{prob}
\label{pr01:exam2021re}
The Lagrangian of a system with a single degree of freedom is given by
\begin{equation}
L(q,\dot q)=\frac12q\dot q^2\;.
\end{equation}
The motion is characterized by positive generalized velocity, total energy $E$ and the initial condition $\lim\limits_{t\to0}q(t)=0$. Find the position $q(t)$ as a function of time $t>0$.
\end{prob}

\begin{prob}
\label{pr01:exam2023}
The dynamics of a system with generalized coordinates $x,y$ is governed by the Lagrangian
\begin{equation}
L=\frac12(x\dot y-y\dot x)-\frac\lambda4(x^2+y^2-v^2)^2\;,
\end{equation}
where $\lambda$ and $v$ are positive constants. Find a solution of the equation of motion that has the lowest possible total energy.
\end{prob}
\chapter{Application: Motion in Central Fields}
\label{chap:centralfields}

\keywords{Two-body problem, effective potential, Binet's equation, Kepler's laws, Laplace--Runge--Lenz vector.}

%%%%%%%%%%%%%%%%%%%%%%%%%%%%%%%%%%%%%%%%%%%%%%%%%%%%%%%%%%%%

\noindent The plan of our course consists of theory {\chaptername}s, intertwined with {\chaptername}s dedicated to specific applications of the general formalism. In this first application \chaptername, we will have a look at a problem that historically provided the key motivation for the development of mechanics: the dynamics of the Solar System. This is an archetype of a system of objects mutually interacting via gravitational forces. Our analysis will be general enough to also encompass other, similar or equivalent problems such as the motion of binary star systems, or the classical limit of the dynamics of atoms and molecules.

We start modeling the planetary dynamics by making several simplifying assumptions, some relatively innocuous but others less so. First, we will treat all the massive objects involved as pointlike; this is a very reasonable approximation for most practical purposes. Second, we will neglect relativistic effects: mostly good until you are into compact stars or black holes. Third, we will assume that the forces between the objects only depend on distance but not on direction. This is certainly correct for Newtonian gravity, but may be somewhat restrictive in the microscopic world, where molecular interactions are often directional. Finally, we will only deal with two massive objects at a time. This is serious. The three- and generally $N$-body problem is tough, the dynamics is very complicated, and only certain special solutions are known. On the other hand, the two-body problem is elementary and still represents a sensible approximation for a variety of systems. Think for example of the motion of a single planet around the Sun, or of a satellite around the Earth.

%%%%%%%%%%%%%%%%%%%%%%%%%%%%%%%%%%%%%%%%%%%%%%%%%%%%%%%%%%%%

\section{Reduction to One-Dimensional Problem}
\label{sec:2bodyreduction}

With the formalities out of the way, I  start right away by writing down the Lagrangian for a two-body system,
\begin{equation}
L=\frac12m_1\dot{\vec r}_1^2+\frac12m_2\dot{\vec r}_2^2-V(\lvert\vec r_2-\vec r_1\rvert)\;.
\label{ch02:Lag2body}
\end{equation}
As indicated by the notation, the potential energy of the interaction between the masses is assumed to only depend on their distance. Do you recall having seen this somewhere? If not, go back to Sect.~\ref{subsec:2body}. Once you have refreshed your memory, you also know how to proceed. Let me repeat the basic steps for your convenience. We trade the position vectors $\vec r_1,\vec r_2$ of the individual masses for the position $\vec R$ of the \emph{center of mass} (CoM) and the relative position $\vec r$, defined by
\begin{equation}
\vec R\equiv\frac{m_1\vec r_1+m_2\vec r_2}{m_1+m_2}\;,\qquad
\vec r\equiv\vec r_2-\vec r_1\;.
\label{ch02:CoMcoordinates}
\end{equation}
In terms of the new variables, the Lagrangian takes the form
\begin{equation}
\boxed{L=\underbrace{\frac12M\dot{\vec R}^2}_{\text{CoM motion}}+\underbrace{\frac12\m\dot{\vec r}^2-V(\abs{\vec r})}_{\text{relative motion}}\;,}
\label{ch02:2bodyLag}
\end{equation}
where $M\equiv m_1+m_2$ is the total mass and $\m\equiv m_1m_2/(m_1+m_2)$ the \emph{reduced mass} of the system. The ensuing \emph{equation of motion} (EoM) for $\vec R$ is trivial, $M\ddot{\vec R}=\vec0$, and guarantees conservation of the total momentum of the system. In the following, I will simply discard the first term in the Lagrangian~\eqref{ch02:2bodyLag} and focus on the relative motion as described by the remaining two terms, depending only on $\vec r$.

\begin{illustration}%
The masses of the Earth and Moon are $\smash{m_{\earth}\approx5.97\times10^{24}\,\text{kg}}$, $\smash{m_{\leftmoon}\approx7.35\times10^{22}\,\text{kg}}$. With the distance of $384{,}000\,\text{km}$, the CoM of the Earth--Moon system is only about $4{,}700\,\text{km}$ off the center of the Earth. The reduced mass is $\smash{\m_{\earth\leftmoon}\approx7.26\times10^{22}\,\text{kg}}$, just $2\,\%$ smaller than the mass of the Moon. The total mass is $\smash{M_{\earth\leftmoon}\approx6.04\times10^{24}\,\text{kg}}$. The mass of the Sun is $m_{\astrosun}\approx1.99\times10^{30}\,\text{kg}$. Given the distance of $151\times10^6\,\text{km}$ from the Earth, the CoM of the Sun--Earth system lies mere $450\,\text{km}$ off the center of the Sun. Thanks to the hierarchy of masses, $\smash{m_{\leftmoon}\ll m_{\earth}\ll m_{\astrosun}}$, the Sun--Earth--Moon system can be reasonably thought of in terms of a pair of two-body problems. It is certainly safe to view the Earth and Moon jointly as a single object of mass $\smash{M_{\earth\leftmoon}}$ revolving around the Sun. One can also think of the Earth--Moon system in terms of revolution of the Moon around the Earth. This approximation is, however, rather crude. The Sun turns out to significantly influence the motion of the Moon. In case of interest, you will find more details including an overview of the history of understanding the three-body problem of the Sun, Earth and Moon in~\cite{Gutzwiller1998}.
\end{illustration}

Mathematically, the relative motion of the original objects with masses $m_1,m_2$ is equivalent to the motion of a single object with mass $\m$, moving in a radial (central) field with a center at the origin of the space of the variable $\vec r$. Any trajectory in such a system always lies in a single plane containing the center of force. This is ultimately because the acceleration has a radial direction. Together with the instantaneous velocity vector at a chosen point of the trajectory, this defines the plane in which the motion occurs. It follows that regardless of the number $n$ of spatial dimensions, it is always possible to orient the Cartesian coordinate frame spanned on the axes $x_1,\dotsc,x_n$ so that the given trajectory lies entirely in the plane of $x_1,x_2$. The dynamics effectively reduces from $\R^n$ to $\R^2$. I will use the more common notation $x,y$ and switch right away to polar coordinates $r,\vp$ by the usual prescription $x=r\cos\vp$, $y=r\sin\vp$. This brings the Lagrangian for the ``particle'' of mass $\m$ to the form
\begin{equation}
L=\frac12\m(\dot r^2+r^2\dot\vp^2)-V(r)\;.
\label{ch02:1particleL}
\end{equation}
The corresponding Lagrange equations for $r,\vp$ are, respectively,
\begin{equation}
\m(\ddot r-r\dot\vp^2)=-V'(r)\;,\qquad
\OD{}t(\m r^2\dot\vp)=0\;.
\label{ch02:2bodyEoM}
\end{equation}
You have also seen these before: cf.~\eqref{ch01:LparticlepolarEoM}. Next we need to figure out how to deal with these equations.

%%%%%%%%%%%%%%%%%%%%%%%%%%%%%%%%%%%%%%%%%%%%%%%%%%%%%%%%%%%%

\subsection{Effective Potential}
\label{subsec:effpot}

The EoM for $\vp$ is trivial, implying that the combination
\begin{equation}
\boxed{J\equiv\m r^2\dot\vp}
\label{ch02:Jdef}
\end{equation}
is a constant of motion. As already mentioned before, this has the physical interpretation as the angular momentum of the particle. We can use it to eliminate $\dot\vp$ from the EoM for $r$, thus bringing it to the form
\begin{equation}
\m\ddot r=-V'(r)+\frac{J^2}{\m r^3}\;.
\label{ch02:EoMradial}
\end{equation}
The reason for grouping both terms not containing a time derivative on the right-hand side is that we can now rewrite the equation further as $\m\ddot r=-V'_\mathrm{eff}(r)$, where
\begin{equation}
\boxed{V_\mathrm{eff}(r)\equiv V(r)+\frac{J^2}{2\m r^2}}
\label{ch02:effpot}
\end{equation}
is the so-called \emph{effective potential}.

What is this good for? Treating $J$ as an a priori free parameter, there is no trace of the azimuthal motion left in our new EoM, $\m\ddot r=-V'_\mathrm{eff}(r)$. We can view it as describing purely one-dimensional motion of a particle with mass $\m$ and potential energy $V_\mathrm{eff}(r)$. The effective one-dimensional Lagrangian is $L=(1/2)\m\dot r^2-V_\mathrm{eff}(r)$. Even more interesting is the total energy, which is by construction conserved,
\begin{equation}
E=\frac12\m\dot r^2+V_\mathrm{eff}(r)=\frac12\m\dot r^2+V(r)+\frac{J^2}{2\m r^2}\;.
\label{ch02:Energy}
\end{equation}
I have used the symbol $E$ for the energy first because it is more common, and second to distinguish it from the Hamilton function of the original, two-dimensional single-particle problem~\eqref{ch02:1particleL}.\footnote{This itself differs from the Hamiltonian corresponding to the full Lagrangian~\eqref{ch02:2bodyLag} of the two-body system by a contribution of the CoM motion, $(1/2)M\dot{\vec R}^2$.} The latter is by means of~\eqref{ch01:Hamiltonian} equal to
\begin{equation}
H=\frac12\m(\dot r^2+r^2\dot\vp^2)+V(r)\;.
\end{equation}
Eliminating $\dot\vp$ using~\eqref{ch02:Jdef} of course recovers~\eqref{ch02:Energy}.

\begin{figure}[t]
\sidecaption[t]
\includegraphics[width=2.9in]{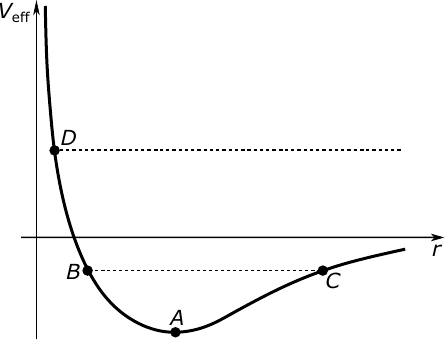}
\caption{Sketch of the effective potential~\eqref{ch02:effpot} in case of a Newton-like potential energy $V(r)=-k/r$ with a positive constant $k$. The dots indicate points on various trajectories (dashed lines) where the radial velocity $\dot r$ vanishes.}
\label{fig02:effpot}
\end{figure}

You certainly recall from your introductory mechanics classes that energy conservation is very useful, especially for one-dimensional problems where it can be used to understand the motion without solving any differential equations. We are now going to exploit this feature. For the sake of illustration, we need a specific potential $V(r)$. Everything said so far holds in principle for any choice of potential. However, keeping in mind that this whole \chaptername{} is motivated by celestial mechanics, we take a potential corresponding to an attractive force decreasing with the inverse square of distance, as appropriate for Newtonian gravity in $\R^3$. We thus set
\begin{equation}
V(r)=-\frac kr\quad\Rightarrow\quad
V_\mathrm{eff}(r)=-\frac kr+\frac{J^2}{2\m r^2}\;,
\end{equation}
where $k$ is a constant positive parameter. This effective potential as a function of $r>0$ is qualitatively sketched in Fig.~\ref{fig02:effpot}. For $r\to0$, the effective potential is dominated by the ``centrifugal'' term $J^2/(2\m r^2)$ that prevents the particle from getting too close to the origin. On the other hand, for $r\to\infty$, the effective potential is dominated by the actual potential energy, $-k/r$.

Qualitatively, much about the motion of the particle can be inferred from the graph of $V_\mathrm{eff}(r)$ and the fact that during the motion, the total energy~\eqref{ch02:Energy} remains constant. It helps to think of literally placing the mass $\m$ in the valley defined by the graph of $V_\mathrm{eff}(r)$, and letting it move therein with certain energy. For our choice of effective potential, there is a unique (global) minimum at point $A$. Should the total energy $E$ exactly correspond to the value of the potential at the minimum, $V_\mathrm{eff}(r_A)$, then by~\eqref{ch02:Energy}, the radial velocity $\dot r$ must vanish at all times. The minimum $A$ therefore corresponds to circular trajectories (orbits) with fixed $r=r_A$ and, by~\eqref{ch02:Jdef}, fixed angular velocity $\dot\vp=J/(\m r_A^2)$. Minimizing the effective potential explicitly gives
\begin{equation}
r_A=\frac{J^2}{k\m}\quad\text{with}\quad
V_\mathrm{eff}(r_A)=-\frac{k^2\m}{2J^2}=-\frac{k}{2r_A}\;.
\label{ch02:minVeff}
\end{equation}
Combining this with~\eqref{ch02:Jdef}, we get the angular velocity $\dot\vp=\sqrt{k/(\m r_A^3)}$. As an exercise, check that this agrees with the properties of circular orbits, obtained in the Newtonian way by balancing the gravitational attraction against the centrifugal force acting on the mass!

The total energy $E$ can never be below $V_\mathrm{eff}(r_A)$. Suppose now that it is higher than that, but still negative. This is illustrated in Fig.~\ref{fig02:effpot} by the dashed line connecting the points $B,C$. We see that such an orbit is bound to the center: the particle can only get as far as $r_C$ from the origin. On the other hand, the centrifugal part of the effective potential does not allow it to get closer than $r_B$. The exact values of the distance $r_B$ of the nearest point of the orbit (\emph{pericenter}), and $r_C$ of the furthest point of the orbit (\emph{apocenter}), are obtained by setting $\dot r=0$ and solving $V_\mathrm{eff}(r)=E$ as a quadratic equation for $1/r$.

It remains to discuss trajectories whose total energy $E$ is non-negative. A typical example is shown in Fig.~\ref{fig02:effpot} as the dashed line emanating from point $D$. This corresponds to orbits that are unbound. The particle get nearest the center of force at $D$. The precise value of $r_D$ is given by the sole positive solution of the quadratic equation $V_\mathrm{eff}(r)=E$ for $1/r$. Having ``bounced off'' at $D$, the particle has sufficient energy to escape the potential well and runs away to infinity.

The analysis of orbits based on the graph of the effective potential is only qualitative. However, it can be turned into a fully quantitative derivation of the orbit using the following steps. The first step is to think of~\eqref{ch02:Energy} as a first-order \emph{ordinary differential equation} (ODE) for the radial distance as a function of time. Upon isolating $\dot r^2$ on one side and taking the square root, the equation can be integrated to give $t$ as a function of $r$,
\begin{equation}
t(r)=\sqrt{\frac\m2}\int\frac{\D r}{\sqrt{E-V(r)-\frac{J^2}{2\m r^2}}}\;.
\label{ch02:integrater}
\end{equation}
This involves a single integration constant that is conventionally fixed by setting $t=0$ at the pericenter or apocenter. The second step is to invert the obtained function to get $r(t)$, radial distance as a function of time. Having done that, we proceed to step three: integrate~\eqref{ch02:Jdef} to find $\vp$ as a function of time,
\begin{equation}
\vp(t)=\frac J\m\int\frac{\D t}{[r(t)]^2}\;.
\label{ch02:integratevp}
\end{equation}
This also involves an integration constant, most conveniently fixed by setting $\vp(0)=0$. This completes the algorithm for finding the orbit, reducing the problem to the two one-dimensional integrations indicated in~\eqref{ch02:integrater} and~\eqref{ch02:integratevp}.

\begin{watchout}%
Keeping in mind that the trajectory in any central field is necessarily planar, and the motion is thus effectively restricted to two dimensions, we expect the solution to involve four integration constants. Two of these were already mentioned, and enter through the integration in~\eqref{ch02:integrater} and~\eqref{ch02:integratevp}. The remaining two integration constants are hidden, since we used conservation of angular momentum and energy to reduce the Lagrange equations to first order. Fixing these remaining integration constants is equivalent to choosing particular values of $J$ and $E$.
\end{watchout}

As straightforward as the above procedure appears, it is usually not easy to do the integrals explicitly. It is certainly possible for the Newtonian potential $V(r)=-k/r$, but even there it takes some grit. We will bypass the problem by following an alternative, less streamlined but more elegant approach. This will not give us explicitly the trajectory in terms of position as a function of time. However, it will still tell us everything about the geometric shape of the orbit, which is for many practical purposes quite sufficient.

%%%%%%%%%%%%%%%%%%%%%%%%%%%%%%%%%%%%%%%%%%%%%%%%%%%%%%%%%%%%

\subsection{Geometry of Orbits}

Suppose we do not care about the precise dependence of $r$ and $\vp$ on time, but are only interested in the shape of the orbit as a geometric curve. In other words, we would like to know how $r$ depends on $\vp$. We thus need to eliminate time as the independent variable in favor of $\vp$. Using~\eqref{ch02:Jdef}, we can convert derivatives with respect to $t$ to derivatives with respect to $\vp$,
\begin{equation}
\OD{}t=\dot\vp\OD{}\vp=\frac{J}{\m r^2}\OD{}\vp\;.
\end{equation}
This turns the second time derivative in the radial EoM~\eqref{ch02:EoMradial} into
\begin{equation}
\ddot r=\frac{J}{\m r^2}\OD{}\vp\left(\frac{J}{\m r^2}\OD r\vp\right)=-\frac{J^2u^2}{\m^2}\frac{\D^2u}{\D\vp^2}\;,
\end{equation}
where $u\equiv 1/r$ is a new, more convenient radial variable. Using in addition that $V'(r)=-u^2\Od{V(1/u)}{u}$, and putting all the pieces together, the radial EoM~\eqref{ch02:EoMradial} becomes a second-order ODE for the function $u(\vp)$,
\begin{equation}
\boxed{\frac{\D^2u}{\D\vp^2}+u+\frac{\m}{J^2}\OD{}uV(1/u)=0\;.}
\label{ch02:Binet}
\end{equation}
This is known as \emph{Binet's equation}.

\begin{figure}[t]
\sidecaption[t]
\includegraphics[width=2.9in]{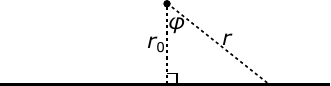}
\caption{Parameterization of the trajectory (thick solid line) of a free particle using polar coordinates. The trajectory is a straight line, described in polar coordinates as $r(\vp)=r_0/\cos\vp$ up to a shift of $\vp$.}
\label{fig02:Binetfree}
\end{figure}

\begin{illustration}%
\label{ex02:Binetfree}%
As a sanity check, let us see what the Binet equation tells us about possible trajectories of a free particle. Setting $V\to0$, \eqref{ch02:Binet} reduces to the equation of a harmonic oscillator, $\D^2 u/\D\vp^2+u=0$. This is solved by
\begin{equation}
\frac{1}{r(\vp)}=u(\vp)=A\cos\vp+B\sin\vp\;,
\end{equation}
where $A,B$ are integration constants. One of these can be fixed by choosing the coordinate system so that $\vp=0$ corresponds to the point on the trajectory nearest the origin. This amounts to requiring that $\Od u\vp=0$ at $\vp=0$, which implies $B=0$. Changing finally the notation for the remaining integration constant as $A=1/r_0$, we arrive at the solution $r(\vp)=r_0/\cos\vp$. Clearly, $r_0$ has the interpretation as the distance from the origin to the nearest point on the trajectory. That the inverse cosine actually represents a straight line, as expected for a free particle, is shown in Fig.~\ref{fig02:Binetfree}.
\end{illustration}

%%%%%%%%%%%%%%%%%%%%%%%%%%%%%%%%%%%%%%%%%%%%%%%%%%%%%%%%%%%%

\section{Kepler Problem}
\label{sec:Keplerproblem}

Equipped with the Binet equation, we can now address the problem of finding the shape of the orbits for the potential $V(r)=-k/r$. This is known as the Kepler problem, regardless of the physical origin of the force and the sign of $k$. A gravitational force is of course always attractive, but the electrostatic Coulomb force, satisfying the same inverse-square law, can be both attractive ($k>0$) and repulsive ($k<0$). With $V(r)=-k/r=-ku$, the Binet equation reduces to
\begin{equation}
\frac{\D^2 u}{\D\vp^2}+u=\frac{k\m}{J^2}\;.
\end{equation}
With a constant right-hand side, this is a trivial modification of the problem we already solved in~\refex{ex02:Binetfree}. Setting again $\vp=0$ at the pericenter, the solution is $u(\vp)=(k\m/J^2)+A\cos\vp$. It is conventional to trade the integration constant $A$ for a dimensionless parameter $e$ called \emph{eccentricity}. Our final result for the Keplerian orbit in polar coordinates is then
\begin{equation}
\boxed{\frac{1}{r(\vp)}=u(\vp)=\frac{k\m}{J^2}(1+e\cos\vp)\;.}
\label{ch02:Keplerorbitgeneral}
\end{equation}

In order that $\vp=0$ actually corresponds to a pericenter where $u$ is maximum, the eccentricity should always be non-negative. It can be explicitly computed in terms of the two integration constants $J,E$. To see how, evaluate the energy~\eqref{ch02:Energy} by looking at the pericenter where $\dot r$ is guaranteed to vanish. Setting $r\to r_0$, we thus get
\begin{equation}
E=-\frac{k}{r_0}+\frac{J^2}{2\m r_0^2}\;.
\end{equation}
On other hand, \eqref{ch02:Keplerorbitgeneral} evaluates at $\vp=0$ to $1/r_0=(k\m/J^2)(1+e)$. Using this to eliminate $r_0$ from the expression for the energy, we find $E=[k^2\m/(2J^2)](e^2-1)$, which inverts to
\begin{equation}
\boxed{e=\sqrt{1+\frac{2EJ^2}{k^2\m}}=\sqrt{1-\frac{E}{V_\mathrm{eff}(r_A)}}\;.}
\label{ch02:eccentricity}
\end{equation}
In the last step, I used~\eqref{ch02:minVeff} to express the eccentricity in terms of the ratio of the actual energy $E$ and the minimum possible energy in an attractive potential.\footnote{For a repulsive potential ($k<0$), the effective potential $V_\mathrm{eff}(r)$ does not have a minimum but is always positive, and so is therefore the total energy~\eqref{ch02:Energy}. In that case, one should always use the first expression for the eccentricity in~\eqref{ch02:eccentricity}.}

The geometric curves described by~\eqref{ch02:Keplerorbitgeneral} correspond to conic sections. Here is a brief overview, we will have a closer look at some of the possibilities later:
\begin{itemize}
\item\emph{Circle} ($e=0$). Here the radial coordinate $r$ is constant. By~\eqref{ch02:eccentricity}, the eccentricity can only vanish when the total energy $E$ is negative and tuned to the minimum of the effective potential $V_\mathrm{eff}$. This requires an attractive potential ($k>0$).
\item\emph{Ellipse} ($0<e<1$). This is a broader class of bound orbits, corresponding to the $BC$ line in Fig.~\ref{fig02:effpot}. Just like circular orbits, elliptic orbits require an attractive potential and negative total energy.
\item\emph{Parabola} ($e=1$). This is a boundary case, corresponding to vanishing total energy, again only possible for attractive potentials.
\item\emph{Hyperbola} ($e>1$). This is the generic case for unbound orbits, corresponding to positive total energy. While possible for both signs of the constant $k$, it is the only option for repulsive potentials ($k<0$).
\end{itemize}

%%%%%%%%%%%%%%%%%%%%%%%%%%%%%%%%%%%%%%%%%%%%%%%%%%%%%%%%%%%%

\subsection{Elliptic Orbits}
\label{subsec:ellipticorbits}

\begin{figure}[t]
\sidecaption[t]
\includegraphics[width=2.9in]{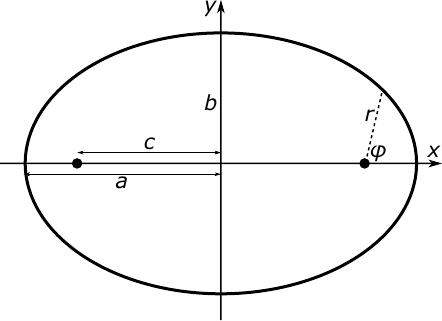}
\caption{The geometry of an ellipse. The focal points are indicated with dots. The semi-major axis $a$ and the semi-minor axis $b$ are shown along with the focal distance $c=\sqrt{a^2-b^2}$. The radial distance $r$ is measured from one of the foci, and the polar angle $\vp$ is taken with respect to the major axis.}
\label{fig02:ellipse}
\end{figure}

Let us now focus on elliptic orbits, which is the case relevant for the motion of planets around the Sun. I would first like to convince you that~\eqref{ch02:Keplerorbitgeneral} with $e<1$ indeed does correspond to an ellipse. To that end, we start from a more familiar definition of an ellipse as the set of points in plane with a given sum of distances from two fixed points (\emph{foci}). Using Cartesian coordinates, we place the foci on the $x$-axis at $x=\pm c$ where $c$ is a positive constant. The definition of an ellipse then translates to
\begin{equation}
\sqrt{(x+c)^2+y^2}+\sqrt{(x-c)^2+y^2}=2a\;,
\end{equation}
where $a$ is the \emph{semi-major axis} of the ellipse. See Fig.~\ref{fig02:ellipse} for a geometric interpretation of the different parameters of the ellipse. If you now keep squaring until you have removed all square roots, you will eventually arrive at the better-known implicit equation for an ellipse,
\begin{equation}
\frac{x^2}{a^2}+\frac{y^2}{b^2}=1\;,
\label{ch02:ellipseCartesian}
\end{equation}
where $b\equiv\sqrt{a^2-c^2}$ is the \emph{semi-minor axis} of the ellipse. This expression shows that the ellipse can be thought of as first constructing a circle of radius $a$ and then rescaling it by $b/a$ in the $y$-direction.

Next we switch to polar coordinates. We place the origin from which the radial distance $r$ is measured at the $x=c$ focus. Taking the polar angle $\vp$ with respect to the $x$-axis as usual then corresponds to the point transformation $x=c+r\cos\vp$, $y=r\sin\vp$. Upon inserting this in~\eqref{ch02:ellipseCartesian} and solving for $r$, we find that
\begin{equation}
\frac ar=\frac{1+e\cos\vp}{1-e^2}\;,
\label{ch02:ellipsepolar}
\end{equation}
where $e\equiv c/a$. It is conventional to take the semi-major axis $a$ and eccentricity $e$ as the primary parameters defining the shape of the ellipse, in terms of which the other parameters can be expressed. Thus, $b=a\sqrt{1-e^2}$ and $c=ea$.

The parameterization~\eqref{ch02:ellipsepolar} of the ellipse in polar coordinates precisely matches our previous result~\eqref{ch02:Keplerorbitgeneral}, obtained by solving the Binet equation. Comparing the prefactors tells us that $a(1-e^2)=J^2/(k\m)$. Using finally~\eqref{ch02:eccentricity} to eliminate the eccentricity gives us a very convenient expression for the semi-major axis of the orbit solely in terms of the energy,
\begin{equation}
\boxed{a=-\frac{k}{2E}\;.}
\label{ch02:ellipseafromE}
\end{equation}

%%%%%%%%%%%%%%%%%%%%%%%%%%%%%%%%%%%%%%%%%%%%%%%%%%%%%%%%%%%%

\subsection{Kepler's Laws}

We are now at a point where we can formulate and prove the three famous laws of planetary motion due to Kepler. We will take them one at a time and prove them right away. I will phrase the laws in terms of the general problem of motion of a point particle in an attractive potential, $V(r)=-k/r$ with $k>0$. The translation to the language of astronomy is obvious.

\runinhead{First Law} \emph{All bounded orbits of the particle are ellipses} (including the degenerate case with $e=0$, that is circles), \emph{with the center of force placed at one of the foci}. We have already proven this, there is nothing more to be done here.

\runinhead{Second Law} \emph{The line connecting the particle to the center of force sweeps equal areas during equal intervals of time.} Let us see where this comes from. Upon infinitesimal motion along the orbit, corresponding to the increment $\D\vp$ in the polar angle, the position vector of the particle sweeps an infinitesimally thin triangle with area $\D A=(1/2)\abs{\vec r\times\D\vec r}=(1/2)r^2\D\vp$. Consequently, the ``area velocity,'' that is the rate at which the position vector covers the area in the orbit plane, is
\begin{equation}
\dot A=\frac12r^2\dot\vp=\frac{J}{2\m}\;,
\label{ch02:areavelocity}
\end{equation}
where we used~\eqref{ch02:Jdef} to eliminate the angular velocity $\dot\vp$. As we see, the area velocity $\dot A$ is constant as claimed as a consequence of the conservation of angular momentum. This (and only this) Kepler law is therefore valid for any choice of the central potential $V(r)$, not just for $V(r)=-k/r$.

\runinhead{Third Law} \emph{The ratio of the square of the orbital period of the particle and the cube of the semi-major axis of its orbit is a constant, independent of the choice of orbit}. To prove this, we use the concept of area velocity just introduced. Upon completing the entire orbit, the position vector of the particle covers the area of the ellipse, $\pi ab$. Hence, $\dot A=\pi ab/T$, where $T$ is the period of orbit. Taking a square and combining this with our previous results, we obtain
\begin{equation}
\begin{split}
\frac{J^2}{4\m^2}&\overset{\eqref{ch02:areavelocity}}{=}\dot A^2=\left(\frac{\pi ab}{T}\right)^2=\frac{\pi^2a^4}{T^2}(1-e^2)\overset{\eqref{ch02:eccentricity}}{=}\frac{\pi^2a^4}{T^2}\left(-\frac{2EJ^2}{k^2\m}\right)\\
&\overset{\eqref{ch02:ellipseafromE}}{=}\frac{\pi^2J^2}{k\m}\frac{a^3}{T^2}\;.
\label{ch02:3rdKepleraux}
\end{split}
\end{equation}
Upon canceling common factors on both sides of the equation, we finally arrive at
\begin{equation}
\boxed{\frac{a^3}{T^2}=\frac{k}{4\pi^2\m}\;.}
\label{ch02:3rdKepler}
\end{equation}

\begin{illustration}%
In a binary gravitational system, we have $k=Gm_1m_2$, where $G$ is the gravitational constant and $m_1,m_2$ the masses of the two objects in the system. Together with an expression for the reduced mass, $\m=m_1m_2/(m_1+m_2)$, this makes the right-hand side of~\eqref{ch02:3rdKepler} evaluate to $G(m_1+m_2)/(4\pi^2)$. Let us see what this implies for the Solar System. The masses of all the terrestrial planets are five or more orders of magnitude smaller than the mass of the Sun. For these planets, $a^3/T^2$ should therefore be to a high accuracy the same and equal to $Gm_{\astrosun}/(4\pi^2)$. On the other hand, Jupiter, the heaviest of the planets, has mass $m_{\jupiter}\approx1.90\times10^{27}\,\text{kg}$. For Jupiter, the ratio $a^3/T^2$ should accordingly be larger than for the Earth by the factor
\begin{equation}
\frac{m_{\astrosun}+m_{\jupiter}}{m_{\astrosun}+m_{\earth}}\approx1+\frac{m_{\jupiter}}{m_{\astrosun}}\approx1+9.5\times10^{-4}\;.
\end{equation}
This is in an excellent agreement with experimental data. As a side remark, it was known already to Kepler that his third law also applies (with high accuracy) to the four \emph{Galilean moons}: Io, Europa, Ganymede and Callisto. Nowadays, the validity of Kepler's third law is experimentally verified even for some exoplanets, for instance for the seven known planets orbiting around the TRAPPIST-1 red dwarf. Check for yourself using the data on the \href{https://en.wikipedia.org/wiki/TRAPPIST-1}{Wikipedia page!}
\end{illustration}

\begin{figure}[t]
\sidecaption[t]
\includegraphics[width=2.9in]{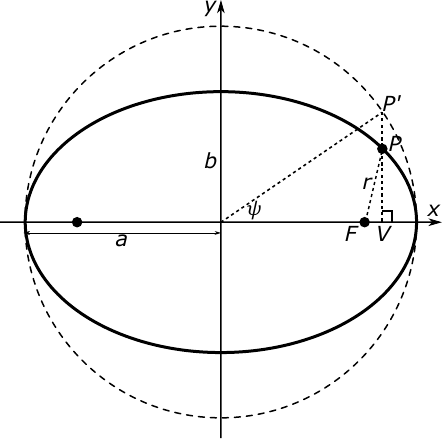}
\caption{Geometric construction of the eccentric anomaly $\psi$. A vertical cord $VP$ passing through a point $P$ on the ellipse is extended until it intersects the large circle of radius $a$ centered at the origin, at a new point $P'$. Using the fact that $\abs{VP}/\abs{VP'}=b/a$, the eccentric anomaly $\psi$ is related to the polar angle $\vp$ and radial distance $r$ defined at the focus $F$ by $b\sin\psi=r\sin\vp$.}
\label{fig02:ellipse2}
\end{figure}

I will close the discussion of elliptic orbits by sketching an alternative approach to the time dependence of the motion along the orbit that does not require integration, unlike the method outlined in Sect.~\ref{subsec:effpot}. The first step is to introduce another angular variable that parameterizes the ellipse, called \emph{eccentric anomaly}, $\psi$. This is measured at the center of the ellipse rather than the focus, as shown in Fig.~\ref{fig02:ellipse2}. From the geometric construction, it follows that the eccentric anomaly parameterizes the Cartesian coordinates of a point $P$ on the ellipse through $x=a\cos\psi$, $y=b\sin\psi$. This makes it easy to compute the radial distance from the center of force,
\begin{equation}
r=\sqrt{(x-c)^2+y^2}=a(1-e\cos\psi)\;,
\label{ch02:rfrompsi}
\end{equation}
where I used that $b=a\sqrt{1-e^2}$ and $c=ea$ to simplify the expression under the square root. The polar angle $\vp$ can be recovered from the eccentric anomaly through
\begin{equation}
r\sin\vp=b\sin\psi\;.
\label{ch02:vpfrompsi}
\end{equation}
This follows directly from Fig.~\ref{fig02:ellipse2}, where $\abs{VP}=r\sin\vp$, $\abs{VP'}=a\sin\psi$, and finally $\abs{VP'}/\abs{VP}=b/a$, since the ellipse can be thought of as a circle vertically rescaled by $b/a$.

To get access to the time dependence of the motion, we take the time derivative of the basic equation for the orbit~\eqref{ch02:Keplerorbitgeneral}. This gives, upon some manipulation,
\begin{equation}
ae\dot\psi\sin\psi\overset{\eqref{ch02:rfrompsi}}{=}\dot r\overset{\eqref{ch02:Keplerorbitgeneral}}{=}\frac{k\m e}{J^2}r^2\dot\vp\sin\vp\overset{\eqref{ch02:Jdef}}{=}\frac{ke}{J}\sin\vp\overset{\eqref{ch02:vpfrompsi}}{=}\frac{keb}{Jr}\sin\psi\;.
\end{equation}
Canceling common factors on the two sides of the equation, we end up with $\dot\psi=kb/(Jar)$. To simplify this further, we note that
\begin{equation}
\frac{kb}{Ja^2}\overset{\eqref{ch02:areavelocity}}{=}\frac{kb}{2\m\dot Aa^2}\overset{\eqref{ch02:3rdKepler}}{=}\frac{2\pi^2ab}{\dot AT^2}=\frac{2\pi}{T}\;.
\end{equation}
Putting all the pieces together, we find that $\dot\psi(1-e\cos\psi)=2\pi/T$. This can be immediately integrated. Choosing an integration constant so that at time $t=0$, we have $\psi=\vp=0$, the solution is
\begin{equation}
\boxed{\psi-e\sin\psi=\frac{2\pi t}{T}\;,}
\label{ch02:Keplereq}
\end{equation}
known as \emph{Kepler's equation}. This determines $\psi$ implicitly as a function of time. Once the eccentric anomaly is known, the time dependence of the polar coordinates $r,\vp$ can be recovered using the relations
\begin{equation}
r=a(1-e\cos\psi)\;,\qquad
\frac1{1-e\cos\psi}=\frac{1+e\cos\vp}{1-e^2}\;.
\end{equation}

%%%%%%%%%%%%%%%%%%%%%%%%%%%%%%%%%%%%%%%%%%%%%%%%%%%%%%%%%%%%

\subsection{Hyperbolic Orbits}

\begin{figure}[t]
\sidecaption[t]
\includegraphics[width=2.9in]{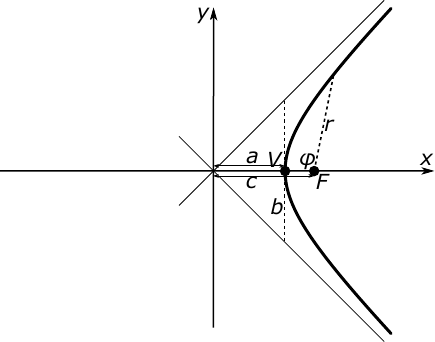}
\caption{The geometry of a hyperbola. Just like for the ellipse in Fig.~\ref{fig02:ellipse}, there are two focal points, located on the $x$-axis at $x=\pm c$. The picture only shows one of them, $F$, and one of the two branches of the hyperbola. The semi-major axis $a$ measures the distance from the origin to the vertex of the hyperbola, $V$. The polar angle $\vp$ is defined so that the vertex of the hyperbola corresponds to $\vp=0$.}
\label{fig02:hyperbola}
\end{figure}

I will now briefly describe the changes necessary to account for the hyperbolic orbits. For the sake of concreteness, I will only consider the case of an attractive potential ($k>0$), where the orbit coincides with the branch of the hyperbola, adjacent to the center of force. For repulsive forces ($k<0$), we would have to deal with the other branch of the hyperbola. With this qualification, the rest of the mathematical setup is almost identical to that for elliptic orbits in Sect.~\ref{subsec:ellipticorbits}.

There are two foci, placed on the $x$-axis at $x=\pm c$. Geometrically, the hyperbola is defined as the set of points with fixed \emph{difference} of distances from the foci,
\begin{equation}
\Bigl\lvert\sqrt{(x+c)^2+y^2}-\sqrt{(x-c)^2+y^2}\Bigr\rvert=2a\;.
\end{equation}
Upon removing the square roots, this can be rewritten in the more familiar form
\begin{equation}
\frac{x^2}{a^2}-\frac{y^2}{b^2}=1\;,
\label{ch02:hyperbolaCartesian}
\end{equation}
where now $b\equiv\sqrt{c^2-a^2}$; see Fig.~\ref{fig02:hyperbola} for a geometric interpretation of the various parameters. The next step is to introduce polar coordinates with origin centered at the focus $F$. We choose to measure the angle $\vp$ with respect to the $x$-axis so that the vertex $V$ of the hyperbola corresponds to $\vp=0$. This establishes a connection between the Cartesian and polar descriptions of the hyperbola through $x=c-r\cos\vp$, $y=r\sin\vp$. It is now a simple exercise to convert the Cartesian equation~\eqref{ch02:hyperbolaCartesian} to
\begin{equation}
\frac ar=\frac{1+e\cos\vp}{e^2-1}\;.
\label{ch02:hyperbolapolar}
\end{equation}
The eccentricity is defined, as before, by $e\equiv c/a$, and is now therefore greater than one. Our polar parameterization~\eqref{ch02:hyperbolapolar} of the hyperbola precisely matches the orbit equation~\eqref{ch02:Keplerorbitgeneral}. Comparing the coefficients, we find that the semi-major axis $a$ only depends on the total energy $E$ through
\begin{equation}
a=\frac{k}{2E}\;.
\end{equation}
This is the correct modification of~\eqref{ch02:ellipseafromE}, appropriate for hyperbolic trajectories.

%%%%%%%%%%%%%%%%%%%%%%%%%%%%%%%%%%%%%%%%%%%%%%%%%%%%%%%%%%%%

\subsection{Laplace--Runge--Lenz Vector}

There is a very useful heuristics in physics whereby simple and beautiful results do not appear out of the blue, but usually have a deeper underlying reason. You may have wondered why the shapes of Keplerian orbits exactly correspond to conic sections. In a randomly chosen potential, the orbits would not only not be simple, but they would not even be closed.\footnote{According to \href{https://en.wikipedia.org/wiki/Bertrand's_theorem}{Bertrand's theorem}, the only central potentials such that all the bound orbits are closed curves, are $V(r)\propto-1/r$ and $V(r)\propto r^2$. Both of these play a prominent role in physics.} I have a good news for you. There is an elegant way to derive the orbit equation~\eqref{ch02:Keplerorbitgeneral} that does not require any integration, or solution of a differential equation to that matter.

The starting point is the observation that the Kepler problem with $V(r)=-k/r$ possesses an exotic vector-valued constant of motion, called \emph{Laplace--Runge--Lenz} (LRL) \emph{vector},
\begin{equation}
\boxed{\vec R\equiv\vekt pJ-k\m\frac{\vec r}{r}\;,}
\label{ch02:Runge}
\end{equation}
where $\vec p\equiv\m\vec r$ is the momentum of the relative motion and $\vec J\equiv\vekt rp$ the corresponding angular momentum. The proof that $\vec R$ is indeed conserved requires some manipulation, but is fairly straightforward,
\begin{equation}
\begin{split}
\dot{\vec R}&=\dot{\vec p}\times\vec J+\vec p\times\xcancel{\dot{\vec J}}-k\m\left(\frac{\dot{\vec r}}{r}-\frac{\dot r\vec r}{r^2}\right)=\vekt FJ-\frac{k\m}r\dot{\vec r}+\frac{k\m}{r^3}\vec r\underbrace{r\dot r}_{=\vec r\cdot\dot{\vec r}}\\
&=-\frac{k\m}{r^3}\vec r\times(\vec r\times\dot{\vec r})-\frac{k\m}r\dot{\vec r}+\frac{k\m}{r^3}\vec r\vec r\cdot\dot{\vec r}=\vec0\;.
\end{split}
\end{equation}
In the intermediate steps, I used the fact that the angular momentum is conserved, the expression $\vec F=-k\vec r/r^3$ for the force due to the potential $V(r)=-k/r$, and finally the standard formula for the vector triple product of three vectors.

To see what the LRL vector is good for, take the scalar product of~\eqref{ch02:Runge} with $\vec r$ and use the cyclic property of the scalar triple product, $\vec r\cdot(\vekt pJ)=\vec J\cdot(\vekt rp)=\vec J^2$. Dividing by $J^2r$ then leads to
\begin{equation}
\frac1r=\frac{k\m}{J^2}\left(1+\frac{R}{k\m}\cos\vp\right)\;,
\end{equation}
where $R\equiv\abs{\vec R}$ and $\vp$ is the angle between $\vec r$ and $\vec R$. This is nothing but~\eqref{ch02:Keplerorbitgeneral}, where the eccentricity is now given in terms of the magnitude of the LRL vector, $e=R/(k\m)$. To see the connection to our previous relation between eccentricity and energy, just take the square of~\eqref{ch02:Runge},
\begin{align}
(k\m e)^2&=\vec R^2=(\vekt pJ)^2-\frac{2k\m}{r}\vec r\cdot(\vekt pJ)+k^2\m^2\\
\notag
&=\vec p^2\vec J^2-\frac{2k\m J^2}{r}+k^2\m^2=2\m J^2\left(\frac{\vec p^2}{2\m}-\frac kr\right)+k^2\m^2=2\m EJ^2+k^2\m^2\;.
\end{align}
This is equivalent to~\eqref{ch02:eccentricity}.

\begin{watchout}%
The utility of the LRL vector does not end here. First, you will meet it again in the exercise problems. Perhaps even more importantly, the LRL vector can be promoted to an operator and used in quantum mechanics. The gravitational Kepler problem is of course hardly relevant there. However, the electrostatic Coulomb force, which dominates the physics of atoms and molecules, satisfies the same inverse-square law as Newtonian gravity. The presence of an additional vector constant of motion allows, among others, for an exact, purely algebraic solution of the quantum-mechanical hydrogen atom problem.
\end{watchout}

%%%%%%%%%%%%%%%%%%%%%%%%%%%%%%%%%%%%%%%%%%%%%%%%%%%%%%%%%%%%

\section*{\probsec}
\addcontentsline{toc}{section}{\probsec}

\begin{prob}
\label{pr02:shaft}
Imagine that you drill a straight vertical shaft passing through the center of the Earth and surfacing on the opposite side. If you then throw a ball of mass $m$ into the shaft, how long will it take the ball to reach the antipodal point on the surface? You can assume that the Earth is a perfect sphere of radius $R$ and constant density $\vr$, and that the effects of air friction and Earth's rotation can be neglected. Find the numerical value of your result.
\end{prob}

\begin{prob}
\label{pr02:2body}
Show that the relative motion can still be separated from the motion of the center of mass if we add to the two-body Lagrangian~\eqref{ch02:Lag2body} the potential energy due to an external uniform gravitational field, and at the same time the harmonic potential $(1/2)m_1\o^2\vec r_1^2+(1/2)m_2\o^2\vec r_2^2$.
\end{prob}

\begin{prob}
\label{pr02:hydrogenBfield}
Using the result of~\refpr{pr01:particleinEMfield}, write down the Lagrangian for a system of two particles with masses $m_1,m_2$ and electric charges $q_1,q_2$ in a uniform external magnetic field $\vec B$. Do not forget the Coulomb interaction between the particles. What condition should the electric charges satisfy so that you can still separate the relative and center-of-mass motion following the approach of Sect.~\ref{sec:2bodyreduction}? Use the Poincar\'e gauge where $\vec A(\vec r)=(1/2)\vekt Br$.
\end{prob}

\begin{prob}
\label{pr02:effpot}
Sketch the effective potential~\eqref{ch02:effpot} for
\begin{equation}
V(r)=-\frac kr-\frac\ell{r^3}\;,
\end{equation}
where both $k$ and $\ell$ are positive. Discuss qualitatively the various types of orbits present in the system. How does your answer depend on the magnitude of $\ell$?
\end{prob}

\begin{prob}
\label{pr02:perturbation}
The orbits of the planets in the Solar System are not perfectly elliptic due to their mutual gravitational interactions. As a simple model of such a perturbation, consider the potential
\begin{equation}
V(r)=-\frac kr+\frac\eps{r^2}\;,
\end{equation}
where the parameter $\eps$ is assumed to be small. Use Binet's equation to analyze the effect of the perturbation. Show that it can be interpreted in terms of slow precession of the elliptic orbit, and find the angle of precession per period. How does your result depend on the sign of $\eps$?
\end{prob}

\begin{prob}
\label{pr02:periapo}
Use the effective potential to calculate the distance of a particle on an elliptic Keplerian orbit from the center of force at the pericenter and apocenter. Check your result using the geometric properties of the ellipse, reviewed in Sect.~\ref{sec:Keplerproblem}.
\end{prob}

\begin{prob}
\label{pr02:Halley}
Halley's comet orbits around the Sun with a period of about $75\,\text{yr}$. At perihelion, the comet reaches as near as $89$ million kilometers to the Sun. Find the eccentricity of the orbit and the speed of the comet at perihelion.
\end{prob}

\begin{prob}
\label{pr02:collapse}
Consider a uniform spherical cloud of cosmic dust of radius $R$ and total mass $M$. While initially at rest, the cloud starts shrinking as a result of its own gravity and eventually collapses to a point. Find the time it takes for the cloud to collapse. Hint: since the gravitational field of a spherically symmetric mass distribution is equal to that of a point charge, the problem is equivalent to finding the time of a free fall of a test particle from distance $R$ to a pointlike center of force of mass $M$. This latter problem can be solved using Kepler's laws.
\end{prob}

\begin{prob}
\label{pr02:hodograph}
Use the Laplace--Runge--Lenz vector to show that as a particle traverses an elliptic Keplerian orbit, its momentum vector traces out a circle. Find the center of this circle in momentum space and its radius. Hint: work out the cross product $\vekt JR$ using the definition~\eqref{ch02:Runge}.
\end{prob}
\chapter{Hamiltonian Mechanics}
\label{chap:Hammechanics}

\keywords{Generalized momenta, phase space, Hamiltonian, Hamilton equations, phase portrait, Legendre transform, cyclic coordinates, Routhian and Routh equations.}

%%%%%%%%%%%%%%%%%%%%%%%%%%%%%%%%%%%%%%%%%%%%%%%%%%%%%%%%%%%%

\noindent Having demonstrated the power of analytical mechanics by using it to solve the Kepler problem, we shall now make further progress on the theoretical frontier. We already have enough experience to appreciate the fact that first-order \emph{ordinary differential equations} (ODEs) are much easier to solve than second-order ones. This is after all what makes conservation laws so useful: they allow us to reduce the order of the equations of motion from second to first. In this~\chaptername, we will develop an alternative approach to mechanics due to Hamilton, where all the equations of motion are first-order by construction. As we will see, this requires introducing new dynamical variables. A second-order Lagrangian system with $n$ independent generalized coordinates requires $2n$ initial conditions to determine the solution to the equations of motion uniquely. The same freedom to choose a solution can be maintained with $2n$ first-order equations of motion. The price to pay for working solely with first-order ODEs is therefore the doubling of the number of dynamical variables (and equations).

For most practical purposes, the Lagrangian formalism remains the first method of choice due to its conceptual simplicity. However, the Hamiltonian formalism is more powerful in several regards. Dynamical systems of first-order ODEs have been studied in great detail. In case of interest, you will find an introduction in~\cite{Percival1982}. In addition, Hamiltonian mechanics is conceptually more fundamental in that it requires one to introduce less geometric structure than Lagrangian mechanics; I will touch upon this point briefly in~\chaptername~\ref{chap:geometryclassmech}. Finally and most importantly for our course, Hamiltonian mechanics offers deeper insight into the nature of conservation laws and makes it easier to exploit their existence for a reduction of the system of equations of motion.

%%%%%%%%%%%%%%%%%%%%%%%%%%%%%%%%%%%%%%%%%%%%%%%%%%%%%%%%%%%%

\section{Basic Setup}

We start with a brief reminder of the Lagrangian formalism. This is based on the action functional,
\begin{equation}
S[q]\equiv\int_{t_1}^{t_2}\D t\,L(q(t),\dot q(t),t)\;,
\label{ch03:actionLag}
\end{equation}
where $q_i$ is a set of $n$ generalized coordinates, parameterizing the configuration of the system. For most mechanical systems, the Lagrange function $L$ equals the difference of the system's kinetic and potential energy. Regardless of whether this is actually the case or not, the physical trajectory of the system is a stationary point of $S[q]$ in the space of trajectories with fixed boundary conditions, $q_i=q_{i1}$ at $t=t_1$ and $q_i=q_{i2}$ at $t=t_2$. This implies the set of Lagrange equations,
\begin{equation}
\frac{\udelta S}{\udelta q_i}=\PD{L}{q_i}-\OD{}t\PD{L}{\dot q_i}=0\;.
\end{equation}

%%%%%%%%%%%%%%%%%%%%%%%%%%%%%%%%%%%%%%%%%%%%%%%%%%%%%%%%%%%%

\subsection{Hamiltonian Form of the Variational Principle}

Hamilton's formulation of mechanics requires two sets of dynamical variables: the already familiar generalized coordinates $q_i$, and the same number $n$ of \emph{generalized momenta} $p_i$. Together, these completely specify the instantaneous kinematical state of the system. This is in contrast to the generalized coordinates alone, which only determine the static configuration of the system. It follows that the generalized momenta carry information about the velocity of motion. The set of all kinematical states of the system, parameterized jointly by $q_i$ and $p_i$, is called the \emph{phase space}.

\begin{illustration}%
\label{ex03:phasespace}%
We know from~\refex{ex01:kinematics} that the configuration space of the system of $N$ particles in an $n$-dimensional Euclidean space is $R^{N\times n}$. Since we need a generalized momentum for each of the generalized coordinates, the phase space of this system is accordingly $R^{2(N\times n)}$.

In order that you do not get the impression that the phase space is always some Euclidean space, let me mention a couple of less trivial examples. The motion of a simple pendulum (Sect.~\ref{subsec:pendulum}) is restricted to a circle. Mathematically, the circle, $S^1$, therefore constitutes the configuration space of the pendulum. The (generalized) velocity of the pendulum, on the other hand, is not constrained and can take arbitrary values. Thus, the phase space of the pendulum is $S^1\times\R$, that is a cylinder. Similarly, the configuration space of the spherical pendulum introduced in~\refpr{pr01:sphericalpendulum} is $S^2$ and the phase space is accordingly $S^2\times\R^2$.
\end{illustration}

Next, we introduce a functional that plays exactly the same role for the Hamiltonian for\-ma\-lism as the action~\eqref{ch03:actionLag} in the Lagrangian formalism,
\begin{equation}
\boxed{S[q,p]\equiv\int_{t_1}^{t_2}\D t\,\biggl[\sum_i\dot q_i(t)p_i(t)-H(q(t),p(t),t)\biggr]\;.}
\label{ch03:actionHam}
\end{equation}
Somewhat unfortunately, this is also called \emph{action}, and is only distinguished from \eqref{ch03:actionLag} by explicitly indicating the arguments. The \emph{Hamilton function} (or simply \emph{Hamiltonian}) $H(q,p,t)$ is a function of the generalized coordinates and momenta, possibly also depending explicitly on time.

\begin{watchout}%
This is another moment where a mathematician becomes desperate. Why do we physicists use the same symbol for the two apparently unrelated functionals~\eqref{ch03:actionLag} and~\eqref{ch03:actionHam}? What does the function $H(q,p,t)$ in~\eqref{ch03:actionHam} have to do with the previously defined Hamiltonian~\eqref{ch01:Hamiltonian}, which is a function of the generalized coordinates and velocities, not momenta? Although I did use the analogy with the Lagrangian formalism to set up the notion of phase space, it helps if you clear your mind and think of~\eqref{ch03:actionHam} as a mere definition for the time being. We will work our way backwards and understand the precise relationship between the Lagrange and Hamilton formulations of mechanics in Sect.~\ref{subsec:relationLagHam}.
\end{watchout}

We are now in the position to formulate the basic dynamical principle of Hamiltonian mechanics. Here you go. The actual, physical trajectory of the system in the phase space is a stationary point of the action $S[q,p]$. (Are you surprised?) It is required that the allowed trajectories satisfy a fixed boundary condition,
\begin{equation}
q_i(t_1)=q_{i1}\;,\qquad
q_i(t_2)=q_{i2}\;.
\end{equation}
No boundary condition is necessary for $p_i$. Try to understand why!

%%%%%%%%%%%%%%%%%%%%%%%%%%%%%%%%%%%%%%%%%%%%%%%%%%%%%%%%%%%%

\subsection{Hamilton Equations of Motion}

Having formulated the variational principle for Hamiltonian mechanics, we infer at once that the dynamical equations of motion take the form $\udelta S/\udelta q_i(t)=\udelta S/\udelta p_i(t)=0$. Using the explicit form~\eqref{ch03:actionHam} of the action, this translates to the \emph{Hamilton equations},
\begin{equation}
\boxed{\dot q_i=\PD{H}{p_i}\;,\qquad
\dot p_i=-\PD{H}{q_i}\;.}
\label{ch03:Hamiltoneq}
\end{equation}
The Hamilton equations constitute a set of $2n$ coupled first-order ODEs for the $2n$ dynamical variables $q_i,p_i$. Their solution is therefore uniquely determined by an initial condition whereby the values of all the $q_i,p_i$ at a chosen time point are specified. In other words, the evolution of the system is determined by specifying a single point: its initial position in the phase space. This allows for a neat graphical representation of the trajectories of the system as curves in the phase space. Take a little break to ponder how this differs from the Lagrangian formalism, where merely drawing the trajectory as a curve in the configuration space does not convey any information about the velocity of the system.

\begin{figure}[t]
\sidecaption[t]
\includegraphics[width=2.9in]{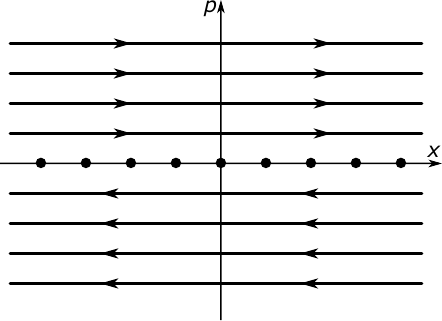}
\caption{Phase portrait of a free particle. The oriented solid lines represent possible trajectories of the particle, moving respectively to the left or right. The dots indicate degenerate trajectories where the particle is at rest.}
\label{fig03:portraitfree}
\end{figure}

\begin{illustration}%
\label{ex03:Ham1d}%
To illustrate the above general setup on an example that is as simple as possible, think of a particle of mass $m$ moving in one dimension. The configuration space is $\R$, and the phase space $\R^2$. Using the Cartesian coordinate $x$ as the sole generalized coordinate, I will call the corresponding generalized momentum simply $p$. Denoting the potential energy of the particle with respect to a force acting on it as $V(x)$, it is easy to guess the Hamiltonian,
\begin{equation}
H(x,p)=\frac{p^2}{2m}+V(x)\;.
\label{ch03:exHam1d}
\end{equation}
The corresponding Hamilton equations are
\begin{equation}
\dot x=\PD{H}{p}=\frac pm\;,\qquad
\dot p=-\PD{H}{x}=-V'(x)\;.
\label{ch03:exHamEoM}
\end{equation}
The first of these is just the usual relation between momentum and velocity, whereas the latter reproduces Newton's second law. In the special case of a free particle, $V(x)=0$, some typical trajectories are sketched graphically in Fig.~\ref{fig03:portraitfree}. This type of diagram is known as the \emph{phase portrait}. The first of the Hamilton equations can also be used to eliminate the momentum $p$ from the second equation. Mathematically, this amounts to a conversion of a pair of first-order ODEs to a single second-order ODE. We thus recover the Lagrange equation, $m\ddot x=-V'(x)$.
\end{illustration}

\begin{figure}[t]
\sidecaption[t]
\includegraphics[width=2.9in]{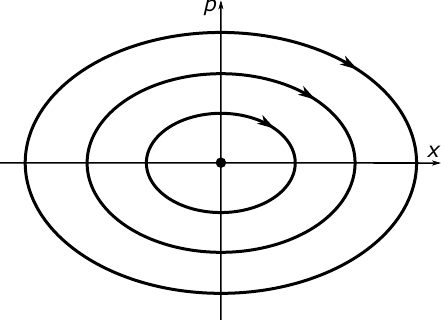}
\caption{Phase portrait of a linear harmonic oscillator. There is one degenerate trajectory, indicated by the dot, corresponding to a particle at rest. All the other trajectories are oriented ellipses with the semi-axis $\sqrt{2E/(m\o^2)}$ in the $x$-direction and $\sqrt{2mE}$ in the $p$-direction, where $E$ is the total energy of the oscillator.}
\label{fig03:portraitoscillator}
\end{figure}

\begin{illustration}%
For a somewhat less trivial illustration, take the linear harmonic oscillator. For a particle with mass $m$, this amounts to choosing the potential as $V(x)=(1/2)m\o^2x^2$, where $\o$ is the characteristic frequency of the oscillator. The full Hamiltonian is still given by~\eqref{ch03:exHam1d}. The Hamilton equations~\eqref{ch03:exHamEoM} now become
\begin{equation}
\dot x=\frac pm\;,\qquad
\dot p=-m\o^2x\;.
\end{equation}
Eliminating the momentum $p$ recovers the standard oscillator equation, $\ddot x=\dot p/m=-\o^2x$. Jumping a little ahead, it is easy to prove using the Hamilton equations~\eqref{ch03:exHamEoM} that the Hamiltonian~\eqref{ch03:exHam1d} is a constant of motion. In the harmonic oscillator case, this guarantees that the trajectories in the phase space are ellipses defined by setting $p^2/(2m)+(1/2)m\o^2x^2$ equal to a constant. The phase portrait of a linear harmonic oscillator is sketched in Fig.~\ref{fig03:portraitoscillator}.
\end{illustration}

%%%%%%%%%%%%%%%%%%%%%%%%%%%%%%%%%%%%%%%%%%%%%%%%%%%%%%%%%%%%

\subsection{Relation Between Lagrangian and Hamiltonian Formalisms}
\label{subsec:relationLagHam}

We saw in the examples worked out above that the generalized momentum $p$ could be algebraically eliminated in favor of the generalized velocity $\dot x$ using one of the Hamilton equations. This can be done fairly generally under moderate technical assumptions. To see what the conditions might be, focus on the first of the Hamilton equations~\eqref{ch03:Hamiltoneq}, $\dot q_i=\Pd{H}{p_i}$. The right-hand side of this is a function of the generalized coordinates and momenta. By the implicit function theorem, the equation therefore defines the generalized momenta in terms of coordinates and velocities,
\begin{equation}
p_i=p_i(q,\dot q,t)\;,
\label{ch03:pfromqdot}
\end{equation}
provided the set of functions $\Pd{H}{p_i}$ has an invertible Jacobian with respect to the variables $p_j$. This translates into the condition that the \emph{Hessian} of the Hamilton function with respect to its momentum variables, $\de^2H/\de p_i\de p_j$, is invertible. I will assume that this is the case whenever we use the Hamiltonian formalism.

With the formalities out of the way, we can eliminate the generalized momenta using~\eqref{ch03:pfromqdot}. However, instead of doing so in the Hamilton equations, we will remove the momenta as the variables of the action. This turns~\eqref{ch03:actionHam} into
\begin{equation}
S[q]=\int_{t_1}^{t_2}\D t\,\biggl[\sum_i\dot q_ip_i(q,\dot q,t)-H(q,p(q,\dot q,t),t)\biggr]\;.
\label{ch03:actionLagfromHam}
\end{equation}
Our new functional only depends on the generalized coordinates as variables. Moreover, the integrand is a function of the coordinates $q_i$, velocities $\dot q_i$, and possibly of time. Should the predictions of the Lagrangian and Hamiltonian formalisms coincide, \eqref{ch03:actionLagfromHam} must agree with the Lagrangian action~\eqref{ch03:actionLag}. This gives an unambiguous algorithm how to switch from the Hamiltonian to the Lagrangian.

\begin{watchout}%
\vspace{-2.5ex}
\runinhead{From the Hamiltonian to the Lagrangian} Solve the equations $\dot q_i=\Pd{H}{p_i}$ algebraically for $p_i$ in terms of the generalized coordinates and velocities. This defines a set of functions $p_i(q,\dot q,t)$. Inserting these in the Hamiltonian $H(q,p,t)$, the corresponding equivalent Lagrangian equals
\begin{equation}
L(q,\dot q,t)=\sum_i\dot q_ip_i(q,\dot q,t)-H(q,p(q,\dot q,t),t)\;.
\label{ch03:LagfromHam}
\end{equation}
This conversion of the Hamiltonian to the Lagrangian as well as its reverse detailed below, which gives the Hamiltonian from the known Lagrangian, is an example of a \emph{Legendre transformation}. You might have met it previously in thermodynamics.
\end{watchout}

\begin{illustration}%
Recall the class of Hamiltonians~\eqref{ch03:exHam1d} describing a particle in one dimension, introduced in~\refex{ex03:Ham1d}. In this case, the conversion of momentum to velocity is trivial, $p=m\dot x$. Inserting this in~\eqref{ch03:LagfromHam}, we find the corresponding Lagrangian,
\begin{equation}
L=m\dot x^2-\biggl[\frac{(m\dot x)^2}{2m}+V(x)\biggr]=\frac12m\dot x^2-V(x)\;,
\end{equation}
which agrees with our expectation.
\end{illustration}

This was the easier bit. Unfortunately, in practice it is more useful to be able to derive the Hamiltonian starting from a known Lagrangian. This is because of the simpler structure of the Lagrangian, which only depends on a single set of dynamical variables. In general, it is not easy to guess the Hamiltonian as we did in~\refex{ex03:Ham1d}, since the generalized momentum may not have a simple physical interpretation one could relate to the configuration of the system. Instead of eliminating the generalized momenta by using the first half of the Hamilton equations~\eqref{ch03:Hamiltoneq}, we therefore need to find a way to introduce momentum variables in the Lagrangian formalism.

To that end, recall that we already did once define a set of momenta starting from a Lagrangian; cf.~Sect.~\ref{sec01:conservationlaws}. While the discussion therein was limited to the context of cyclic coordinates and the associated conservation laws, let us see whether~\eqref{ch01:conjugatemomentum} could be a viable definition of generalized momentum, applicable to any Lagrangian. To check this, we take the Lagrangian~\eqref{ch03:LagfromHam} and find $\Pd{L}{\dot q_i}$ using the chain rule,
\begin{equation}
\PD{L}{\dot q_i}=p_i+\sum_j\dot q_j\PD{p_j}{\dot q_i}-\sum_j\PD{H}{p_j}\PD{p_j}{\dot q_i}=p_i+\sum_j\left(\dot q_j-\PD{H}{p_j}\right)\PD{p_j}{\dot q_i}\;.
\end{equation}
The expression in the parentheses vanishes by means of the first of the Hamilton equations~\eqref{ch03:Hamiltoneq}. This verifies that setting $\Pd{L}{\dot q_i}$ equal to $p_i$ provides the desired inverse of~\eqref{ch03:pfromqdot} that allows us to switch back from generalized velocities to generalized momenta. Let us again put together the basic algorithm for translating from the Lagrangian to the Hamiltonian.

\begin{watchout}%
\vspace{-2.5ex}
\runinhead{From the Lagrangian to the Hamiltonian} Solve the equation $p_i=\Pd{L}{\dot q_i}$ algebraically for $\dot q_i$ in terms of the generalized coordinates and momenta. Mathematically, a unique solution exists provided the Hessian $\de^2L/\de\dot q_i\de\dot q_j$ is invertible. This defines a set of functions $\dot q_i(q,p,t)$. Inserting these in the Lagrangian $L(q,\dot q,t)$, the corresponding equivalent Hamiltonian equals
\begin{equation}
H(q,p,t)=\sum_ip_i\dot q_i(q,p,t)-L(q,\dot q(q,p,t),t)\;.
\label{ch03:HamfromLag}
\end{equation}
This guarantees that the Lagrangian action~\eqref{ch03:actionLag}, once expressed in terms of $q_i$ and $p_i$ as independent dynamical variables, reproduces the Hamiltonian action~\eqref{ch03:actionHam}.
\end{watchout}

\begin{figure}[t]
\sidecaption[t]
\includegraphics[width=2.9in]{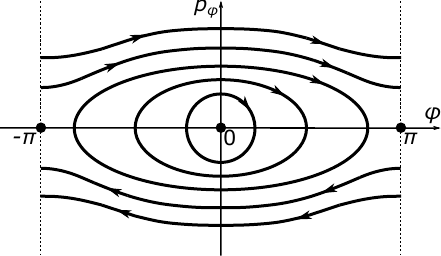}
\caption{Phase portrait of a simple pendulum. There are degenerate trajectories (dots) with $\vp=n\pi$ and any $n\in\Z$. In addition, there are closed elliptic-like trajectories, describing periodic oscillations of the pendulum. Finally, there are open trajectories, corresponding to a ``winding'' motion of the pendulum. The phase portrait is periodic in $\vp$ with period $2\pi$.}
\label{fig03:portraitpendulum}
\end{figure}

\begin{illustration}%
To have at least one nontrivial example of the Legendre transformation from the Lagrangian to the Hamiltonian, recall the simple pendulum, analyzed in Sect.~\ref{subsec:pendulum}. The Lagrangian is given by~\eqref{ch01:pendulumLag}. The sole generalized coordinate is the polar angle $\vp$, and the corresponding generalized momentum is thus $p_\vp=\Pd{L}{\dot\vp}=mL^2\dot\vp$. Inverting this to $\dot\vp=p_\vp/(mL^2)$, we get the Hamiltonian via~\eqref{ch03:HamfromLag},
\begin{equation}
H=p_\vp\frac{p_\vp}{mL^2}-\left[\frac12mL^2\left(\frac{p_\vp}{mL^2}\right)^2+mgL\cos\vp\right]=\frac{p_\vp^2}{2mL^2}-mgL\cos\vp\;.
\end{equation}
The corresponding Hamilton equations are
\begin{equation}
\dot\vp=\PD{H}{p_\vp}=\frac{p_\vp}{mL^2}\;,\qquad
\dot p_\vp=-\PD{H}{\vp}=-mgL\sin\vp\;.
\end{equation}
Of course, by eliminating the generalized momentum, we recover the basic pendulum equation~\eqref{ch01:pendulumEoM}. However, here I would like to use both of the Hamilton equations directly to illustrate the phase portrait of the pendulum; see Fig.~\ref{fig03:portraitpendulum} for a sketch. Unlike the phase portraits of the free particle in Fig.~\ref{fig03:portraitfree} and of the linear harmonic oscillator (Fig.~\ref{fig03:portraitoscillator}), it has a very nontrivial structure. This underlies the utility of a graphical representation of the solutions to the Hamilton equations. In the present case, we find two static trajectories, represented in the figure by dots. However, only of them is stable, namely that corresponding to the lowest possible position, $\vp=0$. The stability is readily observed in the picture: if we increase the energy of the pendulum slightly, it will merely move to one of the nearby quasi-elliptic trajectories, but will keep oscillating around the stable equilibrium. On the other hand, the $\vp=\pi$ solution is unstable. Indeed, giving the pendulum that sits at $\vp=\pi$ a little kick to increase its momentum, it will move to a nearby trajectory that is open and corresponds to perpetual ``winding'' around the circle rather than periodic oscillations. Keep in mind (\refex{ex03:phasespace}) that the phase space of the pendulum is $S^1\times\R$. It is therefore sufficient to draw the phase portrait on any interval of $\vp$ of length $2\pi$, as in Fig.~\ref{fig03:portraitpendulum}. It is however quite common to stretch the graph over several periods, which allows one to track individual open trajectories in terms of a continuously changing variable $\vp$.
\end{illustration}

%%%%%%%%%%%%%%%%%%%%%%%%%%%%%%%%%%%%%%%%%%%%%%%%%%%%%%%%%%%%

\section{Conservation Laws Revisited}

We still have not answered the question what the Hamiltonian entering the action~\eqref{ch03:actionHam} has to do with the eponymous function~\eqref{ch01:Hamiltonian}, defined in Sect.~\ref{sec01:conservationlaws} as the constant of motion representing the energy of the system. To that end, recall that the construction of the latter was based on the assumption that the Lagrangian does not depend explicitly on time. This translates in the Hamiltonian formalism to the assumption that the Hamiltonian, $H(q,p)$, does not depend explicitly on time. With the help of~\eqref{ch00:firstintenergy}, we find that the corresponding first integral of the Hamiltonian action~\eqref{ch03:actionHam} is, up to the overall sign,
\begin{equation}
\cancel{\sum_i\dot q_ip_i}-\biggl[\cancel{\sum_i\dot q_ip_i}-H(q,p)\biggr]=H(q,p)\;.
\end{equation}
The same result follows from the Lagrangian action~\eqref{ch03:actionLag} upon using the relation $\Pd{L}{\dot q_i}=p_i$ and the identification~\eqref{ch03:LagfromHam}. This verifies what you must have suspected all along: the Hamilton function introduced in Sect.~\ref{sec01:conservationlaws} and the Hamiltonian entering~\eqref{ch03:actionHam} indeed are one and the same object, once expressed in terms of appropriate variables. Should the Hamiltonian not be time-independent, as we assumed, we would find using directly the Hamilton equations~\eqref{ch03:Hamiltoneq} and the chain rule that
\begin{equation}
\OD Ht=\sum_i\biggl(\PD{H}{q_i}\underbrace{\dot q_i}_{=\Pd{H}{p_i}}+\PD{H}{p_i}\underbrace{\dot p_i}_{=-\Pd{H}{q_i}}\biggr)+\PD Ht=\PD Ht\;.
\end{equation}
When evaluated on solutions of the Hamilton equations, the variation of the Hamiltonian with time is therefore completely determined by its explicit time dependence as a function of three variables, $H(q,p,t)$. This is a classical predecessor of the quantum \href{https://en.wikipedia.org/wiki/Hellmann%E2%80%93Feynman_theorem}{Feynman--Hellmann theorem}.

%%%%%%%%%%%%%%%%%%%%%%%%%%%%%%%%%%%%%%%%%%%%%%%%%%%%%%%%%%%%

\subsection{Cyclic Coordinates}
\label{subsec:Hamcyclic}

In addition to the Hamiltonian expressing energy, we found in Sect.~\ref{sec01:conservationlaws} that whenever the Lagrangian does not depend explicitly on some generalized coordinate $q_i$, the corresponding conjugate momentum, defined by~\eqref{ch01:conjugatemomentum}, is a constant of motion. This feature is readily reproduced by the Hamiltonian formalism. Namely, it follows directly from the Hamilton equations~\eqref{ch03:Hamiltoneq} that whenever the Hamiltonian does not depend on one of the generalized coordinates, $q_i$, then $\dot p_i=-\Pd{H}{q_i}=0$. The corresponding generalized momentum is a constant of motion.

In spite of the close analogy, there is an important difference in how cyclic coordinates and the corresponding conservation laws feature in the equations of motion of the Lagrangian and Hamiltonian formalism. In the Lagrangian formalism, the natural dynamical variables are the generalized coordinates and velocities, neither of which is itself a constant of motion. All the presence of a cyclic coordinate does is therefore to reduce a system of $n$ second-order ODEs to $n-1$ second-order ODEs and one first-order ODE. The reduction requires a single integration constant, supplied by the value of the conserved conjugate momentum, $p_i$. On the other hand, the Hamiltonian formalism allows us to ignore the cyclic coordinate $q_i$ and the corresponding generalized momentum $p_i$ altogether, thus reducing the system of $2n$ first-order ODEs to $2n-2$ first-order ODEs for the remaining generalized coordinates and momenta.

Let us see a bit more explicitly how this works. Suppose that the Hamiltonian does not depend explicitly on one of the generalized coordinates, $q_j$. We can thus write it as $H=H(q_1,\dotsc,q_{j-1},q_{j+1},\dotsc,q_n,p_1,\dotsc,p_n,t)$. To remind us of the fact that $p_j$ is then guaranteed to be a constant of motion, we replace $p_j\to c$ and write the Hamiltonian instead as
\begin{equation}
H=H(q_1,\dotsc,q_{j-1},q_{j+1},\dotsc,q_n,p_1,\dotsc,p_{j-1},p_{j+1},\dotsc,p_n,c,t)\;.
\end{equation}
The Hamilton equations~\eqref{ch03:Hamiltoneq} applied to all $q_i,p_i$ with $i\neq j$ then constitute a set of coupled $2n-2$ first-order ODEs, where $c$ enters as a mere constant parameter. This is in principle the end of the line. Should we be interested in the time dependence of the generalized velocity $\dot q_j$, we can recover this \emph{after} having solved the Hamilton equations through
\begin{equation}
\dot q_j=\PD Hc\;,
\end{equation}
which is just the first of the Hamilton equations~\eqref{ch03:Hamiltoneq} applied to $q_j$.

\begin{illustration}%
\label{ex03:centralfield}%
For an illustration of the difference in how the Lagrangian and Hamiltonian formalisms deal with cyclic coordinates, recall the problem of motion of a particle in a central conservative field. We previously reduced this to the two-dimensional Lagrangian~\eqref{ch02:1particleL} in polar coordinates,
\begin{equation}
L=\frac12\m(\dot r^2+r^2\dot\vp^2)-V(r)\;.
\label{ch03:2bodyLagpolar}
\end{equation}
The corresponding second-order Lagrange equations are displayed in~\eqref{ch02:2bodyEoM}. However, one of these trivially reduces to a first-order ODE, $J=\m r^2\dot\vp$ with constant $J$, thanks to the fact that $\vp$ is a cyclic coordinate. We thus end up with a second-order \emph{equation of motion} (EoM) for $r$ and a first-order EoM for $\vp$, which are coupled to each other. To make further progress, we have to manually eliminate $\dot\vp$ by replacing $\dot\vp\to J/(\m r^2)$, which gives us a single second-order EoM for $r$ alone,
\begin{equation}
\m\ddot r=-V'(r)+\frac{J^2}{\m r^3}\;.
\label{ch03:2bodyradialEoM}
\end{equation}
For the concrete system at hand, this is a trivial step. It may however get involved for other, more nontrivial problems.

The Hamiltonian formalism makes the reduction automatic. First, the Hamiltonian corresponding to the Lagrangian~\eqref{ch03:2bodyLagpolar} reads
\begin{equation}
H=\frac{p_r^2}{2\m}+\frac{p_\vp^2}{2\m r^2}+V(r)\;,
\end{equation}
with obvious notation for the generalized momenta $p_r,p_\vp$. The fact that the angular variable $\vp$ is cyclic now trivializes two of the Hamilton equations,
\begin{equation}
\dot\vp=\frac{p_\vp}{\m r^2}\;,\qquad
\dot p_\vp=0\;.
\end{equation}
The generalized momentum $p_\vp$ is constant, being nothing but the angular momentum $J$. This leaves us with a mere pair of first-order ODEs for $r$ and $p_r$,
\begin{equation}
\dot r=\frac{p_r}{\m}\;,\qquad
\dot p_r=-V'(r)+\frac{p_\vp^2}{\m r^3}\;.
\end{equation}
These are completely decoupled from the angular motion. If desired, they can be converted into a single second-order EoM for $r$,
\begin{equation}
\m\ddot r=-V'(r)+\frac{p_\vp^2}{\m r^3}\;,
\label{ch03:2bodyradialEoMHamilton}
\end{equation}
equivalent to the radial EoM~\eqref{ch03:2bodyradialEoM} obtained from the Lagrangian formalism. Once this EoM is solved, we can recover the angular velocity of the orbital motion by simply substituting the solution $r(t)$ in the relation $\dot\vp=p_\vp/(\m r^2)$.
\end{illustration}

\begin{watchout}%
You can always eliminate a cyclic coordinate $q_i$ by expressing the velocity $\dot q_i$ in terms of the constant of motion $p_i$ and inserting the result in the remaining Lagrange equations, as we did in the above example. However, you should never do this at the level of the Lagrangian! I urge you to check that by inserting $\dot\vp=J/(\m r^2)$ back into~\eqref{ch03:2bodyLagpolar}, the resulting ``one-dimensional Lagrangian'' gives a wrong EoM for $r$, which differs from~\eqref{ch03:2bodyradialEoM} by the sign of the $J^2/(\m r^3)$ term. To make things even more confusing, note that what I am now warning you against looks very similar to what we did in Sect.~\ref{subsec:relationLagHam}. There, we used the variational EoM to solve for $p_i$ and inserted the result back into the action. What is going on here? Think this through carefully! Hint: recall that we are looking for stationary points of an action functional. This can be done sequentially, by solving a subset of the Euler--Lagrange equations (or even one equation at a time) and inserting the result back into the functional. That however requires the solution to be uniquely determined by the remaining dynamical variables, so that we get a well-defined reduced functional of the~latter.
\end{watchout}

It follows from our discussion that the Hamiltonian formalism is advantageous in the presence of cyclic coordinates. On the other hand, the Lagrangian formalism remains more elementary and compact otherwise. This dichotomy has motivated the development of \emph{Routhian mechanics}, which in a certain well-defined sense interpolates between the Lagrangian and Hamiltonian ones. I will conclude the~\chaptername{} with an outline of Routh's framework.

%%%%%%%%%%%%%%%%%%%%%%%%%%%%%%%%%%%%%%%%%%%%%%%%%%%%%%%%%%%%

\subsection{Routh's Formalism}

In Sect.~\ref{subsec:relationLagHam}, the Hamiltonian action~\eqref{ch03:actionHam} proved useful for understanding the precise relationship between the Lagrangian and Hamiltonian formulations of the same problem. We will now follow the same line of thought, merely expanding the freedom of choice between generalized velocities and momenta. Take the $n$ pairs of variables $q_i,p_i$ and split them into two groups, corresponding to $i=1,\dotsc,N$ and $i=N+1,\dotsc,n$ with $0\leq N\leq n$. Suppose we would like to keep the first $N$ pairs intact, but convert the remaining $n-N$ ones into a ``Lagrangian'' form. This would seem to be a smart thing to do in case all the $q_1,\dotsc,q_N$ are cyclic. However, what follows is entirely independent of such a motivation, and constitutes a general formulation of mechanics, alternative (or rather complementary) to that of Lagrange and Hamilton.

In order to make the distinction between the ``Hamilton-like'' and ``Lagrange-like'' variables explicit, I will attach to them the subscript ``H'' or ``L,'' respectively. In other words, I will identify
\begin{equation}
q_i\equiv q_{\mathrm{H}i}\quad\text{for }i=1,\dotsc,N\;,\qquad
q_i\equiv q_{\mathrm{L}i}\quad\text{for }i=N+1,\dotsc,n\;,
\end{equation}
and analogously for the generalized momenta $p_i$. We will now utilize the Hamilton equations $\dot q_{\mathrm Li}=\Pd{H}{p_{\mathrm Li}}$ to eliminate the generalized momenta $p_{\mathrm Li}$ in terms of the generalized coordinates and velocities,
\begin{equation}
p_{\mathrm Li}=p_{\mathrm Li}(q_\mathrm{H},q_\mathrm{L},\dot q_\mathrm{L},t)\;,\qquad
i=N+1,\dotsc,n\;.
\label{ch03:routhpL}
\end{equation}
Note how the notation helps us understand the technical details: the eliminated momenta may depend on all the generalized coordinates, but by construction only a subset of the generalized velocities. Upon inserting this into the action functional~\eqref{ch03:actionHam}, we can rewrite it as
\begin{equation}
\boxed{S[q_\mathrm{H},p_\mathrm{H},q_\mathrm{L}]=\int_{t_1}^{t_2}\D t\,\biggl[\sum_{i=1}^N\dot q_{\mathrm Hi}p_{\mathrm Hi}-R(q_\mathrm{H},p_\mathrm{H},q_\mathrm{L},\dot q_\mathrm{L},t)\biggr]\;,}
\label{ch03:actionRouth}
\end{equation}
where
\begin{equation}
\begin{split}
R(q_\mathrm{H},p_\mathrm{H},q_\mathrm{L},\dot q_\mathrm{L},t)\equiv{}& H(q_\mathrm{H},p_\mathrm{H},q_\mathrm{L},p_\mathrm{L}(q_\mathrm{H},q_\mathrm{L},\dot q_\mathrm{L},t),t)\\
&-\sum_{i=N+1}^n\dot q_{\mathrm Li}p_{\mathrm Li}(q_\mathrm{H},q_\mathrm{L},\dot q_\mathrm{L},t)\;,
\end{split}
\end{equation}
and $R$ is the \emph{Routh function} (or simply \emph{Routhian}). By construction, the Routhian action~\eqref{ch03:actionRouth} takes the same values as the original Hamiltonian action~\eqref{ch03:actionHam} upon the identification~\eqref{ch03:routhpL}. It is a simple exercise in variational calculus to derive the corresponding dynamical equations of motion, defined by $\udelta S/\udelta q_{\mathrm Hi}=\udelta S/\udelta p_{\mathrm Hi}=0$ for $i=1,\dotsc,N$ and $\udelta S/\udelta q_{\mathrm Li}=0$ for $i=N+1,\dotsc,n$. I recommend that you work out the details yourself. The final result is
\begin{align}
\label{ch03:RouthEoMHam}
\dot q_{\mathrm Hi}=\PD{R}{p_{\mathrm Hi}}\;,\qquad
\dot p_{\mathrm Hi}&=-\PD{R}{q_{\mathrm Hi}}\;,
&\text{for }i&=1,\dotsc,N\;,\\
\label{ch03:RouthEoMLag}
\OD{}t\PD{R}{\dot q_{\mathrm Li}}&=\PD{R}{q_{\mathrm Li}}\;,
&\text{for }i&=N+1,\dotsc n\;.
\end{align}
This is easy to remember: the equations~\eqref{ch03:RouthEoMHam} for the Hamilton-like variables are Hamilton-like, and those for the Lagrange-like variables~\eqref{ch03:RouthEoMLag} are Lagrange-like.

Taken together, these are equivalent to the Hamilton equations~\eqref{ch03:Hamiltoneq} for the action~\eqref{ch03:actionHam} we started with. First, $\Pd{R}{p_{\mathrm Hi}}=\Pd{H}{p_{\mathrm Hi}}$ since the generalized momenta $p_{\mathrm Hi}$ only enter the Routhian through the Hamiltonian. This verifies the first of the Hamilton-like equations~\eqref{ch03:RouthEoMHam}. The second follows by applying the chain rule to the definition of the Routhian,
\begin{equation}
\begin{split}
\PD{R}{q_{\mathrm Hi}}&=\PD{H}{q_{\mathrm Hi}}+\sum_{j=N+1}^n\PD{H}{p_{\mathrm Lj}}\PD{p_{\mathrm Lj}}{q_{\mathrm Hi}}-\sum_{j=N+1}^n\dot q_{\mathrm Lj}\PD{p_{\mathrm Lj}}{q_{\mathrm Hi}}\\
&=\PD{H}{q_{\mathrm Hi}}+\sum_{j=N+1}^n\biggl(\PD{H}{p_{\mathrm Lj}}-\dot q_{\mathrm Lj}\biggr)\PD{p_{\mathrm Lj}}{q_{\mathrm Hi}}\overset{\eqref{ch03:Hamiltoneq}}{=}\PD{H}{q_{\mathrm Hi}}\overset{\eqref{ch03:Hamiltoneq}}{=}-\dot p_{\mathrm Hi}\;.
\end{split}
\end{equation}
It remains to verify the Lagrange-like equations~\eqref{ch03:RouthEoMLag} of the Routhian formalism. We do so by evaluating separately both sides of the equations. On the one hand,
\begin{equation}
\begin{split}
\PD{R}{\dot q_{\mathrm Li}}&=\sum_{j=N+1}^n\PD{H}{p_{\mathrm Lj}}\PD{p_{\mathrm Lj}}{\dot q_{\mathrm Li}}-p_{\mathrm Li}-\sum_{j=N+1}^n\dot q_{\mathrm Lj}\PD{p_{\mathrm Lj}}{\dot q_{\mathrm Li}}\\
&=-p_{\mathrm Li}+\sum_{j=N+1}^n\biggl(\PD{H}{p_{\mathrm Lj}}-\dot q_{\mathrm Lj}\biggr)\PD{p_{\mathrm Lj}}{\dot q_{\mathrm Li}}\overset{\eqref{ch03:Hamiltoneq}}{=}-p_{\mathrm Li}\;.
\end{split}
\end{equation}
On the other hand,
\begin{equation}
\begin{split}
\PD{R}{q_{\mathrm Li}}&=\PD{H}{q_{\mathrm Li}}+\sum_{j=N+1}^n\PD{H}{p_{\mathrm Lj}}\PD{p_{\mathrm Lj}}{q_{\mathrm Li}}-\sum_{j=N+1}^n\dot q_{\mathrm Lj}\PD{p_{\mathrm Lj}}{q_{\mathrm Li}}\\
&=\PD{H}{q_{\mathrm Li}}+\sum_{j=N+1}^n\biggl(\PD{H}{p_{\mathrm Lj}}-\dot q_{\mathrm Lj}\biggr)\PD{p_{\mathrm Lj}}{q_{\mathrm Li}}\overset{\eqref{ch03:Hamiltoneq}}{=}\PD{H}{q_{\mathrm Li}}\overset{\eqref{ch03:Hamiltoneq}}{=}-\dot p_{\mathrm Li}\;.
\end{split}
\end{equation}
This completes the proof that the Routh equations of motion~\eqref{ch03:RouthEoMHam} and~\eqref{ch03:RouthEoMLag} have the same physical content as the original Hamilton equations~\eqref{ch03:Hamiltoneq}.

By the way, setting $N=0$ corresponds to a special case where all the generalized coordinates are Lagrange-like. In this case, the Routhian equals minus the Lagrangian as defined in terms of the Hamiltonian in~\eqref{ch03:LagfromHam}. Our above derivation of the Routh equations therefore includes as a special case a proof of equivalence of the Lagrange and Hamilton equations of motion, which I previously skipped for the sake of brevity. The opposite extreme special case, $N=n$, is trivial. Here we do not trade any momentum variables $p_i$ for the corresponding generalized velocities $\dot q_i$ by~\eqref{ch03:routhpL}. The Routhian is identical to the Hamiltonian and the dynamical equations of motion reduce to~\eqref{ch03:RouthEoMHam}.

It is also possible to arrive at the Routhian formulation of mechanics by starting from the Lagrangian and following the reverse procedure as outlined in Sect.~\ref{subsec:relationLagHam}. For the record, let me briefly sketch the main steps. We start by computing the conjugate momenta $p_i$ for the generalized coordinates $q_i$ with $i=1,\dotsc,N$ via $p_{\mathrm Hi}=\Pd{L}{\dot q_{\mathrm Hi}}$. This defines implicitly the generalized velocities $\dot q_{\mathrm Hi}$ in terms of the other variables,
\begin{equation}
\dot q_{\mathrm Hi}=\dot q_{\mathrm Hi}(q_\mathrm{H},p_\mathrm{H},q_\mathrm{L},t)\;.
\end{equation}
Upon inserting this back into the Lagrangian, the Routhian is obtained as
\begin{equation}
R(q_\mathrm{H},p_\mathrm{H},q_\mathrm{L},\dot q_\mathrm{L},t)\equiv\sum_{i=1}^Np_{\mathrm Hi}\dot q_{\mathrm Hi}(q_\mathrm{H},p_\mathrm{H},q_\mathrm{L},t)-L(q_\mathrm{H},\dot q_\mathrm{H}(q_\mathrm{H},p_\mathrm{H},q_\mathrm{L},t),q_\mathrm{L},\dot q_\mathrm{L},t)\;.
\end{equation}
This recovers the Routhian action~\eqref{ch03:actionRouth}. It is now of course possible to prove the Routh equations~\eqref{ch03:RouthEoMHam} and~\eqref{ch03:RouthEoMLag} as a consequence of the original Lagrange equations~\eqref{ch01:Lagrangeeq}. I will leave this as a straightforward exercise to those interested.

\begin{illustration}%
For a simple illustration, let us get back to the problem of motion in a central field. In \refex{ex03:centralfield}, we compared the Lagrangian and Hamiltonian approaches to this problem. The Routhian approach lies halfway between the two. Given that the polar angle $\vp$ is a cyclic coordinate, we choose it to be treated in a Hamilton-like fashion. Eliminating the angular velocity $\dot\vp$ in terms of the conjugate momentum, $p_\vp=\Pd{L}{\dot\vp}=\m r^2\dot\vp$, we arrive at the Routhian
\begin{equation}
R=p_\vp\dot\vp-L=p_\vp\frac{p_\vp}{\m r^2}-\frac12\m\dot r^2-\frac12\m r^2\left(\frac{p_\vp}{\m r^2}\right)^2+V(r)=\frac{p_\vp^2}{2\m r^2}-\frac12\m\dot r^2+V(r)\;.
\end{equation}
This defines the Routhian action~\eqref{ch03:actionRouth} as a functional of $\vp$, $p_\vp$ and $r$. For the first two of the variables, we get Hamilton-like equations,
\begin{equation}
\dot\vp=\PD{R}{p_\vp}=\frac{p_\vp}{\m r^2}\;,\qquad
\dot p_\vp=-\PD{R}{\vp}=0\;.
\end{equation}
The latter of these allows us to treat $p_\vp$ as a constant parameter. The former then gives us explicitly the angular velocity $\dot\vp$ once the radial coordinate $r$ is known as a function of time. This in turns follows by solving the Lagrange-like EoM for $r$,
\begin{equation}
0=\PD Rr-\OD{}t\PD{R}{\dot r}=m\ddot r+V'(r)-\frac{p_\vp^2}{\m r^3}\;,
\end{equation}
in full agreement with our previous result~\eqref{ch03:2bodyradialEoMHamilton}.
\end{illustration}

%%%%%%%%%%%%%%%%%%%%%%%%%%%%%%%%%%%%%%%%%%%%%%%%%%%%%%%%%%%%

\section*{\probsec}
\addcontentsline{toc}{section}{\probsec}

\begin{prob}
\label{pr03:sphericalpendulum}
A particle of mass $m$ is constrained to move on the surface of a sphere of radius $R$. There are no other forces acting on the particle; you can therefore think of it as the spherical pendulum of~\refpr{pr01:sphericalpendulum} in the limit of vanishing gravity. Starting from the known Lagrangian for the particle in spherical angles $\t,\vp$, deduce the corresponding Hamiltonian. Derive the Hamilton equations and reduce them to an effectively one-dimensional problem by using the fact that $\vp$ is a cyclic coordinate.
\end{prob}

\begin{prob}
\label{pr03:oldexam}
The Hamiltonian of a system is given by
\begin{equation}
H(q,p)=\frac12qp^2\;.
\end{equation}
Find the generalized coordinate $q$ as a function of time, given the total energy $E$ and the initial condition $q(t=0)=q_0$.
\end{prob}

\begin{prob}
\label{pr03:Hamflow}
Suppose that we are given two functions $q(t)$ and $p(t)$ that satisfy the first-order differential equations
\begin{equation}
\dot q=f(q,p)\;,\qquad
\dot p=g(q,p)\;,
\end{equation}
where $f(q,p)$ and $g(q,p)$ are some functions of two variables. A natural question is whether there is a Hamiltonian $H(q,p)$ that reproduces these equations with $q$ being the generalized coordinate and $p$ the generalized momentum. Show that the necessary condition for the existence of a Hamiltonian is that
\begin{equation}
\PD fq+\PD gp=0\;.
\end{equation}
Find the Hamiltonian in case that $f(q,p)=p-q^2$ and $g(q,p)=2qp+q^2$.
\end{prob}

\begin{prob}
\label{pr03:phaseportrait}
In~\refpr{pr01:freefall}, you constructed a Lagrangian for a particle of mass $m$, moving vertically in a uniform gravitational field. Derive the corresponding Hamiltonian and sketch the phase portrait of the system.
\end{prob}

\begin{prob}
\label{pr03:particleinEMfield}
Starting from the Lagrangian given in~\refpr{pr01:particleinEMfield}, show that the Hamiltonian describing the motion of a charged particle in an external electromagnetic field is
\begin{equation}
H(\vec r,\vec p,t)=\frac{[\vec p-q\vec A(\vec r,t)]^2}{2m}+q\p(\vec r,t)\;.
\end{equation}
Show that the ensuing Hamilton equations correctly reproduce the Lorentz force, that is, are equivalent to the Newtonian equation of motion $m\ddot{\vec r}=q(\vec E+\dot{\vec r}\times\vec B)$.
\end{prob}

\begin{prob}
\label{pr03:puzzle}
Here is something to puzzle over. Consider a system whose Lagrangian is given in terms of two dynamical variables $x,y$ as
\begin{equation}
L=\frac12(x\dot y-y\dot x)-\frac\l2(x^2+y^2)\;,
\label{ch03:Lagpuzzle}
\end{equation}
where $\l$ is a positive parameter. On the one hand, it is easy to compute the Hamilton function of this system via~\eqref{ch01:Hamiltonian}; work out the details. On the other hand, the Hessian of the Lagrangian with respect to $\dot x,\dot y$ is not invertible, so it does not seem possible to derive the Hamiltonian via a Legendre transformation as outlined in Sect.~\ref{subsec:relationLagHam}. It looks like our system does not have a Hamiltonian formulation. How is that possible? Remark: the Lagrangian~\eqref{ch03:Lagpuzzle} is not as artificial as it might seem. The term linear in time derivatives arises naturally if we take the Lagrangian~\eqref{ch01:LagEMfield} for a particle in a uniform magnetic field along the $z$-axis and go to the limit $m\to0$. 
\end{prob}

\begin{prob}
\label{pr03:Routhian}
In~\refpr{pr03:sphericalpendulum}, you derived the Lagrangian and Hamiltonian for a particle, moving freely on the surface of a sphere. Since the azimuthal angle $\vp$ is a cyclic coordinate, this looks like a good candidate for an application of the Routhian formalism. Find the Routhian, checking that you get the same result whether you start from the Lagrangian or from the Hamiltonian. Deduce the Routh equation of motion for the polar angle $\t$. Can you find any concrete solutions?
\end{prob}
\chapter{Application: Oscillations of Mechanical Systems}
\label{chap:oscillations}

\keywords{Simple harmonic oscillator, normal modes, normal frequencies.}

%%%%%%%%%%%%%%%%%%%%%%%%%%%%%%%%%%%%%%%%%%%%%%%%%%%%%%%%%%%%

\noindent So far we have focused mostly on developing techniques for construction of the \emph{equation of motion} (EoM) for mechanical systems. As far as finding concrete solutions of the EoM goes, we restricted ourselves to a few exactly solvable problems. However, in physics this is an exception rather than a rule. We thus need new tools that would allow us to find approximate solutions of the EoM whenever an exact solution is out of reach. In this \chaptername, we will develop the simplest yet physically most useful of such approximation methods, which we will heavily rely on in the field theory part of the course. For some further, more advanced methods, see for instance~\cite{Goldstein2013a,Lemos2018,Percival1982}.

The main idea is as follows. Suppose we are able to find exactly one concrete, simple solution of the EoM, $q_{0i}(t)$. This makes it possible to search for other, ``nearby'' solutions that differ from $q_{0i}(t)$ very little in a well-defined sense. Mathematically, one parameterizes the putative new solution $q_i(t)$ of the EoM in terms of its deviation from $q_{0i}(t)$, $\eta_i(t)$, that is,
\begin{equation}
q_i(t)\equiv q_{0i}(t)+\eta_i(t)\;.
\label{ch04:smalldeviation}
\end{equation}
Since $\eta_i$ is supposed to be small, it is reasonable to expand the EoM in its powers and, in the first approximation, drop all powers of $\eta_i$ higher than one. This process is known as the \emph{linearization} of the EoM. The ensuing approximate linear EoM can then be dealt with using standard techniques for linear \emph{ordinary differential equations} (ODEs). In most cases, an exact solution is now possible.

In this~\chaptername, we will be mostly concerned with applications where $q_{0i}$ is a stable equilibrium state of a system. In such cases, a small perturbation leads generally to oscillations around the equilibrium. The simple pendulum is a prototype of such a system; we saw already in Sect.~\ref{subsec:pendulum} how the linearization of the EoM reduces the pendulum to a linear harmonic oscillator. The harmonic approximation turns out to be very powerful. As we will show, it reduces the dynamics of any mechanical system to an equivalent set of independent one-dimensional oscillators.

%%%%%%%%%%%%%%%%%%%%%%%%%%%%%%%%%%%%%%%%%%%%%%%%%%%%%%%%%%%%

\section{Linear Harmonic Oscillator}

Given the central role the one-dimensional (linear) harmonic oscillator plays in the theory of near-equilibrium dynamics of mechanical systems, it will not hurt to start with a little reminder of the basics. To settle the terminology first, I will mean by a linear, or simple, harmonic oscillator any system whose EoM takes the form
\begin{equation}
\boxed{\ddot q+\o^2q=0\;,}
\label{ch04:simpleLHO}
\end{equation}
where $\o$ is a real positive constant. The concrete physical realization of this EoM is irrelevant. It can be a mass oscillating on a spring where $q$ stands for the distance of the mass from its equilibrium position. It can be the simple pendulum in the small-angle approximation, where $q$ would naturally correspond to the angular deviation from equilibrium. However, the same oscillator equation~\eqref{ch04:simpleLHO} can also be realized in a strictly nonmechanical manner. Think for instance of an  $LC$ circuit with an alternating current. Regardless of the physical interpretation, the solution of~\eqref{ch04:simpleLHO}, specified uniquely by the initial conditions $q(0),\dot q(0)$ at time $t=0$, is
\begin{equation}
\boxed{q(t)=q(0)\cos\o t+\frac{\dot q(0)}{\o}\sin\o t\;.}
\label{ch04:simpleLHOsol}
\end{equation}
The oscillator equation~\eqref{ch04:simpleLHO} and its solution~\eqref{ch04:simpleLHOsol} are so pervasive in physics that you should make sure to remember them.

In practice, macroscopic physical systems usually do not satisfy~\eqref{ch04:simpleLHO} exactly due to dissipation effects. The latter can have different realizations depending on the nature of the oscillator itself. In mechanical systems, dissipation arises from friction that converts mechanical energy into internal, thermal energy. In electromagnetic oscillators such as the plain $LC$ circuit, it is the resistance that is responsible for the dissipation of electromagnetic energy into heat. In addition to dissipation, the oscillating system may be naturally \emph{driven} by some external agent. This is particularly relevant for applications in electromagnetism, since electric circuits usually have one or more active elements: sources of current or voltage.

With the above in mind, I will now generalize~\eqref{ch04:simpleLHO} to
\begin{equation}
\ddot q(t)+2\g\dot q(t)+\o^2q(t)=F(t)\;.
\label{ch04:LHOdampeddriven}
\end{equation}
The positive constant parameter $\g$ models the effects of damping of the oscillations due to dissipation. The right-hand side, $F(t)$, takes into account possible driving of the oscillator. It is assumed to be a fixed, known function of time. The new equation~\eqref{ch04:LHOdampeddriven} is certainly more challenging than~\eqref{ch04:simpleLHO} we started with. However, it preserves the key feature: linearity. We can therefore deal with it by finding separately a general solution of the corresponding homogeneous equation, and a single solution of the full, inhomogeneous problem.

To solve the homogeneous equation, $\ddot q+2\g\dot q+\o^2 q=0$, we use the complex ansatz $q\propto\E^{\I\l t}$, which reduces the second-order ODE to a quadratic equation for the characteristic frequency $\l$, $-\l^2+2\I\g\l+\o^2=0$. This is solved by
\begin{equation}
\l_\pm=\I\g\pm\sqrt{\o^2-\g^2}\;.
\label{ch04:dampeddrivenlambdapm}
\end{equation}
Let us first assume that the damping is not too strong, $\g<\o$. Keeping in mind that the generalized coordinate $q(t)$ must be real, we can write the general solution to the homogeneous problem as
\begin{equation}
q_\mathrm{hom}(t)=\E^{-\g t}\Bigl[c_1\cos\Bigl(\sqrt{\o^2-\g^2}\,t\Bigr)+c_2\sin\Bigl(\sqrt{\o^2-\g^2}\,t\Bigr)\Bigr]\;,
\label{ch04:drivenLHOsolhom}
\end{equation}
where $c_{1,2}$ are two real integration constants. These cannot be determined until we have also found a particular solution of the full, inhomogeneous problem.

To that end, we represent the driving ``force'' $F(t)$ by its Fourier transform $\tilde F(\n)$, where $\n$ is a variable with the dimension of angular frequency, dual to time,
\begin{equation}
F(t)=\int_{-\infty}^{+\infty}\frac{\D\n}{2\pi}\tilde F(\n)\E^{-\I\n t}\quad\Leftrightarrow\quad
\tilde F(\n)=\int_{-\infty}^{+\infty}\D t\,F(t)\E^{\I\n t}\;.
\end{equation}
Upon transforming in the same way the dependent variable $q(t)$, we again arrive at an algebraic equation, $(-\n^2-2\I\g\n+\o^2)\tilde q(\n)=\tilde F(\n)$. This in turn gives a particular solution of the damped driven oscillator problem,
\begin{equation}
q_\mathrm{part}(t)=\int_{-\infty}^{+\infty}\frac{\D\n}{2\pi}\frac{\tilde F(\n)\E^{-\I\n t}}{\o^2-\n^2-2\I\g\n}\;.
\label{ch04:drivenLHOsolpart}
\end{equation}
The general solution thereof is given by the sum of~\eqref{ch04:drivenLHOsolhom} and~\eqref{ch04:drivenLHOsolpart}. If desired, a specific solution can now be pinned down by relating the integration constants $c_{1,2}$ to the initial conditions for $q(t)$.

\begin{illustration}%
To check the consistency of the particular solution~\eqref{ch04:drivenLHOsolpart}, let us consider some special cases where its validity is easily verified. First, take a constant driving force, $F(t)=F_0$. (Think for instance of a mass suspended on a spring in a uniform gravitational field.) This corresponds to $\tilde F(\n)=2\pi F_0\d(\n)$, which reduces~\eqref{ch04:drivenLHOsolpart} to $q_\mathrm{part}(t)=F_0/\o^2$. This certainly is a solution to~\eqref{ch04:LHOdampeddriven}.

Another natural choice of the driving force is one that oscillates with certain fixed nonzero frequency $\O$, $F(t)=F_0\cos\O t$. This is reproduced by setting $\tilde F(\n)=\pi F_0[\d(\n+\O)+\d(\n-\O)]$. Inserting that in~\eqref{ch04:drivenLHOsolpart} we obtain, upon a bit of manipulation,
\begin{equation}
q_\mathrm{part}(t)=F_0\frac{(\o^2-\O^2)\cos\O t+2\g\O\sin\O t}{(\o^2-\O^2)^2+(2\g\O)^2}\;.
\end{equation}
Also this is easily checked to satisfy~\eqref{ch04:LHOdampeddriven}. The interesting feature of periodic driving is that it can lead to resonance: the magnitude of the oscillating solution $q_\mathrm{part}(t)$ has a peak at $\O=\o$, i.e.~when the frequency of the driving is tuned to the intrinsic frequency of the oscillator. It is also noteworthy that the particular solution $q_\mathrm{part}(t)$ oscillates with the frequency imposed by the driving force. This is in contrast with the solution~\eqref{ch04:drivenLHOsolhom} of the homogeneous problem, which oscillates with the intrinsic frequency $\o$ of the system. After a time of the order of $1/\g$, these intrinsic oscillations die away due to the damping, and only the forced oscillations remain.
\end{illustration}

The above solution of the damped driven oscillator problem was based on the assumption that the damping is not too strong, $\g<\o$. The opposite case, $\g>\o$, is usually referred to as overdamped. The particular solution~\eqref{ch04:drivenLHOsolpart} of the inhomogeneous problem remains valid even in this case. However, the homogeneous solution~\eqref{ch04:drivenLHOsolhom} needs to be modified owing to the fact that both solutions~\eqref{ch04:dampeddrivenlambdapm} of the characteristic equation are now purely imaginary. We thus have to replace~\eqref{ch04:drivenLHOsolhom} with
\begin{equation}
q_\mathrm{hom}(t)=\E^{-\g t}\Bigl[c_1\exp\Bigl(\sqrt{\g^2-\o^2}\,t\Bigr)+c_2\exp\Bigl(-\sqrt{\g^2-\o^2}\,t\Bigr)\Bigr]\;.
\end{equation}
This type of solution does not involve any oscillations, but rather only damping at two different characteristic rates.

%%%%%%%%%%%%%%%%%%%%%%%%%%%%%%%%%%%%%%%%%%%%%%%%%%%%%%%%%%%%

\section{Small Oscillations in Single-Component Systems}

Now that we know the simple harmonic oscillator inside out, we are ready to tackle the problem of near-equilibrium dynamics in more complicated systems. In order to underline the main idea of our line of attack, we will start with systems with a single degree of freedom, represented by a generalized coordinate $q$. We will consider the broad class of Lagrangians of the type
\begin{equation}
L(q,\dot q)=\frac12m(q)\dot q^2-V(q)\;.
\label{ch04:Lgen1comp}
\end{equation}
The notation suggests that the function $m(q)$ captures the inertial properties of the system, but it does not have to be a mere constant mass parameter. All I will require is that $m(q)$ is positive. The Hamilton function corresponding to~\eqref{ch04:Lgen1comp} is $H=(1/2)m(q)\dot q^2+V(q)$. Hence, $V(q)$ can be interpreted as the potential energy, and I will assume it to be bounded from below.

Let $q_0$ denote the minimum of $V(q)$. Then, $q(t)\equiv q_0$ is a solution of the Lagrange equation for~\eqref{ch04:Lgen1comp},
\begin{equation}
\OD{}t[m(q)\dot q]=\frac12m'(q)\dot q^2-V'(q)\;.
\label{ch04:Lgen1compEoM}
\end{equation}
Physically, it corresponds to a stable equilibrium configuration of the system. Following the philosophy outlined at the beginning of the~\chaptername, we will now search for other solutions $q(t)$ that deviate from the equilibrium by $\eta(t)$, as defined in~\eqref{ch04:smalldeviation}. Expanding everything in~\eqref{ch04:Lgen1compEoM} to first order in $\eta$, we deduce an approximate, linearized EoM for the latter,\footnote{I will use the symbol $\approx$ throughout the entire lecture notes to indicate a linear approximation to the EoM, or the corresponding quadratic approximation to the Lagrangian.}
\begin{equation}
m(q_0)\ddot\eta\approx-V''(q_0)\eta\;.
\label{ch04:Lgen1compEoMlin}
\end{equation}
This describes harmonic oscillations with frequency
\begin{equation}
\boxed{\o=\sqrt{\frac{V''(q_0)}{m(q_0)}}\;.}
\end{equation}
Using~\eqref{ch04:simpleLHOsol}, it is easy to write down the corresponding approximate solution to the original EoM~\eqref{ch04:Lgen1compEoM},
\begin{equation}
q(t)\approx q_0+\eta(0)\cos\o t+\frac{\dot\eta(0)}{\o}\sin\o t\;,
\end{equation}
where $\eta(0)\equiv q(0)-q_0$ and $\dot\eta(0)=\dot q(0)$.

\begin{watchout}%
Instead of expanding the EoM to the first order in $\eta$, one can alternatively expand the Lagrangian to the second order in $\eta$. Since such a quadratic Lagrangian gives a linear EoM, these two approaches are physically entirely equivalent. In case of~\eqref{ch04:Lgen1comp}, we thus get
\begin{equation}
L\approx\frac12m(q_0)\dot\eta^2-\xcancel{V(q_0)}-\xcancel{V'(q_0)\eta}-\frac12V''(q_0)\eta^2\;.
\end{equation}
The term $V(q_0)$ can be discarded due to being a mere constant. In addition, the term linear in $\eta$ vanishes since $V'(q_0)=0$ by the definition of a minimum. The resulting equivalent Lagrangian, $(1/2)m(q_0)\dot\eta^2-(1/2)V''(q_0)\eta^2$, recovers, not surprisingly, the linearized EoM~\eqref{ch04:Lgen1compEoMlin}.

It is obviously easier to expand the EoM to the first order in $\eta$ than to expand the Lagrangian to the second order. So why would one ever bother doing the latter? The answer to this question is that once we proceed to systems with multiple degrees of freedom, it is actually more transparent to deal with the Lagrangian than with a system of coupled linearized ODEs. Moreover, there is a computational cost in first deriving the exact EoM, only to power-expand it subsequently. For these reasons, I will mostly use the Lagrangian approach in what follows. However, I will occasionally remind you of the EoM linearization approach, since it is good to keep in mind as a viable alternative.
\end{watchout}

%%%%%%%%%%%%%%%%%%%%%%%%%%%%%%%%%%%%%%%%%%%%%%%%%%%%%%%%%%%%

\subsection{Example: Ball in a Bowl}

\begin{figure}[t]
\sidecaption[t]
\includegraphics[width=2.9in]{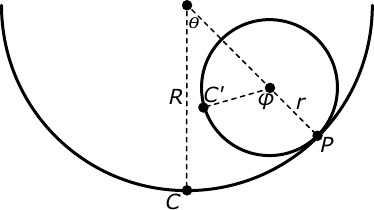}
\caption{A homogeneous ball of radius $r$, rolling without slipping inside a spherical bowl of radius $R$. It is assumed that the motion happens in a single vertical plane. The angular variables $\t$ and $\vp$ are related by $R\t=r\vp$. We are to find the frequency of small oscillations around equilibrium in terms of $r,R$ and the gravitational acceleration $g$.}
\label{fig04:rollingball}
\end{figure}

Consider the motion of a ball that rolls without slipping in a spherical bowl, under the influence of a uniform gravitational field. I will assume that the motion is restricted to a single vertical plane. The more general problem of a fully three-dimensional motion is tricky, and the resulting dynamics can be quite surprising. (In case of interest, search for the keyword ``golf ball paradox.'') See Fig.~\ref{fig04:rollingball} for a visualization of our simplified setup. There are two natural angular variables that can be used to parameterize the configuration of the ball: the angular deviation $\t$ of its \emph{center of mass} (CoM) from the position of lowest potential energy, and the angle of rotation $\vp$ with respect to the point of contact $P$ between the ball and the bowl. These two angles are related by the requirement that the arcs $CP$ and $C'P$ have equal lengths, as follows from the no-slip condition. This is expressed as $R\t=r\vp$.

To construct the kinetic energy, we recall the expression for the moment of inertia of a ball with respect to its CoM, $I=(2/5)mr^2$, where $m$ is the mass of the ball. The angular velocity of the ball with respect to the inertial reference frame attached to the ground is $\dot\vp-\dot\t$. By the same token, the (magnitude of the) linear velocity of its CoM is $r(\dot\vp-\dot\t)$. By putting together the kinetic energies of the translational motion of the CoM and of the rotation with respect to the CoM, we find the total kinetic energy, $(7/10)mr^2(\dot\vp-\dot\t)^2$. The potential energy is easier to calculate, as it is determined by the position of the CoM. Choosing the potential energy to vanish at the lowest possible position, corresponding to $\t=0$, we find that $V=mg(R-r)(1-\cos\t)$. Finally, we need to choose one of the angles $\t,\vp$ as the sole generalized coordinate for the problem. Eliminating $\vp$ in favor of $\t$ by $\vp=(R/r)\t$, the total Lagrangian for the ball reads
\begin{equation}
\begin{split}
L&=\frac7{10}m(R-r)^2\dot\t^2-mg(R-r)(1-\cos\t)\\
&\approx\frac7{10}m(R-r)^2\dot\t^2-\frac12mg(R-r)\t^2\;,
\end{split}
\end{equation}
where the second line shows the relevant part of the Lagrangian upon power expansion in $\t$. We infer that for small deviations from equilibrium, the ball undergoes approximately harmonic oscillations with angular frequency
\begin{equation}
\o=\sqrt{\frac{5g}{7(R-r)}}\;.
\end{equation}
How do you interpret the fact that the frequency diverges in the limit $r\to R$?

%%%%%%%%%%%%%%%%%%%%%%%%%%%%%%%%%%%%%%%%%%%%%%%%%%%%%%%%%%%%

\subsection{Example: Rocking Cylinder}

\begin{figure}[t]
\sidecaption[t]
\includegraphics[width=2.9in]{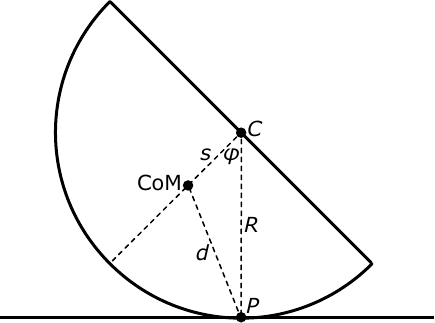}
\caption{A homogeneous half-cylinder of radius $R$ and length $L$, rolling without slipping on a horizontal plane. The task is to find the frequency of small oscillations in terms of $R$ and the gravitational acceleration~$g$.}
\label{fig04:halfcylinder}
\end{figure}

For another example that is technically somewhat more involved, consider a half-cylinder rocking back and forth on a horizontal plane; see Fig.~\ref{fig04:halfcylinder} for a cross-section of the setup by a vertical plane. Here we have only one natural generalized coordinate, $\vp$, measuring the deviation from equilibrium in terms of the angle by which the half-cylinder is tilted. To be able to construct the Lagrangian, we need to know where the CoM of the half-cylinder is located. Its distance from the center of curvature $C$ turns out to be $s=4R/(3\pi)$. If you are unsure how to prove this, you should read about the \href{https://en.wikipedia.org/wiki/Pappus%27s_centroid_theorem}{Pappus centroid (Guldinus) theorem}. The potential energy, again normalized to zero in the equilibrium position $\vp=0$, is then $mgs(1-\cos\vp)$, where $m$ is the mass of the half-cylinder.

The next step is to construct the kinetic energy. The moment of inertia with respect to the axis of revolution of the half-cylinder, passing through $C$, is $I_C=(1/2)mR^2$. This is the same formula as for a full solid cylinder. (Try to understand why.) From here, we can extract the moment of inertia with respect to an axis passing through the CoM via the \href{https://en.wikipedia.org/wiki/Parallel_axis_theorem}{parallel axis (Steiner) theorem}, $I_\mathrm{CoM}=I_C-ms^2$. In addition, the (magnitude of the) linear velocity of the CoM is $d\dot\vp$, where the distance $d$ of the CoM from the point of contact $P$ between the half-cylinder and the ground is given by the law of cosines, $d=\sqrt{R^2+s^2-2Rs\cos\vp}$. The total kinetic energy is then composed out of the translational kinetic energy of the CoM and the energy of rotation with respect to the CoM,
\begin{equation}
\begin{split}
\frac12m(d\dot\vp)^2+\frac12I_\mathrm{CoM}\dot\vp^2&=\frac12m(d^2-s^2)\dot\vp^2+\frac12I_C\dot\vp^2\\
&=\left(\frac34-\frac{4}{3\pi}\cos\vp\right)mR^2\dot\vp^2\;.
\end{split}
\end{equation}
Putting all the pieces together, we arrive at the Lagrangian
\begin{equation}
\begin{split}
L&=\left(\frac34-\frac{4}{3\pi}\cos\vp\right)mR^2\dot\vp^2-\frac4{3\pi}mgR(1-\cos\vp)\\
&\approx\left(\frac34-\frac{4}{3\pi}\right)mR^2\dot\vp^2-\frac2{3\pi}mgR\vp^2\;.
\end{split}
\end{equation}
From here, we get the angular frequency of the rocking motion of the half-cylinder,
\begin{equation}
\o=\sqrt{\frac{8}{9\pi-16}\frac gR}\;.
\end{equation}
In case you find the above solution too complicated, just try to imagine how you would go about the problem using Newtonian mechanics and forces.

%%%%%%%%%%%%%%%%%%%%%%%%%%%%%%%%%%%%%%%%%%%%%%%%%%%%%%%%%%%%

\section{General Theory for Multicomponent Systems}
\label{sec:multicomponent}

It is fairly straightforward the extend the analysis of small oscillations from a single generalized coordinate to systems with multiple degrees of freedom. The general theory however requires a moderate use of linear algebra. In case you have not refreshed your linear algebra background at least in the extent corresponding to~\appendixname~\ref{app:linalg} yet, you should do so now.

Instead of a single generalized coordinate $q$, we will now have $n$ of them, $q_i$. The Lagrangian~\eqref{ch04:Lgen1comp} is thus generalized to
\begin{equation}
L(q,\dot q)=\frac12m_{ij}(q)\dot q_i\dot q_j-V(q)\;.
\label{ch04:Lgenmulti}
\end{equation}
Note that I am using the Einstein summation convention: this will make many algebraic expressions less cluttered. The kinetic energy is now given by a (by assumption positive-definite) $q$-dependent quadratic form in the generalized velocities $\dot q_i$. In other words, the matrix $m_{ij}(q)$ is symmetric and positive-definite. We again take $q_{0i}$ to indicate the minimum of the potential energy $V(q)$; setting $q_i(t)\equiv q_{0i}$ then automatically satisfies the Lagrange equations descending from the Lagrangian~\eqref{ch04:Lgenmulti}. Upon redefining the generalized coordinates as~\eqref{ch04:smalldeviation} and expanding to second order in the deviation variables $\eta_i$, we get
\begin{equation}
L\approx\frac12T_{ij}\dot\eta_i\dot\eta_j-\frac12V_{ij}\eta_i\eta_j\;,
\label{ch04:Lgenmultieta}
\end{equation}
where
\begin{equation}
T_{ij}\equiv m_{ij}(q_0)\;,\qquad
V_{ij}\equiv\frac{\de^2V(q)}{\de q_i\de q_j}\biggr\rvert_{q=q_0}\;.
\label{ch04:TVdef}
\end{equation}
In the process, we dropped the constant contribution $V(q_0)$ and used the fact that all the first derivatives $\Pd{V}{q_i}$ vanish at $q_i=q_{0i}$.

%%%%%%%%%%%%%%%%%%%%%%%%%%%%%%%%%%%%%%%%%%%%%%%%%%%%%%%%%%%%

\subsection{Normal Modes}

The approximate Lagrangian~\eqref{ch04:Lgenmultieta} is completely specified by two quadratic forms, $T_{ij}$ and $V_{ij}$. The first of these is, as already stressed, by assumption positive-definite. The second, $V_{ij}$, is positive-semidefinite by construction, since it corresponds to the Hessian of $V(q)$ at its minimum. The next step is crucial: we use the fact that such a pair of quadratic forms can be simultaneously diagonalized by a suitable linear transformation of the generalized coordinates $\eta_i$. Here is an algorithmic procedure how to do so, following the reasoning outlined at the end of~\appendixname~\ref{app:linalg}:
\begin{enumerate}
\item[(1)] Diagonalize the matrix $T$ by an orthogonal transformation $P_1$,
\begin{equation}
T\to T'\equiv P_1^TTP_1\;,\qquad
V\to V'\equiv P_1^TVP_1\;.
\end{equation}
\item[(2)] Rescale the new generalized coordinates in order to make $T'$ into a unit matrix,
\begin{equation}
T'\to T''\equiv P_2^TT'P_2=\un\;,\qquad
V'\to V''\equiv P_2^TV'P_2\;,
\end{equation}
where $P_2=\diag(1/\sqrt{\l_1},\dotsc,1/\sqrt{\l_n})$ and $\l_1,\dotsc,\l_n$ are the eigenvalues of $T$.
\item[(3)] Diagonalize $V''$ by an additional orthogonal transformation $P_3$,
\begin{equation}
T''\to P_3^TT''P_3=P_3^TP_3=\un\;,\qquad
V''\to P_3^TV''P_3\equiv\O^2\;.
\end{equation}
\end{enumerate}
This series of transformations can be encoded in a single transformation matrix, $P\equiv P_1P_2P_3$. In terms of this matrix, we have $\O^2=P^TVP$. At the same time, the generalized coordinates $\eta_i$ are transformed to $\x_i$ such that
\begin{equation}
\eta_i=P_{ij}\x_j\quad\Leftrightarrow\quad
\x_i=(P^{-1})_{ij}\eta_j\;.
\label{ch04:normalmode}
\end{equation}
In these new variables, the approximate quadratic Lagrangian~\eqref{ch04:Lgenmultieta} takes the form
\begin{equation}
\boxed{L\approx\frac12\sum_{i=1}^n\bigl(\dot\x_i^2-\o_i^2\x_i^2\bigr)\;,}
\label{ch04:Lnormalmode}
\end{equation}
where $\o_i^2$ are the eigenvalues of $\O^2$ (and thus also of $V$). In the standard terminology, each of the $\x_i$s is called a \emph{normal mode}, $\o_i$ being the corresponding \emph{normal frequency}.

Our new Lagrangian~\eqref{ch04:Lnormalmode} is a sum of completely decoupled contributions, one for each normal mode. Each of these has its own separate EoM of the type~\eqref{ch04:simpleLHO}. This brings us to a general algorithm how to find the unique approximate solution of the Lagrange equations for~\eqref{ch04:Lgenmulti}, given the initial conditions $q_i(0)$ and $\dot q_i(0)$:
\begin{enumerate}
\item[(1)] Find the minimum $q_{0i}$ of the potential $V(q)$. Use it to identify the matrices $T_{ij}$ and $V_{ij}$ via~\eqref{ch04:TVdef}.
\item[(2)] Find the transformation $P$ that brings the matrices $T,V$ to the form $P^TTP=\un$ and $P^TVP=\O^2$ with diagonal $\O^2$.
\item[(3)] Fix the initial conditions for $\x_i$ using~\eqref{ch04:normalmode},
\begin{equation}
\x_i(0)=(P^{-1})_{ij}[q_j(0)-q_{0j}]\;,\qquad
\dot\x_i(0)=(P^{-1})_{ij}\dot q_j(0)\;.
\end{equation}
\item[(4)] Solve the Lagrange equations for the normal modes in terms of the initial conditions; cf.~\eqref{ch04:simpleLHOsol},
\begin{equation}
\begin{aligned}
\x_i(t)&=\x_i(0)\cos\o_it+\frac{\dot\x_i(0)}{\o_i}\sin\o_i t\qquad&\text{for }&\o_i\neq0\;,\\
\x_i(t)&=\x_i(0)+\dot\x_i(0)t\qquad&\text{for }&\o_i=0\;.
\end{aligned}
\label{ch04:generalxit}
\end{equation}
\item[(5)] Translate back to the original variables,
\begin{equation}
q_i(t)=q_{0i}+P_{ij}\x_j(t)\;.
\end{equation}
\end{enumerate}

It is easy to modify the above algorithm in case the system is subject to external forces. To include the effect of driving, we take the Lagrange equations for the quadratic Lagrangian~\eqref{ch04:Lgenmultieta} and add a right-hand side following the template~\eqref{ch04:LHOdampeddriven},
\begin{equation}
T_{ij}\ddot\eta_j(t)+V_{ij}\eta_j(t)=F_i(t)\;.
\label{ch04:drivenmulti}
\end{equation}
Here $F_i$ is a set of fixed functions of time, one for each generalized coordinate. Upon the change of variables from $\eta_i$ to $\x_i$, this becomes
\begin{equation}
\ddot\x_i(t)+\o_i^2\x_i(t)=G_i(t)\;,\quad\text{where }G_i(t)\equiv P_{ji}F_j(t)\;.
\end{equation}
A particular solution to these inhomogeneous equations is given by~\eqref{ch04:drivenLHOsolpart}. A general solution then follows by adding the solution~\eqref{ch04:generalxit} of the corresponding homogeneous problem.

What is not easy to is incorporate dissipation effects. That would require adding another matrix term to~\eqref{ch04:drivenmulti}, linear in the generalized velocities $\dot\eta_i$. The problem with doing so is that it is then no longer possible to decouple the resulting set of second-order ODEs into independent normal modes. That would amount to simultaneous diagonalization of three quadratic forms, which is generally not possible.

%%%%%%%%%%%%%%%%%%%%%%%%%%%%%%%%%%%%%%%%%%%%%%%%%%%%%%%%%%%%

\subsection{Example: Coupled Oscillators}

\begin{figure}[t]
\sidecaption[t]
\includegraphics[width=2.9in]{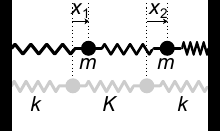}
\caption{Geometric setup of the coupled oscillator problem. The system has a left--right symmetry in that the two masses are equal, as are the spring constants of the two springs by which the masses are attached to the walls. For comparison, the equilibrium state is shown in gray.}
\label{fig04:coupledoscillators}
\end{figure}

I will now illustrate the general theory of normal modes on two simple examples. We start with a one-dimensional system of two equal masses connected by a spring $K$, each of which is in addition attached to a nearby wall by a spring $k$. See Fig.~\ref{fig04:coupledoscillators} for a sketch. We assume that in equilibrium, all the three springs are relaxed (not stretched). What the actual lengths of the springs at rest are is irrelevant. It seems convenient to use as generalized coordinates the displacements $x_1,x_2$ of the two masses from equilibrium. The Lagrangian of the system is then
\begin{equation}
L=\frac12m(\dot x_1^2+\dot x_2^2)-\frac12[k(x_1^2+x_2^2)+K(x_2-x_1)^2]\;,
\label{ch04:coupledoscLag}
\end{equation}
where we used that the extension of the middle spring is $x_2-x_1$. Our Lagrangian is quadratic in the generalized coordinates and there is therefore no need to make any approximations. The system is expected to undergo perfectly harmonic oscillations.

To find the normal modes, we first identify the matrices $T$ and $V$ by matching the Lagrangian to~\eqref{ch04:Lgenmultieta},
\begin{equation}
T=\begin{pmatrix}
m & 0\\
0 & m
\end{pmatrix}=m\un\;,\qquad
V=\begin{pmatrix}
k+K & -K\\
-K & k+K
\end{pmatrix}\;.
\end{equation}
The matrix $T$ already is diagonal and proportional to the unit matrix. We just need to get rid of the mass factor $m$ by rescaling $x_{1,2}$ with $1/\sqrt m$. To diagonalize $V$, note that it can be written as $V=(k+K)\un-K\tau_1$, where $\tau_1$ is one of the Pauli matrices. Using what we know about the eigenvectors and eigenvalues of the Pauli matrices (cf.~\refex{exapp:Pauli}), we conclude that the normal modes of the system are
\begin{equation}
\x_\pm\equiv\sqrt{\frac m2}(x_1\pm x_2)\quad\Leftrightarrow\quad
x_1=\frac1{\sqrt{2m}}(\x_++\x_-)\;,\hspace{1ex}
x_2=\frac1{\sqrt{2m}}(\x_+-\x_-)\;.
\label{ch04:coupledosc_normal}
\end{equation}
In terms of these, the Lagrangian takes the form
\begin{equation}
L=\frac12(\dot\x_+^2+\dot\x_-^2)-\frac12\frac km\x_+^2-\frac12\frac{k+2K}m\x_-^2\;.
\end{equation}
The corresponding normal frequencies are therefore
\begin{equation}
\o_+=\sqrt{\frac km}\;,\qquad
\o_-=\sqrt{\frac{k+2K}{m}}\;.
\end{equation}

The physical nature of the normal modes is most easily understood with the help of~\eqref{ch04:coupledosc_normal}. Setting $\x_-\to0$, we find that $x_1=x_2\propto\x_+$. In other words, the $\x_+$ normal mode amounts to synchronized oscillations of the two masses. Their distance remains constant during such motion, which explains why the normal frequency $\o_+$ does not depend on the spring constant $K$. Setting, on the other hand, $\x_+\to0$, we find that $x_1=-x_2\propto\x_-$. The other normal mode, $\x_-$, therefore corresponds to oscillations of the two masses with opposite phase.

\begin{watchout}%
The Lagrangian~\eqref{ch04:coupledoscLag} does not change if we swap the generalized coordinates, $x_1\leftrightarrow x_2$. This operation can be viewed literally as mirroring the spring chain in Fig.~\ref{fig04:coupledoscillators} with respect to a vertical axis passing through the center of the chain. It is a simple example of a discrete symmetry, often called \emph{parity}. Note how the normal modes of the system reflect this symmetry. Under $x_1\leftrightarrow x_2$, $\x_+$ remains unchanged, whereas $\x_-$ changes sign. The two normal modes are therefore eigenstates of the symmetry operation with eigenvalues $\pm1$.
\end{watchout}

%%%%%%%%%%%%%%%%%%%%%%%%%%%%%%%%%%%%%%%%%%%%%%%%%%%%%%%%%%%%

\subsection{Example: Triatomic Molecule}
\label{subsec04:triatomic}

\begin{figure}[t]
\sidecaption[t]
\includegraphics[width=2.0in]{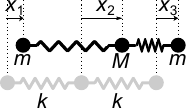}% This figure does not include so many details to warrant a full width of 2.9 in!
\caption{Geometry of a linear triatomic molecule. Given the equal masses of the two atoms at the edges and the equal spring constants of the two springs, the system has a left--right symmetry under which the two edge molecules are swapped. For comparison, the equilibrium state is shown in gray.}
\label{fig04:triatomic}
\end{figure}

Here is one more, somewhat more involved example. Consider the system of three masses, mutually connected by two springs, as shown in Fig.~\ref{fig04:triatomic}. For the sake of simplicity, we assume that the system is linear and the masses can only move in the direction along the springs. If you wish, you can think of this system as an oversimplified classical model of the molecule of carbon dioxide. It is again convenient to use as generalized coordinates the displacements of the masses from equilibrium, in which the springs are relaxed (not stretched). This time, we however need three coordinates, denoted as $x_1$, $x_2$ and $x_3$. The Lagrangian is then
\begin{equation}
L=\frac12m(\dot x_1^2+\dot x_3^2)+\frac12M\dot x_2^2-\frac12k[(x_2-x_1)^2+(x_3-x_2)^2]\;.
\end{equation}
The kinetic energy matrix $T_{ij}$ is diagonal but not proportional to $\un$ due to the unequal masses. We bring it to the desired form by introducing new coordinates $\eta_1,\eta_2,\eta_3$ via
\begin{equation}
\eta_1=x_1\sqrt m\;,\qquad
\eta_2=x_2\sqrt M\;,\qquad
\eta_2=x_3\sqrt m\;.
\end{equation}
This turns the Lagrangian into
\begin{equation}
L=\frac12\dot\eta_i\dot\eta_i-\frac12V_{ij}\eta_i\eta_j\;,
\end{equation}
where
\begin{equation}
V=\begin{pmatrix}
\o^2 & -\o\O & 0\\
-\o\O & 2\O^2 & -\o\O\\
0 & -\o\O & \o^2
\end{pmatrix}\;,\qquad
\o\equiv\sqrt{\frac{k}{m}}\;,\quad
\O\equiv\sqrt{\frac{k}{M}}\;.
\end{equation}

The next step is to find an orthogonal transformation that diagonalizes the matrix $V$. This is a standard problem in linear algebra, and I will thus skip details. The eigenvalues of $V$ are
\begin{equation}
\o_+^2=\o^2+2\O^2=k\left(\frac1m+\frac2M\right)\;,\qquad
\o_-^2=\o^2=\frac km\;,\qquad
\o_0^2=0\;.
\end{equation}
The corresponding real, normalized eigenvectors are, up to a sign,
\begin{equation}
\vec v_+=\frac{1}{\sqrt{2\o^2+4\O^2}}\begin{pmatrix}
\o\\
-2\O\\
\o
\end{pmatrix}\;,\quad
\vec v_-=\frac1{\sqrt2}\begin{pmatrix}
1\\
0\\
-1
\end{pmatrix}\;,\quad
\vec v_0=\frac{1}{\sqrt{\o^2+2\O^2}}\begin{pmatrix}
\O\\
\o\\
\O
\end{pmatrix}\;.
\end{equation}
These eigenvectors span the columns of the orthogonal matrix $P$ that diagonalizes $V$ via the transformation $V\to P^TVP$. This gives us an explicit expression for the original generalized coordinates $x_1,x_2,x_3$ in terms of the normal modes $\x_+,\x_-,\x_0$,
\begin{equation}
\begin{split}
x_1&=\frac{1}{\sqrt m}\left(\frac{\o\x_+}{\sqrt{2\o^2+4\O^2}}+\frac{\x_-}{\sqrt2}+\frac{\O\x_0}{\sqrt{\o^2+2\O^2}}\right)\;,\\
x_2&=\frac{1}{\sqrt M}\left(-\frac{2\O\x_+}{\sqrt{2\o^2+4\O^2}}+\frac{\o\x_0}{\sqrt{\o^2+2\O^2}}\right)\;,\\
x_3&=\frac{1}{\sqrt m}\left(\frac{\o\x_+}{\sqrt{2\o^2+4\O^2}}-\frac{\x_-}{\sqrt2}+\frac{\O\x_0}{\sqrt{\o^2+2\O^2}}\right)\;.
\end{split}
\label{ch04:triatomicxfromxi}
\end{equation}
If desired, the general solution of the Lagrange equations for our system is obtained by combining~\eqref{ch04:triatomicxfromxi} with~\eqref{ch04:generalxit}.

Let us take apart the physical meaning of the three modes. For the $\x_+$ mode, we set $\x_-,\x_0\to0$, upon which
\begin{equation}
(x_1,x_2,x_3)\propto(1/m,-2/M,1/m)\x_+\;.
\end{equation}
This amounts to synchronized oscillations of the two masses on the edges. The middle mass oscillates with an opposite phase so as to maintain the CoM of the system fixed, which is guaranteed by the relation $mx_1+Mx_2+mx_3=0$. As to the $\x_-$ mode, we set $\x_+,\x_0\to0$, which leads to
\begin{equation}
(x_1,x_2,x_3)\propto(1,0,-1)\x_-\;.
\end{equation}
This is easy to interpret: the two masses on the edges oscillate with the opposite phase, while the mass in the middle remains still. In other words, the molecule as a whole periodically stretches and shrinks. The fact that the middle mass is static explains why the normal frequency $\o_-$ does not depend on $M$. Finally, we set $\x_+,\x_-\to0$, which gives us
\begin{equation}
(x_1,x_2,x_3)\propto(1,1,1)\x_0\;.
\end{equation}
This describes linear translational motion of the whole molecule without stretching the springs between the masses. The $\x_0$ mode is therefore not oscillatory, which is confirmed by the above calculated normal frequency, $\o_0=0$.

\begin{watchout}%
As for the coupled oscillator system in Fig.~\ref{fig04:coupledoscillators}, the qualitative properties of the normal modes $\x_+,\x_-,\x_0$ reflect the symmetries of the triatomic molecule. First, we can again swap the two masses on the edges and the corresponding coordinates $x_1,x_3$ without changing the Lagrangian. The notation chosen for the normal modes $\x_+,\x_-$ reminds us that these modes are eigenstates of this parity operation with respective eigenvalues $\pm1$. The zero mode $\x_0$ is even more interesting. Its presence reflects the invariance of the dynamics of the system under simultaneous translations of all the three masses. Mathematically, this is encoded in the fact that $\x_0$ without time derivative does not appear in the diagonalized Lagrangian~\eqref{ch04:Lnormalmode} thanks to $\o_0=0$: $\x_0$ is a cyclic coordinate. As a consequence, the conjugate momentum $\Pd{L}{\dot\x_0}$ is a constant of motion, expressing the conservation of the total momentum of the system. Why is the total momentum not conserved in the coupled oscillator system in Fig.~\ref{fig04:coupledoscillators}?
\end{watchout}

%%%%%%%%%%%%%%%%%%%%%%%%%%%%%%%%%%%%%%%%%%%%%%%%%%%%%%%%%%%%

\section*{\probsec}
\addcontentsline{toc}{section}{\probsec}

\begin{figure}[t]
\sidecaption[t]
\includegraphics[width=2.9in]{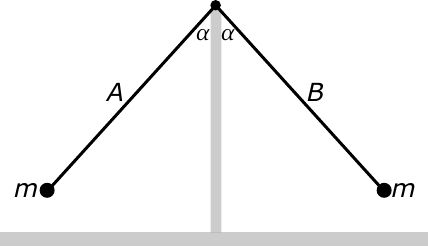}
\caption{Illustration for~\refpr{pr04:swing}. A model of a swing, constructed out of two rigid arms, making a fixed angle $2\a$ with each other. Two children on the swing are modeled as point objects of equal mass $m$.}
\label{fig04:swing}
\end{figure}

\begin{prob}
\label{pr04:criticaldamping}
Find the general solution of the EoM~\eqref{ch04:LHOdampeddriven} for a damped driven oscillator in the special case of \emph{critical damping} where $\g=\o$.
\end{prob}

\begin{prob}
\label{pr04:swing}
A playground swing is constructed out of two rigid arms $A,B$ of equal length $L$, connected together so that they form a fixed angle $2\a$, and suspended in the middle on a vertical pole. In the equilibrium position, both arms form the angle $\a$ with respect to the vertical (see Fig.~\ref{fig04:swing}). The swing can pivot around the point of suspension in a vertical plane. Suppose that we model two children sitting on the swing as point masses $m$ attached to the ends of the arms $A,B$, and that we can neglect the mass of the construction of the swing. Determine the period of small oscillations of this system around equilibrium.
\end{prob}

\begin{figure}[t]
\sidecaption[t]
\includegraphics[width=2.0in]{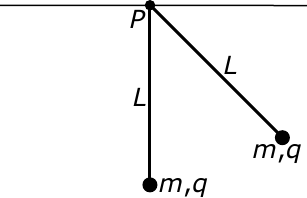}% This figure does not include so many details to warrant a full width of 2.9 in!
\caption{Illustration for~\refpr{pr04:chargedpendulum}. Two particles of equal mass $m$ and electric charge $q$ are attached to two strings of equal length $L$. One of the particles is fixed, whereas the other is free to move in a vertical plane.}
\label{fig04:chargedpendulum}
\end{figure}

\begin{prob}
\label{pr04:chargedpendulum}
Consider a system consisting of two point particles of equal mass $m$ and electric charge $q$. Both are suspended on a massless string of length $L$, attached to the same pivot point $P$ (see Fig.~\ref{fig04:chargedpendulum}). One of the particles is fixed at such a position that its string is vertical. The other particle is free to swing in a vertical plane. The uniform gravitational field $\vec g$ points vertically downwards. Find the frequency of small oscillations of the mobile particle around its equilibrium position. Assume that the actual values of the parameters are such that in equilibrium, the angle between the two strings is smaller than $\pi/2$.
\end{prob}

\begin{prob}
\label{pr04:triatomic}
Suppose that the triatomic molecule in Fig.~\ref{fig04:triatomic} is initially in equilibrium and at rest. Then, the middle molecule receives an instantaneous kick that gives it a nonzero initial velocity $\dot x_2(0)$. Find the positions of all the three molecules as a function of time, given this initial condition.
\end{prob}

\begin{figure}[t]
\sidecaption[t]
\includegraphics[width=2.0in]{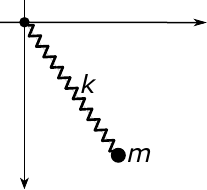}% This figure does not include so many details to warrant a full width of 2.9 in!
\caption{Illustration for~\refpr{pr04:springypendulum}: a ``springy'' pendulum, where a string of fixed length is replaced with a spring of known rest length and spring constant.}
\label{fig04:springypendulum}
\end{figure}

\begin{prob}
\label{pr04:springypendulum}
An interesting variation on the simple pendulum is shown in Fig.~\ref{fig04:springypendulum}. The string of a fixed length is replaced with a spring of rest (unstretched) length $L_0$ and spring constant $k$. Assuming that all motion happens in a single vertical plane, find suitable generalized coordinates for the system and construct a Lagrangian. Then, assuming small oscillations, find the normal modes and normal frequencies of the system.
\end{prob}

\begin{figure}[t]
\sidecaption[t]
\includegraphics[width=2.0in]{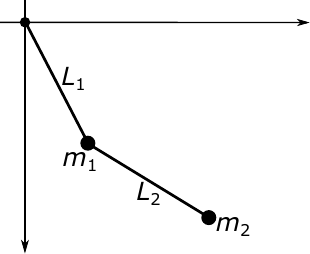}% This figure does not include so many details to warrant a full width of 2.9 in!
\caption{Illustration for~\refpr{pr04:doublependulum}: a double pendulum.}
\label{fig04:doublependulum}
\end{figure}

\begin{prob}
\label{pr04:doublependulum}
A \emph{double pendulum} is a system where the point mass of a simple pendulum is used as the pivot point for another pendulum, attached to it. See Fig.~\ref{fig04:doublependulum} for a sketch. Assuming that all motion happens in a single vertical plane, find suitable generalized coordinates for the system and construct a Lagrangian. Then, assuming small oscillations, find the normal frequencies of the system for the special case $L_1=L_2\equiv L$ and $m_1=m_2\equiv m$.
\end{prob}

\begin{prob}
\label{pr04:chargedparticle}
A particle of mass $m$ and electric charge $q$ is exposed to a combination of a three-dimensional harmonic oscillator potential, $(1/2)m\o^2\vec r^2$, and a static uniform magnetic field $\vec B$. Show that the motion of the particle can be decomposed into a linear combination of normal modes and find the corresponding normal frequencies. Hint: in this problem, it is better to find the normal modes by inspecting the equations of motion rather than the Lagrangian.
\end{prob}
\chapter{Application: Relativistic Mechanics}
\label{chap:relmechanics}

\keywords{Minkowski space and metric, Lorentz transformation, rapidity, covariant and contravariant four-vectors, four-velocity, four-momentum.}

%%%%%%%%%%%%%%%%%%%%%%%%%%%%%%%%%%%%%%%%%%%%%%%%%%%%%%%%%%%%

\noindent Already when we first met the Lagrangian formalism, I advertised that it can accommodate both Newtonian and relativistic mechanics. It is time to keep the promise. I will not repeat the basics of relativistic physics, which you are well familiar with from you introductory-level physics courses. Instead, I will largely focus on the geometry of special relativity. This will naturally lead us to the four-vector formalism that we will heavily rely on in the field theory part of the course. In this~\chaptername, it will help us find a geometric interpretation for the action of a relativistic particle.

%%%%%%%%%%%%%%%%%%%%%%%%%%%%%%%%%%%%%%%%%%%%%%%%%%%%%%%%%%%%

\section{Spacetime Geometry}
\label{sec:spacetimegeometry}

Geometry as a domain of human knowledge is thousands of years old. However, it was not until the advent of special relativity that space and time began to be viewed as parts of a single geometric structure. Since this structure is not hardwired into our everyday experience, it is useful to introduce it following the mathematical analogy with the much more familiar geometry of Euclidean space.

%%%%%%%%%%%%%%%%%%%%%%%%%%%%%%%%%%%%%%%%%%%%%%%%%%%%%%%%%%%%

\subsection{Euclidean Versus Minkowski Space}

Let us start with the $n$-dimensional Euclidean space, $\R^n$. This is a real vector space, endowed with a scalar product. The scalar product of two vectors $\vec u,\vec v\in\R^n$ is defined in terms of their (Cartesian) components $u_i,v_i$ as\footnote{As in the previous \chaptername{}s, I will use the dot notation, common in physics. In the notation of \appendixname~\ref{appsec:scalarproduct}, the scalar product would be $\scal{\vec u}{\vec v}$.}
\begin{equation}
\skal uv\equiv\sum_{i=1}^nu_iv_i\;.
\label{ch05:euclskal}
\end{equation}
Having the scalar product at hand allows one to compute the length of a single vector, $\abs{\vec v}\equiv\sqrt{\skal vv}$. This in turn determines the distance of two points $A,B$ in Euclidean space, represented by their position vectors $\vec r_A,\vec r_B$, as $\abs{\vec r_B-\vec r_A}$. It is common to restrict to infinitesimally separated points so that $\vec r_B-\vec r_A\equiv\D\vec r$. All information about measuring distance in the Euclidean space is then encoded in the \emph{Euclidean metric}, expressing the squared distance of the nearby points,
\begin{equation}
\D s^2\equiv\D\vec r\cdot\D\vec r\;.
\label{ch05:euclmetric}
\end{equation}
Why bother spelling out explicitly the special case of infinitesimally separated points if we can easily compute the distance of any two points in space? The answer is that the concept of a metric can be generalized to geometric structures that do not constitute a vector space. You are going to need the more general notion of distance if you choose to study general relativity. I will return to this briefly in \chaptername~\ref{chap:geometryclassmech}.

We are now ready to introduce the \emph{Minkowski space} (or \emph{Minkowski spacetime}). This is a vector space, $\R^{1,n}$, isomorphic to $\R^{n+1}$. The difference from Euclidean space is in how the ``scalar product,'' and in turn the ``distance'' of two points, is defined. But first thing first. In the notation common in high-energy physics, elements of Minkowski space are denoted using a plain italic font, such as $v\in\R^{1,n}$. Individual components of Minkowski vectors are indicated with the help of a lowercase Greek superscript. The letters $\m,\n$ are most common. However, any other Greek letter may be used if needed for disambiguation of multiple indices. Hence $v^\m$, $v^\l$ and $v^\a$ may all represent the same object. To make things more confusing, the superscript $\m$ on $v^\m$ does not run from $1$ to $n+1$ but rather from $0$ to $n$. The individual components of the Minkowski vector are thus assembled as
\begin{equation}
v^\m\equiv(v^0,\vec v)\;.
\label{ch05:fourvectordef}
\end{equation}
We continue using $\vec v$ for the part of $v^\m$ that collects the components $v^1,\dotsc,v^n$. The notation suggests that $\R^n$ is a natural subspace of $\R^{1,n}$. The Minkowski space is therefore most appropriately interpreted as being isomorphic to $\R\oplus\R^n$. The first part, $\R$, represents time and carries the ``temporal part'' of the Minkowski vector, $v^0$. The second part, $\R^n$, represents space and carries the ``spatial part'' of the Minkowski vector, $\vec v$ or $v^i$ with $i=1,\dotsc,n$.

After the long notational prelude, we next introduce the \emph{Minkowski dot product}, utilizing the Einstein summation convention,
\begin{equation}
u\cdot v\equiv g_{\m\n}u^\m v^\n\equiv-u^0v^0+\skal uv\;.
\label{ch05:minkskal}
\end{equation}
The constant matrix $g_{\m\n}\equiv\diag(-1,+1,\dotsc,+1)$ is the \emph{Minkowski metric}. A special case of a Minkowski dot product of two spacetime vectors is the Minkowski square of a single vector, $v^2\equiv g_{\m\n}v^\m v^\n$. With the Minkowski vectors $x^\m_A$ and $x^\m_B$, indicating the positions in space and time of two points (\emph{events}) $A,B$, the separation of the latter is encoded in the quantity $(x_B-x_A)^2$. Specifically, the separation of two nearby points such that $x^\m_B-x^\m_A\equiv\D x^\m$ is measured by the \emph{spacetime interval},
\begin{equation}
\boxed{\D s^2\equiv\D x\cdot\D x=g_{\m\n}\D x^\m\D x^\n\;.}
\label{ch05:spacetimeinterval}
\end{equation}

\begin{watchout}%
The notation is chosen to highlight as much as possible the similarity between the Euclidean and Minkowski space. It is therefore useful to explicitly point out some important differences. First, the dot product~\eqref{ch05:minkskal} is not a scalar product in the sense of~\appendixname~\ref{appsec:scalarproduct}. This is because the bilinear form defined by $g_{\m\n}$ is symmetric but not positive-definite. Likewise, the square of a nonzero Minkowski vector, $v^2=v\cdot v=g_{\m\n}v^\m v^\n$, can be positive but also zero or even negative. We will see later that these different options have distinct physical interpretations. By the same token, the spacetime interval~\eqref{ch05:spacetimeinterval} should generally not be thought of as measuring a ``distance,'' even though it reduces to the Euclidean metric~\eqref{ch05:euclmetric} in case $\D x^0=0$. Mathematically, the Minkowski space is a special case of a \emph{pseudo-Euclidean space}, $\R^{m,n}\cong\R^m\oplus\R^n$, which has $m$ ``time-like'' and $n$ ``space-like'' coordinates.

Let me append a remark that has no implications for physics yet is highly relevant in practice. There is a natural isomorphism connecting the pseudo-Euclidean spaces $\R^{1,n}$ and $\R^{n,1}$, which allows both of them to be viewed as the same Minkowski space. This gives rise to two different conventions which are, sadly, both commonly used in physics. In this course, I will use consistently the ``mostly-plus'' convention~\eqref{ch05:minkskal} for the Minkowski dot product, which has the advantage that it includes the positive-definite Euclidean scalar product as a special case. This is the choice common in the general theory of relativity, and generally speaking in gravitational physics. The opposite, ``mostly-minus'' convention where~\eqref{ch05:minkskal} is replaced with $u\cdot v\equiv u^0v^0-\skal uv$, is common in particle physics. It is good to be aware that many authors choose, without spelling it out explicitly, a convention pertinent to their area of expertise.
\end{watchout}

%%%%%%%%%%%%%%%%%%%%%%%%%%%%%%%%%%%%%%%%%%%%%%%%%%%%%%%%%%%%

\subsection{Isometry Group}

With the basic definitions out of the way, let us now delve deeper into the structure of the Euclidean and Minkowski spaces. We observe that the Euclidean metric as defined by~\eqref{ch05:euclmetric} does not change when the Cartesian coordinates $r_i$ are transformed in a specific manner. We can translate the coordinates as $\vec r\to\vec r'\equiv\vec r+\vec a$ where $\vec a$ is an arbitrary constant vector. We can also rotate the coordinates as $r_i\to r_i'\equiv R_{ij}r_j$, where $R_{ij}$ is an arbitrary orthogonal $n\times n$ matrix. Both proper and improper rotations are allowed. Any transformation, composed of a sequence of successive translations or rotations, will likewise leave the Euclidean metric unchanged.

What we have just discovered is an example of a mathematical structure known as a \emph{group}. This is any nonempty set $G$ endowed with a binary operation (``multiplication'') $\circ:G\times G\to G$, satisfying the following axioms:
\begin{itemize}
\item The multiplication is associative: $g_1\circ(g_2\circ g_3)=(g_1\circ g_2)\circ g_3$ for all $g_1,g_2,g_3\in G$.
\item There is a unit (identity) element $e\in G$ such that $e\circ g=g\circ e=g$ for all $g\in G$.
\item Each element $g\in G$ has an inverse, usually denoted $g^{-1}$, such that $g\circ g^{-1}=g^{-1}\circ g=e$.
\end{itemize}
The above axioms automatically guarantee that the unit element is unique, as is the inverse $g^{-1}$ of any $g\in G$.

\begin{illustration}%
A \emph{finite group} is a group that, as a set, has a finite number of elements. The simplest, though trivial, example of a finite group is one consisting solely of the unit element, $\{e\}$. The only possible choice of the group operation is $e\circ e=e$, which implies that $e$ is its own inverse. The smallest nontrivial group is $\Z_2=\{e,a\}$ with the multiplication rules
\begin{equation}
e\circ e=e\;,\qquad
e\circ a=a\circ e=a\;,\qquad
a\circ a=e\;.
\end{equation}
Another way to view $\Z_2$ is as the set $\{0,1\}$ where the group operation is defined by addition modulo $2$. The $\Z_2$ is an example of a \emph{cyclic group}, $\Z_n=\{e,a,a^2,\dotsc,a^{n-1}\}$ with positive integer $n$, where the powers of $a$ are defined by iterated multiplication and $a^n$ is required to equal $e$. This is equivalent to the set $\{0,1,\dotsc,n-1\}$ where the group operation is defined by addition modulo $n$.

All the groups mentioned so far are \emph{Abelian}: their group operation is commutative, that is $g_1\circ g_2=g_2\circ g_1$ for all $g_1,g_2\in G$. The smallest \emph{non-Abelian} group is the \href{https://en.wikipedia.org/wiki/Symmetric_group}{symmetric group} $S_3$, the group of all permutations of three elements. This can also be viewed as the group of symmetries (rotations and reflections) of an equilateral triangle, whereby it constitutes a special case, $D_3$, of a \href{https://en.wikipedia.org/wiki/Dihedral_group}{dihedral group}, $D_n$.
\end{illustration}

A transformation that preserves a metric is called an \emph{isometry}. The set of all isometries of the Euclidean metric is a group where the group operation is defined by composition of the isometries as maps. The associativity of this operation follows from the equivalent general property of maps. The unit element corresponds to the identity map, $\vec r\to\vec r$. Finally, any translation or rotation can be inverted, hence any combination of translations and rotations has an inverse as well. The \emph{isometry group} of the Euclidean space $\R^n$ is infinite for any $n$. It has the structure of a \emph{Lie group}, which roughly speaking means that the elements of the isometry group can be parameterized smoothly by a finite number of parameters, and that the operations of group multiplication and inverse are smooth as well.

\begin{illustration}%
There are several important examples of Lie groups that are worth being aware of. The set of all orthogonal $n\times n$ matrices constitutes the \emph{orthogonal group}, $\gr{O}(n)$. The subset consisting of matrices with a unit determinant, corresponding to proper rotations in $\R^n$, is itself a Lie group: the \emph{special orthogonal group} $\gr{SO}(n)$. The isometries of the Euclidean metric span the \emph{Euclidean group} $\gr{E}(n)$, also often referred to as $\gr{ISO}(n)$. The subgroup consisting of proper rotations and translations and their combinations is known as the \emph{special Euclidean group}, $\gr{SE}(n)$.

The concept of an isometry can be promoted to complex spaces: just think of transformations preserving the scalar product $\scal{\vec u}{\vec v}=\sum_{i=1}^nu_i^*v_i$ in $\C^n$. In complex spaces, the notion of an orthogonal map is replaced with a unitary map (see \appendixname~\ref{appsec:scalarproduct}). This leads to the \emph{unitary group} $\gr{U}(n)$, the set of all unitary $n\times n$ matrices. Other, related structures follow the same pattern as in real spaces. Thus, the \emph{special unitary group} $\gr{SU}(n)$ only includes matrices with a unit determinant. Similarly, adding to the unitary group all translations in $\C^n$ and their combinations with unitary transformations, we arrive at the group $\gr{ISU}(n)$.
\end{illustration}

Equipped with all the new terminology, we can now switch from Euclidean to Minkowski space. What are the isometries of the Minkowski metric, encoded in the spacetime interval~\eqref{ch05:spacetimeinterval}? These can again be classified into translations and ``rotations'' and their arbitrary combinations. The translations amount to a mere shift $x^\m\to x'^\m\equiv x^\m+a^\m$ with an arbitrary constant Minkowski vector $a^\m$. The ``rotations'' correspond to linear transformations of the spacetime coordinates $x^\m$, usually expressed as
\begin{equation}
\boxed{x^\m\to x'^\m\equiv\Lambda^\m_{\phantom\m\n}x^\n\;,\quad\text{where}\quad
g_{\m\n}\Lambda^\m_{\phantom\m\a}\Lambda^\n_{\phantom\n\b}=g_{\a\b}\;.}
\label{ch05:Lorentztransfodef}
\end{equation}
The latter condition follows from the requirement of invariance of the spacetime interval, or equivalently the invariance of $g_{\m\n}$. It can be written compactly in a matrix form as $\Lambda^Tg\Lambda=g$. Matrices $\Lambda^\m_{\phantom\m\n}$ with this property are called \emph{Lorentz transformations}, and they span the \emph{Lorentz group}, $\gr{O}(1,n)$. Adding back the spacetime translations, we arrive at the complete isometry group of the Minkowski metric, known as the \emph{Poincar\'e group}, $\gr{ISO}(1,n)$. Our next task will be to understand in more detail the structure and physical interpretation of Lorentz transformations. Before, let me however insert a brief historical remark.

I have followed quite closely the historical development, leading from an explicit construction of the (Euclidean and Minkowski) geometry to the observation that it possesses a distinguished set of symmetries. One can however reverse the reasoning and argue that the geometry of space(time) is largely characterized by its set of isometries. Thus, all possible types of geometries can be enumerated by classifying possible isometry groups. This is the essence of the \emph{Erlangen program}, spelled out by Felix Klein in 1872. Klein's philosophy underlies much of our modern understanding of the fundamental laws of nature, whereby specifying the symmetry is often the starting point of a construction of new physical theories.

%%%%%%%%%%%%%%%%%%%%%%%%%%%%%%%%%%%%%%%%%%%%%%%%%%%%%%%%%%%%

\subsection{Lorentz Transformations}

Much about the structure of the Lorentz group can be learned using a simple trick, going back to the early days of relativity theory. While not actively used by practitioners anymore, it is still reflected in the so-called Wick rotation technique in quantum field theory. Suppose that we trade the time coordinate $x^0$ for $x^{n+1}\equiv\I x^0$. This turns the spacetime interval~\eqref{ch05:spacetimeinterval} into
\begin{equation}
\D s^2=(\D x^1)^2+\dotsb+(\D x^n)^2+(\D x^{n+1})^2\;,
\end{equation}
and thus the Minkowski metric in $\R^{1,n}$ into the Euclidean metric in $\R^{n+1}$. The change does not affect the spatial coordinates $x^i$. Accordingly, the Lorentz group $\gr{O}(1,n)$ contains the group of purely spatial rotations $\gr{O}(n)$ as a subgroup. The nontrivial part of the Lorentz group are the ``rotations'' connecting space and time.

To see what these might be, we go to $\R^{n+1}$ and consider a rotation by angle $\vp$ in the plane of $x^i$ (with fixed $i$) and $x^{n+1}$. This has the matrix form
\begin{equation}
\begin{pmatrix}
x'^{n+1}\\
x'^i
\end{pmatrix}=
\begin{pmatrix}
\cos\vp & -\sin\vp\\
\sin\vp & \cos\vp
\end{pmatrix}
\begin{pmatrix}
x^{n+1}\\
x^i
\end{pmatrix}\;.
\end{equation}
All we have to do now is replace $x^{n+1}\to\I x^0$. Since the Lorentz transformation should be real, we also substitute $\vp\to\I w$ with real $w$. Using the identities $\sin\I w=\I\sinh w$ and $\cos\I w=\cosh w$, our Euclidean rotation becomes
\begin{equation}
\begin{pmatrix}
x'^0\\
x'^i
\end{pmatrix}=
\begin{pmatrix}
\cosh w & -\sinh w\\
-\sinh w & \cosh w
\end{pmatrix}
\begin{pmatrix}
x^0\\
x^i
\end{pmatrix}\;.
\label{ch05:Lorentztransfomatrix}
\end{equation}
The parameter $w$ is called \emph{rapidity}. Another, more familiar form of the Lorentz transformation is obtained by trading the rapidity for the dimensionless velocity parameter $\b$ through
\begin{equation}
\b=\tanh w\quad\Leftrightarrow\quad
w=\log\sqrt{\frac{1+\b}{1-\b}}\;.
\label{ch05:wbeta}
\end{equation}
Then, $\cosh w=1/\sqrt{1-\b^2}\equiv\g$ and $\sinh w=\b\g$ so that~\eqref{ch05:Lorentztransfomatrix} becomes
\begin{equation}
\begin{pmatrix}
x'^0\\
x'^i
\end{pmatrix}=
\begin{pmatrix}
\g & -\b\g\\
-\b\g & \g
\end{pmatrix}
\begin{pmatrix}
x^0\\
x^i
\end{pmatrix}\;.
\label{ch05:Lorentztransfomatrixphysical}
\end{equation}

\begin{watchout}%
This is the right place to elaborate on the actual physical interpretation of the Minkowski spacetime and Lorentz transformations. Upon the identification $x^0\equiv ct$, the spacetime coordinates $x^\m=(ct,\vec x)$ carry information about the time $t$ and place $\vec x$ of an event. The geometric transformation~\eqref{ch05:Lorentztransfomatrix} corresponds to a change of \emph{inertial reference frame} (IMF). Denote as IRF and IRF' the frames attached respectively to the coordinates $x^\m$ and $x'^\m$. The origin of IRF' is at $\vec x'=\vec0$, which corresponds to $t=t'\cosh w$ and $x^i=ct'\sinh w$. Hence, the origin of IRF' moves with respect to IRF with velocity $x^i/t=c\tanh w=\b c$ along the $i$-th coordinate axis. This gives the dimensionless parameter $\b$ a concrete physical content: it is the speed of the Lorentz \emph{boost}~\eqref{ch05:Lorentztransfomatrixphysical} measured in units of the speed of light. Denoting the relative velocity of the two IRFs as $u\equiv\b c$, the Lorentz boost acquires the familiar form
\begin{equation}
t'=\g\left(t-\frac{ux^i}{c^2}\right)\;,\qquad
x'^i=\g(x^i-ut)\;.
\end{equation}
The other spatial coordinates, $x^j$ with $j\neq i$, remain unaffected. Finally, the invariance of the spacetime interval~\eqref{ch05:spacetimeinterval}, $\D s^2=-c^2\D t^2+\D\vec x\cdot\D\vec x$, under Lorentz transformations encodes, among others, the relativistic postulate of constancy of the speed of light.
\end{watchout}

The expression~\eqref{ch05:Lorentztransfomatrixphysical} for the Lorentz boost connecting different IRFs assumes that the relative motion of the IRFs is aligned with one of the coordinate axes. It is a special case of a general formula for the Lorentz boost with velocity $\vec u\equiv\vec\b c$ of an arbitrary direction, which I show here for your interest without proof,
\begin{equation}
\boxed{ct'=\g(ct-\skal\b x)\;,\qquad
\vec x'=\vec x+\frac{\g-1}{\vec\b^2}(\skal\b x)\vec\b-\vec\b\g ct\;,}
\end{equation}
where $\g\equiv1/\sqrt{1-\vec u^2/c^2}=1/\sqrt{1-\vec\b^2}$. The same transformation can be represented in terms of the matrix elements $\Lambda^\m_{\phantom\m\n}$,
\begin{equation}
\Lambda^0_{\phantom{0}0}=\g\;,\qquad
\Lambda^0_{\phantom{0}i}=\Lambda^i_{\phantom{i}0}=-\g u_i/c\;,\qquad
\Lambda^i_{\phantom ij}=\d_{ij}+(\g-1)\frac{u_iu_j}{\vec u^2}\;.
\end{equation}
In the specific case of $n=3$ spatial dimensions, the entire matrix $\Lambda^\m_{\phantom\m\n}$ reads
\begin{equation}
\Lambda=\begin{pmatrix}
\g & -\g{u_x}/c & -\g{u_y}/c & -\g{u_z}/c\\
-\g{u_x}/c & 1+(\g-1){u_x^2}/{\vec u^2} & (\g-1){u_xu_y}/{\vec u^2} & (\g-1){u_xu_z}/{\vec u^2}\\
-\g{u_y}/c & (\g-1){u_yu_x}/{\vec u^2} & 1+(\g-1){u_y^2}/{\vec u^2} & (\g-1){u_yu_z}/{\vec u^2}\\
-\g{u_z}/c & (\g-1){u_zu_x}/{\vec u^2} & (\g-1){u_zu_y}/{\vec u^2} & 1+(\g-1){u_z^2}/{\vec u^2}
\end{pmatrix}\;.
\end{equation}

%%%%%%%%%%%%%%%%%%%%%%%%%%%%%%%%%%%%%%%%%%%%%%%%%%%%%%%%%%%%

\section{Four-Vectors in Relativistic Physics}

The case of $n=3$ spatial dimensions is obviously of most practical relevance. This lies behind the term \emph{four-vector} for vectors in the four-dimensional Minkowski space $\R^{1,3}$. Due to a lack of an appropriate, brief name for vectors in $\R^{1,n}$, I will from now on use consistently the same term, even though most of the material in this~\chaptername{} is valid for Minkowski spacetimes of arbitrary dimension.

The importance of vectors is stressed early on in mathematics and physics education. We know that expressing any mathematical identity or physical law in terms of vectors guarantees their validity independently of the choice of a coordinate frame. In special relativity, this observation naturally extends to the invariance of physical laws under a change of IRF. To make the latter manifest, we need to express physical laws in terms of four-vectors (or higher-rank tensors under Lorentz transformations). To that end, we must be able to pair vector and scalar observables and fold them into four-vectors, that is objects of the type~\eqref{ch05:fourvectordef} whose components transform under a change of IRF as in~\eqref{ch05:Lorentztransfodef}. This is in practice best done case by case.

%%%%%%%%%%%%%%%%%%%%%%%%%%%%%%%%%%%%%%%%%%%%%%%%%%%%%%%%%%%%

\subsection{Relativistic Kinematics}

\begin{figure}[t]
\sidecaption[t]
\includegraphics[width=2.9in]{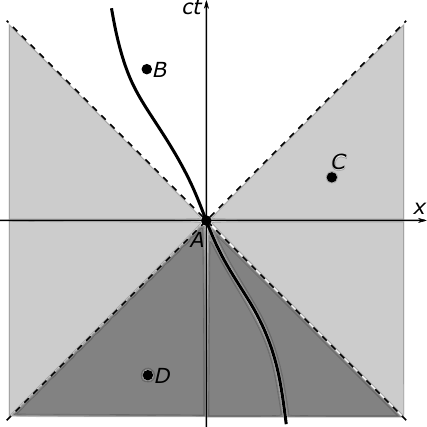}
\caption{Illustration of the separation of Minkowski spacetime into distinct regions with specific properties under orthochronous Lorentz transformations. The color coding is such that the past is dark, the present is light gray and the future is bright. The thick solid curve shows a possible worldline of an object (particle) passing through the event $A$ located at the origin of spacetime.}
\label{fig05:spacetime}
\end{figure}

We are already familiar with one concrete example of a four-vector: the coordinate four-vector $x^\m=(ct,\vec x)$, which is the basic parameter of Minkowski space. While the values of the individual components of this four-vector change upon switching the IRF, the change is constrained by the requirement that the Minkowski square $x^2\equiv x\cdot x$ remains intact. Moreover, it is easy to check that a Lorentz transformation of the type~\eqref{ch05:Lorentztransfomatrix} does not change the sign of $x^0$ provided $x^2<0$. Such time-direction-preserving transformations are called \emph{orthochronous}. The combination of the sign of $x^2$ and (in case of $x^2<0$) the sign of $x^0$ divides the spacetime into three distinct regions. These are indicated in Fig.~\ref{fig05:spacetime} by color and by representative events $B,C,D$. The Minkowski square $x^2_B,x^2_C,x^2_D$ defines the spacetime interval separating these events from the event $A$ placed at the origin of spacetime. Likewise, the sign of $x^0_B,x^0_C,x^0_D$ indicates the temporal order of $B,C,D$ with respect to $A$.

For the event $B$, we find $x_B^2<0$ and at the same time $x^0_B>0$. These two properties together carve out a part of the spacetime that is preserved by orthochronous Lorentz transformations. The characteristic feature of this domain is that it is possible to change frame to an IRF' such that $\vec x'_B=\vec0$. In this frame, the events $A$ and $B$ happen at the same place in a fixed order: first $A$, then $B$. Accordingly, the domain represented by $B$ is referred to as the \emph{future}.

For the event $C$, we have $x_C^2>0$. Here it is always possible to change frame to an IRF' such that $t'_C=0$. In this frame, the events $A$ and $C$ happen at the same time, that is are simultaneous. What is typical is that the temporal order of the events can be changed by a suitable choice of IRF. Accordingly, the events $A$ and $C$ cannot be in any causal connection to each other. For these reasons, the entire region represented by $C$ is called the \emph{present}.

Finally, for the event $D$, we find $x_D^2<0$ but $x^0_D<0$. These two properties are again preserved by orthochronous Lorentz transformations. It is possible to find an IRF' in which the events $A$ and $D$ take place at the same location, but this time the order is opposite than for $B$: first $D$, and only then $A$. Accordingly, the domain represented by $D$ is called the \emph{past}.

Suppose now that we track the motion of an object such as a particle, passing through $A$. With the exception of strictly massless particles such as photons, the velocity of motion must be strictly smaller than the speed of light. The trajectory (\emph{worldline}) of the particle is then \emph{timelike}, $\D s^2<0$, and must lie entirely in the past and future with respect to $A$. It is common to parameterize the worldline by the \emph{proper time} $\tau$, that is time measured in an instantaneous rest frame of the particle. By construction, $\D s^2=-c^2\D\tau^2$. The invariance of the spacetime interval under Lorentz transformation guarantees that the proper time is a tautological \emph{Lorentz scalar} that takes the same value in all IRFs.

In any other IRF, the particle moves with some velocity $\vec v=\Od{\vec x}{t}$. This implies
\begin{equation}
-c^2\D\tau^2=\D s^2=-c^2\D t^2+\D\vec x^2=-c^2\D t^2/\g_{\vec v}^2\;,
\end{equation}
where $\g_{\vec v}\equiv1/\sqrt{1-\vec v^2/c^2}$ is the $\g$-factor associated with the speed of the particle, not to be confused with the $\g$-factor in~\eqref{ch05:Lorentztransfomatrixphysical} of a Lorentz transformation connecting two different IRFs. We find that the proper time is related to time measured in any IRF by \emph{time dilation}, $\D t=\g_{\vec v}\D\tau$. This allows us to construct the \emph{four-velocity}
\begin{equation}
\boxed{u^\m\equiv\OD{x^\m}{\tau}=\g_{\vec v}(c,\vec v)\;.}
\label{ch05:fourvelocity}
\end{equation}
The four-velocity carries complete information about the kinematics of the particle, and by construction transforms as a four-vector. It is normalized as
\begin{equation}
u^2=\g_{\vec v}^2(-c^2+\vec v^2)=-c^2\;.
\label{ch05:fourvelocitysquared}
\end{equation}
With the four-velocity at hand, we can proceed towards a four-vector formulation of relativistic dynamics.

%%%%%%%%%%%%%%%%%%%%%%%%%%%%%%%%%%%%%%%%%%%%%%%%%%%%%%%%%%%%

\subsection{Relativistic Dynamics}

We know that in case the Lagrangian of a particle does not depend explicitly on its position, the corresponding conjugate momentum is conserved. In addition, assuming the Lagrangian does not depend explicitly on time implies conservation of energy. However, in relativity, position and time are folded together into a single four-vector, $x^\m$. We therefore expect that also the corresponding constants of motion, momentum $\vec p$ and energy $E$, should together span a single four-vector. The only problem is that we do not have a Lagrangian for relativistic motion yet. We bypass this problem by constructing a new four-vector and \emph{defining} its components to be energy and momentum. This leads to the notion of \emph{four-momentum},
\begin{equation}
\boxed{p^\m\equiv mu^\m\equiv(E/c,\vec p)\;,}
\label{ch05:fourmomentum}
\end{equation}
where $m$ is the mass of the object (particle) in its rest frame, also called the \emph{rest mass}. Just like the proper time, this is a tautological Lorentz scalar, which guarantees that $p^\m\equiv mu^\m$ indeed is a four-vector. The second equality in~\eqref{ch05:fourmomentum} serves as a definition of $E$ and $\vec p$; the extra factor of $1/c$ is inserted to maintain standard units.

By matching~\eqref{ch05:fourmomentum} to~\eqref{ch05:fourvelocity}, we deduce how the energy and momentum are related to the particle's velocity,
\begin{equation}
E=\g_{\vec v}mc^2\;,\qquad
\vec p=\g_{\vec v}m\vec v\;.
\label{ch05:relenergymomentum}
\end{equation}
Taking a ratio then gives us a useful inverse expression for the velocity, $\vec v=\vec pc^2/E$. Finally, \eqref{ch05:fourvelocitysquared} implies that $p^2=-m^2c^2$, which is equivalent to
\begin{equation}
E=\sqrt{\vec p^2c^2+m^2c^4}\;.
\label{ch05:dispersion}
\end{equation}
Note how all the standard relations of relativistic dynamics immediately follow once we implement Lorentz invariance through four-vectors!

\begin{illustration}%
We know that photons always propagate at the speed of light, and having all the $\g$-factors around could give rise to concerns as to whether the four-vector formalism is suitable for photons at all. Here comes a positive surprise: having reached~\eqref{ch05:dispersion}, it is no problem to take the limit $m\to0$ therein, so that
\begin{equation}
E=\abs{\vec p}c\quad\Rightarrow\quad
\vec v=\frac{\vec p}{\abs{\vec p}}c\;.
\end{equation}
This proves that, indeed, any massless particle necessarily moves at the speed of light. The relation~\eqref{ch05:dispersion} is also very useful for massive particles since it allows us to dispense with the velocity altogether and work solely with the physically more important concepts of energy and momentum. After all, at modern colliders, even relatively heavy particles are accelerated to speeds extremely close to $c$, or equivalently $\g$-factors much higher than one. In such a kinematical regime, the velocity is really not very useful. In case a purely kinematical quantity is needed, the rapidity is much more convenient.
\end{illustration}

So far, I have used four-vectors largely as a conceptual device. However, four-vectors are also a great computational tool suitable for solving concrete problems. I will now illustrate this on a couple of explicit examples. The purpose is not to offer a representative sample of relativistic physics, but rather to demonstrate how the possibility to change perspective by switching a reference frame, baked into the coordinate-free language of four-vectors, sheds light on some rather surprising aspects of relativistic kinematics.

\begin{illustration}%
\label{ex05:fixedtarget}%
Suppose we want to study head-on collisions of two particles with equal rest mass. There are two standard ways of doing so. In a ``fixed-target'' experiment, one of the particles (target) is initially at rest and the other (projectile) hits it. In a ``collider'' experiment, the two particles are accelerated to opposite velocities and then collided. The two setups are related by a Lorentz transformation. I will use the label LAB (laboratory frame) for the fixed-target setup, and CoM (center-of-mass frame) for the collider setup. Let us relate the total four-momentum $P^\m$ of the two-particle system in the two frames. On the one hand,
\begin{equation}
P^\m_\mathrm{LAB}=(E/c+mc,\vec p)\;,
\end{equation}
where $E$ and $\vec p$ are the energy and momentum of the projectile in the LAB frame. This implies
\begin{equation}
P_\mathrm{LAB}^2=-[(E/c)^2+2mE+m^2c^2]+\vec p^2=-2m(E+mc^2)\;,
\end{equation}
where I used~\eqref{ch05:dispersion}. On the other hand, $P_\mathrm{CoM}^\m=(2E'/c,\vec0)$, where $E'$ is the energy of either of the particles in the CoM frame. This leads to $P_\mathrm{CoM}^2=-4E'^2/c^2$. However, the square of a four-vector is invariant under Lorentz transformations, that is independent of the choice of IRF. Comparing our two results then gives $2E'^2=mc^2(E+mc^2)$. The kinematics of the transformation between the LAB and CoM frames is best highlighted by expressing the energies in terms of the respective $\g$-factors as $E=\g_\mathrm{LAB}mc^2$ and $E'=\g_\mathrm{CoM}mc^2$. The rest mass then drops out and we arrive at a purely kinematical relation,
\begin{equation}
2\g_\mathrm{CoM}^2=\g_\mathrm{LAB}+1\;.
\end{equation}
To see how the choice of setup for particle experiments matters, consider a proton at LHC, accelerated to roughly $\g_\mathrm{CoM}\approx7000$. If you wanted to study the same physics using a fixed-target experiment, you would have to accelerate the projectile protons to a whopping $\g_\mathrm{LAB}\approx98\times10^6$. Surely, you would only need to accelerate a single proton instead of two, but that would still increase the energy costs of accelerating the particles by a factor of $\g_\mathrm{LAB}/(2\g_\mathrm{CoM})\approx\g_\mathrm{CoM}\approx7000$.
\end{illustration}

\begin{illustration}%
The spectrum of ultra-high-energy cosmic radiation observed on Earth is expected to be strongly suppressed at energies above certain threshold. This can be explained by inelastic scattering of the radiation on photons of the \emph{cosmic microwave background} (CMB). Consider a high-energy proton $p$ that collides head-on with a background photon $\g$, thus producing the lightest hadron, the neutral pion $\pi^0$, through the process $p+\g\to p+\pi^0$. We would like to find the minimum energy of the proton required for the process to be kinematically viable. We will use the same trick as in~\refex{ex05:fixedtarget}: evaluate the squared total four-momentum of the system in two different ways and compare the results.

For a head-on collision, we can always orient the coordinate frame so that the momentum of both particles is aligned with the $x$-axis. We can therefore write the four-momenta of the proton and the photon before the collision as
\begin{equation}
\begin{split}
p^\m_p&=(E_p/c,p_p,0,0)\;,\\
p^\m_\g&=(E_\g/c,p_\g,0,0)=(E_\g/c,-E_\g/c,0,0)\;.
\end{split}
\end{equation}
From here, we find the square of the total four momentum,
\begin{equation}
(p_p+p_\g)^2=-m_p^2c^2-\frac{2E_\g}{c}\left(\frac{E_p}{c}+p_p\right)\approx-m_p^2c^2-\frac{4E_pE_\g}{c^2}\;,
\end{equation}
where I used that the proton is ultrarelativistic so that $p_p\approx E_p/c$.

To determine the threshold for pion production, we switch to the CoM frame. The minimum initial proton energy required is such that, in this frame, the total momentum after the collision vanishes. Hence, the threshold corresponds to $P^\m_\mathrm{CoM}=((m_p+m_\pi)c,\vec0)$. Taking the square and comparing to our previous result for $P^2$, we can extract the proton energy,
\begin{equation}
E_p\approx\frac{m_\pi(m_\pi+2m_p)c^4}{4E_\g}\;.
\end{equation}
You can guess that the threshold is going to be fairly high from the fact that $E_p$ is inversely proportional to the energy of the CMB photon. Even then, the concrete numerical value is stunning. We need the rest masses, $m_pc^2\approx938\,\text{MeV}$ and $m_\pi c^2\approx135\,\text{MeV}$. In addition, from the CMB temperature of $T_\mathrm{CMB}\approx2.7\,\text{K}$, we estimate $E_\g=k_\mathrm{B}T_\mathrm{CMB}\approx2.3\times10^{-4}\,\text{eV}$. This gives in the end $E_p\approx3\times10^{20}\,\text{eV}$ or, if you wish, $E_p/(m_\pi c^2)\approx2\times10^{12}$ using the rest energy of the pion as a unit. This is another example of extreme relativistic kinematics. The above-found threshold for the energy of cosmic radiation is known as the \href{https://en.wikipedia.org/wiki/Greisen%E2%80%93Zatsepin%E2%80%93Kuzmin_limit}{Greisen--Zatsepin--Kuzmin limit}.
\end{illustration}

%%%%%%%%%%%%%%%%%%%%%%%%%%%%%%%%%%%%%%%%%%%%%%%%%%%%%%%%%%%%

\section{Lagrangian Formulation of Relativistic Mechanics}

Having covered thoroughly the mathematical framework of special relativity, we will now turn to our central question: how to formulate relativistic dynamics using the Lagrangian formalism. In order to keep contact with~\chaptername~\ref{chap:Lagmechanics}, I will temporarily revert to the notation used therein, whereby the position of a particle and its velocity are, respectively, $\vec r$ and $\dot{\vec r}$. Suppose the particle moves in an external field characterized by a potential energy $V(\vec r)$. We are thus looking for a Lagrangian $L(\vec r,\dot{\vec r})$ such that:
\begin{itemize}
\item The conjugate momentum $\Pd{L}{\dot{\vec r}}$ corresponds to the physical momentum $\vec p=\g_{\vec v}m\vec v$ in accord with~\eqref{ch05:relenergymomentum}.
\item The Hamilton function $H=\dot{\vec r}\cdot\vec p-L$ gives the physical energy $E=\g_{\vec v}mc^2+V(\vec r)$, likewise in accord with~\eqref{ch05:relenergymomentum}.
\end{itemize}
This problem has a unique solution,
\begin{equation}
\boxed{L(\vec r,\dot{\vec r})=-mc^2\sqrt{1-\frac{\dot{\vec r}^2}{c^2}}-V(\vec r)\;.}
\label{ch05:Lagrangian}
\end{equation}
Note that the Lagrangian no longer amounts to the difference of kinetic and potential energy. However, it does reproduce the familiar Lagrangian of a nonrelativistic particle upon power expansion in velocity,
\begin{equation}
L=-mc^2+\frac12m\dot{\vec r}^2-V(\vec r)+\bigO(\dot{\vec r}^4)\;.
\end{equation}
With the Lagrangian at hand, we can derive the corresponding \emph{equation of motion} (EoM) using the general Lagrange equation~\eqref{ch01:Lagrangeeq}. This leads to the expected result
\begin{equation}
\OD{\vec p}{t}=\PD{L}{\vec r}=-\grad V(\vec r)\;.
\label{ch05:EoMrelativistic}
\end{equation}

\begin{watchout}%
One a first encounter with special relativity, one is usually served a more or less elaborate argument explaining why Newton's $\vec F=m\ddot{\vec r}$ no longer holds, yet $\vec F=\Od{\vec p}{t}$ still does. Look how helpful it is to change perspective. From the Lagrangian point of view, the fact that the EoM contains $\Od{\vec p}{t}$ is obvious; we do not even need to know what exactly the Lagrangian looks like. That the right-hand side of~\eqref{ch05:EoMrelativistic} is nothing but force is now a mere matter of definition. We can simply postulate that force equals $-\grad V$, or equivalently, that the work done by the force along some path gives the difference of potential energy at the endpoints of the path. All this of course does not mean that the laws of special relativity boil down to a bunch of definitions. Physics is an empirical science, remember? The actual content of special relativity is encoded in the empirically justified requirement of Lorentz invariance and in the concrete form of the Lagrangian~\eqref{ch05:Lagrangian}. This was in turn deduced from the assumption that energy and momentum together constitute a four-vector.
\end{watchout}

We have not done all the work in the previous sections to now just dump the four-vector formalism and go back to $\vec r$ and $\dot{\vec r}$ forever. The relativistic action has a neat geometric interpretation that manifests itself in the four-vector language. Consider the simplest possible system: a free particle. Physically, this guarantees full Lorentz invariance; the presence of a time-independent potential energy $V(\vec r)$ would require choosing a fixed reference frame. Thinking of the trajectory of the particle as a geometric curve in $\R^{1,n}$, we parameterize it as $x^\m(s)$ with an arbitrarily chosen parameter $s$, only subject to a fixed boundary condition at the endpoints of the worldline. The action corresponding to the Lagrangian~\eqref{ch05:Lagrangian} with $V(\vec r)=0$ can then be written as
\begin{equation}
S=-mc\int_{s_1}^{s_2}\sqrt{-g_{\m\n}\D x^\m\D x^\n}=-mc\int_{s_1}^{s_2}\D s\,\sqrt{-g_{\m\n}\OD{x^\m}{s}\OD{x^\n}{s}}\;.
\label{ch05:actionfree}
\end{equation}
This might ring a bell: our relativistic action copies verbatim the problem of finding a path of shortest distance that we dealt with in Sect.~\ref{subsec:shortestdistance}. The only difference is that we have replaced the Euclidean metric with $g_{\m\n}$ and added a minus sign to account for the fact that the worldline is timelike. Geometrically, the variational problem for a free relativistic particle is therefore equivalent to the problem of finding a path that minimizes the proper time elapsed during the motion. I encourage you do take up the exercise of deriving the EoM from the action~\eqref{ch05:actionfree}. You should find that it is equivalent to conservation of relativistic momentum. I will skip the details since we will now deal with a more general problem, of which the free particle is~a~special~case.

%%%%%%%%%%%%%%%%%%%%%%%%%%%%%%%%%%%%%%%%%%%%%%%%%%%%%%%%%%%%

\subsection{Particle in Electromagnetic Field}
\label{subsec:particleEMfield}

We previously dropped the potential energy $V(\vec r)$ since it requires a choice of reference frame and thus obscures the Lorentz invariance of relativistic dynamics. Even in nonrelativistic mechanics, the presence of an external field violates momentum conservation, and Galilei invariance of the dynamical laws to that matter. There, the solution to the conundrum is simple though. One recalls that in practice, an ``external field'' models interaction with a source of the field whose dynamics is of no interest to us. It can thus always be eliminated in favor of a pair interaction between dynamical objects as we did in Sect.~\ref{subsec:2body}. This leads to a setup where total momentum of a system is conserved as it should be, and the EoM satisfies, although I did not stress this back then, the Galilei principle of relativity.

In special relativity, this quick fix does not help. The problem is that a potential energy of the type $V(\vec r_2-\vec r_1)$ represents an instantaneous interaction between particles located at $\vec r_1$ and $\vec r_2$ at the same time. Since the notion of simultaneity depends on the choice of IRF, such kind of ``action at distance'' is not allowed, as it would require a signal that propagates faster than light. The only systematic and consistent way of describing interactions compatible with the principles of special relativity is by making them local. This brings us back to a particle interacting with an external field, except that the field cannot be static, and the interaction must in some way depend on the motion of the particle.

From introductory physics courses, we know one example of an interaction that is consistent with relativity: electromagnetism. We shall now therefore see how to construct an action for a relativistic particle moving in an external electromagnetic field. In~\refpr{pr01:particleinEMfield}, we already saw the Lagrangian for a \emph{non}relativistic particle in a background electromagnetic field. We can reuse it here, but two changes are necessary: we have to modify the kinetic term and rewrite the action using four-vectors. The first step is simple. Following~\eqref{ch05:Lagrangian} as a guide, we try
\begin{equation}
L(\vec r,\dot{\vec r},t)=-mc^2\sqrt{1-\frac{\dot{\vec r}^2}{c^2}}-q\p(\vec r,t)+q\dot{\vec r}\cdot\vec A(\vec r,t)\;.
\label{ch05:LagrangianEM}
\end{equation}
Next, we observe that if we group the scalar and vector potential $\p$ and $\vec A$ into the \emph{four-potential},
\begin{equation}
A^\m\equiv(\p/c,\vec A)\;,
\end{equation}
then the two terms in the Lagrangian depending on the electromagnetic field can be written jointly as $q\dot x\cdot A$, where the dot still indicates a time derivative. This allows us to write the action in a Lorentz-invariant form by adding the interaction term to~\eqref{ch05:actionfree},
\begin{equation}
\begin{split}
S&=\int_{s_1}^{s_2}\Bigl(-mc\sqrt{-g_{\m\n}\D x^\m\D x^\n}+qg_{\m\n}\D x^\m A^\n\Bigr)\\
&=\int_{s_1}^{s_2}\D s\,\biggl(-mc\sqrt{-g_{\m\n}\OD{x^\m}{s}\OD{x^\n}{s}}+qg_{\m\n}\OD{x^\m}{s}A^\n\biggr)\;.
\end{split}
\label{ch05:actionEM}
\end{equation}
The expression on the first line highlights the geometric nature of the action: the first term gives, up to an overall factor, the total proper time elapsed along the worldline, whereas the second term represents a line integral of the four-potential along the worldline. The second line of~\eqref{ch05:actionEM} gives an expression suitable for a derivation of the EoM.

\begin{watchout}%
Before we tackle the EoM, I need to introduce some additional terminology and notation that will make the calculation much more transparent. So far, I have labeled four-vectors consistently with a superscript, as in $v^\m$. This kind of four-vector is called \emph{contravariant} in the physics literature. By taking a product with the metric and summing over one index, it can be turned into the corresponding \emph{covariant} four-vector, $v_\m\equiv g_{\m\n}v^\n$. This makes it possible to write the Minkowski dot product of two four-vectors in multiple ways,
\begin{equation}
u\cdot v=g_{\m\n}u^\m v^\n=u_\m v^\m=u^\m v_\m=g^{\m\n}u_\m v_\n\;,
\end{equation}
where $g^{\m\n}$ is the matrix inverse of $g_{\m\n}$. Just like the Minkowski metric $g_{\m\n}$ can be used to ``lower'' the index of contravariant four-vectors (or other contravariant tensors), the inverse $g^{\m\n}$ serves to raise the index of covariant four-vectors (or other covariant tensors). We thus have a lot of freedom to move dummy indices up or down at will. All we have to remember is that within the Einstein summation convention, the pair of indices summed over must always consist of one superscript and one subscript.

While the process of raising and lowering indices seems pretty innocuous, it is useful to stress that from the point of view of mathematics, covariant and contravariant vectors (or tensors) are entirely different objects. A contravariant vector is an element of the Minkowski space, $\R^{1,n}$. The metric is a symmetric bilinear form, that is a map $\R^{1,n}\times\R^{1,n}\to\R$. The covariant vector $v_\m=g_{\m\n}v^\n$ is obtained by filling one of the inputs of the metric with $v\in\R^{1,n}$, and is therefore a map $\R^{1,n}\to\R$. Thus, a covariant vector is an element of the dual space $\R^{1,n*}$. The lowering of the index on $v\in\R^{1,n}$ establishes a natural isomorphism between $\R^{1,n}$ and $\R^{1,n*}$. The existence of such an isomorphism is a general feature of vector spaces equipped with a scalar product.
\end{watchout}

We now move on to derive the EoM for the action~\eqref{ch05:actionEM}. Its general form is
\begin{equation}
\OD{}s\PD{L}{x'^\m}=\PD{L}{x^\m}\;,
\label{ch05:EoMEMaux}
\end{equation}
where the prime indicates a derivative with respect to $s$. Keep in mind that since the worldline is parameterized by $s$, all of $x^\m$ are treated as dependent variables, that is generalized coordinates. This is in contrast to the usual parameterization of trajectories by time, where $t$ is the independent variable and $x^i$ are its functions. We evaluate the two sides of~\eqref{ch05:EoMEMaux} separately, starting with the left-hand side
\begin{equation}
\OD{}s\PD{L}{x'^\m}=\OD{}{s}\biggl(mc\frac{x'_\m}{\sqrt{-x'^\n x'_\n}}+qA_\m\biggr)=\OD{}{s}\left(m\OD{x_\m}{\tau}+qA_\m\right)\;.
\end{equation}
The right-hand side of~\eqref{ch05:EoMEMaux} is simply $qx'^\n\Pd{A_\n}{x^\m}$. Combining the two results and converting the outermost derivative with respect to $s$ into one with respect to $\tau$, we get an EoM that is independent of the choice of the parameter $s$,
\begin{equation}
q\OD{x^\n}{\tau}\PD{A_\n}{x^\m}=\OD{}\tau\left(m\OD{x_\m}{\tau}+qA_\m\right)=m\frac{\D^2x_\m}{\D\tau^2}+q\OD{x^\n}{\tau}\PD{A_\m}{x^\n}\;.
\end{equation}
This can be finally reassembled as
\begin{equation}
\boxed{m\frac{\D^2x^\m}{\D\tau^2}=\OD{p^\m}{\tau}=qF^{\m\n}u_\n\;,}\qquad\text{where}\quad
F^{\m\n}\equiv\PD{A^\n}{x_\m}-\PD{A^\m}{x_\n}
\label{ch05:EoMEM}
\end{equation}
is the so-called \emph{field-strength tensor}.

I hope you appreciate the elegance of starting with the manifestly Lorentz-invariant action~\eqref{ch05:actionEM}, and thence deriving the four-vector EoM~\eqref{ch05:EoMEM}. The manifest independence of the choice of IRF is the strength of our argument. Its weakness is that the physical content of~\eqref{ch05:EoMEM} is not entirely transparent. I invite you to work out the details. You should find that the temporal and spatial parts of~\eqref{ch05:EoMEM} correspond, respectively, to
\begin{equation}
\OD{E}{t}=q\dot{\vec r}\cdot\vec E\;,\qquad
\OD{\vec p}{t}=q(\vec E+\dot{\vec r}\times\vec B)\;.
\end{equation} 
The first of these relations says that the energy $E$ of the particle as defined in~\eqref{ch05:fourmomentum} is changed by the work done on it by the electric field $\vec E$. The second relation is the actual dynamical law for motion under the effect of the Lorentz force.

%%%%%%%%%%%%%%%%%%%%%%%%%%%%%%%%%%%%%%%%%%%%%%%%%%%%%%%%%%%%

\section*{\probsec}
\addcontentsline{toc}{section}{\probsec}

\begin{prob}
\label{pr05:Poincaregroup}
A general element of the Poincar\'e group corresponds to the affine map
\begin{equation}
x^\m\to x'^\m\equiv\Lambda^\m_{\phantom\m\n}x^\n+a^\m\;,
\label{ch05:poincare}
\end{equation}
where $a^\m$ is a constant four-vector and the matrix $\smash{\Lambda^\m_{\phantom\m\n}}$ preserves the Minkowski metric, that is satisfies the condition in~\eqref{ch05:Lorentztransfodef}. Verify explicitly that the set of all transformations of the type~\eqref{ch05:poincare} satisfies the axioms of a group.
\end{prob}

\begin{prob}
\label{pr05:boostcomposition}
Verify that the composition of two Lorentz boosts in the same direction amounts to addition of their rapidities, $(w_1,w_2)\to w_1+w_2$, or equivalently to the relation
\begin{equation}
(\b_1,\b_2)\to\frac{\b_1+\b_2}{1+\b_1\b_2}
\end{equation}
for the velocity parameters.
\end{prob}

\begin{prob}
\label{pr05:Thomasprecession}
Consider a Lorentz boost $\Lambda_{\vec u}$ with velocity $\vec u$ pointing along the $x$-axis, and a boost $\Lambda_{\vec v}$ with velocity $\vec v$ pointing along the $y$-axis. Suppose the spacetime coordinates are first transformed with $\Lambda_{\vec u}$ and then with $\Lambda_{\vec v}$. Alternatively, we may transform the coordinates first with $\Lambda_{\vec v}$ and only then with $\Lambda_{\vec u}$. Show that if both $\vec u$ and $\vec v$ are small compared to the speed of light, then the results of performing the boosts in the two different orders only differ by a rotation in the $xy$-plane. This is known as the \emph{Thomas precession}. Hint: the easiest way to compare the matrix products $\Lambda_{\vec u}\Lambda_{\vec v}$ and $\Lambda_{\vec v}\Lambda_{\vec u}$ is to multiply the latter by the inverse of the former (or vice versa). With this in mind, calculate the matrix product $\Lambda_{\vec v}^{-1}\Lambda_{\vec u}^{-1}\Lambda_{\vec v}\Lambda_{\vec u}$, expanding in powers of $\vec u,\vec v$ to the first order in both.
\end{prob}

\begin{prob}
\label{pr05:velocitytransfo}
Using the fact that the four-velocity $u^\m=\g_{\vec v}(c,\vec v)$ is a four-vector, derive a transformation rule for the components of the velocity $\vec v$ under a Lorentz boost with speed $u$ in the $x$-direction.
\end{prob}

\begin{prob}
\label{pr05:scattering}
Consider a scattering process involving four particles with four-momenta $p_1^\m,\dotsc,p_4^\m$. Assuming that both the initial and the final state involves two of the particles, energy and momentum conservation require that $p^\m_1+p^\m_2=p^\m_3+p^\m_4$. To characterize the kinematics of the process in a manner independent of the choice of IRF, we can form the dot products $p_i\cdot p_j$ with $i\neq j$. Show that out of the six possible dot products, only two are independent, and the others can be expressed in terms of these. Can you guess how many independent Lorentz-invariant kinematical variables there will be for an $n$-particle scattering process with any $n>4$?
\end{prob}

\begin{prob}
\label{pr05:spinor}
The components of the four-momentum $p^\m$ of a particle of mass $m$ (in $n=3$ spatial dimensions) can be encoded in a complex $2\times 2$ matrix,
\begin{equation}
\Pi\equiv\begin{pmatrix}
p^0+p^3 & p^1-\I p^2\\
p^1+\I p^2 & p^0-p^3
\end{pmatrix}=p^0\un+\skal p\tau\;,
\end{equation}
where $\vec\tau$ is the vector of Pauli matrices. Noting that $\det\Pi=-p^2=m^2c^2$, find the eigenvalues of $\Pi$ for a massless particle. Show that in this case, the matrix $\Pi$ can be written as $\Pi=\vec\x\he{\vec\x}$ for some complex vector $\vec\x\in\C^2$. Remark: this representation of the four-momentum of massless particles in terms of a complex \emph{spinor} is heavily used in modern studies of relativistic scattering.
\end{prob}

\begin{prob}
\label{pr05:Hamiltonian}
Starting from the Lagrangian~\eqref{ch05:Lagrangian} for a relativistic particle of rest mass $m$ moving in the potential $V(\vec r)$, derive the Hamiltonian. Check that the corresponding Hamilton equations have the expected form.
\end{prob}

\begin{prob}
\label{pr05:umupmu}
The fact that the field-strength tensor is antisymmetric in its two indices implies that $F^{\m\n}u_\m u_\n=0$. It then follows from the EoM~\eqref{ch05:EoMEM} that $u_\m\Od{p^\m}{\tau}=0$. Show that this is true for any kind of motion as a consequence of the definitions of four-velocity and four-momentum. Explain the physical content of this identity.
\end{prob}
\chapter{Geometry of Classical Mechanics}
\label{chap:geometryclassmech}

\keywords{Riemannian metric on configuration space, geodesic equation, symplectic form on phase space, Darboux coordinates, canonical transformation, Poisson bracket.}

%%%%%%%%%%%%%%%%%%%%%%%%%%%%%%%%%%%%%%%%%%%%%%%%%%%%%%%%%%%%

\noindent In this \chaptername, we will explore some of the deeper structural aspects of mechanics. The main benefit of doing so will be an improved understanding of the geometrical origin of the Hamiltonian formalism. We will see to what extent generalized coordinates and momenta, while being independent variables in the phase space, are related to each other. We will also learn how to choose these variables directly without having to start from a Lagrangian. Finally, we will gain new insight into the nature of conservation laws and see, for the first time, how they are related to symmetries. In order to motivate the formal development of the Hamiltonian formalism, we will however first revisit the conceptually more accessible Lagrangian formalism.

%%%%%%%%%%%%%%%%%%%%%%%%%%%%%%%%%%%%%%%%%%%%%%%%%%%%%%%%%%%%

\section{Lagrangian Formalism: Geometry of Configuration Space}
\label{sec:geomlag}

To be concrete, let us start with the problem of a single (nonrelativistic) particle of mass $m$, moving in $n$-dimensional Euclidean space under the effect of a potential, $V(\vec r)$. The Lagrangian of the particle can be written in vector form as
\begin{equation}
L=\frac12m\dot{\vec r}^2-V(\vec r)\;.
\label{ch06:lagRn}
\end{equation}
For many purposes, the most natural choice of concrete generalized coordinates for this system is to take the standard Cartesian coordinates $r^i$, $i=1,\dotsc,n$. In the Cartesian coordinates, the squared velocity is simply $\dot{\vec r}^2=\d_{ij}\dot r^i\dot r^j$. Suppose however that we would like to switch to a different set of coordinates in $\R^n$, which I will denote as $q^i$. In terms of these, the length of an infinitesimal line segment in Euclidean space can be expressed using the chain rule as
\begin{equation}
\D\vec r^2=\d_{kl}\D r^k\D r^l=\d_{kl}\PD{r^k}{q^i}\PD{r^l}{q^j}\D q^i\D q^j\equiv g_{ij}(q)\D q^i\D q^j\;,
\end{equation}
where $g_{ij}(q)$ is the Euclidean metric in the new coordinates.

\begin{watchout}
The use of superscripts to label coordinates is common in differential geometry and I will stick to this convention. It gives us additional control over the structure of algebraic expressions and their manipulations. Among others, the Einstein summation convention will, in this \chaptername, apply exclusively to pairs of indices out of which one is a subscript and one a superscript.
\end{watchout}

\begin{illustration}%
\label{ex06:polarcoords}%
In the Euclidean plane $\R^2$, it is common to trade the Cartesian coordinates $x,y$ for the polar coordinates $r,\vp$ via $x=r\cos\vp$, $y=r\sin\vp$. The corresponding line element~is
\begin{equation}
\D\vec r^2=\D r^2+r^2\D\vp^2\;.
\end{equation}
This shows that the components of the Euclidean metric in the polar coordinates are $g_{rr}=1$, $g_{\vp\vp}=r^2$ and $g_{r\vp}=g_{\vp r}=0$. It is easy to promote this result to three dimensions, where one introduces spherical coordinates $r,\t,\vp$ via $x=r\sin\t\cos\vp$, $y=r\sin\t\sin\vp$ and $z=r\cos\t$. The corresponding metric is then
\begin{equation}
\D\vec r^2=\D r^2+r^2\D\t^2+r^2\sin^2\t\,\D\vp^2\;.
\end{equation}
The metric is still represented by a diagonal matrix, but its nonzero elements are now $g_{rr}=1$, $g_{\t\t}=r^2$ and $g_{\vp\vp}=r^2\sin^2\t$.
\end{illustration}

Provided that the relation between the original Cartesian coordinates $r^i$ and the new coordinates $q^i$ is time-independent, the Lagrangian~\eqref{ch06:lagRn} turns to
\begin{equation}
\boxed{L=\frac12mg_{ij}(q)\dot q^i\dot q^j-V(q)\;.}
\label{ch06:lagRnq}
\end{equation}
We shall now make a large conceptual step. Imagine a particle that does not live in the Euclidean space, but rather is confined to a possibly curved (hyper)surface $\S$. This can of course be thought of as a result of constraints, just like a pendulum can be viewed as a particle in $\R^2$ forced to move on a circle. This interpretation is however irrelevant. For all practical purposes, it is not necessary to think of $\S$ as a surface in some higher-dimensional space, but as a ``space'' in its own right that may be curved. In mathematics, such a curved ``space'' is called a \emph{manifold}; I will adopt this terminology for the sake of brevity.

The manifold $\S$ now plays the role of the configuration space for our particle of mass $m$. The variables $q^i$ are coordinates on $\S$ that uniquely specify the position of the particle. Finally, the \emph{Riemannian metric} $g_{ij}(q)$ describes the intrinsic geometry of the manifold. Once the configuration space is not flat, there may not be a preferred set of coordinates such as the Cartesian coordinates in $\R^n$. This underlines the basic property of the Lagrangian formalism that any (generalized) coordinates are equally good. Should we at any point decide to switch from coordinates $q^i$ to new coordinates $\tilde q^i$, all we have to do is to insert in~\eqref{ch06:lagRnq} appropriately modified metric, $\tilde g_{ij}(\tilde q)$. This follows by comparing the expressions for the line element in the two coordinate sets,
\begin{multline}
\D s^2=g_{kl}(q)\D q^k\D q^l=g_{kl}(q(\tilde q))\PD{q^k}{\tilde q^i}\PD{q^l}{\tilde q^j}\D\tilde q^i\D\tilde q^j\ifeq\tilde g_{ij}(\tilde q)\D\tilde q^i\D\tilde q^j\\
\Rightarrow\quad\tilde g_{ij}(\tilde q)=g_{kl}(q(\tilde q))\PD{q^k}{\tilde q^i}\PD{q^l}{\tilde q^j}\;.
\label{ch06:metrictransfo}
\end{multline}

\begin{illustration}%
Two simple examples of curved manifolds can be extracted from \refex{ex06:polarcoords} by restricting to constant radial variable, $r=L$. Doing so in the plane $\R^2$ leads to a one-dimensional manifold $\S$ which is simply a circle of radius $L$ centered at the origin. The corresponding line element is $\D s^2=L^2\D\vp^2$ and the Lagrangian~\eqref{ch06:lagRnq} boils down to $L=(1/2)mL^2\dot\vp^2-V(\vp)$. This generalizes the simple pendulum to an arbitrary potential $V(\vp)$.

Similarly, restricting to constant $r=L$ in $\R^3$ makes $\S$ into a sphere with the metric expressed in spherical coordinates as $\D s^2=L^2\D\t^2+L^2\sin^2\t\,\D\vp^2$. A particle moving on a sphere in the gravitational field is called a spherical pendulum; cf.~\refpr{pr01:sphericalpendulum}. In a generic potential $V(\t,\vp)$, its Lagrangian reads
\begin{equation}
L=\frac12mL^2(\dot\t^2+\dot\vp^2\sin^2\t)-V(\t,\vp)\;.
\end{equation}
\end{illustration}

Since the Lagrangian~\eqref{ch06:lagRnq} takes the same form for any choice of coordinates $q^i$, we would expect the same to be true for the Lagrange \emph{equation of motion} (EoM). Let us see how to express the EoM in terms of the geometric properties of the configuration space $\S$. As the first step, evaluate
\begin{equation}
\OD{}t\PD L{\dot q^i}-\PD L{q^i}=\OD{}t(mg_{ij}\dot q^j)-\frac12m(\de_ig_{jk})\dot q^j\dot q^k+\PD V{q^i}\;,
\end{equation}
where I introduced the shorthand notation $\de_ig_{jk}\equiv\Pd{g_{jk}}{q^i}$. In the next step, we use the chain rule to rewrite $\Od{(g_{ij}\dot q^j)}t=g_{ij}\ddot q^j+(\de_kg_{ij})\dot q^j\dot q^k$. Collecting the terms proportional to $\dot q^j\dot q^k$, the EoM can now be cast as
\begin{equation}
g_{ij}\ddot q^j+\frac12(\de_kg_{ij}+\de_jg_{ik}-\de_ig_{jk})\dot q^j\dot q^k+\frac1m\PD V{q^i}=0\;.
\end{equation}
The last step is to multiply this set of equations with the matrix inverse of the metric $g_{ij}(q)$, indicated by raising the indices, $g^{ij}(q)$. The final form of the EoM is then
\begin{equation}
\boxed{\ddot q^i+\Gamma^i_{jk}\dot q^j\dot q^k+\frac{g^{ij}}m\PD V{q^j}=0\;,}
\label{ch06:lagEoM}
\end{equation}
where $\smash{\Gamma^i_{jk}\equiv(1/2)g^{il}(\de_jg_{lk}+\de_kg_{jl}-\de_lg_{jk})}$ are the \emph{Christoffel symbols} of the metric $g_{ij}$. Equation~\eqref{ch06:lagEoM} is a universal EoM describing the motion of a particle under any potential $V(q)$ on any Riemannian manifold $\S$, in any coordinates.

\begin{watchout}
The Christoffel symbols only depend on the metric and thus characterize the geometry of the configuration space. All the components $\smash{\Gamma^i_{jk}}$ can be computed once the coordinates $q^i$ on $\S$ have been chosen and the metric is known. Importantly, the mass $m$ of the particle only appears in the last, potential term of~\eqref{ch06:lagEoM}. The EoM of a free particle (no potential) is purely geometric, $\ddot q^i+\Gamma^i_{jk}\dot q^j\dot q^k=0$. This is the \emph{geodesic equation}, describing ``straight lines'' on the manifold $\S$.
\end{watchout}

%%%%%%%%%%%%%%%%%%%%%%%%%%%%%%%%%%%%%%%%%%%%%%%%%%%%%%%%%%%%

\section{Hamiltonian Formalism: Geometry of Phase Space}
\label{sec:geomHam}

A geometric reformulation of the Hamiltonian formalism can be developed following the same inductive approach as above. We first write down what we know about the dynamics of a single particle in $\R^n$. Switching to possibly curvilinear coordinates will then highlight certain geometric structure of the phase space. Finally, we will declare this structure to be a general feature of the Hamiltonian formalism, valid also for other systems than a single particle in $\R^n$.

Let us see this program through to the end. For a single particle in $\R^n$, we have a set of generalized coordinates and momenta, $q^i$ and $p_i$, where $i=1,\dotsc,n$. Once we know the Hamiltonian $H(q,p,t)$, we can express the action functional as
\begin{equation}
S[q,p]=\int_{t_1}^{t_2}\D t\,[\dot q^ip_i-H(q,p,t)]\simeq\int_{t_1}^{t_2}\D t\left[\frac12(\dot q^ip_i-q^i\dot p_i)-H(q,p,t)\right]\;.
\label{ch06:hamaction}
\end{equation}
In the second step we integrated by parts, which is indicated by the symbol $\simeq$. We shall now group all the \emph{canonical variables} $q^i,p_i$ into a single $2n$-component object, $\vec\x\equiv(q^1,\dotsc,q^n,p_1,\dotsc,p_n)^T$. Introducing the $2n\times2n$ matrix $\O$ consisting of four $n\times n$ blocks,
\begin{equation}
\O\equiv\begin{pmatrix}
0 & -\un\\
+\un & 0
\end{pmatrix}\;,
\label{ch06:OmegaRn}
\end{equation}
the action~\eqref{ch06:hamaction} can be rewritten as
\begin{equation}
S[\x]=\int_{t_1}^{t_2}\D t\left[\frac12\O_{ij}\x^i\dot\x^j-H(\x,t)\right]\;,
\label{ch06:hamaction2}
\end{equation}
where the indices $i,j,\dotsc$ now run from $1$ to $2n$.

\begin{illustration}%
By taking the functional derivative with respect to $\x^i$, one easily finds the corresponding EoM,
\begin{equation}
\O_{ij}\dot\x^j=\PD H{\x^i}\quad\Leftrightarrow\quad
\dot\x^i=\O^{ij}\PD H{\x^j}\;,
\end{equation}
where $\O^{ij}$ are the elements of the matrix inverse of $\O$. This set of $2n$ equations is nothing but the usual Hamilton equations~\eqref{ch03:Hamiltoneq} in disguise,
\begin{equation}
\dot q^i=\PD H{p_i}\;,\qquad
\dot p_i=-\PD H{q^i}\;.
\label{ch06:Hameqconventional}
\end{equation}
\end{illustration}

%%%%%%%%%%%%%%%%%%%%%%%%%%%%%%%%%%%%%%%%%%%%%%%%%%%%%%%%%%%%

\subsection{Symplectic Formulation of Hamiltonian Mechanics}
\label{sec:geomsymplectic}

The next step is to switch to a new set of coordinates, $\tilde\x^i(\x)$, in the phase space. This turns the first term under the integral in~\eqref{ch06:hamaction2} to $\smash{(1/2)\O_{ij}\x^i(\tilde\x)(\Pd{\x^j}{\tilde\x^k})\dot{\tilde\x}^k}$. What we find is a possibly complicated expression, depending on the precise choice of the new canonical variables. It is however always linear in the time derivatives $\smash{\dot{\tilde\x}^k}$. This is the sought general feature of the Hamiltonian formalism, which guarantees the ensuing dynamical equations to always be first-order in time derivatives.

We now define a general dynamical system in the Hamiltonian formalism by the action in terms of a set of $2n$ canonical coordinates $\smash{\x^i}$,
\begin{equation}
\boxed{S[\x]=\int_{t_1}^{t_2}\D t\,[\o_i(\x(t))\dot\x^i(t)-H(\x(t),t)]\;.}
\label{ch06:sympaction}
\end{equation}
The $2n$ coordinates $\x^i$ take values from a manifold $\S$ that may not be flat like the Euclidean space we started with. Nevertheless, we still call $\S$ the phase space of the system. The Hamiltonian $H(\x,t)$ is a function on $\S$, possibly also depending explicitly on time. It characterizes the dynamics of the system. On the other hand, the set of functions $\o_i(\x)$ characterizes the geometry of the phase space independently of the Hamiltonian. Accordingly, $\o_i(\x)$ is assumed not to depend explicitly on time. This is somewhat analogous to our previous observation that the metric $g_{ij}(q)$ in~\eqref{ch06:lagRnq} describes the geometry of the configuration space, independent of the potential function $V(q)$. It is common to call $\o_i(\x)$ the \emph{symplectic potential} of the phase~space.

It is a simple exercise to work out the functional derivative of the action~\eqref{ch06:sympaction} and check that the corresponding EoM reads
\begin{equation}
\boxed{\O_{ij}\dot\x^j=\PD H{\x^j}\;,}\quad\text{where}\quad
\O_{ij}\equiv\PD{\o_j}{\x^i}-\PD{\o_i}{\x^j}\;.
\label{ch06:sympEoM}
\end{equation}
The antisymmetric matrix function $\O_{ij}(\x)$ on $\S$ is called the \emph{symplectic form}. I will often use the shorthand notation $\de_i$ for $\Pd{}{\x^i}$ so that the symplectic form becomes $\O_{ij}=\de_i\o_j-\de_j\o_i$, and the general Hamilton EoM reduces to $\O_{ij}\dot\x^j=\de_iH$.

\begin{illustration}%
Let us check that the notation used in our general Hamiltonian action~\eqref{ch06:sympaction} agrees with the special case of a single particle we started from. There, the action was~\eqref{ch06:hamaction2} and the matrix $\O_{ij}$ was constant as defined by~\eqref{ch06:OmegaRn}. We can always go back from~\eqref{ch06:sympaction} to~\eqref{ch06:hamaction2} by setting $\o_i(\x)=-(1/2)\O_{ij}\x^j$. This leads to $\de_i\o_j-\de_j\o_i=\O_{ij}$, which is consistent with the general definition of the symplectic form in~\eqref{ch06:sympEoM}.
\end{illustration}

The symplectic form defines some kind of geometric structure on the phase space. It is however a structure that we have no prior experience with. Let us list its basic properties and then comment on their mathematical and physical significance:
\begin{enumerate}
\item[(a)] The symplectic form $\O_{ij}(\x)$ establishes an antisymmetric bilinear form on (the tangent space to) the phase space $\S$.
\item[(b)] Under a change of coordinates, $\x^i\to\tilde\x^i(\x)$, the components of the symplectic form transform by
\begin{equation}
\tilde\O_{ij}(\tilde\x)=\O_{kl}(\x(\tilde\x))\PD{\x^k}{\tilde\x^i}\PD{\x^l}{\tilde\x^j}\;.
\label{ch06:sympformtransfo}
\end{equation}
\item[(c)] The symplectic form as a function on $\S$ satisfies the \emph{Bianchi identity}
\begin{equation}
\de_k\O_{ij}+\de_i\O_{jk}+\de_j\O_{ki}=0\;.
\label{ch06:Bianchi}
\end{equation}
\item[(d)] The symplectic form is nondegenerate, that is, the matrix $\O_{ij}(\x)$ is invertible everywhere on the phase space $\S$.
\end{enumerate}
The property (a) says, at our level of sophistication, merely that $\O_{ij}$ is an antisymmetric matrix, which follows from its definition in~\eqref{ch06:sympEoM}. The property (b) says that $\O_{ij}(\x)$ behaves as a tensor field on $\S$, and is analogous to the transformation property of the metric~\eqref{ch06:metrictransfo} under a charge of coordinates. To check this property, one first observes using~\eqref{ch06:sympaction} that the symplectic potential $\o_i(\x)$ is a tensor of rank one, that is, $\tilde\o_i(\tilde\x)=\o_j(\x(\tilde\x))\Pd{\x^j}{\tilde\x^i}$. The rest then follows again from the definition of $\O_{ij}(\x)$ in~\eqref{ch06:sympEoM}. The latter together with the symmetry of second partial derivatives also verifies the Bianchi identity (c). Finally, the \emph{assumption} (d) that the symplectic form is nondegenerate is necessary to invert the Hamiltonian EoM~\eqref{ch06:sympEoM} to
\begin{equation}
\dot\x^i=\O^{ij}\PD H{\x^j}\;,
\label{ch06:sympEoM2}
\end{equation}
where $\O^{ij}(\x)$ is again the matrix inverse of $\O_{ij}(\x)$. In this form, the EoM ensures that the initial value problem for the canonical coordinates $\x^i(t)$ is well-defined and allows for a unique solution. In other words, the action~\eqref{ch06:sympaction} actually completely determines the dynamics of our Hamiltonian system.

\begin{illustration}%
For a single particle moving in $n$-dimensional Euclidean space, the phase space is $\smash{\S=\R^n\times\R^n=\R^{2n}}$. While the Hamilton EoM~\eqref{ch06:sympEoM} takes the same form for any choice of coordinates on $\S$, the choice $\vec\x=(q^i,p_i)$ is special in that $\O_{ij}$ is constant and given by~\eqref{ch06:OmegaRn}. Such coordinates on the phase space are called \emph{Darboux coordinates}; they generalize the notion of Cartesian coordinates in Euclidean space.

It turns out that for any phase space $\S$, one can always introduce Darboux coordinates that are well-defined in some open set on $\S$. This is the content of the \emph{Darboux theorem}. It guarantees that on any phase space, one can introduce generalized coordinates $q^i$ and their conjugate momenta $p_i$ such that the Hamilton equations take the conventional form~\eqref{ch06:Hameqconventional}. The catch, and the very reason why I make the effort to set up the symplectic formalism for Hamiltonian mechanics, is that the Darboux coordinates may not exist globally on the whole phase space. This is illustrated by the next example.
\end{illustration}

\begin{illustration}%
\label{ex06:sympS2}%
The unit sphere $S^2$ is an example of a curved manifold on which one can define a symplectic structure. To see how, recall that it is possible to parameterize the sphere using a unit vector variable in $\R^3$, which I will denote as $\vec n$. The three components of $\vec n$ are of course not mutually independent, but rather are related by the constraint $\abs{\vec n}=1$. Thinking of the unit vector as a function of two independent coordinates $\x^i$ on the sphere, $\vec n(\x)$, there is a natural nondegenerate, antisymmetric bilinear form that respects the rotational symmetry of the sphere,
\begin{equation}
\O_{ij}(\x)\equiv\a\vec n(\x)\cdot[\de_i\vec n(\x)\times\de_j\vec n(\x)]\;.
\label{ch06:sympS2}
\end{equation}
The (nonzero) constant $\a$ fixes the overall normalization of this symplectic form. Thinking of $\vec n$ as representing the spin degrees of freedom of a particle, the value of $\a$ can be determined from the properties of angular momentum; cf.~\refpr{pr06:sympS2}.

Let us see what the corresponding EoM looks like. The left-hand side of~\eqref{ch06:sympEoM} evaluates to $\O_{ij}\dot\x^j=\a\vec n\cdot(\de_i\vec n\times\dot{\vec n})=\alpha\de_i\vec n\cdot(\dot{\vec n}\times\vec n)$. Assuming that the Hamiltonian only depends on the canonical variables $\x^i$ through $\vec n(\x)$, the right-hand side of the EoM becomes $\Pd H{\x^i}=(\Pd H{\vec n})\cdot\de_i\vec n$. The EoM is thus equivalent to $(\a\dot{\vec n}\times\vec n-\Pd{H}{\vec n})\cdot\de_i\vec n=0$. Using the fact that $\de_i\vec n$ with $i=1,2$ is a basis of linearly independent tangent vectors to the sphere, this can be expressed as
\begin{equation}
\a\dot{\vec n}\times\vec n-\PD H{\vec n}=\l\vec n\;,
\end{equation}
where $\l$ can be interpreted as a Lagrange multiplier for the constraint $\abs{\vec n}=1$. This form of the EoM does not depend on $\x^i$ or the derivatives of $\vec n$ with respect to $\x^i$, and we can therefore think of it as an equation for $\vec n$ itself. Taking finally a cross-product with $\vec n$ to eliminate $\l$, we obtain an explicit first order evolution equation,
\begin{equation}
\dot{\vec n}=\frac1\a\vec n\times\PD H{\vec n}\;.
\label{ch06:LandauLifshitz}
\end{equation}
This is known as the \emph{Landau--Lifshitz equation}, and describes the dynamics of spin in presence of external forces.

We have not used any specific choice of canonical coordinates on the sphere, and made the point that the EoM~\eqref{ch06:LandauLifshitz} is independent of such a choice. Sometimes, it may nevertheless be convenient to trade the implicit parameterization of the sphere by the unit vector $\vec n$ for concrete coordinates. The natural choice are the spherical angles $\t,\vp$, in terms of which $\vec n=(\sin\t\cos\vp,\sin\t\sin\vp,\cos\t)$. A quick calculation using~\eqref{ch06:sympS2} shows that in these coordinates, the sole independent offdiagonal element of the symplectic form is $\O_{\t\vp}=\a\sin\t$. The other offdiagonal element is fixed by antisymmetry, $\O_{\vp\t}=-\a\sin\t$. From here it is just a small step to find Darboux coordinates on the sphere. Indeed, we can take for instance $\x^1=\cos\t$ and $\x^2=\a\vp$. It is easy to check with the help of~\eqref{ch06:sympformtransfo} that in these coordinates, $\O_{ij}(\x)=-\ve_{ij}$ in accord with~\eqref{ch06:OmegaRn}. We see, however, that such Darboux coordinates are only meaningful away from the poles of the sphere, at which the the polar angle $\vp$ is ill-defined. This is not mere bad luck in the choice of coordinates. The sphere is an example of a symplectic manifold on which a set of globally well-defined Darboux coordinates does not exist. This mathematical subtlety has profound consequences for the physical properties of spin systems such as ferromagnets.
\end{illustration}

%%%%%%%%%%%%%%%%%%%%%%%%%%%%%%%%%%%%%%%%%%%%%%%%%%%%%%%%%%%%

\subsection{Canonical Transformations}

We know by now how to set up the action using arbitrary coordinates in the phase space. Should we wish to change the coordinates, we have~\eqref{ch06:sympformtransfo} that tells us how to convert the components of the symplectic form. The Hamiltonian remains the same, although in the new coordinates, it may of course be represented by a different function. Yet, it is sometimes convenient to restrict oneself to transformations of coordinates that preserve the already known structure of the phase space. To motivate what comes below, just think of translations and rotations in Euclidean space. These have the property that they map one set of Cartesian coordinates to another, equally valid set of Cartesian coordinates. In mathematical terms, translations and rotations constitute isometries of the Euclidean metric, as we already saw in Sect.~\ref{sec:spacetimegeometry}.

This concept of isometry can be readily generalized to any manifold $\S$ endowed with a Riemannian metric $g_{ij}(q)$. We say that the smooth and invertible map $q^i\to\tilde q^i(q)$ is an isometry of the metric if the new metric $\tilde g_{ij}$ given by~\eqref{ch06:metrictransfo} coincides with $g_{ij}$. More precisely, we require that
\begin{equation}
\tilde g_{ij}(\tilde q)=g_{ij}(\tilde q)\quad\Leftrightarrow\quad
g_{kl}(q(\tilde q))\PD{q^k}{\tilde q^i}\PD{q^l}{\tilde q^j}=g_{ij}(\tilde q)\;.
\end{equation}
Motivated by the analogy with isometries of metrics on Riemannian manifolds, we shall now briefly look at the properties of transformations of canonical coordinates that preserve the symplectic form via~\eqref{ch06:sympformtransfo},
\begin{equation}
\tilde\O_{ij}(\tilde\x)=\O_{ij}(\tilde\x)\quad\Leftrightarrow\quad
\boxed{\O_{kl}(\x(\tilde\x))\PD{\x^k}{\tilde\x^i}\PD{\x^l}{\tilde\x^j}=\O_{ij}(\tilde\x)\;.}
\label{ch06:canontransfo}
\end{equation}
In physics, such maps are called \emph{canonical transformations}. The corresponding mathematical term, which I will not use here, is \emph{symplectomorphism}. In terms of the Darboux coordinates $q^i,p_i$, the distinguishing feature of canonical transformations, equivalent to~\eqref{ch06:canontransfo}, is that they preserve the form of the Hamilton EoM~\eqref{ch06:Hameqconventional}. In other words, upon transforming from $\vec\x=(q^i,p_i)$ to $\tilde{\vec\x}=(\tilde q^i,\tilde p_i)$, the Hamilton equations become
\begin{equation}
\dot{\tilde q}^i=\PD{\tilde H}{\tilde p_i}\;,\qquad
\dot{\tilde p}_i=-\PD{\tilde H}{\tilde q^i}\;,
\end{equation}
where the new Hamiltonian is obtained by mere substitution of variables, $\tilde H(\tilde q,\tilde p,t)=H(q(\tilde q,\tilde p),p(\tilde q,\tilde p),t)$.

\begin{watchout}%
The parallel between isometries of a metric and symplectomorphisms of a symplectic form does not go very far. The isometries of a Riemannian manifold span a \emph{Lie group}. This is itself a finite-dimensional manifold. In other words, the number of ``independent'' types of symmetry transformations of a metric (think translations and rotations in $\R^n$) is always finite. In contrast, the freedom to set up a canonical transformation is so huge that the space of all possible symplectomorphisms turns out to be infinite-dimensional. I shall not develop the formal properties of canonical transformations further, but instead work out a couple of illustrative examples.
\end{watchout}

\begin{illustration}%
\label{ex06:symplinear}%
Take the familiar example of a particle in $\R^n$ along with a set of Darboux coordinates, $\vec\x=(q^1,\dotsc,q^n,p_1,\dotsc,p_n)$. The generalized coordinates $q^i$ can be chosen arbitrarily; important is that $p_i$ are the corresponding conjugate momenta. However, if it helps, you can think of $q^i$ simply as the Cartesian coordinates in space. To get basic insight into the nature of canonical transformations, consider a transformation of coordinates that is \emph{linear}, $\x^i=P^i_{\phantom ij}\tilde\x^j$, where $P$ is an invertible real $2n\times2n$ matrix. The condition~\eqref{ch06:canontransfo} that the transformation be canonical then reduces to
\begin{equation}
\O_{kl}P^k_{\phantom ki}P^l_{\phantom lj}=\O_{ij}\quad\Leftrightarrow\quad
P^T\O P=\O\;,
\label{ch06:Sp2n}
\end{equation}
where $\O$ is the constant matrix~\eqref{ch06:OmegaRn}. The set of all matrices $P$ that satisfy~\eqref{ch06:Sp2n} constitutes the \emph{symplectic group} $\gr{Sp}(2n,\R)$. This is a real Lie group of dimension $n(2n+1)$. For a very specific example, take the simplest case of $n=1$ and consider a rotation of the canonical coordinates $q,p$,
\begin{equation}
\left.\begin{aligned}
q&=\tilde q\cos\a+\tilde p\sin\a\\
p&=-\tilde q\sin\a+\tilde p\cos\a
\end{aligned}\right\}\quad\Leftrightarrow\quad
P=\begin{pmatrix}
\cos\a & \sin\a\\
-\sin\a & \cos\a
\end{pmatrix}\;.
\label{ch06:pqalpha}
\end{equation}
It is easy to check that the rotation matrix $P$ satisfies~\eqref{ch06:Sp2n} for any choice of the angle $\a$. We can choose in particular $\a=\pi/2$, which leads to $q=\tilde p$ and $p=-\tilde q$. This is a very special case of a canonical transformation: the Hamilton equations~\eqref{ch06:Hameqconventional} remain valid if we swap the generalized coordinates and momenta and change the sign of the former.
\end{illustration}

The matrix form~\eqref{ch06:Sp2n} of the condition for canonical transformations is useful even if the transformation is not linear. Just note that $\smash{J^i_{\phantom ij}\equiv\Pd{\x^i}{\tilde\x^j}}$ is the Jacobian matrix of the inverse transformation $\smash{\tilde\x^i\to\x^i(\tilde\x)}$. The condition~\eqref{ch06:canontransfo} is then equivalent to $J(\tilde\x)^T\O(\x(\tilde\x))J(\tilde\x)=\O(\tilde\x)$, where $\O(\x)$ is now the matrix with elements $\O_{ij}(\x)$. Here is a simple example.

\begin{illustration}%
\label{ex06:canonLHO}%
Consider a particle moving in one dimension with its phase space $\S=\R\times\R=\R^2$. The transformation
\begin{equation}
\left.\begin{aligned}
q&=\sqrt{\frac{2\tilde p}{m\o}}\sin\tilde q\\
p&=\sqrt{2m\o\tilde p}\cos\tilde q
\end{aligned}\right\}\quad\Leftrightarrow\quad
J=\begin{pmatrix}
\sqrt{\frac{2\tilde p}{m\o}}\cos\tilde q & \frac1{\sqrt{2m\o\tilde p}}\sin\tilde q\\
-\sqrt{2m\o\tilde p}\sin\tilde q & \sqrt{\frac{m\o}{2\tilde p}}\cos\tilde q
\end{pmatrix}\;,
\end{equation}
where $m$ and $\o$ are positive parameters, is canonical. It is tailored to simplify the solution of the EoM for a linear harmonic oscillator.
\end{illustration}

%%%%%%%%%%%%%%%%%%%%%%%%%%%%%%%%%%%%%%%%%%%%%%%%%%%%%%%%%%%%

\subsection{Poisson Bracket}
\label{subsec:Poisson}

I will now introduce yet another structure on the phase space, which will eventually naturally guide us to revisit conservation laws. To motivate what is coming, consider the problem of finding the time dependence of a function $F(\x)$ on the phase space when evaluated on a trajectory $\x^i(t)$ of a system. The trajectory itself is determined as the unique solution to the EoM~\eqref{ch06:sympEoM2} with given initial conditions. Using the chain rule, we find the time derivative of $F(\x(t))$,
\begin{equation}
\dot F=\PD F{\x^i}\dot\x^i=\O^{ij}\PD F{\x^i}\PD H{\x^j}\;.
\label{ch06:evolF}
\end{equation}
The right-hand side can be thought of more abstractly as a bilinear operation, acting on a pair of functions on the phase space. For any two functions $F,G$ on the phase space, we thus introduce the shorthand notation
\begin{equation}
\boxed{\{F,G\}\equiv\O^{ij}\PD F{\x^i}\PD G{\x^j}\;.}
\label{ch06:poissonbracket}
\end{equation}
This is the \emph{Poisson bracket} of $F$ and $G$. Choosing $G$ as the Hamiltonian recovers the time evolution of $F(\x(t))$ we started with in the form $\dot F=\{F,H\}$. Another important special case of the Poisson bracket arises from taking for $F,G$ a pair of canonical coordinates,
\begin{equation}
\boxed{\{\x^i,\x^j\}=\O^{ij}(\x)\;.}
\end{equation}
The Poisson bracket therefore encodes complete information about the symplectic structure of the phase space.

\begin{illustration}%
\label{ex06:canonicalPoisson}%
Consider once again a system living in $\R^n$, whose phase space is $\S=\R^n\times\R^n=\R^{2n}$. Inverting the constant matrix~\eqref{ch06:OmegaRn}, we find that in terms of the Darboux coordinates $q^i,p_i$, the Poisson bracket can be expressed as
\begin{equation}
\{F,G\}=\PD F{q^i}\PD G{p_i}-\PD F{p_i}\PD G{q^i}\;.
\label{ch06:poissonconventional}
\end{equation}
A special case of this are the Poisson brackets of the Darboux coordinates themselves,
\begin{equation}
\{q^i,q^j\}=\{p_i,p_j\}=0\;,\qquad
\{q^i,p_j\}=\d^i_j\;.
\label{ch06:PoissonDarboux}
\end{equation}
Finally, the Hamilton equations~\eqref{ch06:Hameqconventional} now acquire a simple symmetric form, $\dot q^i=\Pd H{p_i}=\{q^i,H\}$ and $\dot p_i=-\Pd H{q^i}=\{p_i,H\}$.
\end{illustration}

The Poisson bracket has several important properties through which it establishes an elegant mathematical structure on the algebra of functions on the phase space:
\begin{enumerate}
\item[(a)] Bilinearity. Namely, for any triplet of functions $A_1,A_2,B$ on the phase space and any two constants $c_1,c_2$,
\begin{equation}
\{c_1A_1+c_2A_2,B\}=c_1\{A_1,B\}+c_2\{A_2,B\}\;.
\end{equation}
\item[(b)] Antisymmetry. For any two functions $A,B$ on the phase space,
\begin{equation}
\{A,B\}=-\{B,A\}\;.
\end{equation}
\item[(c)] For any triplet of functions $A,B,C$ on the phase, the double Poisson bracket satisfies the \emph{Jacobi identity},
\begin{equation}
\{A,\{B,C\}\}+\{B,\{C,A\}\}+\{C,\{A,B\}\}=0\;.
\label{ch06:Jacobi}
\end{equation}
\item[(d)] Finally, the Poisson bracket satisfies an analog of the Leibniz (product) rule,
\begin{equation}
\{A,BC\}=B\{A,C\}+C\{A,B\}\;,
\label{ch06:productrule}
\end{equation}
for any triplet of functions $A,B,C$ on the phase space.
\end{enumerate}
The properties (a), (b) and (c) together imply that the space of functions on the phase space is a \emph{Lie algebra}. Adding the product rule (d) endows the space of functions on the phase space with the structure of a \emph{Poisson algebra}. The bilinearity and antisymmetry follow immediately from the definition~\eqref{ch06:poissonbracket} of the Poisson bracket. Likewise, the product rule~\eqref{ch06:productrule} for the Poisson bracket is a simple consequence of that for the partial derivatives of functions. The only property that takes some effort to prove is the Jacobi identity~\eqref{ch06:Jacobi}. I will skip details, but in case you would like to try your luck, you will need the Bianchi identity~\eqref{ch06:Bianchi} and patience.

The above already makes it clear that the simple shorthand notation~\eqref{ch06:poissonbracket} reveals a lot of structure that would be hard to notice otherwise. Here is another bit. The introduction of the Poisson bracket above was motivated by the evolution equation~\eqref{ch06:evolF} for the value of a function $F$ along a trajectory on the phase space. We now replace the Hamiltonian with an arbitrary (time-independent) function $Q$ on the phase space. This still defines a set of ``trajectories'' on the phase space via the map $\x^i_0\to\x^i(t)$, where $\x^i(t)$ is the solution to the equation
\begin{equation}
\dot\x^i(t)=\at{\{\x^i,Q\}}{\x=\x(t)}\quad\text{such that }\x^i(0)=\x^i_0\;.
\label{ch06:flow}
\end{equation}
In mathematical terminology, the function $Q$ generates a \emph{flow} on the phase space via~\eqref{ch06:flow}. It is often more practical to work with infinitesimally small time increments $\udelta t$, which results in a shift of the canonical coordinates, directly reflecting the Poisson bracket,
\begin{equation}
\udelta\x^i=\udelta t\{\x^i,Q\}\;.
\label{ch06:flowinfty}
\end{equation}
Here comes the magic: for \emph{any} choice of $Q$ and any fixed $t$, the map $\x^i_0\to\x^i(t)$ generated by $Q$ is a canonical transformation! The proof is technical and relies on both the Bianchi identity~\eqref{ch06:Bianchi} for the symplectic form and the Jacobi identity~\eqref{ch06:Jacobi} for the Poisson bracket. I will thus skip details and instead work out a simple example.

\begin{illustration}%
In~\refex{ex06:symplinear}, we saw an example of a canonical transformation depending on a continuous parameter. Inverting~\eqref{ch06:pqalpha} gives
\begin{equation}
\tilde q(\a)=q\cos\a-p\sin\a\;,\qquad
\tilde p(\a)=q\sin\a+p\cos\a\;.
\label{ch06:flowexample}
\end{equation}
We now want to find a function $Q$ that generates the map $(q,p)\to(\tilde q(\a),\tilde p(\a))$.  This is most easily identified with the help of the infinitesimal transformation~\eqref{ch06:flowinfty}, which suggests that
\begin{equation}
\udelta q=-p\udelta\a\ifeq\udelta\a\{q,Q\}=\udelta\a\PD Qp\;,\qquad
\udelta p=q\udelta\a\ifeq\udelta\a\{p,Q\}=-\udelta\a\PD Qq\;.
\end{equation}
It is easy to see that $Q(q,p)=-(q^2+p^2)/2$ does the job. We should still make sure that~\eqref{ch06:flowexample} actually represents the exact flow generated by our $Q$. This means that $\tilde q(\a)$ and $\tilde p(\a)$ should satisfy, for all $\a$, the differential equation~\eqref{ch06:flow}, which in the present case takes the form
\begin{equation}
\OD{\tilde q(\a)}\a=\{q,Q\}\Bigr\rvert_{\substack{q=\tilde q(\a)\\p=\tilde p(\a)}}=-\tilde p(\a)\;,\quad
\OD{\tilde p(\a)}\a=\{p,Q\}\Bigr\rvert_{\substack{q=\tilde q(\a)\\p=\tilde p(\a)}}=\tilde q(\a)\;.
\end{equation}
These flow equations are indeed solved by~\eqref{ch06:flowexample} with the initial condition $\tilde q(0)=q$ and $\tilde p(0)=p$.
\end{illustration}

%%%%%%%%%%%%%%%%%%%%%%%%%%%%%%%%%%%%%%%%%%%%%%%%%%%%%%%%%%%%

\subsection{Symmetries and Conservation Laws}
\label{subsec:Hamsymcons}

The discussion of flow generated by the Poisson bracket with a given function on the phase space leads to the last topic of our exploration of the geometry of mechanics. It follows from the evolution equation~\eqref{ch06:evolF} that a time-independent function $F$ on the phase space is a constant (integral) of motion if and only if $\{F,H\}=0$. Here is another piece of magic: given two constants of motion $F,G$, their Poisson bracket $\{F,G\}$ is also a constant of motion. The proof is very simple this time, and relies on the Jacobi identity~\eqref{ch06:Jacobi} in the form
\begin{equation}
\{\{F,G\},H\}=-\{H,\{F,G\}\}=\{F,\{G,H\}\}+\{G,\{H,F\}\}=0\;.
\end{equation}
The Poisson bracket allows us to construct new constants of motion out of those we already know. That is however not all. Any constant of motion $Q$, as any other function on the phase space, defines a flow $\x^i_0\to\x^i(t)$ via~\eqref{ch06:flow}. Under this flow, the Hamiltonian $H(\x(t))$ satisfies the evolution equation
\begin{equation}
\begin{split}
\dot H(\x(t))&=\at{\PD H{\x^i}}{\x=\x(t)}\dot\x^i(t)=\at{\PD H{\x^i}}{\x=\x(t)}\at{\{\x^i,Q\}}{\x=\x(t)}\\
&=\at{\O^{ij}\PD H{\x^i}\PD Q{\x^j}}{\x=\x(t)}=\at{\{H,Q\}}{\x=\x(t)}=0\;,
\end{split}
\end{equation}
where in the last step I used the assumption that $Q$ is conserved. The Hamiltonian does not change under the flow generated by any constant of motion. The flow generated by a constant of motion $Q$ therefore has two features:
\begin{itemize}
\item It is a canonical transformation that preserves the symplectic form $\O_{ij}$.
\item It is a symmetry of the Hamiltonian of the system.
\end{itemize}
What we have discovered is a very general statement that any constant of motion gives rise via~\eqref{ch06:flow} to a symmetry of the system. We will see in~\chaptername~\ref{chap:symmetries} that the opposite is also true: knowing a priori a symmetry of the system allows us to construct the corresponding constant of motion, or conservation law. Symmetries and conservation laws are in a one-to-one correspondence.

\begin{illustration}%
We already know (cf.~Sect.~\ref{subsec:Hamcyclic}) that if the Hamiltonian does not depend on one specific generalized coordinate $q^i$, then the corresponding conjugate momentum $p_i$ is a constant of motion. This agrees with our new formulation of conservation laws, since according to~\eqref{ch06:poissonconventional}, $\{p_i,H\}=-\Pd H{q^i}=0$. With the Poisson brackets~\eqref{ch06:PoissonDarboux} for the Darboux coordinates, one finds that the conserved momentum $p_i$ generates a canonical transformation on the phase space, solving the differential equations $\dot{\tilde q}^j(t)=\d^j_i$ and $\dot{\tilde p}_j(t)=0$. With the initial condition $\smash{\tilde q^j(0)=q^j}$ and $\smash{\tilde p_j(0)=p_j}$, the unique solution is $\tilde q^j(t)=\smash{q^j+\d^j_it}$ and $\tilde p_j(t)=p_j$. In other words, the conserved momentum $p_i$ generates translations of the conjugate coordinate $q^i$, while leaving all the other canonical variables unchanged.
\end{illustration}

\begin{illustration}%
\label{ex06:angularmomentum}%
The advantage of using the Poisson bracket is that it allows us to identify constants of motion without having to switch variables so that one of the generalized coordinates is cyclic. For an example, consider a particle in $\R^3$ in a central field. The Hamiltonian~is
\begin{equation}
H=\frac{\vec p^2}{2m}+V(r)\;.
\end{equation}
Here the canonical variables are $r^i,p_i$, and $\smash{r\equiv\sqrt{\d_{ij}r^ir^j}}$ is the radial distance. In this system, the vector of angular momentum, $\vec J\equiv\vekt rp$, is a constant of motion. To prove that $\vec J$ is conserved, it is sufficient to check that $\{J^i,\vec p^2\}=\{J^i,V(r)\}=0$ separately for each component of $\vec J$. This can be verified using the product rule~\eqref{ch06:productrule}. Let us show how the calculation goes in case of $J^1=r^2p_3-r^3p_2$. First,
\begin{equation}
\{J^1,\vec p^2\}=\PD{J^1}{r^i}\PD{\vec p^2}{p_i}=2\sum_{i=2}^3\PD{J^1}{r^i}p_i=2(p_3p_2-p_2p_3)=0\;.
\end{equation}
The Poisson bracket with the potential is equally simple,
\begin{equation}
\begin{split}
\{J^1,V(r)\}&=-\PD{J^1}{p_i}\PD{V(r)}{r^i}=-\sum_{i=2}^3\PD{J^1}{p_i}\frac{r^i}rV'(r)\\
&=-\frac{V'(r)}r(-r^3r^2+r^2r^3)=0\;.
\end{split}
\end{equation}
The same argument applies to $J^2$ and $J^3$ by mere permutation of indices, which shows that the whole vector $\vec J$ is conserved.

What are the canonical transformations that the three components of angular momentum generate? Using again $J^1$ as an example, the infinitesimal transformation~\eqref{ch06:flowinfty} amounts to
\begin{equation}
\begin{aligned}
\udelta r^1&=0\;,\qquad&
\udelta r^2&=-r^3\udelta t\;,\qquad&
\udelta r^3&=+r^2\udelta t\;,\\
\udelta p_1&=0\;,\qquad&
\udelta p_2&=-p_3\udelta t\;,\qquad&
\udelta p_3&=+p_2\udelta t\;.
\end{aligned}
\end{equation}
This makes it clear that $J^1$ generates rotations of both $\vec r$ and $\vec p$ in the plane of the second and third component, that is around the first coordinate axis. Repeating the same steps for $J^2$ and $J^3$ shows that they likewise generate rotations of $\vec r$ and $\vec p$ around the second and third axis, respectively.
\end{illustration}

%%%%%%%%%%%%%%%%%%%%%%%%%%%%%%%%%%%%%%%%%%%%%%%%%%%%%%%%%%%%

\section*{\probsec}
\addcontentsline{toc}{section}{\probsec}

\begin{prob}
\label{pr06:lagsphere}
Write down the Riemannian metric on a unit sphere $S^2$ in the spherical coordinates $\t,\vp$. Calculate the Christoffel symbols and convince yourself that the solutions of the geodesic equation $\smash{\ddot q^i+\Gamma^i_{jk}\dot q^j\dot q^k=0}$ are great circles. Hint: you do not need to solve the geodesic equation in full generality. By a suitable choice of orientation of the Cartesian coordinate axes with respect to which the spherical angles $\t,\vp$ are defined, one can always assume the geodesic to satisfy a simple initial condition such as $\t(0)=\pi/2$ and $\dot\t(0)=0$.
\end{prob}

\begin{prob}
\label{pr06:canon}
Consider a system with one degree of freedom with the standard Darboux coordinates $q,p$. We now wish to redefine these coordinates to
\begin{equation}
\tilde q=f(q)\;,\qquad
\tilde p=g(q,p)\;,
\end{equation}
where $f$ is a function of a single variable whereas $g$ depends on two variables. Given the function $f(q)$, what are the allowed choices of $g(q,p)$ so that the transformation is canonical? Give some explicit examples.
\end{prob}

\begin{prob}
\label{pr06:LeviCivita}
In this and the following problems, we will simplify the notation by writing all indices as subscripts. Now the \emph{Levi-Civita symbol} $\ve_{ijk}$ with $i,j,k\in\{1,2,3\}$ is defined to equal $+1$ if $(i,j,k)$ is an even permutation of $(1,2,3)$ and $-1$ if $(i,j,k)$ is an odd permutation of $(1,2,3)$. In all other cases, $\ve_{ijk}=0$. Prove the following identities for a product of two Levi-Civita symbols, summed over one or two indices,
\begin{equation}
\ve_{ijm}\ve_{klm}=\d_{ik}\d_{jl}-\d_{il}\d_{jk}\;,\qquad
\ve_{ikl}\ve_{jkl}=2\d_{ij}\;.
\label{ch06:epseps}
\end{equation}
\end{prob}

\begin{prob}
\label{pr06:PoissonJrP}
Write the Cartesian components of angular momentum of a particle in three dimensions as $J_i=\ve_{ijk}r_jp_k$. Using the results of~\refpr{pr06:LeviCivita}, show that
\begin{equation}
\{J_i,r_j\}=\ve_{ijk}r_k\;,\qquad
\{J_i,p_j\}=\ve_{ijk}p_k\;,\qquad
\{J_i,J_j\}=\ve_{ijk}J_k\;.
\end{equation}
\end{prob}

\begin{prob}
\label{pr06:sympS2}
Using the components of the symplectic form on a sphere in spherical coordinates, calculated in~\refex{ex06:sympS2}, show that the Cartesian components of the unit vector $\vec n$ satisfy the Poisson bracket
\begin{equation}
\{n_i,n_j\}=\b\ve_{ijk}n_k\;.
\label{ch06:PoissonS2}
\end{equation}
How is the parameter $\b$ related to the overall scale $\a$ of the symplectic form~\eqref{ch06:sympS2}? Check that this Poisson bracket reproduces the Landau--Lifshitz equation~\eqref{ch06:LandauLifshitz} via $\dot{\vec n}=\{\vec n,H\}$. Now think of the unit vector $\vec n$ as describing a particle with spin of magnitude $J$ via $\vec n=\vec J/J$. Using the Poisson bracket for angular momentum, derived in~\refpr{pr06:PoissonJrP}, fix $\beta$ in terms of $J$. The dynamics of spin in an external magnetic field $\vec B$ is governed by the Hamiltonian $H=-\m\skal JB$, where the constant $\m$ is the magnetic moment of the particle. Solve the Landau--Lifshitz equation for this Hamiltonian and show that its solution describes \emph{Larmor precession} of the spin.
\end{prob}

\begin{prob}
\label{pr06:poiEM}
For a charged particle in an electromagnetic field, the momentum $\vec p$ conjugate to the position vector $\vec r$ is $\vec p=m\dot{\vec r}+q\vec A(\vec r,t)$, where $m,q$ are the mass and charge of the particle and $\vec A$ is the vector potential of the external field. Since the conjugate momentum depends on the vector potential, it is not physical, that is measurable. Therefore, one sometimes uses instead the physical, kinematical momentum, $\vec\pi\equiv m\dot{\vec r}=\vec p-q\vec A(\vec r,t)$. Using the standard Poisson brackets $\{r_i,r_j\}=\{p_i,p_j\}=0$ and $\{r_i,p_j\}=\d_{ij}$, show that the components of the kinematical momentum satisfy
\begin{equation}
\{r_i,\pi_j\}=\d_{ij}\;,\qquad
\{\pi_i,\pi_j\}=q\ve_{ijk}B_k\;,
\end{equation}
where $\vec B=\rot\vec A$ is the magnetic field. Writing the Hamiltonian of the particle as $H=\vec\pi^2/(2m)+q\p(\vec r,t)$, where $\p$ is the scalar potential of the electromagnetic field, derive the EoM for $\vec\pi$ and check that it has the expected form. Hint: you have to take into account the explicit time dependence of the kinematical momentum through the vector potential. This modifies the evolution equation for $\vec\pi$ to $\dot{\vec\pi}=\{\vec\pi,H\}-q\Pd{\vec A}t$.
\end{prob}

\begin{prob}
\label{pr06:isoosc}
Consider a particle of mass $m$ moving in $n$ dimensions in the isotropic potential $V(\vec r)=(1/2)m\o^2\vec r^2$. Show that the following quantities are conserved for any $i,j$,
\begin{equation}
A_{ij}\equiv\frac{p_ip_j}{2m}+\frac12m\o^2r_ir_j\;.
\end{equation}
Remark: together with the angular momentum tensor $J_{ij}\equiv r_ip_j-r_jp_i$, this gives us altogether $n^2$ constants of motion. There must obviously be numerous relationships between them since a general solution to the EoM only depends on $2n$ integration constants. Try to think what these relations might be.
\end{prob}
\chapter{Application: Dynamics of Rigid Bodies}
\label{chap:rigidbody}

\keywords{Rigid body, body frame and space frame, angular velocity, tensor of inertia, parallel axes theorem, principal axes and moments of inertia, Euler equations, precession, intermediate axis theorem.}

%%%%%%%%%%%%%%%%%%%%%%%%%%%%%%%%%%%%%%%%%%%%%%%%%%%%%%%%%%%%

\noindent Most of the applications of analytical mechanics we have discussed so far concerned the dynamics of systems consisting of point masses. The few examples of massive objects of a nonzero size were limited to simple shapes such as a sphere, where we could rely on our previous experience from introductory mechanics. In this \chaptername, we shall at last deal with the dynamics of extended objects systematically. The presentation of the material in this \chaptername{} is inspired by Chap.~8 of~\cite{Jose1998a}. See also~\cite{Leyvraz2015} for a more elementary treatment, very similar to ours, which stresses the matrix formulation of rotation kinematics.

Let me start with a few basic clarifications to set up the stage. First and foremost, we will only be concerned with finite \emph{rigid} bodies. The position and orientation of a rigid body may change with time, for instance as a consequence of the action of external forces. However, the shape is assumed to be fixed, as is the mass distribution inside the object. The rigidity assumption is reasonable for solid bodies, as long as the external forces are negligible compared to the internal elastic forces that would be induced by changing the shape of the body appreciably. Second, we will always assume one point of the body to have a fixed position in space. This is just a matter of practical convenience. We know that for nonrelativistic dynamics, the motion of the \emph{center of mass} (CoM) can be separated from the ``internal motion'' in relative coordinates. We can therefore think that the CoM motion has already been solved for, and focus on the rotation of the body around the CoM. In this case, the CoM would obviously be our fixed point. However, one can also imagine situations where any chosen point in the body is physically attached to a fixed position, as in a pendulum. The analysis of the rigid body problem in this \chaptername{} is valid regardless of the choice of the fixed point. Finally, unlike in \chaptername~\ref{chap:geometryclassmech}, I shall restrict from the outset to three spatial dimensions, and indicate all spatial indices with subscripts.

%%%%%%%%%%%%%%%%%%%%%%%%%%%%%%%%%%%%%%%%%%%%%%%%%%%%%%%%%%%%

\section{Kinematics of a Rigid Body}

\begin{figure}[t]
\sidecaption[t]
\includegraphics[width=2.9in]{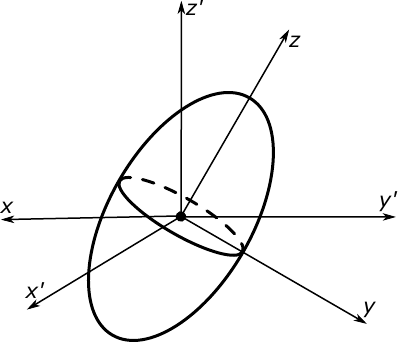}
\caption{Visualization of the body frame, attached to the rigid body, and the space frame, fixed in the laboratory IRF. The two Cartesian frames share the origin, placed at the fixed point of the body.}
\label{fig07:spacebodyframe}
\end{figure}

We will analyze the dynamics of a rigid body using the Lagrangian formalism. To that end, we first need to identify the configuration space. We shall use a Cartesian coordinate frame, dubbed the \emph{body frame}, attached to the rigid body. The origin of this Cartesian frame is chosen to coincide with the fixed point of the body. The position of any other point of the rigid body is then specified by its coordinates $r_i$ in the body frame. The rigidity requirement is equivalent to saying that the coordinates $r_i$ of any point of the body are constants, independent of time. The orientation of the rigid body in space is described using another Cartesian coordinate frame: the \emph{space frame}. This is fixed to an \emph{inertial reference frame} (IRF), and its origin coincides with that of the body frame. I will distinguish the body and space frames by using primes for the latter. See Fig.~\ref{fig07:spacebodyframe} for an illustration of the two frames.

I will denote as $\vec e_i$ the basis of unit vectors pointing along the coordinate axes of the body frame, and analogously $\vec e_i'$ for the space frame. The basis vectors and the components of the position vector $\vec r$ in the two frames are related by mutually inverse linear transformations,
\begin{equation}
\vec e_i=\vec e_j'R_{ji}\;,\qquad
r_i=(R^{-1})_{ij}r_j'\;,
\label{ch07:eidef}
\end{equation}
so that $\vec r=r_i\vec e_i=r_i'\vec e_i'$ is independent of the choice of frame. Since both $\vec e_i$ and $\vec e_i'$ form an orthonormal basis, the matrix $R$ must be orthogonal,
\begin{equation}
R^TR=RR^T=\un\quad\Leftrightarrow\quad
R_{ki}R_{kj}=R_{ik}R_{jk}=\d_{ij}\;.
\label{ch07:OGconstraint}
\end{equation}
I will in addition always assume that the two coordinate frames have the same (right-handed) orientation, which implies that $R$ is a proper rotation and accordingly has a unit determinant. The trajectory of the rigid body is completely specified by giving the orthonormal basis $\vec e_i$ of the body frame, or equivalently the rotation matrix $R$, as a function of time. Our first conclusion therefore is that the configuration space of the rigid body consists of the set of proper rotations.

\begin{illustration}%
\label{ex07:2drotations}%
For the sake of illustration, let us restrict the motion of the rigid body to two dimensions. This amounts to identifying the $z$- and $z'$-axes of the body and space frames, and only allowing rotation in the $xy$ plane. Such rotation can be described by a $2\times2$ orthogonal matrix $R$, parameterized by a single angle $\vp$ such that
\begin{equation}
R=\begin{pmatrix}
\cos\vp & -\sin\vp\\
\sin\vp & \cos\vp
\end{pmatrix}\;.
\label{ch07:rotation2d}
\end{equation}
This implies the following, mutually dual transformations of the basis vectors and the corresponding coordinates,
\begin{equation}
\begin{pmatrix}
\vec e_x,\vec e_y
\end{pmatrix}=
\begin{pmatrix}
\vec e_x',\vec e_y'
\end{pmatrix}
\raisebox{-1.3ex}{$\begin{pmatrix}
\cos\vp & -\sin\vp\\
\sin\vp & \cos\vp
\end{pmatrix}$}\;,\qquad
\raisebox{-1.3ex}{$\begin{pmatrix}
x\\
y
\end{pmatrix}$}=
\raisebox{-1.3ex}{$\begin{pmatrix}
\cos\vp & \sin\vp\\
-\sin\vp & \cos\vp
\end{pmatrix}$}
\raisebox{-1.3ex}{$\begin{pmatrix}
x'\\
y'
\end{pmatrix}$}\;.
\end{equation}
Mathematically, the configuration space for rotations of a rigid body in two dimensions, that is around a fixed axis, corresponds to a circle, $S^1$, parameterized by the angle $\vp$. This is the same configuration space as, for instance, for a pendulum swinging in a fixed vertical plane.
\end{illustration}

\begin{watchout}%
Guided by Sect.~\ref{sec:geomlag}, we can conclude that the motion of a free rigid body traces a geodesic on the manifold of proper rotations, which is the Lie group $\gr{SO}(3)$. However, while certainly true, this does not really help the intuition. First, we do not know how such a configuration space looks like as a manifold. Second, we do not even know how to construct a Riemannian metric on it. I will therefore content myself with this remark on the geometry of the motion of rigid bodies, and follow a more elementary approach closer to physics.
\end{watchout}

Next, we need to understand how to describe the kinematics of the rigid body in the IRF to which the space frame is attached. This is characterized by the time dependence of $\vec e_i$, or equivalently $R_{ji}$, in~\eqref{ch07:eidef}. Taking a time derivative and expressing the result in the body frame, we get
\begin{equation}
\dot{\vec e}_i=\vec e_k'\dot R_{ki}=\vec e_j(R^{-1})_{jk}\dot R_{ki}\equiv\vec e_j\O_{ji}\;,
\label{ch07:dotei}
\end{equation}
where we introduced the \emph{angular velocity} matrix $\O\equiv R^{-1}\dot R$. Using the fact that $R$ is orthogonal so that $R^{-1}=R^T$ and $\Od{(R^TR)}{t}=\dot R^TR+R^T\dot R=0$, we easily deduce that $\O$ is antisymmetric,
\begin{equation}
\O^T=(R^T\dot R)^T=\dot R^TR=-R^T\dot R=-\O\;.
\end{equation}
It follows that $\O$ can be parameterized by a vector $\vec\o=\o_i\vec e_i$,
\begin{equation}
\O=\begin{pmatrix}
0 & -\o_3 & +\o_2\\
+\o_3 & 0 & -\o_1\\
-\o_2 & +\o_1 & 0
\end{pmatrix}\quad\Leftrightarrow\quad
\O_{ij}=-\ve_{ijk}\o_k\;.
\end{equation}
Finally, we use the fact that the orthonormal basis $\vec e_i$ is right-handed so that $\vec e_i\times\vec e_j=\ve_{ijk}\vec e_k$. This allows us to rewrite~\eqref{ch07:dotei} as
\begin{equation}
\dot{\vec e}_i=-\vec e_j\ve_{jik}\o_k=\o_k\ve_{kij}\vec e_j=\o_k\vec e_k\times\vec e_i\quad\Rightarrow\quad
\boxed{\dot{\vec e}_i=\vekt\o e_i\;.}
\label{ch07:dotei2}
\end{equation}
This is a familiar expression for the rate of change of a vector under rotation with angular velocity $\vec\o$.

\begin{illustration}%
\label{ex07:2drotations2}%
Let us briefly return to the simpler case of two-dimensional rotations, introduced in \refex{ex07:2drotations}. With the $2\times2$ rotation matrix~\eqref{ch07:rotation2d}, we find that the angular velocity matrix is
\begin{equation}
\O=R^T\dot R=\begin{pmatrix}
0 & -\dot\vp\\
\dot\vp & 0
\end{pmatrix}\;,
\label{ch07:Omega2d}
\end{equation}
which encodes the scalar angular velocity of rotations in a plane in a trivial manner. Moreover, the rate of change of the body frame~\eqref{ch07:dotei2} boils down to $\dot{\vec e}_i=\dot\vp\ve_{ij}\vec e_j$, or equivalently $\dot{\vec e}_x=\dot\vp\vec e_y$ and $\dot{\vec e}_y=-\dot\vp\vec e_x$.
\end{illustration}

Before we move on, let us quickly summarize the two most important results we have found so far:
\begin{itemize}
\item The configuration space of a rigid body moving around a fixed point consists of proper rotations, or equivalently orthogonal matrices $R$ with unit determinant.
\item The kinematical state of the rigid body is specified by the configuration matrix $R$ and the angular velocity matrix $\O=R^{-1}\dot R$ as a function of time. The matrix $\O$ is related to the more familiar angular momentum vector $\vec\o$ by $\O_{ij}=-\ve_{ijk}\o_k$, where $\o_i$ are the components of $\vec\o$ in the body frame.
\end{itemize}

%%%%%%%%%%%%%%%%%%%%%%%%%%%%%%%%%%%%%%%%%%%%%%%%%%%%%%%%%%%%

\section{Dynamics of a Rigid Body}

Following the philosophy of the Lagrangian formalism, we now have to construct the Lagrangian, and then derive the corresponding \emph{equation of motion} (EoM). Altogether, this is not a difficult task. However, owing to the nontrivial geometry of the configuration space, the calculation involves certain amount of algebraic manipulations. We will therefore take it one step at a time. 

%%%%%%%%%%%%%%%%%%%%%%%%%%%%%%%%%%%%%%%%%%%%%%%%%%%%%%%%%%%%

\subsection{Kinetic Energy and Tensor of Inertia}

We start by writing down the kinetic energy of the rotating rigid body as observed in the space frame. The velocity of a chosen point in the body is obtained by inverting the relation between $r_i$ and $r_i'$ in~\eqref{ch07:eidef} and taking a derivative, $\dot r_i'=\dot R_{ij}r_j$. (Remember that the body coordinates $r_i$ of any point in the rigid body are fixed.) Introducing the mass density $\vr(\vec r)$ of the rigid body and denoting as $V$ the domain in the body frame occupied by the body, the total kinetic energy is then simply
\begin{equation}
T=\frac12\int_V\D^3\!\vec r\,\vr(\vec r)\dot r_k'\dot r_k'=\frac12\dot R_{ki}\dot R_{kj}\int_V\D^3\!\vec r\,\vr(\vec r)r_ir_j\;.
\end{equation}
The integral on the right-hand side is independent of the motion of the rigid body. It describes the second moment of the mass distribution in the body; I will use the shorthand notation $K_{ij}$ for it. The kinetic energy can then be written compactly as
\begin{equation}
T=\frac12K_{ij}\dot R_{ki}\dot R_{kj}=\frac12\tr(\dot RK\dot R^T)\;,\qquad
K_{ij}\equiv\int_V\D^3\!\vec r\,\vr(\vec r)r_ir_j\;.
\label{ch07:kinenergy}
\end{equation}

We will use~\eqref{ch07:kinenergy} below when we write down the full Lagrangian for the rigid body. However, to get more physical insight, it is useful to rewrite the kinetic energy in terms of the angular velocity vector $\vec\o$. To that end, note that $\dot R^T\dot R=\dot R^TRR^T\dot R=\O^T\O=\O\O^T$. This allows us to express the kinetic energy as
\begin{equation}
T=\frac12K_{kl}\O_{km}\O_{lm}=\frac12K_{kl}\ve_{kmi}\ve_{lmj}\o_i\o_j=\frac12K_{kl}(\d_{ij}\d_{kl}-\d_{il}\d_{kj})\o_i\o_j\;,
\end{equation}
where we used the first identity in~\eqref{ch06:epseps}. Contracting the indices in the parentheses with $K_{kl}$ finally brings the kinetic energy to the form
\begin{equation}
\boxed{T=\frac12I_{ij}\o_i\o_j\;,\qquad
I_{ij}\equiv\d_{ij}\tr K-K_{ij}=\int_V\D^3\!\vec r\,\vr(\vec r)(\d_{ij}\vec r^2-r_ir_j)\;.}
\label{ch07:kinenergy2}
\end{equation}
By going from~\eqref{ch07:kinenergy} to~\eqref{ch07:kinenergy2}, we have effectively traded the matrix $K_{ij}$ of second moments of the mass distribution for the \emph{tensor of inertia} $I_{ij}$.

\begin{illustration}%
In introductory mechanics, one usually deals with rotations around a fixed axis. Suppose the direction of the axis is given by a unit vector $\vec n$. The rotation can then be parameterized by a scalar angular velocity $\o$ in terms of which $\vec\o=\o\vec n$, or $\o_i=\o n_i$. The kinetic energy~\eqref{ch07:kinenergy2} now becomes
\begin{equation}
T=\frac12I_{\vec n}\o^2\;,\qquad
I_{\vec n}\equiv I_{ij}n_in_j=
\int_V\D^3\!\vec r\,\vr(\vec r)[\vec r^2-(\skal nr)^2]\;.
\label{ch07:In}
\end{equation}
Here $I_{\vec n}$ is the moment of inertia for rotations around the fixed axis $\vec n$. In practice, one seldom needs to calculate the moment of inertia from scratch though. There are large tables that compile moments of inertia for a number of different geometric shapes; see for instance the relevant \href{https://en.wikipedia.org/wiki/List_of_moments_of_inertia
}{Wikipedia page}.
\end{illustration}

\begin{watchout}%
As briefly hinted above, the kinetic energy~\eqref{ch07:kinenergy} establishes a metric on the Lie group $\gr{SO}(3)$ of proper rotations. This metric is not uniquely determined by the geometry of spatial rotations, but rather depends on the mass distribution in the rigid body. Its properties reflect the symmetries of both the rigid body and the ambient space. On the one hand, one can always change the space frame by an orthogonal transformation $\smash{\vec e_i'\to\tilde{\vec e}_i'\equiv\vec e_j'P_{ji}}$. This can be incorporated by replacing $R$ with $\smash{P^TR}$ everywhere. Such a change of basis does not affect the kinetic energy~\eqref{ch07:kinenergy}, which is therefore invariant under any rotation of the ambient space. On the other hand, one may also want to change the body frame, which would amount to the orthogonal transformation $\vec e_i\to\tilde{\vec e}_i\equiv\vec e_j P_{ji}$. This is equivalent to replacing $R$ with $RP$. In the kinetic energy~\eqref{ch07:kinenergy}, the change of body frame therefore manifests effectively by replacing $K$ with $PKP^T$. Obviously, not every change of the body frame leaves the kinetic energy intact. The allowed transformations of the body frame correspond to symmetries of the $K_{ij}$ matrix, or of the tensor of inertia itself.
\end{watchout}

Before we proceed, it is useful to mention some properties of the tensor of inertia. First, it is obvious from~\eqref{ch07:In} that $I_{\vec n}$ is always non-negative, since $\vec r^2-(\skal nr)^2$ is just the squared distance of $\vec r$ from the axis of rotation. This is sufficient to conclude that $I_{ij}$ as a quadratic form is positive-semidefinite. In fact, it is strictly positive-definite for any body that has a nonzero size in all three dimensions. Second, the tensor of inertia can be diagonalized by a suitable choice of the basis $\vec e_i$, in which
\begin{equation}
I=\begin{pmatrix}
I_1 & 0 & 0\\
0 & I_2 & 0\\
0 & 0 & I_3
\end{pmatrix}\quad\Rightarrow\quad
T=\frac12(I_1\o_1^2+I_2\o_2^2+I_3\o_3^2)\;.
\end{equation}
The eigenvalues $I_i$ of the tensor of inertia are called the \emph{principal moments of inertia}, and the corresponding basis vectors $\vec e_i$ define the \emph{principal axes of inertia}.

\begin{illustration}%
For a uniform solid sphere of mass $m$ and radius $R$, the moment of inertia is $(2/5)mR^2$ with respect to any axis passing through the center. Accordingly, the tensor of inertia is proportional to the unit matrix, $I_{ij}=(2/5)\d_{ij}mR^2$. Any set of Cartesian coordinate axes passing through the center can be taken as the principal axes of inertia. The freedom to choose the principal axes at will reflects the symmetry of the sphere under rotations around the center.

For a uniform solid cylinder of mass $m$, radius $R$ and height $H$, the moment of inertia with respect to the axis of symmetry is $(1/2)mR^2$. The moment of inertia with respect to any axis passing through the CoM and perpendicular to the axis of symmetry is $(1/12)m(3R^2+H^2)$. Here the principal axes of inertia can be chosen as any set of Cartesian coordinate axes passing through the CoM, one of which coincides with the symmetry axis of the cylinder. The freedom to choose the principal axes again reflects the symmetry of the cylinder, which is however lower than that of the sphere. Another manifestation of the symmetry of the cylinder is that two of the three principal moments of inertia are equal to each other.

For a rigid body such that all the three principal moments of inertia are different, the principal axes of inertia are uniquely determined by the eigenvectors of the tensor of inertia.
\end{illustration}

\begin{figure}[t]
\sidecaption[t]
\includegraphics[width=2.0in]{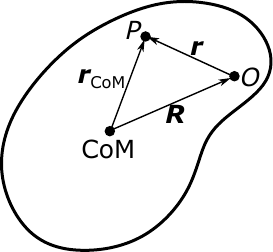}% This figure does not include so many details to warrant a full width of 2.9 in!
\caption{Schematic illustration of the geometry involved in shifting the position of the fixed point from the CoM to the point $O$. The symbol $P$ refers to the variable point whose location one integrates over to determine the tensor of inertia via~\eqref{ch07:kinenergy2}.}
\label{fig07:parallelaxes}
\end{figure}

Third, should we decide to change the point in the rigid body that we keep fixed, we do not need to recalculate the whole tensor of inertia every time. Suppose we choose as the reference point the CoM, and then switch to the point $O$ (see Fig.~\ref{fig07:parallelaxes}). The tensor of inertia with respect to $O$ can be written as
\begin{equation}
I_{ij}=\int_V\D^3\!\vec r\,\vr(\vec r)[\d_{ij}(\vec r_\mathrm{CoM}-\vec R)^2-(r_{\mathrm{CoM},i}-R_i)(r_{\mathrm{CoM},j}-R_j)]\;.
\end{equation}
By expanding the expression in square brackets in powers of $\vec R$, we find that the terms linear in $\vec R$ drop by the definition of CoM. We then get a simple relation between the tensors of inertia with respect to $O$ and the CoM, respectively,
\begin{equation}
\boxed{I_{ij}=I_{ij}^\mathrm{CoM}+m(\d_{ij}\vec R^2-R_iR_j)\;,}
\end{equation}
where $m$ is the total mass of the rigid body. A special case of this tensor identity follows by projecting it to the direction defined by a unit vector $\vec n$,
\begin{equation}
I_{\vec n}=I_{\vec n}^\mathrm{CoM}+m[\vec R^2-(\skal nR)^2]\;.
\end{equation}
This is known as the \emph{parallel axis} (or \emph{Steiner}) \emph{theorem}.

\begin{illustration}%
Consider a thin uniform rod of mass $m$ and length $L$. The moment of inertia with respect to an axis passing through the center of the rod and perpendicular to it is $I^\mathrm{CoM}_\perp=(1/12)mL^2$. By the Steiner theorem, the moment of inertia with respect to an axis passing through one of the ends of the rod, and perpendicular to the rod, is
\begin{equation}
I^\mathrm{end}_\perp=I^\mathrm{CoM}_\perp+m\left(\frac L2\right)^2=\frac13mL^2\;.
\end{equation}
\end{illustration}

%%%%%%%%%%%%%%%%%%%%%%%%%%%%%%%%%%%%%%%%%%%%%%%%%%%%%%%%%%%%

\subsection{Lagrangian of a Rigid Body}

We have finally reached the point where we can write down the Lagrangian for a rigid body using the rotation matrix $R$ as the generalized coordinate. One subtlety we have to be careful about is that the components $R_{ij}$ are not mutually independent due to the relations~\eqref{ch07:OGconstraint}. We deal with these constraints by introducing a symmetric matrix $\Lambda$ of Lagrange multipliers and adding to the Lagrangian the term $\Lambda_{ij}(R_{ki}R_{kj}-\d_{ij})$. Finally, we need to add a potential energy $V(R)$, which depends on the orientation of the body in some a priori unknown manner. Altogether, the Lagrangian becomes
\begin{equation}
\boxed{L=\frac12\tr(\dot RK\dot R^T)-V(R)+\tr(R\Lambda R^T-\Lambda)\;.}
\label{ch07:lagrangian}
\end{equation}

We can now derive the Lagrange EoM in the usual manner by treating the matrix elements $R_{ij}$ as independent dynamical variables. To streamline the derivation of the EoM, let us prepare two pieces of algebra beforehand. First, recall that the trace of a product of two matrices can be written as $\tr(AB)=A_{ij}B_{ji}$. Taking the derivative of the trace with respect to a matrix element of one of the matrices can thus be expressed in two equivalent ways,
\begin{equation}
\PD{[\tr(AB)]}{B_{ji}}=A_{ij}\quad\Leftrightarrow\quad
\PD{[\tr(AB)]}{B^T}=A\;.
\end{equation}
Second, we will need an expression for the second time derivative of $R$. To that end, we rewrite the definition $\O=R^T\dot R$ as $\dot R=R\O$ and differentiate,
\begin{equation}
\ddot R=\dot R\O+R\dot\O\quad\Rightarrow\quad
R^T\ddot R=\dot\O+\O^2\;.
\label{ch07:ddotR}
\end{equation}

With these preparations, we now start by taking the variation of the action based on~\eqref{ch07:lagrangian} with respect to $R$,
\begin{equation}
\ddot RK=2R\Lambda-\PD VR\;.
\end{equation}
In order to eliminate the Lagrangian multiplier matrix $\Lambda$, we multiply this equation from the left by $R^T$ and take the antisymmetric part. Using subsequently~\eqref{ch07:ddotR}, we arrive at the final matrix form of the EoM,
\begin{equation}
\boxed{(\dot\O K+K\dot\O)+(\O^2K-K\O^2)=\PD{V}{R^T}R-R^T\PD VR\;.}
\label{ch07:EoMmatrix}
\end{equation}
As we will show below, the left-hand side of~\eqref{ch07:EoMmatrix} has a simple interpretation in terms of the principal moments of inertia and the angular velocity. The right-hand side encodes the action of external forces on the rigid body through the corresponding torque. We do not need to work out the correspondence in detail, since we will mostly deal with the dynamics of free rigid bodies without external forces acting upon them.

\begin{illustration}%
Let us illustrate the content of the EoM~\eqref{ch07:EoMmatrix} on the simple case of two-dimensional rotations, discussed previously in~\refex{ex07:2drotations} and~\refex{ex07:2drotations2}. Here the EoM is an antisymmetric $2\times2$ equation and thus only has one independent component. With the expression~\eqref{ch07:Omega2d} for the angular velocity matrix, one easily checks that~\eqref{ch07:EoMmatrix} boils down to
\begin{equation}
I_\mathrm{2d}\ddot\vp=R_{i1}\PD V{R_{i2}}-R_{i2}\PD V{R_{i1}}\;,
\label{ch07:EoMmatrix2d}
\end{equation}
where $I_\mathrm{2d}\equiv\tr I=\tr K$. The right-hand side can be further rewritten using~\eqref{ch07:rotation2d} as
\begin{equation}
R_{i1}\PD V{R_{i2}}-R_{i2}\PD V{R_{i1}}=-\PD V{R_{ij}}\PD{R_{ij}}\vp\;.
\end{equation}
Thinking of the potential as a function of the angle $\vp$, this is simply $-V'(\vp)$ expressed through the matrix elements $R_{ij}$ via the chain rule. The EoM for two-dimensional rotations thus takes an extremely simple form,
\begin{equation}
I_\mathrm{2d}\ddot\vp=-V'(\vp)\;.
\label{ch07:EoM2d}
\end{equation}
The same result can be obtained even more easily by rewriting the Lagrangian~\eqref{ch07:lagrangian} in terms of the unconstrained variable $\vp$. Namely, using~\eqref{ch07:rotation2d} gives $L=(1/2)I_\mathrm{2d}\dot\vp^2-V(\vp)$, which in turn reproduces the EoM~\eqref{ch07:EoM2d}.
\end{illustration}

%%%%%%%%%%%%%%%%%%%%%%%%%%%%%%%%%%%%%%%%%%%%%%%%%%%%%%%%%%%%

\subsection{Euler Equations}

The matrix EoM~\eqref{ch07:EoMmatrix} is very compact, but also not very transparent. In order to appreciate its content, we switch to the body frame defined by the principal axes of inertia. In this frame, $K_{ij}$ is diagonal and we denote its diagonal elements as $K_i$, so that $K_{ij}=\d_{ij}K_i$ (no summation over $i$). These are related to the principal moments of inertia by~\eqref{ch07:kinenergy2},
\begin{equation}
K=\begin{pmatrix}
K_1 & 0 & 0\\
0 & K_2 & 0\\
0 & 0 & K_3
\end{pmatrix}\;,\qquad
I=\begin{pmatrix}
K_2+K_3 & 0 & 0\\
0 & K_1+K_3 & 0\\
0 & 0 & K_1+K_2
\end{pmatrix}\;.
\end{equation}
The first term on the left-hand side of~\eqref{ch07:EoMmatrix} then becomes
\begin{equation}
(\dot\O K+K\dot\O)_{ij}=-\sum_k\ve_{ijk}\dot\o_k(K_i+K_j)=-\sum_k\ve_{ijk}I_k\dot\o_k\;.
\end{equation}
Moreover, $\O^2$ is a symmetric matrix whose matrix elements can be computed with the help of the first of the identities in~\eqref{ch06:epseps}, $(\O^2)_{ij}=\O_{ik}\O_{kj}=-\ve_{ikm}\ve_{jkn}\o_m\o_n=-\d_{ij}\vec\o^2+\o_i\o_j$. This makes it possible to rewrite the second term on the left-hand side of~\eqref{ch07:EoMmatrix} as
\begin{equation}
(\O^2K-K\O^2)_{ij}=\o_i\o_j(K_j-K_i)=\o_i\o_j(I_i-I_j)\;.
\end{equation}
Putting all the pieces together, \eqref{ch07:EoMmatrix} is equivalent to a set of first-order \emph{ordinary differential equations} (ODEs) for the components of the angular velocity vector,
\begin{equation}
\boxed{\begin{aligned}
I_1\dot\o_1-(I_2-I_3)\o_2\o_3&=\tau_1\;,\\
I_2\dot\o_2-(I_3-I_1)\o_3\o_1&=\tau_2\;,\\
I_3\dot\o_3-(I_1-I_2)\o_1\o_2&=\tau_3\;,
\end{aligned}}
\label{ch07:Eulereq}
\end{equation}
where $\tau_i$ are the components of the torque due to external forces in the body frame. These are the so-called \emph{Euler equations}.

\begin{watchout}%
The Euler equations are highly implicit. Remember that $\o_i$ are the components of the angular velocity vector in the body frame, whose motion with respect to the IRF spanned on the space frame is as yet unknown. Indeed, the evolution of the basis vectors $\vec e_i$ of the body frame depends on the angular velocity itself through~\eqref{ch07:dotei2}. This looks quite hopeless. Luckily, we can take advantage of angular momentum, whose time dependence in the IRF is under control.
\end{watchout}

Building on the definition of angular momentum in introductory mechanics as the product of moment of inertia and angular velocity, we define the angular momentum vector $\vec J$ component-wise with respect to the principal axes of inertia,
\begin{equation}
\vec J\equiv\sum_iI_i\o_i\vec e_i\;,\qquad\text{or }\vec J\equiv(I_1\o_1,I_2\o_2,I_3\o_3)\text{ in the body frame}\;.
\label{ch07:angularmomentum}
\end{equation}
Writing the Euler equations~\eqref{ch07:Eulereq} collectively as $I_i\dot\o_i=\tau_i+\sum_{j,k}\ve_{ijk}I_j\o_j\o_k$, we can compute the time derivative of angular momentum with the help of~\eqref{ch07:dotei2},
\begin{equation}
\begin{split}
\vec{\dot J}&=\sum_iI_i(\dot\o_i\vec e_i+\o_i\dot{\vec e}_i)\\
&=\sum_i\Bigl(\tau_i+\cancel{\sum_{j,k}\ve_{ijk}I_j\o_j\o_k}\Bigr)\vec e_i+\cancel{\sum_iI_i\o_i\sum_{j,k}\ve_{ijk}\vec e_j\o_k}=\tau_i\vec e_i\;.
\end{split}
\end{equation}
This says that the rate of change of angular momentum is determined by the torque of the external forces. In the special of a free rigid body that is not subject to any forces, the angular momentum is conserved. This is of course well-known. Here I have just shown how the conservation of angular momentum can be recovered within the Lagrangian approach to rigid bodies. Note that the Lagrangian formalism does not \emph{require} us to introduce the concept of angular momentum, since the EoM follows directly from the Lagrangian in terms of the generalized coordinates $R_{ij}$. However, knowing that the vector $\vec{J}$ is time-independent will be very useful below when we search for concrete solutions to the Euler equations.

%%%%%%%%%%%%%%%%%%%%%%%%%%%%%%%%%%%%%%%%%%%%%%%%%%%%%%%%%%%%

\section{Sample Applications}

The Euler equations~\eqref{ch07:Eulereq} represent three coupled nonlinear first-order ODEs for the angular velocity. As such, they are not easy to solve in full generality. Instead, I will illustrate their utility on a few special cases. We will always assume that there are no external forces, or at least that the torque of the external forces with respect to the chosen fixed point of the rigid body is zero.\footnote{This weaker assumption is satisfied for instance for rigid bodies subject to a uniform gravitational field, provided the fixed point is chosen as the CoM. This makes it possible to apply the results of this \chaptername{} to rigid bodies freely falling in the Earth's gravitational field.}

%%%%%%%%%%%%%%%%%%%%%%%%%%%%%%%%%%%%%%%%%%%%%%%%%%%%%%%%%%%%

\subsection{Free Spherical Top}

We start with the simplest special case where all the three principal moments of inertia are equal, $I_1=I_2=I_3\equiv \bar I$. Such a rigid body is referred to as the \emph{spherical top}. An example of a spherical top would indeed be a sphere or a solid ball rotating around its CoM. However, there are other examples that do not necessarily have full spherical symmetry. For instance, a solid cube has equal moments of inertia with respect to all three axes passing through the CoM and parallel to some of the edges of the cube. Cubic symmetry is sufficient to ensure full isotropy of the tensor of inertia. Namely, $I_{ij}=\d_{ij}\bar I$ implies by~\eqref{ch07:In} that $I_{\vec n}=\bar I$ for any orientation of the axis $\vec n$ passing through the CoM.

With the above comments out of the way, let us see what the dynamics of a spherical top looks like. Thanks to the fact that all the principal moments of inertia are equal, the angular momentum is simply $\vec J=\bar I\vec\o$. Conservation of angular momentum then implies that the angular velocity is a constant vector. The spherical top rotates at a constant rate around a fixed axis, defined by the direction of $\vec\o$. The same conclusion can be obtained directly from the Euler equations~\eqref{ch07:Eulereq}.

%%%%%%%%%%%%%%%%%%%%%%%%%%%%%%%%%%%%%%%%%%%%%%%%%%%%%%%%%%%%

\subsection{Free Symmetric Top}
\label{subsec:symtop}

A \emph{symmetric top} is a rigid body whose tensor of inertia is not completely isotropic, but rather has a reduced, axial symmetry. This is the case when two of the principal moments of inertia equal each other but differ from the third one. Examples of a symmetric top include a cylinder or a cone, or indeed any homogeneous solid body with axial symmetry.

I will use the convention that $I_1=I_2\neq I_3$. The Euler equations~\eqref{ch07:Eulereq} in the absence of external torque then reduce to
\begin{equation}
\begin{split}
I_1\dot\o_1-(I_1-I_3)\o_2\o_3&=0\;,\\
I_1\dot\o_2-(I_3-I_1)\o_3\o_1&=0\;,\\
I_3\dot\o_3&=0\;.
\end{split}
\end{equation}
The third equation guarantees that $\o_3$ is constant. In order to see what the first two equations tell us, we introduce a shorthand notation for the constant combination
\begin{equation}
\bar\o\equiv\frac{I_3-I_1}{I_1}\o_3=\left(\frac{I_3}{I_1}-1\right)\o_3\;.
\label{ch07:baromega}
\end{equation}
Then the first two Euler equations acquire the simple form
\begin{equation}
\dot\o_1+\bar\o\o_2=0\;,\qquad
\dot\o_2-\bar\o\o_1=0\;.
\end{equation}
These equations describe harmonic oscillations, as one easily checks by taking an additional time derivative and combining them to $\ddot\o_1+\bar\o^2\o_1=\ddot\o_2+\bar\o^2\o_2=0$. Putting all the pieces together, we can write the general solution of the Euler equations for a free symmetric top as
\begin{equation}
\o_1(t)=\o_0\cos(\bar\o t+\a)\;,\qquad
\o_2(t)=\o_0\sin(\bar\o t+\a)\;,\qquad
\o_3(t)=\o_3\;.
\label{ch07:symtopsol}
\end{equation}
Here the amplitudes $\o_0$ and $\o_3$ and the phase $\a$ act as integration constants. These three parameters together uniquely specify the solution to the Euler equations as three first-order ODEs for the functions $\o_i(t)$.

There is a fly in the ointment though. Namely, $\o_i$ are the components of angular velocity in the body frame, whose dependence on time we do not know. To get some insight in how the motion of the symmetric top looks in the space frame, we take advantage of the conservation of angular momentum. To that end, note that for the solution~\eqref{ch07:symtopsol}, both 
\begin{equation}
\abs{\vec\o}^2=\o_0^2+\o_3^2\;,\qquad
\skal J\o=I_1(\o_1^2+\o_2^2)+I_3\o_3^2=I_1\o_0^2+I_3\o_3^2
\end{equation}
are constant. As a consequence, the angle between the time-dependent vector $\vec\o$ and the fixed vector $\vec J$ also remains constant. Using the definitions $\vec J=(I_1\o_1,I_2\o_2,I_3\o_3)$ and $\vec \o=(\o_1,\o_2,\o_3)$ in the body frame, we find moreover that
\begin{equation}
\vec\o=\frac{\vec J}{I_1}+\left(1-\frac{I_3}{I_1}\right)\o_3\vec e_3=\frac{\vec J}{I_1}-\bar\o\vec e_3\;.
\label{ch07:symtopcoplanar}
\end{equation}
As a consequence, the three vectors $\vec J$, $\vec\o$ and $\vec e_3$ are coplanar. Furthermore, \eqref{ch07:symtopcoplanar} together with~\eqref{ch07:dotei2} gives
\begin{equation}
\dot{\vec e}_3=\vekt\o e_3=\frac1{I_1}\vekt Je_3\;.
\end{equation}
This describes \emph{precession} of the vector $\vec e_3$, defining the axis of symmetry of the rigid body, around the fixed vector $\vec J$ with angular velocity $\o_\mathrm{prec}={\abs{\vec J}}/{I_1}$.

\begin{figure}[t]
\sidecaption[t]
\includegraphics[width=2.9in]{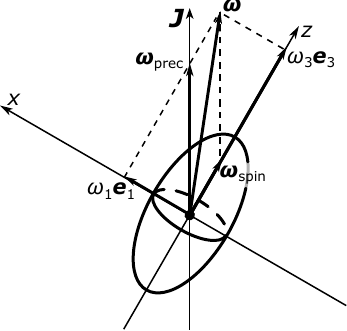}
\caption{Snapshot of the kinematics of a free symmetric top in the case $I_1=I_2>I_3$. The angular momentum vector $\vec J$ is oriented vertically. For simplicity, we assume that at the moment the snapshot is taken, the $\o_2$ component of angular velocity $\vec\o$ vanishes. The $x$-axis of the body frame then lies in the plane defined by the vectors $\vec J$ and $\vec\o$.}
\label{fig07:symmetrictop}
\end{figure}

We now have enough information to completely characterize the rotational motion of the symmetric top; see Fig.~\ref{fig07:symmetrictop} for a visualization. By~\eqref{ch07:symtopcoplanar}, the instantaneous angular velocity vector $\vec\o$ can be split as $\vec\o=\vec\o_\mathrm{prec}+\vec\o_\mathrm{spin}$. Accordingly, the motion can be viewed as a composition of two separate rotations. One of these is the above-mentioned precession with constant angular velocity $\vec\o_\mathrm{prec}=\vec J/I_1$ around the fixed axis defined by the angular momentum vector $\vec J$. The other is a spinning motion of the top with angular velocity $\vec\o_\mathrm{spin}=-\bar\o\vec e_3$ around its symmetry axis. Matching this to the alternative decomposition of the angular velocity into mutually orthogonal components, $\vec\o=\o_i\vec e_i$, requires that $\o_3=\o_\mathrm{prec}\cos\t+\o_\mathrm{spin}$, where $\t$ is the fixed angle between $\vec J$ and $\vec e_3$. Inserting $\o_\mathrm{spin}=-\bar\o$ finally leads to a simple expression for the precession rate that only depends on the kinematics of the motion,
\begin{equation}
\o_\mathrm{prec}=\frac{I_3\o_3}{I_1\cos\t}\;.
\label{ch07:symtopprecession}
\end{equation}

\begin{illustration}%
Consider a uniform thin homogeneous disk of mass $m$ and radius $R$. Choosing the axis of symmetry to be $\vec e_3$, the principal moments of inertia with respect to axes passing through the CoM are $I_1=I_2=(1/4)mR^2$ and $I_3=(1/2)mR^2$. For rotational motion such that the axis of symmetry $\vec e_3$ is nearly aligned with the direction of angular momentum $\vec J$ (that is the angle $\t$ is very small), we get
\begin{equation}
\o_\mathrm{prec}\approx2\o_3\;,\qquad
\o_\mathrm{spin}=-\bar\o=-\o_3\;.
\end{equation}
The precession rate equals approximately twice the rate of the spin of the disk around its symmetry axis. For this concrete geometry, one can observe the precession by throwing a plate in the air and watching its ``wobbling.'' The factor of two relating the wobble and spin rates can be understood in elementary terms using the balance of forces in the rotating plate; see~\cite{Tuleja2007} for details.
\end{illustration}

%%%%%%%%%%%%%%%%%%%%%%%%%%%%%%%%%%%%%%%%%%%%%%%%%%%%%%%%%%%%

\subsection{Stability of a Free Asymmetric Top}

Let us finally address, if only briefly and superficially, the generic case where all the principal moments of inertia are different. We are not going to attempt a general solution of the Euler equations~\eqref{ch07:Eulereq}. Instead, we note that there are special solutions that describe, even in this general case, uniform rotation at constant angular velocity. In fact, it is obvious from~\eqref{ch07:Eulereq} that such solutions have to satisfy the relations $\o_1\o_2=\o_2\o_3=\o_3\o_1=0$. Hence, only one of the components of $\vec\o$ can be nonzero. Without loss of generality, we can assume that $\o_3\neq0$ whereas $\o_1=\o_2=0$. The physical parameters of the corresponding uniform rotation are then
\begin{equation}
\vec\o=(0,0,\o_3)\;,\qquad
\vec J=(0,0,I_3\o_3)\;.
\label{ch07:uniformrotation}
\end{equation}
Conservation of angular momentum ensures that the third body axis points in a fixed direction in space, and so does therefore the angular velocity vector $\vec\o=\vec J/I_3$.

Suppose now that we perturb the motion of the rigid body mildly, for instance by applying a small torque that slightly deflects the body from its uniform rotation. The effect this has on the rigid body can be qualitatively understood by linearizing the Euler equations in small deviations from the uniform rotation~\eqref{ch07:uniformrotation}. Upon writing
\begin{equation}
\o_1(t)=\udelta\o_1(t)\;,\qquad
\o_2(t)=\udelta\o_2(t)\;,\qquad
\o_3(t)=\o_3+\udelta\o_3(t)\;,
\end{equation}
and expanding~\eqref{ch07:Eulereq} to first order in the deviations $\udelta\o_i$, we get the approximate linearized EoM,
\begin{equation}
\begin{split}
I_1\udelta\dot\o_1-(I_2-I_3)\o_3\udelta\o_2&\approx0\;,\\
I_2\udelta\dot\o_2-(I_3-I_1)\o_3\udelta\o_1&\approx0\;,\\
I_3\udelta\dot\o_3&\approx0\;.
\end{split}
\label{ch07:Eulerlinearized}
\end{equation}
The last equation says merely that we can increase or decrease the speed of uniform rotation around the third axis with impunity. The first two equations in~\eqref{ch07:Eulerlinearized} are more interesting though. Combining them leads to two separated second-order ODEs for $\udelta\o_1$ and $\udelta\o_2$,
\begin{equation}
\udelta\ddot\o_{1,2}+\frac{(I_3-I_1)(I_3-I_2)}{I_1I_2}\o_3^2\udelta\o_{1,2}\approx0\;.
\end{equation}
This has two linearly independent solutions for $\udelta\o_{1,2}$, proportional to $\exp(\pm\I\o_\mathrm{osc}t)$, with ``frequency'' $\o_\mathrm{osc}$ given by
\begin{equation}
\o_\mathrm{osc}=\sqrt{\frac{(I_3-I_1)(I_3-I_2)}{I_1I_2}}\,\o_3\;.
\end{equation}

Should $I_3$ be the largest or the smallest of the three principal moments of inertia, $\o_\mathrm{osc}$ is real and the solutions correspond to oscillations of angular velocity around the unperturbed value, $\o_3\vec e_3$. The rotation around the third body axis we started with is stable. Should, on the other hand, $I_3$ be the intermediate of the three principal moments of inertia (that is $I_1<I_3<I_2$ or $I_1>I_3>I_2$), then $\o_\mathrm{osc}$ becomes purely imaginary. In this case, $\udelta\o_{1,2}$ will grow exponentially with time until our linearized approximation to the EoM breaks down; the original rotation around the third body axis is unstable. This result is known as the \emph{intermediate axis theorem}. It may lead to rather striking behavior of freely rotating objects, manifested by the \emph{Dzhanibekov effect}, observed by a Soviet cosmonaut in space in 1985.

\begin{watchout}%
The stability analysis changes dramatically in case the rotational motion of the body allows for dissipation of kinetic energy. This happens for instance for bodies with movable solid parts, or objects containing fluids. To the extent that one can still define the tensor of inertia, the free rotation of such a body will eventually settle to rotation around the principal axis with maximum moment of inertia. This has a simple explanation. Denoting the components of angular momentum with respect to the body frame spanned on the principal axes of inertia as $J_i=I_i\o_i$, the kinetic energy can be written as
\begin{equation}
T=\frac12(I_1\o_1^2+I_2\o_2^2+I_3\o_3^2)=\frac{J_1^2}{2I_1}+\frac{J_2^2}{2I_2}+\frac{J_3^2}{2I_3}\;.
\end{equation}
In a free rotation, the angular momentum $\vec J$ is still conserved. This implies that its magnitude, $\smash{\abs{\vec J}=\sqrt{\hbox to0pt{\phantom{$J^1$}}\smash{J_1^2+J_2^2+J_3^2}}}$, remains fixed although the individual components $J_i$ in the body frame may change with time. The presence of dissipation will eventually bring the body to a state with lowest kinetic energy at fixed $\vec J$. This is acquired when $\vec J$ points along the principal axis with the largest moment of inertia, that is $T_\mathrm{min}=\vec J^2/(2I_\mathrm{max})$.

Somewhat surprisingly, people only learned this fairly recently, and did so the hard way. Namely, in 1958, the U.S.~released the \emph{Explorer 1} satellite to an orbit around the Earth. The satellite was designed to spin around its long axis (minimum moment of inertia), but failed to do so, ending up in a precession motion. In hindsight, this was caused by the presence of flexible antennas that dissipated kinetic energy.
\end{watchout}

%%%%%%%%%%%%%%%%%%%%%%%%%%%%%%%%%%%%%%%%%%%%%%%%%%%%%%%%%%%%

\section*{\probsec}
\addcontentsline{toc}{section}{\probsec}

\begin{prob}
\label{pr07:rotvector}
In three dimensions, the relationship between the body frame and the space frame can always be described as a single rotation by angle $\vp$ around an axis, defined by a unit vector $\vec n$. Show that the result of such a rotation on any of the basis vectors of the space frame can be written as
\begin{equation}
\vec e_i=\vec e_i'\cos\vp+\vec n(\skal ne_i')(1-\cos\vp)+(\vekt ne_i')\sin\vp\;.
\end{equation}
Find the corresponding rotation matrix $R$, and check that in the special case of a rotation in the $xy$ plane, your matrix is consistent with~\eqref{ch07:rotation2d}. Verify that the vector $\vec n$ is an eigenvector of $R$ with unit eigenvalue, and that the rotation angle $\vp$ can be recovered from the matrix $R$ via the relation $\tr R=1+2\cos\vp$.
\end{prob}

\begin{prob}
\label{pr07:rotframe}
The relation between the descriptions of the same position vector in the space and body frames allows us to understand the nature of inertial forces that appear in a noninertial, rotating reference frame. An observer attached to the body frame will describe the position of any object using the Cartesian coordinates $r_i$, its velocity in terms of $\dot r_i$, and acceleration in terms of $\ddot r_i$. Find the relationship between the acceleration $\ddot{\vec r}_\mathrm{body}=\ddot r_i\vec e_i$, observed in the rotating (body) frame, and the acceleration $\ddot{\vec r}=\ddot r_i'\vec e_i'$, observed in the inertial (space) frame.
\end{prob}

\begin{prob}
\label{pr07:inertia}
A homogeneous tetrahedron of mass $m$ is placed in a Cartesian coordinate frame so that one of its vertices lies at the origin and the other three on the respective positive coordinate semiaxes, each at distance $L$ from the origin. Calculate the tensor of inertia of the tetrahedron with respect to the origin. What are the principal axes and moments of inertia?
\end{prob}

\begin{prob}
\label{pr07:pendulum}
A \emph{physical pendulum} is a solid object, suspended at a fixed point so that it can swing around a given horizontal axis. Find the frequency of small oscillations of the pendulum in a uniform gravitational field, given its mass $m$, moment of inertia $I$ with respect to the axis of rotation, and the distance $d$ of its CoM from the axis. Check that your answer gives as a special case the frequency of small oscillations of a simple pendulum, consisting of a point mass $m$ on a string of length $L$.
\end{prob}

\begin{prob}
\label{pr07:wobble}
Recall the motion of a free symmetric top, discussed in Sect.~\ref{subsec:symtop}. Denote as $\a$ the angle between the angular velocity $\vec\o$ and the axis of symmetry $\vec e_3$ of the top. Show that this angle is related to the angle $\t$ between the angular momentum $\vec J$ and the axis of symmetry by
\begin{equation}
\tan\a=\frac{I_3}{I_1}\tan\t\;.
\end{equation}
\end{prob}

\begin{prob}
\label{pr07:asymtop}
The set of Euler equations~\eqref{ch07:Eulereq} can be obtained in an alternative, and arguably simpler, manner if we know a priori that the angular momentum $\vec J=J_i\vec e_i$ and torque $\vec\tau=\tau_i\vec e_i$ are related by $\vec{\dot J}=\vec\tau$. Derive the Euler equations by combining this with the equation~\eqref{ch07:dotei2} for the rate of change of the basis vectors of the body frame.
\end{prob}
\chapter{Lagrangian and Hamiltonian Mechanics of~Continuous~Systems}
\label{chap:LagHamcont}

\keywords{Fields, Lagrangian density, Klein--Gordon theory, Schr\"odinger theory, Hamiltonian density, quasiparticles, mass matrix, Nambu--Goldstone and Higgs boson.}

%%%%%%%%%%%%%%%%%%%%%%%%%%%%%%%%%%%%%%%%%%%%%%%%%%%%%%%%%%%%

\noindent With this~\chaptername, our course on analytical mechanics and field theory enters its second part. In the mechanics part, we developed an arsenal of tools that allowed us to deal with many applications. Much of this development was however subject to the simplifying assumption that material objects can be treated as point masses, or at worst as rigid bodies of a finite size. Their kinematics was accordingly described in terms of a finite number of parameters such as mass, (linear or angular) velocity and (linear or angular) momentum.

The next level of sophistication is based on a refined framework where properties of matter are described using local observables that may change with position as well as time. Thus, for instance, the mass density or temperature of matter may vary from place to place. Both of these are examples of \emph{scalar fields}, defined by a single value as a function of position and time. Similarly, the velocity of a medium such as a fluid may take different values at different points in space. This is an example of a \emph{vector field}. What is common to all these local observables is that the corresponding fields are assumed to be continuous (and in practice mostly even smooth) functions. This limits the applicability of classical field theory to resolution scales much larger than the average distance of atoms or molecules in the given material. Physical observables that, directly or indirectly, probe the atomic structure of matter require a quantum-mechanical treatment.

Operationally, the major difference between field theory and mechanics is that the \emph{equation of motion} (EoM) will now be a \emph{partial differential equation} (PDE). This is the price we have to pay. What we gain is access to much more detailed properties of matter, including phenomena whose description within mechanics would not have been possible at all.

%%%%%%%%%%%%%%%%%%%%%%%%%%%%%%%%%%%%%%%%%%%%%%%%%%%%%%%%%%%%

\section{Fields in the Lagrangian Formalism}
\label{sec:fieldLagrange}

Having clarified the importance of fields, we can start right away with the formal setup, which is a straightforward generalization of that we used in mechanics. One of the main changes is that spatial coordinates will no longer play the role of generalized coordinates defining the dynamical variables of the system. Instead, they will merely parameterize the actual dynamical variables, namely the fields themselves. From now on, I will thus use the symbol $x^{\m,\n,\dotsc}$ for spacetime coordinates and, if needed separately, $x^{i,j,\dotsc}$ for spatial coordinates.\footnote{Although the basic formalism of field theory can be used without change for any choice of spacetime coordinates, I will always implicitly assume that the spatial coordinates $x^i$ are Cartesian. The Greek index in $x^\m$ is a mere umbrella notation for spatial and temporal coordinates, and does \emph{not} imply relativistic physics without further qualifications. However, in case the field theory at hand is relativistic, $x^\m$ will stand for the standard Minkowski coordinates.} The collection of all spatial coordinates, defining a point in space, will be indicated with $\vec x$. A generic set of field variables will be denoted as $\p^{A,B,\dotsc}$. Unless explicitly stated otherwise, the superscript $A$ may label different scalar fields, but also merely distinguish different components of vector (or generally tensor) fields. The collection of all fields in a given system will sometimes be indicated with plain $\p$ without any index.

The set of generalized coordinates of a field theory now consists of $\p^A(\vec x)$ for all relevant choices of $A$ and all points $\vec x$ in the physical space. Suppose that the fields take values from some set $\S_\mathrm{Lag}$, which may be a vector space but also a curved surface, or manifold. When needed, I will refer to $\S_\mathrm{Lag}$ as the \emph{target space} of the theory. The configuration space of a field theory in $d$ spatial dimensions then formally consists of all maps $\p:\R^d\to\S_\mathrm{Lag}$.

%%%%%%%%%%%%%%%%%%%%%%%%%%%%%%%%%%%%%%%%%%%%%%%%%%%%%%%%%%%%

\subsection{Variational Formulation of Lagrangian Field Theory}

We are now ready to formulate Lagrangian field theory in terms of its action functional. This is defined by integration of a \emph{Lagrangian density} $\La$ over spacetime,
\begin{equation}
\boxed{S[\p]\equiv\int_\O\D^D\!x\,\La(\p,\de\p,\de\de\p,\dotsc,x)\;.}
\label{ch08:action}
\end{equation}
The Lagrangian density $\La$ is assumed to be a function of the fields $\p^A$ and of a finite number of their derivatives with respect to the spacetime coordinates, denoted symbolically as $\de\p,\de\de\p,\dotsc$. It may also depend explicitly, that is directly, on the spacetime coordinates themselves. The integration domain $\O$ can in principle be chosen arbitrarily, although in concrete applications one usually integrates over the entire spacetime of dimension $D\equiv d+1$.

Just like in mechanics, the basic dynamical principle of classical field theory now dictates that the actual, physical field configurations $\p^A(x)$ are stationary points of the action functional $S[\p]$. Thus, the EoM for the fields amounts to evaluating the functional derivatives $\udelta S/\udelta\p^A(x)$ and setting them to zero. This can be done using the techniques of \chaptername~\ref{chap:mathintro}. In the process, we need to assume suitable boundary conditions that ensure that boundary terms arising from integration by parts drop out. For instance, should the Lagrangian density only depend on the first derivatives of $\p^A$, which is sufficient for most of our course, we need to assume that the values of $\p^A$ at the boundary $\de\O$ of $\O$ are fixed. Using~\eqref{ch00:functionaldergeneral} or~\eqref{ch00:functionalderhigher}, we then find that the EoM takes the form
\begin{equation}
\boxed{\frac{\udelta S}{\udelta\p^A}=\PD\La{\p^A}-\de_\m\PD\La{(\de_\m\p^A)}+\dotsb=0\;,}
\label{ch08:ELequation}
\end{equation}
where $\de_\m$ is a shorthand notation for $\Pd{}{x^\m}$. The ellipsis indicates additional terms arising from the dependence of $\La$ on higher derivatives of the fields. For the sake of brevity, I will not spell these terms out explicitly, and instead refer you to~\refpr{pr08:Schr2ndorder} and~\refpr{pr08:KdV} for examples of higher-derivative field theories.

\begin{watchout}%
The fixed boundary condition for the fields may seem to have been engineered with the sole purpose not to have to bother about the contribution of surface terms. While that may indeed be the case, it is a fact that changing the boundary condition would not have affected the validity of the EoM~\eqref{ch08:ELequation} inside $\O$. It could however impose additional restrictions on the fields at the boundary $\de\O$. In~\chaptername~\ref{chap:fluid}, we will see an example of how we can modify the boundary condition so that it works for us.
\end{watchout}

%%%%%%%%%%%%%%%%%%%%%%%%%%%%%%%%%%%%%%%%%%%%%%%%%%%%%%%%%%%%

\subsection{Examples}
\label{subsec:fieldexamples}

To illustrate the use of the variational formulation of field theory and the corresponding EoM~\eqref{ch08:ELequation}, we next work out a couple of examples where the EoM can be exactly solved. In both examples, the number of spatial dimensions $d$ can be arbitrary and the target space $\S_\mathrm{Lag}$ is an actual vector space. Our first example is a theory of a \emph{real scalar field}, which is a colloquial way of saying that the target space is $\S_\mathrm{Lag}=\R$.

\begin{illustration}%
\label{ex08:KG}%
The \emph{Klein--Gordon} (KG) \emph{theory} is a relativistic theory of a single real scalar field $\p$, defined by the Lagrangian density
\begin{equation}
\La=\frac1{2c^2}(\de_t\p)^2-\frac12(\grad\p)^2-\frac{m^2c^2}{2\hbar^2}\p^2\;,
\label{ch08:KGlag}
\end{equation}
where $\de_t$ is just a different notation for a time derivative, $\de_t\p\equiv\dot\p$. This will oftentimes be more convenient than the dot notation familiar from mechanics, especially when we deal with a linear EoM in the form of a differential operator acting on the field. The EoM following from the Lagrangian density~\eqref{ch08:KGlag} is the KG equation,
\begin{equation}
\left(\frac1{c^2}\de_t^2-\grad^2+\frac{m^2c^2}{\hbar^2}\right)\p=0\;.
\label{ch08:KGeq}
\end{equation}
This is a homogeneous linear second-order PDE. As such, its solution can be sought as a linear combination of plane waves of the type $\p(\vec x,t)\propto\exp[\I(\skal kx-\o t)]$. Here $\vec k$ is the wave vector and $\o$ the frequency of the wave. Inserting this in~\eqref{ch08:KGeq}, we find that the frequency and wave vector are related by the \emph{dispersion relation}
\begin{equation}
\o^2=c^2\vec k^2+\frac{m^2c^4}{\hbar^2}\;.
\label{ch08:KGdisp}
\end{equation}
Under the quantum-mechanical particle--wave duality, the plane wave corresponds to a particle with momentum $\vec p=\hbar\vec k$ and energy $E=\hbar\o$. The dispersion relation~\eqref{ch08:KGdisp} is then equivalent to the relativistic relation~\eqref{ch05:dispersion} between energy and momentum, $\smash{E=\sqrt{\vec p^2c^2+m^2c^4}}$, for particles of rest mass $m$. In the following, I will often use the particle--wave duality to give a (quasi)particle interpretation to plane-wave solutions to the EoM, without spelling out all the details of the argument every time.

The KG theory describes relativistic particles. It should therefore be possible to present it in a way that makes its relativistic nature manifest. To that end, I shall adopt the set of units common in high-energy physics, in which $\hbar=c=1$. In these units, there is no difference between energy and frequency, or between momentum and the wave vector. Moreover, the zeroth component of the coordinate four-vector, $x^0$, is simply the time $t$. Accordingly, the time derivative $\de_t$ is identified with $\de_0\equiv\Pd{}{x^0}$. The field $\p$ is a scalar with respect to all Poincar\'e transformations~\eqref{ch05:poincare} between different inertial reference frames in the sense that
\begin{equation}
\p'(x')=\p(x)\quad\text{where }
x'^\m=\Lambda^\m_{\phantom\m\n}x^\n+a^\m\;.
\end{equation}
The matrix $\Lambda$ is required to preserve the Minkowski metric, i.e.~satisfy $g_{\m\n}\Lambda^\m_{\phantom\m\a}\Lambda^\n_{\phantom\n\b}=g_{\a\b}$, whereas $a^\m$ can be an arbitrary constant four-vector. With these qualifications, the Lagrangian density~\eqref{ch08:KGlag} can be written compactly as
\begin{equation}
\La=-\frac12(\de_\m\p)^2-\frac12m^2\p^2\;,
\end{equation}
and is itself a scalar function of spacetime coordinates. I have used the standard shorthand notation $\smash{(\de_\m\p)^2\equiv g^{\m\n}\de_\m\p\de_\n\p=-(\de_0\p)^2+(\grad\p)^2}$. This does not strictly speaking follow our convention for squares of four-vectors in Minkowski spacetime. It is however so ingrained in the high-energy physics literature that it would not be reasonable to modify it. The EoM~\eqref{ch08:KGeq} now acquires an equally compact form,
\begin{equation}
(\Box+m^2)\p=0\;,
\end{equation}
where $\Box\equiv -g^{\m\n}\de_\m\de_\n=-\de_\m\de^\m$ is the \emph{d'Alembert operator}. Accordingly, the dispersion relation~\eqref{ch08:KGdisp} is reduced to $\smash{E=\sqrt{\vec p^2+m^2}}$, which is the form most commonly employed in particle physics.
\end{illustration}

\begin{illustration}%
\label{ex08:Schr}%
For another illustrative example we take a \emph{complex scalar field} $\psi$, corresponding to $\S_\mathrm{Lag}=\C$, and consider the Lagrangian density
\begin{equation}
\La=\frac{\I\hbar}2(\psi^*\de_t\psi-\psi\de_t\psi^*)-\frac{\hbar^2}{2m}\grad\psi^*\cdot\grad\psi-V\psi^*\psi\;,
\label{ch08:SchrLag}
\end{equation}
where $V$ is a fixed function of the spatial coordinates, $\vec x$. This is sometimes referred to as the \emph{Schr\"odinger theory} for reasons that will become clear shortly. Note that while the field $\psi$ itself is complex, the Lagrangian density, hence the action, is real. This ensures the existence of a single, unique EoM representing the requirement that its solutions are stationary points of the action.

The EoM for the Lagrangian density~\eqref{ch08:SchrLag} can be written as
\begin{equation}
\I\hbar\de_t\psi=\left(-\frac{\hbar^2\grad^2}{2m}+V\right)\psi\;.
\label{ch08:Schreq}
\end{equation}
This is the Schr\"odinger equation for the wave function of a particle of mass $m$ moving in the potential $V(\vec x)$. In the absence of the potential, the EoM~\eqref{ch08:Schreq} has plane-wave solutions $\psi(\vec x,t)\propto\exp[\I(\skal kx-\o t)]$ with the frequency and wave vector satisfying the dispersion relation $\hbar\o=\hbar^2\vec k^2/(2m)$. Using the particle--wave duality, this corresponds to particles with energy $E=\vec p^2/(2m)$.
\end{illustration}

\begin{watchout}
The EoM~\eqref{ch08:Schreq} was obtained by treating $\psi$ and $\psi^*$ formally as independent variables. This may sound inappropriate given that I insisted on the importance of the Lagrangian density, hence action, being real. What I did was take a simple shortcut that is common when dealing with complex fields. Namely, one can always think of a complex field in terms of its real and imaginary parts, $\psi=\psi_1+\I\psi_2$. This amounts to viewing the target space $\S_\mathrm{Lag}$ as $\C$ or $\R\times\R$, respectively. Using the chain rule, one then finds that
\begin{equation}
\frac{\udelta S}{\udelta\psi}=\frac12\left(\frac{\udelta S}{\udelta\psi_1}-\I\frac{\udelta S}{\udelta\psi_2}\right)\;,\qquad
\frac{\udelta S}{\udelta\psi^*}=\frac12\left(\frac{\udelta S}{\udelta\psi_1}+\I\frac{\udelta S}{\udelta\psi_2}\right)\;.
\end{equation}
This makes it clear that the formal functional derivatives $\udelta S/\udelta\psi$ and $\udelta S/\udelta\psi^*$ vanish if and only if $\udelta S/\udelta\psi_1$ and $\udelta S/\udelta\psi_2$ do. The requirement that the action be real ensures that $\udelta S/\udelta\psi^*$ is the complex conjugate of $\udelta S/\udelta\psi$, and so we only have one independent (complex) EoM for the complex field $\psi$.

The same kind of argument shows that, more generally, redefining the set of fields $\p^A$ by a linear transformation only mixes the corresponding equations of motion by analogous linear combinations. Thus, setting $\p^A=P^A_{\phantom AB}\tilde\p^B$ with a constant invertible matrix $P$ and using the chain rule, we get
\begin{equation}
\frac{\udelta S}{\udelta\tilde\p^A}=\frac{\udelta S}{\udelta\p^B}\PD{\p^B}{\tilde\p^A}=\frac{\udelta S}{\udelta\p^B}P^B_{\phantom BA}\;.
\label{ch08:functionalchainrule}
\end{equation}
Of course, we are free to redefine the fields inside the action by \emph{any} (nonsingular) nonlinear transformation, should we wish to do so. This leads to a further generalization of~\eqref{ch08:functionalchainrule} to a chain rule for functional derivatives. I shall however not spell the chain rule out explicitly, for we will not need it.
\end{watchout}

What is common to both of our sample field theories~\eqref{ch08:KGlag} and~\eqref{ch08:SchrLag} is that they describe \emph{free}, that is noninteracting, particles. With this in mind, it is common to refer to the fields $\p$ and $\psi$ themselves as free or noninteracting. One feature that is worth stressing is our observation that the state with a single particle of given momentum maps to a particular plane-wave solution to the EoM. However, thanks to the linearity of both~\eqref{ch08:KGeq} and~\eqref{ch08:Schreq}, we can form arbitrary linear combinations of such plane-wave solutions. This ultimately points to the significance of field theory upon quantization: it describes quantum systems with an arbitrary number of particles. We will return to this interpretation below in Sect.~\ref{sec:fieldQuasi}, where we will see how the Lagrangian formalism also naturally incorporates interactions among plane waves or particles.

%%%%%%%%%%%%%%%%%%%%%%%%%%%%%%%%%%%%%%%%%%%%%%%%%%%%%%%%%%%%

\section{Fields in the Hamiltonian Formalism}
\label{sec:fieldHamilton}

Before we dive deeper into the physical properties of classical field theory, we take a step aside and clarify how to rephrase what we already know in the Hamiltonian language. It would in principle be possible to develop a symplectic approach to field theory by closely following Sect.~\ref{sec:geomHam}. Here I will however prioritize simplicity over generality and elegance. Instead of introducing the symplectic form and then following it wherever it takes us, we will follow the low-key approach of~\chaptername~\ref{chap:Hammechanics}. This amounts to restricting the choice of field variables to the generalized coordinates and momenta, that is the Darboux coordinates in the phase space. We will take the Lagrangian formulation of field theory as the starting point, assuming that the Lagrangian density $\La$ does not depend on higher than first derivatives of the fields.

We have already identified the generalized coordinates as the set of fields $\p^A$ entering the action~\eqref{ch08:action}. Similarly to mechanics, the corresponding conjugate momenta are defined by taking a derivative of the Lagrangian density with respect to the generalized velocities,
\begin{equation}
\pi_A\equiv\PD\La{\dot\p^A}=\PD{\La}{(\de_t\p^A)}\;.
\label{ch08:Legendre}
\end{equation}
It is assumed that this relation has a unique solution for the velocities $\de_t\p^A$ in terms of the momenta $\pi_A$, the fields $\p^A$, and possibly their spatial derivatives. Using this to eliminate the velocities, we then define the \emph{Hamiltonian density} by a field-theoretic generalization of the Legendre transformation~\eqref{ch03:HamfromLag},
\begin{equation}
\boxed{\Ha=\pi_A\de_t\p^A-\La\;.}
\label{ch08:Hamdef}
\end{equation}
This is a function of the generalized coordinates $\p^A$, generalized momenta $\pi_A$ and their spatial derivatives. It represents the energy density of the system.

We are now ready to give a definition of a Hamiltonian field theory. Its canonical variables consist of the generalized coordinates $\p^A(\vec x)$ and the generalized momenta $\pi_A(\vec x)$, for all possible values of $A$ and at all points $\vec x$ in space. Formally, the phase space of the theory is thus a collection of maps $(\p,\pi):\R^d\to\S_\mathrm{Ham}$, where $\S_\mathrm{Ham}$ is the target space of the theory in the Hamiltonian formulation. This is in most cases different from the target space $\S_\mathrm{Lag}$ of the Lagrangian version of the theory.

%%%%%%%%%%%%%%%%%%%%%%%%%%%%%%%%%%%%%%%%%%%%%%%%%%%%%%%%%%%%

\subsection{Variational Formulation of Hamiltonian Field Theory}

Having set up the phase space, we proceed to define the dynamics of a Hamiltonian field theory by its action as a functional of the phase space variables. Motivated by the action~\eqref{ch08:action} in the Lagrangian formalism and the relation~\eqref{ch08:Hamdef} between the Lagrangian and Hamiltonian densities, we write the Hamiltonian action as
\begin{equation}
\boxed{S[\p,\pi]=\int_{t_1}^{t_2}\D t\int_{\vec\O}\D^d\!\vec x\,[\pi_A\de_t\p^A-\Ha(\p,\pi,\grad\p,\grad\pi)]\;,}
\label{ch08:actionHam}
\end{equation}
where $\vec\O$ is the domain in space, $\R^d$, on which the canonical variables $\p^A(\vec x)$ and $\pi_A(\vec x)$ are defined. This generalizes the Hamiltonian form~\eqref{ch03:actionHam} of the action in mechanics. The time integration is done over an arbitrarily chosen interval, at the endpoints of which the boundary condition is specified. Working out the functional derivatives of the action with respect to $\p^A$ and $\pi_A$ gives the field-theoretic equivalent of Hamilton's equations~\eqref{ch03:Hamiltoneq},
\begin{equation}
\boxed{\begin{split}
\de_t\p^A&=\PD\Ha{\pi_A}-\divg\PD\Ha{(\grad\pi_A)}\;,\\
\de_t\pi_A&=-\PD\Ha{\p^A}+\divg\PD\Ha{(\grad\p^A)}\;.
\end{split}}
\label{ch08:Hameq}
\end{equation}

\begin{watchout}%
As in mechanics, the Hamiltonian and Lagrangian formulations of the same theory, including the corresponding equations of motion, are equivalent to each other. One disadvantage of Hamiltonian field theory is that it explicitly separates space from time, as is clearly visible in both the action~\eqref{ch08:actionHam} and the EoM~\eqref{ch08:Hameq}. This is an issue in relativistic field theory, where maintaining manifest invariance under Poincar\'e transformations is desirable. On the other hand, the Hamiltonian formulation of field theory is often used in nonrelativistic applications such as condensed-matter or atomic physics.
\end{watchout}

Let us check the validity of the Hamiltonian setup on the two examples worked out in Sect.~\ref{subsec:fieldexamples}. We start with the KG theory of~\refex{ex08:KG}.

\begin{illustration}%
\label{ex08:KGham}%
We simplify the Lagrangian~\eqref{ch08:KGlag} by switching to the high-energy physics units but without using four-vectors,
\begin{equation}
\La=\frac12(\de_t\p)^2-\frac12(\grad\p)^2-\frac12m^2\p^2\;.
\end{equation}
This gives us the corresponding conjugate momentum, $\pi=\Pd\La{(\de_t\p)}=\de_t\p$. The Hamiltonian density follows as
\begin{equation}
\Ha=\frac12\pi^2+\frac12(\grad\p)^2+\frac12m^2\p^2\;.
\label{ch08:KGham}
\end{equation}
The first equation in~\eqref{ch08:Hameq} reproduces the relation between generalized velocity and momentum, $\de_t\p=\pi$. The second equation therein is then equivalent to the KG equation~\eqref{ch08:KGeq}, $\de_t\pi=\grad^2\p-m^2\p$. Note that the target space of the Hamiltonian formulation of the KG theory is $\S_\mathrm{Ham}=\R\times\R$ as compared to $\S_\mathrm{Lag}=\R$, owing to the doubled number of variables in the Hamiltonian formalism.
\end{illustration}

\begin{illustration}%
\label{ex08:SchrHam}%
To apply the Hamiltonian formalism to the Schr\"odinger theory~\eqref{ch08:SchrLag}, we have to be a bit careful. Treating $\psi$ and $\psi^*$ naively as independent generalized coordinates as we did in~\refex{ex08:Schr}, we would find from~\eqref{ch08:Legendre} that the corresponding generalized momenta are, up to normalization, again $\psi$ and $\psi^*$. This indicates that the phase space cannot contain four independent field variables. One way around this problem would be to introduce two constraints implementing the relationships between the generalized coordinates and momenta implied by~\eqref{ch08:Legendre}. Here we will take a more down-to-earth approach, based on the observation that the action of the Schr\"odinger theory already is in the Hamiltonian form~\eqref{ch08:actionHam}. To make this explicit, we parameterize $\psi$ by its real and imaginary parts, $\psi=\psi_1+\I\psi_2$. Upon integration by parts, the action as defined by the Lagrangian density~\eqref{ch08:SchrLag} through~\eqref{ch08:action} reads
\begin{equation}
S[\psi_{1,2}]\simeq\int\D^D\!x\left\{2\hbar\psi_2\de_t\psi_1-\frac{\hbar^2}{2m}[(\grad\psi_1)^2+(\grad\psi_2)^2]-V(\psi_1^2+\psi_2^2)\right\}\;.
\end{equation}
To bring this into the form~\eqref{ch08:actionHam}, we identify $\p\equiv\psi_1$ and $\pi\equiv2\hbar\psi_2$. The Hamiltonian density is accordingly
\begin{equation}
\Ha=\frac{\hbar^2}{2m}(\grad\p)^2+\frac1{8m}(\grad\pi)^2+V\left(\p^2+\frac{\pi^2}{4\hbar^2}\right)\;.
\label{ch08:SchrHam}
\end{equation}
This makes it clear that the target spaces of the Lagrangian and Hamiltonian formulations of the Schr\"odinger theory coincide, $\S_\mathrm{Lag}=\S_\mathrm{Ham}=\R\times\R=\C$. The Hamilton equations~\eqref{ch08:Hameq} constitute a pair of real PDEs of first order in time derivatives. It is easy to check that they can be combined into a single complex EoM, equivalent to the Schr\"odinger equation~\eqref{ch08:Schreq} .

A simpler and mathematically more elegant way to obtain the same result is to write the action of the Schr\"odinger theory in the complex form,
\begin{equation}
S[\psi,\psi^*]\simeq\int\D^D\!x\left(\I\hbar\psi^*\de_t\psi-\frac{\hbar^2}{2m}\grad\psi^*\cdot\grad\psi-V\psi^*\psi\right)\;.
\end{equation}
We now identify $\psi$ as the sole generalized coordinate. Using~\eqref{ch08:Legendre}, we find the conjugate momentum $\pi=\I\hbar\psi^*$. The fact that these two are not mutually independent as complex variables reconfirms that the target space is merely $\S_\mathrm{Ham}=\C$. Applying~\eqref{ch08:Hamdef} then gives the Hamiltonian density
\begin{equation}
\Ha=\frac{\hbar^2}{2m}\grad\psi^*\cdot\grad\psi+V\psi^*\psi=-\frac\I\hbar\left(\frac{\hbar^2}{2m}\grad\pi\cdot\grad\psi+V\pi\psi\right)\;.
\label{ch08:SchrHam2}
\end{equation}
This in turn reproduces the Schr\"odinger equation as the Hamiltonian EoM~\eqref{ch08:Hameq}. In hindsight, the subtleties we encountered when developing the Hamiltonian description of the Schr\"odinger theory arise from insisting that the action takes the form~\eqref{ch08:actionHam} with explicitly separated generalized coordinates and momenta. Had we instead followed the general symplectic form~\eqref{ch06:sympaction} of the action, the derivation of the EoM would have been unproblematic. Here the target space is clearly $\S_\mathrm{Ham}=\C$ and there is no reason to suspect the presence of constraints.
\end{illustration}

%%%%%%%%%%%%%%%%%%%%%%%%%%%%%%%%%%%%%%%%%%%%%%%%%%%%%%%%%%%%

\subsection{Canonical Structure of Hamiltonian Field Theory}

We saw in Sect.~\ref{subsec:Poisson} that the Poisson bracket encodes in a very compact manner the symplectic structure of Hamiltonian mechanics. This feature survives the transition from mechanics to field theory. Since we have limited our treatment of Hamiltonian field theory to the Darboux coordinates $\p^A(\vec x)$ and $\pi_A(\vec x)$, we can follow~\refex{ex06:canonicalPoisson}. Guided by~\eqref{ch06:poissonconventional}, we define the Poisson bracket of two \emph{functionals} $F,G$ on the phase space of the Hamiltonian field theory by
\begin{equation}
\boxed{\{F,G\}\equiv\int_{\vec\O}\D^d\!\vec x\left[\frac{\udelta F}{\udelta\p^A(\vec x)}\frac{\udelta G}{\udelta\pi_A(\vec x)}-\frac{\udelta F}{\udelta\pi_A(\vec x)}\frac{\udelta G}{\udelta\p^A(\vec x)}\right]\;.}
\label{ch08:Poissonfields}
\end{equation}
In the special case that $F,G$ are chosen from the canonical variables themselves, we get a field-theoretic version of~\eqref{ch06:PoissonDarboux},
\begin{align}
\{\p^A(\vec x),\p^B(\vec y)\}=\{\pi_A(\vec x),\pi_B(\vec y)\}&=0\;,\\
\notag
\{\p^A(\vec x),\pi_B(\vec y)\}&=\d^A_B\d^d(\vec x-\vec y)\;,
\end{align}
where $\d^d(\vec x-\vec y)$ is the $d$-dimensional Dirac $\d$-function.

The Poisson bracket allows us to cast the Hamilton equations~\eqref{ch08:Hameq} in a neat and concise form, formally identical to mechanics. To that end, we first introduce the integral Hamiltonian, representing the total energy of the system,
\begin{equation}
H[\p,\pi]=\int_{\vec\O}\D^d\!\vec x\,\Ha(\p,\pi,\grad\p,\grad\pi)\;.
\end{equation}
This is a functional on the phase space. Applying the definition~\eqref{ch08:Poissonfields} of the Poisson bracket makes~\eqref{ch08:Hameq} equivalent to
\begin{equation}
\boxed{\begin{split}\de_t\p^A(\vec x)&=\frac{\udelta H}{\udelta\pi_A(\vec x)}=\{\p^A(\vec x),H\}\;,\\
\de_t\pi_A(\vec x)&=-\frac{\udelta H}{\udelta\p^A(\vec x)}=\{\pi_A(\vec x),H\}\;.\end{split}}
\end{equation}
In general, the time evolution of any functional $F$ on the phase space that does not explicitly depend on time, along a solution to the Hamilton equations~\eqref{ch08:Hameq}, is given by $\de_tF=\{F,H\}$.

\begin{illustration}%
In~\refex{ex08:SchrHam} we gave two different Hamiltonian formulations of the Schr\"odinger theory. The second of these allows for an elegant formulation of the Poisson bracket in terms of the complex field $\psi$ and its conjugate $\psi^*$. Namely, by using that the generalized momentum conjugate to $\psi$ is $\I\hbar\psi^*$, we can write the Poisson bracket~\eqref{ch08:Poissonfields} of any two functionals on the phase space of the Schr\"odinger theory as
\begin{equation}
\{F,G\}=-\frac\I\hbar\int_{\vec\O}\D^d\!\vec x\,\left[\frac{\udelta F}{\udelta\psi(\vec x)}\frac{\udelta G}{\udelta\psi^*(\vec x)}-\frac{\udelta F}{\udelta\psi^*(\vec x)}\frac{\udelta G}{\udelta\psi(\vec x)}\right]\;.
\end{equation}
This implies as a special case the Poisson brackets for $\psi$ and $\psi^*$ themselves,
\begin{equation}
\{\psi(\vec x),\psi(\vec y)\}=\{\psi^*(\vec x),\psi^*(\vec y)\}=0\;,\qquad
\{\psi(\vec x),\psi^*(\vec y)\}=-\frac\I\hbar\d^d(\vec x-\vec y)\;.
\end{equation}
By the same token, using the first expression for the Hamiltonian density in~\eqref{ch08:SchrHam2}, the Schr\"odinger equation~\eqref{ch08:Schreq} can be cast simply as $\I\hbar\de_t\psi=\I\hbar\{\psi,H\}=\udelta H/\udelta\psi^*$. 
\end{illustration}

%%%%%%%%%%%%%%%%%%%%%%%%%%%%%%%%%%%%%%%%%%%%%%%%%%%%%%%%%%%%

\section{Small Oscillations as Quasiparticles}
\label{sec:fieldQuasi}

The few examples of field theories we have worked out so far describe free fields. Operationally, the Lagrangian density of a theory of noninteracting fields can be recognized by being strictly bilinear (quadratic) in the fields. The corresponding EoM is then linear, and thus satisfies the principle of superposition. As pointed out before, the importance of free theories is not only in our ability to solve their EoM exactly, but also in the fact that when quantized, they describe noninteracting~particles.

Going beyond the free field paradigm is nontrivial, as it requires us to deal with nonlinear PDEs. As a consequence, it is rare that the EoM of an interacting field theory would be exactly solvable. However, there is a general strategy that allows one to analyze the physical content of an interacting field theory following the philosophy of \chaptername~\ref{chap:oscillations}. The basic workflow that parallels the setup of Sect.~\ref{sec:multicomponent} is as follows:
\begin{enumerate}
\item[(1)] Find a stationary state of the theory. Typically but not inevitably, this is the state of lowest energy: the \emph{ground state}.
\item[(2)] Expand the Lagrangian density (or action) to second order in deviations of the fields from their values in the stationary state.
\item[(3)] Diagonalize the resulting bilinear Lagrangian (or action).
\end{enumerate}
The last step identifies the normal modes of the theory with respect to the chosen stationary state. As in our previous examples, the EoM for normal modes is linear and its plane-wave solutions can be interpreted in terms of free particles. In condensed-matter physics, it is common to denote a normal mode and the corresponding particle respectively as a \emph{collective mode} and a \emph{quasiparticle}. The higher-order contributions to the Lagrangian, obtained by the expansion in step (2), generate interactions among the quasiparticles. While it is possible to take these interactions systematically into account, here we will focus on identifying the spectrum of normal modes.

%%%%%%%%%%%%%%%%%%%%%%%%%%%%%%%%%%%%%%%%%%%%%%%%%%%%%%%%%%%%

\subsection{Mass Spectrum of Relativistic Scalar Fields}

The task is easiest to deal with in relativistic theories of scalar fields. Consider the following class of Lagrangian densities for a set of real scalar fields $\p^A$,\footnote{In some applications, it is natural to also consider complex scalar fields. These can however always be represented by their real and imaginary parts, and are therefore covered by our present analysis.}
\begin{equation}
\La=-\frac12\d_{AB}\de_\m\p^A\de^\m\p^B-V(\p)\;.
\label{ch08:lagmulti}
\end{equation}
The first term in~\eqref{ch08:lagmulti} is called \emph{kinetic term}. Its form hints that we have assumed it to be already diagonalized in the field labels $A,B$. This can always be ensured by a linear transformation of the fields. The overall sign is fixed by the requirement that the derivative part of the Hamiltonian density is positive-definite; cf.~\eqref{ch08:KGham}. The term $V(\p)$ in~\eqref{ch08:lagmulti} represents potential energy of interactions among the fields, and is assumed to be bounded from below for the same reason.

\begin{watchout}
Equation~\eqref{ch08:lagmulti} is certainly not the most general classical Lagrangian density one could write down. We could add for instance terms with a higher power of derivatives of $\p^A$, or higher derivatives thereof. We could also multiply the kinetic term by an arbitrary function of the fields without further derivatives. All these possibilities are relevant in the context of \emph{effective field theory}, which dictates that one includes in the Lagrangian density \emph{all} terms compatible with the assumed symmetries of the theory. However, there is a well-defined context in which the class of scalar theories~\eqref{ch08:lagmulti} is singled out as relevant in a technical sense. Namely, in quantum field theory, field theories that are ``renormalizable'' play a distinguished role. It turns out that all ``perturbatively renormalizable'' relativistic theories of scalar fields in any number $d\geq2$ of spatial dimensions span a subset of the class~\eqref{ch08:lagmulti}.
\end{watchout}

A straightforward generalization of~\refex{ex08:KGham} shows that the generalized momenta $\pi_A$ conjugate to the generalized coordinates $\p^A$ are $\pi_A=\d_{AB}\de_t\p^B$. The Hamiltonian density of the theory~\eqref{ch08:lagmulti} is then
\begin{equation}
\Ha=\frac12\d^{AB}\pi_A\pi_B+\frac12\d_{AB}\grad\p^A\cdot\grad\p^B+V(\p)\;.
\end{equation}
The simple form of the derivative part of the Hamiltonian density ensures that in the ground state, $\pi_A=0$ and $\p^A$ are coordinate-independent constants, determined by minimizing the potential $V(\p)$. Denoting the values of the fields minimizing the potential as $\smash{\p^A_0}$ and parameterizing the fields as $\smash{\p^A=\p^A_0+\vp^A}$, the potential can be expanded in powers of the deviations $\vp^A$ from the minimum as
\begin{equation}
V(\p)=V(\p_0)+\smash{\xcancel{\at{\PD V{\p^A}}{\p=\p_0}}}\vp^A+\frac12\underbrace{\at{\frac{\de^2V}{\de\p^A\de\p^B}}{\p=\p_0}}_{\equiv m^2_{AB}}\vp^A\vp^B+\bigO(\vp^3)\;.
\end{equation}

The matrix $\smash{m^2_{AB}}$ is called the \emph{mass matrix} of the theory~\eqref{ch08:lagmulti}. Being by construction symmetric and positive-semidefinite, it can be diagonalized by an orthogonal transformation of the fields $\vp^A$. Denoting its eigenvalues as $\smash{m_A^2}$ and the corresponding linear combinations of fields (eigenvectors of the mass matrix) as $\smash{\x^A}$, the Lagrangian density~\eqref{ch08:lagmulti} is turned into
\begin{equation}
\La\approx-V(\p_0)-\frac12\sum_A\left[(\de_\m\x^A)^2+m_A^2(\x^A)^2\right]\;.
\end{equation}
As in \chaptername~\ref{chap:oscillations}, I use the symbol $\approx$ to indicate a quadratic approximation to the Lagrangian, or the linearized version of the EoM. The fields $\x^A$ are our normal modes. They represent a set of relativistic particles with respective masses $m_A$. This is our main result with widespread applications in particle physics. Let us illustrate it on a couple of examples.

\begin{illustration}%
We start with a trivial example. The Lagrangian density
\begin{equation}
\La=-\frac12(\de_\m\p)^2-\frac12m^2\p^2-\frac\l{4!}\p^4
\label{ch08:lagphi4}
\end{equation}
defines the \emph{$\p^4$-theory}. The parameter $\l$ is assumed to be positive; the denominator factor of $4!$ is just conventional. The potential energy of the theory obviously has its minimum at $\p=0$. Accordingly, the Lagrangian density~\eqref{ch08:lagphi4} already has the form of a series expansion. We therefore conclude that the $\phi^4$-theory describes interacting relativistic particles of mass $m$.

The $\p^4$-theory has a complexified version with a complex scalar field $\p$, which is usually written as
\begin{equation}
\La=-\de_\m\p^*\de^\m\p-m^2\p^*\p-\frac\l4(\p^*\p)^2\;.
\end{equation}
This can be cast in a real form possessing a kinetic term with a standard normalization by parameterizing $\p$ by two real fields $\p^{1,2}$ as $\p\equiv(\p^1+\I\p^2)/\sqrt2$. The Lagrangian density of the complex $\p^4$-theory can then be expressed as
\begin{equation}
\La=-\frac12\d_{AB}\de_\m\p^A\de^\m\p^B-\frac12m^2\d_{AB}\p^A\p^B-\frac\l{16}(\d_{AB}\p^A\p^B)^2\;,
\label{ch08:lagphi4complex}
\end{equation}
where all the dummy indices run over the set $\{1,2\}$. In this form, both the kinetic term and the mass matrix are diagonal. This shows that the complex $\p^4$-theory describes a pair of interacting particles with mass $m$.

There is another straightforward generalization of our original $\p^4$-theory~\eqref{ch08:lagphi4}, whose Lagrangian density is identical to~\eqref{ch08:lagphi4complex} except for a different range of indices, $A,B=1,\dotsc,n$. This theory is usually referred to as the \emph{$\gr{O}(n)$ linear sigma model}. In various incarnations, this model plays a distinguished role across physical disciplines, from the Higgs mechanism in particle physics to the theory of phase transitions.
\end{illustration}

\begin{illustration}%
Having settled the basics, let us have a look at a more interesting example. Consider a theory of a complex scalar field $\p$ defined by the Lagrangian density
\begin{equation}
\La=-\de_\m\p^*\de^\m\p-\frac\l4(\p^*\p-v^2)^2\;,
\label{ch08:lagMexicanhat}
\end{equation}
where $v$ is a real parameter, assumed to be positive. The potential of this theory is $V(\p)=(\l/4)(\p^*\p-v^2)^2$. It is obviously minimized by any $\p_0$ such that $\abs{\p_0}=v$. The minimum of the potential is not unique! This is actually a feature, not a bug. The existence of multiple degenerate ground states is a hallmark of \emph{spontaneous symmetry breaking}. In our case, the Lagrangian density~\eqref{ch08:lagMexicanhat} is symmetric with respect to the transformation $\smash{\p\to\E^{\I\eps}\p}$ for any phase $\eps$. This symmetry is not preserved by any of the minima of the potential. Instead, changing the phase of the field transforms one minimum into another.

To find the spectrum of normal modes, we need to choose one specific minimum and expand the Lagrangian density in deviations from this minimum. We thus set $\p_0=v$ and introduce real fields $\vp^1,\vp^2$ by
\begin{equation}
\p=v+\frac1{\sqrt2}(\vp^1+\I\vp^2)\;.
\end{equation}
Inserting this in the Lagrangian density~\eqref{ch08:lagMexicanhat} leads to
\begin{equation}
\La\approx-\frac12\d_{AB}\de_\m\vp^A\de^\m\vp^B-\frac12\l v^2(\vp^1)^2\;.
\end{equation}
The Lagrangian density is now diagonal, hence $\vp^1,\vp^2$ are the sought normal modes. The $\vp^2$ mode does not have any mass term. It corresponds to a massless relativistic particle, known as the \emph{Nambu--Goldstone boson}. The presence of a massless particle is a direct consequence of spontaneous symmetry breaking. On the other hand, the particle described by $\vp^1$ carries a mass $\sqrt\l v$. This is a prototype of a \emph{Higgs boson}.
\end{illustration}

%%%%%%%%%%%%%%%%%%%%%%%%%%%%%%%%%%%%%%%%%%%%%%%%%%%%%%%%%%%%

\subsection{Oscillations Around Inhomogeneous Stationary States}

The stationary state one expands around does not necessarily have to be constant. This is especially true in nonrelativistic field theories, but can also happen in relativistic ones. Even there it is possible to find stationary states that are stable with respect to small fluctuations of the fields, yet do not minimize the energy of the system. Instead of a general discussion, I will work out in detail one illustrative example.

The \emph{sine-Gordon theory} is a relativistic theory of a real scalar field $\p$ defined by the Lagrangian density
\begin{equation}
\La=-\frac12(\de_\m\p)^2-m^2v^2\left(1-\cos\frac\p v\right)\;,
\label{ch08:SGlag}
\end{equation}
where $m$ and $v$ are positive parameters. The EoM of this theory,
\begin{equation}
\Box\p+m^2v\sin\frac\p v=0\;,
\label{ch08:SGeq}
\end{equation}
has a static inhomogeneous solution that we already met in~\refpr{pr00:SGkink} as the \emph{sine-Gordon kink}. It depends on a single spatial coordinate which I shall call $z$,
\begin{equation}
\p_0(z)=4v\arctan\E^{mz}\;.
\label{ch08:SGkink}
\end{equation}
This is not the absolute minimum of the Hamiltonian of the sine-Gordon theory. (That would be $\p_0/v=0$ up to an integer multiple of $2\pi$.) However, it is still a local minimum. It minimizes the energy on the subset of fields satisfying the boundary conditions $\p(z=-\infty)=0$ and $\p(z=+\infty)=2\pi v$.

The next step would be to shift the original field as $\p=\p_0+\vp$ and expand the action of the sine-Gordon theory to the second order in $\vp$. However, in this case it is technically easier to follow the alternative approach of linearizing the EoM. Expanding~\eqref{ch08:SGeq} to first order in $\vp$ gives
\begin{equation}
\left[\Box+m^2\cos\frac{\p_0(z)}v\right]\vp(x)=\left[\Box+m^2\left(1-\frac2{\cosh^2mz}\right)\right]\vp(x)\approx0\;,
\label{ch08:SGeqlin}
\end{equation}
where $x$ without indices still stands for the collection of all spacetime coordinates. The linearized EoM~\eqref{ch08:SGeqlin} is solved by functions that behave as plane waves in all the spacetime coordinates but $z$. Introducing the shorthand notation $\vec x_\perp$ for the spatial coordinates other than $z$, we can parameterize the plane-wave solutions as $\vp(x)=\hat\vp(z)\exp[\I(\vec k_\perp\cdot\vec x_\perp-\o t)]$. This transforms~\eqref{ch08:SGeqlin} into an ordinary differential equation for the profile function $\hat\vp(z)$, 
\begin{equation}
\left(-\de_z^2-\frac{2m^2}{\cosh^2mz}\right)\hat\vp(z)\approx(\o^2-\vec k_\perp^2-m^2)\hat\vp(z)\;.
\label{ch08:SGeq1d}
\end{equation}

This linear one-dimensional EoM has a remarkable interpretation. Up to a choice of normalization, the differential operator on the left-hand side, $-\de_z^2-2m^2/\cosh^2mz$, corresponds to the Hamiltonian of a particle moving in a potential well. Given that the coefficient on the right-hand side, $\o^2-\vec k_\perp^2-m^2$, is constant, \eqref{ch08:SGeq1d} is nothing but the eigenvalue problem for this one-dimensional Hamiltonian where $\o^2-\vec k_\perp^2-m^2$ is the corresponding ``energy.'' The mentioned effective one-dimensional Hamiltonian $-\de_z^2-2m^2/\cosh^2mz$ is well-known under the name \emph{P\"oschl--Teller Hamiltonian}, and known are also its eigenstates and eigenvalues. The derivation of these eigenstates and eigenvalues is a beautiful problem in quantum mechanics that can be dealt with elegantly using the technique of annihilation and creation operators. However, here I shall content myself with a summary of the results for the spectrum of the P\"oschl--Teller Hamiltonian along with the corresponding normal modes of the sine-Gordon theory on the kink background~\eqref{ch08:SGkink}:
\begin{itemize}
\item There is a single bound state with $\hat\vp(z)\propto1/\cosh mz$ and eigenvalue $-m^2$. This corresponds to solutions to the linearized EoM~\eqref{ch08:SGeqlin} with dispersion relation $\o^2=\vec k_\perp^2$. These solutions are localized on the kink~\eqref{ch08:SGkink} in the $z$-direction, but propagate freely as massless particles in the other directions.
\item There is a continuum of eigenstates parameterized by quasimomentum $k_z$, with $\hat\vp(z)\propto(-\I k_z+m\tanh mz)\exp(\I k_zz)$ and eigenvalue $k_z^2$. This amounts to solutions of~\eqref{ch08:SGeqlin} with dispersion relation $\smash{\o^2=\vec k_\perp^2+k_z^2+m^2}$. These propagate in the entire space, and far away from the kink~\eqref{ch08:SGkink} behave as free particles of mass~$m$.
\end{itemize}

The sine-Gordon kink~\eqref{ch08:SGkink} separates two domains in space, where the value of $\p_0$ lies near two neighboring minima of the potential of the sine-Gordon theory, $m^2v^2[1-\cos(\p/v)]$. It is thus an example of a \emph{domain wall}, which is an interface dividing two domains in space, in both of which the system resides in one of a discrete set of degenerate ground states. The normal modes that are localized in the $z$-direction can be interpreted as surface waves on the domain wall. 

%%%%%%%%%%%%%%%%%%%%%%%%%%%%%%%%%%%%%%%%%%%%%%%%%%%%%%%%%%%%

\section*{\probsec}
\addcontentsline{toc}{section}{\probsec}

\begin{prob}
\label{pr08:Schr2ndorder}
Integrate the action of the Schr\"odinger theory~\eqref{ch08:SchrLag} by parts to bring it in the~form
\begin{equation}
S[\psi,\psi^*]\simeq\int\D^D\!x\,\psi^*\left(\I\hbar\de_t+\frac{\hbar^2\vec\nabla^2}{2m}-V\right)\psi\;.
\label{ch08:Schraction}
\end{equation}
This new action contains second derivatives of $\psi$. Use it to derive the EoM, and check that the EoM is equivalent to the Schr\"odinger equation~\eqref{ch08:Schreq}.
\end{prob}

\begin{prob}
\label{pr08:KdV}
For another example of a higher-derivative field theory, consider the following Lagrangian density for a real scalar field $\p$ living in one spatial dimension,
\begin{equation}
\La=\frac12\de_t\p\de_x\p+(\de_x\p)^3-\frac12(\de_x^2\p)^2\;,
\end{equation}
where $\de_x\equiv\Pd{}{x}$ is a derivative with respect to the sole spatial coordinate. Show that the corresponding EoM is the \emph{Korteweg--de Vries equation}
\begin{equation}
\de_t\psi+6\psi\de_x\psi+\de_x^3\psi=0\;,\quad\text{where }\psi\equiv\de_x\p\;.
\end{equation}
This equation is known to describe surface waves in shallow water.
\end{prob}

\begin{prob}
\label{pr08:nlsm}
The target space of a field theory does not need to be a vector space. Consider the relativistic theory defined by the following Lagrangian density,
\begin{equation}
\La=-\frac12\de_\mu\vec n\cdot\de^\mu\vec n\;.
\end{equation}
Here $\vec n$ is a triplet of scalar fields satisfying the constraint $\skal nn=1$. This is a Lagrangian field theory with the target space $\S_\mathrm{Lag}=S^2\subset\R^3$. For reasons of purely historical interest, it is called the \emph{nonlinear sigma model}. Suppose that we choose a stationary state of the theory as $\vec n_0=(0,0,1)$. Show that small oscillations around this state correspond to two massless particles. Hint: you can do this in two different ways. Either expand the Lagrangian density to second order in deviations of $\vec n$ from $\vec n_0$, or expand the EoM to first order in the deviations. If you choose the latter approach, remember that the variable $\vec n$ is constrained so you need a Lagrange multiplier when deriving the EoM.
\end{prob}

\begin{prob}
\label{pr08:LandauLifshitz}
To follow up on \refpr{pr06:sympS2}, imagine that we have a collection of unit-vector spin variables $\vec n(\vec x)$, one at each point in space. One can think of this as a continuum model of ferromagnets. Spins at different points are mutually independent. This motivates the following local generalization of the Poisson bracket~\eqref{ch06:PoissonS2},
\begin{equation}
\{n^A(\vec x),n^B(\vec y)\}=\frac1M\ve^{AB}_{\phantom{AB}C}n^C(\vec x)\d^d(\vec x-\vec y)\;;
\end{equation}
the constant $M$ is interpreted as the average spin density in the ferromagnetic ground state. The simplest Hamiltonian for the spin variable $\vec n$ one can write down is
\begin{equation}
H[\vec n]=\frac{\vr_\mathrm{s}}2\int_{\vec\O}\D^d\!\vec x\,\d_{AB}\grad n^A\cdot\grad n^B\;,
\end{equation}
where $\vr_\mathrm{s}$ is a material constant known as the spin stiffness. Show that the ensuing Hamilton equation for $\vec n$ can be written as
\begin{equation}
\de_t\vec n=\frac{\vr_\mathrm{s}}M\vec n\times\vec\nabla^2\vec n\;.
\end{equation}
This is the local differential version of the Landau--Lifshitz equation~\eqref{ch06:LandauLifshitz}.
\end{prob}

\begin{prob}
\label{pr08:spinwave}
The uniform ferromagnetic ground state can be described by the constant field $\vec n_0=(0,0,1)$. Local deviations of the spin configuration from this ground state can be parameterized by a vector field $\vec\eta$ such that $\vec n=\vec n_0+\vec\eta$. Expand the EoM derived in \refpr{pr08:LandauLifshitz} to first order in $\vec\eta$ and show that it has plane-wave solutions with dispersion relation $\o=\vr_\mathrm{s}\vec k^2/M$. These are the ferromagnetic \emph{spin waves}. Hint: the field $\vec n$ is a unit vector, which implies that the three components of $\vec\eta$ are not all mutually independent. It is convenient to use $\eta^1,\eta^2$ as the independent variables.
\end{prob}

\begin{prob}
\label{pr08:auxfield}
A relativistic theory of a real scalar field $\p$ and a four-vector field $A^\m$ is defined by the Lagrangian density
\begin{equation}
\La=\frac12 A_\m A^\m-A^\m\de_\m\p\;.
\end{equation}
Use the EoM to find the normal modes of this theory.
\end{prob}

\begin{prob}
\label{pr08:sineGordon}
Starting from the static sine-Gordon kink~\eqref{ch08:SGkink}, use Lorentz invariance of the sine-Gordon theory~\eqref{ch08:SGlag} to find a solution to the sine-Gordon equation~\eqref{ch08:SGeq} that corresponds to a kink moving at constant velocity $v$ along the $z$-axis.
\end{prob}

\begin{prob}
\label{pr08:superfluid}
A theory of a single real scalar field $\p$ is defined by the Lagrangian density
\begin{equation}
\La=P\bigl(\de_t\p-(\grad\p)^2/(2m)\bigr)\;,
\label{ch08:EFTsuperfluid}
\end{equation}
where $P$ is a given, fixed function and $m$ a constant positive parameter. First convince yourself that $\p_0(x)=\m t$, where $\m$ is a constant parameter, solves the EoM of this theory. Then, find the normal modes of the theory with respect to the stationary state $\p_0$ and show that they correspond to plane waves with dispersion relation
\begin{equation}
\o^2=\frac{P'(\m)}{mP''(\m)}\vec k^2\;.
\end{equation}
Remark: the Lagrangian density~\eqref{ch08:EFTsuperfluid} describes a (nonrelativistic) superfluid whose constituents have mass $m$. The parameter $\m$ represents the chemical potential of the superfluid and the function $P(\m)$ its pressure in thermodynamic equilibrium.
\end{prob}
\chapter{Application: Elasticity of Solids}
\label{chap:elasticity}

\keywords{Continuum limit, Young modulus, body (material) coordinates, strain tensor, Lam\'e coefficients, bulk and shear modulus, Cauchy stress tensor, Navier--Cauchy equation, longitudinal and transverse sound.}

%%%%%%%%%%%%%%%%%%%%%%%%%%%%%%%%%%%%%%%%%%%%%%%%%%%%%%%%%%%%

\noindent As the first application of the field theory machinery, developed in~\chaptername~\ref{chap:LagHamcont}, we will have a look at elasticity of solids. This is a classic part of physics whose mathematical foundations were laid in the early 19\textsuperscript{th} century largely by Cauchy and Poisson. In a certain sense, this choice of subject is naturally motivated by our study of rigid bodies in~\chaptername~\ref{chap:rigidbody}. Therein, we focused entirely on the motion of solid objects as a whole, and neglected the internal dynamics of the material they are made of. Here we will fill the gap and investigate the basic macroscopic material properties of solids. This will be our first encounter with \emph{continuum mechanics}. Borrowing some of the concepts from our discussion of rigid bodies, we will develop a general language for description of continuous media. Apart from applying this language immediately to elastic solids, we will also use it later in~\chaptername~\ref{chap:fluid} when we talk about fluids.

We will start with a detailed discussion of a naive one-dimensional model of an elastic solid as a collection of point masses connected by springs. While certainly physically inadequate in the detail, this will help us develop an intuitive understanding of the Lagrangian formalism for continuous media. The introduction is followed by the core material of the~\chaptername. We build a Lagrangian field theory of (nonrelativistic) elastic solids and interpret its physical content in terms of strain and stress of the elastic medium. Then, we apply the constructed field theory to small oscillations of solids: elastic sound waves. Parts of this~\chaptername{} are inspired by~\cite{Landau1986}, to which I refer the reader for further details.

%%%%%%%%%%%%%%%%%%%%%%%%%%%%%%%%%%%%%%%%%%%%%%%%%%%%%%%%%%%%

\section{Spring Model of a Solid}
\label{sec:springmodel}

\begin{figure}[t]
\sidecaption[t]
\includegraphics[width=2.9in]{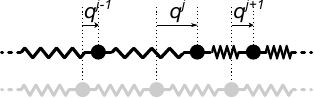}
\caption{One-dimensional spring chain model of a solid. All the masses equal $m$, all the springs have the same spring constant $k$. The distance between nearest neighbors in equilibrium is $a$. For comparison, the equilibrium state is indicated in gray.}
\label{fig09:springmodel}
\end{figure}

Neglecting the microscopic quantum-mechanical nature of matter, one can think of the properties of a solid as arising from elastic forces between neighboring atoms. The simplest one-dimensional classical model of a solid is shown in Fig.~\ref{fig09:springmodel}. A set of point objects, all of mass $m$, is connected by linear springs, all with the same spring constant $k$. This is a discrete system, which however already has an infinite number of degrees of freedom. Let us set up a coordinate $x$ along the spring chain so that in equilibrium, the position of the $j$-th object is $x^j_0=ja$, where $a$ is a constant. We will describe the configuration of the system using a set of generalized coordinates $q^j$, measuring the displacement of the masses from equilibrium, $q^j\equiv x^j-x^j_0=x^j-ja$.

The extension of the spring connecting the $j$-th and $(j+1)$-th mass is $x^{j+1}-x^j-a=q^{j+1}-q^j$. The Lagrangian of the spring chain therefore takes the form
\begin{equation}
L=\frac12m\sum_j(\dot q^j)^2-\frac12k\sum_j(q^{j+1}-q^j)^2\;.
\label{ch09:chainLag}
\end{equation}
The corresponding \emph{equation of motion} (EoM) reads
\begin{equation}
\ddot q^j-\O^2(q^{j+1}-2q^j+q^{j-1})=0\;,\qquad
\O\equiv\sqrt{\frac km}\;.
\label{ch09:chainEoM}
\end{equation}
This set of linear differential equations can be solved by using an exponential ansatz,
\begin{equation}
q^j(t)=\hat q\E^{\I(\k j-\o t)}\;,
\label{ch09:chainansatz}
\end{equation}
where $\hat q$ is a complex amplitude and $\k$ a discrete equivalent of a wave number. This reduces~\eqref{ch09:chainEoM} to an algebraic equation for $\o$ as a function of $\k$, the solution being
\begin{equation}
\o^2=4\O^2\sin^2\frac\k2\;.
\label{ch09:chaindisp}
\end{equation}
Although the spring chain is discrete, we observe a substantial similarity with plane-wave solutions to the equations of motion for free fields we dealt with in~\chaptername~\ref{chap:LagHamcont}. In order to make the analogy explicit, we need to take a continuum limit of the discrete spring chain.

%%%%%%%%%%%%%%%%%%%%%%%%%%%%%%%%%%%%%%%%%%%%%%%%%%%%%%%%%%%%

\subsection{Continuum Limit}
\label{subsec:springmodelcontinuous}

Suppose we can only resolve the motion of the chain at length scales much longer than $a$. This is a sensible assumption: the lattice spacing of real solids is way too small to be observed macroscopically. Unless we look at the solid at the resolution of an electron microscope, it is a good approximation to treat it as a continuous medium. To see how to make the transition from the discrete spring chain to a continuous elastic string, we rewrite the EoM~\eqref{ch09:chainEoM} as
\begin{equation}
\frac ma\ddot q^j-ka\frac{q^{j+1}-2q^j+q^{j-1}}{a^2}=0\;.
\label{ch09:chainEoMcontlimit}
\end{equation}
We now trade the discrete label $j$ for a continuous variable $\p$ indicating a position along the chain by replacing $j\to\p/a$. Likewise, the local displacement of the chain from its equilibrium position becomes a function of $\p$, that is $q^j\to q(\p)$. The large fraction in~\eqref{ch09:chainEoMcontlimit} is then seen to be a discrete approximation to the second derivative $\de_\p^2q$. The ratio $m/a$ indicates the mass per unit length of the chain, and we will treat it as the mass density of the continuous string, $\vr_0$. Finally, the combination $ka$ indicates the force in the string induced by its unit relative extension. This is called \emph{Young modulus} and denoted as $E$. With all the new notation, we can now take the formal limit $a\to0$, which converts the discrete EoM~\eqref{ch09:chainEoMcontlimit} into a second-order \emph{partial differential equation} (PDE) for the displacement $q$ as a function of $\p$ and $t$,
\begin{equation}
\vr_0\de^2_tq-E\de^2_\p q=\vr_0\de^2_tx-E\de^2_\p x=0\;.
\label{ch09:stringwaveeq}
\end{equation}
The second, equivalent form is based on the fact that in equilibrium, the position of a point of the string (with fixed $\p$) is $x_0=\p$, and its displacement from equilibrium is therefore $q=x-\p$. What we found is a wave equation with plane-wave solutions
\begin{equation}
q(\p,t)\propto\E^{\I(p\p-\o t)}\;,\qquad
\o^2=\frac E{\vr_0}p^2\;.
\label{ch09:springplanewave}
\end{equation}

In our analysis above, we chose to take the continuum limit at the level of the EoM. However, we could have done so already at the level of the Lagrangian. To that end, simply rewrite~\eqref{ch09:chainLag} as
\begin{equation}
L=a\sum_j\biggl[\frac12\frac ma(\dot q^j)^2-\frac12ka\biggl(\frac{q^{j+1}-q^j}{a}\biggr)^2\biggr]\;.
\end{equation}
This reveals the familiar combinations $m/a$ and $ka$. Moreover, $(q^{j+1}-q^j)/a$ is just a discrete approximation to the first derivative, $\de_\p q=\de_\p x-1$. Finally, $a\sum_j$ can be viewed as a Riemann sum that, in the limit $a\to0$, becomes an integral over the string. This leads immediately to an action as a functional of the field $x(\p,t)$,
\begin{equation}
S[x]=\int\D\p\,\D t\,\La(x,\de x)\;,\qquad
\La\equiv\frac12\vr_0(\de_tx)^2-\frac12E(\de_\p x-1)^2\;.
\label{ch09:chainaction}
\end{equation}
This action reproduces~\eqref{ch09:stringwaveeq} as its variational EoM.

Yet another way to approach the continuum limit is by inspecting directly the solution~\eqref{ch09:chainansatz}. The latter can be identified with the continuous plane wave~\eqref{ch09:springplanewave} if we set $\k=pa$, since then $\k j=(pa)(\p/a)=p\p$. Simultaneously, the parameter $\O$ can be rewritten as
\begin{equation}
\O=\sqrt{\frac km}=\frac1a\sqrt{\frac{ka}{m/a}}=\frac1a\sqrt{\frac E{\vr_0}}\;.
\end{equation}
Taking the continuum limit $a\to0$ then brings~\eqref{ch09:chaindisp} to the form
\begin{equation}
\o^2=\lim_{a\to0}4\O^2\sin^2\frac\k2=\lim_{a\to0}\frac{4E}{a^2\vr_0}\sin^2\frac{pa}2=\frac E{\vr_0}p^2\;,
\end{equation}
in accord with~\eqref{ch09:springplanewave}. In practice, \eqref{ch09:springplanewave} will be a good approximation of the wave on  the discrete spring chain as long as $\k\ll2\pi$, that is, as long as the wavelength $2\pi/p$ is much longer than the scale $a$.

%%%%%%%%%%%%%%%%%%%%%%%%%%%%%%%%%%%%%%%%%%%%%%%%%%%%%%%%%%%%

\subsection{Space and Body Coordinates}

We have achieved a detailed understanding of the discrete spring chain in Fig.~\ref{fig09:springmodel} and its continuum limit. We found that in this limit, the chain behaves like an elastic medium that supports \emph{longitudinal} compression waves, whereby the individual elements of the medium move in the direction of propagation of the wave. We have thus developed a primitive model of \emph{sound} in solids. Moreover, from~\chaptername~\ref{chap:LagHamcont}, we have at hand a machinery that allows us to deal with such continuous waves.

One detail is disturbing though. The notation we used suggests that the variable $\p$ should be treated as a field that depends on the spacetime coordinates $x$ and $t$. However, the action~\eqref{ch09:chainaction} is explicitly spelled out as a functional of $x(\p,t)$. The time $t$ plays the role of a parameter that allows us to track the evolution of the system. As such, it will always be one of the independent variables in the variational problem. On the other hand, $x$ and $\p$ are to some extent interchangeable, and it is therefore important to understand the difference between these two ``coordinates.'' Fixing the value of $x$ defines a specific point in space, regardless of the motion of the spring chain in the space. This is the good old spatial coordinate that we have been using as one of the independent variables parameterizing field theory. Fixing the value of $\p$ defines a specific point on the spring chain, as is obvious from the origin of $\p$ in the label $j$ via the replacement $j\to\p/a$. We will call $\p$ the body coordinate.

The function $x(\p,t)$ that enters the action~\eqref{ch09:chainaction} therefore gives the position of a specified element of the chain in space, at given time $t$. The collection of $x(\p,t)$ as a function of time for different choices of $\p$ corresponds to the set of trajectories of the individual elements of the spring chain. It is not surprising that this is a natural building block for the construction of the action. On the other hand, the function $\p(x,t)$ identifies the element of the chain that happens to be at the point $x$ in space, at time $t$. This has a less immediate physical interpretation than $x(\p,t)$. However, as long as the evolution of the spring chain is nonsingular, both descriptions in terms of $x(\p,t)$ and $\p(x,t)$ are equivalent. Mathematically, the functions $x(\p,t)$ and $\p(x,t)$ for fixed time $t$ are mutually inverse to each other. In the following, it will be more convenient to use $\p(x,t)$ and treat it as a scalar field in spacetime.

%%%%%%%%%%%%%%%%%%%%%%%%%%%%%%%%%%%%%%%%%%%%%%%%%%%%%%%%%%%%

\section{Kinematics of Continuous Media}
\label{sec:solidkinematics}

We are now ready to promote the above observations to a general framework for description of classical matter. The latter is assumed to be a continuous medium that fills the $d$-dimensional Euclidean space, $\R^d$. Our main assumption is that it is possible to uniquely label the individual elements of the medium with a set of labels $\p^A$, $A=1,\dotsc,d$ in a way that allows us to track their motion. The labels $\p^A$ are the \emph{body} (or sometimes \emph{material}) \emph{coordinates}, which remain attached to the chosen element of the medium. This idea is borrowed from our discussion of rigid body dynamics in~\chaptername~\ref{chap:rigidbody}, except that deformations of the medium now obviously make it impossible to set up a rigid Cartesian body frame, attached to the material of the medium. The functions $\p^A(\vec x,t)$ constitute a set of scalar fields in spacetime, whose values determine which element of the medium is located at position $\vec x$ at time $t$. This description of matter is often referred to as \emph{Eulerian}.

\begin{watchout}%
There is a considerable ambiguity in the way that the labels $\p^A$ are assigned to the elements of the medium. Any prediction for the observable physical properties of matter must be independent of how the assignment is done. While the choice of assignment is in principle arbitrary, in practice it is common to require that the values of $\p^A$ coincide with the spatial coordinates $x^i$ in a chosen reference state of the medium, indicated with a subscript $0$, that is
\begin{equation}
\p_0^A(\vec x,t)=\d^A_i x^i\;.
\label{ch09:equilibrium}
\end{equation}
Equivalently, one can say that the position of the element with labels $\p^A$ in the reference state is $x_0^i=\d^i_A\p^A$. I will assume that the reference state is a time-independent stationary (equilibrium) state of the system under fixed external conditions. This is exactly the choice we made above in case of the one-dimensional spring chain.
\end{watchout}

\begin{figure}[t]
\sidecaption[t]
\includegraphics[width=2.9in]{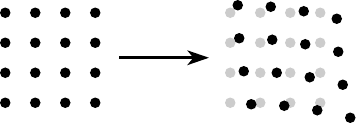}
\caption{Visualization of the motion of a medium in the Lagrangian picture. Displayed are the positions of a fixed set of medium elements at two different times. In order to highlight the displacement of the medium over this time interval, the original positions of the medium elements are indicated using gray dots in the later snapshot.}
\label{fig09:bodycoords}
\end{figure}

Another possibility how to describe the motion of the medium is by following the trajectories of its individual elements. Thus, the functions $x^i(\p,t)$ specify the position of a given element of the medium at time $t$. This description of matter is known as \emph{Lagrangian}. The Lagrangian picture makes it somewhat easier to visualize the motion of the medium; see Fig.~\ref{fig09:bodycoords} for a simple illustration. The assumption that the labeling of the medium elements is unique at any time $t$ requires that the Eulerian map $x^i\to\p^A(\vec x,t)$ at fixed time $t$ is invertible. The inverse is precisely the Lagrangian map $\p^A\to x^i(\p,t)$. Assuming absence of singularities, we find that the configuration space of a medium filling the entire space $\R^d$ consists of all smooth invertible maps $x^i\to\p^A(\vec x)$. This defines the \emph{diffeomorphism group}, $\gr{Diff}(\R^d)$.

%%%%%%%%%%%%%%%%%%%%%%%%%%%%%%%%%%%%%%%%%%%%%%%%%%%%%%%%%%%%

\subsection{Matter Density and Current}
\label{subsec:mattercurrent}

Now that we know how to define a configuration of a medium at fixed time, we would like to be able to describe its motion. In the Lagrangian picture, this is simple. The local velocity of an element of the medium is given by a mere time derivative, $\de_t x^i(\p,t)$, at fixed $\p^A$. We are however going to use the Eulerian picture defined by the maps $\p^A(\vec x,t)$. Let us denote the trajectory of a fixed medium element by $\vec x(t)$, and the velocity we would like to calculate by $\dot{\vec x}(t)$. Taking a total time derivative of $\p^A(\vec x(t),t)$, which is by assumption constant along the trajectory, we get
\begin{equation}
0=\OD{\p^A}t=\PD{\p^A}{x^i}\dot x^i+\PD{\p^A}t\equiv M^A_{\phantom Ai}\dot x^i+\PD{\p^A}t\;,
\end{equation}
where $M^A_{\phantom Ai}\equiv\de_i\p^A$ is the Jacobian matrix of the map $x^i\to\p^A(\vec x,t)$. Since this transformation is by assumption invertible, so is the Jacobian, and we conclude that the local Eulerian velocity of the medium can be expressed in terms of derivatives of the scalar fields $\p^A(\vec x,t)$ as
\begin{equation}
\boxed{\dot x^i=-(M^{-1})^i_{\phantom iA}\de_t\p^A\;.}
\label{ch09:Eulervelocity}
\end{equation}

The basic kinematical properties of a medium are given by its velocity~\eqref{ch09:Eulervelocity} and mass density, $\vr$. It turns out that the two can be encoded rather elegantly in a single mathematical object. I will now briefly outline how to do so, although we will only fully benefit from the construction in~\chaptername~\ref{chap:fluid} when we discuss fluid dynamics. The reason for this digression is that the argument naturally builds upon our derivation of the Eulerian velocity~\eqref{ch09:Eulervelocity}. Feel free to skip the rest of the section on a first reading, and return to it at a later point.

Let us make the simplifying assumption that the equilibrium state defining the labels $\p^A$ on the medium elements via~\eqref{ch09:equilibrium} is spatially uniform. Denoting the constant mass density in this uniform state as $\vr_0$, we next introduce the \emph{matter~current}
\begin{equation}
\boxed{J^\m\equiv\frac{\vr_0}{d!}\ve^{\m\n_1\dotsb\n_d}\ve_{A_1\dotsb A_d}\de_{\n_1}\p^{A_1}\dotsb\de_{\n_d}\p^{A_d}\;.}
\label{ch09:mattercurrent}
\end{equation}
Let us unpack this very compact expression. First, $J^\m$ is formally a vector in spacetime, although our medium is not necessarily relativistic. You can think of $J^\m$ as a mere shorthand notation for a ``density'' $J^0$ and ``current'' $\vec J$ with components $J^i$. It will oftentimes be convenient to use the index $0$ for the time component of (possibly only formal) spacetime tensors, whether the medium at hand actually is relativistic or not. Furthermore, the $d$-dimensional \emph{Levi-Civita symbol} $\ve_{A_1\dotsb A_d}$ is defined to equal respectively $+1$ or $-1$ if $(A_1,\dotsc,A_d)$ is an even or odd permutation of $(1,\dotsc,d)$, and zero otherwise. Likewise, the spacetime Levi-Civita symbol $\ve^{\m_1\dotsb\m_D}$ is defined by the sign of $(\m_1,\dotsc,\m_D)$ as a permutation of the spacetime indices $(0,1,\dotsc,d)$.

Having defined all the notation, we next use the \href{https://en.wikipedia.org/wiki/Leibniz_formula_for_determinants}{Leibniz formula} for the determinant of a $d\times d$ matrix to observe that $J^0$ is proportional to the determinant of the Jacobian matrix $M^A_{\phantom Ai}$,
\begin{equation}
J^0=\frac{\vr_0}{d!}\ve^{0i_1\dotsb i_d}\ve_{A_1\dotsb A_d}M^{A_1}_{\phantom{A_1}i_1}\dotsb M^{A_d}_{\phantom{A_d}i_d}=\vr_0\det M\;.
\label{ch09:matterdensity}
\end{equation}
Now $\vr_0$ was defined to be the density of the medium in equilibrium, or by~\eqref{ch09:equilibrium} equivalently the density in the material coordinates. It follows that $J^0$ is the density of the medium in the space coordinates, $\vr$.

The matter current has two additional simple properties, both as a consequence of the antisymmetry of the Levi-Civita symbol. First, it satisfies the conservation law $\de_\m J^\m=0$. (We will get back to conservation laws in general, and this one in particular, in~\chaptername~\ref{chap:symmetries}.) Second, the current is orthogonal to the gradient of all the material coordinates, $J^\m\de_\m\p^A=0$. Spelling this relation out in time and space indices and using~\eqref{ch09:Eulervelocity}, we find
\begin{equation}
J^0\de_0\p^A+J^iM^A_{\phantom Ai}=0\quad\Rightarrow\quad
J^i=-J^0(M^{-1})^i_{\phantom iA}\de_0\p^A=\vr\dot x^i\;.
\end{equation}
Altogether, we thus have $J^\m=(\vr,\vr\dot x^i)$. Being equal to density times velocity justifies calling $J^i$ a current.

\begin{illustration}%
\label{ex09:mattercurrent1d}%
In order to demystify the formal manipulations surrounding the concept of matter current, let us check the special case of $d=1$. This should match our continuous description of the one-dimensional spring chain in Sect.~\ref{subsec:springmodelcontinuous}. Here we have no need for the index $A$, and~\eqref{ch09:mattercurrent} reduces to
\begin{equation}
J^\m=\vr_0\ve^{\m\n}\de_\n\p=(\vr_0\de_x\p,-\vr_0\de_t\p)\;,
\end{equation}
where $x$ stands for the sole spatial coordinate. This makes the properties $\de_\m J^\m=0$ and $J^\m\de_\m\p=0$ manifest. Moreover, the Jacobian matrix $M^A_{\phantom Ai}$ boils down to $M=\de_x\p$. With the help of~\eqref{ch09:Eulervelocity}, we can then cast the matter current as
\begin{equation}
J^\m=\vr_0\de_x\p(1,-\de_t\p/\de_x\p)=\vr_0\de_x\p(1,\dot x)\;.
\end{equation}
The last thing to clarify is the meaning of the prefactor $\vr_0\de_x\p$. The constant $\vr_0$ was defined in Sect.~\ref{subsec:springmodelcontinuous} as the mass per unit section of the spring chain. Then $\vr_0\D\p$ is the mass contained in the segment $\D\p$ of the spring chain, and accordingly $\vr_0\de_x\p$ the mass per unit length in the coordinate space, that is, the density $\vr$.
\end{illustration}

%%%%%%%%%%%%%%%%%%%%%%%%%%%%%%%%%%%%%%%%%%%%%%%%%%%%%%%%%%%%

\section{Lagrangian Field Theory of Solids}
\label{sec:solidLagrangian}

We have reached a point where we can construct a Lagrangian density in terms of the generalized coordinates $\p^A$. While our discussion of kinematics in Sect.~\ref{sec:solidkinematics} was fairly general and applies equally to both solids and fluids, relativistic or not, we shall now focus on nonrelativistic solids. This will allow us to be more concrete. We will get back to fluids in~\chaptername~\ref{chap:fluid}. On the contrary, intrinsically relativistic solids are of marginal importance in nature, and we will therefore ignore them altogether.

With these qualifications, we can search for the Lagrangian density of a solid as the difference of densities of kinetic and potential energy,
\begin{equation}
\La=\Ta-\Va\;.
\end{equation}
Note that the Lagrangian density cannot depend on $\p^A$ without derivatives. Namely, translation invariance of Euclidean space and of the reference equilibrium state dictates that any shift of the labels $\p^A$ by a constant can be compensated by a corresponding shift of the spatial coordinates $x^i$; cf.~\eqref{ch09:equilibrium}. For the Lagrangian density to be insensitive to such constant shifts of $\p^A$, every field $\p^A(\vec x,t)$ inside it must carry at least one derivative. We will only consider the simplest possibility that every $\p^A(\vec x,t)$ carries \emph{exactly} one derivative.\footnote{For weak deformations of the solid in the sense that $\p^A-\p^A_0=\p^A-\d^A_ix^i$ varies sufficiently slowly in space, adding terms with higher derivatives will have a negligible effect.}

We conclude that our Lagrangian density should be some function of $\de_\m\p^A$ only, that is a function of $\de_0\p^A$ and $M^A_{\phantom Ai}$. We can however be even more precise. Namely, for the Lagrangian density to be a scalar under spatial rotations, all the spatial indices on the various factors of $M^A_{\phantom Ai}$ must be paired and summed over. Thus, $M^A_{\phantom Ai}$ can only enter through the combination
\begin{equation}
\Xi^{AB}\equiv\d^{ij}M^A_{\phantom Ai}M^B_{\phantom Bj}=\grad\p^A\cdot\grad\p^B\;,\quad\text{or equivalently }
\Xi\equiv MM^T\;.
\label{ch09:Xidef}
\end{equation}
Altogether, the Lagrangian density should therefore be a function of $\de_0\p^A$ and $\Xi^{AB}$. The kinetic energy density can, in fact, be written down at once: it should equal a half of the mass density times the squared velocity. The latter can be expressed in terms of our variables $\de_0\p^A$ and $\Xi^{AB}$ with the help of~\eqref{ch09:Eulervelocity},
\begin{equation}
\dot{\vec x}^2=\d_{ij}(M^{-1})^i_{\phantom iA}(M^{-1})^j_{\phantom jB}\de_0\p^A\de_0\p^B=(\Xi^{-1})_{AB}\de_0\p^A\de_0\p^B\;.
\end{equation}
Using finally~\eqref{ch09:matterdensity} and $\det M=\sqrt{\det\Xi}$, we get
\begin{equation}
\boxed{\Ta=\frac{\vr_0}2\sqrt{\det\Xi}(\Xi^{-1})_{AB}\de_t\p^A\de_t\p^B\;.}
\label{ch09:solidLagkin}
\end{equation}

\begin{watchout}%
This looks like a very idiosyncratic way of expressing such a simple concept as kinetic energy. That is the price we have to pay for formulating the dynamics of elastic solids via a variational principle with $\p^A$ as the generalized coordinates. We will see later, with the hindsight gained in~\chaptername~\ref{chap:symmetries}, that the variational EoM descending from~\eqref{ch09:solidLagkin} and the potential energy density $\Va$ is equivalent to local conservation of energy and momentum. Within this~\chaptername, we will content ourselves with analyzing just the linearized version of the EoM. The real advantage of the Lagrangian approach is that it makes it straightforward to describe elasticity of solids beyond the linear approximation.
\end{watchout}

\begin{illustration}%
To follow up on~\refex{ex09:mattercurrent1d}, let us see what~\eqref{ch09:solidLagkin} reduces to in the special case of $d=1$. Here we have $M=\de_x\p$ and accordingly $\Xi=(\de_x\p)^2$. Using the facts that $\dot x=-\de_t\p/\de_x\p$ and $\vr=\vr_0\de_x\p$, the kinetic energy density can then be written as
\begin{equation}
\Ta=\frac{\vr_0}2\sqrt{(\de_x\p)^2}\frac1{(\de_x\p)^2}(\dot x\de_x\p)^2=\frac12\vr_0(\de_x\p)\dot x^2=\frac12\vr\dot x^2\;,
\end{equation}
as expected. As a minor side remark, I have used that $\de_x\p$ is positive. This follows from the fact that in equilibrium, $\p_0=x$ so that $\de_x\p_0=1$, and from the assumption that the map $x\to\p(x,t)$ is invertible at all times and continuous as a function of $t$.
\end{illustration}

Next we turn our attention to the potential energy density $\Va$. We implicitly assume that $\Va$ is well-defined and only depends on the local, instantaneous deformation of the solid. In other words, $\Va$ is some as yet unspecified function of $\Xi^{AB}$. To stress the physical content of this assumption, let me point out possible effects that it excludes. Thus, the potential energy should not depend on the motion of the medium, that is the generalized velocities $\de_t\p^A$. By the same token, the potential energy of the configuration of the medium should not have a ``memory,'' that is depend on the past states. Finally, we exclude dissipative effects, which are notoriously difficult to take into account in the variational formulation of both mechanics and field theory. The dynamical regime constrained by our fundamental assumption is called \emph{hyperelasticity}. It provides a reasonable, numerically accurate description of most real solids, at least within a certain range of deformations.

The deformation of the material is measured by the deviation of $\Xi^{AB}$ from its value in equilibrium, $\smash{\Xi_0^{AB}=\d^{AB}}$. In the literature, $\Xi^{AB}$ is sometimes referred to as the \emph{Finger strain tensor}.\footnote{The matrix inverse, $(\Xi^{-1})_{AB}$, called the \emph{Cauchy strain tensor}, is more common. However, in our approach based on the fields $\p^A(\vec x,t)$ as the generalized coordinates, the Finger strain tensor is the natural object to work with.} Before moving on, let us first convince ourselves that $\Xi^{AB}$ is indeed only sensitive to actual deformations of shape, and not to rigid motions of the material, that is translations and rotations. Rigid translations only shift the spatial coordinates $x^i$ by a constant, and thus leave $M^A_{\phantom Ai}$ and by extension $\Xi^{AB}$ itself intact. As to rigid rotations, consider a linear transformation of $x^i$ to new coordinates $x'^i$ via $\smash{x^i=R^i_{\phantom ij}x'^j}$, where $\smash{R^i_{\phantom ij}}$ is an orthogonal matrix. This changes $\smash{M^A_{\phantom Ai}}$ to
\begin{equation}
M'^A_{\phantom {'A}i}=\de_j\p^A\PD{x^j}{x'^i}=M^A_{\phantom Aj}R^j_{\phantom ji}\;.
\end{equation}
In matrix form, this amounts to $M'=MR$. It follows from the orthogonality of the transformation matrix, $RR^T=\un$, that $\Xi'=M'M'^T=MRR^TM^T=MM^T=\Xi$. As expected, $\Xi^{AB}$ cannot be changed by any translation or rotation of the medium as a whole. It is therefore a good measure of local deformations of the material.

Since the potential energy density does not contain any time derivatives $\de_t\p^A$, its contribution to the variational EoM takes the form
\begin{equation}
\de_i\PD\Va{(\de_i\p^A)}=\de_i\PD\Va{M^A_{\phantom Ai}}=2\d^{ij}\de_i\left(\PD\Va{\Xi^{AB}}M^B_{\phantom Bj}\right)\;.
\end{equation}
Except for the overall sign, this should have the physical interpretation as the density of net force on an element of the solid. The expression in the parentheses itself, or equivalently $\Pd\Va{\Xi^{AB}}$, represents \emph{stress} in the medium. This generalizes the concept of pressure known from fluids in a way we will clarify in more detail below. Here I just point out that the net force on a solid element arises from the gradient of stress inside the material.

The precise dependence of the potential energy density on $\Xi^{AB}$ is constrained by the symmetries of the solid. Most solids have a crystalline structure, whose symmetry dictates what kind of combinations of the components $\Xi^{AB}$ can enter the Lagrangian density. Answering this question even for simple crystalline structures requires heavy use of tensor calculus; some (though by far not exhaustive) details can be found in Sect.~10 of~\cite{Landau1986}. Here we focus instead on the idealized case of \emph{isotropic} solids.

%%%%%%%%%%%%%%%%%%%%%%%%%%%%%%%%%%%%%%%%%%%%%%%%%%%%%%%%%%%%

\section{Isotropic Solids}
\label{sec:isotropicsolids}

A single monocrystal of a solid material is never perfectly isotropic. However, many solids appear in nature in the form of polycrystals, with the individual crystalline grains being randomly oriented. Such a polycrystal can behave as effectively isotropic when observed at length scales much longer than the typical size of a grain.

Complete isotropy of the material requires that $\Va$ should not change upon an orthogonal transformation of the body coordinates, $\p^A=R^A_{\phantom AB}\p'^B$, which changes the matrix $\Xi$ to $\Xi'=R^T\Xi R$. Being symmetric, $\Xi$ can always be diagonalized by such an orthogonal transformation. It follows that the potential energy density $\Va$ can only depend on the $d$ eigenvalues of $\Xi$. An immediate corollary is that there can be at most $d$ algebraically independent functions $\Va(\Xi)$. These can be chosen arbitrarily, but it is conventional to take as the set of independent rotationally-invariant functions the traces of powers of $\Xi$, that is $\tr\Xi^n$ with $n=1,\dotsc,d$. The most general potential energy density of an isotropic solid can then be expressed as a function of these,
\begin{equation}
\boxed{\Va=\Va(\tr\Xi,\dotsc,\tr\Xi^d)\;.}
\label{ch09:solidLagpot}
\end{equation}
An alternative choice of the $d$ algebraically independent invariants that is sometimes used is given by the coefficients of the characteristic equation $\Xi$ satisfies as per the \href{https://en.wikipedia.org/wiki/Cayley-Hamilton_theorem}{Cayley--Hamilton theorem}. For instance, for $d=3$ these would be $\tr\Xi$, $[(\tr\Xi)^2-\tr\Xi^2]/2$ and $\det\Xi$, which equals $[(\tr\Xi)^3-3\tr\Xi\tr\Xi^2+2\tr\Xi^3]/6$.

%%%%%%%%%%%%%%%%%%%%%%%%%%%%%%%%%%%%%%%%%%%%%%%%%%%%%%%%%%%%

\subsection{Linear Elasticity}
\label{subsec:linearelasticity}

\begin{figure}[t]
\sidecaption[t]
\includegraphics[width=2.9in]{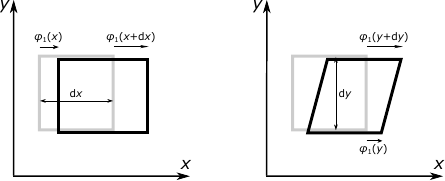}
\caption{Examples of deformations of an element of the medium, corresponding to different sectors of the infinitesimal strain tensor~\eqref{ch09:inftystrain}. Left panel: expansion in a single direction, measured by $e_{11}$. Right panel: shear deformation due to nonzero $\de_2\vp_1$, captured by $e_{12}$.}
\label{fig09:shear}
\end{figure}

Together, \eqref{ch09:solidLagkin} and~\eqref{ch09:solidLagpot} completely specify the dynamics of an isotropic hyperelastic medium. However, the corresponding EoM is not easy to solve due to being highly nonlinear. To illustrate the basic physical properties of elastic solids, we shall from now on restrict the discussion to small deviations from the equilibrium~\eqref{ch09:equilibrium}. To that end, we will utilize the standard workflow for small oscillations of fields as outlined in Sect.~\ref{sec:fieldQuasi}. The first step is to parameterize the deviations of the medium from equilibrium by a set of (by assumption small) fields $\vp^A$ such that
\begin{equation}
\p^A=\p_0^A-\vp^A=\d^A_ix^i-\vp^A\;.
\label{ch09:displacement}
\end{equation}
Note the opposite sign as compared to our previously used convention for field derivatives. This is to ensure that $\vp^A$ can be interpreted literally in terms of the physical displacement of the medium from equilibrium, that is as $\d^A_i(x^i-x^i_0)=\d^A_ix^i-\p^A$. Upon inserting~\eqref{ch09:displacement} in the definition~\eqref{ch09:Xidef} of $\Xi^{AB}$, we find
\begin{equation}
\Xi^{AB}=\d^{ij}(\d^A_i-\de_i\vp^A)(\d^B_j-\de_j\vp^B)=\d^{AB}-2\d^{Ai}\d^{Bj}e_{ij}+\grad\vp^A\cdot\grad\vp^B\;,
\label{ch09:Xiapprox}
\end{equation}
where I introduced the shorthand notation
\begin{equation}
e_{ij}\equiv\frac12(\de_i\vp_j+\de_j\vp_i)\;,\qquad
\vp_i\equiv\d_{iA}\vp^A\;.
\label{ch09:inftystrain}
\end{equation}
The symmetric matrix $e_{ij}$ is called the \emph{infinitesimal strain tensor}. It parameterizes small deformations of the medium in the following way (see Fig.~\ref{fig09:shear} for illustration):
\begin{itemize}
\item There is no strain if the medium as a whole undergoes an infinitesimal translation (constant $\vp_i$) or rotation ($\vp_i=\O_{ij}x^j$ with antisymmetric $\O_{ij}$).
\item The diagonal elements $e_{11}=\de_1\vp_1$ etc.~measure relative expansion or compression along one of the coordinate axes.
\item As a consequence, the trace of the infinitesimal strain tensor, $\tr e$, measures relative change of volume of a medium element.
\item The offdiagonal elements such as $e_{12}=(\de_1\vp_2+\de_2\vp_1)/2$ measure shear deformation of the material.
\end{itemize}

The next step of our general workflow for small oscillations is to expand the Lagrangian density to second order in the deviation fields, here $\vp^A$. In case of the kinetic energy density~\eqref{ch09:solidLagkin}, this amounts to approximating
\begin{equation}
\boxed{\Ta\approx\frac{\vr_0}2\d_{AB}\de_t\vp^A\de_t\vp^B=\frac{\vr_0}2\d^{ij}\de_t\vp_i\de_t\vp_j\;.}
\label{ch09:solidLagkinapprox}
\end{equation}
In case of the potential energy density, this requires writing down the most general second-order polynomial in $e_{ij}$ consistent with isotropy. There are only two possibilities, and accordingly the bilinear part of the potential energy density depends on two a priori unknown material constants. It can be written alternatively using the index or matrix notation as
\begin{equation}
\boxed{\Va\approx\frac\l2(\d^{ij}e_{ij})^2+\m\d^{ik}\d^{jl}e_{ij}e_{kl}=\frac\l2(\tr e)^2+\m\tr e^2\;,}
\label{ch09:solidLagpotapprox1}
\end{equation}
where the parameters $\l,\m$ are known as \emph{Lam\'e coefficients}. In practice, a somewhat different basis of quadratic invariants is usually employed. Using the fact that
\begin{equation}
\tr e^2=\tr\left(e-\frac{\tr e}d\un+\frac{\tr e}d\un\right)^2=\tr\left(e-\frac{\tr e}d\un\right)^2+\frac{(\tr e)^2}d\;,
\end{equation}
the potential energy density can be rewritten as
\begin{equation}
\boxed{\Va\approx\frac K2(\tr e)^2+\m\tr\left(e-\frac{\tr e}d\un\right)^2\;,}
\label{ch09:solidLagpotapprox2}
\end{equation}
where $K\equiv\l+(2\m/d)$. The advantage of this parameterization is that it explicitly separates changes in overall volume from deformations of shape. Namely, $\tr e$ measures, as pointed out above, the relative change in volume of a medium element. The coefficient $K$ is dubbed \emph{bulk modulus}. On the other hand, the combination $e-(\tr e/d)\un$ has a vanishing trace and thus is not sensitive to changes in volume. It parameterizes solely the change of shape of the medium element. The coefficient $\m$ is called \emph{shear modulus}.

The complete bilinear Lagrangian density for small deformations of an isotropic solid is given by~\eqref{ch09:solidLagkinapprox} together with~\eqref{ch09:solidLagpotapprox1} or~\eqref{ch09:solidLagpotapprox2}. In order to derive the corresponding EoM, we need to evaluate, among others, the derivative of the Lagrangian density with respect to a spatial derivative of $\vp_i$. Let us prepare this beforehand. Using the chain rule, we get
\begin{equation}
\PD\La{(\de_i\vp_j)}\approx\PD\La{e_{kl}}\PD{e_{kl}}{(\de_i\vp_j)}=\frac12\PD{\La}{e_{kl}}(\d^i_k\d^j_l+\d^i_l\d^j_k)=\PD{\La}{e_{ij}}\;.
\end{equation}
Equipped with this identity, it takes a little effort to obtain the linearized EoM,
\begin{equation}
\boxed{\vr_0\ddot\vp_i\approx\de^j\s_{ij}\;,}\qquad\text{where }
\s_{ij}\equiv\d_{ik}\d_{jl}\PD{\Va}{e_{kl}}\approx\l\d_{ij}\tr e+2\m e_{ij}\;.
\label{ch09:solidEoMstress}
\end{equation}
The symmetric matrix $\s_{ij}$ is the \emph{Cauchy stress tensor}. Physically, $\s_{ij}$ measures the $i$-th component of force per unit area on a surface whose normal vector points along the $j$-th coordinate axis. With this interpretation, \eqref{ch09:solidEoMstress} is seen to simply encode Newton's second law, with mass time acceleration on the left-hand side and force on the right-hand side.

\begin{illustration}%
\label{ex09:fluid}%
Ideal fluids can be viewed as a special limit of elastic solids in which there is no shear stress. The only internal forces that ideal fluids experience are those of pressure, $P$. Accordingly, the linearized EoM for the displacement vector $\vec\vp$ in an ideal fluid should take the form $\vr_0\ddot{\vec\vp}\approx-\grad P$. Matching this to~\eqref{ch09:solidEoMstress} requires that $\s_{ij}=-P\d_{ij}$. The physical content of this relation is obvious. In ideal fluids, internal forces of pressure always act along the normal to a surface, and are isotropic. The minus sign indicates that pressure pushes against any surface in the fluid.

Using the decomposition $e_{ij}=\d_{ij}(\tr e/d)+(e_{ij}-\d_{ij}\tr e/d)$, the relation $\s_{ij}=-P\d_{ij}$ is seen to be satisfied only if $\m=0$ and simultaneously $P=-K\tr e$. The vanishing of the shear modulus confirms that ideal fluids do not experience any shear stress. The relation between pressure and the trace of the strain tensor moreover shows that $1/K$ measures the relative change in the fluid volume with increasing pressure. Thus, $1/K$ can be identified with the \emph{compressibility} of the fluid.
\end{illustration}

Using the definition~\eqref{ch09:inftystrain} of the infinitesimal strain tensor allows one to rewrite the divergence of the Cauchy stress tensor as
\begin{equation}
\begin{split}
\de^j\s_{ij}&=\l\de_i\tr e+2\m\de^je_{ij}=\l\de_i\de^j\vp_j+\m(\de^j\de_i\vp_j+\de^j\de_j\vp_i)\\
&=(\l+\m)\de_i\divg\vec\vp+\m\grad^2\vp_i\;.
\end{split}
\end{equation}
This brings the EoM~\eqref{ch09:solidEoMstress} to the vector form
\begin{equation}
\boxed{\vr_0\ddot{\vec\vp}\approx(\l+\m)\grad(\divg\vec\vp)+\m\grad^2\vec\vp\;,}
\label{ch09:solidEoMNavierCauchy}
\end{equation}
called the \emph{Navier--Cauchy equation}. This is the basic dynamical equation of linear elasticity. It can be used in multiple ways. On the one hand, setting the left-hand side of~\eqref{ch09:solidEoMNavierCauchy} to zero leads to a condition for local static equilibrium that determines the distribution of strain, and via~\eqref{ch09:solidEoMstress} also of stress, in an elastic medium. An example of this kind of application of the Navier--Cauchy equation appears in~\refpr{pr09:sphere}. On the other hand, \eqref{ch09:solidEoMNavierCauchy} is a convenient starting point for the analysis of small oscillations of the medium and their propagation. This will be our next target.

%%%%%%%%%%%%%%%%%%%%%%%%%%%%%%%%%%%%%%%%%%%%%%%%%%%%%%%%%%%%

\subsection{Elastic Waves}
\label{subsec:elasticwaves}

Equation~\eqref{ch09:solidEoMNavierCauchy} is a homogeneous second-order linear PDE and as such is ideally suited for the analysis of elastic waves in solids. We proceed as usual by adopting an exponential ansatz,
\begin{equation}
\vec\vp(\vec x,t)=\hat{\vec\vp}\E^{\I(\skal kx-\o t)}\;,
\end{equation}
where $\vec k$ is the wave vector and $\o$ the frequency of the wave. Now that our dynamical variable $\vec\vp$ is a vector, we also need a vector amplitude, $\hat{\vec\vp}$. The exponential plane-wave ansatz reduces the EoM~\eqref{ch09:solidEoMNavierCauchy} to an algebraic equation,
\begin{equation}
\vr_0\o^2\hat{\vec\vp}\approx(\l+\m)\vec k(\vec k\cdot\hat{\vec\vp})+\m\vec k^2\hat{\vec\vp}\;.
\label{ch09:elasticwavesdisp}
\end{equation}
To proceed, we decompose the amplitude into components parallel (\emph{longitudinal}) and perpendicular (\emph{transverse}) to the wave vector, $\hat{\vec\vp}=\hat{\vec\vp}_\mathrm{L}+\hat{\vec\vp}_\mathrm{T}$. Decomposing in the same way~\eqref{ch09:elasticwavesdisp} itself shows that there are two different solutions for the frequency, corresponding to the longitudinal and transverse components of the wave,
\begin{equation}
\boxed{\o_\mathrm{L}^2=\frac{\l+2\m}{\vr_0}\vec k^2\;,\qquad
\o_\mathrm{T}^2=\frac\m{\vr_0}\vec k^2\;.}
\label{ch09:elasticwavesdisp2}
\end{equation}

We have found that small perturbations of strain and stress in an isotropic solid propagate in the form of elastic waves, which are nothing else but \emph{sound}. There are two different basic types of sound waves in isotropic solids:
\begin{itemize}
\item\emph{Longitudinal waves.} For these, the oscillations of the medium elements are in the direction of propagation of the wave. This is the type of compression wave that we found in the one-dimensional spring model of Sect.~\ref{sec:springmodel}. Longitudinal sound waves exist in solids and fluids alike.
\item\emph{Transverse waves.} In this case, the medium elements oscillate in one of the directions perpendicular to the direction of propagation of the wave. In $d$ spatial dimensions, transverse waves can therefore acquire $d-1$ different polarizations, as compared to a single type of longitudinal wave. As is clear from~\eqref{ch09:elasticwavesdisp2}, the existence of transverse waves critically relies on the presence of shear stress. Transverse sound waves therefore exist in solids but not in fluids.
\end{itemize}
Both types of sound waves have a linear dispersion relation, with frequency proportional to the magnitude of the wave vector (that is the wave number). The respective phase velocities of the longitudinal and transverse waves are
\begin{equation}
c_\mathrm{L}=\sqrt{\frac{\l+2\m}{\vr_0}}=\sqrt{\frac{K+2\m(1-1/d)}{\vr_0}}\;,\qquad
c_\mathrm{T}=\sqrt{\frac\m{\vr_0}}\;.
\label{ch09:cLcT}
\end{equation}

%%%%%%%%%%%%%%%%%%%%%%%%%%%%%%%%%%%%%%%%%%%%%%%%%%%%%%%%%%%%

\section*{\probsec}
\addcontentsline{toc}{section}{\probsec}

\begin{prob}
\label{pr09:soundspeed}
Explain why the bulk modulus $K$ and the shear modulus $\m$ (but not necessarily the Lam\'e coefficient $\l$) must be positive. Use the fact that $K>0$ to deduce a constraint on possible values of the ratio $c_\mathrm{L}/c_\mathrm{T}$ of sound velocities. Check that your constraint is satisfied by the following data for selected solid materials (in $d=3$ dimensions):
\begin{center}
\begin{tabular}{p{2.0cm}p{2.0cm}p{1.2cm}}
\hline\noalign{\smallskip}
Material & $c_\mathrm{L}\text{ [m/s]}$ & $c_\mathrm{T}\text{ [m/s]}$ \\
\noalign{\smallskip}\svhline\noalign{\smallskip}
Aluminum & 6420 & 3040\\
Copper & 5010 & 2270\\
Iron & 4994 & 2809\\
Nickel & 6040 & 3000\\
Titanium & 6070 & 3125\\
Zinc & 4210 & 2440\\
\noalign{\smallskip}\hline\noalign{\smallskip}
\end{tabular}
\end{center}
\end{prob}

\begin{prob}
\label{pr09:sigmatoe}
In Sect.~\ref{sec:springmodel}, we characterized the elastic properties of a one-dimensional string by the Young modulus $E$. A general definition of $E$ is as follows. Suppose we ``pull'' the solid in a way that creates purely one-dimensional stress so that the only nonvanishing component of the Cauchy stress tensor is $\s_{11}$. The Young modulus is then introduced as a proportionality factor between $\s_{11}$ and the  component of the infinitesimal strain tensor corresponding to expansion in the direction of the pull, $e_{11}=\s_{11}/E$. However, the expansion of a bulk material in one direction is usually accompanied by a contraction in the transverse directions. The amount of transverse contraction is characterized by the \emph{Poisson ratio} $\n$ through $e_{22}=\dotsb=e_{dd}=-\n e_{11}$. Use this special case of stress to argue that the infinitesimal strain tensor and Cauchy stress tensor are generally related by
\begin{equation}
e_{ij}=\frac1E[(1+\n)\s_{ij}-\n\d_{ij}\tr\s]\;.
\label{ch09:eintermsofsigma}
\end{equation}
\end{prob}
Comparing with~\eqref{ch09:solidEoMstress}, find the relation between the Young modulus $E$ and Poisson ratio $\n$, and the Lam\'e coefficients $\l,\m$. You can check your result against the table of relations among different elastic coefficients on the corresponding \href{https://en.wikipedia.org/wiki/Elastic_modulus}{Wikipedia page}.

\begin{prob}
\label{pr09:Poissonratio}
Find an expression for the Poisson ratio in terms of the bulk modulus $K$ and the shear modulus $\mu$. Use this to show that the Poisson ratio must lie in the range
\begin{equation}
-1<\n<\frac1{d-1}\;.
\end{equation}
How would you physically characterize the limit $\n\to1/(d-1)$?
\end{prob}

\begin{prob}
\label{pr09:sphericalstrain}
Suppose that a three-dimensional isotropic solid medium is strained in a perfectly spherically symmetric manner. This can be described by the displacement vector $\vec\vp(\vec x)=f(r)\vec n$, where $r\equiv\abs{\vec x}$ is the radial distance from the origin, and $\vec n\equiv\vec x/r$ a radial unit vector. Calculate the (Cartesian) components of the infinitesimal strain tensor. You should find that
\begin{equation}
e_{ij}(\vec x)=\frac{f(r)}r\d_{ij}+\left[\frac{f'(r)}{r^2}-\frac{f(r)}{r^3}\right]x_ix_j\;.
\end{equation}
\end{prob}

\begin{prob}
\label{pr09:sphere}
Consider a solid spherical shell with inner radius $a$ and outer radius $b$. The shell is filled with a gas of pressure $P_a$, whereas the ambient pressure is $P_b$. Using the result of~\refpr{pr09:sphericalstrain}, find the radial displacement of the material due to the pressure. Hint: first show that the radial displacement vector $\vec\vp$ necessarily satisfies $\smash{\grad(\divg\vec\vp)=\grad^2\vec\vp}$ as a consequence of having a vanishing curl. This turns the Navier--Cauchy equation into the condition of constant divergence of $\vec\vp$. Find a general spherically symmetric solution of this condition. Finally, impose the boundary condition that the projection of the Cauchy stress tensor into the radial direction, $\s_{ij}n^in^j$, equals minus the outside pressure ($P_a$ at the inner surface and $P_b$ at the outer surface). Give a physical interpretation of your result in the special case $P_a=P_b$.
\end{prob}

\begin{prob}
\label{pr09:NavierCauchy}
In our derivation of the Navier--Cauchy equation, we did not take into account any other forces but the stress caused by the strain of the medium. Forces acting only on the surface of the material can be taken into account by a suitable boundary condition as in~\refpr{pr09:sphere}. On the other hand, volume forces have to be included by modifying the EoM~\eqref{ch09:solidEoMNavierCauchy}. To be concrete, suppose that the medium is exposed to a constant gravitational field $\vec g$. Add a suitable term representing the effect of gravity to the potential energy density~\eqref{ch09:solidLagpot}, and thence derive the corrected EoM. Check your result by using it to calculate the pressure distribution in an incompressible liquid in hydrostatic equilibrium.
\end{prob}
\chapter{Symmetries and Conservation Laws}
\label{chap:symmetries}

\keywords{Local conservation law, continuous symmetry, Noether theorem, Noether current and charge, energy--momentum tensor, angular momentum tensor.}

%%%%%%%%%%%%%%%%%%%%%%%%%%%%%%%%%%%%%%%%%%%%%%%%%%%%%%%%%%%%

\noindent We have already encountered the concept of a conservation law on various occasions throughout the course. At a general level, our discussion of conservation laws was however limited to mechanics, where they manifest themselves through functions on the configuration or phase space that are constant along the trajectory of the system. So far, our only tool to uncover the presence of such constants of motion has been based on the assumption that the Lagrangian or Hamiltonian of the system does not depend explicitly on time or one of the generalized coordinates.

In this \chaptername, we will start by promoting the concept of a conservation law to a local (differential) law. This comes along with the notion of a local density and current of a conserved charge. Then, we introduce the seemingly unrelated concept of a symmetry of a dynamical system. The core of the \chaptername{} is devoted to the Noether theorem, showing that a symmetry of a system automatically guarantees the presence of a conservation law. A separate subsection is dedicated to the omnipresent example of such a correspondence, relating the conservation of energy and momentum to the invariance of physical laws under translations in time and space. At the end, we will return to mechanics as a special case of field theory. In this restricted setting, we will be able to use our experience gathered in \chaptername~\ref{chap:geometryclassmech} to show that the correspondence between symmetries and conservation laws is bijective, that is one-to-one.

%%%%%%%%%%%%%%%%%%%%%%%%%%%%%%%%%%%%%%%%%%%%%%%%%%%%%%%%%%%%

\section{Local Conservation Laws}
\label{sec:conservationlaws}

As a motivation, let us recall a couple of examples of local conservation laws that one encounters early on in the undergraduate physics curriculum. For instance, in electromagnetism, one shows that electric charge is conserved as a consequence of Maxwell's equations. The conservation law takes the local form
\begin{equation}
\PD\vr t+\divg\vec J=0\;,
\label{ch10:consEMcharge}
\end{equation}
where $\vr$ is the density of electric charge and $\vec J$ the density of electric current. In addition, it follows from Maxwell's equations that in the absence of electric charges, the energy of the electromagnetic field is also locally conserved,\footnote{In the presence of charged matter, the energy of the electromagnetic field alone is not conserved anymore, although the total energy of the field and matter still is.}
\begin{equation}
\PD ut+\divg\vec S=0\;.
\label{ch10:consEMenergy}
\end{equation}
Here $u\equiv(1/2)\eps_0\vec E^2+\vec B^2/(2\m_0)$ is the energy density of the electromagnetic field and $\vec S\equiv\vekt EB/\m_0$ the Poynting vector, representing the energy flux.

\begin{figure}[t]
\sidecaption[t]
\includegraphics[width=2.0in]{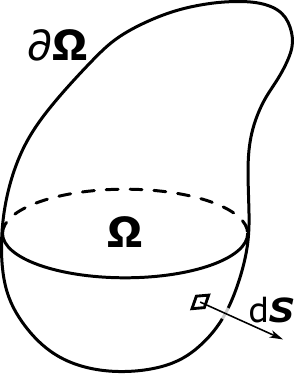}% This figure does not include so many details to warrant a full width of 2.9 in!
\caption{Visualization of the geometry involved in the calculation of the total charge~\eqref{ch10:QinOmega} contained in domain $\vec\O$ and the flux of charge~\eqref{ch10:fluxoutofOmega} through the boundary $\de\vec\O$. The area element $\D\vec S$ on the boundary is oriented outwards.}
\label{fig10:conservation}
\end{figure}

Both examples~\eqref{ch10:consEMcharge} and~\eqref{ch10:consEMenergy} of local conservation laws take the same mathematical form. This is not a coincidence. Suppose there is an extensive quantity $Q$ that we know to be conserved. I will generally refer to $Q$ as a \emph{charge} except for specific examples where a different terminology is established. Being extensive, the total charge contained in a domain $\vec\O$ in space can be determined by integration of the local \emph{charge density} $\vr$,
\begin{equation}
Q_{\vec\O}(t)=\int_{\vec\O}\D^d\!\vec x\,\vr(\vec x,t)\;.
\label{ch10:QinOmega}
\end{equation}
If the domain $\vec\O$ were completely isolated from the outside world, we would expect $Q_{\vec\O}$ to be time-independent, that is conserved. In general, the amount of charge inside $\vec\O$ can change, but only if some charge ``flows'' in or out through the surface $\de\vec\O$. The flow is parameterized by the \emph{current density} $\vec J$, which represents the amount of charge per second that passes through a unit area perpendicular to $\vec J$. The flux of charge through the boundary $\de\vec\O$, that is the amount of charge per second that leaves the domain $\vec\O$, is then given by integration over $\de\vec\O$,
\begin{equation}
\Phi_{\de\vec\O}=\int_{\de\vec\O}\D\vec S\cdot\vec J\;.
\label{ch10:fluxoutofOmega}
\end{equation}
The outward direction of the flow is guaranteed by the choice of orientation of the surface $\de\vec\O$ so that the oriented area element $\D\vec S$ points out of $\vec\O$. (See Fig.~\ref{fig10:conservation} for a visualization of the geometry.) The fact that no charge can appear in $\vec\O$ or disappear from it without having passed through the surface is encoded in the condition $\Od{Q_{\vec\O}}t=-\Phi_{\de\vec\O}$. Using the divergence theorem to convert the surface integral into a volume one, this boils down to
\begin{equation}
\int_{\vec\O}\D^d\!\vec x\,\biggl(\PD\vr t+\divg\vec J\biggr)=0\quad\Rightarrow\quad
\boxed{\PD\vr t+\divg\vec J=0\;.}
\label{ch10:consgeneral}
\end{equation}
For the rigorously minded, I have assumed that the domain $\vec\O$ is chosen as time-independent but otherwise arbitrary. This guarantees that the integrand in~\eqref{ch10:consgeneral} has to vanish point by point in space. The local conservation law~\eqref{ch10:consgeneral} has an important corollary. Suppose that the current density $\vec J$ vanishes on the boundary $\de\vec\O$. This may be natural, for instance, if we stretch $\vec\O$ to the whole space $\R^d$ and we know that all fields in the system decay sufficiently rapidly at spatial infinity. Then, the integral charge $Q_{\vec\O}$ is actually time-independent, that is conserved.

\begin{illustration}%
\label{ex10:continuityequation}%
An important example of a local conservation law that one meets in the undergraduate curriculum is the continuity equation for a (nonrelativistic) fluid,
\begin{equation}
\PD\vr t+\divg(\vr\vec v)=0\;,
\label{ch10:continuityeq}
\end{equation}
where $\vr(\vec x,t)$ is the mass density of the fluid and $\vec v(\vec x,t)$ its local (Eulerian) velocity. Matching this to the general conservation law~\eqref{ch10:consgeneral}, we see that the continuity equation represents conservation of mass in nonrelativistic continuum mechanics. The combination $\vr\vec v$ can accordingly be interpreted as the density of mass current.
\end{illustration}

\begin{illustration}%
\label{ex10:consprobability}%
For another example from a familiar context, consider the wave function $\psi(\vec x,t)$ of a quantum-mechanical particle of mass $m$ moving in a potential $V(\vec x)$. Here setting $\vr(\vec x,t)\equiv\abs{\psi(\vec x,t)}^2$ defines the probability density for finding the particle at a given position. It is known that the total probability, that is $\int_{\R^d}\D^d\!\vec x\,\vr(\vec x,t)$, is conserved by the unitary time evolution of quantum mechanics. Accordingly, the probability density satisfies a local conservation law of the type~\eqref{ch10:consgeneral}, where
\begin{equation}
\vec J=\frac\hbar{2\I m}(\psi^*\grad\psi-\psi\grad\psi^*)
\end{equation}
is the probability current. The local conservation law can be proven by calculating the time derivative of $\vr(\vec x,t)$ using the Schr\"odinger equation and manipulating the result into the form of a gradient of a vector function.
\end{illustration}

Before we conclude the introductory account of conservation laws, let me mention that it is common to merge the charge density $\vr$ and current density $\vec J$ into a single object, $J^\m\equiv(\vr,\vec J)$. Here I follow the convention introduced in Sect.~\ref{sec:solidkinematics}, using the Greek index $\m$ to refer jointly to the density $J^0\equiv\vr$ and current $\vec J$ even for systems that are not relativistic. The point of doing so is that for many formal manipulations, the distinction between temporal and spatial components is unimportant. Our general local conservation law~\eqref{ch10:consgeneral} now acquires a particularly compact form
\begin{equation}
\de_\m J^\m=0\;.
\label{ch10:consgeneralrel}
\end{equation}

In all the examples mentioned until now, we could have either guessed the existence of a local conservation law by intuition, or derived it by an explicit manipulation of the \emph{equation of motion} (EoM). This is however not satisfactory. We need a more systematic way of checking whether a given system possesses a conservation law, and finding out what form this conservation law takes. That is where the concept of symmetry plays an invaluable role.

%%%%%%%%%%%%%%%%%%%%%%%%%%%%%%%%%%%%%%%%%%%%%%%%%%%%%%%%%%%%

\section{Continuous Symmetries in Field Theory}
\label{sec:continuoussymmetries}

While the term ``symmetry'' is often used in a colloquial way, in the context of the variational formulation of field theory (or its special case, mechanics), it has a precise meaning. We say that a theory of a set of fields, $\p^A$, has a continuous symmetry if its action $S[\p]$ is invariant under an \emph{infinitesimal transformation}
\begin{equation}
\boxed{\udelta\p^A\equiv\eps\df^A(\p,\de\p,\dotsc,x)\;.}
\label{ch10:inftytransfo}
\end{equation}
Here $\eps$ is a constant parameter, while $\df^A$ is a set of functions that may depend on the fields and their derivatives as well as explicitly on the spacetime coordinates. The invariance of the action is understood as the vanishing of $S[\p+\eps\df]-S[\p]$ to first order in $\eps$, that is as
\begin{equation}
S[\p+\eps\df]-S[\p]=\bigO(\eps^2)\;.
\label{ch10:invariance}
\end{equation}
This is often abbreviated as $\udelta S[\p]=0$. In the language of~\chaptername~\ref{chap:mathintro}, the condition of invariance amounts to the vanishing of the Gateaux derivative~\eqref{ch00:Gateauxder} of the action $S[\p]$ in the direction defined by $\df^A$.

\begin{illustration}%
\label{ex10:KG}%
Consider the theory of a free massless \emph{Klein--Gordon} (KG) field (cf.~\refex{ex08:KG}), defined by the Lagrangian density
\begin{equation}
\La=\frac12(\de_0\p)^2-\frac12(\grad\p)^2=-\frac12(\de_\m\p)^2\;.
\label{ch10:LagKGmassless}
\end{equation}
Since the Lagrangian density only depends on the derivatives of the scalar field $\p$, the action of this theory is obviously invariant under shifts of the field by a constant,
\begin{equation}
\udelta\p=\eps\quad\Leftrightarrow\quad
F(\p,\de\p,\dotsc,x)=1\;.
\label{ch10:shiftsymmetry}
\end{equation}
\end{illustration}

In the above example, we did not actually have to assume the parameter $\eps$ to be infinitesimal. The equality $S[\p+\eps F]=S[\p]$ for any finite value of $\eps$ is apparent. This is not a rule. In most cases, one has to carefully evaluate the shift of the action under~\eqref{ch10:inftytransfo} to first order in $\eps$. There are however exceptional examples where it is easier to consider a \emph{finite transformation},
\begin{equation}
\boxed{\p^A\to\p'^A\equiv\DF^A(\p,\de\p,\dotsc,x,\eps)\;,}
\label{ch10:finitetransfo}
\end{equation}
which is a function of $\eps\in\R$ such that $\DF^A(\p,\de\p,\dotsc,x,0)=\p^A$. Should the action be left unchanged by~\eqref{ch10:finitetransfo} for all $\eps$, it is automatically guaranteed to be invariant under the infinitesimal version of the transformation,
\begin{multline}
\p'^A-\p^A=\at{\PD{\DF^A(\p,\de\p,\dotsc,x,\eps)}{\eps}}{\eps=0}\eps+\bigO(\eps^2)\\
\Rightarrow\quad
\df^A(\p,\de\p,\dotsc,x)=\at{\PD{\DF^A(\p,\de\p,\dotsc,x,\eps)}{\eps}}{\eps=0}\;.
\end{multline}

\begin{illustration}%
\label{ex10:Schr}%
Recall the Schr\"odinger theory, defined in~\refex{ex08:Schr} by the Lagrangian density
\begin{equation}
\La=\frac{\I\hbar}2(\psi^*\de_t\psi-\psi\de_t\psi^*)-\frac{\hbar^2}{2m}\grad\psi^*\cdot\grad\psi-V\psi^*\psi\;.
\label{ch10:LagSchr}
\end{equation}
Since every factor of $\psi$ in the Lagrangian is accompanied by $\psi^*$, it follows at once that the Lagrangian density (and hence also the action) is invariant under the finite transformation
\begin{equation}
\psi\to\psi'=\E^{\I\eps}\psi\;,\qquad
\psi^*\to\psi'^*=\E^{-\I\eps}\psi^*\;.
\end{equation}
In this case, checking invariance under the corresponding infinitesimal transformation, $\udelta\psi=\I\eps\psi$ and $\udelta\psi^*=-\I\eps\psi^*$, would be unnecessarily complicated.
\end{illustration}

%%%%%%%%%%%%%%%%%%%%%%%%%%%%%%%%%%%%%%%%%%%%%%%%%%%%%%%%%%%%

\subsection{Noether Theorem}
\label{subsec:Noethertheorem}

Having defined the concept of symmetry, we now utilize an ingeniously simple trick. Instead of the infinitesimal symmetry transformation~\eqref{ch10:inftytransfo}, let us shift the fields $\p^A$ by a localized transformation,
\begin{equation}
\udelta\p^A(x)\equiv\eps(x)\df^A(\p,\de\p,\dotsc,x)\;,
\label{ch10:localshift}
\end{equation}
which only differs from~\eqref{ch10:inftytransfo} by allowing the parameter $\eps(x)$ to depend on the coordinates in an arbitrary manner. This will in general change the action. To first order in $\eps$, the shift of the action can be written schematically as
\begin{equation}
\udelta S=\int\D^D\!x\,\left(G\eps+G^\m\de_\m\eps+G^{\m\n}\de_\m\de_\n\eps+\dotsb\right)\;,
\label{ch10:localshiftS}
\end{equation}
where the tensors $G^{\m\n\dotsb}$ are some local functions of the fields, their derivatives and the spacetime coordinates. Choosing $\eps$ to actually be constant reduces~\eqref{ch10:localshift} to a symmetry of the action. Then~\eqref{ch10:localshiftS} should vanish, which requires that $\smash{\int\D^D\!x\,G=0}$. This is only possible if $G$ is a mere surface term, that is $G=\de_\m\tilde G^\m$ for some vector function $\smash{\tilde G^\m}$. Using integration by parts as appropriate, the shift of the action~\eqref{ch10:localshiftS} under coordinate-dependent $\eps(x)$ can then be brought to the form\footnote{I do not indicate the domain of integration, and ignore whatever surface terms might arise from integration by parts. This is because the validity of the ensuing local conservation law ultimately relies only on the bulk EoM. That itself is likewise unaffected by boundary terms, which may at most require a particular boundary condition for the fields.}
\begin{equation}
\boxed{\udelta S\simeq\int\D^D\!x\,J^\m\de_\m\eps}\qquad
\text{(definition of $J^\m$)}\;,
\label{ch10:GellMannLevy}
\end{equation}
where $J^\m$ is a local vector function of the fields, their derivatives, and the coordinates.

It now takes just a little extra effort to see what all this has to do with conservation laws. Suppose we choose the fields $\p^A$ as functions on the spacetime so that they satisfy the EoM of the theory. For the sake of brevity, I will often use the particle physics terminology and refer to such fields as \emph{on-shell}. Then, by construction, the shift $\udelta S$ should vanish for \emph{any} choice of $\udelta\p^A$. In the terminology of~\chaptername~\ref{chap:mathintro}, the on-shell condition requires vanishing of the functional derivative $\udelta S/\udelta\p^A$, which is equivalent to the vanishing of the Gateaux derivative $D_FS[\p]$ in all ``directions'' $F^A$. Integrating~\eqref{ch10:GellMannLevy} by parts, we find that $\smash{\int\D^D\!x\,\eps\de_\m J^\m}$ should vanish for any choice of $\eps(x)$ provided the fields $\smash{\p^A(x)}$ are on-shell. This is only possible if
\begin{equation}
\boxed{\de_\m J^\m=0}\qquad
\text{(on-shell)}\;.
\label{ch10:Noethertheorem}
\end{equation}

\begin{watchout}%
Let us reexamine carefully what we have done here. We started by assuming that the action $S[\p]$ possesses a continuous symmetry. This led us to~\eqref{ch10:GellMannLevy}, which is required to hold for any $\p^A$ and $\udelta\p^A=\eps\df^A$. Equation~\eqref{ch10:GellMannLevy} defines the \emph{Noether current} $J^\m$. Then we specialized to fields that satisfy the EoM, that is, required $\udelta S=0$ just for on-shell $\p^A$ but for any choice of $\udelta\p^A$. This finally gave us~\eqref{ch10:Noethertheorem}: a continuous symmetry of the action ensures the existence of a local conservation law. This statement is known as the \emph{Noether theorem}. This existence theorem is complemented by an explicit algorithm for construction of the Noether current via~\eqref{ch10:GellMannLevy}. The Noether current is guaranteed by~\eqref{ch10:Noethertheorem} to be conserved for on-shell fields.
\end{watchout}

\begin{illustration}%
\label{ex10:KG2}%
To follow up on \refex{ex10:KG}, let us calculate the shift of the action of the free massless scalar field under $\udelta\p(x)=\eps(x)$,
\begin{equation}
\udelta S=\int\D^D\!x\,\left(\de_0\p\de_0\eps-\grad\phi\cdot\grad\eps\right)=-\int\D^D\!x\,\de^\m\p\de_\m\eps\;.
\end{equation}
Using~\eqref{ch10:GellMannLevy}, we readily identify the Noether current,
\begin{equation}
J^0=\de_0\p\text{ and }
\vec J=-\grad\p\;,\qquad\text{or }
J^\m=-\de^\m\p\;.
\end{equation}
In this case, the conservation of the Noether current is equivalent to the EoM, that is the KG equation, $\de_\m J^\m=-\de_\m\de^\m\p=0$.
\end{illustration}

\begin{illustration}%
\label{ex10:SchrU(1)}%
For another example, we revisit the Schr\"odinger theory of~\refex{ex10:Schr}. Here we are interested in the shift of the action under the local infinitesimal transformation $\udelta\psi(x)=\I\eps(x)\psi(x)$ and $\udelta\psi^*=-\I\eps(x)\psi^*(x)$. Upon some manipulation, the shift of the action can be brought to the form required by~\eqref{ch10:GellMannLevy},
\begin{equation}
\udelta S=\int\D^D\!x\,\biggl[-\hbar\psi^*\psi\de_t\eps+\frac{\I\hbar^2}{2m}(\psi^*\grad\psi-\psi\grad\psi^*)\cdot\grad\eps\biggr]\;,
\end{equation}
which leads to the identification of the Noether charge density and current as
\begin{equation}
J^0=-\hbar\psi^*\psi\;,\qquad
\vec J=\frac{\I\hbar^2}{2m}(\psi^*\grad\psi-\psi\grad\psi^*)\;.
\label{ch10:currentSchr}
\end{equation}
This result agrees with what we observed in~\refex{ex10:consprobability}, except for the overall normalization of the Noether current $J^\m$, which carries an extra factor of $-\hbar$ here. The moral lesson is that if the current $J^\m$ satisfies the conservation law~\eqref{ch10:Noethertheorem}, then so does its multiple $\a J^\m$ for any $\a\in\R$. The ambiguity in the normalization of the Noether current cannot be fixed from the assumed symmetry of the action alone.
\end{illustration}

So far our only tool to identify the Noether current, whose existence is guaranteed by the Noether theorem, has been~\eqref{ch10:GellMannLevy}. However, computing the shift of the action under the localized transformation~\eqref{ch10:localshift} is not always the easiest approach. There is an alternative that is often more convenient. The price to pay is that, to keep things simple, we have to make stronger assumptions on the form of the Lagrangian density. Suppose that the latter only depends on the fields and their first derivatives and possibly also on the spacetime coordinates, i.e.~it is of the form $\La(\p,\de\p,x)$. Then the shift of the action can be computed explicitly using the chain rule,
\begin{equation}
\begin{split}
\udelta S&=\int\D^D\!x\,\biggl[\PD\La{\p^A}\eps\df^A+\PD\La{(\de_\m\p^A)}\de_\m(\eps\df^A)\biggr]\\
&=\int\D^D\!x\,\biggl\{\biggl[\PD\La{\p^A}\df^A+\PD{\La}{(\de_\m\p^A)}\de_\m\df^A\biggr]\eps+\biggl[\PD{\La}{(\de_\m\p^A)}\df^A\biggr]\de_\m\eps\biggr\}\;.
\end{split}
\label{ch10:Noetheraux}
\end{equation}
In order that this vanishes for constant $\eps$, so that the transformation~\eqref{ch10:inftytransfo} actually is a symmetry of the action, the expression in the first square brackets on the second line must be a surface term. Hence, there must be a local vector function $K^\m$ so that
\begin{equation}
\boxed{\PD\La{\p^A}\df^A+\PD{\La}{(\de_\m\p^A)}\de_\m\df^A\ifeq\de_\m K^\m\;.}
\label{ch10:invariancecondition}
\end{equation}
I will call this the \emph{invariance condition}. Inserting this in~\eqref{ch10:Noetheraux} and integrating by parts brings the shift of the action under the localized transformation~\eqref{ch10:localshift} into the form~\eqref{ch10:GellMannLevy}, whence we readily identify the Noether current,
\begin{equation}
\boxed{J^\m=\PD{\La}{(\de_\m\p^A)}\df^A-K^\m\;.}
\label{ch10:Noethercurrent}
\end{equation}
The invariance condition~\eqref{ch10:invariancecondition} together with the explicit expression~\eqref{ch10:Noethercurrent} for the Noether current will in practice be our main tool when dealing with symmetries and the corresponding conservation laws.

\begin{illustration}%
Let us illustrate the use of~\eqref{ch10:invariancecondition} and~\eqref{ch10:Noethercurrent} on the examples of a free massless relativistic scalar field and the Schr\"odinger field, respectively. In the former case, it is obvious that the Lagrangian density~\eqref{ch10:LagKGmassless} satisfies~\eqref{ch10:invariancecondition} trivially with $F=1$ and $K^\m=0$. From~\eqref{ch10:Noethercurrent} we then get immediately $J^\m=\Pd\La{(\de_\m\p)}=-\de^\m\p$, in accord with what we found previously.

In case of the Schr\"odinger theory defined by the Lagrangian density~\eqref{ch10:LagSchr}, we find likewise that~\eqref{ch10:invariancecondition} is satisfied with $F_\psi=\I\psi$, $F_{\psi^*}=-\I\psi^*$ and $K^\m=0$. This leads via~\eqref{ch10:Noethercurrent} in turn to
\begin{equation}
J^\m=\I\psi\PD{\La}{(\de_\m\psi)}-\I\psi^*\PD{\La}{(\de_\m\psi^*)}\;.
\end{equation}
This is equivalent to~\eqref{ch10:currentSchr}.
\end{illustration}

It is possible to generalize the above argument to Lagrangian densities depending on arbitrarily high derivatives of fields. This however involves certain amount of combinatorics, and I therefore merely quote the final result. The generalization of the invariance condition~\eqref{ch10:invariancecondition} itself is straightforward,
\begin{equation}
\sum_{n=0}^\infty\PD{\La}{(\de_{\m_1}\dotsb\de_{\m_n}\p^A)}\de_{\m_1}\dotsb\de_{\m_n}F^A\ifeq\de_\m K^\m\;.
\end{equation}
What requires some work is to convert this into the Noether current,
\begin{equation}
J^\m=-K^\m+\sum_{n=1}^\infty\sum_{k=1}^n(-1)^{k+1}\biggl[\de_{\a_2}\dotsb\de_{\a_k}\PD{\La}{(\de_{\a_2}\dotsb\de_{\a_n}\de_\m\p^A)}\biggr]\de_{\a_{k+1}}\dotsb\de_{\a_n}F^A\;.
\label{ch10:Noethercurrentgeneral}
\end{equation}
The index $n$ counts the order of derivatives of fields in the Lagrangian density, and our previous expression~\eqref{ch10:Noethercurrent} corresponds to the special case of $n=1$.

%%%%%%%%%%%%%%%%%%%%%%%%%%%%%%%%%%%%%%%%%%%%%%%%%%%%%%%%%%%%

\subsection{Energy--Momentum Tensor}
\label{subsec:EMtensor}

Among the many examples of symmetries and the associated conservation laws, there is one that deserves a separate section for being present in any reasonable field theory. Indeed, the conservation laws for energy and momentum belong to the most fundamental laws of physics that are valid across physical disciplines. Let us see how they can be recovered from the general formalism of the Noether theorem.

We consider a translation of spacetime coordinates, $x^\m\to x'^\m=x^\m+\eps^\m$. In $D$ spacetime dimensions, this amounts to $D$ independent transformations, one for each coordinate. We therefore expect a set of $D$ Noether currents. Under the coordinate translation, the fields $\p^A$ transform via
\begin{equation}
\p'^A(x')=\p^A(x)\quad\Leftrightarrow\quad
\p'^A(x)=\p^A(x-\eps)\;.
\end{equation}
The first form of the field transformation simply says that the set of $n$ fields $\p^A$, interpreted as a hypersurface in a $(D+n)$-dimensional space, remains unchanged while the coordinate canvas under it shifts. This is usually called ``passive transformation.'' A simple Taylor expansion now gives $\p'^A(x)=\p^A(x)-\eps^\m\de_\m\p^A(x)+\bigO(\eps^2)$. This matches~\eqref{ch10:inftytransfo} with
\begin{equation}
\udelta\p^A=-\eps^\m\de_\m\p\;,\qquad
F^A_\m=-\de_\m\p^A\;.
\end{equation}

In general, the action of a theory will be invariant under spacetime translations if the Lagrangian density does not depend explicitly on the spacetime coordinates.\footnote{This rule of thumb can also be applied to a subset of coordinates. One expects conservation of energy for Lagrangian densities that do not depend on time, and conservation of the $i$-th component of momentum provided the Lagrangian density does not depend explicitly on $x^i$.} We will however restrict the discussion to Lagrangian densities that do not contain higher than first derivatives of fields, so that we can take advantage of the explicit expression~\eqref{ch10:Noethercurrent} for the Noether current. Picking a translation along a specific coordinate axis, defined by index $\n$, the invariance condition~\eqref{ch10:invariancecondition} requires that
\begin{equation}
\PD{\La}{\p^A}F^A_\n+\PD{\La}{(\de_\m\p^A)}\de_\m F^A_\n=-\PD{\La}{\p^A}\de_\n\p^A-\PD{\La}{(\de_\m\p^A)}\de_\m\de_\n\p^A\ifeq\de_\m K^\m_\n\;.
\end{equation}
This is solved by $K^\m_\n=-\d^\m_\n\La$. The corresponding set of Noether currents, called the \emph{energy--momentum} (EM) \emph{tensor} and denoted $T^\m_{\phantom\m\n}$, is given by~\eqref{ch10:Noethercurrent}. It is however conventional to define it with an opposite overall sign for reasons that will become clear shortly,
\begin{equation}
\boxed{T^\m_{\phantom\m\n}=\PD{\La}{(\de_\m\p^A)}\de_\n\p^A-\d^\m_\n\La\;.}
\label{ch10:EMtensor}
\end{equation}
In case of interest, here is a generalization of the EM tensor to Lagrangian densities depending on arbitrarily high field derivatives, which follows directly from~\eqref{ch10:Noethercurrentgeneral},
\begin{equation}
T^\m_{\phantom\m\n}=-\d^\m_\n\La+\sum_{n=1}^\infty\sum_{k=1}^n(-1)^{k+1}\biggl[\de_{\a_2}\dotsb\de_{\a_k}\PD{\La}{(\de_{\a_2}\dotsb\de_{\a_n}\de_\m\p^A)}\biggr]\de_{\a_{k+1}}\dotsb\de_{\a_n}\de_\n\p^A\;.
\end{equation}

Let us inspect in detail the physical content of the EM tensor. To that end, it is best to separate the discussion of invariance under translations of space and time.

\runinhead{Invariance Under Time Translations} In this case, we set in~\eqref{ch10:EMtensor} $\n=0$. This gives a single Noether current $T^\m_{\phantom\m0}$, whose temporal and spatial parts are
\begin{equation}
T^0_{\phantom00}=\PD{\La}{(\de_0\p^A)}\de_0\p^A-\La\;,\qquad
T^i_{\phantom i0}=\PD{\La}{(\de_i\p^A)}\de_0\p^A\;.
\end{equation}
The $T^0_{\phantom00}$ component is recognized as the Hamiltonian density $\Ha$; cf.~\eqref{ch08:Hamdef}. This verifies that the ensuing conservation law,
\begin{equation}
\de_\m T^\m_{\phantom\m0}=\de_0T^0_{\phantom00}+\de_iT^i_{\phantom i0}=0\;,
\label{ch10:consenergy}
\end{equation}
represents local energy conservation. Accordingly, $T^i_{\phantom i0}$ has the physical meaning of energy flux.

\runinhead{Invariance Under Space Translations} Here we get a set of $d$ Noether currents $T^\m_{\phantom\m j}$ with $j=1,\dotsc,d$. Their temporal and spatial parts are
\begin{equation}
T^0_{\phantom0j}=\PD{\La}{(\de_0\p^A)}\de_j\p^A\;,\qquad
T^i_{\phantom ij}=\PD{\La}{(\de_i\p^A)}\de_j\p^A-\d^i_j\La\;.
\end{equation}
These are interpreted respectively as the momentum density and the momentum flux. The corresponding conservation law,
\begin{equation}
\de_\m T^\m_{\phantom\m j}=\de_0T^0_{\phantom0j}+\de_iT^i_{\phantom ij}=0\;,
\label{ch10:consmomentum}
\end{equation}
represents local momentum conservation. The conservation laws for energy and momentum, \eqref{ch10:consenergy} and~\eqref{ch10:consmomentum}, can be written jointly in the very compact form
\begin{equation}
\de_\m T^\m_{\phantom\m\n}=0\;.
\end{equation}

\begin{illustration}%
\label{ex10:EMSchr}%
Let us illustrate the general construction of the EM tensor on the Schr\"odinger theory, defined by~\eqref{ch10:LagSchr}. A brief manipulation gives the energy and momentum currents,
\begin{align}
T^0_{\phantom00}&=\frac{\hbar^2}{2m}\grad\psi^*\cdot\grad\psi+V\psi^*\psi\;,\quad&
T^i_{\phantom i0}&=-\frac{\hbar^2}{2m}(\de^i\psi^*\de_0\psi+\de_0\psi^*\de^i\psi)\;,\\
\notag
T^0_{\phantom0j}&=\frac{\I\hbar}2(\psi^*\de_j\psi-\psi\de_j\psi^*)\;,&
T^i_{\phantom ij}&=-\frac{\hbar^2}{2m}(\de^i\psi^*\de_j\psi+\de_j\psi^*\de^i\psi)-\d^i_j\La\;.
\end{align}
The $\smash{T^0_{\phantom00}}$ component equals, unsurprisingly, the Hamiltonian density of the Schr\"odinger theory. The momentum density $\smash{T^0_{\phantom0j}}$ matches the probability current we found in~\refex{ex10:consprobability}, up to overall normalization. The precise relation, $\smash{T^0_{\phantom0i}=-mJ^i}$, turns out to be enforced by the Galilei invariance of the Schr\"odinger theory.
\end{illustration}

Most of our general arguments use systematically Greek spacetime indices without relying on a particular type of spacetime geometry. In relativistic theories, it is however common to use as $x^\m$ the Minkowski coordinates, and to raise the second index of the EM tensor $T^\m_{\phantom\m\n}$ using the Minkowski metric. This gives a rank-two contravariant tensor $T^{\m\n}$, which for Lagrangian densities independent of higher than first derivatives of fields takes the form
\begin{equation}
T^{\m\n}=\PD{\La}{(\de_\m\p^A)}\de^\n\p^A-g^{\m\n}\La\;.
\label{ch10:EMtensorcontravariant}
\end{equation}
The corresponding covariant tensor, $T_{\m\n}$, is the EM tensor that enters the Einstein equation for the gravitational field in the general theory of relativity. The physical interpretation of $T^{\m\n}$ of course follows that of $T^\m_{\phantom\m\n}$.

\begin{illustration}%
Consider a modification of the KG theory, endowed with a potential energy term for the scalar field,
\begin{equation}
\La=-\frac12(\de_\m\p)^2-\frac12m^2\p^2-V(\p)\;.
\end{equation}
It follows immediately from~\eqref{ch10:EMtensorcontravariant} that
\begin{equation}
T^{\m\n}=-\de^\m\p\de^\n\p-g^{\m\n}\La\;.
\end{equation}
This has a feature that one has no a priori reason to expect based on the derivation of the EM tensor from invariance under spacetime translation. Namely, our contravariant EM tensor $T^{\m\n}$ is symmetric in its two indices. This property turns out to be valid generally for relativistic theories of scalar fields. Its physical content is that, as a consequence of Lorentz invariance, there is a close relation between the energy flux and momentum density.
\end{illustration}

In addition to invariance under spacetime translations, there is another fundamental symmetry of space(time), namely that under rotations. This can be used to derive the corresponding set of Noether currents in close parallel with the derivation of the EM tensor. Here we will however take a shortcut. Given that $\smash{T^0_{\phantom0i}}$ represents momentum density, we can simply define the density of angular momentum by following the analogy with mechanics, that is as $\smash{M^0_{\phantom0ij}=x_iT^0_{\phantom0j}-x_jT^0_{\phantom0i}}$. This definition can then be extended to the tensor object $\smash{M^\m_{\phantom\m ij}}$, which includes both angular momentum density and angular momentum flux. We will not follow this line of reasoning in full generality, but instead focus on relativistic field theories. Here one defines the \emph{angular momentum tensor} $M^{\m\n\l}$ by
\begin{equation}
M^{\m\n\l}\equiv x^\n T^{\m\l}-x^\l T^{\m\n}\;.
\label{ch10:angularmomentum}
\end{equation}
A short calculation using the conservation of energy and momentum shows that, for on-shell fields,
\begin{equation}
\begin{split}
\de_\m M^{\m\n\l}&=\de_\m(x^\n T^{\m\l})-\de_\m(x^\l T^{\m\n})\\
&=g^\n_\m T^{\m\l}+\xcancel{x^\n\de_\m T^{\m\l}}-g^\l_\m T^{\m\n}-\xcancel{x^\l\de_\m T^{\m\n}}=T^{\n\l}-T^{\l\n}\;.
\end{split}
\end{equation}
The angular momentum tensor satisfies a local conservation law, but only if the EM tensor $T^{\m\n}$ is symmetric. That is, as pointed out above, in turn only true for theories of scalar fields. This makes physical sense. The $M^{\m\n\l}$ as defined by~\eqref{ch10:angularmomentum} only takes into account the orbital angular momentum of the fields, not their intrinsic angular momentum (spin). It is not surprising that the conservation of such orbital momentum is restricted to scalar fields, which correspond to spinless particles. It is possible to generalize the definition of the angular momentum tensor to particles of arbitrary spin, represented by tensor (or spinor) fields. That would however take us well beyond the scope of the course, and I will therefore not dwell on details.

%%%%%%%%%%%%%%%%%%%%%%%%%%%%%%%%%%%%%%%%%%%%%%%%%%%%%%%%%%%%

\section{Noether Theorem in Mechanics}
\label{sec:symmetryinmechanics}

Everything we have done in this~\chaptername{} so far applies to field theories in an arbitrary number of spatial dimensions $d$. This includes the extreme case of $d=0$, which is nothing but mechanics. We therefore get the Noether theorem for mechanics for free as a special of that for field theory. Below, I reiterate the main results in a notation adapted to that we used in the first half of the course for mechanics. The main difference is that the fields $\p^A(\vec x,t)$ now become the generalized coordinates $q^i(t)$. Derivatives with respect to the sole independent variable, that is the time $t$, are indicated with a dot. The concepts of Lagrangian and Lagrangian density coincide, and I will use the standard notation $L$. Finally, the ``Noether current'' $J^\m$ now has a single component $J^0$, which I will denote as $Q$ and call the \emph{Noether charge}.

The starting point is again the assumption that the action $S[q]$ is invariant under a continuous infinitesimal transformation, now in the form
\begin{equation}
\udelta q^i=\eps F^i(q,\dot q,\dotsc,t)\;,
\label{ch10:inftytransfomechanics}
\end{equation}
descending from~\eqref{ch10:inftytransfo}. The set of functions $F^i$ may depend on the generalized coordinates and their time derivatives, as well as explicitly on time. Next, we promote the constant parameter $\eps$ in~\eqref{ch10:inftytransfomechanics} to a time-dependent function, $\eps(t)$. Such a generalized transformation will in general modify the action. However, the assumed invariance of the action in the limit of constant $\eps$ ensures that its shift under the localized transformation can be written in a form analogous to~\eqref{ch10:GellMannLevy},
\begin{equation}
\udelta S\simeq\int\D t\,Q\dot\eps\;.
\label{ch10:GellMannLevymechanics}
\end{equation}
This can be used to extract the Noether charge $Q$. For $q^i(t)$ satisfying the EoM, that is along any trajectory of the system, $Q$ is then guaranteed to be conserved, as follows from~\eqref{ch10:Noethertheorem}.

The above algorithm is universally applicable to any mechanical system. However, in case the Lagrangian only depends on the generalized coordinates $q^i$ and their first derivatives, the generalized velocities $\dot q^i$, one can be more explicit. The statement of symmetry is then equivalent to the existence of a function $K$ that satisfies an invariance condition analogous to~\eqref{ch10:invariancecondition},
\begin{equation}
\boxed{\PD{L}{q^i}F^i+\PD{L}{\dot q^i}\dot F^i\ifeq\dot K\;.}
\label{ch10:invarianceconditionmechanics}
\end{equation}
The corresponding Noether charge follows as a special case of~\eqref{ch10:Noethercurrent},
\begin{equation}
\boxed{Q=\PD{L}{\dot q^i}F^i-K\;.}
\label{ch10:Noethercharge}
\end{equation}

\begin{watchout}%
The importance of~\eqref{ch10:Noethercharge} in combination with~\eqref{ch10:invarianceconditionmechanics} is that it bypasses the need to search for cyclic coordinates. As long as we know that the action of a mechanical system has a symmetry, we can now compute the corresponding Noether charge in any generalized coordinates we wish. This is not just a hypothetical advantage, as the following example shows.
\end{watchout}

\begin{illustration}%
\label{ex10:centralfield}%
The Lagrangian of a particle in a central field in three dimensions, $L=(1/2)m\dot{\vec r}^2-V(r)$, is invariant under proper rotations, $\vec r\to\vec r'=R\vec r$, where $R$ is an orthogonal matrix with unit determinant. The infinitesimal form of a proper rotation can be written as
\begin{equation}
\udelta\vec r=\vekt\a r\;,\qquad\text{or }
\udelta r^i=\ve^i_{\phantom ijk}\a^jr^k\;,
\end{equation}
where $\vec\a$ is an infinitesimal vector of parameters, specifying the axis and angle of the rotation. Thus, a rotation around the $j$-th Cartesian axis corresponds to the function $\smash{F^i_j=\ve^i_{\phantom ijk}r^k}$. From~\eqref{ch10:Noethercharge}, we then get immediately the vector of Noether charges,
\begin{equation}
Q_j=\PD{L}{\dot r^i}F^i_j=p_i\ve^i_{\phantom ijk}r^k=\ve_{jk}^{\phantom{jk}i}r^kp_i\;,
\end{equation}
where $\vec p=\Pd{L}{\dot{\vec r}}=m\dot{\vec r}$ is the vector of conjugate momentum. Our set of Noether charges corresponds to the vector $\vec Q=\vekt rp$, which is nothing but angular momentum. Note that we were able to compute the whole vector of angular momentum at once. The alternative would have been to rewrite the Lagrangian in spherical coordinates, and then notice that it does not depend on the azimuthal angle. This would guarantee the conservation of \emph{one} component of angular momentum. The conservation of the whole angular momentum vector would then follow by repeating the argument with different orientations of the coordinate axes. The derivation of angular momentum conservation based on the Noether theorem has the advantage that it preserves, and exploits, the full rotational symmetry of the system.
\end{illustration}

%%%%%%%%%%%%%%%%%%%%%%%%%%%%%%%%%%%%%%%%%%%%%%%%%%%%%%%%%%%%

\subsection{Correspondence of Symmetries and Conservation Laws}

While I have phrased the Noether theorem in the Lagrangian language, everything said applies to the Hamiltonian formalism as well. The identification of the Noether charge (or Noether current in field theory) relies only on the assumed invariance of the action, regardless of what variables the action as a functional may depend on. This is best illustrated by following up on~\refex{ex10:centralfield}.

\begin{illustration}%
The Hamiltonian form of the action of a particle in a central field is
\begin{equation}
S[\vec r,\vec p]=\int_{t_1}^{t_2}\D t\,[\dot{\vec r}\cdot\vec p-H(\vec r,\vec p)]\equiv\int_{t_1}^{t_2}\D t\,L_\mathrm{H}(\vec r,\vec p,\dot{\vec r},\dot{\vec p})\;,
\end{equation}
where the Hamiltonian is $H(\vec r,\vec p)=\vec p^2/(2m)+V(r)$. The action is invariant under a simultaneous rotation of the coordinate vector $\vec r$ and the momentum $\vec p$,
\begin{equation}
\udelta r^i\equiv\a^jF^i_{(\vec r)j}=\a^j(\ve^i_{\phantom{i}jk}r^k)\;,\qquad
\udelta p_i\equiv\a^jF_{(\vec p)ij}=\a^j(\ve_{ij}^{\phantom{ij}k}p_k)\;.
\end{equation}
When applied to the \emph{Hamiltonian Lagrangian} $L_\mathrm{H}$, the invariance condition~\eqref{ch10:invarianceconditionmechanics}  is manifestly satisfied with $K=0$, and~\eqref{ch10:Noethercharge} then gives
\begin{equation}
Q_j=\PD{L_\mathrm{H}}{\dot r^i}F^i_{(\vec r)j}+\PD{L_\mathrm{H}}{\dot p_i}F_{(\vec p)ij}=p_iF^i_{(\vec r)j}=\ve_{jk}^{\phantom{jk}i}r^kp_i\;.
\end{equation}
This is the same result we found before starting from the Lagrangian form of action.
\end{illustration}

One important feature that is particular to the Hamiltonian formalism is that one can reverse the line of argument and reconstruct the symmetry from the known Noether charge. This is something we already saw in Sect.~\ref{subsec:Hamsymcons}, including~\refex{ex06:angularmomentum} that showed how spatial rotations are generated by angular momentum. Now we have all the tools needed to tie up the loose ends.

Adapting the language of Sect.~\ref{sec:geomHam} to our current purposes, any function $Q$ on the phase space generates a transformation of the canonical variables $\x^i$, defined by
\begin{equation}
\udelta\x^i\equiv\eps\{\x^i,Q\}\quad\Leftrightarrow\quad
F^i(\x)=\{\x^i,Q\}=\O^{ij}\PD Q{\x^j}\;,
\label{ch10:symptransfo}
\end{equation}
where $\O^{ij}$ is the matrix inverse of the symplectic form of the system, $\O_{ij}$. Now the general form of action in the Hamiltonian formalism is
\begin{equation}
S[\x]=\int_{t_1}^{t_2}\D t\,[\o_i(\x)\dot\x^i-H(\x,t)]\equiv\int_{t_1}^{t_2}\D t\,L_\mathrm{H}(\x,\dot\x,t)\;,
\end{equation}
where $\o_i(\x)$ is the symplectic potential. Let us evaluate the shift of the Hamiltonian Lagrangian $L_\mathrm{H}$ under~\eqref{ch10:symptransfo}. This requires a series of steps, and I simplify the algebraic expressions by using the shorthand notation $\de_i\equiv\Pd{}{\x^i}$,
\begin{align}
\notag
(1/\eps)\udelta L_\mathrm{H}&=[(\de_j\o_i)\O^{jk}\de_kQ]\dot\x^i+\o_i\de_t(\O^{ij}\de_j Q)-(\de_iH)\O^{ij}\de_jQ\\
&=(\O_{ji}+\de_i\o_j)\O^{jk}(\de_kQ)\dot\x^i+\o_i\de_t(\O^{ij}\de_j Q)-\{H,Q\}\\
\notag
&=-\O_{ij}\O^{jk}(\de_kQ)\dot\x^i+(\dot\x^k\de_k\o_i)\O^{ij}\de_jQ+\o_i\de_t(\O^{ij}\de_j Q)-\{H,Q\}\\
\notag
&=\de_t(-Q+\o_i\O^{ij}\de_jQ)-\{H,Q\}\;.
\end{align}
In the intermediate steps, I used the relation between the symplectic form and the symplectic potential, $\O_{ij}=\de_i\o_j-\de_j\o_i$, and the chain rule. Also, in order to be able to put $Q$ under a time derivative in the last step, I assumed that it does not depend on time explicitly, but only implicitly through the canonical variables $\x^i$. The same assumption applies to the symplectic potential and in turn the symplectic form.

At this point, we assume that the generator $Q$ of the transformation~\eqref{ch10:symptransfo} has a vanishing Poisson bracket with the Hamiltonian, that is, it is conserved. Then, we see that the Hamiltonian Lagrangian satisfies the invariance condition~\eqref{ch10:invarianceconditionmechanics} with
\begin{equation}
K=-Q+\o_i\O^{ij}\PD Q{\x^j}\;.
\end{equation}
This means that the transformation~\eqref{ch10:symptransfo} of the canonical variables, induced by $Q$, indeed is a symmetry of the action as we expected. From~\eqref{ch10:Noethercharge}, we then finally obtain the corresponding Noether charge,
\begin{equation}
\PD{L_\mathrm{H}}{\dot\x^i}F^i-K=\o_iF^i-K=\o_i\O^{ij}\PD{Q}{\x^j}-K=Q\;.
\end{equation}
This closes the circle, showing that the relation between symmetries of the action and conserved charges is an equivalence. The Noether theorem proven here and the idea of flow generated by conserved charges on the phase space, developed in Sect.~\ref{sec:geomHam}, constitute the two mutually inverse directions of the equivalence.

%%%%%%%%%%%%%%%%%%%%%%%%%%%%%%%%%%%%%%%%%%%%%%%%%%%%%%%%%%%%

\subsection{Moral Lessons}

Much of the apparatus of analytical mechanics and field theory, including its basic applications, was developed in the period between mid-18\textsuperscript{th} and mid-19\textsuperscript{th} century. It is therefore quite surprising that the relationship between symmetries and conservation laws was not appreciated until 1915 when Emmy Noether proved her eponymous theorem. While formulated as a theorem in variational calculus, this breakthrough was clearly motivated by physical applications.

In spite of most of classical physics having been already established by 1915, Noether's theorem constitutes one of the deepest and most profound physical laws. It ensures that some conservation laws must be realized in any sensible physical theory, whatever form the latter might take. Thus, for instance, it is empirically established that the laws of nature do not change with time and location. This leads to the universal laws of conservation of energy and momentum. The development of field theory, first classical and then quantum, has however borne countless examples of further applications of Noether's theorem. In fact, the theorem has helped to make symmetry a central concept on which all of our current theories of fundamental physics are based. More than a hundred years after Noether's discovery, the study of symmetries in physics keeps providing new insights, and the importance of symmetries for our understanding of physics keeps growing.

%%%%%%%%%%%%%%%%%%%%%%%%%%%%%%%%%%%%%%%%%%%%%%%%%%%%%%%%%%%%

\section*{\probsec}
\addcontentsline{toc}{section}{\probsec}

\begin{prob}
\label{pr10:phi4real}
A relativistic theory of $n$ real scalar fields $\phi^A$ is defined by
\begin{equation}
\La=-\frac12\d_{AB}\de_\m\p^A\de^\m\p^B-\frac12m^2\d_{AB}\p^A\p^B-\frac\l{16}(\d_{AB}\p^A\p^B)^2\;,
\end{equation}
where $\l$ is a positive constant. Consider a linear transformation of the fields, $\p^A=P^A_{\phantom AB}\p'^B$, where $P$ is a nonsingular matrix. What kind of condition does $P$ have to satisfy so that this transformation is a symmetry of the theory? An infinitesimal version of the symmetry transformation can be represented by $P\approx\un+\eps Q$, where $\eps$ is a parameter. What kind of condition does $Q$ have to satisfy? For given $Q$, find the corresponding Noether current.
\end{prob} 

\begin{prob}
\label{pr10:phi4complex}
A theory of a single \emph{complex} scalar field $\p$ is defined by the Lagrangian density
\begin{equation}
\La=-\de_\m\p^*\de^\m\p-m^2\p^*\p-\frac\l4(\p^*\p)^2\;.
\end{equation}
This theory is invariant under the phase transformation $\p\to\E^{\I\eps}\p$. Find the corresponding Noether current. By writing the complex field in terms of two real fields, $\p=(\phi^1+\I\phi^2)/\sqrt2$, check that your result for the Noether current agrees with the special case $n=2$ of~\refpr{pr10:phi4real}.
\end{prob}

\begin{prob}
\label{pr10:SchrGalilei}
Recall the Lagrangian density of the free Schr\"odinger theory,
\begin{equation}
\La=\frac{\I\hbar}2(\psi^*\de_t\psi-\psi\de_t\psi^*)-\frac{\hbar^2}{2m}\vec\nabla\psi^*\cdot\vec\nabla\psi\;.
\end{equation}
Show that the action of the theory is invariant under the Galilei transformation
\begin{equation}
\vec x'=\vec x+\vec vt\;,\qquad
\psi'(\vec x',t)=\exp\left[\frac\I\hbar m\left(\skal vx+\frac12\vec v^2t\right)\right]\psi(\vec x,t)\;,
\end{equation}
where $\vec v$ is the velocity parameter of the Galilei boost. Find the corresponding set of Noether currents. Hint: you should find, after some amount of manipulation, that the Lagrangian density is shifted by the boost by a surface term, $\udelta\La=-\divg(\vec vt\La)$.
\end{prob}

\begin{prob}
\label{pr10:Schrgauged}
Suppose that the otherwise free Schr\"odinger field is exposed to fixed, coordinate-dependent external fields $A_t$ and $\vec A$. Its Lagrangian density is thus modified to
\begin{equation}
\La=\frac{\I\hbar}2(\psi^*D_t\psi-\psi D_t\psi^*)-\frac{\hbar^2}{2m}\vec D\psi^*\cdot\vec D\psi\;,
\end{equation}
where $D_t\psi\equiv(\de_t-\I A_t)\psi$ and $\vec D\psi\equiv(\vec\nabla-\I\vec A)\psi$ are the \emph{covariant derivatives} of the field. Show that the Lagrangian density is invariant under a simultaneous transformation of $\psi$ and the external fields,
\begin{equation}
\psi\to\E^{\I\eps}\psi\;,\qquad
A_t\to A_t+\de_t\eps\;,\qquad
\vec A\to\vec A+\vec\nabla\eps\;,
\label{ch10:gaugetransfo}
\end{equation}
where $\eps$ is allowed to depend on coordinates in an arbitrary manner. Check that the Noether current of the free Schr\"odinger theory without external fields can be recovered by taking a derivative of $\La$ with respect to $A_t$ and $\vec A$ and then setting both fields to zero. Why do you think this trick to find the Noether current works?
\end{prob}

\begin{prob}
\label{pr10:idealfluid}
Irrotational motion of an ideal fluid of local density $\vr(\vec x,t)$ and Eulerian velocity $\vec v(\vec x,t)$ is described by the action
\begin{equation}
S[\vr,\vec v,\p]=\int\D^D\!x\,\biggl\{\frac12\vr\vec v^2-u(\vr)+\p\biggl[\PD\vr t+\vec\nabla\cdot(\vr\vec v)\biggr]\biggr\}\;,
\end{equation}
where $u(\vr)$ is the internal energy density of the fluid and $\p(\vec x,t)$ a Lagrange multiplier for the constraint that the continuity equation~\eqref{ch10:continuityeq} be satisfied. Find the EoM for all the three indicated variables and give their physical interpretation. Then use the EoM for $\vec v$ to eliminate the latter in favor of $\vr,\phi$, thus getting a new action $\tilde S[\vr,\phi]$. Use this new action to find the EM tensor of the fluid. \end{prob}

\begin{prob}
\label{pr10:topological}
Consider a theory of a real scalar field $\p$ in one spatial dimension. In such a theory, one can define the current $J^\m\equiv\ve^{\m\n}\de_\n\p$, 
where $\ve^{\m\n}$ is the two-dimensional Levi-Civita symbol. Show that the current $J^\m$ is conserved \emph{identically}, that is independently of the EoM for $\p$. Suppose now that the theory has a Lagrangian density
\begin{equation}
\La=-\frac12(\de_\m\p)^2-V(\p)\;,
\end{equation}
where the potential function $V(\p)$ has a discrete set of minima $\p_i$ at which $V(\p_i)=0$. Find possible values of the integral charge, $Q=\int_{-\infty}^{+\infty}J^0\,\D x^1$, for field configurations whose total energy is finite. 
\end{prob}

\begin{prob}
\label{pr10:mattercurrent}
Another example of an identically conserved current is the matter current~\eqref{ch09:mattercurrent} in a continuous medium. Show that in this case, the equation $\de_\m J^\m=0$ expresses the conservation of mass of the medium. Why do you think the current is conserved identically rather than on-shell as a consequence of a symmetry of the action?
\end{prob}

\begin{prob}
\label{pr10:helix}
An infinitely long thin wire is bent into a helix, parameterized in cylindrical coordinates $(\vr,\vp,z)$ by $z=\a\vp$ and $\vr=R$, where $\a$ and $R$ are positive constants. The wire is charged with a constant linear density of electric charge $\l$. Argue that the dynamics of a charged particle, moving in the electrostatic field of the wire, possesses a constant of motion (in addition to energy). Find an expression for this constant of motion in terms of the particle's position and momentum.
\end{prob}

\begin{prob}
\label{pr10:LaplaceRungeLenz}
Recall that a particle of mass $m$, moving in the central field $V(r)=-k/r$ with constant $k$, has another vector constant of motion in addition to angular momentum $\vec J$, namely the Laplace--Runge--Lenz vector
\begin{equation}
\vec R\equiv\vekt pJ-km\frac{\vec r}r\;.
\end{equation}
Find the Lagrangian symmetry for which $\vec R$ is the corresponding Noether charge. You may want to use the results of~\refpr{pr06:PoissonJrP}.
\end{prob}
\chapter{Application: Fluid Dynamics}
\label{chap:fluid}

\keywords{Continuity equation, pressure, Euler equation, hydrodynamic sound, potential (irrotational) flow, velocity potential, Bernoulli equation, gravity waves, shear and bulk viscosity, Navier--Stokes equation.}

%%%%%%%%%%%%%%%%%%%%%%%%%%%%%%%%%%%%%%%%%%%%%%%%%%%%%%%%%%%%

\noindent Equipped with a thorough understanding of conservation laws, we shall now turn our attention to an important sector of classical field theory: fluid dynamics. This is a vast area of science with a broad range of applications from very large scales (cosmology, astrophysics, atmospheric and ocean science) to smaller but still macroscopic scales (including aerodynamics, laboratory-scale hydrodynamics, hydraulics and biomechanics) and finally to microscopic scales (as for instance in the quark--gluon plasma created in collisions of relativistic heavy ions). The content of this~\chaptername{} should be understood as a brief primer on the theoretical foundations underlying the above applications. To goal is to give you the background necessary for further independent study of more advanced (and typically also more specialized) literature. Along the way, I will however also take the opportunity to illustrate some further general aspects of classical field theory. These include the discussion of equivalence of conservation laws and the dynamical \emph{equation of motion} (EoM), the generalization of Noether's theorem to approximate symmetries, and the use of a variational principle to generate a natural boundary condition for the fields.

The core of the~\chaptername{} constitutes Sect.~\ref{sec:idealhydro}, which addresses various aspects of nonrelativistic ideal (perfect) fluids. Selected applications however require going beyond the nonrelativistic limit, or beyond the assumption of absence of dissipation. In Sect.~\ref{sec:hydromisc}, I outline how one can extend the basic formalism to include the effects of viscosity, or to account for intrinsically relativistic dynamics of fluids. Those of you who are interested will find much more information (including references for further reading) in the \href{http://www.damtp.cam.ac.uk/user/tong/fluids.html}{Lectures on Fluid Mechanics} by David Tong.

%%%%%%%%%%%%%%%%%%%%%%%%%%%%%%%%%%%%%%%%%%%%%%%%%%%%%%%%%%%%

\section{Lagrangian Field Theory of Ideal Hydrodynamics}
\label{sec:idealhydro}

Let us initiate the exploration of fluids by asking ourselves what we expect to find. Recall that a system that is in thermodynamic equilibrium has no memory of initial conditions or of the process via which it reached equilibrium. The equilibrium can be viewed as statistically the most random state consistent with given constraints. During the relaxation towards equilibrium, all information about the detailed microstate of the system is dissipated except for the data that are protected by conservation laws, such as the total mass, energy, or momentum of the system. The equilibrium state is thus fully determined by whatever conserved charges the system might possess.

Hydrodynamics, or fluid dynamics, provides an effective macroscopic description of systems near thermodynamic equilibrium. It is assumed that the microscopic processes responsible for equilibration are fast enough so that the system may be expected to be in equilibrium \emph{locally}. The intensive thermodynamic variables characterizing equilibrium may however slowly vary with both time and position. In this sense, the collective hydrodynamic behavior of matter should be completely fixed by a combination of equilibrium thermodynamics and local conservation laws.

The hydrodynamic description of matter may take different forms, depending on what symmetries (and accordingly what conservation laws) one assumes. For most of this \chaptername, we will consider ordinary (nonrelativistic) matter, where the relevant conservation laws are those for Newtonian mass, energy and momentum. We will rely heavily on the language for description of continuous media, developed in \chaptername~\ref{chap:elasticity}. Here the local conservation of mass is automatically incorporated through the identical conservation of the matter current~\eqref{ch09:mattercurrent} (see also the related~\refpr{pr10:mattercurrent}). As a consequence, the basic dynamical equations for fluids should descend from local conservation laws for energy and momentum alone.

We learned in \chaptername~\ref{chap:symmetries} that the conservation of Noether currents is generally a consequence of the dynamical EoM. However, here I claim that in case of fluids, the EoM should be \emph{equivalent} to local conservation of energy and momentum. This is an interesting twist in the story of symmetries and conservation laws that deserves a more detailed discussion.

%%%%%%%%%%%%%%%%%%%%%%%%%%%%%%%%%%%%%%%%%%%%%%%%%%%%%%%%%%%%

\subsection{Equation of Motion from Conservation Laws}
\label{subsec:EoMfromconservation}

Let us temporarily put aside the description of continuous media in terms of material coordinates, and consider a generic theory of a set of fields $\p^A$. We initially assume that the Lagrangian density does not contain higher than first derivatives of the fields. Combining the expression~\eqref{ch10:Noethercurrent} with the invariance condition~\eqref{ch10:invariancecondition}, we find
\begin{equation}
\de_\m J^\m=\biggl[\de_\m\PD{\La}{(\de_\m\p^A)}-\PD{\La}{\p^A}\biggr]F^A\;.
\label{ch11:currentEoMeq1}
\end{equation}
This identity should be satisfied for any field configuration $\p^A(x)$, not just for those solving the EoM. It is easy to generalize to Lagrangians containing arbitrarily high derivatives of the fields. Indeed, the shift of the action $S[\p]$ under the localized transformation~\eqref{ch10:localshift} can be expressed semi-explicitly using the chain rule as
\begin{equation}
\udelta S=\int\D^D\!x\,\frac{\udelta S}{\udelta\p^A(x)}\udelta\p^A(x)=\int\D^D\!x\,\frac{\udelta S}{\udelta\p^A(x)}\eps(x)F^A(\p,\de\p,\dotsc,x)\;.
\end{equation}
This should match~\eqref{ch10:GellMannLevy} for any choice of $\eps(x)$, which is only possible if
\begin{equation}
\boxed{\de_\m J^\m=-\frac{\udelta S}{\udelta\p^A}F^A\;.}
\label{ch11:currentEoMeq2}
\end{equation}
For Lagrangian densities only depending on the fields and their first derivatives, this obviously reduces to~\eqref{ch11:currentEoMeq1}.

Equation~\eqref{ch11:currentEoMeq2} holds by construction for an arbitrary field configuration. For fields satisfying the EoM (which we refer to as \emph{on-shell}), the right-hand side vanishes and we recover current conservation as expected. It is however interesting to ask whether the implication can be reversed. Suppose that we are given a field configuration $\p^A$ for which $\de_\m J^\m=0$ with the current $J^\m$ defined by~\eqref{ch10:GellMannLevy}. Is this sufficient to conclude that $\udelta S/\udelta\p^A=0$, that is, the fields are on-shell? If yes, then the current conservation indeed would be equivalent to the EoM. Mathematically speaking, it appears that a possible proof of the equivalence relies on our ability to ``divide'' \eqref{ch11:currentEoMeq2} by the factor $F^A$.

\begin{illustration}%
Imagine a theory of a field $\p$ whose action is invariant under a constant shift of the field, $\p\to\p+\eps$, that is $F=1$. Then~\eqref{ch11:currentEoMeq2} tells us that $\de_\m J^\m=-\udelta S/\udelta\p$, whatever the Noether current actually looks like. In this case, the equivalence of current conservation and the EoM is manifest. We already observed this equivalence in the special case of a free massless Klein--Gordon field in~\refex{ex10:KG2}.
\end{illustration}

\begin{illustration}%
What is really of interest to us is the case where the fields $\p^A$ with $A=1,\dotsc,d$ are the body (material) coordinates of a continuous medium. We focus on the Noether currents associated with invariance under spacetime translations, spanning the \emph{energy--momentum} (EM) \emph{tensor} $T^\m_{\phantom\m\n}$. The spacetime translation $x^\m\to x^\m+\eps^\m$ is equivalent to $\udelta\p^A=-\eps^\m\de_\m\p^A\equiv\eps^\m F^A_\m$, so that $F^A_\m=-\de_\m\p^A$. From~\eqref{ch11:currentEoMeq2} we then find that\footnote{The right-hand side of the equation seems to have a wrong sign, yet it is correct. This is because when defining in Sect.~\ref{subsec:EMtensor} the EM tensor via~\eqref{ch10:GellMannLevy}, we chose an opposite overall sign.}
\begin{equation}
\de_\m T^\m_{\phantom\m\n}=-\frac{\udelta S}{\udelta\p^A}\de_\n\p^A\;.
\end{equation}
Recall now that the body coordinates $\p^A$ are by construction related to the spatial coordinates $x^i$ by a smooth invertible map, $x^i\to\p^A(\vec x,t)$. This requires that the Jacobian matrix of the transformation,  $M^A_{\phantom Ai}\equiv\de_i\p^A$, be invertible. As a consequence, the EoM for the medium can be recovered from local momentum conservation alone,
\begin{equation}
\frac{\udelta S}{\udelta\p^A}=-(M^{-1})^i_{\phantom iA}\de_\m T^\m_{\phantom\m i}\;.
\end{equation}
We did not actually need to use the conservation of energy. Instead, energy conservation follows from the EoM, which itself follows from conservation of momentum. This is consistent with our experience from Newtonian mechanics, where conservation of mechanical energy is a secondary result, the primary law being the momentum balance as expressed by Newton's second law of motion.
\end{illustration}

\begin{watchout}
The above example demonstrates that if we manage to formulate hydrodynamics as a field theory with the material coordinates $\p^A$ as the basic variables, its dynamics will be identical to local conservation of momentum (the local conservation of energy being a consequence). What, then, is the point of constructing the action for the fields $\p^A$ at all? The answer is that this makes it easier to find the EM tensor, and to relate it to the equilibrium thermodynamics of the fluid. It is however good to keep in mind that this is a convenience, not a necessity. Including the effects of viscosity, or any other form of dissipation, is notoriously difficult in Lagrangian field theory. In Sect.~\ref{subsec:viscousfluids}, we will do so by modifying directly the momentum conservation law.
\end{watchout}

%%%%%%%%%%%%%%%%%%%%%%%%%%%%%%%%%%%%%%%%%%%%%%%%%%%%%%%%%%%%

\subsection{Energy--Momentum Tensor of Nonrelativistic Fluids}

Having clarified what we expect to find, the next step is to construct the Lagrangian density for a fluid, and then find the corresponding EM tensor. The kinetic energy density of a fluid is the same as that of a solid, and is given by~\eqref{ch09:solidLagkin}. The potential energy due to internal forces in the fluid is of course different. Ideal fluids do not experience any shear stress, only pressure due to overall compression or expansion. As such, the potential energy density can only depend on the local mass density~\eqref{ch09:matterdensity} of the fluid. The latter is in turn fixed by the determinant of the matrix $\smash{\Xi^{AB}\equiv\d^{ij}M^A_{\phantom Ai}M^B_{\phantom Bj}=\grad\p^A\cdot\grad\p^B}$. It follows that the Lagrangian density of a nonrelativistic ideal fluid can be written as
\begin{equation}
\La=\frac{\vr_0}2\sqrt{\abs\Xi}(\Xi^{-1})_{AB}\de_t\p^A\de_t\p^B-\Va(\abs\Xi)\;,
\label{ch11:LagfluidNR}
\end{equation}
where I used the shorthand notation $\abs\Xi\equiv\det\Xi$. The function $\Va(\abs\Xi)$ can be interpreted as the internal energy density of the local equilibrium. As such, it is fixed by the equilibrium \emph{equation of state} of the fluid, which may be evaluated separately before one has to deal with the macroscopic flow of the fluid.

\begin{watchout}%
From thermodynamics, we are used to the fact that the equation of state of a single chemical compound in equilibrium is determined by two independent thermodynamic variables. One can choose these for instance as the mass density and entropy. On the other hand, we expect that our Lagrangian field theory~\eqref{ch11:LagfluidNR} will only be able to capture macroscopic processes in the fluid that are nondissipative, and thus do not involve local entropy production. The fact that the entropy of each element of the fluid remains constant means that the thermodynamic state of the element only depends on a sole independent variable, which can be chosen as the mass density. This is reflected by the notation used in~\eqref{ch11:LagfluidNR}.
\end{watchout}

In order to derive the EM tensor corresponding to the Lagrangian density~\eqref{ch11:LagfluidNR}, we need to be able to evaluate the derivatives of $\La$ with respect to $\de_\m\p^A$. The dependence on time derivatives is simple. On the other hand, spatial derivatives of $\p^A$ only enter the Lagrangian density through the inverse of $\Xi=MM^T$, and through $\sqrt{\abs\Xi}=\abs M$. To make further progress, we use the following algebraic identities, whose proof is relegated to an exercise (see~\refpr{pr11:auxiliaryidentities}),
\begin{equation}
\PD{\abs{M}}{M^A_{\phantom Ai}}=\abs M(M^{-1})^i_{\phantom iA}\;,\qquad
\PD{\smash{(M^{-1})^j_{\phantom jB}}}{M^A_{\phantom Ai}}=-(M^{-1})^i_{\phantom iB}(M^{-1})^j_{\phantom jA}\;.
\label{ch11:auxiliaryidentities}
\end{equation}
Using these gives, upon some manipulation, the useful auxiliary formula
\begin{equation}
\PD{\La}{M^A_{\phantom Ai}}=\frac{\vr_0}2\sqrt{\abs\Xi}(M^{-1})^i_{\phantom iA}\vec v^2-\vr_0\sqrt{\abs\Xi}(M^{-1})^j_{\phantom jA}v^iv_j-2\abs\Xi\Va'(\abs\Xi)(M^{-1})^i_{\phantom iA}\;,
\end{equation}
where the prime on $\Va$ indicates a derivative with respect to the displayed argument, and $v^i\equiv-(M^{-1})^i_{\phantom iA}\de_t\p^A$ is the local (Eulerian) velocity of the fluid; cf.~Sect.~\ref{subsec:mattercurrent}, in particular~\eqref{ch09:Eulervelocity}. It now takes little effort to assemble all the elements of the EM tensor~\eqref{ch10:EMtensor} as functions of the local velocity $\vec v$ and mass density $\vr=\vr_0\sqrt{\abs\Xi}$,
\begin{equation}
\begin{aligned}
T^0_{\phantom00}&=\frac12\vr\vec v^2+\Va(\abs\Xi)\;,\qquad&
T^i_{\phantom i0}&=\biggl[\frac12\vr\vec v^2+2\abs\Xi\Va'(\abs\Xi)\biggr]v^i\;,\\
T^0_{\phantom0j}&=-\vr v_j\;,\qquad&
T^i_{\phantom ij}&=\bigl[\Va(\abs\Xi)-2\abs\Xi\Va'(\abs\Xi)\bigr]\d^i_j-\vr v^iv_j\;.
\end{aligned}
\end{equation}
The combination $\smash{\Va(\abs\Xi)-2\abs\Xi\Va'(\abs\Xi)}$ in $\smash{T^i_{\phantom ij}}$ has a simple physical interpretation. Namely, using that $\smash{\vr=\vr_0\sqrt{\abs\Xi}}$, we can write $\smash{2\abs\Xi\Va'(\abs\Xi)=\vr\Va'(\vr)}$. Consequently, 
\begin{equation}
2\abs\Xi\Va'(\abs\Xi)-\Va(\abs\Xi)=\vr\Va'(\vr)-\Va(\vr)=-\OD{[\Va(\vr)/\vr]}{(1/\vr)}\equiv P
\label{ch11:PfromV}
\end{equation}
is minus the derivative of specific internal energy (internal energy per unit mass) with respect to the specific volume. At fixed entropy, this is nothing but the thermodynamic pressure as a consequence of the first law of thermodynamics. This leads us to the final result for the EM tensor of the fluid, expressed as a function of the mass density, pressure and the local velocity,
\begin{equation}
\boxed{\begin{aligned}
T^0_{\phantom00}&=\frac12\vr\vec v^2+\Va(\vr)\;,\qquad&
T^i_{\phantom i0}&=\biggl[\frac12\vr\vec v^2+\Va(\vr)+P\biggr]v^i\;,\\
T^0_{\phantom0j}&=-\vr v_j\;,\qquad&
T^i_{\phantom ij}&=-P\d^i_j-\vr v^iv_j\;.
\end{aligned}}
\label{ch11:EMtensorfluid}
\end{equation}

We already know that the EoM of our Lagrangian field theory of ideal fluids is equivalent to local conservation of energy and momentum. The energy conservation takes the form
\begin{equation}
\de_tT^0_{\phantom00}+\de_iT^i_{\phantom i0}=\de_tT^0_{\phantom00}+\divg[(T^0_{\phantom{0}0}+P)\vec v]=0\;,
\end{equation}
where $T^0_{\phantom00}$ has the interpretation as the total energy density of the moving fluid. What is more interesting is the local conservation of momentum, which by our discussion in Sect.~\ref{subsec:EoMfromconservation} should be the actual fundamental dynamical law for fluids. Using the continuity equation~\eqref{ch10:continuityeq}, reproduced for convenience here,
\begin{equation}
\PD\vr t+\divg(\vr\vec v)=0\;,
\label{ch11:continuityeq}
\end{equation}
to rewrite $\de_t(\vr\vec v)=\vr\de_t\vec v-\vec v\divg(\vr\vec v)$, the momentum conservation condition $\de_\m T^\m_{\phantom\m j}=0$ can be cast as
\begin{equation}
\boxed{\PD{\vec v}t+(\skal v\nabla)\vec v=-\frac{\grad P}\vr\;.}
\label{ch11:Eulereq}
\end{equation}
This is known as the \emph{Euler equation}. It has a very simple Newtonian interpretation. Namely, the left-hand side of~\eqref{ch11:Eulereq} is the acceleration of the fluid element, obtained from the velocity $\vec v(\vec x(t),t)$ by taking a total time derivative along the trajectory $\vec x(t)$ of the element. The right-hand side of~\eqref{ch11:Eulereq} is then the net force due to pressure per unit mass. The continuity equation~\eqref{ch11:continuityeq} and Euler equation~\eqref{ch11:Eulereq} together constitute the fundamental dynamical laws of ideal hydrodynamics.

\begin{illustration}%
To illustrate the utility of the continuity and Euler equations, consider an equilibrium solution where the density $\vr_0$ and pressure $P_0$ are constant and there is no motion, $\vec v_0=\vec0$. As the next step, we introduce the deviations of all the dynamical variables from equilibrium via
\begin{equation}
\vr\equiv\vr_0+\udelta\vr\;,\qquad
P\equiv P+\udelta P\;,\qquad
\vec v\equiv\vec v_0+\udelta\vec v\;,
\end{equation}
and linearize the equations~\eqref{ch11:continuityeq} and~\eqref{ch11:Eulereq} in $\udelta\vr$, $\udelta P$ and $\udelta\vec v$,
\begin{equation}
\PD{\udelta\vr}{t}+\vr_0\divg\udelta\vec v\approx0\;,\qquad
\PD{\udelta\vec v}{t}\approx-\frac{\grad\udelta P}{\vr_0}\;.
\label{ch11:soundaux}
\end{equation}
Taking the time derivative of the linearized continuity equation and the divergence of the linearized Euler equation, we can eliminate the velocity variable $\udelta\vec v$, thus getting a single linear second-order \emph{partial differential equation} (PDE),
\begin{equation}
\frac{\de^2\udelta\vr}{\de t^2}-\grad^2\udelta P\approx0\;.
\end{equation}
We have already assumed that the equation of state of the fluid is given by a function of a single variable, such as the density $\vr$. We can therefore write $\udelta P=P'(\vr)\udelta\vr\approx P'(\vr_0)\udelta\vr$, where the derivative $P'(\vr)=\Od P\vr$ is taken at constant entropy. This is physically sensible, since the mechanical oscillations of pressure and density around their equilibrium values are usually too fast for any heat transfer to take place. Having used the thermodynamic equation of state, we have now arrived at a single second-order PDE for a single variable $\udelta\vr$,
\begin{equation}
\frac{\de^2\udelta\vr}{\de t^2}-P'(\vr_0)\grad^2\udelta\vr\approx0\;.
\end{equation}
This is an ordinary wave equation that has plane-wave solutions, $\udelta\vr\propto\exp[\I(\skal kx-\o t)]$, whose frequency $\o$ is related to the wave vector $\vec k$ by $\smash{\o^2=P'(\vr_0)\vec k^2}$. This is hydrodynamic sound, with $\smash{\sqrt{P'(\vr_0)}}$ playing the role of speed of propagation. Note that by~\eqref{ch11:soundaux}, the local velocity of the fluid in the plane wave is proportional to $\vec k$. This means that sound waves in fluids are longitudinal. There are no transverse sound waves in fluids due to the absence of shear stress.

For a concrete illustration, consider an ideal gas with the equation of state $P=\vr RT$, where $R$ is the specific gas constant and $T$ the thermodynamic temperature. In a reversible adiabatic process, the ratio $P/\vr^\k$ with $\k$ being the adiabatic index, remains constant. Then, $P'(\vr)=\k P/\vr$ so that the speed of sound reads $\sqrt{\k RT}$.
\end{illustration}

%%%%%%%%%%%%%%%%%%%%%%%%%%%%%%%%%%%%%%%%%%%%%%%%%%%%%%%%%%%%

\subsection{Effect of External Fields}

In many applications, the fluid is also subject to other forces than those of pressure. Of particular relevance are conservative external forces that act on the whole volume of the fluid, such as gravity. With this important example in mind, I shall denote the external force per unit mass of the fluid as $\vec g$. The assumption that the external field be conservative translates into the existence of a potential $\psi$ such that $\vec g=-\grad\psi$. The effect of the external field is then taken into account by adding to the Lagrangian density the term
\begin{equation}
\La_{\vec g}=-\vr\psi=-\vr_0\sqrt{\abs\Xi}\psi\;.
\label{ch11:Lagextfield}
\end{equation}
In order to see how the external field affects the motion of the fluid, let us return to the basics. We know from before that the EoM for the fluid should be equivalent to the conservation law for its EM tensor. The latter in turn arises from invariance of the action of the fluid under translations in space and time. However, the external field makes the Lagrangian density depend explicitly on the coordinates. This breaks the assumed translation invariance, and so can be expected to lead to violations of the conservation of energy and momentum.

To see concretely how the conservation laws are violated, recall the trick that we used in Sect.~\ref{subsec:Noethertheorem} to prove the Noether theorem. We again subject the action to the localized translation $x^\m\to x'^\m=x^\m+\eps^\m(x)$. However, with the external field present, we no longer expect the shift of the action to take the form~\eqref{ch10:GellMannLevy}. Instead, we find
\begin{equation}
\udelta S=\int\D^D\!x\,(-T^\m_{\phantom\m\n}\de_\m\eps^\n+\eps^\n\de_\n^\mathrm{exp}\La)\;.
\label{ch11:GellMannexplicit}
\end{equation}
The EM tensor (including the conventional minus sign) can still be extracted from the shift of the action induced by the implicit dependence on coordinates through the fields $\p^A$. The new term in~\eqref{ch11:GellMannexplicit} accounts for the explicit dependence of the Lagrangian density on the coordinates. The notation reminds us that the derivative $\de_\n^\mathrm{exp}$ only acts on the coordinates that appear in the Lagrangian density explicitly through the external field, not on the dynamical fields $\p^A$. For on-shell fields, the shift of the action should still vanish for any choice of $\eps(x)$. Upon integration by parts, this gives a generalized conservation law in the form
\begin{equation}
\boxed{\de_\m T^\mu_{\phantom\m\n}=-\de_\n^\mathrm{exp}\La\;.}
\label{ch11:EMconsapproximate}
\end{equation}
The same result can be obtained by evaluating the divergence of the EM tensor as given by~\eqref{ch10:EMtensor} and using the EoM, under the assumption that the Lagrangian density does not contain higher than first derivatives of the fields.

We now have what we need to continue the discussion of the effect of external fields on fluid dynamics. All that needs to be done is to find the modification of the EM tensor due to~\eqref{ch11:Lagextfield}, and the corresponding generalized conservation law. We assume that the external field is static (time-independent). The effect of the external field then turns out to be twofold:
\begin{itemize}
\item The energy current is modified to $\smash{T^\m_{\phantom\m0}+\psi J^\m}$, where $\smash{T^\m_{\phantom\m\n}}$ denotes the original EM tensor~\eqref{ch11:EMtensorfluid} and $\smash{J^\m=(\vr,\vr\vec v)}$ is the matter current defined by~\eqref{ch09:mattercurrent}. Since the external field is assumed to be static, local energy conservation survives, and can be expressed alternatively as
\begin{equation}
\de_\m(T^\m_{\phantom\m0}+\psi J^\m)=0\quad\Leftrightarrow\quad
\de_\m T^\m_{\phantom\m0}=-J^\m\de_\m\psi\;.
\end{equation}
The latter form has a particularly clear physical interpretation, whereby the right-hand side, equal to $\skal Jg$, gives the power of the external field per unit volume, that is the rate at which the field supplies energy to the fluid.
\item The momentum current $T^\m_{\phantom\m j}$ remains unchanged. However, the local momentum conservation law is modified in accord with~\eqref{ch11:EMconsapproximate},
\begin{equation}
\de_\m T^\m_{\phantom\m j}=\vr\de_j\psi=-\vr g_j\;.
\end{equation}
This in turn modifies the Euler equation~\eqref{ch11:Eulereq} to
\begin{equation}
\boxed{\PD{\vec v}t+(\skal v\nabla)\vec v=-\frac{\grad P}\vr+\vec g\;.}
\label{ch11:Eulereqmod}
\end{equation}
This could have been expected from the Newtonian interpretation of the Euler equation. It is however reassuring to see the same result automatically reproduced by our field-theoretic setup.
\end{itemize}

%%%%%%%%%%%%%%%%%%%%%%%%%%%%%%%%%%%%%%%%%%%%%%%%%%%%%%%%%%%%

\subsection{Potential Flow}

The continuity equation~\eqref{ch11:continuityeq} and Euler equation~\eqref{ch11:Eulereqmod} together constitute a set of coupled nonlinear PDEs for the density $\vr$ and velocity $\vec v$. (The remaining variable in the Euler equation, namely pressure, is assumed to be a known function of density.) Solving such a set of equations is a hard problem. It can however be significantly simplified for certain special cases of flow. Suppose that the motion of the fluid does not involve any circulation. Technically, this amounts to the vanishing of the integral $\oint\vec v\cdot\D\vec x$ over any closed loop in space. Locally, it requires that the velocity field $\vec v$ has a vanishing curl, or equivalently $\de_iv_j=\de_jv_i$. Under this assumption, $\vec v$ is a conservative field, and there must be a \emph{velocity potential} $\p(\vec x,t)$ such that
\begin{equation}
\vec v=\grad\p\;.
\label{ch11:velocitypotential}
\end{equation}
A fluid flow without circulation is accordingly called potential, or \emph{irrotational}.

\begin{illustration}%
A potential flow becomes particularly simple when the fluid is incompressible. In this limit, the density $\vr$ is a mere constant and the continuity equation~\eqref{ch11:continuityeq} reduces to $\grad^2\p=0$. The velocity potential can then be found using the techniques developed in mathematics for solving the Laplace equation. In the special case of a flow that is effectively two-dimensional, one can use insight from complex calculus. This is often employed to analyze the flow of incompressible fluids around obstacles. See~\refpr{pr11:conformal} for an example.
\end{illustration}

Let us now see how the assumption that the flow of the fluid is potential simplifies the Euler equation. First, the nonlinear term on the left-hand side of~\eqref{ch11:Eulereqmod} can be rewritten as
\begin{equation}
[(\skal v\nabla)\vec v]^i=v^j\de_jv^i=v^j\de^iv_j=\frac12\de^i\vec v^2=\frac12(\grad\vec v^2)^i\;.
\end{equation}
Furthermore, \eqref{ch11:PfromV} implies that
\begin{equation}
\grad P=\cancel{(\grad\vr)\Va'(\vr)}+\vr\Va''(\vr)\grad\vr-\cancel{\Va'(\vr)\grad\vr}\quad\Rightarrow\quad
\frac{\grad P}\vr=\grad[\Va'(\vr)]\;.
\end{equation}
Putting everything together, the Euler equation~\eqref{ch11:Eulereqmod} including the effect of the external field $\vec g$ can be cast as
\begin{equation}
\grad\biggl[\PD\p t+\frac12\vec v^2+\Va'(\vr)+\psi\biggr]=\vec0\;.
\end{equation}
The argument of the gradient must therefore not depend on the spatial coordinates. Any dependence on time alone can however be absorbed into a redefinition of the velocity potential $\p$ that does not affect the velocity, $\vec v=\grad\p$. We conclude that, up to the choice of velocity potential, the Euler equation for potential flow amounts to
\begin{equation}
\boxed{\frac12\vec v^2+\Va'(\vr)+\psi+\PD\p t=0\;.}
\label{ch11:Bernoullieq}
\end{equation}
This is a fairly general version of the \emph{Bernoulli equation}, valid for an unsteady but irrotational flow.

The basic dynamical equations describing potential flow are now the continuity equation~\eqref{ch11:continuityeq} and the Bernoulli equation~\eqref{ch11:Bernoullieq}, along with the relation~\eqref{ch11:velocitypotential} between the velocity and its potential. Remarkably, these three PDEs can be recovered from a simplified variational principle where $\vr,\vec v,\p$ rather than the material coordinates $\p^A$ are the dynamical variables. The action is defined by
\begin{equation}
\boxed{S[\vr,\vec v,\p]\equiv\int\D^D\!x\,\biggl\{\frac12\vr\vec v^2-\Va(\vr)-\vr\psi+\p\biggl[\PD\vr t+\divg(\vr\vec v)\biggr]\biggr\}\;.}
\label{ch11:actionLin}
\end{equation}
Except for the presence of the potential $\psi$ and a different notation for the internal energy, we already saw this in~\refpr{pr10:idealfluid}. The continuity equation is now implemented as a constraint with the velocity potential $\p$ playing the role of the Lagrange multiplier. The EoM for $\vec v$ descending from~\eqref{ch11:actionLin} is equivalent to~\eqref{ch11:velocitypotential}. Inserting $\vec v=\grad\p$ back into the action and integrating by parts, we obtain a modified action for $\vr$ and $\p$ alone,
\begin{equation}
\tilde S[\vr,\p]\simeq-\int\D^D\!x\,\biggl[\frac12\vr(\grad\p)^2+\Va(\vr)+\vr\psi+\vr\PD\p t\biggr]\;.
\label{ch11:actionLin2}
\end{equation}

\begin{watchout}%
In spite of leading to the same dynamical equations, the action~\eqref{ch11:actionLin} cannot be obtained from the Lagrangian~\eqref{ch11:LagfluidNR} by some clever change of variables. This is because in the description based on the material coordinates $\p^A$, mass conservation is already automatically built in. Equivalently, the matter current~\eqref{ch09:mattercurrent} is identically conserved without having to impose the EoM for the fields. On the other hand, in the formulation~\eqref{ch11:actionLin}, the continuity equation encoding local mass conservation is one of the Lagrange equations of motion. In other words, when computing the functional derivative of~\eqref{ch11:actionLin}, one also allows for variations of density and velocity that violate mass conservation. The two variational principles for ideal fluids are therefore fundamentally different. Those interested will find a discussion of the connection between these two formulations of hydrodynamics (in the special case of incompressible fluids) in Sect.~2.5 of the above-mentioned \href{http://www.damtp.cam.ac.uk/user/tong/fluids.html}{Lectures on Fluid Mechanics}.
\end{watchout}

\begin{figure}[t]
\sidecaption[t]
\includegraphics[width=2.9in]{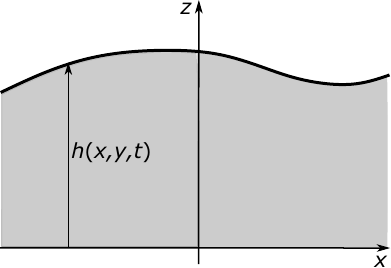}
\caption{Two-dimensional cross-section of the geometry involved in the discussion of gravity waves on the free surface of water.}
\label{fig11:gravitywave}
\end{figure}

I shall now exemplify the use of our new variational principle by applying it to a remarkable everyday phenomenon: waves on the free surface of water, also called \emph{gravity waves}. The discussion below is motivated by and closely follows Sect.~1.4 of~\cite{Stone2009a}. Apart from the interesting physics involved, the calculation also illustrates how a variational principle can be utilized to generate a natural boundary condition for the fields.

Having in mind water in a uniform gravitational field, we will make the underlying assumptions that we are dealing with an incompressible liquid of density $\vr_0$, and that the potential of the external field is $\psi=-\skal gx$ with constant~$\vec g$. We imagine a three-dimensional geometry whose two-dimensional cross-section is shown in Fig.~\ref{fig11:gravitywave}. The Cartesian $z$-axis is oriented vertically against the direction of the gravitational field $\vec g$. The other two axes, $x$ and $y$, are horizontal. We assume that the fluid is bounded from below by a flat surface (seabed) at $z=0$. The free surface at height $h(x,y,t)$ is in contact with an outside atmosphere whose pressure is fixed and constant. With these assumptions, the action~\eqref{ch11:actionLin2} reduces to one for $\p$ and $h$,
\begin{equation}
\tilde{\tilde S}[\p,h]=\int_\O\D^4\!x\,\biggl[-\frac12\vr_0(\grad\p)^2+\vr_0\skal gx-\vr_0\PD\p t\biggr]\;.
\end{equation}
I have dropped the constant contribution of internal energy, $\Va(\vr_0)$. The integration domain $\O$ is the infinite slab defined by $0\leq z\leq h(x,y,t)$ with $t,x,y\in\R$. What makes this problem nontrivial is that the height function $h(x,y,t)$ is a priori unknown. It therefore has to be treated as an additional variable of the variational problem.

Let us now carefully evaluate the shift of the action under a small change of the velocity potential $\udelta\p$ and of the free surface $\udelta h$. Keeping track of surface terms generated by partial integration, we find
\begin{align}
\notag
\udelta \tilde{\tilde S}={}&\int_\O\D^4\!x\,\biggl[-\vr_0\grad\p\cdot\grad\udelta\p-\vr_0\PD{\udelta\p}{t}\biggr]\\
\notag
&+\at{\int\D t\,\D x\,\D y\,\udelta h\biggl[-\frac12\vr_0(\grad\p)^2+\vr_0\skal gx-\vr_0\PD\p t\biggr]}{z=h(x,y,t)}\\
={}&\int_\O\D^4\!x\,\vr_0\udelta\p\grad^2\p-\int_{\de\O}\vr_0\udelta\p(1,\de_x\p,\de_y\p,\de_z\p)\cdot\D\vec S\\
\notag
&-\int\D t\,\D x\,\D y\,\udelta h\at{\biggl[\frac12\vr_0(\grad\p)^2+\vr_0 gz+\vr_0\de_t\p\biggr]}{z=h(x,y,t)}
\;.
\end{align}
Here $\D\vec S$ is an oriented area element on the boundary $\de\O$ of the integration domain. At the seabed, the area element is proportional to $(0,0,0,-1)$. At the free surface, the direction of the normal vector to the surface is most easily obtained by taking the gradient of the function defining the surface via $z-h(x,y,t)=0$, This gives
\begin{equation}
\D\vec S\propto(-\de_th,-\de_ xh,-\de_yh,1)\quad\text{at the free surface}\;.
\end{equation}
We now demand that the shift of the action vanishes for any choice of $\udelta\p$ and $\udelta h$. This gives first of all the EoM $\grad^2\p=0$ that should be satisfied in the bulk domain $\O$. However, it also enforces the following boundary conditions,
\begin{equation}
\begin{aligned}
\de_z\p&=0\qquad&&\text{at the seabed, }z=0\;,\\
\de_z\p-\de_th-\de_xh\de_x\p-\de_yh\de_y\p&=0\qquad&&\text{at the free surface, }z=h\;,\\
\smash{\tfrac12(\grad\p)^2}+gz+\de_t\p&=0\qquad&&\text{at the free surface, }z=h\;.
\end{aligned}
\label{ch11:freesurface}
\end{equation}
These boundary conditions together with the Laplace equation for $\p$ inside $\O$ define the mathematical problem that we next have to solve.

We will not even attempt a general solution of the problem. Instead, we will search for approximate plane-wave solutions corresponding to small-amplitude oscillations, so that we can linearize the equations wherever possible. Since the liquid is bounded in the vertical direction, we expect to find waves that propagate horizontally. We therefore use the ansatz
\begin{equation}
\p(\vec x,t)=\hat\p(z)\exp[\I(\vec k_\perp\cdot\vec x_\perp-\o t)]\;,
\label{ch11:gravitywave}
\end{equation}
where the subscript $\perp$ indicates two-dimensional vectors in the horizontal $xy$ plane. Inserting our ansatz in the Laplace equation, we find that the $z$-dependent amplitude $\hat\p(z)$ should satisfy the second order ordinary differential equation $\hat\p''(z)-\vec k_\perp^2\p(z)=0$. The boundary condition $\de_z\p=0$ at $z=0$ selects a unique solution up to an overall scale, $\hat\p(z)=\hat\p_0\cosh kz$, where $k\equiv\abs{\vec k_\perp}$.

So far we have not committed any approximation. However, the difficult, nonlinear part of the problem is hidden in the two boundary conditions to be satisfied at the free surface. Taking the time derivative of the last line in~\eqref{ch11:freesurface}, these boundary conditions can be combined to
\begin{equation}
\frac12\de_t(\grad\p)^2+g(\de_z\p-\grad_\perp h\cdot\grad_\perp\p)+\de^2_t\p=0\qquad\text{at }z=h(\vec x_\perp,t)\;.
\end{equation}
Now we use the assumption that the oscillations of the liquid have a small amplitude. This allows us to drop the $(\grad\p)^2$ and $\grad_\perp h\cdot\grad_\perp\p$ terms. The ensuing linearized boundary condition, $g\de_z\p+\de_t^2\p\approx0$, leads to an algebraic condition for the frequency $\o$ and wave vector $\vec k_\perp$ of the plane wave~\eqref{ch11:gravitywave},
\begin{equation}
\o^2=gk\tanh kh_0\;,
\label{ch11:gravitywavedisp}
\end{equation}
where $h_0$ is the position of the free surface (i.e.~the depth of the water) in equilibrium.

Let us consolidate our results. The exponential ansatz~\eqref{ch11:gravitywave} was convenient for the sake of finding a concrete solution. However, the velocity potential of the actual flow of the water must be real. We can take either the real or the imaginary part of~\eqref{ch11:gravitywave}; these are two linearly independent solutions, differing only by their overall phase. For the sake of illustration, I choose the imaginary part. The velocity potential and the corresponding velocity field of the wave are then
\begin{equation}
\begin{split}
\p(\vec x,t)&=\hat\p_0\sin(\vec k_\perp\cdot\vec x_\perp-\o t)\cosh kz\;,\\
\vec v_\perp(\vec x,t)&=\hat\p_0\vec k_\perp\cos(\vec k_\perp\cdot\vec x_\perp-\o t)\cosh kz\;,\\
v_z(\vec x,t)&=\hat\p_0k\sin(\vec k_\perp\cdot\vec x_\perp-\o t)\sinh kz\;.
\end{split}
\end{equation}
One immediate observation is that in the limit $kz\ll1$ (corresponding to distances from the seabed much shorter than the wavelength $\l=2\pi/k$), the transverse velocity $\vec v_\perp$ is much larger than the vertical velocity $v_z$. In this limit, therefore, the motion of the liquid is nearly horizontal. In the opposite limit $kz\gg1$ (distances from seabed much longer than the wavelength $\l=2\pi/k$), we have $\cosh kz\approx\sinh kz\approx\E^{kz}/2$. This implies $\vec v_\perp^2+v_z^2\approx\hat\p_0^2k^2\E^{2kz}/4$. In other words, the local motion of the liquid is nearly circular, with an amplitude that grows with distance from the seabed.

Before closing the discussion of gravity waves, I should stress the fact that the dispersion relation~\eqref{ch11:gravitywavedisp} is nonlinear in the wave number $k$. For such waves, the propagation of a localized wave packet is better characterized by the group velocity,
\begin{equation}
\OD\o k=\frac12\sqrt{\frac{g}{k\tanh kh_0}}\biggl(\tanh kh_0+\frac{kh_0}{\cosh^2kh_0}\biggr)\;.
\label{ch11:gravitywavegroup}
\end{equation}
In the limit $kh_0\ll1$ or equivalently $h_0\ll\l/(2\pi)$, we get $\Od\o k\to\sqrt{gh_0}$. This can be viewed either as waves on shallow water (small $h_0$), or waves of long wavelength (small $k$). In the opposite limit $kh_0\gg1$, or $h_0\gg\l/(2\pi)$, we find the asymptotic behavior $\Od\o k\approx(1/2)\sqrt{g/k}$. This regime is most naturally thought of as waves on deep water. Altogether, the group velocity~\eqref{ch11:gravitywavegroup} is a monotonically decreasing function of the wave number. This means that at fixed depth $h_0$, the long-wavelength component of the wave packet moves the fastest. This agrees with experience. When a heavy object such as a boat passes by on an otherwise still water surface, we observe the long-wavelength waves arrive first, and the short-wavelength oscillations last.

%%%%%%%%%%%%%%%%%%%%%%%%%%%%%%%%%%%%%%%%%%%%%%%%%%%%%%%%%%%%

\section{Miscellaneous Extensions}
\label{sec:hydromisc}

Having developed the basics of hydrodynamics for nonrelativistic ideal fluids, I will now briefly outline some of its extensions. First, we will have a look at how internal friction (viscosity) affects the EoM for the fluid. This is relevant for many practical applications. Then, I will sketch how the entire setup changes if we know from the outset that the dynamics of the fluid is relativistic. This finds applications for instance in relativistic astrophysics and cosmology.

%%%%%%%%%%%%%%%%%%%%%%%%%%%%%%%%%%%%%%%%%%%%%%%%%%%%%%%%%%%%

\subsection{Viscous Fluids}
\label{subsec:viscousfluids}

We will not be able to include the effects of viscosity within the Lagrangian description of the fluid. For once, it pays off to adopt a Newtonian approach. Here the generalized local momentum conservation takes the form of Newton's second law,
\begin{equation}
\de_t(\vr v^i)=\de_jT^{ji}+\vr g^i\;,
\label{ch11:EoMNewton}
\end{equation}
where, according to~\eqref{ch11:EMtensorfluid},
\begin{equation}
T^{ij}=-P\d^{ij}-\vr v^iv^j\;.
\label{ch11:stressideal}
\end{equation}
The spatial part of the EM tensor, $T^{ij}$, also called \emph{stress tensor} in the terminology we used previously for elastic solids, captures momentum transfer in the fluid due to internal forces. While the left-hand side of~\eqref{ch11:EoMNewton} and the $\vr g^i$ term on the right-hand side are dictated by the basic laws of mechanics and Newtonian gravity, it is not given a priori that the stress tensor must take the form~\eqref{ch11:stressideal}. Indeed, we previously deduced this from the assumption that the internal forces in the fluid can be represented by a potential energy that only depends on the density of the fluid.

Let us now try to be agnostic and understand better what form the stress tensor \emph{can} take. Representing the internal forces within the fluid, it should depend somehow on the kinematical state of the fluid. Unless the fluid is an extremely diluted gas in which the mean free path of the molecules is comparable to or longer than their average distance, it is reasonable to assume that the internal forces have a short range. The stress tensor should then depend only on the local density $\vr$ and velocity $\vec v$ of the fluid. Also, in a fluid in uniform motion, there should be no other forces but isotropic pressure. This is because in such a situation, we can always find an \emph{inertial reference frame} (IRF) in which the fluid as a whole is at rest. The pressure itself is taken care of by the $-P\d^{ij}$ contribution to the stress tensor. The additional term in~\eqref{ch11:stressideal}, $-\vr v^iv^j$, is purely inertial and is fixed by Galilei invariance of nonrelativistic mechanics; this is indirectly touched upon in~\refpr{pr11:EulerGalileiinvar}. Any other contributions to the stress tensor must contain derivatives of the velocity field.

Inspired by the above observation, we can actually view~\eqref{ch11:stressideal} as the leading contribution to a systematic expansion of $T^{ij}$ in derivatives of the velocity field. The derivative contributions to the stress tensor are sensitive to differential motion of the fluid, and are therefore naturally expected to capture the effects of viscosity. (There cannot be any friction forces in a fluid moving as a whole with uniform velocity.) Symbolically, the ideal-fluid stress tensor~\eqref{ch11:stressideal} is thus generalized to
\begin{equation}
T^{ij}=-P\d^{ij}-\vr v^iv^j+\ve^{ij}\;,
\label{ch11:stressviscous}
\end{equation}
where $\ve^{ij}$ is the so-called \emph{viscous stress tensor}. Isotropy dictates that at the lowest order in derivatives and velocity, it takes the generic form
\begin{equation}
\ve^{ij}=\a\biggl(\PD{v^i}{x_j}+\PD{v^j}{x_i}\biggr)+\b\d^{ij}\PD{v^k}{x^k}+\g\biggl(\PD{v^i}{x_j}-\PD{v^j}{x_i}\biggr)\;,
\end{equation}
where $\a,\b,\g$ are material constants that may depend on the density. The $\g$-term is only relevant for fluids in which a significant transfer of angular momentum takes place. This is the case for instance for ferromagnetic fluids where the fluid elements carry spin. I will however assume that the $\g$ parameter is negligible, which is the case for ``ordinary'' fluids one encounters in nature. The viscous stress tensor is then parameterized by two material constants in the conventional form
\begin{equation}
\ve^{ij}=\eta\biggl(\PD{v^i}{x_j}+\PD{v^j}{x_i}-\frac2d\d^{ij}\PD{v^k}{x^k}\biggr)+\zeta\d^{ij}\PD{v^k}{x^k}\;.
\end{equation}
The second term measures the rate of local changes in the fluid volume. The parameter $\zeta$ is called \emph{bulk viscosity}. The first term is insensitive to volume changes, and instead measures the rate of shear strain of the fluid. The coefficient $\eta$ is accordingly called \emph{shear viscosity}. Both viscosity coefficients are required to be non-negative by the second law of thermodynamics.

It is now a matter of simple algebra to implement the effects of viscosity in the EoM by combining~\eqref{ch11:EoMNewton} with~\eqref{ch11:stressviscous}. The result can be written in a vector form, generalizing~\eqref{ch11:Eulereqmod},
\begin{equation}
\boxed{\vr\biggl[\PD{\vec v}t+(\skal v\nabla)\vec v\biggr]=-\grad P+\vr\vec g+\eta\grad^2\vec v+\biggl[\zeta+\biggl(1-\frac2d\biggr)\eta\biggr]\grad(\divg\vec v)\;.}
\label{ch11:NavierStokes}
\end{equation}
This is the celebrated \emph{Navier--Stokes equation}. Understanding the global existence and smoothness properties of solutions to this equation is considered one of the most important unresolved problems in mathematics with immense practical implications.

%%%%%%%%%%%%%%%%%%%%%%%%%%%%%%%%%%%%%%%%%%%%%%%%%%%%%%%%%%%%

\subsection{Relativistic Ideal Hydrodynamics}

Before we wrap up our discussion of fluids, I will briefly sketch the changes necessary in case the dynamics of the fluid is intrinsically relativistic. To keep things simple, we will return to the assumption that the fluid is ideal and thus can be described in the language of Lagrangian field theory. In this framework, the transition from nonrelativistic fluids to relativistic ones is simple. Namely, Lorentz invariance requires that the derivatives of the material coordinates $\p^A$ only enter the Lagrangian density through the combination
\begin{equation}
\Xi^{AB}\equiv g^{\m\n}\de_\m\p^A\de_\n\p^B\;.
\label{ch11:Xidefrel}
\end{equation}
This is the relativistic version of~\eqref{ch09:Xidef}. The rest of the argument is the same as for nonrelativistic fluids. Namely, one can always switch to an IRF, also called the \emph{local rest frame}, in which a chosen fluid element at a given moment does not move. In this frame, $\Xi^{AB}$ reduces to $\grad\p^A\cdot\grad\p^B$, and we know from before that the Lagrangian density can only depend on the determinant of $\Xi$, $\abs\Xi$. Altogether, the action for the relativistic ideal fluid then takes the Lorentz-invariant form
\begin{equation}
S[\p]=\int\D^D\!x\,\Fa(\abs\Xi)\;,
\label{ch11:actionrelfluid}
\end{equation}
where $\Fa$ is some as yet unspecified function that should be related to the pressure and internal energy of the fluid.

\begin{watchout}
Note that unlike in the nonrelativistic case, the ``kinetic'' and ``potential'' parts of the Lagrangian density are subsumed into a single term. This is a consequence of Lorentz invariance. Should the form of the Lagrangian density be independent of the choice of IRF, it is not possible to separate the energy of the motion of the fluid from the energy due to its static compression or expansion. It is nevertheless possible to show that in the limit of motion much slower than the speed of light, the relativistic action~\eqref{ch11:actionrelfluid} reproduces our previous result~\eqref{ch11:LagfluidNR}. The internal energy density $\Va$ turns out to be determined by $-\Fa$ upon subtracting the relativistic rest energy of the fluid. I will not work out the details at this stage, since it will prove easier to understand the nonrelativistic limit at the level of the EM tensor.
\end{watchout}

The next step is to work out the EM tensor descending from the action~\eqref{ch11:actionrelfluid}. Using the auxiliary identity
\begin{equation}
\PD{\abs\Xi}{(\de_\m\p^A)}=2\abs\Xi(\Xi^{-1})_{AB}\de^\m\p^B\;,
\end{equation}
which is proven analogously to the first relation in~\eqref{ch11:auxiliaryidentities}, one arrives quickly at
\begin{equation}
T^{\m\n}=2\abs\Xi\Fa'(\abs\Xi)(\Xi^{-1})_{AB}\de^\m\p^A\de^\n\p^B-g^{\m\n}\Fa(\abs\Xi)\;.
\label{ch11:Tmunuaux}
\end{equation}
This can be further simplified by recalling the matter current~\eqref{ch09:mattercurrent}. The current is proportional to $(1,v^i)$, and thus also to the local four-velocity $u^\m$ of the fluid. From the property $J^\m\de_\m\p^A=0$ we then infer that all the gradients $\de_\m\p^A$ are orthogonal (in the Minkowski sense) to the four-velocity. Using this fact and the definition~\eqref{ch11:Xidefrel} of $\Xi^{AB}$, we find that the tensor $G^{\m\n}\equiv(\Xi^{-1})_{AB}\de^\m\p^A\de^\n\p^B-u^\m u^\n$ satisfies
\begin{equation}
G^{\m\n}u_\n=u^\m\;,\qquad
G^{\m\n}\de_\n\p^A=\de^\m\p^A\;.
\end{equation}
This means that, in fact, $G^{\m\n}=g^{\m\n}$. As a consequence, the EM tensor~\eqref{ch11:Tmunuaux} of the relativistic ideal fluid can be expressed as
\begin{equation}
T^{\m\n}=2\abs\Xi\Fa'(\abs\Xi)(g^{\m\n}+u^\m u^\n)-g^{\m\n}\Fa(\abs\Xi)\;.
\label{ch11:EMtensorrel}
\end{equation}

We have now eliminated the explicit dependence of the EM tensor on the material coordinates. What remains is to relate the function $\Fa(\abs\Xi)$ to the thermodynamic properties of the fluid. To that end, we switch to the local rest frame. In this IRF, $T^{\m\n}$ should be strictly diagonal as there is no momentum, no energy flux, and no shear stress in a fluid at rest. Moreover, isotropy requires that all the spatial diagonal elements of $T^{\m\n}$ are equal. The EM tensor in the local rest frame therefore takes the simple form $T^{\m\n}=\operatorname{diag}(-\Va,-P,-P,\dotsc)$. Here $\Va$ is the internal energy density and $P$ the pressure, both measured in the local rest frame. These are \emph{scalar} fields, parameterizing the local thermodynamics of the fluid. Using the four-velocity $u^\m$ which itself reduces to $(1,\vec0)$ in the local rest frame, the EM tensor can be expressed in terms of the pressure and internal energy density in a covariant form as
\begin{equation}
\boxed{T^{\m\n}=-Pg^{\m\n}-(\Va+P)u^\m u^\n\;.}
\label{ch11:EMtensorrel2}
\end{equation}

\begin{watchout}
The EM tensor for a relativistic fluid appears in the literature in various forms depending on the chosen set of conventions, and it is fair to admit that our expression~\eqref{ch11:EMtensorrel2} is not the most usual one. Its form is largely dictated by our sign convention for the Minkowski metric. Moreover, we defined the EM tensor so that it is $\smash{T^0_{\phantom00}}$ that has the interpretation as the energy density. This implies that in our conventions, $\smash{T^{00}}$ equals minus the energy density.
\end{watchout}

With the help of the projection $T^{\m\n}u_\m u_\n=-\Va$, it is now easy to match~\eqref{ch11:EMtensorrel2} to~\eqref{ch11:EMtensorrel}, which finally reveals the relationship between $\Fa(\abs\Xi)$ and the thermodynamic properties of the fluid,
\begin{equation}
\Fa(\abs\Xi)=-\Va(\abs\Xi)\;,\qquad
P(\abs\Xi)=2\abs\Xi\Va'(\abs\Xi)-\Va(\abs\Xi)\;.
\end{equation}
We conclude the discussion with the observation that the momentum density and current, entering the relativistic conservation law $\de_\m T^{\m\n}=0$, can be easily matched to their nonrelativistic counterparts in~\eqref{ch11:EMtensorfluid}. To that end, note that in the limit of slow motion, the relativistic internal energy density $\Va$ is dominated by the rest energy, that is $\Va\approx\vr$. Also, the four-velocity is well approximated by $u^\m\approx(1,v^i)$. Within this approximation, we find using~\eqref{ch11:EMtensorrel2} that
\begin{equation}
T^0_{\phantom0j}\approx-\vr v_j\;,\qquad
T^i_{\phantom ij}\approx-P\d^i_j-\vr v^iv_j\;,
\end{equation}
where I in addition used the fact that the contribution of pressure to $\Va+P$ is negligible. This recovers the nonrelativistic momentum density and current in~\eqref{ch11:EMtensorfluid}. The matching of the energy density and current is somewhat more involved as it requires a careful evaluation of the subleading contributions to energy, and I will therefore not go into details.

%%%%%%%%%%%%%%%%%%%%%%%%%%%%%%%%%%%%%%%%%%%%%%%%%%%%%%%%%%%%

\section*{\probsec}
\addcontentsline{toc}{section}{\probsec}

\begin{prob}
\label{pr11:auxiliaryidentities}
Given an invertible square matrix $A^i_{\phantom ij}$, prove the identities
\begin{equation}
\PD{(\det A)}{A^i_{\phantom ij}}=(\det A)(A^{-1})^j_{\phantom ji}\;,\qquad
\PD{\smash{(A^{-1})^i_{\phantom ij}}}{A^k_{\phantom kl}}=-(A^{-1})^i_{\phantom ik}(A^{-1})^l_{\phantom lj}\;.
\end{equation}
Hint: as to the first identity, it is advantageous to use the relation $\log\det A=\tr\log A$. For the second one, simply differentiate the equation $AA^{-1}=\un$ with respect to one of the matrix elements of $A$.
\end{prob}

\begin{prob}
\label{pr11:rotatingbucket}
An ideal incompressible liquid of density $\vr$ rotates steadily and uniformly with angular velocity $\o$ around a vertical axis. Find the pressure distribution in the liquid, assuming that the gravitational field $\vec g$ points vertically downwards and is uniform.
\end{prob}

\begin{prob}
\label{pr11:Eulersolunidirectional}
The flow of a fluid is said to be \emph{unidirectional} if there is a Cartesian coordinate system in which the velocity only has one component. In three spatial dimensions, a unidirectional flow can be represented by the velocity field $\vec v(\vec x,t)=(0,0,v_z(\vec x,t))$. Find all unidirectional solutions to the continuity and Euler equations for an ideal incompressible fluid in presence of a conservative field $\vec g(\vec x)$! Hint: first use the continuity equation to deduce that $v_z$ does not depend on $z$. Then take the curl of the Euler equation and show that $v_z(x,y,t)=f_1(t)+f_2(x,y)$, where $f_1$ and $f_2$ are arbitrary functions of the indicated arguments. Finally, compute the pressure using the Euler equation.
\end{prob}

\begin{prob}
\label{pr11:EulerGalileiinvar}
Show that the continuity and Euler equations are invariant under the following Galilei transformations,
\begin{equation}
\begin{gathered}
\vec x'=\vec x-\vec ut\;,\qquad
\vec v'(\vec x',t)=\vec v(\vec x,t)-\vec u\;,\\
\vr'(\vec x',t)=\vr(\vec x,t)\;,\qquad
P'(\vec x',t)=P(\vec x,t)\;,\qquad
\vec g'(\vec x',t)=\vec g(\vec x,t)\;.
\end{gathered}
\label{ch11:Galilei}
\end{equation}
To be precise, assuming that the velocity $\vec v(\vec x,t)$, density $\vr(\vec x,t)$, pressure $P(\vec x,t)$ and conservative external field $\vec g(\vec x,t)$ satisfy the equations in the unprimed coordinates, show that the functions $\vec v'(\vec x',t)$, $\vr'(\vec x',t)$, $P'(\vec x',t)$ and $\vec g'(\vec x',t)$ defined by~\eqref{ch11:Galilei} satisfy the same equations in the primed coordinates.
\end{prob}

\begin{prob}
\label{pr11:conformal}
The fact that potential (irrotational) flow of incompressible fluids is governed by the Laplace equation $\grad^2\p=0$ where $\p$ is the velocity potential, can be used to generate flow patterns satisfying a host of different boundary conditions. Take for instance the complex function $f(z)\equiv z+(1/z)$ where $z\equiv x+\I y\in\C$. Convince yourself that $\operatorname{Re}f(z)$ as a function of $x,y$ satisfies the Laplace equation in $\R^2$. Show that when restricted to $x^2+y^2\geq1$, this function can be viewed as the velocity potential of a flow around the circle $x^2+y^2=1$ whose velocity converges to a constant at infinity. When viewed as a two-dimensional cross-section of a three-dimensional flow, this describes a steady flow of a fluid around a cylindric obstacle.
\end{prob}

\begin{prob}
\label{pr11:viscousstresstensor}
Show that for a fluid under uniform rotation, the viscous stress tensor vanishes. In other words, a uniformly rotating fluid behaves like ideal. This explains why when stirring your tea or coffee in a cup, the liquid tends to settle to rigid-body-like rotation. Viscosity suppresses differential rotation. A bonus challenge: try to find all time-independent velocity fields describing flow of an \emph{incompressible} fluid such that the viscous stress tensor vanishes. Hint: use the condition of vanishing viscous stress tensor to show that all second partial derivatives of the velocity field vanish.
\end{prob}

\begin{prob}
\label{pr11:viscousflowunidirectional}
Consider a unidirectional steady flow of a viscous incompressible fluid in the $z$-direction, without any external field $\vec g$. Use the continuity and Navier--Stokes equations to obtain a partial differential equation for the single nonvanishing component of the velocity, $v_z$. With the help of your result, show that steady flow of a viscous incompressible fluid through a straight pipe of a circular cross-section has a parabolic velocity profile.
\end{prob}

\begin{figure}[t]
\sidecaption[t]
\includegraphics[width=2.9in]{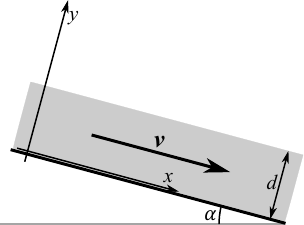}
\caption{Illustration for \refpr{pr11:inclinedplane}. A layer of an incompressible viscous liquid flows down a plane inclined by angle $\a$. The task is to find the speed of the liquid at the free surface.}
\label{fig11:inclinedplane}
\end{figure}

\begin{prob}
\label{pr11:inclinedplane}
An incompressible liquid of density $\vr$ and shear viscosity $\eta$ flows down an inclined plane as shown in Fig.~\ref{fig11:inclinedplane}. We assume that the liquid forms a uniform layer of thickness $d$, and flows steadily along the $x$-axis as indicated in the figure. Find the speed of the liquid at its free surface in terms of the given parameters $\vr,\eta,d$, the inclination angle $\a$ and the gravitational acceleration~$g$. Hint: the velocity $\vec v$ of the liquid should satisfy the boundary conditions $v_x=0$ at the contact with the inclined plane and $\Pd{v_x}{y}=0$ at the free surface.
\end{prob}

\chapter{Application: Electrodynamics}
\label{chap:electrodynamics}

\keywords{Four-current, four-potential, field-strength tensor, gauge invariance, retarded potential, dipole radiation, Rayleigh scattering, nonlinear electrodynamics, Proca theory, Maxwell--Chern--Simons theory, axion electrodynamics.}

%%%%%%%%%%%%%%%%%%%%%%%%%%%%%%%%%%%%%%%%%%%%%%%%%%%%%%%%%%%%

\noindent The first two applications of field theory we discussed in~\chaptername~\ref{chap:elasticity} and~\chaptername~\ref{chap:fluid} were motivated by everyday phenomena, and were largely nonrelativistic in nature. In this last~\chaptername{} of the course, we shall look at another application, which is intrinsically relativistic yet equally or even more important: electromagnetism. The basic phenomenology of electromagnetism is well covered by virtually any undergraduate physics curriculum. If you need a refresher, any standard source such as~\cite{Purcell2013,Feynman2011} will do. Here we will focus on developing a formulation of electromagnetism as a relativistic field theory. This will, among others, help us understand to what extent Maxwell's theory is unique, and how it might be meaningfully modified.

%%%%%%%%%%%%%%%%%%%%%%%%%%%%%%%%%%%%%%%%%%%%%%%%%%%%%%%%%%%%

\section{Electromagnetism as Relativistic Field Theory}

Let us start with a brief overview of the basic laws of electromagnetism. At the undergraduate level, the basic degrees of freedom of electromagnetism are usually taken to be the electric field $\vec E$ and the magnetic field $\vec B$. These two vector fields satisfy the set of Maxwell's equations\footnote{Throughout this entire~\chaptername{} up to a single exception, we will work in the physical space with $d=3$ dimensions. A generalization of Maxwell's equations to other dimensions exists. However, the magnetic field then has to be treated as a rank-two antisymmetric tensor $B_{ij}$ rather than a vector.}
\begin{align}
\label{ch12:maxwell1stpair}
\rot\vec E&=-\PD{\vec B}t\;,&
\divg\vec B&=0\;,\\
\label{ch12:maxwell2ndpair}
\divg\vec E&=\frac\vr{\eps_0}\;,&
\rot\vec B&=\m_0\vec J+\frac1{c^2}\PD{\vec E}t\;.
\end{align}
Here $\vr$ is the density of electric charge that sources the electric field. Likewise, the electric current density $\vec J$ sources magnetic fields.

An important feature of electromagnetism is that the information about the two vector fields $\vec E,\vec B$ can be encoded into a single scalar and a single vector, namely the electromagnetic potentials $\p$ and $\vec A$. The electric and magnetic fields can be uniquely reconstructed from the latter as
\begin{equation}
\vec E=-\grad\p-\PD{\vec A}t\;,\qquad
\vec B=\rot\vec A\;.
\label{ch12:potentials}
\end{equation}
The correspondence between the fields and their potentials is however not one-to-one. Indeed, the potentials can be redefined by the \emph{gauge transformation}
\begin{equation}
\p\to\p-\PD\chi t\;,\qquad
\vec A\to\vec A+\grad\chi
\label{ch12:gaugetransfo}
\end{equation}
with an arbitrary function $\chi$ of spacetime coordinates, without changing the fields. In practice, it is common to fix, or at least narrow down, the ambiguity in the choice of the potentials by imposing an additional constraint that they should satisfy.

The Maxwell equations imply several conservation laws that we will later be able to relate to the  symmetries of the Lagrangian density for electromagnetism. The most important of these is the conservation of electric charge,
\begin{equation}
\PD\vr t+\divg\vec J=0\;.
\label{ch12:chargeconservation}
\end{equation}
We will also revisit the local conservation of energy, expressed in the form
\begin{equation}
\PD ut+\divg\vec S=-\skal JE\;,
\label{ch12:energyconservation}
\end{equation}
where
\begin{equation}
u=\frac12\eps_0\vec E^2+\frac1{2\m_0}\vec B^2\;,\qquad
\vec S=\frac1{\m_0}\vekt EB
\label{ch12:energydensityflux}
\end{equation}
are the energy density and flux, respectively. The latter is usually referred to as the \emph{Poynting vector}. The nonzero right-hand side of~\eqref{ch12:energyconservation} represents the power per unit volume, exerted by the electric field on charged matter, and thus the transfer of energy between the electromagnetic field and matter.

%%%%%%%%%%%%%%%%%%%%%%%%%%%%%%%%%%%%%%%%%%%%%%%%%%%%%%%%%%%%

\subsection{Four-Current, Four-Potential, Field-Strength Tensor}

The first step towards a Lorentz-covariant reformulation of electromagnetism is the observation that the electric charge of any object is Lorentz-invariant, that is, it takes the same value in any \emph{inertial reference frame} (IRF). We will model electric current as motion of a charged fluid, with local charge density $\vr$ and velocity $\vec v$. The current density is then a simple product of the two,
\begin{equation}
\vec J=\vr\vec v\;.
\end{equation}
While the electric charge itself is Lorentz-invariant, its density is not as a consequence of Lorentz contraction. It can however be related to the charge density $\vr_0$ measured in the fluid's local rest frame by $\vr=\g_{\vec v}\vr_0$ where $\smash{\g_{\vec v}\equiv1/\sqrt{1-(\vec v/c)^2}}$. This shows that the charge density and current can be assembled into a single four-vector object,
\begin{equation}
\boxed{J^\m\equiv(c\vr,\vec J)=\vr_0u^\m\;,}
\label{ch12:fourcurrent}
\end{equation}
where $u^\m=\g_{\vec v}(c,\vec v)$ is the local four-velocity of the medium. This is the \emph{four-current}.

Now that we have put the sources of electromagnetic fields into a single Lorentz-covariant object, we must do the same with the electric and magnetic fields themselves. Here a single four-vector will not do, since $\vec E$ and $\vec B$ have altogether six a priori independent components. One might naively expect a pair of four-vectors, one for each of the fields. That would however require adding new degrees of freedom. It turns out to be easier to first covariantize the electromagnetic potentials $\p$ and $\vec A$. These fit neatly into a single four-vector, namely the \emph{four-potential}
\begin{equation}
\boxed{A^\m\equiv(\p/c,\vec A)\;.}
\label{ch12:fourpotential}
\end{equation}

\begin{watchout}%
In case of the four-current, the definition~\eqref{ch12:fourcurrent} was physically justified by our observation that $c\vr$ and $\vec J$ together transform as components of a four-vector, thanks to their proportionality to the components of four-velocity. In case of the four-potential, we do not have such a compelling justification. This is closely related to the freedom to redefine $\p$ and $\vec A$ by the gauge transformation~\eqref{ch12:gaugetransfo}, which can in principle be done independently in any IRF. However, assuming that the same set of potentials can somehow be ``measured'' in different IRFs, then~\eqref{ch12:gaugetransfo} can be rewritten neatly as
\begin{equation}
A_\m\to A_\m+\de_\m\chi\;.
\label{ch12:gaugetransfoAmu}
\end{equation}
This suggests that we are on the right track, and we shall therefore take the definition~\eqref{ch12:fourpotential} for granted.
\end{watchout}

Our next task is to express the electric and magnetic fields in terms of the four-potential. Both fields are given by linear combinations of first partial derivatives of the electromagnetic potentials. From~\eqref{ch12:gaugetransfoAmu} it follows that $\de_\m A_\n\to\de_\m A_\n+\de_\m\de_\n\chi$. The only way to obtain a quantity that is invariant under the gauge transformation~\eqref{ch12:gaugetransfoAmu} from linear combinations of $\de_\m A_\n$ is to antisymmetrize in the indices $\m,\n$ by setting
\begin{equation}
\boxed{F_{\m\n}\equiv\de_\m A_\n-\de_\n A_\m\;.}
\label{ch12:Fmunu}
\end{equation}
Let us inspect what we found. Using the definition~\eqref{ch12:fourpotential} together with~\eqref{ch12:potentials}, the components of the \emph{field-strength tensor} $F_{\m\n}$ are seen to equal
\begin{equation}
F_{\m\n}=\begin{pmatrix}
0 & -E_x/c & -E_y/c & -E_z/c\\
E_x/c & 0 & B_z & -B_y\\
E_y/c & -B_z & 0 & B_x\\
E_z/c & B_y & -B_x & 0
\end{pmatrix}\;.
\label{ch12:Fmunumatrix}
\end{equation}
The six independent components of the antisymmetric tensor $F_{\m\n}$ encode elegantly the electromagnetic fields. This is therefore the desired Lorentz-covariant formulation of the electromagnetic fields, which would have been difficult to guess without using invariance under the gauge transformation~\eqref{ch12:gaugetransfoAmu} as a guide.

Having introduced the basic building blocks of the covariant formulation of electromagnetism, let us briefly digress and comment on their transformation properties under Lorentz transformations. The four-current $J^\m$ and four-potential $A^\m$ are both four-vectors and their transformation thus copies that of the spacetime coordinates,
\begin{equation}
x'^\m=\Lambda^\m_{\phantom\m\n}x^\n\quad\Rightarrow\quad
J'^\m(x')=\Lambda^\m_{\phantom\m\n}J^\n(x)\;,\quad
A'^\m(x')=\Lambda^\m_{\phantom\m\n}A^\n(x)\;.
\end{equation}
The rule for $A^\m(x)$ fixes through~\eqref{ch12:potentials} uniquely the transformation properties of the electric and magnetic fields. For practical applications, it is however more convenient to be able to transform the electromagnetic fields between different IRFs directly, without having to use the potentials in the intermediate steps. To that end, we lift the indices on the field-strength tensor using the Minkowski metric, $F^{\m\n}\equiv g^{\m\a}g^{\n\b}F_{\a\b}$. The new contravariant tensor $F^{\m\n}$ carries two four-vector indices, which indicates the transformation rule
\begin{equation}
F'^{\m\n}(x')=\Lambda^\m_{\phantom\m\a}\Lambda^\n_{\phantom\n\b}F^{\a\b}(x)\;.
\label{ch12:FmunuLorentztransfo}
\end{equation}
This is straightforward to translate into the ordinary (spatial) vector notation in terms of the electric field $\vec E$, the magnetic field $\vec B$, and the velocity $\vec v$ of the Lorentz boost. Denoting with $\parallel$ and $\perp$ the components of the fields parallel and perpendicular to $\vec v$, respectively, \eqref{ch12:FmunuLorentztransfo} is equivalent to
\begin{equation}
\begin{aligned}
\vec E'_\parallel&=\vec E_\parallel\;,\qquad&
\vec E'_\perp&=\g_{\vec v}(\vec E_\perp+\vekt vB)\;,\\
\vec B'_\parallel&=\vec B_\parallel\;,\qquad&
\vec B'_\perp&=\g_{\vec v}[\vec B_\perp-(1/c^2)\vekt vE]\;.
\end{aligned}
\label{ch12:EBtransfo}
\end{equation}

%%%%%%%%%%%%%%%%%%%%%%%%%%%%%%%%%%%%%%%%%%%%%%%%%%%%%%%%%%%%

\subsection{Covariant Formulation of Maxwell's Equations}

The next step is to reformulate the Maxwell equations~\eqref{ch12:maxwell1stpair} and~\eqref{ch12:maxwell2ndpair} in terms of the field-strength tensor. Being a pair of (spatial) vector equations and a pair of scalar equations, the Maxwell equations have altogether eight components. This suggests their relativistic formulation should amount to two four-vector differential equations. Indeed, it is straightforward to check that all the Maxwell equations are subsumed in
\begin{equation}
\boxed{\ve^{\k\l\m\n}\de_\l F_{\m\n}=0\;,\qquad
\de_\n F^{\m\n}=\m_0J^\m\;.}
\label{ch12:maxwellcov}
\end{equation}
The first of these, also equivalently written as
\begin{equation}
\de_\l F_{\m\n}+\de_\m F_{\n\l}+\de_\n F_{\l\m}=0\;,
\label{ch12:Bianchi}
\end{equation}
is equivalent to the pair of sourceless Maxwell equations~\eqref{ch12:maxwell1stpair}. It is often referred to as the \emph{Bianchi identity}. The second equation in~\eqref{ch12:maxwellcov} is equivalent to the pair of Maxwell equations~\eqref{ch12:maxwell2ndpair} containing sources of electromagnetic fields.

\begin{illustration}%
One benefit of expressing the Maxwell equations in the Lorentz-covariant form is that this makes the conservation of electric charge manifest. Indeed, by taking the four-divergence of the second equation in~\eqref{ch12:maxwellcov} and using the antisymmetry of the field-strength tensor, we find immediately that
\begin{equation}
\m_0\de_\m J^\m=\de_\m\de_\n F^{\m\n}=0\;.
\end{equation}
This is equivalent to~\eqref{ch12:chargeconservation}. It is worth noting that the conservation of the four-current $J^\m$ is a consequence of the Maxwell equations alone. It does not require the \emph{equation of motion} (EoM) for the matter that constitutes the current.
\end{illustration}

The principal advantage of working with the electromagnetic potentials is that they make it straightforward to solve the Maxwell equations for a given distribution of electric charge and current. To see this, we insert the definition~\eqref{ch12:Fmunu} of the field-strength tensor into the second equation in~\eqref{ch12:maxwellcov}, getting\footnote{The Bianchi identity is satisfied identically by~\eqref{ch12:Fmunu}. This reflects the definition of electromagnetic potentials~\eqref{ch12:potentials} as a way to resolve the pair of Maxwell equations~\eqref{ch12:maxwell1stpair} without sources.}
\begin{equation}
\de_\n\de^\m A^\n-\de_\n\de^\n A^\m=\m_0J^\m\;.
\end{equation}
It is now convenient to adopt the \emph{Lorenz gauge condition},
\begin{equation}
\de_\m A^\m=\frac1{c^2}\PD\p t+\divg\vec A=0\qquad\text{(Lorenz gauge)}\;.
\label{ch12:Lorenzgauge}
\end{equation}
This brings the EoM for the four-potential to the form
\begin{equation}
-\de_\n\de^\n A^\m\equiv\Box A^\m=\m_0 J^\m\;,
\label{ch12:AmuEoM}
\end{equation}
where $\Box=(1/c^2)\de^2/\de t^2-\grad^2$ is the d'Alembert operator.

\begin{watchout}%
The reduced EoM~\eqref{ch12:AmuEoM} justifies a posteriori treating the four-potential $A^\m$ as a four-vector. Namely, \eqref{ch12:AmuEoM} determines $A^\m$ in terms of the four-vector source $J^\m$ up to the addition of an arbitrary solution to the homogeneous equation $\Box A^\m=0$. The latter ambiguity can be fixed by a judicious choice of boundary conditions. Once we have established that the four-potential $A^\m$ actually transforms as a four-vector, it follows that the Lorenz gauge condition~\eqref{ch12:Lorenzgauge} itself is Lorentz-invariant, that is valid simultaneously in all IRFs. This verifies the consistency of our relativistic formulation of electromagnetism.
\end{watchout}

%%%%%%%%%%%%%%%%%%%%%%%%%%%%%%%%%%%%%%%%%%%%%%%%%%%%%%%%%%%%

\section{Electromagnetic Radiation}

The EoM~\eqref{ch12:AmuEoM} describes both free electromagnetic fields in vacuum (in the special case of $J^\m=0$) and the generation of electromagnetic fields by given sources. Should one want to study the back-reaction of the fields on charged matter, then~\eqref{ch12:AmuEoM} has to be supplemented with an EoM for the latter. (A Lorentz-covariant formulation of the dynamics of a charged particle was worked out in Sect.~\ref{subsec:particleEMfield}.) Here we will focus on a different application though, namely generation of electromagnetic radiation by accelerating charges, treating the motion of charges as known.

In this case, the relevant solution of~\eqref{ch12:AmuEoM}, known as the \emph{retarded potential}, is
\begin{equation}
\boxed{A^\m(\vec x,t)=\frac{\m_0}{4\pi}\int\D^3\!\vec x'\,\frac{J^\m(\vec x',t_\mathrm{ret})}{\abs{\vec x-\vec x'}}\;,}\qquad\text{where }
t_\mathrm{ret}\equiv t-\frac{\abs{\vec x-\vec x'}}{c}\;.
\label{ch12:retardedpotential}
\end{equation}
For a static source, represented by $J^\m(\vec x)$, this clearly reduces to the results for the electrostatic and vector potentials (the latter in the \emph{Coulomb gauge}, $\divg\vec A=0$), well-known from electrostatics and magnetostatics. Inserting the retarded time $t_\mathrm{ret}$ makes sure that~\eqref{ch12:retardedpotential} solves the general time-dependent equation~\eqref{ch12:AmuEoM} in the Lorenz gauge. Physically, the retarded time takes into account the fact that electromagnetic fields propagate with a finite speed, and any change in the source will therefore affect the field at a distance from it with certain delay.

%%%%%%%%%%%%%%%%%%%%%%%%%%%%%%%%%%%%%%%%%%%%%%%%%%%%%%%%%%%%

\subsection{Electric Dipole Radiation}

\begin{figure}[t]
\sidecaption[t]
\includegraphics[width=2.9in]{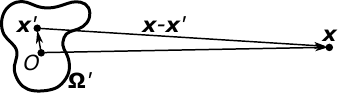}
\caption{Illustration of the geometry involved in the analysis of electric dipole radiation. The source of the radiation is localized to the spatial domain $\vec\O'$. The electromagnetic fields are observed at the point $\vec x$. The origin of coordinates $O$ is chosen so that $\abs{\vec x}\gg\abs{\vec x'}$ for all points $\vec x'\in\vec\O'$.}
\label{fig12:dipoleradiation}
\end{figure}

For the sake of illustration, we will now focus on a common type of electromagnetic radiation that arises from interaction of electromagnetic fields with matter that is overall electrically neutral, but its charge distribution changes with time. This serves among others as a classical model for oscillations of electric charge in atoms or molecules. Our main assumption is that the sources of electromagnetic fields are localized to a finite spatial domain, $\vec\O'$; see Fig.~\ref{fig12:dipoleradiation} for a visualization of the situation. We place the origin of coordinates somewhere inside $\vec\O'$ and observe the electromagnetic fields at a point $\vec x$ far away from the sources, so that $\abs{\vec x}$ can be treated as much larger than $\abs{\vec x'}$ for any $\vec x'\in\vec\O'$. Under this assumption, the retarded potential~\eqref{ch12:retardedpotential} can be approximated with
\begin{equation}
A^\m(\vec x,t)\approx\frac{\m_0}{4\pi r}\int_{\vec\O'}\D^3\!\vec x'\,J^\m(\vec x',t-r/c)\;,\qquad\text{where }r\equiv\abs{\vec x}\;.
\label{ch12:dipoleAmu}
\end{equation}

Our next strategy will be as follows. Since the sources of the electromagnetic fields are localized to a small domain, it should be possible to express the integral over $\vec\O'$ in terms of the multipole moments of the charge distribution. Thanks to our assumption of overall neutrality, the leading multipole, that is the total charge, vanishes. The electromagnetic fields far away from the sources should therefore be dominated by the contribution of the electric dipole moment $\vec p$. We will start by evaluating the vector potential $\vec A$ in terms of the dipole moment, for which we only need to integrate over the spatial current $\vec J$. This is sufficient to determine the magnetic field $\vec B$. In the next step, we will find the electric field $\vec E$ from the magnetic field by using Maxwell's equations. Once we have both fields, we will be able to compute the power of the radiation emitted by the source using the Poynting vector.

Let us now see this program through to the end. In order to relate the spatial integral of $\vec J$ to the dipole moment, we use the charge conservation~\eqref{ch12:chargeconservation} to rewrite $\smash{J^i=\de_j(x^iJ^j)-x^i\de_jJ^j=\de_j(x^iJ^j)+x^i\de_t\vr}$. Integrating over space and using the fact that the current by assumption vanishes outside of $\vec\O'$, we then get
\begin{equation}
\int_{\vec\O'}\D^3\!\vec x'\,J^i(\vec x',t-r/c)=\int_{\vec\O'}\D^3\!\vec x'\,x'^i\de_t\vr(\vec x',t-r/c)\;.
\end{equation}
The last integral is nothing but the time derivative of the dipole moment of the charge distribution, which brings the vector potential to the form
\begin{equation}
\vec A(\vec x,t)\approx\frac{\m_0}{4\pi r}\vec{\dot p}(t-r/c)\;.
\label{ch12:dipoleA}
\end{equation}
The next step is to compute the magnetic field. This follows straightforwardly as
\begin{equation}
\vec B(\vec x,t)\approx\frac{\m_0}{4\pi}\rot\frac{\vec{\dot p}(t-r/c)}r=-\frac{\m_0}{4\pi}\biggl[\frac{\vec x}{r^3}\times\vec{\dot p}(t-r/c)+\frac{\vec x}{cr^2}\times\vec{\ddot p}(t-r/c)\biggr]\;;
\end{equation}
I converted the derivative of $\vec{\dot p}(t-r/c)$ with respect to $r$ to a derivative with respect to time using the chain rule. Far away from the sources, the term proportional to $\vec{\ddot p}$ will be dominant, being suppressed by one less inverse power of distance. The asymptotic behavior of the magnetic field at large distance will therefore be dominated by
\begin{equation}
\vec B(\vec x,t)\approx-\frac{\m_0}{4\pi cr^2}\vec x\times\vec{\ddot p}(t-r/c)\;.
\label{ch12:dipoleB}
\end{equation}

Our next task is to find the electric field. Here we take advantage of the Amp\`ere law, which far away from the sources where there is no current reduces to $\Pd{\vec E}{t}=c^2\rot\vec B$. Based on the experience from the calculation of $\vec B$, we expect the large-distance behavior of the electric field to be dominated by the term where the curl in $\rot\vec B$ acts on $\vec{\ddot p}$. Using that $\de_i[\vec{\ddot p}(t-r/c)]=-(1/c)(x_i/r)\vec{\dddot p}(t-r/c)$ and subsequently integrating over time simply by removing one of the dots on the dipole moment, we arrive at
\begin{equation}
\vec E(\vec x,t)\approx\frac{\m_0}{4\pi r^3}\vec x\times[\vec x\times\vec{\ddot p}(t-r/c)]\;.
\label{ch12:dipoleE}
\end{equation}

\begin{watchout}%
What we have done is not quite innocuous. In bypassing explicit time integration, we tacitly assumed that there are no contributions to the electric field that would be static and thus drop out of the Amp\`ere law. In other words, when integrating the equation $\Pd{\vec E}{t}=c^2\rot\vec B$, we neglected the possibility that there might be a time-independent but space-dependent integration ``constant.'' Such a static field might arise from the monopole part of the multipole expansion, should our charge distribution carry nonzero net charge.
\end{watchout}

The dominant parts of the electric and magnetic fields~\eqref{ch12:dipoleE} and~\eqref{ch12:dipoleB} satisfy the simple orthogonality relation $\vec E=-c\vekt nB$, where $\vec n\equiv\vec x/r$ is the radial unit vector. This means that the vectors $\vec n$, $\vec E$ and $\vec B$ form a right-handed orthogonal system, which suggests that the electromagnetic fields propagate radially outwards in the form of a spherical wave. This is confirmed by computing the Poynting vector, $\vec S=\vekt EB/\m_0=(c/\m_0)\vec B^2\vec n$. Inserting the magnetic field from~\eqref{ch12:dipoleB}, we get
\begin{equation}
\boxed{\vec S(\vec x,t)\approx\frac{\m_0}{16\pi^2cr^2}\abs{\vec n(\vec x)\times\vec{\ddot p}(t-r/c)}^2\vec n(\vec x)\;.}
\label{ch12:dipoleS}
\end{equation}
This is a fairly general expression for the energy flux radiated by a time-dependent, localized charged distribution that is overall electrically neutral. We see that the energy flux decreases with the second power of the radial distance $r$. As a consequence, upon integration over all directions, we find that there is nonzero energy transfer away from the source towards spatial infinity. At the same time, the Poynting vector is proportional to the squared second time derivative of the dipole moment. This leads to the conclusion that accelerated charges emit electromagnetic radiation.

The result~\eqref{ch12:dipoleS} also justifies a posteriori all the approximations we have made. Namely, we dropped contributions to the electric and magnetic fields which decrease with higher powers of distance $r$. These would generate corrections to our Poynting vector decreasing faster than $1/r^2$, and would therefore not affect the energy transfer.

\begin{illustration}%
\label{ex12:dipoleradiation}%
For a concrete illustration, suppose that the motion of the sources is such that the dipole moment $\vec p$ has a fixed direction and only its magnitude varies with time. This special case includes for instance the oscillations of electric charge, enforced by a passing linearly polarized electromagnetic wave. The cross product in~\eqref{ch12:dipoleS} now simplifies to $\vec n\times\vec{\ddot p}=\ddot p\sin\t$, where $p(t)$ is the magnitude of the dipole moment and $\t$ the angle between $\vec p$ and $\vec x$. The Poynting vector then becomes
\begin{equation}
\vec S(\vec x,t)\approx\frac{\m_0 [\ddot p(t-r/c)]^2\sin^2\t}{16\pi^2cr^2}\vec n(\vec x)\;.
\end{equation}
The radiation of a linearly oscillating dipole has a very simple directional nature, the maximum power being emitted in the plane perpendicular to $\vec p$. The total radiated power is obtained by integration over the full solid angle,
\begin{equation}
P(r,t)=\int_0^\pi2\pi r^2\abs{\vec S(\vec x,t)}\sin\t\,\D\t\approx\frac{\m_0[\ddot p(t-r/c)]^2}{6\pi c}\;.
\label{ch12:dipoleP}
\end{equation}
\end{illustration}

%%%%%%%%%%%%%%%%%%%%%%%%%%%%%%%%%%%%%%%%%%%%%%%%%%%%%%%%%%%%

\subsection{Rayleigh Scattering}

As a particular application of our analysis of electromagnetic radiation, we will now look at elastic scattering of an incident electromagnetic wave by atoms or molecules. An atom that is not ionized carries zero net electric charge. Moreover, suppose that it initially does not carry any electric dipole moment. An external electric field $\vec E$ will then polarize the atom, inducing a dipole moment that is, in the first approximation, proportional to the field,
\begin{equation}
\vec p=\a\vec E\;.
\label{ch12:polarizability}
\end{equation}
The constant $\a$ is the \emph{polarizability} of the atom. This linear approximation assumes that the electric field is not too strong. Also, in case the field is time-dependent, it is meaningful to assume that the induced dipole moment will be proportional to the instantaneous electric field only if the characteristic frequency of the latter is much lower than the frequency of transitions between different energy levels of the atom. This assumption is reasonable for the interaction of electromagnetic waves in the visible part of the spectrum with atoms or simple molecules.

To be concrete, we now assume that the electric field $\vec E$ belongs to a linearly polarized electromagnetic wave of (angular) frequency $\o$. Then $\vec{\ddot p}=-\o^2\vec{p}=-\a\o^2\vec E$. Using this in~\eqref{ch12:dipoleP} leads to
\begin{equation}
P(r,t)\approx\frac{\m_0\a^2\o^4[\vec E(t-r/c)]^2}{6\pi c}\;,
\end{equation}
where $\vec E(t)$ is the electric field at the location of the atom. Averaging over the fast oscillations of the field gives $\langle[\vec E(t)]^2\rangle=(1/2)\vec E_0^2$, where $\vec E_0$ is the amplitude of the electromagnetic wave. The time-averaged power of the radiation, emitted by the atom, is then
\begin{equation}
\langle P(r,t)\rangle\approx\frac{\m_0\a^2\o^4\vec E_0^2}{12\pi c}\;.
\label{ch12:Rayleigh}
\end{equation}
The regime of scattering of light by matter, described by the classical linear model of polarization~\eqref{ch12:polarizability}, is called \emph{Rayleigh scattering}.

\begin{illustration}%
Our result~\eqref{ch12:Rayleigh} describes for instance the power loss (attenuation) of a light beam passing through a gas or liquid. It is clearly proportional to the intensity of the incident light beam, which scales with the squared amplitude $\vec E_0^2$. The relative attenuation rate depends, apart from the universal constants such as $\m_0$ and $c$, only on the polarizability $\a$ and the frequency $\o$. In particular the proportionality to the fourth power of frequency is notable. The two edges of the spectrum of visible light have frequencies that differ fairly accurately by a factor of two. This translates to the ratio of 16 in the attenuation rates for the violet and red ends of the visible spectrum. The fact that the violet end of the spectrum is scattered by matter much more strongly than the red end explains why the sky is blue.
\end{illustration}

%%%%%%%%%%%%%%%%%%%%%%%%%%%%%%%%%%%%%%%%%%%%%%%%%%%%%%%%%%%%

\section{Lagrangian Formulation of Electromagnetism}

Many of the basic properties of Maxwell's electrodynamics can be reproduced elegantly starting from a variational principle. Apart from the practical utility, however, the Lagrangian formulation also sheds light on the structure of electromagnetism as a physical theory. We will be finally able to understand why Maxwell's equations essentially \emph{have to} look the way they do.

%%%%%%%%%%%%%%%%%%%%%%%%%%%%%%%%%%%%%%%%%%%%%%%%%%%%%%%%%%%%

\subsection{Lagrangian Density and Equation of Motion}

The essential part of the Lagrangian setup is the choice of generalized coordinates. Guided by the observation that the Maxwell equations~\eqref{ch12:maxwellcov} are of first order in derivatives of the electromagnetic fields and thus of second order in derivatives of the four-potential $A^\m$, we will take $A^\m$ itself as the set of independent generalized coordinates. This choice is not without subtleties. The most obvious of these stems from the fact that Maxwell's equations only depend on the field-strength tensor $F^{\m\n}$ (or the fields $\vec E,\vec B$), and not explicitly on the four-potential. This reflects the invariance of Maxwell's equations under the gauge transformation~\eqref{ch12:gaugetransfoAmu}, which must be mirrored by the Lagrangian density of the theory. Furthermore, we know that Maxwell's equations are linear in $F^{\m\n}$ and thus satisfy the superposition principle. This requires that the Lagrangian density must be bilinear in the four-potential. Altogether, the Lagrangian density of electromagnetism is required to feature the following properties:
\begin{enumerate}
\item[(a)] Lorentz invariance.
\item[(b)] Gauge invariance.
\item[(c)] Superposition principle. 
\end{enumerate}
Condition (a) requires that the Lagrangian density is built using solely the Lorentz-covariant objects $A^\m$, $F^{\m\n}$ and $J^\m$. The condition (b) essentially implies that the part of the Lagrangian density describing electromagnetic fields in absence of sources is built out of $F^{\m\n}$ alone (see~\refpr{pr12:ChernSimons} for an exception); by condition (c) this part must be quadratic in $F^{\m\n}$. Finally, the part of the Lagrangian density that captures the coupling between electromagnetic fields and matter should be linear in $A^\m$ so as to provide a source term in the EoM.

The most general Lagrangian density consistent with the above requirements takes the form
\begin{equation}
\La=uF_{\m\n}F^{\m\n}+v\ve^{\m\n\a\b}F_{\m\n}F_{\a\b}+wJ^\m A_\m\;,
\label{ch12:Lagansatz}
\end{equation}
where $u,v,w$ are some constants. Note that the $w$-term can be consistent with gauge invariance in spite of appearances. Namely, upon the gauge transformation~\eqref{ch12:gaugetransfoAmu}, the Lagrangian density shifts by $wJ^\m\de_\m\chi$, which equals $-w\chi\de_\m J^\m$ up to a surface term. The action will therefore be gauge-invariant provided the source four-current $J^\m$ is conserved, which is something we expect and desire. To constrain the parameters $u,v,w$, we compute the functional derivative of the action based on~\eqref{ch12:Lagansatz},
\begin{equation}
\frac{\udelta S}{\udelta A_\m}=wJ^\m+4u\de_\n F^{\m\n}\;.
\end{equation}
The main surprise is that the $v$-term does not contribute at all to the EoM and therefore can be dropped; this is the subject of~\refpr{pr12:Pontryagin}. The EoM $\udelta S/\udelta A_\m=0$ takes the form of the Maxwell equations with source, that is the second equation in~\eqref{ch12:maxwellcov}. A precise matching requires that $w/(4u)=-\m_0$. The overall normalization of the coefficients $u,w$ can be fixed by requiring that the Lagrangian density gives a correctly normalized \emph{energy--momentum} (EM) \emph{tensor}. We will verify in Sect.~\ref{subsec:EMEMtensor} that the correct choice is $w=1$ and $u=-1/(4\m_0)$. We thus arrive at our final result for the Lagrangian density of electrodynamics,
\begin{equation}
\boxed{\La=-\frac1{4\m_0}F_{\m\n}F^{\m\n}+J^\m A_\m\;.}
\label{ch12:LagEM}
\end{equation}
The Euler--Lagrange equation associated with this Lagrangian density is the second equation in~\eqref{ch12:maxwellcov}. The form of this equation is uniquely fixed by the above constraints (a)--(c). The only input that is left undetermined are the values of the constants that enter the EoM, that is the vacuum permeability $\m_0$ and the speed of light $c$. This is a major achievement of our relativistic formulation of electromagnetism. Such rigidity of Maxwell's equations would be completely invisible in their nonrelativistic formulation as shown in~\eqref{ch12:maxwell1stpair} and~\eqref{ch12:maxwell2ndpair}.

\begin{watchout}%
By constructing the Lagrangian density, we managed to reproduce one half of Maxwell's equations, namely those that contain the sources of electromagnetic fields. How about the first pair of Maxwell equations~\eqref{ch12:maxwell1stpair}, or equivalently the first relation in~\eqref{ch12:maxwellcov} that we dubbed Bianchi identity? This does not have the status of a Lagrangian EoM. Namely, as the name suggests, the Bianchi identity is an \emph{identity}, valid for any configuration of the four-potential $A^\m(x)$. It is a direct consequence of the mathematical redundancy, involved in parameterizing the electromagnetic fields by the scalar and vector potential. The true EoM following from the Lagrangian density~\eqref{ch12:LagEM}, on the other hand, is not satisfied by just any field configuration. This distinction between the Bianchi identity and the dynamical EoM will be important in Sect.~\ref{sec:EMextensions} when we consider possible modifications of electromagnetism. Namely, no matter how we change the Lagrangian, the Bianchi identity will remain valid, only the EoM will change.
\end{watchout}

%%%%%%%%%%%%%%%%%%%%%%%%%%%%%%%%%%%%%%%%%%%%%%%%%%%%%%%%%%%%

\subsection{Energy--Momentum Tensor}
\label{subsec:EMEMtensor}

The energy and momentum carried by electromagnetic fields play an important role in understanding the physics of electromagnetic radiation. The Lagrangian formalism makes it easy to find what the energy and momentum look like. In order to keep the discussion simple, we will from now on only consider electromagnetic fields in the absence of sources, that is, we set $J^\m=0$. An extension of the local conservation laws of energy and momentum in presence of electric charge and current is the subject of~\refpr{pr12:EMconservation}.

Using the fact that for the Lagrangian density~\eqref{ch12:LagEM}, $\Pd\La{(\de_\m A_\n)}=-F^{\m\n}/\m_0$, the EM tensor of the free electromagnetic field is easily extracted with the help of the general formula~\eqref{ch10:EMtensorcontravariant},
\begin{equation}
T^{\m\n}=-\frac1{\m_0}F^{\m\a}\de^\n A_\a+\frac1{4\m_0}g^{\m\n}F_{\a\b}F^{\a\b}\;.
\label{ch12:EMtensorcan}
\end{equation}
This \emph{canonical EM tensor} however has a serious flaw: it depends explicitly on the four-potential. We certainly do not want the energy and momentum density of the electromagnetic fields to depend on the choice of gauge for $A^\m$.

This issue has a history that is nearly as long as that of Noether's theorem itself. Without going into details, I just note that the resolution of the puzzle lies in the ambiguity of the procedure we used in Sect.~\ref{subsec:Noethertheorem} to extract the Noether current from a given action and symmetry. The first thing to notice is that one can add to the Noether current any vector field whose divergence identically vanishes; such a modification cannot be detected at the level of the action in~\eqref{ch10:GellMannLevy}, and does not affect the resulting local conservation law. The canonical EM tensor~\eqref{ch12:EMtensorcan} can thus be redefined to
\begin{align}
\Theta^{\m\n}&\equiv T^{\m\n}+\frac1{\m_0}\de_\a(F^{\m\a}A^\n)\\
\notag
&=-\frac1{\m_0}F^{\m\a}\de^\n A_\a+\frac1{4\m_0}g^{\m\n}F_{\a\b}F^{\a\b}+\frac1{\m_0}(\underline{A^\n\de_\a F^{\m\a}}+F^{\m\a}\de_\a A^\n)\;.
\end{align}
This is because of the antisymmetry of $F^{\m\a}$ under the exchange of its indices, which guarantees that $\de_\m\Theta^{\m\n}=\de_\m T^{\m\n}$ identically. Moreover, the underlined term vanishes by the EoM in the absence of sources. The rest of the ``improved'' EM tensor can be folded into the neat expression
\begin{equation}
\boxed{\Theta^{\m\n}=-\frac1{\m_0}F^{\m\a}F^\n_{\phantom\n\a}+\frac1{4\m_0}g^{\m\n}F_{\a\b}F^{\a\b}\;.}
\label{ch12:EMtensorHilbert}
\end{equation}
This \emph{Hilbert EM tensor} solves our problem with gauge invariance. Moreover, it has the additional feature that it is manifestly symmetric. While inconsequential in our current setting, this is important when one tries to couple the electromagnetic field to dynamical gravity as in the general theory of relativity.

It is an easy exercise to work out the physical content of the Hilbert EM tensor in terms of the electric and magnetic fields using~\eqref{ch12:Fmunumatrix}. We find that:
\begin{itemize}
\item The energy density is determined by
\begin{equation}
\Theta^{00}=-\frac12\eps_0\vec E^2-\frac1{2\m_0}{\vec B^2}\;.
\end{equation}
This combination of electric and magnetic fields is expected from undergraduate electromagnetism. The opposite overall sign stems from our conventions whereby it is $\Theta^0_{\phantom00}$ that is the energy density.
\item The momentum density and energy flux both correspond to
\begin{equation}
\Theta^{0i}=\Theta^{i0}=-c\eps_0(\vekt EB)^i\;.
\end{equation}
More precisely, the energy current reads $-c\Theta^{i0}=(\vekt EB)^i/\m_0$, which is nothing but the Poynting vector $\vec S$ in~\eqref{ch12:energydensityflux}. On the other hand, the momentum density is $-\Theta^{0i}/c=\eps_0(\vekt EB)^i$.
\item The momentum flux stems from from the \emph{Maxwell stress tensor}
\begin{equation}
\Theta^{ij}=\eps_0E^iE^j+\frac1{\m_0}B^iB^j-\d^{ij}\biggl(\frac12\eps_0\vec E^2+\frac1{2\m_0}\vec B^2\biggr)\;.
\end{equation}
\end{itemize}

%%%%%%%%%%%%%%%%%%%%%%%%%%%%%%%%%%%%%%%%%%%%%%%%%%%%%%%%%%%%

\section{Modifications of Electromagnetism}
\label{sec:EMextensions}

Throughout the whole course, I have tried to emphasize the conceptual unity of classical mechanics and field theory by showing how a relatively small set of basic ideas can be successfully applied in many different contexts. Most of the time, we however had to take a given Lagrangian or action for granted. In this \chaptername, we have changed the point of view and demonstrated how the form of the action can be severely constrained using a priori expected symmetries. We shall now pursue this approach to its logical conclusion and show what kinds of other dynamics are \emph{possible} if one starts from Maxwell's electromagnetism and drops some of the assumptions we made previously. We will thus conclude the \chaptername{} and the whole course by offering an insight into the theoretically driven ``model-building'' mindset, which is particularly common in high-energy physics.

The list of possible modifications of electromagnetism compiled below is certainly not exhaustive. On the other hand, every single theory included in the list has proven to be relevant in some corner of physics, sometimes in multiple ones. This underlines the universality of local field theory: given the set of degrees of freedom and required symmetries, there is only a finite number of options how to construct the action. It is then not so surprising that the same type of physical theory may appear in vastly different contexts.

\runinhead{Nonlinear Electrodynamics} The modification of electromagnetism that is in a certain sense least dramatic is based on dropping the superposition principle. In the absence of sources and under the assumption that the EoM remains of first order in the derivatives of electromagnetic fields, the Lagrangian density should be a Lorentz-scalar function of the field-strength tensor, without any further derivatives. From before, we know two simple Lorentz-scalar objects one can build out of $F^{\m\n}$, namely $F_{\m\n}F^{\m\n}$ and $F_{\m\n}G^{\m\n}$, where
\begin{equation}
G^{\m\n}\equiv\frac12\ve^{\m\n\a\b}F_{\a\b}
\label{ch12:dualtensor}
\end{equation}
is the \emph{dual} electromagnetic tensor. It turns out that there are no other algebraically independent scalars; a detailed proof of this not-quite-trivial statement is left to~\refpr{pr12:nonlinear}. Taking it for the time being for granted, the Lagrangian density will be some generic function of $F_{\m\n}F^{\m\n}$ and $F_{\m\n}G^{\m\n}$. This describes a class of theories that go under the umbrella name \emph{nonlinear electrodynamics}. There are numerous physically inspired choices of the Lagrangian density; in case of interest, you will find more information in~\cite{Sorokin2022}. Here I mention at least one, which is motivated by quantum electrodynamics. In the latter, the presence of virtual electrons and positrons induces a feeble self-interaction among photons that is absent in classical electromagnetism. This interaction can be encoded in a classical action for the electromagnetic field itself. The dominant effect corresponds to a term in the Lagrangian density that is of fourth order in $F^{\m\n}$ and describes photon--photon scattering. It is captured by the \emph{Euler--Heisenberg Lagrangian} density,
\begin{equation}
\La=-\frac1{4\m_0}F_{\m\n}F^{\m\n}+\frac{\hbar^3\a^2}{90\m_0^2m_e^4c^5}\biggl[(F_{\m\n}F^{\m\n})^2+\frac74(F_{\m\n}G^{\m\n})^2\biggr]+\dotsb\;,
\end{equation}
where $m_e$ is the electron mass and $\a\equiv e^2/(4\pi\eps_0\hbar c)$ the fine-structure constant. The ellipsis indicates further contributions of higher order in the field-strength tensor.

\runinhead{Axion Electrodynamics} In our derivation of the Lagrangian density~\eqref{ch12:LagEM}, we rushed to discard the term proportional to $F_{\m\nu}G^{\m\n}$ as being irrelevant for the EoM. Yet, this term can still lead to observable effects. In turns out that under reasonable assumptions and for reasons related to topology, the parameter $v$ in~\eqref{ch12:Lagansatz} can only take two different values. In the vacuum, we have $v=0$. However, there are materials called \emph{topological insulators} where $v$ can be nonzero. This has no effect on the propagation of electromagnetic fields inside such materials. The difference in the values of $v$ becomes relevant at an interface between a topological insulator and vacuum. Intriguingly, the presence of the $v$-term in the Lagrangian density leads to the effective appearance of a \emph{magnetic monopole} as a mirror image of a physical electric charge placed in the vacuum near the interface. See~\cite{Qi2009} for a more detailed but accessible discussion of this phenomenon.

A remarkable twist in the story happens if we promote the constant $v$ to a dynamical degree of freedom. This is done by adding to the electromagnetic field $A^\m$ a (pseudo)scalar field $\p$ and extending the Lagrangian density~\eqref{ch12:LagEM} to\footnote{From now on I resort to the high-energy physics units in which the speed of light as well as the vacuum permeability $\m_0$ are set to one.}
\begin{equation}
\La=-\frac12(\de_\m\p)^2-\frac12m^2\p^2-\frac14F_{\m\n}F^{\m\n}-\frac C8\p\ve^{\m\n\a\b}F_{\m\n}F_{\a\b}\;,
\label{ch12:Lagaxion}
\end{equation}
where $m$ is the mass of the scalar mode and $C$ a constant, measuring the strength of the interaction between the scalar and the photon. This theory is called \emph{axion electrodynamics}. It describes electromagnetic interactions of the axion, a hypothetical particle that might constitute one of the components of dark matter. The same type of theory also describes the electromagnetic interactions of the neutral pion $\pi^0$, which is an electrically neutral pseudoscalar meson with a lifetime of about $8.5\times10^{-17}\,\mathrm{s}$. In this case, the field $\p$ represents the neutral pion and the coefficient $C$ is analytically calculable. The axion-like coupling of $\pi^0$ to the electromagnetic field is responsible for the anomalous decay of the neutral pion into a pair of photons. Some further properties of axion electrodynamics are discussed in~\refpr{pr12:axion}.

\runinhead{Proca Theory} Another way to modify electromagnetism is to give up the otherwise sacred requirement of gauge invariance. This opens the possibility to add many new terms to the Lagrangian density. However, maintaining the assumption of linearity of the EoM singles out a term of the type $A_\m A^\m$ as the simplest possible modification of Maxwell's theory. This leads to the \emph{Proca theory},
\begin{equation}
\La=-\frac14F_{\m\n}F^{\m\n}-\frac12m^2A_\m A^\m\;.
\label{ch12:Proca}
\end{equation}
This theory turns out to describe massive particles, with mass equal to $m$; cf.~\refpr{pr12:Proca}. Their physics is quite different from that of the photon, that is the quantum of the electromagnetic field. Unlike the photon which can only have two different polarizations (both transverse), a plane wave described by the Proca theory can also be polarized longitudinally. Accordingly, the Proca particle has three degrees of freedom, as appropriate for massive particles of spin one.

\runinhead{Maxwell--Chern--Simons Theory} For our final example of modified electromagnetism, we switch to two spatial (or three spacetime) dimensions. Here we do not need to give up any of our requirements on the Lagrangian density to find something interesting. Namely, it is possible to add a new term to the Lagrangian density that is Lorentz-invariant, bilinear in $A^\m$, and preserves gauge invariance up to a surface term. The result is conventionally written as
\begin{equation}
\La=-\frac14F_{\m\n}F^{\m\n}+\frac{k}{4\pi}\ve^{\l\m\n}A_\l\de_\m A_\n\;,
\label{ch12:ChernSimons}
\end{equation}
where $k$ is a constant and $\ve^{\l\m\n}$ is the three-dimensional Levi-Civita symbol, normalized so that $\ve^{012}=1$. This is the \emph{Maxwell--Chern--Simons theory}. Intriguingly, the new term in the Lagrangian density makes the excitations described by the $A^\m$ field massive without breaking gauge invariance as in the Proca theory. The basic properties of the Maxwell--Chern--Simons theory are worked out in~\refpr{pr12:ChernSimons} and~\refpr{pr12:ChernSimonsspectrum}. An even more remarkable modification of electromagnetism is obtained by removing the standard kinetic term $-(1/4)F_{\m\n}F^{\m\n}$ altogether and only keeping the second term in~\eqref{ch12:ChernSimons}. This is the \emph{Chern--Simons theory} which, among others, describes the low-energy physics of the \emph{fractional quantum Hall effect}.

%%%%%%%%%%%%%%%%%%%%%%%%%%%%%%%%%%%%%%%%%%%%%%%%%%%%%%%%%%%%

\section*{\probsec}
\addcontentsline{toc}{section}{\probsec}

\begin{prob}
\label{pr12:invariants}
The combinations $F_{\m\n}F^{\m\n}$ and $\ve^{\m\n\a\b}F_{\m\n}F_{\a\b}$ should by construction be invariant under Lorentz transformations. Rewrite them in terms of the electric and magnetic fields $\vec E$ and $\vec B$, and check their Lorentz invariance explicitly using the transformation rules~\eqref{ch12:EBtransfo} for the latter. What are the values of these two invariants for a linearly polarized electromagnetic plane wave in vacuum?
\end{prob}

\begin{prob}
\label{pr12:EBfields}
Suppose that the electric and magnetic fields $\vec E$ and $\vec B$ are both constant, that is independent of both space and time. Show that one can then always find an IRF in which (at least) one of the following conditions is satisfied:
\begin{itemize}
\item Both $\vec E$ and $\vec B$ are zero.
\item One of $\vec E$ and $\vec B$ is zero, and the other is aligned along the positive $x$-semiaxis.
\item The $\vec E$ and $\vec B$ fields are perpendicular to each other and aligned respectively with the positive $x$ and $y$ semiaxes.
\item The $\vec E$ and $\vec B$ fields are parallel to each other and aligned along the $x$-axis.
\end{itemize}
\end{prob}

\begin{prob}
\label{pr12:Pontryagin}
Show by a direct calculation that adding a term proportional to $\ve^{\m\n\a\b}F_{\m\n}F_{\a\b}$ to the Lagrangian density of the electromagnetic field does not change its EoM. This suggests that we are dealing with a surface term. There should be a ``current'' $K^\mu$ such that $\ve^{\m\n\a\b}F_{\m\n}F_{\a\b}=\de_\m K^\m$. Try to guess what this current looks like.
\end{prob}

\begin{prob}
\label{pr12:EMconservation}
Calculate the divergence of the Hilbert EM tensor~\eqref{ch12:EMtensorHilbert} in presence of charged matter with four-current $J^\m$. Give a physical interpretation of your result. Hint: you should find that on the solutions of the EoM, $\de_\m\Theta^{\m\n}$ is proportional to $F^{\n\a}J_\a$. To bring the divergence to this form, you will need to use the Bianchi identity.
\end{prob}

\begin{prob}
\label{pr12:nonlinear}
In this problem, you will show in two different ways that any Lorentz-scalar  function built out of $F^{\m\n}$ can be written in terms of the fundamental invariants $F_{\m\n}F^{\m\n}$ and $F_{\m\n}G^{\m\n}$, where $G^{\m\n}$ is the dual tensor~\eqref{ch12:dualtensor}. You may want to start by checking that this is correct for some obvious candidate Lorentz scalars, such as
\begin{equation}
G_{\m\n}G^{\m\n}=-F_{\m\n}F^{\m\n}=2\biggl(\frac{\vec E^2}{c^2}-\vec B^2\biggr)\quad\text{or}\quad
\det F^{\m\n}=\frac{(\skal EB)^2}{c^2}\;.
\end{equation}
By the way, the last relation is an example of the general identity $\det A=(\operatorname{pf}A)^2$, valid for any antisymmetric matrix $A$, where $\operatorname{pf}A$ is its \href{https://en.wikipedia.org/wiki/Pfaffian}{Pfaffian}. If you are ambitious, you may also try something more advanced such as
\begin{equation}
F^{\k\l}F_{\l\m}F^{\m\n}F_{\n\k}=2\biggl(\frac{\vec E^2}{c^2}-\vec B^2\biggr)^2+4\frac{(\skal EB)^2}{c^2}\;.
\end{equation}
Why does this work? The first proof is indirect. Explain why, taking into account the freedom to choose an IRF at will, the values of the electric and magnetic fields $\vec E$ and $\vec B$ are completely fixed by merely two parameters. You can use the results of~\refpr{pr12:EBfields}. This shows that there can be at most two independent Lorentz-invariant combinations of $\vec E$ and $\vec B$. For the second proof, show that upon a Lorentz transformation $x'^\m=\Lambda^\m_{\phantom\m\n}x^\n$, the tensor $F^\m_{\phantom\m\n}=F^{\m\l}g_{\l\n}$, treated as a $4\times4$ matrix $F$, transforms by the similarity transformation $F'(x')=\Lambda F(x)\Lambda^{-1}$. This means that the eigenvalues of $F^\m_{\phantom\m\n}$ are Lorentz-invariant, and any Lorentz-scalar quantity built out of $F^\m_{\phantom\m\n}$ depends only on these eigenvalues. By evaluating the characteristic equation for $F^\m_{\phantom\m\n}$, convince yourself that its eigenvalues in turn only depend on the two fundamental invariants $F_{\m\n}F^{\m\n}$ and $F_{\m\n}G^{\m\n}$. Remark: at a somewhat advanced level, there is a direct and constructive proof that any Lorentz-invariant polynomial built of $F^{\m\n}$ can be reduced to the above two fundamental invariants. More details can be found in~\cite{Escobar2014}.
\end{prob}

\begin{prob}
\label{pr12:axion}
Recall the axion electrodynamics as defined by the Lagrangian density~\eqref{ch12:Lagaxion}. Derive the EoM for $\p$ and for the electromagnetic field. In particular, show that the coupling between $\p$ and $F_{\m\n}$ defined by the last term in the Lagrangian density acts as an additional electric charge density proportional to $\skal B\nabla\p$, and additional electric current density proportional to $\vec B\dot\p-\vekt E\nabla\p$.
\end{prob}

\begin{prob}
\label{pr12:Proca}
Show that the Proca theory~\eqref{ch12:Proca} describes three independent degrees of freedom, all corresponding to a particle with mass $m$. These can be interpreted as the polarizations of a massive spin-1 particle. Hint: use the EoM to deduce that for any nonzero $m$, the vector field $A_\m$ must satisfy the condition $\de_\m A^\m=0$.
\end{prob}

\begin{prob}
\label{pr12:ChernSimons}
Convince yourself that the Lagrangian density~\eqref{ch12:ChernSimons} of the Maxwell--Chern--Simons theory is not gauge-invariant, but rather changes upon a gauge transformation by a surface term. Show that the corresponding EoM nevertheless is gauge-invariant, and takes the form
\begin{equation}
\de_\n F^{\m\n}=\frac k{4\pi}\ve^{\m\a\b}F_{\a\b}\;.
\label{ch12:MaxwellChernSimonsEoM}
\end{equation}
\end{prob}

\begin{prob}
\label{pr12:ChernSimonsspectrum}
Show that the Maxwell--Chern--Simons theory describes relativistic particles of mass~$|k|/(2\pi)$. Hint: use the Lorenz gauge and search for plane-wave solutions of the form $A^\m(x)=\hat A^\m\E^{\I p\cdot x}$, where $\hat A^\m$ is an amplitude and $p^\m$ the three-momentum of the wave. You may need the identity $\smash{\ve^{\m\a\b}\ve_{\m\g\d}=-(\d^\a_\g\d^\b_\d-\d^\a_\d\d^\b_\g)}$; the overall minus sign comes from the signature of the Minkowski metric.
\end{prob}
\appendix
\chapter{Solutions of Exercise Problems}
\label{app:solutions}

%%%%%%%%%%%%%%%%%%%%%%%%%%%%%%%%%%%%%%%%%%%%%%%%%%%%%%%%%%%%

\section{Problems in \chaptername~\ref*{chap:mathintro}}

\begin{sol}{pr00:Lagmech}
The EL equation of the given functional $S_\mathrm{L}$ reads
\begin{equation}
m\ddot{\vec r}+\grad V(\vec r)=\vec0\;.
\end{equation}
This is nothing but Newton's second law for a particle of mass $m$, acted upon by the force $\vec F=-\grad V$ of the external field.
\end{sol}

\begin{sol}{pr00:Hammech}
The EL equations for $\vec r$ and $\vec p$ are, respectively,
\begin{equation}
\dot{\vec p}+\grad V(\vec r)=\vec0\;,\qquad
\dot{\vec r}-\frac{\vec p}m=\vec0\;.
\end{equation}
The former equation takes the form of Newton's second law with $\vec p$ being the momentum of the particle. The relation between the momentum and velocity of the particle is fixed by the latter equation. Since the density of the functional $S_\mathrm{H}$ does not depend on the time derivative of momentum, no integration by parts is required when deriving the EL equation for $\vec p$. This means that we do not have impose any specific boundary condition on $\vec p(t)$ at $t_1$ and $t_2$.
\end{sol}

\begin{sol}{pr00:SGkink}
The density of the given functional does not depend explicitly on $x$, we can therefore use the first integral~\eqref{ch00:firstintenergy},
\begin{equation}
I_x=2\sin^2\frac\p2-\frac12(\p')^2\;.
\end{equation}
According to the boundary condition, $\p$ converges to a constant value for which $\sin\p/2=0$ for both $x\to\pm\infty$. This ensures that $I_x=0$. The latter condition is a first-order ODE for $\p$ that can be directly integrated, leading to
\begin{equation}
\p(x)=4\arctan\E^{x-x_0}\;,
\label{ch00:SGkink}
\end{equation}
where $x_0$ is an integration constant indicating the position of the center of the sine-Gordon kink. It remains to see why our solution is an actual minimum of the functional $F[\p]$. To that end, we rewrite the functional as
\begin{equation}
\begin{split}
F[\p]&=\int_{-\infty}^{+\infty}\D x\,\biggl[\frac12(\p')^2+2\sin^2\frac\p2\biggr]\\
&=\int_{-\infty}^{+\infty}\D x\,\biggl[\frac12\biggl(\p'-2\sin\frac\p2\biggr)^2+2\p'\sin\frac\p2\biggr]\\
&=\frac12\int_{-\infty}^{+\infty}\D x\,\biggl(\p'-2\sin\frac\p2\biggr)^2+2\int_0^{2\pi}\D\p\,\sin\frac\p2\\
&=8+\frac12\int_{-\infty}^{+\infty}\D x\,\biggl(\p'-2\sin\frac\p2\biggr)^2\;.
\end{split}
\end{equation}
Note how we used the boundary condition to convert the second part of the integral to a constant independent on the concrete profile $\p(x)$, interpolating between the two boundary values. The remaining integral is obviously \emph{minimized} by a profile such that $\p'=2\sin(\p/2)$, which is equivalent to the condition $I_x=0$ we previously used to derive the result~\eqref{ch00:SGkink}. 
\end{sol}

\begin{sol}{pr00:resistance}
Let us start with the solution utilizing an explicit constraint. Since the constraint $\rot\vec E=\vec0$ is local, we need a vector of Lagrange multiplier functions $\vec\l(\vec x)$. Putting this together with the resistive power density, we arrive at the functional
\begin{equation}
F_\l[\vec J,\vec\l]\equiv\int\D^3\vec x\,[\skal JE-\vec\l\cdot(\rot\vec E)]=\int\D^3\vec x\,\biggl[\frac{\vec J^2}\s-\vec\l\cdot\biggl(\rot\frac{\vec J}\s\biggr)\biggr]\;,
\end{equation}
where we used the Ohm law to eliminate the electric field in favor of the current density. The ensuing EL equation for $\vec J$ reads
\begin{equation}
\vec J=\frac12\rot\vec\l\;,
\end{equation}
which implies $\divg\vec J=0$. What is the physical interpretation? Demanding that $\vec E$ is conservative is tantamount to Kirchhoff's voltage law. The result of solving the variational problem for the resistive power, that is the condition on $\vec J$ having vanishing divergence, is in turn equivalent to Kirchhoff's current law.

An alternative way to arrive at the same result is to realize that a conservative electrostatic field can be represented by its potential $\p$ through $\vec E=-\grad\p$. We now do not need any Lagrange multiplier, but rather define the functional to be minimized directly as the total power dissipated by the medium,
\begin{equation}
F[\p]\equiv\int\D^3\vec x\,\skal JE=\int\D^3\vec x\,\s(\grad\p)^2\;.
\end{equation}
The corresponding EL equation is $\divg(\s\grad\p)=-\divg\vec J=0$, giving us the same result for the current distribution as before.
\end{sol}

\begin{sol}{pr00:isoperimetric}
Let us start with the approach based on polar coordinates. With $r=f(\vp)$, the area enclosed by the curve $\Gamma$ is given by the functional
\begin{equation}
A[f]\equiv\frac12\int_0^{2\pi}\D\vp\,[f(\vp)]^2\;.
\end{equation}
With the infinitesimal line element $\D s=\sqrt{\D r^2+r^2\D\vp^2}=\D\vp\sqrt{[f(\vp)]^2+[f'(\vp)]^2}$, we also readily construct a functional representing the length of the curve,
\begin{equation}
L[f]\equiv\int_0^{2\pi}\D\vp\,\sqrt{[f(\vp)]^2+[f'(\vp)]^2}\;.
\end{equation}
We are dealing with a global constraint on the value of $L[f]$. We therefore need to introduce a constant Lagrange multiplier $\l$, which gives rise to the modified area functional $A_\l[f]\equiv A[f]-\l(L[f]-\ell)$, where $\ell$ is the prescribed length of $\Gamma$. The corresponding EL equation for $f(\vp)$ is
\begin{equation}
f-\frac{\l f}{\sqrt{f^2+(f')^2}}+\l\OD{}\vp\frac{f'}{\sqrt{f^2+(f')^2}}=0\;.
\end{equation}
This is obviously solved by a constant $f(\vp)=\l$, which suggests that the optimal choice of shape of $\Gamma$ is a circle. However, within the present approach using polar coordinates, it is not easy to understand why the constant is the only relevant solution of the EL equation. Let us therefore switch gears.

The alternative approach to the problem is based on a parametric description of the curve $\Gamma$ in terms of Cartesian coordinates, $(x(t),y(t))$. We choose the parameterization so that going around $\Gamma$ amounts to taking $t$ from $0$ to $1$. The area and length functionals can now be expressed as (check this in detail!)
\begin{equation}
A[x,y]\equiv-\int_0^1\D t\,\dot x(t)y(t)\;,\qquad
L[x,y]\equiv\int_0^1\D t\,\sqrt{[\dot x(t)]^2+[\dot y(t)]^2}\;.
\end{equation}
Again, we need to introduce a Lagrange multiplier $\l$, upon which we are searching for stationary points of the modified area functional $A_\l[x,y]\equiv A[x,y]-\l(L[x,y]-\ell)$. In this formulation, $x$ is a cyclic coordinate, which gives the first integral
\begin{equation}
I_x=-y-\frac{\l\dot x}{\sqrt{\dot x^2+\dot y^2}}\;.
\end{equation}
In addition, we have the EL equation for $y$,
\begin{equation}
-\dot x+\l\OD{}t\frac{\dot y}{\sqrt{\dot x^2+\dot y^2}}=0\;.
\end{equation}
This is a total time derivative and so can be immediately integrated once. Using the first integral $I_x$ to eliminate the square root in the denominator of the result, we end up with a single first-order ODE,
\begin{equation}
\dot x(x+c)+\dot y(y+I_x)=0\;,
\end{equation}
where $c$ is an integration constant. This is equivalent to
\begin{equation}
\OD{}t\bigl[(x+c)^2+(y+I_x)^2\bigr]=0\;,
\end{equation}
which describes a circle centered at the point $(x,y)=(-c,-I_x)$. The radius of the circle is, naturally, $\ell/(2\pi)$.
\end{sol}

\begin{sol}{pr00:snell}
We parameterize the path between $A$ and $B$ as $\vec r(t)$ so that the endpoints are at $t=0$ and $t=1$. With the time to traverse an infinitesimal segment of the path equal $\D t=\D s/(c/n)$, the total time of travel between $A$ and $B$ is given by the~functional
\begin{equation}
T[\vec r]\equiv\frac1c\int_0^1\D t\,n(\vec r(t))\abs{\dot{\vec r}(t)}=\frac1c\int_0^1\D t\,n(\vec r(t))\sqrt{[\dot{\vec r}(t)]^2}\;.
\end{equation}
The corresponding EL equation takes the form
\begin{equation}
\OD{}t\frac{n\dot{\vec r}}{\abs{\dot{\vec r}}}=\abs{\dot{\vec r}}\grad n\;.
\end{equation}
We can eliminate the dependence on the arbitrary parameter $t$ by going back to the line segment $\D s=\abs{\dot{\vec r}}\D t$ and realizing that $\dot{\vec r}/\abs{\dot{\vec r}}\equiv\vec\tau$ is the unit tangent vector to the trajectory of the light ray. This allows us to rewrite the EL equation as
\begin{equation}
\OD{(n\vec\tau)}{s}=\grad n\;.
\end{equation}
This differential equation governs the geometry of the trajectory actually followed by light in an inhomogeneous optical medium.

Suppose now that the refraction index only varies along one Cartesian coordinate, say $x$. This implies that $\grad n$ must be parallel to $\vec e_x$, the basis unit vector in the $x$-direction. Taking a cross product of the EL equation with $\vec e_x$, we find that
\begin{equation}
\OD{}s(n\vec e_x\times\vec\tau)=\vec0\;.
\end{equation}
Thus, $n\vec e_x\times\vec\tau$ is constant along the trajectory. It follows that the light ray propagates in a single plane containing $\vec e_x$. Moreover, the combination $n\sin\t$ where $\t$ is the angle between $\vec e_x$ and $\vec\tau$ remains constant. This is exactly the content of Snell's law.
\end{sol}

%%%%%%%%%%%%%%%%%%%%%%%%%%%%%%%%%%%%%%%%%%%%%%%%%%%%%%%%%%%%

\section{Problems in \chaptername~\ref*{chap:Lagmechanics}}

\begin{sol}{pr01:atwood}
Let us choose as the sole generalized coordinate $q$ the (vertical) displacement of mass $m_1$ from an arbitrarily chosen reference point. We will use the convention that positive $q$ means upward motion and negative $q$ downward motion. The change of potential energy of the system with respect to the reference state is then $m_1gq-m_2gq$. Altogether, the Lagrangian reads
\begin{equation}
L=\frac12(m_1+m_2)\dot q^2-(m_1-m_2)gq\;.
\end{equation}
The EoM is thus
\begin{equation}
\ddot q=\frac{m_2-m_1}{m_2+m_1}g\;.
\end{equation}
This agrees with the prediction of Newton's laws.
\end{sol}

\begin{sol}{pr01:sphericalpendulum}
Since the pendulum is forced by the string to move on the surface of a sphere with radius $L$, it is natural to use the spherical angles $\t,\vp$ as the generalized coordinates. To conform with standard conventions, let us use $\t$ for the angular deviation of the pendulum from the vertical direction, and $\vp$ as the azimuthal angle in the horizontal plane. The potential energy is then $-mgL\cos\t$, and the entire Lagrangian reads
\begin{equation}
L=\frac12mL^2(\dot\t^2+\dot\vp^2\sin^2\t)+mgL\cos\t\;.
\end{equation}
The Lagrange equations for $\t,\vp$ are, respectively,
\begin{equation}
\begin{split}
mL^2\ddot\t&=mL^2\dot\vp^2\sin\t\cos\t-mgL\sin\t\;,\\
\OD{}t(mL^2\dot\vp\sin^2\t)&=0\;.
\end{split}
\end{equation}
Uniform circular motion in a horizontal plane corresponds to constant $\t$ and $\vp$ that is a linear function of time, that is constant $\dot\vp$. The EoM for $\vp$ is identically satisfied for any choice of the constant $\t$ and $\dot\vp$. On the other hand, the EoM for $\t$ requires that
they are related by
\begin{equation}
\dot\vp^2=\frac{g}{L\cos\t}\;.
\end{equation}
This fixes the angular velocity of the horizontal motion up to a choice of orientation.
\end{sol}

\begin{sol}{pr01:pulley}
We introduce two generalized coordinates: the (upward) vertical displacement $z$ of the block $m$ from a chosen reference position, and the extension $x$ of the spring. In terms of these variables, the linear velocities of the two blocks are, respectively, $\dot z$ and $\dot z+\dot x$. Adding the contributions of the gravitational potential energy and the elastic energy of the spring, we get the Lagrangian,
\begin{equation}
L=\frac12m\dot z^2+\frac12M(\dot z+\dot x)^2-mgz-\frac12kx^2\;.
\end{equation}
The corresponding Lagrange equations for $z$ and $x$ are
\begin{equation}
\begin{split}
m\ddot z+M(\ddot z+\ddot x)&=-mg\;,\\
M(\ddot z+\ddot x)&=-kx\;.
\end{split}
\end{equation}
The first of the equations can be used to eliminate $\ddot z$ in favor of $\ddot x$,
\begin{equation}
\ddot z=-\frac{mg}{M+m}-\frac{M\ddot x}{M+m}\;.
\label{ch01:zfromx}
\end{equation}
Inserting this in turn in the second of the Lagrange equations, we arrive at an EoM for $x$ alone,
\begin{equation}
\m\ddot x+kx=\m g\;,
\end{equation}
where $\m\equiv Mm/(M+m)$ is the reduced mass of the system. This describes harmonic oscillations of $x$ around the mean value $\m g/k$ with squared angular frequency $k/\m$. By~\eqref{ch01:zfromx}, the oscillations also propagate to the vertical coordinate $z$, where they are superimposed on motion with constant acceleration $-mg/(M+m)$.
\end{sol}

\begin{sol}{pr01:exam2021}
The potential energy does not depend explicitly on time, which guarantees conservation of total energy as expressed by the Hamilton function,
\begin{equation}
H=\frac12m\dot x^2+V(x)=\frac12m\dot x^2-c(x^2-a^2)^2\;.
\end{equation}
From the given boundary condition, both $\dot x$ and $V(x)$ vanish at $t\to\pm\infty$, hence the total energy must be zero. By setting $H=0$, we find
\begin{equation}
\dot x=\sqrt{\frac{2c}m}(a^2-x^2)\;.
\end{equation}
We used the fact that by the boundary condition, the velocity $\dot x$ must be positive in the region $-a<x<+a$. This equation is easily solved by direction integration,
\begin{equation}
\frac1a\operatorname{arctanh}\frac xa=\sqrt{\frac{2c}m}(t-t_0)\;,
\end{equation}
where the integration constant $t_0$ indicates the time at which the particle passes through the origin, $x=0$. Expressing finally $x$ as a function of time, we get
\begin{equation}
x(t)=a\tanh\biggl[\sqrt{\frac{2c}m}\,a(t-t_0)\biggr]\;.
\end{equation}
\end{sol}

\begin{sol}{pr01:freefall}
The Lagrangian is obviously
\begin{equation}
L=\frac12m\dot z^2-mgz\;,
\end{equation}
and the corresponding EoM reads
\begin{equation}
m\ddot z=-mg\;.
\end{equation}
This describes, hardly surprisingly, vertical motion with constant acceleration $-g$. Upon translation to the new variable $Z$ defined by $z\equiv Z-(1/2)gt^2$, we get
\begin{equation}
\begin{split}
L&=\frac12m(\dot Z-gt)^2-mg\left(Z-\frac12gt^2\right)\\
&=\frac12m\dot Z^2+\OD{}t\left(-mgZt+\frac13mg^2t^3\right)\;.
\end{split}
\end{equation}
We know that adding a total derivative to the Lagrangian does not change the EoM, so our Lagrangian is equivalent to that of a free particle, $(1/2)m\dot Z^2$. The physical interpretation of this result is that by replacing $z$ with $Z$, we have effectively switched into a freely falling reference frame. In this frame, there is a new inertial force that exactly cancels the force of gravity, hence the particle behaves as free.
\end{sol}

\begin{sol}{pr01:particleinEMfield}
Let us start by rewriting the known EoM of the charged particle subject to the Lorentz force in terms of the electromagnetic potentials $\p,\vec A$. Recalling that
\begin{equation}
\vec E=-\grad\p-\PD{\vec A}t\;,\qquad
\vec B=\rot\vec A\;,
\end{equation}
the EoM can be cast as
\begin{equation}
\begin{split}
m\ddot{\vec r}&=q(\vec E+\dot{\vec r}\times\vec B)=-q\left(\grad\p+\PD{\vec A}t\right)+q\dot{\vec r}\times(\rot\vec A)\\
&=-q\left(\grad\p+\PD{\vec A}t\right)+q[\grad(\dot{\vec r}\cdot\vec A)-(\dot{\vec r}\cdot\grad)\vec A]\;,
\end{split}
\label{ch01:Lorentz}
\end{equation}
where the gradient operator $\grad$ is understood to act only on the $\vec r$-dependence of the potentials, not on $\dot{\vec r}$.

Next, let us see whether we can recover this result using the given Lagrangian. We start by working out the derivatives of the Lagrangian with respect to the position and velocity vectors,
\begin{equation}
\PD{L}{\vec r}=-q\grad\p+q\grad(\dot{\vec r}\cdot\vec A)\;,\qquad
\PD{L}{\dot{\vec r}}=m\dot{\vec r}+q\vec A\;.
\end{equation}
The Lagrange equation for $\vec r$ is then equivalent to
\begin{equation}
\begin{split}
0=\OD{}t\PD{L}{\dot{\vec r}}-\PD{L}{\vec r}&=m\ddot{\vec r}+q\left[\PD{\vec A}t+(\dot{\vec r}\cdot\grad)\vec A\right]-[-q\grad\p+q\grad(\dot{\vec r}\cdot\vec A)]\\
&=m\ddot{\vec r}+q\left(\grad\p+\PD{\vec A}t\right)-q[\grad(\dot{\vec r}\cdot\vec A)-(\dot{\vec r}\cdot\grad)\vec A]\;,
\end{split}
\end{equation}
which agrees with~\eqref{ch01:Lorentz}.

The Hamilton function corresponding to the Lagrangian is readily evaluated as
\begin{equation}
H=\frac12m\dot{\vec r}^2+q\p\;.
\end{equation}
Unlike the Lagrangian itself, this can be interpreted as the kinetic energy of the particle plus its potential energy with respect to the electric field. The magnetic field does not contribute to the energy of the particle. This is in accord with the fact that the magnetic part of the Lorentz force is perpendicular to the velocity, and thus cannot do any work.
\end{sol}

\begin{sol}{pr01:exam2021re}
The conserved energy of the system is given by the Hamiltonian,
\begin{equation}
E=H=\dot q\PD L{\dot q}-L=\frac12q\dot q^2\;.
\end{equation}
Given that we know the velocity to be positive, this has a unique solution for $\dot q$,
\begin{equation}
\dot q=\sqrt{\frac{2E}q}\;.
\end{equation}
This is a first-order ODE for $q$, which is easily solved by direct integration. With the help of the given initial condition on $q$, we find that
\begin{equation}
q(t)=\left(\frac{9E}2\right)^{1/3}t^{2/3}\;.
\end{equation}
\end{sol}

\begin{sol}{pr01:exam2023}
The energy of the system is given by the Hamilton function,
\begin{equation}
H=\dot x\PD L{\dot x}+\dot y\PD L{\dot y}-L=\frac\l4(x^2+y^2-v^2)^2\;.
\end{equation}
This shows that the energy of the system is bounded from below, and solutions of lowest possible energy should satisfy $x^2+y^2=v^2$ at all times. To see what these solutions look like, we use the Lagrangian to find the equations of motion,
\begin{equation}
\dot x=-\l y(x^2+y^2-v^2)\;,\qquad
\dot y=+\l x(x^2+y^2-v^2)\;.
\end{equation}
For $x^2+y^2=v^2$, both generalized velocities are zero. The trajectories with the lowest possible energy therefore correspond to any constant $x_0$ and $y_0$ such that $x_0^2+y_0^2=v^2$.
\end{sol}

%%%%%%%%%%%%%%%%%%%%%%%%%%%%%%%%%%%%%%%%%%%%%%%%%%%%%%%%%%%%

\section{Problems in \chaptername~\ref*{chap:centralfields}}

\begin{sol}{pr02:shaft}
At radial distance $r$ from the Earth's center, the force acting on the ball of mass $m$ comes from the sphere of radius $r$ and density $\rho$. Denoting the mass of this sphere as $M(r)=(4/3)\pi r^3\rho$, the radial force is therefore
\begin{equation}
\vec F(r)=-\frac{GmM(r)}{r^2}\vec n\;,
\end{equation}
where $G$ is the gravitational constant and $\vec n$ a unit radial vector. By a simple manipulation using the fact that $M(r)=M(R)r^3/R^3$, the force can be rewritten as
\begin{equation}
\vec F(r)=-\frac{GmM(R)}{r^2}\frac{r^3}{R^3}\vec n=-\frac{GmM(R)r}{R^3}\vec n=-mg\frac rR\vec n\;,
\end{equation}
where $g=GM(R)/R^2$ is the gravitational acceleration at the surface. We find that the force acting on the ball grows linearly with distance from the Earth's center. It can be encoded in the potential energy $V(r)=mgr^2/(2R)$. Altogether, the motion of the ball in the shaft is governed by the effective one-dimensional Lagrangian
\begin{equation}
L=\frac12m\dot r^2-\frac12m\frac gRr^2\;.
\end{equation}
The corresponding EoM, $m\ddot r=-(mg/R)r$, describes harmonic oscillations with angular frequency $\omega=\sqrt{g/R}$. The trajectory connecting the point on the surface where the ball was thrown into the shaft with its antipode is a half of the period of the oscillator. The time needed to reach the antipodal point is therefore
\begin{equation}
t=\frac\pi\omega=\pi\sqrt{\frac Rg}\;.
\end{equation}
Numerically, this evaluates to about $42\,\text{min}$.
\end{sol}

\begin{sol}{pr02:2body}
Denoting the uniform external gravitational field as $\vec g$, the complete Lagrangian including the pair interaction and the gravitational and harmonic potentials reads
\begin{equation}
L=\frac12m_1\dot{\vec r}_1^2+\frac12m_2\dot{\vec r}_2^2-V(\lvert\vec r_2-\vec r_1\rvert)+m_1\skal gr_1+m_2\skal gr_2-\frac12m_1\o^2\vec r_1^2-\frac12m_2\o^2\vec r_2^2\;.
\end{equation}
Upon switching to the CoM and relative coordinates~\eqref{ch02:CoMcoordinates}, the Lagrangian becomes
\begin{equation}
L=\underbrace{\frac12M\dot{\vec R}^2+M\skal gR-\frac12M\o^2\vec R^2}_{\text{CoM motion}}+\underbrace{\frac12\m\dot{\vec r}^2-V(\abs{\vec r})-\frac12\m\o^2\vec r^2}_{\text{relative motion}}\;.
\end{equation}
This makes the separation of the CoM and relative motion manifest.
\end{sol}

\begin{sol}{pr02:hydrogenBfield}
Dropping right away the scalar potential $\p$ and using the Poincar\'e gauge for the vector potential, the Lagrangian for our two charges in a uniform magnetic field is
\begin{equation}
L=\frac12m_1\dot{\vec r}_1^2+\frac12m_2\dot{\vec r}_2^2-V(\lvert\vec r_2-\vec r_1\rvert)+\frac12q_1\vec B\cdot(\vec r_1\times\dot{\vec r}_1)+\frac12q_2\vec B\cdot(\vec r_2\times\dot{\vec r}_2)\;,
\end{equation}
where I used the cyclic property of the scalar triple product of vectors to factor out the magnetic field. Also, $V(\lvert\vec r_2-\vec r_1\rvert)$ stands for the Coulomb interaction of the charges. When expressed in terms of the CoM and relative coordinates~\eqref{ch02:CoMcoordinates}, the relevant part of the magnetic term in the Lagrangian becomes
\begin{equation}
\begin{split}
q_1\vec r_1\times\dot{\vec r}_1+q_2\vec r_2\times\dot{\vec r}_2={}&(q_1+q_2)\vec R\times\dot{\vec R}+\frac1{M^2}(q_1m_2^2+q_2m_1^2)\vec r\times\dot{\vec r}\\
&+\frac1M(q_2m_1-q_1m_2)(\vec R\times\dot{\vec r}+\vec r\times\dot{\vec R})\;.
\end{split}
\end{equation}
For the CoM and relative motion to separate the last term has to vanish, which is only possible if
\begin{equation}
\frac{q_1}{m_1}=\frac{q_2}{m_2}\;.
\end{equation}
Provided this condition is satisfied, the full Lagrangian can be simplified to
\begin{equation}
L=\underbrace{\frac12M\dot{\vec R}^2+\frac12Q\vec B\cdot(\vec R\times\dot{\vec R})}_{\text{CoM motion}}+\underbrace{\frac12\m\dot{\vec r}^2-V(\abs{\vec r})+\frac12\bar q\vec B\cdot(\vec r\times\dot{\vec r})}_{\text{relative motion}}\;,
\end{equation}
where $Q\equiv q_1+q_2$ is the total charge and $\bar q$ the reduced charge, defined by $\bar q/\m=q_1/m_1=q_2/m_2$.
\end{sol}

\begin{figure}[t]
\includegraphics[width=\textwidth]{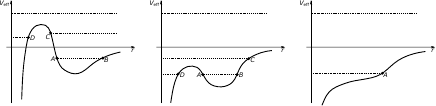}
\caption{Illustration for the solution of~\refpr{pr02:effpot}. The three panels show three qualitatively different regimes of the effective potential~\eqref{ch02:effpotprsol}, depending on the magnitude of the parameter~$\ell$. Left panel: $\ell<J^4/(16\m^2k)$. Middle panel: $J^4/(16\m^2k)<\ell<J^4/(12\m^2k)$. Right panel: $J^4/(12\m^2k)<\ell$.}
\label{fig02:solution:effpot}
\end{figure}

\begin{sol}{pr02:effpot}
The effective potential is
\begin{equation}
V_\mathrm{eff}(r)=-\frac{k}{r}-\frac{\ell}{r^3}+\frac{J^2}{2\m r^2}\;.
\label{ch02:effpotprsol}
\end{equation}
For $r\to0$, the effective potential is dominated by the $\ell$-term and goes to minus infinity. For $r\to\infty$, on the other hand, it is dominated by the $k$-term, and approaches zero from below. The qualitative behavior of the effective potential between these two limits depends on the magnitude of $\ell$. For very small $\ell$, the effective potential will mostly resemble Fig.~\ref{fig02:effpot}, but as $r$ approaches zero, the negative $\ell$-term will eventually win. The effective potential will therefore have a local maximum with positive energy and a local minimum with negative energy, as shown in the left panel of Fig.~\ref{fig02:solution:effpot}. There will be bound orbits with negative energy ($AB$) and unbound orbits with positive energy (such as that emanating from $C$). There will also be bound orbits (such as that emanating from $D$), whose energy can be both positive and negative, which will eventually collapse to the origin. Finally, there will be unbound orbits (such as the dashed line above the local maximum) that may both collapse to the origin and run away to infinity, depending on the direction of motion.

With increasing $\ell$, the energy of the local maximum decreases until at some point, it crosses the horizontal axis and becomes negative; see the middle panel of Fig.~\ref{fig02:solution:effpot}. This happens at $\ell=\ell_1$ for which the discriminant of $V_\mathrm{eff}(r)=0$ as a quadratic equation for $r$ vanishes. An explicit calculation gives $\ell_1=J^4/(16\m^2k)$. For $\ell>\ell_1$, we still find bound orbits with negative energy ($AB$). There will also be two types of orbits that are bound and eventually collapse to the origin (dashed lines emanating from $C$ and $D$). Finally, we will find unbound orbits that may both collapse to the origin and run away to infinity (dashed line above the horizontal axis).

Increasing $\ell$ further, the local minimum and maximum of the effective potential will converge to each other until they eventually coalesce at $\ell=\ell_2$. To find this point, we evaluate the discriminant of $V_\mathrm{eff}'(r)=0$ as a quadratic equation for $r$ and set it to zero. This gives $\ell_2=J^4/(12\m^2k)$. For $\ell>\ell_2$, the effective potential does not have any local maximum or minimum; see the right panel of Fig.~\ref{fig02:solution:effpot} for an illustration. In this case, there are only two qualitatively different types of orbits. Those with negative energy (dashed line emanating from $A$) are bound and will eventually collapse to the origin. Orbits with positive energy (upper dashed line), on the other hand, may both collapse to the origin and run away to infinity.
\end{sol}

\begin{sol}{pr02:perturbation}
For the given potential, the Binet equation evaluates to
\begin{equation}
\frac{\D^2u}{\D\vp^2}+\left(1+\frac{2\eps\m}{J^2}\right)u=\frac{k\m}{J^2}\;.
\end{equation}
In other words, the $\eps$-perturbation in the potential changes the ``frequency'' of the cosine term in the orbit equation~\eqref{ch02:Keplerorbitgeneral}, modifying the latter to
\begin{equation}
u(\vp)=\frac{k\m}{J^2}\biggl[1+e\cos\biggl(\sqrt{1+\frac{2\eps\m}{J^2}}\,\vp\biggr)\biggr]\;.
\end{equation}
For small $\eps$, the effect of the perturbation can be interpreted as slow precession of the orbit. Starting from $\vp=0$, the particle returns to the same radial distance for 
\begin{equation}
\vp=\frac{2\pi}{\sqrt{1+\frac{2\eps\m}{J^2}}}\equiv2\pi+\Delta\vp\;.
\end{equation}
We conclude that the angle of precession per orbit is $\Delta\vp\approx-2\pi\eps\m/J^2$.
\end{sol}

\begin{sol}{pr02:periapo}
For an orbit such as the $BC$ line in Fig.~\ref{fig02:effpot}, the distance at peri- and apocenter is found by solving
\begin{equation}
V_\mathrm{eff}(r)=-\frac kr+\frac{J^2}{2\m r^2}\ifeq E
\end{equation}
as a quadratic equation for $1/r$. This gives
\begin{equation}
\frac1{r_{B,C}}=\frac{k\m}{J^2}\Biggl(1\pm\sqrt{1+\frac{2EJ^2}{k^2\m}}\Biggr)\;.
\end{equation}
This agrees with~\eqref{ch02:Keplerorbitgeneral} evaluated at $\vp=0$ and $\vp=\pi$, with the eccentricity $e$ given by~\eqref{ch02:eccentricity}.
\end{sol}

\begin{sol}{pr02:Halley}
Given the period of orbit $T=75\,\text{yr}$ and distance at perihelion $r_\mathrm{per}=a(1-e)=89\times10^6\,\text{km}$, the eccentricity can be computed using Kepler's third law~\eqref{ch02:3rdKepler} with $k/\m=Gm_{\astrosun}$. This leads to
\begin{equation}
e=1-\left(\frac{4\pi^2r_\mathrm{per}^3}{Gm_{\astrosun}T^2}\right)^{1/3}\approx0.967\;.
\end{equation}
To find the speed at perihelion $v_\mathrm{per}$, we evaluate the area velocity~\eqref{ch02:areavelocity} in two different ways, namely as $\pi ab/T$ and as $(1/2)r_\mathrm{per}v_\mathrm{per}$. Using the relation $b=a\sqrt{1-e^2}$ and inserting our previous result for the eccentricity, we arrive at
\begin{equation}
v_\mathrm{per}=\frac{2\pi r_\mathrm{per}}T\sqrt{\frac{1+e}{(1-e)^3}}\approx54\,\text{km/s}\;.
\end{equation}
\end{sol}

\begin{sol}{pr02:collapse}
The point charge will obviously fall into the center with mass $M$ along a straight trajectory. This can be thought of as a half of an ellipse with eccentricity $e=1$ and semi-major axis $a=R/2$. Using this in Kepler's third law, we find that the time needed for the gravitational collapse (equal to half of the orbit period) is
\begin{equation}
t=\sqrt{\frac{\pi^2R^3}{8GM}}=\sqrt{\frac{3\pi}{32G\vr}}\;.
\end{equation}
As the second expression shows, the time of collapse only depends on the initial density $\vr$ of the cloud.
\end{sol}

\begin{sol}{pr02:hodograph}
The suggested cross-product evaluates to
\begin{equation}
\vekt JR=\vec J\times(\vekt pJ)-k\m\vec J\times\frac{\vec r}{r}=J^2\vec p-k\m\vec J\times\frac{\vec r}{r}\;,
\end{equation}
where I used the fact that the vectors $\vec J$ and $\vec p$ are perpendicular to each other. The second and last step is to express the momentum $\vec p$ in terms of everything else,
\begin{equation}
\vec p=\frac{\vekt JR}{J^2}+\frac{k\m}{J^2}\vec J\times\frac{\vec r}{r}\;.
\end{equation}
The first term on the right-hand side is a constant vector. The second term has a constant magnitude and only its direction changes as the particle traverses its elliptic orbit. We conclude that the momentum vector traces out a circle with center $(\vekt JR)/J^2$ and radius $k\m/J$.
\end{sol}

%%%%%%%%%%%%%%%%%%%%%%%%%%%%%%%%%%%%%%%%%%%%%%%%%%%%%%%%%%%%

\section{Problems in \chaptername~\ref*{chap:Hammechanics}}

\begin{sol}{pr03:sphericalpendulum}
We parameterize the position vector $\vec r$ of the particle by the spherical angles $\t,\vp$ through $\vec r=(R\sin\t\cos\vp,R\sin\t\sin\vp,R\cos\t)$. The Lagrangian consists just of the kinetic energy and equals
\begin{equation}
L=\frac12m\dot{\vec r}^2=\frac12mR^2(\dot\t^2+\dot\vp^2\sin^2\t)\;.
\end{equation}
The conjugate momenta corresponding to $\t$ and $\vp$ are, respectively, $p_\t=mR^2\dot\t$ and $p_\vp=mR^2\dot\vp\sin^2\t$. This leads to the Hamiltonian
\begin{equation}
H=\frac{p_\t^2}{2mR^2}+\frac{p_\vp^2}{2mR^2\sin^2\t}\;.
\end{equation}
The Hamilton equations are accordingly
\begin{equation}
\dot\t=\frac{p_\t}{mR^2}\;,\qquad
\dot\vp=\frac{p_\vp}{mR^2\sin^2\t}\;,\qquad
\dot p_\t=\frac{p_\vp^2\cos\t}{mR^2\sin^3\t}\;,\qquad
\dot p_\vp=0\;.
\end{equation}
Treating $p_\vp$ as a constant parameter, the equations for $\t$ and $p_\t$ govern an equivalent one-dimensional problem. Once $\t$ is known as a function of time, the time dependence of $\vp$ can be found by integrating the equation $\dot\vp=p_\vp/(mR^2\sin^2\t)$.
\end{sol}

\begin{sol}{pr03:oldexam}
The Hamilton equations of motion for this problem are
\begin{equation}
\dot q=\PD Hp=qp,\qquad
\dot p=-\PD Hq=-\frac12p^2\;.
\end{equation}
The latter equation only depends on $p$ and so can be easily integrated to
\begin{equation}
p(t)=\frac{p_0}{1+(1/2) p_0t}\;,
\end{equation}
where $p_0$ is the momentum at time $t=0$. This can in turn be expressed in terms of the given parameters $E$ and $q_0$. Namely, from energy conservation, we deduce that $E=(1/2)q_0p_0^2$, hence $p_0=\sqrt{2E/q_0}$. The last step is to plug the solution for $p(t)$ into the first of the Hamilton equations. This gives an equation for $q$ that is likewise easy to integrate. The final answer is
\begin{equation}
q(t)=q_0\left(1+\frac12p_0t\right)^2=q_0+\sqrt{2Eq_0}\,t+\frac12Et^2\;.
\end{equation}
The same result can be found more easily by using the result for $p(t)$ together with energy conservation, $(1/2)q_0p_0^2=(1/2)q(t)p(t)^2$, so that $q(t)=q_0[p_0/p(t)]^2$.

An alternative solution is to notice that as a consequence of the Hamilton equations, the second time derivative of the coordinate $q$ is particularly simple,
\begin{equation}
\ddot q=\dot qp+q\dot p=qp^2-\frac12qp^2=\frac12qp^2=E\;.
\end{equation}
The motion therefore has constant acceleration $E$. Using elementary mechanics, we can write down the trajectory at once, $q(t)=q_0+v_0t+(1/2)Et^2$, where $v_0=\dot q(0)$. It remains to express $v_0$ in terms of the given parameters $E$ and $q_0$. This follows from the first of the Hamilton equations and energy conservation, $v_0=q_0p_0=q_0\sqrt{2E/q_0}=\sqrt{2Eq_0}$. Plugging this back into our solution then gives the final answer for $q(t)$, in accord with the solution obtained by direct integration of Hamilton's equations.

Yet another way to solve the problem is to use energy conservation to eliminate $p$ in favor of $q$, $p=\sqrt{2E/q}$. Together with the first of the Hamilton equations, this gives a first-order differential equation for $q$ alone,
\begin{equation}
\dot q=qp=\sqrt{2Eq}\;.
\end{equation}
This equation is straightforward to solve by direct integration, the result being
\begin{equation}
q=\biggl(\sqrt{q_0}+\sqrt{\frac E2}\,t\biggr)^2=q_0+\sqrt{2Eq_0}\,t+\frac12Et^2\;.
\end{equation}
\end{sol}

\begin{sol}{pr03:Hamflow}
We want to find a Hamiltonian $H(q,p)$ such that
\begin{equation}
\PD{H(q,p)}{q}=-g(q,p)\;,\qquad
\PD{H(q,p)}{p}=f(q,p)\;.
\end{equation}
Using the symmetry of second partial derivatives, this gives the necessary condition
\begin{equation}
\PD fq=\frac{\de^2H}{\de q\de p}=\frac{\de^2H}{\de p\de q}=-\PD gp\;.
\end{equation}
This consistency condition is satisfied by the choice $f(q,p)=p-q^2$ and $g(q,p)=2qp+q^2$. The find the corresponding Hamiltonian, we first integrate the equation $\Pd{H}{p}=f(q,p)=p-q^2$, which leads to
\begin{equation}
H(q,p)=\frac12p^2-q^2p+c(q)\;,
\end{equation}
where the ``integration constant'' is an arbitrary function of $q$. The other Hamilton equation then reduces to
\begin{equation}
-2qp-q^2=-g(q,p)=\PD{H}{q}=-2qp+c'(q)\;.
\end{equation}
This is solved in turn by $c(q)=-q^3/3$ up to an additive constant. The final result for the Hamiltonian is therefore, up to a constant,
\begin{equation}
H(q,p)=\frac12p^2-q^2p-\frac13q^3\;.
\end{equation}
\end{sol}

\begin{figure}[t]
\sidecaption[t]
\includegraphics[width=2.9in]{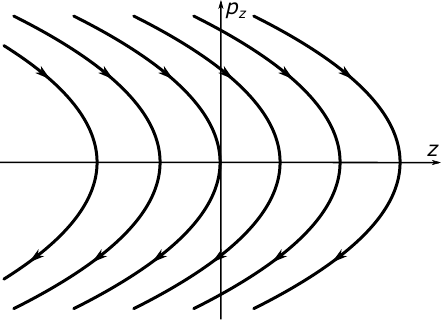}
\caption{Illustration for the solution of~\refpr{pr03:phaseportrait}: phase portrait of vertical motion in a uniform gravitational field.}
\label{fig03:solportrait}
\end{figure}

\begin{sol}{pr03:phaseportrait}
Denoting the Cartesian coordinate in the vertical direction as $z$, the Lagrangian for the particle that is only exposed to a uniform gravitational field is $L=(1/2)m\dot z^2-mgz$. The conjugate momentum is $p_z=m\dot z$, which leads to the Hamiltonian
\begin{equation}
H=\frac{p_z^2}{2m}+mgz\;.
\end{equation}
The corresponding Hamilton equations are
\begin{equation}
\dot z=\frac{p_z}{m}\;,\qquad
\dot p_z=-mg\;.
\end{equation}
The phase portrait defined by these equations is sketched in Fig.~\ref{fig03:solportrait}.
\end{sol}

\begin{sol}{pr03:particleinEMfield}
The momentum vector $\vec p$, conjugate to the position vector $\vec r$, is $\vec p=\Pd{L}{\dot{\vec r}}=m\dot{\vec r}+q\vec A$. As a consequence, the Hamiltonian as a function of $\vec r$ and $\vec p$ is given by
\begin{equation}
\begin{split}
H&=\vec p\cdot\dot{\vec r}-\biggl(\frac12m\dot{\vec r}^2-q\p+q\dot{\vec r}\cdot\vec A\biggr)\\
&=\frac1m\vec p\cdot(\vec p-q\vec A)-\biggl[\frac1{2m}(\vec p-q\vec A)^2-q\p+\frac qm(\vec p-q\vec A)\cdot\vec A\biggr]\\
&=\frac1{2m}(\vec p-q\vec A)^2+q\p\;.
\end{split}
\end{equation}
The first of the Hamilton equations merely reconfirms the relation between momentum and velocity, $\dot{\vec r}=\Pd{H}{\vec p}=(\vec p-q\vec A)/m$. The second of the Hamilton equations is more nontrivial, and we will work out separately its two sides. On the left-hand side, we have
\begin{equation}
\dot{\vec p}=m\ddot{\vec r}+q\OD{\vec A}{t}=m\ddot{\vec r}+q\biggl[\PD{\vec A}{t}+(\dot{\vec r}\cdot\grad)\vec A\biggr]\;.
\end{equation}
On the right-hand side, we find
\begin{equation}
-\PD{H}{\vec r}=q\grad(\dot{\vec r}\cdot\vec A)-q\grad\p\;,
\end{equation}
where the gradient operator $\grad$ is meant to only act on the $\vec r$-dependence of the electromagnetic potentials, not on $\dot{\vec r}$. Combining the two sides of the Hamilton equation together, we get
\begin{equation}
\begin{split}
m\ddot{\vec r}&=q[\grad(\dot{\vec r}\cdot\vec A)-(\dot{\vec r}\cdot\grad)\vec A]+q\biggl(-\grad\p-\PD{\vec A}t\biggr)\\
&=q\dot{\vec r}\times(\rot\vec A)+q\vec E=q(\vec E+\dot{\vec r}\times\vec B)\;,
\end{split}
\end{equation}
which is the standard Lorentz force as expected.
\end{sol}

\begin{sol}{pr03:puzzle}
The source of the confusion is that the given Lagrangian is already linear in velocities, which is a hallmark of the \emph{Hamiltonian} form of action~\eqref{ch03:actionHam}. Indeed, the corresponding Lagrange equations are first-order ODEs for $x,y$. We are more used to the situation where the Lagrangian is quadratic in velocities, which leads to second-order equations of motion. In that case, if we want to make the equations first-order, we need to introduce the additional variables $p_i$, thereby converting the Lagrangian action $S[q]$~\eqref{ch03:actionLag} into the Hamiltonian action $S[q,p]$~\eqref{ch03:actionHam}. Here we have nothing to do. To see the connection of the given Lagrangian to the Hamiltonian action~\eqref{ch03:actionHam}, add to it a total derivative, $\Od{(xy/2)}{t}$, and change the notation from $x,y$ to $p,q$. This brings the corresponding action to the form
\begin{equation}
S[x,y]=\int_{t_1}^{t_2}\D t\,\biggl[x\dot y-\frac\l2(x^2+y^2)\biggr]
\to S[q,p]=\int_{t_1}^{t_2}\D t\,\biggl[p\dot q-\frac\l2(p^2+q^2)\biggr]\;.
\end{equation}
This is nothing but the action of a linear harmonic oscillator in suitably chosen units for length and momentum. The corresponding Hamiltonian is $H(q,p)=(\l/2)(q^2+p^2)=(\l/2)(x^2+y^2)$, in accord with the prediction of~\eqref{ch01:Hamiltonian}.
\end{sol}

\begin{sol}{pr03:Routhian}
The Lagrangian and Hamiltonian are, respectively,
\begin{equation}
L=\frac12mR^2(\dot\t^2+\dot\vp^2\sin^2\t)\;,\qquad
H=\frac{p_\t^2}{2mR^2}+\frac{p_\vp^2}{2mR^2\sin^2\t}\;.
\end{equation}
Starting from the Lagrangian, we trade the azimuthal velocity $\dot\vp$ for the conjugate momentum $p_\vp=\Pd{L}{\dot\vp}=mR^2\dot\vp\sin^2\t$. This gives the Routhian
\begin{equation}
R= p_\vp\dot\vp-L=\frac{p_\vp^2}{2mR^2\sin^2\t}-\frac12mR^2\dot\t^2\;.
\end{equation}
Starting, on the other hand, from the Hamiltonian, we want to eliminate $p_\t$ by using the Hamilton equation $\dot\t=\Pd{H}{p_\t}=p_\t/(mR^2)$. The Routhian should then be $R=H-\dot\t p_\t$, which reproduces the result we obtained previously starting from the Lagrangian. The Lagrange-like Routh equation for $\t$ reads
\begin{equation}
0=\PD{R}{\t}-\OD{}t\PD{R}{\dot\t}=-\frac{p_\vp^2\cos\t}{mR^2\sin^3\t}+mR^2\ddot\t\;.
\end{equation}
At first sight, it is not obvious how to solve this in full generality. We can, however, use a simple trick. For any specific trajectory, we can choose the orientation of the coordinate axes so that one of the initial conditions for the trajectory is $\dot\vp=0$. Since $p_\vp$ is a constant of motion, it follows that $p_\vp=0$ at all times. This reduces the EoM for $\t$ to $\ddot\t=0$. The solution of this equation corresponds to uniform motion along of the meridians of the sphere.
\end{sol}

%%%%%%%%%%%%%%%%%%%%%%%%%%%%%%%%%%%%%%%%%%%%%%%%%%%%%%%%%%%%

\section{Problems in \chaptername~\ref*{chap:oscillations}}

\begin{sol}{pr04:criticaldamping}
The EoM for the critically damped oscillator takes the form
\begin{equation}
\ddot q(t)+2\o\dot q(t)+\o^2q(t)=F(t)\;.
\end{equation}
The characteristic equation of the homogeneous problem has a double root, $\l_\pm=\I\o$. The corresponding linearly independent solutions of the homogeneous EoM are $\E^{-\o t}$ and $t\E^{-\o t}$. The general solution of the EoM is therefore
\begin{equation}
q(t)=\E^{-\o t}(c_1+c_2t)+q_\mathrm{part}(t)\;,
\end{equation}
where $c_1,c_2$ are arbitrary integration constants and the particular solution $q_\mathrm{part}$ of the inhomogeneous equations is still given by~\eqref{ch04:drivenLHOsolpart}.
\end{sol}

\begin{figure}[t]
\sidecaption[t]
\includegraphics[width=2.9in]{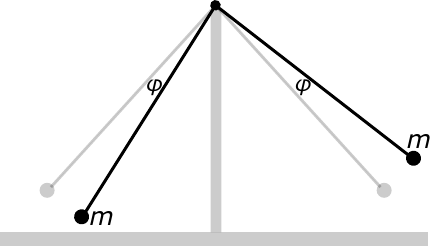}
\caption{Illustration for the solution of~\refpr{pr04:swing}: the generalized coordinate $\vp$ indicates the angular deviation of the swing from equilibrium. The angle between the two arms of the swing is fixed and equals $2\a$. For comparison, the equilibrium configuration is shown in gray.}
\label{fig04:swingsol}
\end{figure}

\begin{sol}{pr04:swing}
The kinetic energy of the system is simply $T=2\times(1/2)m(L\dot\vp)^2=mL^2\dot\vp^2$, where $\vp$ denotes the angular displacement of the swing from the equilibrium position in the counterclockwise direction (see Fig.~\ref{fig04:swingsol}). The angles that the arms $A$ and $B$ form with respect to the vertical are respectively $\a-\vp$ and $\a+\vp$. Setting the zero potential energy level to the pivot point of the swing, the total potential energy of the system is
\begin{equation}
V=-mgL[\cos(\a-\vp)+\cos(\a+\vp)]=-2mgL\cos\a\cos\vp\;.
\end{equation}
To describe small oscillations around the equilibrium, we need to expand the Lagrangian to the second order in powers of $\vp$,
\begin{equation}
L=mL^2\dot\vp^2+2mgL\cos\a\cos\vp\approx2mgL\cos\a+mL^2\dot\vp^2-mgL\vp^2\cos\a\;.
\end{equation}
By taking the ratio of the coefficients of $\vp^2$ and $\dot\vp^2$, we get the squared frequency, and thence the period of oscillations,
\begin{equation}
\o^2=\frac{g\cos\a}L\quad\Rightarrow\quad T=2\pi\sqrt{\frac{L}{g\cos\a}}\;.
\end{equation}
\end{sol}

\begin{sol}{pr04:chargedpendulum}
The distance of the two masses is $r=2L\sin(\vp/2)$ where $\vp$ is the angle between the two strings, which we will use as the generalized coordinate. Setting the level of zero gravitational potential energy to the pivot point $P$, the total (gravitational and electrostatic) potential energy of the mobile particle then is
\begin{equation}
V=-mgL\cos\vp+\frac{q^2}{8\pi\eps_0L\sin(\vp/2)}\;.
\end{equation}
The kinetic energy is $T=(1/2)mL^2\dot\vp^2$. According to the general theory of small oscillations, we now have to expand the Lagrangian to the second order in the displacement from equilibrium. Suppose that the value of $\vp$ in equilibrium is $\vp_0$ and the displacement is denoted as $\t$, so that $\vp=\vp_0+\t$. The potential is then expanded to the second order in $\t$ as
\begin{equation}
\begin{split}
V=&-mgL\left(\cos\vp_0-\t\sin\vp_0-\frac{\t^2}2\cos\vp_0\right)\\
&+\frac{q^2}{8\pi\eps_0L}\biggl[\frac1{\sin(\vp_0/2)}-\frac\t2\frac{\cos(\vp_0/2)}{\sin^2(\vp_0/2)}+\frac{\t^2}8\frac{1+\cos^2(\vp_0/2)}{\sin^3(\vp_0/2)}\biggr]+\mathcal{O}(\t^3)\;.
\end{split}
\end{equation}
The equilibrium position $\vp_0$ can now be determined by requiring that the terms of first order in $\t$ cancel. This leads to the condition
\begin{equation}
\sin^3\frac{\vp_0}2=\frac{q^2}{32\pi\eps_0mgL^2}\;.
\end{equation}
This can be used in turn to simplify the part of the potential quadratic in $\t$. Upon some manipulation, we find the Lagrangian for small oscillations around the equilibrium,
\begin{equation}
L\approx\frac12mL^2\dot\t^2-\frac32mgL\t^2\cos^2\frac{\vp_0}2\;.
\end{equation}
This gives at once the result for the frequency of small oscillations,
\begin{equation}
\omega=\sqrt{\frac{3g}L}\cos\frac{\vp_0}2\;.
\end{equation}
\end{sol}

\begin{sol}{pr04:triatomic}
The given initial conditions are $x_1(0)=x_2(0)=x_3(0)=0$ and
\begin{equation}
\dot x_1(0)=\dot x_3(0)=0\;,\qquad
\dot x_2(0)\equiv v_0\neq0\;.
\end{equation}
These translate via~\eqref{ch04:triatomicxfromxi} to initial conditions for the normal modes, $\x_+(0)=\x_-(0)=\x_0(0)=0$ and
\begin{equation}
\dot\x_+(0)=-\frac{2\sqrt k\,v_0}{\sqrt{2\o^2+4\O^2}}\;,\quad
\dot\x_-(0)=0\;,\quad
\dot\x_0(0)=\sqrt{\frac{kM}{m}}\frac{v_0}{\sqrt{\o^2+2\O^2}}\;.
\end{equation}
These initial conditions give us in turn the time evolution of the normal modes via~\eqref{ch04:generalxit},
\begin{equation}
\x_+(t)=\frac{\dot\x_+(0)}{\o_+}\sin\o_+t\;,\qquad
\x_-(t)=0\;,\qquad
\x_0(t)=\dot\x_0(0)t\;.
\end{equation}
Using once again~\eqref{ch04:triatomicxfromxi} then finally leads to the desired expressions for the displacements $x_1,x_2,x_3$ as a function of time,
\begin{equation}
x_1(t)=x_3(t)=\frac{v_0}{1+2m/M}\left(t-\frac{\sin\o_+t}{\o_+}\right)\;,\quad
x_2(t)=\frac{v_0}{1+2m/M}\left(t+\frac{2m}{M}\frac{\sin\o_+t}{\o_+}\right)\;.
\end{equation}
Let us see how to interpret the result. The terms linear in $t$ indicate that, ignoring the oscillations, the molecule as a whole moves with the velocity $v_0/(1+2m/M)$. This follows from the fact that the initial kick provides the molecule with momentum $Mv_0$, which translates into the velocity $Mv_0/(2m+M)$ of the CoM. The fact that there is no oscillating component corresponding to the $\x_-$ mode stems from the symmetry of the initial conditions under $x_1\leftrightarrow x_3$. As to the contributions of the $\x_+$ mode, there are no $\cos\o_+t$ terms owing to the vanishing initial condition for all the $x_{1,2,3}(0)$. Finally, the coefficients of $\sin\o_+t$ can be fixed from the given initial condition for $\dot x_{1,2,3}(0)$. All in all, we could have therefore pinned the solution down without the translation to the normal modes and back using~\eqref{ch04:triatomicxfromxi}. The advantage of using the normal modes is that it allows us to find the solution by following a clearly outlined algorithm, without having to supply additional physical insight.
\end{sol}

\begin{sol}{pr04:springypendulum}
We use as the generalized coordinates the angular deviation $\vp$ of the pendulum from the vertical axis, and the extension $s$ of the spring. Using Cartesian coordinates as for the simple pendulum in Fig.~\ref{fig01:pendulum}, the position vector of the mass $m$ is
\begin{equation}
\vec r=((L_0+s)\cos\vp,(L_0+s)\sin\vp)\;.
\end{equation}
From here, we get the kinetic energy $(1/2)m\dot{\vec r}^2=(1/2)m[\dot s^2+(L_0+s)^2\dot\vp^2]$. The potential energy of the system consists of the gravitational potential energy and the elastic energy of the spring, and thus equals $(1/2)ks^2-mg(L_0+s)\cos\vp$. Altogether, the Lagrangian of the system reads
\begin{equation}
L=\frac12m[\dot s^2+(L_0+s)^2\dot\vp^2]-\frac12ks^2+mg(L_0+s)\cos\vp\;.
\end{equation}
The minimum of the potential energy lies at $\vp_0=0$ and $s_0=mg/k$; this corresponds to the stretching of the spring at rest due to the gravitational pull on the mass $m$. To find the spectrum of small oscillations, we trade $s$ for $\eta\equiv s-s_0$ and expand the Lagrangian to the second order in $\eta$ and $\vp$,
\begin{equation}
L\approx\frac12m\dot\eta^2+\frac12m(L_0+s_0)^2\dot\vp^2-\frac12k\eta^2-\frac12mg(L_0+s_0)\vp^2\;,
\end{equation}
where we discarded contributions that are constant. In this approximation, the dynamics of $\eta$ and $\vp$ are decoupled; they are therefore the normal modes of the system. The corresponding normal frequencies are
\begin{equation}
\o_\eta=\sqrt{\frac km}\;,\qquad
\o_\vp=\sqrt{\frac{g}{L_0+s_0}}=\sqrt{\frac{g}{L_0+mg/k}}\;.
\end{equation}
\end{sol}

\begin{sol}{pr04:doublependulum}
We need two generalized coordinates to uniquely determine the positions of the masses, and we use to that end the angles $\vp_1,\vp_2$ between the strings and the vertical axis. With the same choice of Cartesian coordinates as in Fig.~\ref{fig01:pendulum}, the position vectors of the two masses are then
\begin{equation}
\vec r_1=(L_1\cos\vp_1,L_1\sin\vp_1)\;,\qquad
\vec r_2=\vec r_1+(L_2\cos\vp_2,L_2\sin\vp_2)\;.
\end{equation}
From here, the kinetic energy of the system follows as $(1/2)m_1\dot{\vec r}_1^2+(1/2)m_2\dot{\vec r}_2^2$. The gravitational potential energy is $-m_1gx_1-m_2gx_2$. Putting all the pieces together, we get after some manipulation the complete Lagrangian,
\begin{equation}
\begin{split}
L={}&\frac12(m_1+m_2)L_1^2\dot\vp_1^2+\frac12m_2L_2^2\dot\vp_2^2+m_2L_1L_2\dot\vp_1\dot\vp_2\cos(\vp_1-\vp_2)\\
&+m_1gL_1\cos\vp_1+m_2g(L_1\cos\vp_1+L_2\cos\vp_2)\;.
\end{split}
\end{equation}
The equilibrium position is $\vp_1=\vp_2=0$ and we can therefore expand the Lagrangian directly in powers of $\vp_1,\vp_2$,
\begin{equation}
\begin{split}
L\approx{}&\frac12(m_1+m_2)L_1^2\dot\vp_1^2+\frac12m_2L_2^2\dot\vp_2^2+m_2L_1L_2\dot\vp_1\dot\vp_2\\
&-\frac12(m_1+m_2)gL_1\vp_1^2-\frac12m_2gL_2\vp_2^2\;,
\end{split}
\end{equation}
up to a constant. This is as far as we get without a massive use of matrices. From now on, we therefore focus on the special case where $L_1=L_2\equiv L$ and $m_1=m_2\equiv m$. This reduces the above approximate quadratic Lagrangian to
\begin{equation}
L\approx\frac12mL^2(2\dot\vp_1^2+\dot\vp_2^2+2\dot\vp_1\dot\vp_2)-\frac12mgL(2\vp_1^2+\vp_2^2)\;.
\end{equation}
This is an example of a Lagrangian where the kinetic energy matrix is not diagonal, although the potential energy matrix is. Our general algorithm based on first diagonalizing the former and then the latter therefore does not seem to be the best choice here. Instead, we will first rescale the angles as $\vp_1=\eta_1/\sqrt2$ and $\vp_2=\eta_2$. This gives a Lagrangian of the type~\eqref{ch04:Lgenmultieta} with
\begin{equation}
T=mL^2\begin{pmatrix}
1 & 1/\sqrt2\\
1/\sqrt2 & 1
\end{pmatrix}\;,\qquad
V=mgL\un\;.
\end{equation}
We can now find the normal frequencies using that $T=mL^2(\un+\tau_1/\sqrt2)$. Indeed, the eigenvalues of $T$ are $mL^2(1\pm1/\sqrt2)$. Dividing $mgL$ by these eigenvalues then gives the (squared) normal frequencies,
\begin{equation}
\o_+^2=\frac1{1+1/\sqrt2}\frac gL=(2-\sqrt2)\frac gL\;,\qquad
\o_-^2=\frac{1}{1-1/\sqrt2}\frac gL=(2+\sqrt2)\frac gL\;.
\end{equation}
\end{sol}

\begin{sol}{pr04:chargedparticle}
The EoM for the particle is given by a combination of the linear restoring force of the harmonic oscillator and the Lorentz force due to the magnetic field,
\begin{equation}
m\ddot{\vec r}=-m\o^2\vec r+q\dot{\vec r}\times\vec B\;.
\end{equation}
Choosing the orientation of the Cartesian coordinate system so that $\vec B$ points along the $z$-axis and using the shorthand notation $\O\equiv q\abs{\vec B}/m$, the EoM takes the component form
\begin{equation}
\ddot x-\O\dot y+\o^2x=0\;,\qquad
\ddot y+\O\dot x+\o^2y=0\;,\qquad
\ddot z+\o^2z=0\;.
\end{equation}
Clearly, the motion along the $z$-axis separates from the motion in the $xy$-plane. Thus, the former is one of the normal modes, oscillating with normal frequency $\o$. The motion in the $xy$-plane can be dealt with by introducing the complex coordinate $w\equiv x+\I y$. This allows one to combine the equations for $x$ and $y$ into a single complex EoM,
\begin{equation}
\ddot w+\I\O\dot w+\o^2w=0\;.
\end{equation}
This is a linear second-order ODE that can be solved in the usual way. We use the ansatz $w\propto\E^{\I\l t}$, which reduces the EoM to an algebraic equation for the normal frequency $\l$, $\l^2+\O\l-\o^2=0$. This is solved by
\begin{equation}
\l_\pm=-\frac\O2\pm\sqrt{\frac{\O^2}{4}+\o^2}\;.
\end{equation}
With the convention that normal frequencies are always positive or at least non-negative, the sought normal frequencies describing the motion of a charged particle in a combination of a magnetic field and a harmonic potential are
\begin{equation}
\sqrt{\o^2+\frac{\O^2}{4}}\pm\frac\O2\quad\text{and}\quad\o\;.
\end{equation}
\end{sol}

%%%%%%%%%%%%%%%%%%%%%%%%%%%%%%%%%%%%%%%%%%%%%%%%%%%%%%%%%%%%

\section{Problems in \chaptername~\ref*{chap:relmechanics}}

\begin{sol}{pr05:Poincaregroup}
Representing the Poincar\'e transformation~\eqref{ch05:poincare} by the pair $(\Lambda,a)$, the composition of transformations $(\Lambda,a)$ and $(\Lambda',a')$ gives
\begin{multline}
x^\m\xrightarrow{(\Lambda,a)}\Lambda^\m_{\phantom\m\n}x^\n+a^\m\xrightarrow{(\Lambda',a')}\Lambda'^\m_{\phantom{,\m}\n}(\Lambda^\n_{\phantom\n\l}x^\l+a^\n)+a'^\m\\
=\Lambda'^\m_{\phantom{,\m}\n}\Lambda^\n_{\phantom\n\l}x^\l+(\Lambda'^\m_{\phantom{,\m}\n}a^\n+a'^\m)\;.
\end{multline}
This is again a transformation of the type~\eqref{ch05:poincare}. Hence, the set of Poincar\'e transformations is closed under composition and the multiplication rule reads
\begin{equation}
(\Lambda',a')\circ(\Lambda,a)=(\Lambda'\Lambda,\Lambda'a+a')\;.
\label{ch05:poincarecomposition}
\end{equation}
With this shorthand notation, it is straightforward to check the associativity axiom; one finds that
\begin{equation}
(\Lambda'',a'')\circ(\Lambda',a')\circ(\Lambda,a)=(\Lambda''\Lambda'\Lambda,\Lambda''\Lambda'a+\Lambda''a'+a'')
\end{equation}
regardless of how the triple product is bracketed. The unit element of the Poincar\'e group is the identity map $x^\m\to x^\m$, represented by $(\un,0)$. Finally, the inverse of $(\Lambda,a)$ with respect to the composition is
\begin{equation}
(\Lambda,a)^{-1}=(\Lambda^{-1},-\Lambda^{-1}a)\;.
\end{equation}
\end{sol}

\begin{sol}{pr05:boostcomposition}
Composition of two Lorentz transformations of the type~\eqref{ch05:Lorentztransfomatrix} amounts to simple matrix multiplication,
\begin{multline}
\begin{pmatrix}
\cosh w_1 & -\sinh w_1\\
-\sinh w_1 & \cosh w_1
\end{pmatrix}
\begin{pmatrix}
\cosh w_2 & -\sinh w_2\\
-\sinh w_2 & \cosh w_2
\end{pmatrix}\\
=\begin{pmatrix}
\cosh w_1\cosh w_2+\sinh w_1\sinh w_2 & -\cosh w_1\sinh w_2-\sinh w_1\cosh w_2\\
-\sinh w_1\cosh w_2-\cosh w_1\sinh w_2 & \sinh w_1\sinh w_2+\cosh w_1\cosh w_2
\end{pmatrix}\\
=\begin{pmatrix}
\cosh(w_1+w_2) & -\sinh(w_1+w_2)\\
-\sinh(w_1+w_2) & \cosh(w_1+w_2)
\end{pmatrix}\;.
\end{multline}
This proves that rapidities simply add. For the second part of the problem, use~\eqref{ch05:wbeta},
\begin{equation}
(\b_1,\b_2)\to\tanh(w_1+w_2)=\frac{\tanh w_1+\tanh w_2}{1+\tanh w_1\tanh w_2}=\frac{\b_1+\b_2}{1+\b_1\b_2}\;.
\end{equation}
\end{sol}

\begin{sol}{pr05:Thomasprecession}
Since only the $xy$-plane is relevant for the problem, we can restrict the $\Lambda^\m_{\phantom\m\n}$ matrices to their upper-left $3\times3$ corner. This way, our solution will hold regardless of the actual number of spatial dimensions, $n\geq2$. Using~\eqref{ch05:Lorentztransfomatrixphysical}, the boost in the $x$-direction can be represented by the matrix
\begin{equation}
\Lambda_{\vec u}=\begin{pmatrix}
\g_{\vec u} & -\b_{\vec u}\g_{\vec u} & 0\\
-\b_{\vec u}\g_{\vec u} & \g_{\vec u} & 0\\
0 & 0 & 1
\end{pmatrix}=
\begin{pmatrix}
1 & -u/c & 0\\
-u/c & 1 & 0\\
0 & 0 & 1
\end{pmatrix}+\bigO(u^2)\;.
\end{equation}
Likewise, for the boost in the $y$-direction, we have
\begin{equation}
\Lambda_{\vec v}=\begin{pmatrix}
\g_{\vec v} & 0 & -\b_{\vec v}\g_{\vec v}\\
0 & 1 & 0\\
-\b_{\vec v}\g_{\vec v} & 0 & \g_{\vec v}
\end{pmatrix}=
\begin{pmatrix}
1 & 0 & -v/c\\
0 & 1 & 0\\
-v/c & 0 & 1
\end{pmatrix}+\bigO(v^2)\;.
\end{equation}
It remains to work out the product $\Lambda_{\vec v}^{-1}\Lambda_{\vec u}^{-1}\Lambda_{\vec v}\Lambda_{\vec u}$. Keeping in mind that inverting a Lorentz transformation amounts to flipping the sign of the velocity, and dropping terms of order $\bigO(u^2,v^2)$ along the way, we arrive rather quickly at the final result,
\begin{equation}
\Lambda_{\vec v}^{-1}\Lambda_{\vec u}^{-1}\Lambda_{\vec v}\Lambda_{\vec u}=\begin{pmatrix}
1 & 0 & 0\\
0 & 1 & -\b_{\vec u}\b_{\vec v}\\
0 & \b_{\vec u}\b_{\vec v} & 1
\end{pmatrix}+\bigO(u^2,v^2)\;.
\end{equation}
This describes an infinitesimal rotation in the $xy$-plane by angle $\b_{\vec u}\b_{\vec v}$, as we wanted to demonstrate.
\end{sol}

\begin{sol}{pr05:velocitytransfo}
The four-velocity $u^\m=\g_{\vec v}(c,\vec v)$ is a four-vector, which means that its change under a Lorentz boost in the $x$-direction can be written in a matrix form as
\begin{equation}
\begin{pmatrix}
\g_{\vec v'}c\\
\g_{\vec v'}v'_x\\
\g_{\vec v'}v'_y
\end{pmatrix}=
\begin{pmatrix}
\g_u & -\b_u\g_u & 0\\
-\b_u\g_u & \g_u & 0\\
0 & 0 & 1
\end{pmatrix}
\begin{pmatrix}
\g_{\vec v}c\\
\g_{\vec v}v_x\\
\g_{\vec v}v_y
\end{pmatrix}\;.
\label{ch05:velocityboost}
\end{equation}
To save space, I have cropped the matrices after the $y$-component; all the velocity components other than $v_x$ will transform under the boost in the same way. The first row of~\eqref{ch05:velocityboost} allows us to extract the $\g$-factor of the transformed velocity,
\begin{equation}
\g_{\vec v'}=\g_u\g_{\vec v}\left(1-\frac{uv_x}{c^2}\right)\;.
\end{equation}
With this at hand, the second and third rows of~\eqref{ch05:velocityboost} then give the transformation rules for the components of the velocity,
\begin{equation}
v'_x=\frac{v_x-u}{1-uv_x/c^2}\;,\qquad
v'_y=\frac{v_y}{\g_u(1-uv_x/c^2)}\;.
\end{equation}
\end{sol}

\begin{sol}{pr05:scattering}
Let us denote the (a priori different) masses of the four particles as $m_1,m_2,m_3,m_4$. Using that $p_i^2=-m_i^2c^2$, we get pairwise relations between the dot products $p_i\cdot p_j$ by placing half of the four-momenta on either side of the energy--momentum conservation and taking the square,
\begin{equation}
\begin{split}
-(m_1^2+m_2^2)c^2+2p_1\cdot p_2=(p_1+p_2)^2&=(p_3+p_4)^2=-(m_3^2+m_4^2)c^2+2p_3\cdot p_4\;,\\
-(m_1^2+m_3^2)c^2-2p_1\cdot p_3=(p_1-p_3)^2&=(p_2-p_4)^2=-(m_2^2+m_4^2)c^2-2p_2\cdot p_4\;,\\
-(m_1^2+m_4^2)c^2-2p_1\cdot p_4=(p_1-p_4)^2&=(p_2-p_3)^2=-(m_2^2+m_3^2)c^2-2p_2\cdot p_3\;.
\end{split}
\end{equation}
This brings the number of independent Lorentz-scalar combinations of the four-momenta down to three, for instance $p_1\cdot p_i$ with $i=2,3,4$. However, these are constrained by one more linear relation, which follows by dotting the energy--momentum conservation condition into $p_1^\m$,
\begin{equation}
p_1\cdot(-p_2+p_3+p_4)=p_1^2=-m_1^2c^2\;.
\end{equation}
At the end of the day, there are therefore only two independent combinations of the four-momenta that characterize the kinematics of the scattering process in a way independent of the choice of IRF.

Let us sketch how the argument generalizes to an arbitrary number $n$ of particles participating in the scattering. To simplify the notation, I will use the convention that all the four-momenta are oriented out of the scattering zone, so that the energy--momentum conservation condition reads
\begin{equation}
p_1^\m+\dotsb+p_n^\m=0\;.
\end{equation}
There are altogether $n(n-1)/2$ different combinations of pairs of four-momenta $p_i\cdot p_j$. However, some of these can be immediately eliminated. For instance, we can use the squared energy--momentum conservation in the form
\begin{equation}
(p_i+p_n)^2=\Bigl(\sum_{\substack{j<n\\ j\neq i}}p_j\Bigr)^2
\end{equation}
to get rid of all $p_i\cdot p_n$ with $i<n$. This leaves us with the $(n-1)(n-2)/2$ variables $p_i\cdot p_j$ with $i,j<n$. One of these can still be eliminated using the relation
\begin{equation}
-m_n^2c^2=p_n^2=(p_1+\dotsb+p_{n-1})^2\;.
\end{equation}
This means that there are at most
\begin{equation}
\frac12(n-1)(n-2)-1=\frac12n(n-3)
\end{equation}
independent combinations of the four-momenta. For $n=4$, this gives mere $2$ independent variables as we found above. For higher $n$, there can be further relations among the remaining variables, depending on the number of spacetime dimensions. These essentially arise from the fact that in an $(n+1)$-dimensional spacetime, there can be at most $n+1$ linearly independent energy--momentum vectors.
\end{sol}

\begin{sol}{pr05:spinor}
The given matrix $\Pi$ satisfies $\det\Pi=-p^2=m^2c^2$ and $\tr\Pi=2p^0$. Recalling that the determinant and trace equal, respectively, the product and sum of the eigenvalues of $\Pi$, we find that for a massless particle, the eigenvalues must be $2p^0=2\abs{\vec p}$ and $0$. Since the matrix $\Pi$ is Hermitian, the corresponding normalized eigenvectors, $\vec e_{2\abs{\vec p}}$ and $\vec e_0$, constitute an orthonormal basis of $\C^2$. It then follows from~\eqref{app:decompose} that the action of $\Pi$ on an arbitrary column vector $\vec v$ is given by
\begin{equation}
\Pi\vec v=2\abs{\vec p}\vec e_{2\abs{\vec p}}\scal{\vec e_{2\abs{\vec p}}}{\vec v}=2\abs{\vec p}\vec e_{2\abs{\vec p}}\vec e_{2\abs{\vec p}}^\dagger\vec v\;.
\end{equation}
The desired representation $\Pi=\vec\x\he{\vec\x}$ then follows by setting $\vec\x\equiv\sqrt{2\abs{\vec p}}\,\vec e_{2\abs{\vec p}}$.
\end{sol}

\begin{sol}{pr05:Hamiltonian}
The conjugate momentum $\vec p$ follows from the Lagrangian,
\begin{equation}
\vec p=\PD{L}{\dot{\vec r}}=\frac{m\dot{\vec r}}{\sqrt{1-\dot{\vec r}^2/c^2}}\quad\Leftrightarrow\quad
\dot{\vec r}=\frac{\vec pc^2}{\sqrt{\vec p^2c^2+m^2c^4}}\;.
\end{equation}
The Hamiltonian then follows as
\begin{equation}
\begin{split}
H=\dot{\vec r}\cdot\vec p-L&=\frac{m\dot{\vec r}^2}{\sqrt{1-\dot{\vec r}^2/c^2}}+mc^2\sqrt{1-\frac{\dot{\vec r}^2}{c^2}}+V(\vec r)\\
&=\frac{mc^2}{\sqrt{1-\dot{\vec r}^2/c^2}}+V(\vec r)=\sqrt{\vec p^2c^2+m^2c^4}+V(\vec r)\;.
\end{split}
\end{equation}
This agrees with the expression~\eqref{ch05:dispersion} for the energy of a relativistic particle as a function of its momentum. The Hamilton equations of motion take the form
\begin{equation}
\dot{\vec r}=\PD{H}{\vec p}=\frac{\vec pc^2}{\sqrt{\vec p^2c^2+m^2c^4}}\;,\qquad
\dot{\vec p}=-\PD{H}{\vec r}=-\grad V(\vec r)\;.
\end{equation}
The first of these just reproduces the rule for conversion between velocity and momentum, whereas the second agrees with the Lagrangian EoM~\eqref{ch05:EoMrelativistic}.
\end{sol}

\begin{sol}{pr05:umupmu}
The key observation is that the Minkowski square of four-velocity is constant, $u_\m u^\m=-c^2$. This implies that $u_\m\Od{u^\m}{\tau}=(1/m)u_\m\Od{p^\m}{\tau}=0$. Using the definitions~\eqref{ch05:fourvelocity} and~\eqref{ch05:fourmomentum} of four-velocity and four-momentum, this translates to
\begin{equation}
\OD{E}{t}=\vec v\cdot\OD{\vec p}{t}=\skal Fv\;.
\end{equation}
We have recovered the well-known fact that the rate of change of energy of an object equals the rate of work, that is power, done by the force acting on it.
\end{sol}

%%%%%%%%%%%%%%%%%%%%%%%%%%%%%%%%%%%%%%%%%%%%%%%%%%%%%%%%%%%%

\section{Problems in \chaptername~\ref*{chap:geometryclassmech}}

\begin{sol}{pr06:lagsphere}
This is basically the same task as the one we dealt with in~\refpr{pr03:Routhian}, just reformulated in a more mathematical language. The metric on the sphere is fixed by the line element, $\D s^2=\D\t^2+\sin^2\t\D\vp^2$. Consequently, the metric as a matrix $\smash{g_{ij}}$ is diagonal with the entries $\smash{g_{\t\t}=1}$ and $\smash{g_{\vp\vp}=\sin^2\t}$. The diagonal entries of the inverse $\smash{g^{ij}}$ are accordingly $\smash{g^{\t\t}=1}$ and $\smash{g^{\vp\vp}=1/\sin^2\t}$. Using the definition of the Christoffel symbols, it turns out that the only nonzero components are
\begin{equation}
\Gamma^\t_{\vp\vp}=-\sin\t\cos\t\;,\qquad
\Gamma^\vp_{\t\vp}=\Gamma^\vp_{\vp\t}=\cot\t\;.
\end{equation}
This gives the geodesic equations in the form
\begin{equation}
\ddot\t-\dot\vp^2\sin\t\cos\t=0\;,\qquad
\ddot\vp+2\dot\t\dot\vp\cot\t=0\;.
\end{equation}
These equations have, among others, a class of solutions of the type $\t(t)=\pi/2$ and $\vp(t)=a+bt$, where $a,b$ are constants. These solutions satisfy the boundary condition $\t(0)=\pi/2$ and $\dot\t(0)=0$, which can always be ensured by a suitable choice of orientation of the Cartesian coordinate axes. The solutions describe motion at constant speed along the equator of the sphere, which is a great circle.
\end{sol}

\begin{sol}{pr06:canon}
In the Darboux coordinates, the components of the symplectic form are $\O_{ij}=-\ve_{ij}$. The given transformation will be canonical if and only if~\eqref{ch06:canontransfo} is satisfied. This leads to the constraint $(\Od fq)(\Pd gp)=1$, whose general solution is
\begin{equation}
g(q,p)=\frac{p}{f'(q)}+\tilde g(q)\;,
\end{equation}
with arbitrary $f(q)$ such that $f'(q)\neq0$, and an arbitrary function $\tilde g(q)$.
\end{sol}

\begin{sol}{pr06:LeviCivita}
The first thing to notice is that the product $\ve_{ijm}\ve_{klm}$ is antisymmetric separately in $i,j$ and $k,l$, and so must vanish whenever $i=j$ or $k=l$. This property is indeed respected by the given right-hand side. We can therefore assume that $i\neq j$ and $k\neq l$. The pairs $i,j$ and $k,l$ must now be the same up to permutation, since otherwise there would be no value of $m$ for which both $\ve_{ijm}$ and $\ve_{klm}$ would be simultaneously nonzero. The terms $\d_{ik}\d_{jl}$ and $\d_{il}\d_{jk}$ on the right-hand side account respectively for the possibilities that $i=k$ and $j=l$, or $i=l$ and $j=k$. Multiplying now the first of the given identities with $\d_{jl}$, we get
\begin{equation}
\ve_{ijm}\ve_{kjm}=\d_{jl}\ve_{ijm}\ve_{klm}=\d_{ik}\underbrace{\d_{jl}\d_{jl}}_{=3}-\underbrace{\d_{il}\d_{lj}\d_{jk}}_{=\d_{ik}}=2\d_{ik}\;.
\end{equation}
This is equivalent to the second identity that was to be proven.
\end{sol}

\begin{sol}{pr06:PoissonJrP}
All we have to do is use multiple times the basic Poisson brackets $\{r_i,r_j\}=\{p_i,p_j\}=0$ and $\{r_i,p_j\}=\d_{ij}$ together with the product rule. First,
\begin{equation}
\{J_i,r_j\}=\{\ve_{ikl}r_kp_l,r_j\}=\ve_{ikl}r_k\{p_l,r_j\}=-\d_{jl}\ve_{ikl}r_k=\ve_{ijk}r_k\;.
\end{equation}
The proof of $\{J_i,p_j\}=\ve_{ijk}p_k$ is entirely equivalent and we can therefore skip the details. Finally, using the two already derived Poisson brackets of $J_i$, we get
\begin{equation}
\begin{split}
\{J_i,J_j\}&=\{J_i,\ve_{jkl}r_kp_l\}=\ve_{jkl}(r_k\{J_i,p_l\}+p_l\{J_i,r_k\})\\
&=\ve_{jkl}(\ve_{ilm}r_kp_m+\ve_{ikm}r_mp_l)\\
&=(\d_{jm}\d_{ik}-\cancel{\d_{ij}\d_{km}})r_kp_m+(\cancel{\d_{ij}\d_{lm}}-\d_{il}\d_{jm})r_mp_l\\
&=r_ip_j-r_jp_i=\ve_{ijk}J_k\;.
\end{split}
\end{equation}
In the intermediate steps, we used repeatedly the first identity for the product of two Levi-Civita symbols from~\refpr{pr06:LeviCivita}.
\end{sol}

\begin{sol}{pr06:sympS2}
In~\refex{ex06:sympS2} we found that in the spherical coordinates, the symplectic form on the sphere is determined by the elements $\O_{\t\vp}=-\O_{\vp\t}=\a\sin\t$. The inverse matrix $\O^{ij}$ therefore has offdiagonal elements $\O^{\t\vp}=-\O^{\vp\t}=-1/(\a\sin\t)$. The Poisson bracket of the components of $\vec n$ can now be calculated directly via~\eqref{ch06:poissonbracket},
\begin{equation}
\{n_i,n_j\}=\O^{\t\vp}(\de_\t n_i\de_\vp n_j-\de_\vp n_i\de_\t n_j)\;.
\end{equation}
With the explicit parameterization of $\vec n$ in terms of the spherical angles, one can check component by component that $\{n_i,n_j\}=-(1/\a)\ve_{ijk}n_k$. This corresponds to $\b=-1/\a$. The Poisson bracket for $\vec n$ gives in turn directly
\begin{equation}
\dot n_i=\{n_i,H\}=\{n_i,n_j\}\PD H{n_j}=-\frac1\a\ve_{ijk}n_k\PD H{n_j}\;,
\end{equation}
which reproduces the Landau--Lifshitz equation~\eqref{ch06:LandauLifshitz}. Finally, by simply rescaling each component by $J$, the Poisson bracket for $\vec n$ is equivalent to $\{J_i,J_j\}=\b J\ve_{ijk}J_k$, which implies that $\b=1/J$, or $\a=-J$. For the specific choice of Hamiltonian, $H=-\m\skal JB$, the Landau--Lifshitz equation reduces to $\dot{\vec n}=\m\vekt nB$. Let us choose the orientation of the magnetic field along the third coordinate axis. Then the evolution equation for $\vec n$ splits component-wise as
\begin{equation}
\dot n_1=\m Bn_2\;,\qquad
\dot n_2=-\m Bn_1\;,\qquad
\dot n_3=0\;.
\end{equation}
In terms of the spherical angles, this amounts to $\dot\t=0$ and $\dot\vp=-\m B$. This describes precession of $\vec n$ around the vector of magnetic field with angular velocity $-\m B$.
\end{sol}

\begin{sol}{pr06:poiEM}
Since all components of $\vec r$ have vanishing Poisson brackets with each other, it follows trivially that $\{r_i,\pi_j\}=\{r_i,p_j\}=\d_{ij}$. The Poisson bracket of two components of the kinematical momentum can be calculated as follows,
\begin{equation}
\begin{split}
\{\pi_i,\pi_j\}&=\{p_i-qA_i,p_j-qA_j\}=-q(\{A_i,p_j\}+\{p_i,A_j\})\\
&=-q\left(\PD{A_i}{r_j}-\PD{A_j}{r_i}\right)=q\ve_{ijk}B_k\;.
\end{split}
\end{equation}
Using this Poisson bracket, one then deduces the evolution equation for $\vec\pi$,
\begin{equation}
\begin{split}
\dot\pi_i&=\{\pi_i,H\}-q\PD{A_i}t=\frac{\pi_j}m\{\pi_i,\pi_j\}+q\{\pi_i,\p\}-q\PD{A_i}t\\
&=\frac1m\ve_{ijk}\pi_jB_k-q\left(\PD\p{r_i}+\PD{A_i}t\right)\;.
\end{split}
\end{equation}
This can be written in the vector form as $\dot{\vec\pi}=q(\vec E+\dot{\vec r}\times\vec B)$, which has the expected form, encoding the Lorentz force.
\end{sol}

\begin{sol}{pr06:isoosc}
We can do even more than we were asked for by introducing the quantity
\begin{equation}
A_M\equiv A_{ij}M_{ij}=\frac1{2m}\vec p^TM\vec p+\frac12m\o^2\vec r^TM\vec r\;,
\label{ch06:AM}
\end{equation}
associated with an arbitrary real symmetric matrix $M$. By repeated use of the product rule for the Poisson bracket, one can then prove the following identity,
\begin{equation}
\{A_M,A_N\}=\frac{\o^2}2[M,N]_{ij}\ve_{ijk}J_k\;,
\end{equation}
where $M,N$ are symmetric matrices and $[M,N]\equiv MN-NM$ their commutator. Now the Hamiltonian of the isotropic oscillator is a special case of~\eqref{ch06:AM} where $M$ is the unit matrix, $H=A_\un$. It follows that $\{A_M,H\}=0$ for any choice of the matrix $M$, which is equivalent to saying that $\{A_{ij},H\}=0$ for all $i,j$.
\end{sol}

%%%%%%%%%%%%%%%%%%%%%%%%%%%%%%%%%%%%%%%%%%%%%%%%%%%%%%%%%%%%

\section{Problems in \chaptername~\ref*{chap:rigidbody}}

\begin{sol}{pr07:rotvector}
In order to avoid visual clutter, we temporarily drop the subscript on the chosen basis vector of the space frame and denote it simply as $\vec e'$. The starting point is to decompose $\vec e'$ into parts parallel with and perpendicular to $\vec n$, $\vec e'=\vec e'_\parallel+\vec e'_\perp$, where
\begin{equation}
\vec e'_\parallel=\vec n(\skal ne')\;,\qquad
\vec e'_\perp=\vec e'-\vec n(\skal ne')\;.
\end{equation}
Upon rotation around the axis $\vec n$ by the angle $\vp$, $\vec e'_\parallel$ remains unchanged, whereas $\vec e'_\perp$ is rotated to $\vec e'_\perp\cos\vp+(\vekt ne'_\perp)\sin\vp$. Altogether, we find that $\vec e'$ is rotated to
\begin{equation}
\begin{split}
\vec e&=\vec e'_\parallel+\vec e'_\perp\cos\vp+(\vekt ne'_\perp)\sin\vp\\
&=\vec n(\skal ne')+[\vec e'-\vec n(\skal ne')]\cos\vp+\{\vec n\times[\vec e'-\xcancel{\vec n(\skal ne')}]\}\sin\vp\\
&=\vec e'\cos\vp+\vec n(\skal ne')(1-\cos\vp)+(\vekt ne')\sin\vp\;.
\end{split}
\end{equation}
The collection of transformations for all the basis vectors $\vec e'_i$ can now be written in a matrix form that matches the definition~\eqref{ch07:eidef} of $R$,
\begin{align}
\label{ch07:rotationmatrix}
&\begin{pmatrix}
\vec e_1,\vec e_2,\vec e_3
\end{pmatrix}=
\begin{pmatrix}
\vec e_1',\vec e_2',\vec e_3'
\end{pmatrix}\cdot\\
\notag
&\begin{pmatrix}
n_1^2+(n_2^2+n_3^2)\cos\vp & n_1n_2(1-\cos\vp)-n_3\sin\vp & n_1n_3(1-\cos\vp)+n_2\sin\vp \\
n_1n_2(1-\cos\vp)+n_3\sin\vp & n_2^2+(n_1^2+n_3^2)\cos\vp & n_2n_3(1-\cos\vp)-n_1\sin\vp \\
n_1n_3(1-\cos\vp)-n_2\sin\vp & n_2n_3(1-\cos\vp)+n_1\sin\vp & n_3^2+(n_1^2+n_2^2)\cos\vp
\end{pmatrix}\;,
\end{align}
where $n_i$ are the components of $\vec n$ in the \emph{space} frame. We also used that $R_{ij}=\vec e'_i\cdot\vec e_j$ and $\vec e'_i\cdot(\vekt ne'_j)=\vec n\cdot(\vec e'_j\times\vec e'_i)=-\ve_{ijk}\skal ne'_k=-\ve_{ijk}n_k$. It is easy to see that the matrix $R$ defined by~\eqref{ch07:rotationmatrix} satisfies
\begin{equation}
R_{ij}n_j=n_i\;,\qquad
\tr R=1+2\cos\vp\;;
\end{equation}
the first of these properties amounts to saying that $\vec n$ is an eigenvector of $R$ with unit eigenvalue. Finally, for a rotation around the $z'$-axis by angle $\vp$, $\vec n=(0,0,1)$. Our general matrix $R$ reduces to
\begin{equation}
R=\begin{pmatrix}
\cos\vp & -\sin\vp & 0\\
\sin\vp & \cos\vp & 0\\
0 & 0 & 1
\end{pmatrix}\;.
\end{equation}
This is consistent with~\eqref{ch07:rotation2d} as expected.
\end{sol}

\begin{sol}{pr07:rotframe}
We start by writing $\vec r=r_i\vec e_i$ and taking two time derivatives, which gives the acceleration as seen in the IRF attached to the space frame,
\begin{equation}
\ddot{\vec r}=\underbrace{\ddot r_i\vec e_i}_{=\ddot{\vec r}_\mathrm{body}}+2\dot r_i\dot{\vec e}_i+r_i\ddot{\vec e}_i\;.
\end{equation}
The second term can be rewritten as $2\dot r_i\vec\o\times\vec e_i=2\vec\o\times\dot{\vec r}_\mathrm{body}$, where $\dot{\vec r}_\mathrm{body}=\dot r_i\vec e_i$ is the velocity observed in the noninertial reference frame attached to the rotating body. Finally, the second time derivative $\ddot{\vec e}_i$ in the third term can be expressed as
\begin{equation}
\ddot{\vec e}_i=\OD{(\vekt\o e_i)}{t}=\dot{\vec\o}\times\vec e_i+\vec\o\times\dot{\vec e}_i=\dot{\vec\o}\times\vec e_i+\vec\o\times(\vekt\o e_i)\;.
\end{equation}
Putting all the pieces together, the acceleration observed in the noninertial rotating frame can be related to the acceleration seen in the inertial space frame by
\begin{equation}
\ddot{\vec r}_\mathrm{body}=\ddot{\vec r}-\dot{\vec\o}\times\vec r-2\vec\o\times\dot{\vec r}_\mathrm{body}-\vec\o\times(\vekt\o r)\;.
\label{ch07:inertial}
\end{equation}
The second term on the right-hand side represents a fictitious force (per unit mass) due to the acceleration of the rotation: the so-called \emph{Euler force}. The third term gives the more familiar \emph{Coriolis force}. Finally, the last term in~\eqref{ch07:inertial} represents the \emph{centrifugal force}.
\end{sol}

\begin{sol}{pr07:inertia}
The matrix of second moments of the mass distribution for the tetrahedron can be expressed in a particularly symmetric form as
\begin{equation}
K_{ij}=\vr\int_{\substack{x,y,z\geq0\\x+y+z\leq L}}\D^3\!\vec r\,r_ir_j\;,
\end{equation}
where $\vr$ is the constant mass density of the tetrahedron. Since the integration domain is symmetric with respect to permutations of the coordinates, it is clear that $K_{ij}$ only has two independent components. Namely, all diagonal elements $K_{ii}$ are equal, as are all offdiagonal elements $K_{ij}$, $i\neq j$. An explicit calculation gives
\begin{equation}
K_{ii}=\frac1{60}{\vr L^5}=\frac1{10}mL^2\;,\qquad
K_{i\neq j}=\frac1{120}{\vr L^5}=\frac1{20}mL^2\;,
\end{equation}
where the total mass is $m=\vr V$ and the volume of the tetrahedron is $V=L^3/6$. As a consequence, the tensor of inertia, $I_{ij}=\d_{ij}\tr K-K_{ij}$, takes the matrix form
\begin{equation}
I=\frac1{20}mL^2\begin{pmatrix}
4 & -1 & -1\\
-1 & 4 & -1\\
-1 & -1 & 4
\end{pmatrix}\;.
\end{equation}
One of the eigenvalues of this matrix is $(1/10)mL^2$, the corresponding eigenvector being any (nonzero) multiple of $(1,1,1)$. The other two eigenvalues are degenerate, both being equal to $(1/4)mL^2$. The corresponding eigenvectors should be orthogonal to $(1,1,1)$, and therefore lie in the plane defined by $x+y+z=0$. Any orthogonal basis in this plane can represent principal axes of inertia. Our tetrahedron is an exotic example of a symmetric top!
\end{sol}

\begin{sol}{pr07:pendulum}
The Lagrangian of the physical pendulum is given by
\begin{equation}
L=\frac12I\dot\vp^2-V(\vp)\;,
\end{equation}
where $\vp$ is the angle parameterizing the deviation of the pendulum from the state of lowest energy (equilibrium). The potential energy only depends on the position of the CoM, and so can be written as $V(\vp)=-mgd\cos\vp$. The EoM corresponding to our Lagrangian is
\begin{equation}
I\ddot\vp+mgd\sin\vp=0\;.
\end{equation}
Upon expansion to first order in $\vp$, we find the frequency of small oscillations,
\begin{equation}
\o=\sqrt{\frac{mgd}{I}}\;.
\end{equation}
In the case of a simple pendulum of length $L$, the moment of inertia is $I=mL^2$ and the distance of the CoM from the axis of rotation is $d=L$. This leads immediately to $\o=\sqrt{g/L}$, which is the correct answer for a simple pendulum.
\end{sol}

\begin{sol}{pr07:wobble}
From $\vec\o=(\o_1,\o_2,\o_3)$, we find that the tangent of the angle $\a$ is
\begin{equation}
\tan\a=\frac{\sqrt{\o_1^2+\o_2^2}}{\o_3}\;.
\end{equation}
On the other hand, from $\vec J=(I_1\o_1,I_2\o_2,I_3\o_3)=(I_1\o_1,I_1\o_2,I_3\o_3)$, we get the tangent of the angle $\t$,
\begin{equation}
\tan\t=\frac{I_1\sqrt{\o_1^2+\o_2^2}}{I_3\o_3}\;.
\end{equation}
A comparison of the two results leads to the desired relation $\tan\a=(I_3/I_1)\tan\t$.
\end{sol}

\begin{sol}{pr07:asymtop}
Writing the angular momentum in the basis of principal axes of inertia as $\vec J=J_i\vec e_i$ gives the rate of change
\begin{equation}
\begin{split}
\vec{\dot J}&=\dot J_i\vec e_i+J_i\dot{\vec e}_i=\dot J_i\vec e_i+J_i\o_k\ve_{kij}\vec e_j=\dot J_i\vec e_i+\ve_{ijk}\vec e_i\o_jJ_k\\
&=(\dot J_i+\ve_{ijk}\o_jJ_k)\vec e_i\;.
\end{split}
\end{equation}
Using now once more the expression for angular momentum in the basis of principal axes of inertia, $J_i=I_i\o_i$ (no summation over $i$ implied), we find that the EoM $\vec{\dot J}=\vec\tau$ is equivalent to
\begin{equation}
I_i\dot\o_i+\sum_{j,k}\ve_{ijk}I_k\o_j\o_k=\tau_i\;,
\end{equation}
which is just a compact way of writing the Euler equations.
\end{sol}

%%%%%%%%%%%%%%%%%%%%%%%%%%%%%%%%%%%%%%%%%%%%%%%%%%%%%%%%%%%%

\section{Problems in \chaptername~\ref*{chap:LagHamcont}}

\begin{sol}{pr08:Schr2ndorder}
The Lagrangian density in~\eqref{ch08:Schraction} does not contain any derivatives of $\psi^*$. Setting $\udelta S/\udelta\psi^*$ to zero therefore gives directly the Schr\"odinger equation in the form
\begin{equation}
\left(\I\hbar\de_t+\frac{\hbar^2\grad^2}{2m}-V\right)\psi=0\;.
\end{equation}
Let us check whether we get the same EoM from $\udelta S/\udelta\psi$. By applying term by term~\eqref{ch00:functionalderhigher}, we find
\begin{equation}
\frac{\udelta S}{\udelta\psi}=-V\psi^*-\I\hbar\de_t\psi^*+\frac{\hbar^2\grad^2}{2m}\psi^*\ifeq0\;.
\end{equation}
This is equivalent to the complex conjugate of the Schr\"odinger equation.
\end{sol}

\begin{sol}{pr08:KdV}
A direct application of~\eqref{ch00:functionalderhigher} to the given Lagrangian density leads to
\begin{equation}
\begin{split}
\frac{\udelta S}{\udelta\p}&=-\de_t\PD\La{(\de_t\p)}-\de_x\PD{\La}{(\de_x\p)}+\de_x\de_x\PD{\La}{(\de_x\de_x\p)}\\
&=-\de_t\de_x\p-3\de_x(\de_x\p)^2-\de_x^2(\de_x^2\p)=-\de_t\psi-6\psi\de_x\psi-\de_x^3\psi\;,
\end{split}
\end{equation}
where $\psi=\de_x\p$. Setting this to zero gives the Korteweg--de Vries equation.
\end{sol}

\begin{sol}{pr08:nlsm}
The deviations of $\vec n$ from the assumed equilibrium state $\smash{\vec n_0}$ can be parameterized by the components $\smash{n^1}$ and $\smash{n^2}$. The last component is then $\smash{n^3=\sqrt{1-(n^1)^2-(n^2)^2}}$, and the equilibrium state corresponds to $n^1=n^2=0$. Inserting this parameterization into the Lagrangian density, we find at once
\begin{equation}
\La\approx-\frac12\sum_{A,B=1}^2\d_{AB}\de_\m n^A\de^\m n^B\;.
\end{equation}
This makes it clear that the deviations of $\vec n$ from equilibrium map to two massless relativistic particles.

Alternatively, one can expand the EoM to first order in the deviations. To derive the EoM, we introduce the Lagrange multiplier $\l(x)$ that implements the constraint $\skal nn=1$ at all spacetime points. The Lagrangian density is thus extended to
\begin{equation}
\La=-\frac12\de_\m\vec n\cdot\de^\m\vec n+\frac\l2(\skal nn-1)\;.
\end{equation}
Now we can take the functional derivative, treating $\vec n$ as a set of three independent scalar fields. This leads to the EoM
\begin{equation}
\Box\vec n=\l\vec n\;.
\label{ch08:nlsmEoM}
\end{equation}
By taking the scalar product with $\vec n$ and using the constraint on $\vec n$, we can isolate the Lagrange multiplier, $\l=\vec n\cdot\Box\vec n$. This makes it clear that $\l$ is of at least second order in the components $n^1,n^2$, parameterizing small deviations of $\vec n$ from the equilibrium. At the linearized level, the EoM~\eqref{ch08:nlsmEoM} is therefore equivalent to
\begin{equation}
\Box n^1\approx\Box n^2\approx0\;.
\end{equation}
This is a pair of KG equations for particles with zero mass.
\end{sol}

\begin{sol}{pr08:LandauLifshitz}
The EoM for $\vec n$ is obtained by taking the Poisson bracket with the Hamiltonian,
\begin{equation}
\begin{split}
\de_tn^A(\vec x)&=\{n^A(\vec x),H[\vec n]\}=\vr_\mathrm{s}\int_{\vec\O}\D^d\!\vec y\,\d_{BC}\{n^A(\vec x),\grad n^B(\vec y)\}\cdot\grad n^C(\vec y)\\
&\simeq-\vr_\mathrm{s}\int_{\vec\O}\D^d\!\vec y\,\d_{BC}\{n^A(\vec x),n^B(\vec y)\}\grad^2n^C(\vec y)\\
&=-\frac{\vr_\mathrm{s}}M\int_{\vec\O}\D^d\!\vec y\,\d_{BC}\ve^{AB}_{\phantom{AB}D}n^D(\vec x)\grad^2n^C(\vec y)\d^d(\vec x-\vec y)\\
&=-\frac{\vr_\mathrm{s}}M\ve^A_{\phantom ACD}n^D(\vec x)\grad^2n^C(\vec x)=\frac{\vr_\mathrm{s}}M\ve^A_{\phantom ABC}n^B(\vec x)\grad^2n^C(\vec x)\;.
\end{split}
\end{equation}
In the second step of the manipulation of the Poisson bracket, we used integration by parts, as indicated by the $\simeq$ symbol. The final result can be written in a vector form as $\de_t\vec n=(\vr_\mathrm{s}/M)\vec n\times\grad^2\vec n$.
\end{sol}

\begin{sol}{pr08:spinwave}
Simply inserting $\vec n=\vec n_0+\vec\eta$ in the EoM derived in~\refpr{pr08:LandauLifshitz}, we get
\begin{equation}
\de_t\vec\eta=\frac{\vr_\mathrm{s}}M(\vec n_0+\vec\eta)\times\grad^2\vec\eta\approx\frac{\vr_\mathrm{s}}M\vec n_0\times\grad^2\vec\eta\;,
\end{equation}
where the $\approx$ symbol denotes a linear approximation. Let us find plane-wave solutions of the linearized EoM. Setting
\begin{equation}
\vec\eta(\vec x,t)=\hat{\vec\eta}\E^{\I(\skal kx-\o t)}\;,
\end{equation}
where $\hat{\vec\eta}$ is a complex amplitude, converts the EoM as a PDE to an algebraic equation,
\begin{equation}
\I\o\hat{\vec\eta}\approx\frac{\vr_\mathrm{s}\vec k^2}M\vec n_0\times\hat{\vec\eta}\;.
\end{equation}
In terms of the components of $\hat{\vec\eta}$, this requires that $\hat\eta^3\approx0$ and moreover
\begin{equation}
\I\o\hat\eta^1\approx-\frac{\vr_\mathrm{s}\vec k^2}M\hat\eta^2\;,\qquad
\I\o\hat\eta^2\approx\frac{\vr_\mathrm{s}\vec k^2}M\hat\eta^1\;.
\end{equation}
This describes a wave with dispersion relation $\o=\vr_\mathrm{s}\vec k^2/M$ whose $\eta^1$ and $\eta^2$ com\-po\-nents have equal magnitudes and a relative phase shift of $\pi/2$. The wave is therefore circularly polarized.
\end{sol}

\begin{sol}{pr08:auxfield}
The EoM for the $\p$ and $A^\m$ field reads respectively
\begin{equation}
\de_\m A^\m=0\;,\qquad
A_\m=\de_\m\p\;.
\end{equation}
The second of these shows that $A^\m$ does not represent an independent dynamical degree of freedom. Combining the two equations gives $\Box\p=0$. The theory therefore includes a sole normal mode, corresponding to a massless relativistic particle.
\end{sol}

\begin{sol}{pr08:sineGordon}
Thanks to Lorentz invariance of the sine-Gordon theory, performing a Lorentz transformation on the kink solution will give another solution to the EoM. We transform into an IRF moving at velocity $-v$ along the $z$-axis. Denoting the coordinates in the new IRF with a prime, the desired Lorentz transformation reads
\begin{equation}
z'=\g(z+vt)\;,\qquad
t'=\g\left(t+\frac{vz}{c^2}\right)\;,
\end{equation}
where $\g\equiv1/\sqrt{1-v^2/c^2}$. By the definition of a scalar field, the solution in the new IRF satisfies $\p_0'(z',t')=\p_0(z)$. When combined with our Lorentz transformation and the explicit expression~\eqref{ch08:SGkink} for the kink solution, this gives
\begin{equation}
\p_0'(z',t')=4v\arctan\E^{m\g(z'-vt')}\;.
\end{equation}
\end{sol}

\begin{sol}{pr08:superfluid}
The EoM of the theory takes the form
\begin{equation}
\de_t\bigl[P'\bigl(\de_t\p-(\grad\p)^2/(2m)\bigr)\bigr]-\frac1m\grad\cdot\bigl[\grad\p\, P'\bigl(\de_t\p-(\grad\p)^2/(2m)\bigr)\bigr]=0\;,
\label{ch08:superfluidEoM}
\end{equation}
where the primes indicate a derivative of $P$ with respect to its argument. For $\p_0=\m t$, this boils down to $\de_t[P'(\m)]=0$, which is certainly satisfied since $P'(\m)$ is a time-independent constant. Next we shift the field by $\p=\p_0+\vp$ and expand the EoM~\eqref{ch08:superfluidEoM} to first order in $\vp$. This gives upon a short manipulation
\begin{equation}
P''(\m)\de_t^2\vp-\frac1mP'(\m)\grad^2\vp\approx0\;.
\end{equation}
Inserting the ansatz for a plane-wave solution, $\vp(\vec x,t)\propto\exp[\I(\skal kx-\o t)]$, finally gives the dispersion relation of the normal mode,
\begin{equation}
\o^2=\frac{P'(\m)}{mP''(\m)}\vec k^2\;.
\label{ch08:superfluiddisp}
\end{equation}

Alternatively, one can expand the Lagrangian density to second order in $\vp$,
\begin{equation}
\La=P(\m)+P'(\m)\left[\de_t\vp-\frac{(\grad\vp)^2}{2m}\right]+\frac12P''(\m)(\de_t\vp)^2+\bigO(\vp^3)\;.
\end{equation}
The constant term can be ignored, and so can be the boundary term $P'(\m)\de_t\vp$. The resulting bilinear Lagrangian density for the normal mode,
\begin{equation}
\La\approx\frac12P''(\m)(\de_t\vp)^2-\frac1{2m}{P'(\m)}(\grad\vp)^2\;,
\end{equation}
reproduces the dispersion relation~\eqref{ch08:superfluiddisp} found previously.
\end{sol}

%%%%%%%%%%%%%%%%%%%%%%%%%%%%%%%%%%%%%%%%%%%%%%%%%%%%%%%%%%%%

\section{Problems in \chaptername~\ref*{chap:elasticity}}

\begin{sol}{pr09:soundspeed}
For the equilibrium state of the solid to be stable, the bilinear part of the potential energy density, shown in~\eqref{ch09:solidLagpotapprox1} or~\eqref{ch09:solidLagpotapprox2}, must be positive-definite. The two terms in the second form, \eqref{ch09:solidLagpotapprox2}, can be set to zero separately and independently by a suitable choice of displacement $\vp_i$. On the one hand, consider an overall expansion or compression, corresponding to $\vp_i\propto x_i$. This gives $e_{ij}\propto\d_{ij}$, which makes the $\m$-term in~\eqref{ch09:solidLagpotapprox2} vanish. Positive definiteness then requires that $K>0$. On the other hand, one can imagine expansion along different coordinate axes by different scale factors, $\vp_i=\eps_ix_i$ (no summation over $i$). This gives a diagonal infinitesimal strain tensor, $e_{ij}=\eps_i\d_{ij}$. Choosing the scale factors $\eps_i$ so that $\sum_i\eps_i=0$ ensures that the volume does not change, that is $\tr e=0$. Positive definiteness then requires that $\m>0$. Note that one cannot apply the same argument to the \eqref{ch09:solidLagpotapprox1} form of the potential energy density. This is because $\tr e^2$ is nonzero and positive for any nonzero $e_{ij}$. As a consequence, one cannot use positive definiteness to argue that $\l>0$.

Using that $K>0$, it follows from the second expression for $c_\mathrm{L}$ in~\eqref{ch09:cLcT} that
\begin{equation}
c_\mathrm{L}>\sqrt{\frac{2\m(1-1/d)}{\vr_0}}=\sqrt{2\left(1-\frac1d\right)}\,c_\mathrm{T}\quad\Rightarrow\quad
\frac{c_\mathrm{L}}{c_\mathrm{T}}>\sqrt{2\left(1-\frac1d\right)}\;.
\end{equation}
For $d=3$, this gives $c_\mathrm{L}/c_\mathrm{T}>\sqrt{4/3}\approx1.15$. This bound is satisfied with a good margin for all the data in the given table.
\end{sol}

\begin{sol}{pr09:sigmatoe}
As a symmetric matrix, the local value of the infinitesimal strain tensor $e_{ij}$ can always be diagonalized by a suitable choice of Cartesian coordinates. It follows from~\eqref{ch09:solidEoMstress} that the local Cauchy stress tensor $\s_{ij}$ is diagonalized in the same coordinate basis. We can now imagine the set of special cases where only a single diagonal element of the stress tensor $\s_{ii}$ (no summation over $i$) is nonzero. This gives $e_{ii}=\s_{ii}/E$ and $e_{jj}=-\n e_{ii}=-(\n/E)\s_{ii}$ for any $j\neq i$. Putting all of these relations for different choices of $i$ together, we get
\begin{equation}
e_{ij}=\frac1E[(1+\n)\s_{ij}-\n\d_{ij}\tr\s]\;.
\label{ch09:sigmatoe}
\end{equation}
Having verified the validity of this tensor relation in the basis where both $e_{ij}$ and $\s_{ij}$ are diagonal is enough to conclude that it holds in any Cartesian coordinate basis.

Now we use the relation $\s_{ij}=\l\d_{ij}\tr e+2\m e_{ij}$ from~\eqref{ch09:solidEoMstress}. Taking the trace gives $\tr\s=(d\l+2\m)\tr e$, which allows us to invert the relation between $e_{ij}$ and $\s_{ij}$ to
\begin{equation}
e_{ij}=\frac1{2\m}\s_{ij}-\frac\l{2\m(d\l+2\m)}\d_{ij}\tr\s\;.
\end{equation}
Comparing this to~\eqref{ch09:sigmatoe} leads to the identification
\begin{equation}
\frac{1+\n}{E}=\frac1{2\m}\;,\qquad
\frac\n E=\frac\l{2\m(d\l+2\m)}\;.
\label{ch09:Poissonaux}
\end{equation}
These equations are easy to solve for $E$ and $\n$ as a function of $\l$ and $\m$,
\begin{equation}
E=2\m\frac{d\l+2\m}{(d-1)\l+2\m}\;,\qquad
\n=\frac\l{(d-1)\l+2\m}\;.
\end{equation}
\label{ch09:Poisson}
\end{sol}

\begin{sol}{pr09:Poissonratio}
One possibility is to use the final result of~\refpr{pr09:sigmatoe} for $\n$ as a function of $\l$ and $\m$ and combine it with the relation $K=\l+(2\m/d)$. A possible alternative is to follow the same line of argument up to~\eqref{ch09:Poissonaux}. Here we take a ratio of the two relations to eliminate the Young modulus, which gives
\begin{equation}
\frac{1+\n}\n=\frac{d\l+2\m}\l=\frac{dK}{K-(2\m/d)}\;.
\end{equation}
This is easily solved, the final result being
\begin{equation}
\n=\frac{dK-2\m}{d(d-1)K+2\m}\;.
\end{equation}
To find the bounds on possible values of the Poisson ratio, we partially decompose the fraction, either as
\begin{equation}
\n=-1+\frac{d^2K}{d(d-1)K+2\m}>-1\;,
\end{equation}
or as
\begin{equation}
\n=\frac1{d-1}\left[1-\frac{2d\m}{d(d-1)K+2\m}\right]<\frac1{d-1}\;.
\end{equation}
Finally, taking the trace of~\eqref{ch09:eintermsofsigma} gives
\begin{equation}
\tr e=\frac{1+(1-d)\n}E\tr\s\;.
\end{equation}
This shows that in the limit $\n\to1/(d-1)$, we also have $\tr e\to0$. There is no volume change in spite of possible local deformations of shape. Therefore, the limit $\n\to1/(d-1)$ can be interpreted as representing a solid that is incompressible.
\end{sol}

\begin{sol}{pr09:sphericalstrain}
This is a simple calculation. The easiest approach is to express the displacement vector as $\vec\vp(\vec x)=\tilde f(r)\vec x$ where $\tilde f(r)\equiv f(r)/r$. Using that $\de_ir=x_i/r$, this gives
\begin{equation}
\de_i\vp_j=\tilde f(r)\d_{ij}+\tilde f'(r)x_j\de_ir=\tilde f(r)\d_{ij}+\frac{\tilde f'(r)}rx_ix_j\;.
\end{equation}
This is already symmetric in the indices $i,j$. Substituting for $\tilde f(r)$ therefore gives directly the desired result for $e_{ij}$.
\end{sol}

\begin{sol}{pr09:sphere}
Any spherically symmetric vector field has a vanishing curl. This applies in particular to our radial displacement $\vec\vp$. The equality of $\grad(\divg\vp)$ and $\grad^2\vp$ follows from the vector calculus identity
\begin{equation}
\rot(\rot\vec\vp)=\grad(\divg\vec\vp)-\grad^2\vec\vp\;.
\end{equation}
Alternatively, one can use the index notation to argue that as a consequence of vanishing curl, $\de_i\vp_j=\de_j\vp_i$ for any choice of $i,j$. Then,
\begin{equation}
[\grad(\divg\vec\vp)]_i=\de_i\de^j\vp_j=\de^j\de_i\vp_j=\de^j\de_j\vp_i=\grad^2\vp_i\;.
\end{equation}
Moving on, the divergence of the radial displacement field, $\vec\vp(\vec x)=f(r)\vec n$, is
\begin{equation}
\divg\vec\vp=f'(r)+\frac{2f(r)}{r}=\frac1{r^2}\de_r[r^2f(r)]\;.
\end{equation}
By the Navier--Cauchy equation~\eqref{ch09:solidEoMNavierCauchy}, this should be constant. Denoting the constant as $\a$, we find the general solution
\begin{equation}
f(r)=\frac13\a r+\frac\b{r^2}\;,
\label{ch09:spherefr}
\end{equation}
where $\b$ is an integration constant. The values of $\a,\b$ are to be fixed using the boundary conditions at the inner and outer surface of our spherical shell. To that end, we need to evaluate the Cauchy stress tensor $\s_{ij}$ via~\eqref{ch09:solidEoMstress}. Using the fact that $\tr e=\divg\vec\vp$, we find
\begin{equation}
\s_{ij}=\l\left[f'(r)+\frac{2f(r)}r\right]\d_{ij}+2\m\left\{\frac{f(r)}r\d_{ij}+\left[\frac{f'(r)}{r^2}-\frac{f(r)}{r^3}\right]x_ix_j\right\}\;.
\end{equation}
Inserting our general solution~\eqref{ch09:spherefr} and projecting the stress tensor to the radial unit vector $\vec n$, we obtain the simple relation
\begin{equation}
\s_{ij}n^in^j=\frac\a3(3\l+2\m)-\frac{4\b\m}{r^3}\;.
\end{equation}
Demanding finally that this equals $-P_a$ and $-P_b$ for $r=a$ and $r=b$, respectively, gives the following values for $\a$ and $\b$,
\begin{equation}
\a=-\frac1K\frac{a^3P_a-b^3P_b}{a^3-b^3}\;,\qquad
\b=\frac1{4\m}\frac{P_a-P_b}{(1/{a^3})-(1/{b^3})}\;.
\end{equation}
Together with~\eqref{ch09:spherefr}, these determine completely the displacement of the spherical shell. In the special case of $P_a=P_b\equiv P$, we find $\a=-P/K$ and $\beta=0$. The corresponding radial shift is given by $f(r)=-Pr/(3K)$. The spherical shell will be strained even if the pressures inside and outside are the same. This is a consequence of the spherical geometry: the outer surface is larger than the inner one. The resulting imbalance of forces from the outside and inside leads to a compression of the shell.
\end{sol}

\begin{sol}{pr09:NavierCauchy}
Keeping the orientation of the Cartesian coordinate axes defining the space coordinates arbitrary, the potential energy density due to the gravitational field can be written as
\begin{equation}
\Va_\mathrm{grav}=-\vr\skal gx=-\vr_0\sqrt{\det\Xi}\,\skal gx\;.
\end{equation}
In the approximation of small deformations, the determinant of $\Xi$ can be expressed with the help of~\eqref{ch09:Xiapprox} as $\det\Xi\approx1-2\tr e$ so that $\Va_\mathrm{grav}\approx-\vr_0(1-\tr e)\skal gx$. The contribution of gravitational potential energy to the EoM is then
\begin{equation}
\de_j\PD{\Va_\mathrm{grav}}{(\de_j\vp_i)}=\de_j\PD{\Va_\mathrm{grav}}{e_{ji}}=\de^i(\vr_0\skal gx)=\vr_0g^i\;.
\end{equation}
In this approximation, the Navier--Cauchy equation is therefore modified to
\begin{equation}
\vr_0\ddot{\vec\vp}\approx(\l+\m)\grad(\divg\vec\vp)+\m\grad^2\vec\vp+\vr_0\vec g\;.
\end{equation}
In the special case of a fluid (cf.~\refex{ex09:fluid}), the condition for static equilibrium reduces to $\grad P\approx\vr_0\vec g$. This recovers the usual expression $P\approx\vr_0\skal gx$. The approximation requires that the density of the fluid does not change appreciably as a result of the pressure, that is, the fluid can be treated as incompressible.
\end{sol}

%%%%%%%%%%%%%%%%%%%%%%%%%%%%%%%%%%%%%%%%%%%%%%%%%%%%%%%%%%%%

\section{Problems in \chaptername~\ref*{chap:symmetries}}

\begin{sol}{pr10:phi4real}
Due to the linearity of the transformation, each term in the Lagrangian density must be separately invariant. It is easiest to inspect the mass term, for which
\begin{equation}
\d_{AB}\p^A\p^B=\d_{AB}P^A_{\phantom AC}P^B_{\phantom BD}\p'^C\p'^D\;.
\end{equation}
For this to equal $\d_{AB}\p'^A\p'^B$, the matrix $P$ must satisfy $\smash{\d_{AB}P^A_{\phantom AC}P^B_{\phantom BD}=\d_{CD}}$, that is be orthogonal, $\smash{PP^T=P^TP=\un}$. Inserting $P\approx\un+\eps Q$ and expanding the orthogonality condition to first order in $\eps$, we find that $Q+Q^T=0$. The generator $Q$ must be antisymmetric. For a given antisymmetric matrix $Q_{AB}$, the invariance condition is satisfied with $K_Q^\m=0$, and the Noether current is proportional to
\begin{equation}
J_Q^\m=Q_{AB}\p^A\de^\m\p^B\;.
\label{ch10:JQ}
\end{equation}
\end{sol}

\begin{sol}{pr10:phi4complex}
The infinitesimal version of the phase transformation corresponds to, respectively, $F_\p=\I\p$ and $F_{\p^*}=-\I\p^*$. The invariance condition is satisfied with $K^\m=0$, and the corresponding Noether current is
\begin{equation}
J^\m=\I(\p^*\de^\m\p-\p\de^\m\p^*)\;.
\end{equation}
In terms of the real fields $\p^{1,2}$ defined by $\p=(\p^1+\I\p^2)/\sqrt{2}$, the current reads
\begin{equation}
J^\m=\p^2\de^\m\p^1-\p^1\de^\m\p^2=-\ve_{AB}\p^A\de^\m\p^B\;.
\end{equation}
This is in accord with the solution of~\refpr{pr10:phi4real}. Namely, setting $\p=(\p^1+\I\p^2)/\sqrt{2}$ makes the Lagrangian densities of the two theories identical. Our present result for the Noether current in terms of $\p^{1,2}$ is recovered from~\eqref{ch10:JQ} by using $Q_{AB}=-\ve_{AB}$ as the sole linearly independent antisymmetric $2\times2$ matrix. 
\end{sol}

\begin{sol}{pr10:SchrGalilei}
The key step is to calculate the infinitesimal transformation of the Schr\"odinger field. The given transformation rule is equivalent to
\begin{equation}
\begin{split}
\psi'(\vec x,t)&=\exp\biggl[\frac\I\hbar m\biggl(\skal vx-\frac12\vec v^2t\biggr)\biggr]\psi(\vec x-\vec v t,t)\\
&=\psi(\vec x,t)+\frac\I\hbar m\skal vx\psi(\vec x,t)-\vec vt\cdot\grad\psi(\vec x,t)+\bigO(\vec v^2)\;.
\end{split}
\end{equation}
This allows to write the infinitesimal transformation of the Schr\"odinger field as
\begin{equation}
\udelta\psi\equiv v^iF_{\psi i}\;,\qquad
F_{\psi i}=\frac\I\hbar mx_i\psi-t\de_i\psi\;.
\end{equation}
The transformation rule for $\psi^*$ follows by taking the complex conjugate. Before we calculate the variation of the Lagrangian density under a Galilei boost, it is useful to prepare the infinitesimal transformations of the derivatives of $\psi$,
\begin{equation}
\begin{split}
\udelta(\de_t\psi)&=\de_t(\udelta\psi)=\frac\I\hbar m\skal vx\de_t\psi-\vec v\cdot\grad\psi-\vec vt\cdot\grad\de_t\psi\;,\\
\udelta(\grad\psi)&=\grad(\udelta\psi)=\frac\I\hbar m\vec v\psi+\frac\I\hbar m\skal vx\grad\psi-\vec vt\cdot\grad(\grad\psi)\;.
\end{split}
\end{equation}
What follows is a somewhat tedious yet straightforward calculation, at the end of which we find that $\udelta\La=-\de_i(v^it\La)$. Hence, the Galilei boost satisfies the invariance condition in the form
\begin{equation}
\udelta\La=v^j\de_\m K^\m_j\;,\qquad
K^\m_j=-\d^\m_jt\La\;.
\end{equation}
This has a simple interpretation. It means that the Lagrangian density as a function of coordinates is itself a scalar with respect to Galilei boosts, and thus transforms as $\La'(\vec x',t)=\La(\vec x,t)$. This corresponds to $\La'(\vec x,t)=\La(\vec x-\vec v t,t)=\La(\vec x,t)-\vec vt\cdot\grad\La(\vec x,t)+\bigO(\vec v^2)$, hence indeed $\udelta\La=-\vec vt\cdot\grad\La$.

It is now easy to extract the $d$ Noether currents corresponding to the $d$ possible directions of the boost velocity $\vec v$. The temporal components of the currents are
\begin{equation}
\begin{split}
J^0_{\phantom 0j}&=\PD\La{(\de_t\psi)}F_{\psi j}+\PD\La{(\de_t\psi^*)}F_{\psi^*j}-K^0_j\\
&=-mx_j\psi^*\psi-\frac{\I\hbar}2t(\psi^*\de_j\psi-\psi\de_j\psi^*)\;.
\end{split}
\end{equation}
The spatial components, on the other hand, turn out to be
\begin{equation}
\begin{split}
J^i_{\phantom 0j}&=\PD\La{(\de_i\psi)}F_{\psi j}+\PD\La{(\de_i\psi^*)}F_{\psi^*j}-K^i_j\\
&=\frac{\I\hbar}2x_j(\psi^*\de^i\psi-\psi\de^i\psi^*)+\frac{\hbar^2t}{2m}(\de^i\psi^*\de_j\psi+\de_j\psi^*\de^i\psi)+\d^i_jt\La\;.
\end{split}
\end{equation}
Interestingly, this set of Noether currents is related to the probability current $J^\m$ of~\refex{ex10:consprobability} and the EM tensor $T^\m_{\phantom\m\n}$ of the Schr\"odinger theory, computed in~\refex{ex10:EMSchr}. It is easy to check that the currents $J^\m_{\phantom\m j}$ owing their existence to Galilei boosts can be expressed as
\begin{equation}
J^\m_{\phantom\m j}=-mx_jJ^\m-tT^\m_{\phantom\m j}\;.
\end{equation}
\end{sol}

\begin{sol}{pr10:Schrgauged}
Introducing the shorthand notation $A_\m\equiv(A_t,\vec A)$, the covariant derivatives can be expressed jointly as $D_\m\psi\equiv(\de_\m-\I A_\m)\psi$. Likewise, the transformation rules for the external fields are subsumed into $A_\m\to A_\m+\de_\m\eps$. The covariant derivative of $\psi$ then transforms as
\begin{equation}
\begin{split}
D_\m\psi\to(D_\m\psi)'&=[\de_\m-\I(A_\m+\de_\m\eps)]\bigl(\E^{\I\eps}\psi\bigr)\\
&=\cancel{(\I\de_\m\eps)\E^{\I\eps}\psi}+\E^{\I\eps}\de_\m\psi-\I(A_\m+\cancel{\de_\m\eps})\E^{\I\eps}\psi=\E^{\I\eps}D_\m\psi\;.
\end{split}
\end{equation}
In other words, upon the change of phase of $\psi$, its covariant derivatives just acquire the same extra phase $\E^{\I\eps}$ as $\psi$ itself. This makes it clear that the given Lagrangian density is invariant under the simultaneous transformation of $\psi$ and $A_\m$. By taking a derivative with respect to the external fields and subsequently setting them to zero, we find
\begin{align}
\at{\PD\La{A_t}}{A=0}&=-\I\psi\at{\PD{\La}{(D_t\psi)}}{A=0}+\I\psi^*\at{\PD{\La}{(D_t\psi^*)}}{A=0}=\hbar\psi^*\psi\;,\\
\notag
\at{\PD\La{\vec A}}{A=0}&=-\I\psi\at{\PD{\La}{(\vec D\psi)}}{A=0}+\I\psi^*\at{\PD{\La}{(\vec D\psi^*)}}{A=0}=-\frac{\I\hbar^2}{2m}(\psi^*\grad\psi-\psi\grad\psi^*)\;.
\end{align}
This agrees with the Noether current as calculated in~\refex{ex10:SchrU(1)}, except for the overall sign. To see why this trick works, just write down the condition that the action in presence of the external fields is invariant under a simultaneous transformation of $\psi$ and $A_\m$ as
\begin{equation}
\udelta S=\int\D^D\!x\,\biggl[\PD{\La}{A_\m}\de_\m\eps+\frac{\udelta S}{\udelta\psi}\eps F_\psi+\frac{\udelta S}{\udelta\psi^*}\eps F_{\psi^*}\biggr]\ifeq0\;.
\end{equation}
We know that upon setting $A_\m\to0$, the sum of the last two terms under the integral must boil down to $J^\m\de_\m\eps$ up to a surface term, where $J^\m$ is the Noether current in the absence of external fields. This gives immediately
\begin{equation}
\udelta S\simeq\int\D^D\!x\,\biggl(\at{\PD\La{A_\m}}{A=0}\de_\m\eps+J^\m\de_\m\eps\biggr)\ifeq0\quad\Rightarrow\quad
J^\m=-\at{\PD{\La}{A_\m}}{A=0}\;.
\end{equation}
We have thus managed to explain why the trick works, including the extra sign.
\end{sol}

\begin{sol}{pr10:idealfluid}
Let us start with the equations of motion. The EoM for $\p$ expresses the constraint $\Pd\vr t+\divg(\vr\vec v)=0$, that is local mass conservation. The EoM for $\vec v$ can be written as $\vec v=\grad\p$. This says that the flow of the fluid is irrotational, and that $\p$ is the velocity potential for $\vec v$. Finally, the EoM for $\vr$ reads
\begin{equation}
\frac12\vec v^2-u'(\vr)-\PD\p t-\vec v\cdot\grad\p=0\quad\Rightarrow\quad
\frac12\vec v^2+u'(\vr)+\PD\p t=0\;,
\end{equation}
where in the second step, we used the already established relation $\vec v=\grad\p$. This is the Bernoulli equation for unsteady irrotational flow. Substituting $\vec v=\grad\p$ into the action and integrating by parts, we find the new Lagrangian density
\begin{equation}
\tilde\La=-\frac12\vr(\grad\p)^2-u(\vr)-\vr\PD\p t\;.
\end{equation}
This form of the Lagrangian density is practically convenient since it only contains $\vr$ without derivatives and $\p$ with derivatives. The individual components of the ensuing EM tensor are
\begin{equation}
\begin{aligned}
T^0_{\phantom00}&=\frac12\vr(\grad\p)^2+u(\vr)\;,\qquad&
T^i_{\phantom i0}&=-\vr\de^i\p\de_0\p\;,\\
T^0_{\phantom0j}&=-\vr\de_j\p\;,\qquad&
T^i_{\phantom ij}&=-\vr\de^i\p\de_j\p-\d^i_j\tilde\La\;.
\end{aligned}
\end{equation}
The $T^0_{\phantom00}$ component, $\smash{(1/2)\vr\vec v^2+u(\vr)}$, obviously has the meaning of energy density, as we expected. Similarly, the $\smash{T^0_{\phantom0j}}$ component, $-\vr v_j$, represents momentum density up to the overall sign.
\end{sol}

\begin{sol}{pr10:topological}
The conservation of the current $J^\m=\ve^{\m\n}\de_\n\p$ follows directly from the antisymmetry of the Levi-Civita symbol, without having to use the EoM for the field $\p$. Denoting the sole spatial coordinate as $x$, the density of the conserved charge is $J^0=\ve^{01}\de_1\p=\de_x\p$. The integral charge $Q$ is then determined by the boundary condition on the field at spatial infinity,
\begin{equation}
Q=\int_{-\infty}^{+\infty}\D x\,\de_x\p=\p(+\infty)-\p(-\infty)\;.
\end{equation}
For the field configuration to have a finite total energy, the energy density must drop to zero at spatial infinity. This requires in turn the value of $\p(x,t)$ to converge to one of the minima of the potential for $x\to\pm\infty$. It follows that the possible values of $Q$ for such field configurations are $\p_i-\p_j$ for any choice of the pair of minima $\p_{i,j}$.
\end{sol}

\begin{sol}{pr10:mattercurrent}
As shown in Sect.~\ref{subsec:mattercurrent}, the matter current can be interpreted as $J^\m=\vr(1,\dot x^i)$, where $\vr$ is the density of the medium. The conservation of the matter current therefore takes the form of the continuity equation~\eqref{ch10:continuityeq}. The reason why the matter current is conserved identically is built in the basic description of the medium, whereby it consists of mutually distinguishable elements, each of which carries a fixed amount of mass. The conservation of mass is therefore a property of the configuration space as defined by the maps $x^i\to\p^A(\vec x)$. In an Eulerian description of the medium where the density and velocity may be treated as dynamical variables, the conservation of mass is not automatic and has to be imposed by hand, as we saw in~\refpr{pr10:idealfluid}.
\end{sol}

\begin{sol}{pr10:helix}
The helix is symmetric under a simultaneous translation along the $z$-axis and rotation in the $xy$ plane. (This is sometimes called \emph{screw motion}.) The symmetry transformation can be parameterized by an angle $\eps$ such that
\begin{equation}
\vp\to\vp'=\vp+\eps\;,\qquad
z\to z'=z+\a\eps\;.
\end{equation}
In Cartesian coordinates, this transformation corresponds to
\begin{equation}
x'=x\cos\eps-y\sin\eps\;,\qquad
y'=x\sin\eps+y\cos\eps\;,\qquad
z'=z+\a\eps\;.
\label{ch10:screw}
\end{equation}
We only need the infinitesimal version of the transformation,
\begin{equation}
\udelta x=-\eps y\;,\qquad
\udelta y=\eps x\;,\qquad
\udelta z=\a\eps\;.
\label{ch10:screwinfty}
\end{equation}
The key observation is that the electrostatic potential of a charge distribution must have the same symmetry as its source, which in our case is the helix. Since the kinetic energy of a particle is invariant separately under both translations and rotations, we conclude that the entire Lagrangian of a test particle moving in the field of the charged helix will be invariant under the screw motion~\eqref{ch10:screw}. Using the Noether theorem together with the infinitesimal transformation~\eqref{ch10:screwinfty} then gives the corresponding Noether charge,
\begin{equation}
Q=xp_y-yp_x+\a p_z=J_z+\a p_z\;.
\end{equation}
\end{sol}

\begin{sol}{pr10:LaplaceRungeLenz}
Using temporarily only subscripts, the $j$-th component of the Laplace--Runge--Lenz vector can be written as
\begin{equation}
R_j=\ve_{jkl}p_kJ_l-km\frac{r_j}r\;.
\label{ch10:RungeLenz}
\end{equation}
The symmetry transformation generated by the Laplace--Runge--Lenz vector is
\begin{equation}
\udelta r_i\equiv\eps_jF_{ij}\;,\qquad
F_{ij}=\{r_i,R_j\}\;.
\label{ch10:RungeLenztransfo1}
\end{equation}
Using the Poisson brackets computed in~\refpr{pr06:PoissonJrP}, we get by a series of steps
\begin{equation}
\begin{split}
\{r_i,R_j\}&=\ve_{jkl}(J_l\{r_i,p_k\}+p_k\{r_i,J_l\})=\ve_{jkl}(J_l\d_{ik}-\ve_{lim}r_mp_k)\\
&=-\ve_{ijl}J_l-(\d_{ij}\d_{km}-\d_{ik}\d_{jm})r_mp_k\\
&=-r_ip_j+r_jp_i-\d_{ij}\skal rp+r_jp_i=-r_ip_j+2r_jp_i-\d_{ij}\skal rp\;.
\end{split}
\label{ch10:RungeLenztransfo2}
\end{equation}
In the Hamiltonian formalism, we would have to evaluate separately the Poisson bracket $\{p_i,R_j\}$ to deduce the transformation rule for canonical momentum. However, in the Lagrangian formalism, $\vec p=m\dot{\vec r}$ and the transformation of momentum is induced by that of the coordinate.

This completes the task given in the assignment of the problem. It is however instructive to check explicitly that~\eqref{ch10:RungeLenztransfo2} actually is a symmetry of the action. Recalling that the Lagrangian is
\begin{equation}
L=\frac12m\dot{\vec r}^2+\frac kr\;,
\end{equation}
let us evaluate its shift under the transformation~\eqref{ch10:RungeLenztransfo1},
\begin{equation}
\udelta L=\eps_j\left(m\dot r_i\dot F_{ij}-\frac k{r^3}r_iF_{ij}\right)\;.
\end{equation}
After inserting the above-calculated result for $F_{ij}$ and some rather uninspiring manipulations, we find that $\udelta L=\de_t(\eps_jK_j)$, where
\begin{equation}
K_j=r_j\vec p^2-p_j(\skal rp)+km\frac{r_j}r\;.
\end{equation}
This checks the invariance condition for the transformation generated by the Laplace--Runge--Lenz vector. Having got this far, we can also verify that a subsequent use of the Noether theorem recovers the expression~\eqref{ch10:RungeLenz}. Indeed, what we get is
\begin{align}
\notag
\PD L{\dot r_i}F_{ij}-K_j&=p_i(-r_ip_j+2r_jp_i-\cancel{\d_{ij}\skal rp})-r_j\vec p^2+\cancel{p_j(\skal rp)}-km\frac{r_j}r\\
&=-p_j(\skal rp)+2r_j\vec p^2-r_j\vec p^2-km\frac{r_j}r\\
\notag
&=r_j\vec p^2-p_j(\skal rp)-km\frac{r_j}r=[\vec p\times(\vekt rp)]_j-km\frac{r_j}r\\
\notag
&=(\vekt pJ)_j-km\frac{r_j}r\;.
\end{align}
This verifies that the Laplace--Runge--Lenz vector is the Noether charge associated with the symmetry transformation~\eqref{ch10:RungeLenztransfo1}.
\end{sol}

%%%%%%%%%%%%%%%%%%%%%%%%%%%%%%%%%%%%%%%%%%%%%%%%%%%%%%%%%%%%

\section{Problems in \chaptername~\ref*{chap:fluid}}

\begin{sol}{pr11:auxiliaryidentities}
Thanks to the cyclicity of trace, derivatives of functions of a  matrix variable $A$ inside a trace can be performed as in ordinary calculus. This is easy to check for a simple power of the matrix,
\begin{multline}
\udelta A^n=(\udelta A)A^{n-1}+A(\udelta A)A^{n-2}+\dotsb+A^{n-2}(\udelta A)A+A^{n-1}(\udelta A)\\
\Rightarrow\quad
\tr(\udelta A^n)=n\tr(A^{n-1}\udelta A)\quad\Rightarrow\quad
\PD{(\tr A^n)}{A^i_{\phantom ij}}=n(A^{n-1})^j_{\phantom ji}\;.
\end{multline}
The same kind of rule applies by extension to any function of $A$ that can be expanded in a Taylor series, in particular
\begin{equation}
\PD{(\tr\log A)}{A^i_{\phantom ij}}=(A^{-1})^j_{\phantom ji}\;.
\end{equation}
The first of the given identities then follows by evaluating separately the derivative of $\log\det A$. The second identity follows straightforwardly by computing the variation of $AA^{-1}=\un$ under a change of $A$,
\begin{multline}
(\udelta A)A^{-1}+A(\udelta A^{-1})=0\quad\Rightarrow\quad
\udelta A^{-1}=-A^{-1}(\udelta A)A^{-1}\\
\text{or equivalently}\quad
(\udelta A^{-1})^i_{\phantom ij}=-(A^{-1})^i_{\phantom ik}(\udelta A)^k_{\phantom kl}(A^{-1})^l_{\phantom lj}\;.
\end{multline}
\end{sol}

\begin{sol}{pr11:rotatingbucket}
We use Cartesian coordinates such that the $z$-axis is pointing vertically upwards. The trajectory of a fluid element at vertical position $z$ and distance $r$ from the $z$-axis can then be described as $\vec x(t)=(r\cos\o t,r\sin\o t,z)$. Taking the time derivative gives the velocity field of the fluid,
\begin{equation}
\vec v=\dot{\vec x}=(-r\o\sin\o t,r\o\cos\o t,0)=(-\o y,\o x,0)\;.
\end{equation}
This field has a vanishing divergence and thus satisfies the continuity equation for an incompressible fluid. The Euler equation gives the gradient of pressure via
\begin{equation}
\frac{\grad P}\vr=\vec g-\xcancel{\PD{\vec v}t}-(\skal v\nabla)\vec v=\vec g-\o(x\de_y-y\de_x)\vec v=(\o^2x,\o^2y,-g)\;.
\end{equation}
By integrating this, we find
\begin{equation}
P(\vec x)=P_0-\vr gz+\frac12\vr\o^2(x^2+y^2)\;,
\end{equation}
where $P_0$ is the pressure at the origin. The term linear in $z$ is the usual hydrostatic pressure due to gravity. The part quadratic in $x,y$ is a consequence of the centrifugal force acting on the rotating fluid.
\end{sol}

\begin{sol}{pr11:Eulersolunidirectional}
For an incompressible fluid, the continuity equation reduces to $\divg\vec v=0$, which boils down to $\de_zv_z=0$ in case of a unidirectional flow along the $z$-axis. This means that $v_z$ can only depend on the $x,y$ coordinates and time. As a consequence, the nonlinear term in the Euler equation, $(\skal v\nabla)\vec v=v_z\de_z\vec v$, vanishes. Using the assumptions that the external field $\vec g$ is conservative and the fluid is incompressible (and the density thus constant), the curl of the Euler equation is just
\begin{equation}
\PD{(\rot\vec v)}{t}=\vec0\quad\Rightarrow\quad
\frac{\de^2v_z}{\de t\,\de x}=\frac{\de^2v_z}{\de t\,\de y}=0\;.
\end{equation}
This implies that the derivative $\Pd{v_z}t$ can only depend on time. The most general unidirectional velocity field consistent with the continuity and Euler equations is therefore defined by
\begin{equation}
v_z(x,y,t)=f_1(t)+f_2(x,y)\;,
\end{equation}
where $f_1$ and $f_2$ are arbitrary functions of the indicated arguments. Inserting this back into the Euler equation determines the gradient of pressure via
\begin{equation}
\frac{\grad P}\vr=\vec g-\PD{\vec v}t-(\skal v\nabla)\vec v=(g_x,g_y,g_z-f_1'(t))\;.
\end{equation}
Denoting as $\psi$ the potential of the external field then gives the pressure,
\begin{equation}
P(\vec x)=-\vr[f_1'(t) z+\psi(\vec x)]+P_0\;,
\end{equation}
where $P_0$ is an arbitrary integration constant.
\end{sol}

\begin{sol}{pr11:EulerGalileiinvar}
The key step is to find how derivatives with respect to space and time transform with the change of IRF. Using the chain rule and the given transformation rules for the coordinates, we find
\begin{equation}
\grad_{\vec x'}=\grad_{\vec x}\;,\qquad
\PD{}{t'}=\PD{}{t}+\skal u\nabla_{\vec x}\;.
\end{equation}
With the help of this dictionary, we now evaluate the left-hand side of the continuity equation in the primed coordinates,
\begin{equation}
\begin{split}
\PD{\vr'}{t'}+\grad_{\vec x'}\cdot(\vr'\vec v')&=\left(
\PD{}{t}+\skal u\grad_{\vec x}\right)\vr+\grad_{\vec x}\cdot[\vr(\vec v-\vec u)]\\
&=\PD\vr{t}+\cancel{\skal u\nabla_{\vec x}\vr}+\grad_{\vec x}\cdot(\vr\vec v)-\cancel{(\grad_{\vec x}\vr)\cdot\vec u}\;.
\end{split}
\end{equation}
This confirms that the forms of the continuity equation in the two IRFs are equivalent. The Euler equation is dealt with in the same way. Its left-hand side in the prime coordinates reads
\begin{equation}
\begin{split}
\PD{\vec v'}{t'}+(\vec v'\cdot\grad_{\vec x'})\vec v'&=\left(
\PD{}{t}+\skal u\grad_{\vec x}\right)(\vec v-\vec u)+[(\vec v-\vec u)\cdot\grad_{\vec x}](\vec v-\vec u)\\
&=\left(
\PD{}{t}+\cancel{\skal u\grad_{\vec x}}\right)\vec v+(\skal v\nabla_{\vec x})\vec v-\cancel{(\skal u\nabla_{\vec x})\vec v}\\
&=-\frac{\grad_{\vec x}P}\vr+\vec g=-\frac{\grad_{\vec x'}P'}{\vr'}+\vec g'\;.
\end{split}
\end{equation}
This confirms that the Euler equation in the primed coordinates is a consequence of the Euler equation in the unprimed coordinates.
\end{sol}

\begin{sol}{pr11:conformal}
We know from complex calculus that for any meromorphic function $f(z)$ in the complex plane, both the real part $\operatorname{Re}f$ and the imaginary part $\operatorname{Im}f$ as functions of $x,y$ are harmonic, that is satisfy the Laplace equation. Specifically for
\begin{equation}
f(z)=z+\frac1z=x+\I y+\frac{x-\I y}{x^2+y^2}\;,
\end{equation}
the real part is
\begin{equation}
\p(x,y)\equiv\operatorname{Re}f(z)=x+\frac x{x^2+y^2}\;.
\end{equation}
Treating this as a velocity potential, the corresponding velocity field is $\vec v=\grad\p\equiv(v_x,v_y)$ with
\begin{equation}
v_x=1+\frac{y^2-x^2}{(x^2+y^2)^2}\;,\qquad
v_y=-\frac{2xy}{(x^2+y^2)^2}\;.
\end{equation}
On the unit circle $x^2+y^2=1$, this velocity field reduces to
\begin{equation}
\vec v\bigr\rvert_{x^2+y^2=1}=(2y^2,-2xy)=2y(y,-x)\;.
\end{equation}
This is perpendicular to the position vector $\vec x\equiv(x,y)$. In other words, the velocity field is tangential to the boundary of the circle. At infinity, the velocity field converges to a constant,
\begin{equation}
\lim_{\abs{\vec x}\to\infty}\vec v=(1,0)\;.
\end{equation}
The above properties confirm the interpretation that the field $\vec v$ represents the potential flow of an incompressible fluid around the obstacle given by the unit circle. 
\end{sol}

\begin{sol}{pr11:viscousstresstensor}
The velocity field of a medium rotating uniformly with fixed angular velocity $\vec\o$ is $\vec v=\vekt\o x$. Writing this in components as $v_i=\ve_{ijk}\o_jx_k$, we infer that $\de_jv_i=-\ve_{ijk}\o_k$. The fact that this is antisymmetric with respect to the exchange of $i,j$ guarantees that both parts of the viscous stress tensor, proportional respectively to the shear and bulk viscosity, vanish.

Let us ask generally for what (steady and incompressible) velocity fields the viscous stress tensor vanishes. The assumption of incompressibility implies via the continuity equation that $\divg\vec v=0$. The viscous stress tensor then simplifies to
\begin{equation}
\ve^{ij}=\eta\biggl(\PD{v^i}{x_j}+\PD{v^j}{x_i}\biggr)\ifeq0\;.
\label{ch11:zeroviscousstress}
\end{equation}
Taking another derivative and using the symmetry of second partial derivatives gives
\begin{equation}
\de_j\de_k v_i=-\de_j\de_i v_k=-\de_i\de_j v_k=\de_i\de_k v_j=\de_k\de_i v_j=-\de_k\de_j v_i=-\de_j\de_kv_i\;.
\end{equation}
This means that all second partial derivatives of the velocity field must vanish. The general solution of this condition is $v^i=\a^i+\b^{ij}x_j$ with constant $\a^i$ and $\b^{ij}$. Inserting this back into~\eqref{ch11:zeroviscousstress} leads to the condition that $\b^{ij}$ is antisymmetric. It can then be traded for a dual vector $\vec\o$ such that $\b^{ij}=-\ve^{ijk}\o_k$. The velocity field thus acquires the form $\vec v=\vec\a+\vekt\o x$. In other words, the only steady incompressible velocity fields that have an identically vanishing viscous stress tensor are those of a uniform flow (constant velocity) and uniform rigid rotation (constant angular velocity), and their linear combinations.
\end{sol}

\begin{sol}{pr11:viscousflowunidirectional}
Under the assumption of incompressibility, the continuity equation reduces to $\divg\vec v=0$, which in the case of unidirectional flow along the $z$-axis gives $\de_zv_z=0$. Since the flow is by assumption also steady, that is $\vec v=(0,0,v_z(x,y))$, the left-hand side of the Navier--Stokes equation identically vanishes. The Navier--Stokes equation therefore reduces to
\begin{equation}
\grad P=\eta\grad^2\vec v\;.
\label{ch11:steadyNavierStokes}
\end{equation}
The transverse components of this equation are trivial, saying merely that $\de_xP=\de_yP=0$; the pressure is a function of $z$. On the other hand, we already showed that $v_z$ only depends on $x$ and $y$. The $z$-component of~\eqref{ch11:steadyNavierStokes}, $P'(z)=\eta\grad^2v_z(x,y)$, then requires that both sides are coordinate-independent constants. Calling this constant $\a$, we infer that the pressure varies linearly along the $z$-axis,
\begin{equation}
P(z)=P_0+\a z\;,
\end{equation}
where $P_0$ is an integration constant. The velocity of the flow must in turn satisfy the two-dimensional Poisson equation
\begin{equation}
\grad^2v_z=\frac\a\eta\;.
\label{ch11:viscousflowz}
\end{equation}
For a pipe with a circular cross-section of radius $R$, we are looking for a solution satisfying the boundary condition $v_z=0$ at $x^2+y^2=R^2$. This is easy to guess,
\begin{equation}
v_z(x,y)=\frac\a{4\eta}(x^2+y^2-R^2)\;.
\end{equation}
This is indeed the expected parabolic profile. The linear second-order PDE~\eqref{ch11:viscousflowz} however makes it straightforward to find the velocity profile also for straight pipes of other cross-sections.
\end{sol}

\begin{sol}{pr11:inclinedplane}
Under the assumption of incompressibility (constant density $\vr$), the continuity equation reduces to $\divg\vec v=0$. The Navier--Stokes equation in turn becomes
\begin{equation}
\vr\left[\PD{\vec v}t+(\skal v\nabla)\vec v\right]=-\grad P+\vr\vec g+\eta\grad^2\vec v\;.
\end{equation}
The flow of the liquid is assumed to be steady and unidirectional, so the Cartesian components of the velocity field are $\vec v=(v_x(\vec x),0,0)$. This further reduces the continuity equation to $\de_xv_x=0$, and the Navier--Stokes equation to
\begin{equation}
\grad P=\vr\vec g+\eta\grad^2\vec v\;.
\label{ch11:inclinedplane}
\end{equation}
Let us first focus on the $y$-component of this vector equation. There is no velocity term and we get simply $\de_yP=\vr g_y=-\vr g\cos\a$. This is solved by $P(x,y)=-\vr gy\cos\a+\tilde P(x)$, where $\tilde P(x)$ is an arbitrary function of $x$. The latter can be fixed by observing that at the free surface of the liquid, the pressure should be constant and match the pressure (let us call it $P_0$) of the surrounding air. Choosing the origin of coordinates as in Fig.~\ref{fig11:inclinedplane} so that the free surface is at $y=d$, this fixes $\tilde P(x)=P_0+\vr gd\cos\a$. We now have a complete solution for the pressure~of~the~liquid,
\begin{equation}
P=P_0+\vr g(d-y)\cos\a\;.
\end{equation}
What really matters is actually only that the pressure does not depend on $x$. This reduces the $x$-component of the equation~\eqref{ch11:inclinedplane} to $\eta\grad^2v_x=-\vr g_x=-\vr g\sin\a$. We already know from the continuity equation that $v_x$ does not depend on $x$, and so $\grad^2v_x=\de_y^2v_x$. It follows that the second derivative of $v_x$ with respect to $y$ is constant. The general solution to our second-order ordinary differential equation for $v_x(y)$ is
\begin{equation}
v_x=-\frac{\vr gy^2}{2\eta}\sin\a+c_1y+c_0\;.
\end{equation}
From the boundary condition $v_x=0$ at $y=0$, we fix $c_0=0$. Likewise, from the boundary condition $\de_yv_x=0$ at $y=d$, it follows that $c_1=(\vr gd/\eta)\sin\a$. This fixes completely $v_x$ as a function of $y$, including its value at the free surface of the liquid,
\begin{equation}
v_x(y)=\frac{\vr g}{2\eta}y(2d-y)\sin\a\quad\Rightarrow\quad
v_x(d)=\frac{\vr gd^2}{2\eta}\sin\a\;.
\end{equation}
\end{sol}

%%%%%%%%%%%%%%%%%%%%%%%%%%%%%%%%%%%%%%%%%%%%%%%%%%%%%%%%%%%%

\section{Problems in \chaptername~\ref*{chap:electrodynamics}}

\begin{sol}{pr12:invariants}
On the one hand, we find
\begin{equation}
F_{\m\n}F^{\m\n}=F_{0i}F^{0i}+F_{i0}F^{i0}+F_{ij}F^{ij}=2\left(\vec B^2-\frac{\vec E^2}{c^2}\right)\;.
\end{equation}
On the other hand,
\begin{equation}
\ve^{\m\n\a\b}F_{\m\n}F_{\a\b}=4\ve^{0ijk}F_{0i}F_{jk}=4\ve^{ijk}\left(-\frac{E_i}c\right)\ve_{jkl}B^l=-\frac8c\skal EB\;.
\end{equation}
The invariance of both expressions under the Lorentz transformation~\eqref{ch12:EBtransfo} is straightforward to check. For a linearly polarized electromagnetic wave, both invariants vanish.
\end{sol}

\begin{sol}{pr12:EBfields}
Suppose that, in a given IRF, $\skal EB=0$. By a suitable choice of Cartesian coordinates, one can then always ensure that the only possibly nonzero component of $\vec E$ is along a chosen coordinate axis, whereas the only possibly nonzero component of $\vec B$ is along one of the other coordinate axes. It remains to inspect the case where, in a given IRF, both fields are nonzero and have a nonvanishing scalar product (that is are not mutually orthogonal). We now want to prove that it is possible to find another IRF in which the fields are parallel. Their orientation along the $x$-axis can then be ensured by a simple rotation of the coordinates.

To prove the above claim, we consider a Lorentz boost with velocity $\vec v$ perpendicular to both $\vec E$ and $\vec B$. According to~\eqref{ch12:EBtransfo}, this changes the fields to
\begin{equation}
\vec E'=\g_{\vec v}(\vec E+\vekt vB)\;,\qquad
\vec B'=\g_{\vec v}\left(\vec B-\frac1{c^2}\vekt vE\right)\;.
\end{equation}
Upon the Lorentz transformation, the electric and magnetic fields will be parallel if $\vec E'\times\vec B'=\vec0$. Working out the cross product while using the assumption $\skal vE=\skal vB=0$, the desired velocity of the boost turns out to be determined implicitly by
\begin{equation}
\frac{\vec v}{1+\vec v^2/c^2}=\frac{\vekt EB}{\vec B^2+\vec E^2/c^2}\;.
\label{ch12:EBaux}
\end{equation}
We must demonstrate that a velocity $\vec v$ satisfying this condition actually exists. To that end, note that as the magnitude of the velocity increases from zero to $c$, the magnitude of the left-hand side of~\eqref{ch12:EBaux} increases monotonically from zero to $c/2$. There will therefore be a unique solution for the velocity provided the right-hand side of~\eqref{ch12:EBaux} is smaller than $c/2$. This is easy to prove using our assumption that $\vec E$ and $\vec B$ are nonzero and not perpendicular to each other, so that
\begin{equation}
\abs{\vekt EB}<\abs{\vec E}\abs{\vec B}=c\frac{\abs{\vec E}}c\abs{\vec B}\leq\frac c2\left(\vec B^2+\frac{\vec E^2}{c^2}\right)\;;
\end{equation}
the last inequality being equivalent to $(\abs{\vec B}-\abs{\vec E}/c)^2\geq0$.
\end{sol}

\begin{sol}{pr12:Pontryagin}
The contribution of the $\ve^{\a\b\g\d}F_{\a\b}F_{\g\d}$ term in the Lagrangian density to the EoM for $A_\m$ is
\begin{equation}
\de_\n\PD{(\ve^{\a\b\g\d}F_{\a\b}F_{\g\d})}{(\de_\n A_\m)}=2\ve^{\a\b\g\d}\de_\n\left[F_{\a\b}\PD{(F_{\g\d})}{(\de_\n A_\m)}\right]\\
=4\ve^{\a\b\n\m}\de_\n F_{\a\b}\;,
\end{equation}
which vanishes by the Bianchi identity. By using similar reasoning, we easily convince ourselves that we are dealing with a surface term,
\begin{equation}
\ve^{\m\n\a\b}F_{\m\n}F_{\a\b}=2\ve^{\m\n\a\b}(\de_\m A_\n)F_{\a\b}=\de_\m\bigl(2\ve^{\m\n\a\b}A_\n F_{\a\b}\bigr)\;.
\end{equation}
This shows that $K^\m=2\ve^{\m\n\a\b}A_\n F_{\a\b}$.
\end{sol}

\begin{sol}{pr12:EMconservation}
To calculate the divergence of the Hilbert EM tensor, we will need to evaluate the derivative of the $F_{\a\b}F^{\a\b}$ term. This can be simplified using the Bianchi identity in the form~\eqref{ch12:Bianchi},
\begin{equation}
\de_\n(F_{\a\b}F^{\a\b})=2F^{\a\b}\de_\n F_{\a\b}=-2F^{\a\b}(\de_\a F_{\b\n}+\de_\b F_{\n\a})=-4F^{\a\b}\de_\a F_{\b\n}\;.
\end{equation}
The divergence of the Hilbert EM tensor~\eqref{ch12:EMtensorHilbert} is now easy to compute,
\begin{equation}
\begin{split}
\de_\m\Theta^{\m\n}&=-\frac1{\m_0}(\de_\m F^{\m\a})F^\n_{\phantom\n\a}-\cancel{\frac1{\m_0}F^{\m\a}\de_\m F^\n_{\phantom\n\a}}-\cancel{\frac1{\m_0}F^{\a\b}\de_\a F_\b^{\phantom\b\n}}=F^{\n\a}J_\a\;,
\end{split}
\end{equation}
where in the last step, we used the Maxwell equations. The physical content of this generalized conservation law is as follows:
\begin{itemize}
\item The temporal component, $\de_\m\Theta^{\m0}=F^{0i}J_i=\skal JE/c$, is equivalent to the local energy conservation~\eqref{ch12:energyconservation}.
\item The spatial part evaluates to
\begin{equation}
\de_\m\Theta^{\m j}=F^{j\m}J_\m=(\vr\vec E+\vekt JB)^j\;.
\end{equation}
The right-hand side is nothing but the force per unit volume by which the electromagnetic fields act on charged matter. This measures the transfer of momentum between the fields and charges.
\end{itemize}
\end{sol}

\begin{sol}{pr12:nonlinear}
That the electric and magnetic fields $\vec E$ and $\vec B$ are, modulo choice of IRF, completely determined by just two parameters, follows from~\refpr{pr12:EBfields}. As to the second suggested proof, recall the Lorentz transformation~\eqref{ch12:FmunuLorentztransfo} of the contravariant field-strength tensor $F^{\m\n}$. To deduce a transformation rule for $F^\m_{\phantom\m\n}=F^{\m\l}g_{\l\n}$, we use the fact that the Minkowski metric itself is Lorentz-invariant,
\begin{equation}
\Lambda^\a_{\phantom\a\m}\Lambda^\b_{\phantom\b\n}g_{\a\b}=g_{\m\n}\quad\Leftrightarrow\quad
g_{\a\b}=(\Lambda^{-1})^\m_{\phantom\m\a}(\Lambda^{-1})^\n_{\phantom\n\b}g_{\m\n}\;.
\end{equation}
This leads to
\begin{equation}
\begin{split}
F'^\m_{\phantom{'\m}\n}&=F'^{\m\l}g_{\l\n}=\Lambda^\m_{\phantom\m\a}\Lambda^\l_{\phantom\l\b}F^{\a\b}(\Lambda^{-1})^\g_{\phantom\g\l}(\Lambda^{-1})^\d_{\phantom\d\n}g_{\g\d}\\
&=\Lambda^\m_{\phantom\m\a}(\Lambda^{-1})^\d_{\phantom\d\n}F^{\a\g}g_{\g\d}=\Lambda^\m_{\phantom\m\a}F^\a_{\phantom\a\b}(\Lambda^{-1})^\b_{\phantom\b\n}\;.
\end{split}
\end{equation}
This proves that $F^\m_{\phantom\m\n}$ transforms by conjugation with the matrix $\Lambda$. The characteristic equation for the eigenvalues $\l$ of $F^\m_{\phantom\m\n}$ is best evaluated by using its explicit matrix representation in terms of the electric and magnetic fields, which leads to
\begin{equation}
\l^4+\left(\vec B^2-\frac{\vec E^2}{c^2}\right)\l^2-\frac{(\skal EB)^2}{c^2}=0\;.
\end{equation}
This completes the proof that the eigenvalues of $F^\m_{\phantom\m\n}$, and by extension any Lorentz scalar built out of $F^\m_{\phantom\m\n}$, can be expressed in terms of the two fundamental invariants $F_{\m\n}F^{\m\n}$ and $F_{\m\n}G^{\m\n}$.
\end{sol}

\begin{sol}{pr12:axion}
The EoM for the scalar field $\p$ is a variation on the Klein--Gordon equation,
\begin{equation}
(\Box+m^2)\p+\frac C8\ve^{\m\n\a\b}F_{\m\n}F_{\a\b}=0\;.
\end{equation}
The EoM for the electromagnetic field follows likewise as
\begin{equation}
\de_\n F^{\m\n}+\frac C2\de_\n\bigl(\p\ve^{\m\n\a\b}F_{\a\b}\bigr)=\de_\n F^{\m\n}+\frac C2\ve^{\m\n\a\b}(\de_\n\p)F_{\a\b}=0\;,
\end{equation}
where in the second step, we used the Bianchi identity for $F_{\a\b}$. The temporal and spatial parts of this EoM translate respectively into
\begin{equation}
\begin{split}
\divg\vec E&=-C\vec B\cdot\grad\p\;,\\
\rot\vec B&=\de_t\vec E+C\vec B\de_t\p-C\vec E\times\grad\p\;.
\end{split}
\end{equation}
This shows that the presence of the scalar field $\p$ can be incorporated in the Maxwell equations effectively as an additional charge density $-C\vec B\cdot\grad\p$ and an additional current density $C(\vec B\de_t\p-\vec E\times\grad\p)$.
\end{sol}

\begin{sol}{pr12:Proca}
The EoM implied by the Lagrangian density of the Proca theory is
\begin{equation}
\de_\n F^{\m\n}+m^2A^\m=0\;.
\end{equation}
By taking the divergence of this equation and using the antisymmetry of the field-strength tensor, we find that $m^2\de_\m A^\m=0$. For any nonzero $m$, $\de_\m A^\m$ must therefore vanish. Inserting this back into the EoM, the latter reduces to
\begin{equation}
(\Box+m^2)A^\m=0\;.
\label{ch12:KGProca}
\end{equation}
This is a Klein--Gordon-type equation, which is expected to describe relativistic particles with rest mass $m$. With a plane-wave ansatz
\begin{equation}
A^\m(x)=\hat A^\m\E^{\I p\cdot x}\;,
\end{equation}
where $p^\m$ is the four-momentum, \eqref{ch12:KGProca} leads to the expected on-shell condition $p^2=-m^2$. The above-derived constraint $\de_\m A^\m=0$ reduces to $p_\m\hat A^\m=p\cdot\hat A=0$. This shows that for any fixed four-momentum $p^\m$, there are only three linearly independent plane-wave solutions. In the rest frame of the particle, $p^\m=(m,\vec0)$ and any purely spatial $\hat A^\m$ will be a solution. In a frame where the Proca particle moves, $p^\m=(E,\vec p)$. Here one finds two solutions for $\smash{\hat A^\m}$ that are purely spatial and perpendicular to $\vec p$. These correspond to transversely polarized waves. There is also one more linearly independent possibility for $\smash{\hat A^\m}$ whose spatial part is parallel to $\vec p$. This amounts to a longitudinally polarized wave.
\end{sol}

\begin{sol}{pr12:ChernSimons}
We start by rewriting the Lagrangian density~\eqref{ch12:ChernSimons} as
\begin{equation}
\La=-\frac14F_{\m\n}F^{\m\n}+\frac k{8\pi}\ve^{\l\m\n}A_\l F_{\m\n}\;.
\end{equation}
This makes it clear that upon a gauge transformation $A_\m\to A_\m+\de_\m\chi$, the Lagrangian density shifts by
\begin{equation}
\udelta\La=\frac k{8\pi}\ve^{\l\m\n}(\de_\l\chi)F_{\m\n}=\frac k{8\pi}\de_\l\bigl(\chi\ve^{\l\m\n}F_{\m\n}\bigr)\;,
\end{equation}
where in the last step, we used the Bianchi identity for the field-strength tensor. This confirms that the Lagrangian density is gauge-invariant up to a surface term. The EoM of the Maxwell--Chern--Simons theory is derived by evaluating separately
\begin{equation}
\begin{split}
\PD\La{A_\m}&=\frac k{8\pi}\ve^{\m\n\l}F_{\n\l}\;,\\
\de_\n\PD{\La}{(\de_\n A_\m)}&=\de_\n\left(-F^{\n\m}+\frac k{4\pi}\ve^{\l\n\m}A_\l\right)=\de_\n F^{\m\n}-\frac k{8\pi}\ve^{\m\n\l}F_{\n\l}\;.
\end{split}
\end{equation}
Setting these two expressions equal to each other, we get the final result for the EoM,
\begin{equation}
\de_\n F^{\m\n}=\frac k{4\pi}\ve^{\m\n\l}F_{\n\l}=\frac k{4\pi}\ve^{\m\a\b}F_{\a\b}\;.
\end{equation}
\end{sol}

\begin{sol}{pr12:ChernSimonsspectrum}
In the Lorenz gauge, the EoM of the Maxwell--Chern--Simons theory~\eqref{ch12:MaxwellChernSimonsEoM} reduces to
\begin{equation}
\Box A^\m=\frac k{4\pi}\ve^{\m\a\b}F_{\a\b}=\frac k{2\pi}\ve^{\m\a\b}\de_\a A_\b\;.
\end{equation}
With the plane-wave ansatz $A^\m(x)=\hat A^\m\E^{\I p\cdot x}$ where $p_\m\hat A^\m=0$ by the Lorenz gauge condition, this becomes
\begin{equation}
p^2\hat A^\m=\frac{\I k}{2\pi}\ve^{\m\a\b}p_\a\hat A_\b\;.
\label{ch12:MCSEoMaux}
\end{equation}
Importantly, this makes clear that $p^2=0$ cannot correspond to any physical solution, as it would imply that the plane wave satisfies $F_{\a\b}=0$. Such a solution for $A^\m$ does not involve any oscillations of the electromagnetic field, and is thus a mere gauge artifact. Getting back to~\eqref{ch12:MCSEoMaux}, we can find the allowed values of $p^2$ by iteration in terms of the amplitude $\hat A^\m$,
\begin{equation}
\begin{split}
p^2\hat A^\m&=\frac1{p^2}\left(\frac{\I k}{2\pi}\right)^2\ve^{\m\a\b}p_\a\ve_{\b\g\d}p^\g\hat A^\d\\
&=\frac1{p^2}\left(\frac k{2\pi}\right)^2(\d^\m_\g\d^\a_\d-\d^\m_\d\d^\a_\g)p_\a p^\g\hat A^\d\\
&=\frac1{p^2}\left(\frac k{2\pi}\right)^2[(\xcancel{p\cdot\hat A})p^\m-p^2\hat A^\m]=-\left(\frac k{2\pi}\right)^2\hat A^\m\;.
\end{split}
\end{equation}
This confirms that the Maxwell--Chern--Simons theory describes relativistic particles with mass $\abs k/(2\pi)$.
\end{sol}
\chapter{Introduction to Linear Algebra}
\label{app:linalg}

\keywords{Vector, vector space, linear independence, basis, dimension, operator, eigenvector, eigenvalue, spectrum, trace, determinant, scalar product, Hermitian operator, unitary operator, symmetric operator, orthogonal operator, bilinear form, quadratic form, signature.}

%%%%%%%%%%%%%%%%%%%%%%%%%%%%%%%%%%%%%%%%%%%%%%%%%%%%%%%%%%%%

\noindent In this \appendixname, we give a brief overview of the basics of linear algebra, assuming some familiarity with matrices and their manipulation. The text is by no means complete; refer to any textbook on linear algebra for a more systematic treatment of the subject. The choice of topics covered here is geared towards the two major applications of linear algebra in our course: the kinematics of a rotating rigid body and the dynamics of coupled oscillations. The rest of the material is included largely just to make the \appendixname{} reasonably self-contained.

%%%%%%%%%%%%%%%%%%%%%%%%%%%%%%%%%%%%%%%%%%%%%%%%%%%%%%%%%%%%

\section{Vectors and Vector Spaces}

We start with an informal reminder of the basic definitions. A \emph{vector space} $V$ (also called a \emph{linear space}) is a set of elements called \emph{vectors} that can be manipulated in a specific manner. Thus, a vector can be multiplied by a number (referred to as a \emph{scalar}), and two vectors can be added to each other. Both operations are subsumed into the concept of a \emph{linear combination}: for any two vectors $\vec u,\vec v\in V$ and scalars $a,b$, the combination $a\vec u+b\vec v$ is well-defined and also lies in $V$. These basic operations on vectors have the usual properties of associativity, distributivity and commutativity. In most cases of practical interest, the scalars are either real or complex numbers, and we then speak accordingly of a real or a complex vector space.

\begin{illustration}%
\label{exapp:vectorspace}%
The set of $n$-tuples $(x_1,\dotsc,x_n)$ of real numbers, known as $\R^n$, is a vector space. Multiplication by a scalar and addition are both defined component-wise,
\begin{equation}
\begin{split}
a(x_1,\dotsc,x_n)&\equiv(ax_1,\dotsc,ax_n)\;,\\
(x_1,\dotsc,x_n)+(y_1,\dotsc,y_n)&\equiv(x_1+y_1,\dotsc,x_n+y_n)\;.
\end{split}
\end{equation}
The space $\R^n$ of real $n$-tuples is a special case of a space of $m\times n$ arrays (matrices), denoted as $\R^{m\times n}$. Also here, multiplication by a scalar and addition are defined component-wise. A qualitatively different example of a vector space is that of polynomials of a single variable and a degree of at most $n$. Multiplication of polynomials by a scalar and their addition is defined by the analogous operations on polynomials as functions. Each polynomial can be represented by its coefficients,
\begin{equation}
P(x)\equiv c_0+c_1x+c_2x^2+\dotsb+ c_nx^n\to(c_0,\dotsc,c_n)\;.
\label{app:poly}
\end{equation}
This establishes a one-to-one mapping between the space of polynomials and $\R^{n+1}$ or $\C^{n+1}$, depending on whether the coefficients are real or complex. The space of polynomials itself is then denoted as $P_n(\R)$ or $P_n(\C)$, respectively.
\end{illustration}

Every vector space has a distinguished element dubbed the zero vector, $\vec0$, which can be thought of literally as arising from multiplying any vector with a zero scalar. A set of vectors $\vec v_1,\dotsc,\vec v_n$ is called \emph{linearly independent} if the equation $a_1\vec v_1+\dotsb+a_n\vec v_n=\vec0$ for the scalars $a_1,\dotsc,a_n$ has only the trivial solution $a_1=\dotsb=a_n=0$. Otherwise, the vectors are called \emph{linearly dependent}. A linearly independent set of vectors $\vec e_1,\dotsc,\vec e_n$ such that every other vector $\vec v\in V$ can be written as their linear combination, $\vec v=v_1\vec e_1+\dotsb+v_n\vec e_n$ with some scalars $v_1,\dotsc,v_n$, is called a \emph{basis} of $V$. The number of elements $n$ of the basis denotes the \emph{dimension} of the space.

\begin{illustration}%
In any vector space there are infinitely many bases. However, some bases may naturally suggest themselves depending on the concrete vector space. Thus, a canonical choice of basis in $\R^n$ is
\begin{equation}
\vec e_1=(1,0,0,\dotsc,0,0)\;,\ 
\vec e_2=(0,1,0,\dotsc,0,0)\;,\ \dotsc\;,
\vec e_n=(0,0,0,\dotsc,0,1)\;.
\end{equation}
In this basis, $(x_1,\dotsc,x_n)=x_1\vec e_1+\dotsb+x_n\vec e_n$. Likewise, the natural choice of basis of the $(n+1)$-dimensional space $P_n$ of polynomials of degree up to $n$ (cf.~\refex{exapp:vectorspace})~is
\begin{equation}
\vec e_0=1\;,\ 
\vec e_1=x\;,\ 
\vec e_2=x^2\;,\ \dotsc\;,\ 
\vec e_n=x^n\;.
\end{equation}
The polynomial~\eqref{app:poly} can then be equivalently represented as the linear combination $c_0\vec e_0+\dotsb+c_n\vec e_n$.
\end{illustration}

To bring the abstract notion of a vector space closer to the more elementary and possibly more familiar matrix formalism, we collect the basis vectors in a row matrix and the components $v_1,\dotsc,v_n$ of a vector $\vec v$ in the basis in a column matrix. The vector can then be formally represented by a matrix product,
\begin{equation}
\vec v=\begin{pmatrix}
\vec e_1,\dotsc,\vec e_n
\end{pmatrix}
\raisebox{-3ex}{$\begin{pmatrix}
v_1\\
\vdots\\
v_n
\end{pmatrix}$}\;.
\label{app:matrixrep}
\end{equation}
The same vector can be represented by its components in many different bases. It is therefore important to be able to relate the descriptions of vectors in different bases. Suppose we switch from an ``old'' basis $\vec e_1,\dotsc,\vec e_n$ to a ``new'' basis $\vec e_1',\dotsc,\vec e_n'$. All the elements of the latter can of course be represented by linear combinations of the old basis vectors. Let us write the relation of the two bases as
\begin{equation}
\begin{pmatrix}
\vec e_1',\dotsc,\vec e_n'
\end{pmatrix}=
\begin{pmatrix}
\vec e_1,\dotsc,\vec e_n
\end{pmatrix}
\raisebox{-3ex}{$\begin{pmatrix}
P_{11} & \dotsc & P_{1n}\\
\vdots & \ddots & \vdots\\
P_{n1} & \cdots & P_{nn}
\end{pmatrix}$}=
\begin{pmatrix}
\vec e_1,\dotsc,\vec e_n
\end{pmatrix}
P\;.
\label{app:basischange}
\end{equation}
Using this in~\eqref{app:matrixrep},
\begin{equation}
\vec v=\begin{pmatrix}
\vec e_1,\dotsc,\vec e_n
\end{pmatrix}
PP^{-1}
\raisebox{-3ex}{$\begin{pmatrix}
v_1\\
\vdots\\
v_n
\end{pmatrix}$}=
\begin{pmatrix}
\vec e_1',\dotsc,\vec e_n'
\end{pmatrix}
P^{-1}
\raisebox{-3ex}{$\begin{pmatrix}
v_1\\
\vdots\\
v_n
\end{pmatrix}$}\;,
\end{equation}
we see that the coordinates of a vector in the two bases are related by
\begin{equation}
\raisebox{-3ex}{$\begin{pmatrix}
v_1'\\
\vdots\\
v_n'
\end{pmatrix}$}=P^{-1}
\raisebox{-3ex}{$\begin{pmatrix}
v_1\\
\vdots\\
v_n
\end{pmatrix}$}\;.
\label{app:basis_transfo}
\end{equation}
Apart from its immediate practical utility, the ability to change basis flexibly plays a key role for understanding the action of operators on vector spaces, which we introduce next.

%%%%%%%%%%%%%%%%%%%%%%%%%%%%%%%%%%%%%%%%%%%%%%%%%%%%%%%%%%%%

\section{Operators on Vector Spaces}

An \emph{operator} $\hat A$ is a map $\hat A:V\to V$ on the vector space $V$ that is linear, i.e.~$\hat A(a\vec u+b\vec v)=a(\hat A\vec u)+b(\hat A\vec v)$ for any vectors $\vec u,\vec v\in V$ and scalars $a,b$. Thanks to its linearity, the operator is fully specified by its action on any chosen basis,
\begin{equation}
\begin{pmatrix}
\smash{\hat A\vec e_1,\dotsc,\hat A\vec e_n}
\end{pmatrix}=
\begin{pmatrix}
\vec e_1,\dotsc,\vec e_n
\end{pmatrix}
\raisebox{-3ex}{$\begin{pmatrix}
A_{11} & \dotsc & A_{1n}\\
\vdots & \ddots & \vdots\\
A_{n1} & \cdots & A_{nn}
\end{pmatrix}$}\;.
\end{equation}
This can be expressed equivalently as $\hat A\vec e_i=A_{1i}\vec e_1+\dotsb+A_{ni}\vec e_n$. Note that while the operator itself is defined without a reference to a particular basis, its \emph{matrix elements} $A_{ij}$ depend on the choice of basis. Once the basis is fixed, we can use \eqref{app:matrixrep} to deduce the action of the operator on an arbitrary vector in $V$,
\begin{equation}
\hat A\vec v=\begin{pmatrix}
\smash{\hat A\vec e_1,\dotsc,\hat A\vec e_n}
\end{pmatrix}
\raisebox{-3ex}{$\begin{pmatrix}
v_1\\
\vdots\\
v_n
\end{pmatrix}$}=
\begin{pmatrix}
\vec e_1,\dotsc,\vec e_n
\end{pmatrix}
\raisebox{-3ex}{$\begin{pmatrix}
A_{11} & \dotsc & A_{1n}\\
\vdots & \ddots & \vdots\\
A_{n1} & \cdots & A_{nn}
\end{pmatrix}$}
\raisebox{-3ex}{$\begin{pmatrix}
v_1\\
\vdots\\
v_n
\end{pmatrix}$}\;.
\label{app:Aonv}
\end{equation}
Thus, the $i$-th component of $\hat A\vec v$ is $(\hat A\vec v)_i=A_{i1}v_1+\dotsb+A_{in}v_n$.

\begin{illustration}%
An example of a linear operator that one meets early on in physics education is rotation in the Euclidean plane $\R^2$ around the origin. Using the canonical Cartesian basis $\vec e_1=(1,0)$ and $\vec e_2=(0,1)$, the action of a counterclockwise rotation $\hat R_\vp$ by angle $\vp$ is completely determined by $\hat R_\vp\vec e_1=(\cos\vp,\sin\vp)$ and $\hat R_\vp\vec e_2=(-\sin\vp,\cos\vp)$. Hence, the matrix representation of the rotation in the canonical basis is
\begin{equation}
R_\vp=\begin{pmatrix}
\cos\vp & -\sin\vp\\
\sin\vp & \cos\vp
\end{pmatrix}\;.
\label{app:rotmatrix}
\end{equation}
Accordingly, the action of $\hat R_\vp$ on an arbitrary vector $\vec v=v_1\vec e_1+v_2\vec e_2\in\R^2$ leads to the vector $\hat R_\vp\vec v$ with components
\begin{equation}
(\hat R_\vp\vec v)_1=v_1\cos\vp-v_2\sin\vp\;,\qquad
(\hat R_\vp\vec v)_2=v_1\sin\vp+v_2\cos\vp\;.
\end{equation}
\end{illustration}

\begin{illustration}%
Taking a derivative with respect to the sole variable $x$ defines a linear operator $\hat\de$ on the space $P_n$ of polynomials of degree at most $n$. As always, the operator is completely specified by its action on a basis of the space. In particular for the monomial basis $1,x,x^2,\dotsc,x^n$, we have $\hat\de1=0$ and $\hat\de x^k=kx^{k-1}$ for $k=1,\dotsc,n$. In this basis, the derivative operator is therefore represented by the matrix
\begin{equation}
\de=\begin{pmatrix}
0 & 1 & 0 & \dotsb & 0\\
0 & 0 & 2 & \dotsb & 0\\
0 & 0 & 0 & \dotsb & 0\\[2pt]
\smash{\vdots} & \smash{\vdots} & \smash{\vdots} & \smash{\ddots} & \smash{\vdots}\\
0 & 0 & 0 & \dotsb & 0
\end{pmatrix}\;.
\end{equation}
\end{illustration}

The last thing we need to figure out is how the matrix representation of an operator changes upon a change of basis of the vector space. By combining~\eqref{app:basischange} with~\eqref{app:basis_transfo} and~\eqref{app:Aonv}, we find
\begin{equation}
\hat A\vec v=
\begin{pmatrix}
\vec e_1',\dotsc,\vec e_n'
\end{pmatrix}
\raisebox{-3ex}{$\begin{pmatrix}
A_{11}' & \dotsc & A_{1n}'\\
\vdots & \ddots & \vdots\\
A_{n1}' & \cdots & A_{nn}'
\end{pmatrix}$}
\raisebox{-3ex}{$\begin{pmatrix}
v_1'\\
\vdots\\
v_n'
\end{pmatrix}$}=
\begin{pmatrix}
\vec e_1,\dotsc,\vec e_n
\end{pmatrix}
P
\raisebox{-3ex}{$\begin{pmatrix}
A_{11}' & \dotsc & A_{1n}'\\
\vdots & \ddots & \vdots\\
A_{n1}' & \cdots & A_{nn}'
\end{pmatrix}$}
P^{-1}
\raisebox{-3ex}{$\begin{pmatrix}
v_1\\
\vdots\\
v_n
\end{pmatrix}$}\;.
\label{app:optransfo}
\end{equation}
Denoting the matrix representations of the operator $\hat A$ in the two bases as $A$ and $A'$ (without the hat), we conclude that $A'=P^{-1}AP$.

%%%%%%%%%%%%%%%%%%%%%%%%%%%%%%%%%%%%%%%%%%%%%%%%%%%%%%%%%%%%

\subsection{Functions of Operators, Trace and Determinant}

The details of the matrix representation of an operator $\hat A:V\to V$ depend on the choice of basis of the space $V$. However, there are certain combinations of the matrix elements $A_{ij}$ that take the same value in any basis, and thus give us direct access to the properties of the operator $\hat A$ itself. The first of these is the \emph{trace},
\begin{equation}
\tr A\equiv\sum_{i=1}^nA_{ii}\;,
\label{app:tracedef}
\end{equation}
that is the sum of the diagonal elements of the matrix $A_{ij}$. It is easy to prove that the trace of a product of matrices satisfies the cyclic property
\begin{equation}
\tr(A_1A_2\dotsb A_{k-1}A_k)=\tr(A_kA_1A_2\dotsb A_{k-1})
\end{equation}
for any $k\geq1$. This implies at once that the trace of an operator is indeed well-defined independently of the choice of basis, since $\tr A'=\tr(P^{-1}AP)=\tr(PP^{-1}A)=\tr A$. 

The other combination of the matrix elements $A_{ij}$ we wish to introduce is the \emph{determinant}, defined by
\begin{equation}
\det A\equiv\sum_\s\sgn\s\,A_{1\s(1)}\dotsb A_{n\s(n)}\;.
\label{app:detdef}
\end{equation}
Here the sum runs over all permutations of $n$ elements and $\sgn\s$ is the sign (parity) of the permutation. Importantly, the determinant factorizes over a product of matrices,
\begin{equation}
\det(A_1\dotsb A_k)=(\det A_1)\dotsb(\det A_k)\;.
\end{equation}
Together with the fact that $\det\un=1$, this shows that $\det P^{-1}=1/\det P$ for any invertible matrix $P$. An immediate corollary is that the determinant of an operator is, like the trace, independent of the choice of basis, $\det A'=\det(P^{-1}AP)=(\det P^{-1})(\det A)(\det P)=\det A$.

\begin{illustration}%
The trace of the rotation matrix~\eqref{app:rotmatrix} is $\tr R_\vp=2\cos\vp$. In this case, the trace therefore gives us access to the angle of the rotation regardless of the basis in which we might want to express the operator $\hat R_\vp$. In addition, we find that $\det R_\vp=1$. This is not a coincidence. The determinant of any rotation, also in higher-dimensional Euclidean spaces, always equals one. We will understand why below when we introduce the concept of an orthogonal operator.
\end{illustration}

The mere fact that matrices can be added and multiplied allows one to define a very general class of operations on matrices. Suppose that $f$ is a function of a single variable that is analytic, i.e.~it has a convergent power series. The latter can be used to define the same function on square matrices,
\begin{equation}
f(x)=\sum_{k=0}^\infty\frac{f_k}{k!}x^k\quad\to\quad
f(A)\equiv\sum_{k=0}^\infty\frac{f_k}{k!}A^k\;.
\label{app:functionoperator}
\end{equation}
Thinking of $A$ as a representation of an operator $\hat A$ in a particular basis on the vector space $V$, we have thus also defined the function of the operator itself, $f(\hat A)$. It is obvious from the construction that upon a change of basis such that $A\to P^{-1}AP$, the matrix representation of $f(\hat A)$ changes in the same way, $f(A)\to P^{-1}f(A)P$.

\begin{illustration}%
One function that is worth mentioning explicitly is the exponential. In fact, you are certainly already familiar with one special case of the exponential of an operator, possibly without being aware of it. Indeed, recall the polynomial space $P_n$ and evaluate the exponential of the derivative operator $\hat\de$ through its action on a polynomial $P(x)$ from the space,
\begin{equation}
\exp(a\hat\de)P(x)=\sum_{k=0}^\infty\frac{a^k\hat\de^k}{k!}P(x)=\sum_{k=0}^\infty\frac{a^k}{k!}P^{(k)}(x)=P(x+a)\;.
\end{equation}
This is nothing but the usual Taylor expansion: the exponential of the derivative operator is equivalent to a shift of the argument of the polynomial. Putting this specific example aside, the exponential of an operator $\hat A$ on a vector space $V$ also enters a remarkable identity that connects the trace and determinant of the operator,
\begin{equation}
\exp\tr\hat A=\det\exp\hat A\;.
\label{app:exptr}
\end{equation}
This plays an important role in diverse areas of mathematics and theoretical physics. In order to get an intuitive understanding of where this relation comes from, we need further insight into the structure of operators independent of the choice of basis of the vector space.
\end{illustration}

%%%%%%%%%%%%%%%%%%%%%%%%%%%%%%%%%%%%%%%%%%%%%%%%%%%%%%%%%%%%

\subsection{Eigenvectors and Eigenvalues}

A nonzero vector $\vec v$ that satisfies the equation
\begin{equation}
\hat A\vec v=\l\vec v\;
\label{app:eigenvalue}
\end{equation}
for some scalar $\l$, is called an \emph{eigenvector} of the operator $\hat A$, and $\lambda$ is then the corresponding \emph{eigenvalue}. Thanks to the linearity of~\eqref{app:eigenvalue}, the eigenvector is only determined up to multiplication by a scalar. Loosely speaking, it is the ``direction'' of $\vec v$ that decides whether or not it is an eigenvector of $\hat A$, not its ``magnitude.'' The set of all eigenvalues of the operator is called its \emph{spectrum}. As follows from~\eqref{app:Aonv}, in a given basis on $V$, the definition~\eqref{app:eigenvalue} of an eigenvalue reduces to $A\vec v=\lambda\vec v$.\footnote{Here we are using the same symbol $\vec v$ for a vector in $V$ and the set of its components in a given basis. There is no such an ambiguity for operators, which are distinguished by a hat from the set of their matrix elements.} Upon a change of basis, the components of the eigenvector change to $\vec v'=P^{-1}\vec v$ in line with~\eqref{app:basis_transfo}. Likewise, according to~\eqref{app:optransfo}, the matrix elements of the operator change to $A'=P^{-1}AP$. Hence, in the new basis, $A'\vec v'=P^{-1}APP^{-1}\vec v=P^{-1}A\vec v=\l\vec v'$. This confirms that the notion of an eigenvalue (and by extension that of the spectrum) is well-defined. While the components of the eigenvector obviously depend on the choice of basis, the eigenvalue itself does not.

\begin{illustration}%
\label{exapp:Pauli}%
Recall the Pauli matrices from your basic course on quantum mechanics,
\begin{equation}
\tau_1=\begin{pmatrix}
0 & 1\\
1 & 0
\end{pmatrix}\;,\qquad
\tau_2=\begin{pmatrix}
0 & -\I\\
+\I & 0
\end{pmatrix}\;,\qquad
\tau_3=\begin{pmatrix}
1 & 0\\
0 & -1
\end{pmatrix}\;.
\end{equation}
All the Pauli matrices have the same spectrum, $\{+1,-1\}$. Up to arbitrary normalization, the corresponding eigenvectors are (you might want to check this explicitly)
\begin{equation}
\vec v^{(1)}_+=\begin{pmatrix}1\\1\end{pmatrix}\;,
\vec v^{(1)}_-=\begin{pmatrix}1\\-1\end{pmatrix}\;,\quad
\vec v^{(2)}_+=\begin{pmatrix}1\\\I\end{pmatrix}\;,
\vec v^{(2)}_-=\begin{pmatrix}1\\-\I\end{pmatrix}\;,\quad
\vec v^{(3)}_+=\begin{pmatrix}1\\0\end{pmatrix}\;,
\vec v^{(3)}_-=\begin{pmatrix}0\\-1\end{pmatrix}\;.
\label{app:Paulieigen}
\end{equation}
\end{illustration}

A simple but useful observation is that a set of eigenvectors, corresponding to a different eigenvalue each, is linearly independent. This shows that in a vector space $V$ of dimension $n$, an operator $\hat A$ can have at most $n$ different eigenvalues. If the operator does have $n$ different eigenvalues $\l_i$, $i=1,\dotsc,n$, then the corresponding eigenvectors $\vec v_i$ constitute a distinguished basis of the space $V$. In this basis, the operator is represented by a diagonal matrix,
\begin{equation}
A=\begin{pmatrix}
\l_1 & 0 & 0 & \dotsb & 0\\
0 & \l_2 & 0 & \dotsb & 0\\
0 & 0 & \l_3 & \dotsb & 0\\[2pt]
\smash\vdots & \smash\vdots & \smash\vdots & \smash\ddots & \smash\vdots\\
0 & 0 & 0 & \dotsb & \l_n
\end{pmatrix}\;.
\label{app:diagA}
\end{equation}
To save some writing, it is common to abbreviate this as $A=\diag(\l_1,\dotsc,\l_n)$. For many operators $\hat A$ that have fewer than $n$ different eigenvalues, a basis of $V$ consisting entirely of eigenvectors of $\hat A$ still exists. Such operators are called \emph{diagonalizable}.

\begin{watchout}%
Reducing the set of matrix elements $A_{ij}$ of $\hat A$ to a diagonal matrix is of great practical utility. It makes it much easier to perform algebraic manipulations with the matrix. It also offers an alternative way to define functions of diagonalizable operators. Namely, it is easy to check using~\eqref{app:functionoperator} that in a basis of eigenvectors of $\hat A$, $f(A)=\diag(f(\l_1),\dotsc,f(\l_n))$. Last but not least, for diagonalizable operators, the important concepts of trace and determinant have a very simple relation to the set of eigenvalues,
\begin{equation}
\tr\hat A=\sum_{i=1}^n\l_i\;,\qquad
\det\hat A=\prod_{i=1}^n\l_i\;.
\end{equation}
This provides the ultimate explanation of the origin of the identity~\eqref{app:exptr} as a vast generalization of the basic property of the exponential function, $\E^{x+y}=\E^x\E^y$. For the sake of completeness, let us add that~\eqref{app:exptr} holds for any operator $\hat A$, even if it is not diagonalizable.
\end{watchout}

The above clearly demonstrates the value of knowing the spectrum of an operator and the corresponding eigenvectors. However, we still do not have a systematic way to compute the spectrum. To that end, note that~\eqref{app:eigenvalue} can be rewritten as $(\hat A-\l\hat\un)\vec v=\vec0$, where $\hat\un$ is the identity operator defined by $\hat\un\vec v=\vec v$ for all $\vec v\in V$. In other words, $\vec v$ is an eigenvector of the operator $\hat A-\l\hat\un$ with zero eigenvalue. Now an operator that has zero in its spectrum cannot be invertible: to prove this, assume the opposite and multiply~\eqref{app:eigenvalue} by the supposedly existent inverse $\hat A^{-1}$. We know from matrix algebra (cf.~\href{https://en.wikipedia.org/wiki/Cramer%27s_rule}{Cramer's rule}) that a square matrix is invertible if and only if it has a nonzero determinant. This allows us to rewrite the definition of an eigenvalue of an operator equivalently as the condition
\begin{equation}
\text{$\l$ is an eigenvalue of $\hat A$}\quad\Leftrightarrow\quad\det(\hat A-\l\hat\un)=0\;.
\label{app:secular}
\end{equation}
The latter constitutes the \emph{characteristic equation} of the operator, whose roots are its eigenvalues. Thus, finding the spectrum of an operator in an $n$-dimensional vector space boils down to solving an algebraic equation of degree $n$. From the fundamental theorem of algebra, we know that in the complex domain, every such equation has, counting multiplicity of roots, exactly $n$ solutions. Counting the real roots of real algebraic equations is less straightforwards. We will therefore for the time being focus on complex vector spaces, and return to real vector spaces later on.

Suppose the characteristic equation has $n$ \emph{different} roots; this is the default case barring accidental tuning of the coefficients of the equation. Then the operator $\hat A$ has $n$ different eigenvalues. Using the argument above~\eqref{app:diagA}, we conclude that ``most'' operators are diagonalizable: this is good news! An operator may not be diagonalizable only when its spectrum is degenerate, that is, its characteristic equation has roots with multiplicity greater than one. We give a simple example below. However, in the next section we will introduce further structure, which guarantees that all operators relevant for this course are indeed diagonalizable.

\begin{illustration}%
Recall once more the space $P_n$ of polynomials of degree $n$ or less and the derivative operator $\hat\de$ acting on it. By definition, a nonzero polynomial $P(x)$ is an eigenvector of $\hat\de$ if $\smash{\hat\de P(x)=P'(x)\ifeq\l P(x)}$. This is a simple first-order differential equation that is solved by $P(x)=\E^{\l x}P(0)$. The problem is that our solution is not a polynomial for any nonzero $\l$. Thus, the derivative operator has only one eigenvalue $\l=0$ and one corresponding linearly independent eigenvector, $P(x)=1$. The conclusion is that $\hat\de$ is trivially diagonalizable on the one-dimensional space $P_0$ of polynomials of degree zero (that is constant functions). On $P_n$ with any $n\geq1$, the derivative operator is not diagonalizable.
\end{illustration}

%%%%%%%%%%%%%%%%%%%%%%%%%%%%%%%%%%%%%%%%%%%%%%%%%%%%%%%%%%%%

\section{Spaces with Scalar Product}
\label{appsec:scalarproduct}

The abstract notion of a \emph{scalar product} generalizes the concept familiar from elementary vector algebra. Formally, a scalar product on a complex vector space $V$ is a map $(\cdot,\cdot):V\times V\to\C$ with the following properties:
\begin{itemize}
\item Linearity with respect to addition: $\scal{\vec u}{\vec v+\vec w}=\scal{\vec u}{\vec v}+\scal{\vec u}{\vec w}$ and $\scal{\vec u+\vec v}{\vec w}=\scal{\vec u}{\vec w}+\scal{\vec v}{\vec w}$ for all $\vec u,\vec v,\vec w\in V$.
\item Linearity with respect to multiplication by a scalar $a$: $\scal{\vec u}{a\vec v}=a\scal{\vec u}{\vec v}$ and $\scal{a\vec u}{\vec v}=a^*\scal{\vec u}{\vec v}$. Note that the scalar $a$ is complex-conjugated when it is factored out of the first argument of the scalar product.
\item Symmetry: $\scal{\vec u}{\vec v}=\scal{\vec v}{\vec u}^*$.
\item Positivity: $\scal{\vec v}{\vec v}$ is real and non-negative for all $\vec v\in V$, and $\scal{\vec v}{\vec v}=0$ if and only if $\vec v=\vec 0$. The expression $\sqrt{\scal{\vec v}{\vec v}}$ can be interpreted as the length of the vector $\vec v$.
\end{itemize}
Two vectors whose scalar product is equal to zero are called \emph{orthogonal}. A given basis $\vec e_1,\dotsc,\vec e_n$ of the vector space $V$ is called \emph{orthonormal} if it satisfies $\scal{\vec e_i}{\vec e_j}=\d_{ij}$. In other words, each of the basis vectors has a unit length and every two different vectors are mutually orthogonal.

\begin{illustration}%
In spite of working with complex vector spaces, you can image vectors therein intuitively as ``arrows'' and the scalar product as measuring the ``angle'' between the vectors. It is therefore important to stress that the concept of scalar product has, just like vector spaces themselves, a much broader landscape of applications beyond this simple geometric picture. In particular in quantum mechanics, spaces whose elements are functions play a key role.\footnote{Most function spaces relevant for quantum mechanics are, in fact, \emph{infinite-dimensional}. Here we will sweep this subtlety under the carpet and pretend that all mathematical operations on functions as vectors remain well-defined without further qualifications.} First, consider the space of (smooth) functions on the interval $[-1,+1]$, endowed with the scalar product
\begin{equation}
\scal{f}{g}\equiv\int_{-1}^{+1}f(x)^*g(x)\,\D x\;.
\end{equation}
This space has a distinguished basis, obtained from the basis $1,x,x^2,\dotsc$ by \href{https://en.wikipedia.org/wiki/Gram%E2%80%93Schmidt_process}{Gram--Schmidt orthogonalization}. Its elements are the \emph{Legendre polynomials} that can be most elegantly, if somewhat implicitly, expressed as
\begin{equation}
P_n(x)\equiv\frac1{2^nn!}\frac{\D^n}{\D x^n}(x^2-1)^n\quad\text{for any integer $n\geq0$}\;.
\end{equation}
These polynomials are all mutually orthogonal, but they are not normalized to unity. Their ``length'' is given by $\scal{P_m}{P_n}=2\d_{mn}/(2n+1)$. The Legendre polynomials are closely related to the spherical harmonics, which describe the wave functions of simultaneous eigenstates of the square of the operator of angular momentum and one of its components. They have many fascinating properties; see the relevant \href{https://en.wikipedia.org/wiki/Legendre_polynomials}{Wikipedia page} for further information.

Another infinite-dimensional complex vector space relevant for quantum mechanics is that of (smooth and bounded) functions on $\R$ with the scalar product
\begin{equation}
\scal{f}{g}\equiv\int_{-\infty}^{+\infty}f(x)^*g(x)\E^{-x^2}\D x\;.
\end{equation}
Starting from the same set of monomials $1,x,x^2,\dotsc$, the Gram--Schmidt orthogonalization now leads to a quite different basis, namely that of \emph{Hermite polynomials},
\begin{equation}
H_n(x)\equiv(-1)^n\E^{x^2}\frac{\D^n}{\D x^n}\E^{-x^2}\;,\qquad n\geq0\;.
\end{equation}
These constitute an orthogonal basis of the space, with the ``length'' given by $\scal{H_m}{H_n}=2^nn!\sqrt\pi\d_{mn}$. The Hermite polynomials determine the wave functions of the stationary states of a quantum harmonic oscillator; see the corresponding \href{https://en.wikipedia.org/wiki/Hermite_polynomials}{Wikipedia page} for more information.
\end{illustration}

An orthonormal basis allows us to find the components of a given vector easily using the scalar product,
\begin{equation}
\vec v=\sum_{i=1}^nv_i\vec e_i\;,\quad\text{where }
v_i=\scal{\vec e_i}{\vec v}\;.
\label{app:scalar_product}
\end{equation}
The scalar product of two vectors can then be expressed component-wise as
\begin{equation}
\scal{\vec u}{\vec v}=\sum_{i,j=1}^n\scal{u_i\vec e_i}{v_j\vec e_j}=\sum_{i,j=1}^nu_i^*v_j\scal{\vec e_i}{\vec e_j}=\sum_{i,j=1}^nu_i^*v_j\d_{ij}=\sum_{i=1}^nu_i^*v_i\;.
\label{app:scaluv}
\end{equation}
Likewise, it is easy to find a component expression for the action of an operator $\hat A$ on a vector $\vec v$,
\begin{equation}
(\hat A\vec v)_i=\scal{\vec e_i}{\hat A\vec v}=\sum_{j=1}^nv_j\scal{\vec e_i}{\hat A\vec e_j}\;.
\label{app:decompose}
\end{equation}
A comparison with \eqref{app:Aonv} shows that the matrix elements of $\hat A$ in an orthonormal basis are given by
\begin{equation}
A_{ij}=\scal{\vec e_i}{\hat A\vec e_j}\;.
\label{app:Aij}
\end{equation}

%%%%%%%%%%%%%%%%%%%%%%%%%%%%%%%%%%%%%%%%%%%%%%%%%%%%%%%%%%%%

\subsection{Hermitian and Unitary Operators}

Having at hand the scalar product, one can define the \emph{Hermitian conjugate}, $\he{\hat A}$, of a given operator $\hat A$ through
\begin{equation}
\scal{\vec u}{\he{\hat A}\vec v}=\scal{\hat A\vec u}{\vec v}\quad\text{for all }\vec u,\vec v\in V\;.
\label{app:hermconj}
\end{equation}
To unpack the content of the definition, let us have a look at the matrix elements of an operator in an orthonormal basis. A series of simple steps using~\eqref{app:Aij} gives
\begin{equation}
(\he A)_{ij}=\scal{\vec e_i}{\he{\hat A}\vec e_j}=\scal{\hat A\vec e_i}{\vec e_j}=\scal{\vec e_j}{\hat A\vec e_i}^*=A_{ji}^*\;.
\label{app:heAmatel}
\end{equation}
In terms of matrices, Hermitian conjugation corresponds to a combination of complex conjugation and transposition. It is therefore not surprising Hermitian conjugation has some simple properties that mimic those of transposition,
\begin{equation}
\he{(\he{\hat A})}=\hat A\;,\qquad
\he{(\hat A\hat B)}=\he{\hat B}\he{\hat A}\;.
\label{app:hermconj2}
\end{equation}
Both of these can be easily proven directly from the definition~\eqref{app:hermconj} and the properties of the scalar product, independently of a choice of basis.

The concept of Hermitian conjugation naturally leads to a special class of operators with distinguished properties. Thus, an operator is called \emph{Hermitian} if it is equal to its Hermitian conjugate, $\he{\hat A}=\hat A$. Let us see what this implies. We start with a single eigenvector $\vec v$ of a Hermitian operator $\hat A$. Using the definition~\eqref{app:hermconj}, we find
\begin{equation}
\l\scal{\vec v}{\vec v}=\scal{\vec v}{\hat A\vec v}=\scal{\he{\hat A}\vec v}{\vec v}=\scal{\hat A\vec v}{\vec v}=\l^*\scal{\vec v}{\vec v}\;.
\end{equation}
This implies that all eigenvalues of Hermitian operators are necessarily real. Next, consider two eigenvectors $\vec v_1,\vec v_2$ with eigenvalues $\l_1\neq\l_2$. Here we find following the same reasoning that
\begin{equation}
\l_1\scal{\vec v_1}{\vec v_2}=\scal{\hat A\vec v_1}{\vec v_2}=\scal{\vec v_1}{\hat A\vec v_2}=\l_2\scal{\vec v_1}{\vec v_2}\;.
\end{equation}
This is only possible if eigenvectors corresponding to different eigenvalues are mutually orthogonal. In fact, it is not difficult to prove, although we will skip details, that Hermitian operators are diagonalizable and it is always possible to construct an orthonormal basis of the vector space out of their eigenvectors. This is the content of the \emph{spectral theorem}. To summarize:
\begin{enumerate}
\item[(H.1)] All eigenvalues of a Hermitian operator are real.
\item[(H.2)] Eigenvectors of a Hermitian operator corresponding to different eigenvalues are mutually orthogonal.
\item[(H.3)] In the vector space on which a Hermitian operator acts, there is an orthonormal basis composed of its eigenvectors.
\end{enumerate}

\begin{illustration}%
In \refex{exapp:Pauli}, we reviewed the spectrum and eigenvectors of the Pauli matrices. All the Pauli matrices are obviously Hermitian. Check that the pairs of eigenvectors as given in~\eqref{app:Paulieigen} are all mutually orthogonal. The set of all $n\times n$ Hermitian matrices, that is matrices for which $A^*_{ji}=A_{ij}$, is a \emph{real} vector space of dimension $n^2$. In the special case of $n=2$, the four matrices $\un,\tau_1,\tau_2,\tau_3$ define an orthogonal basis of this space with respect to the scalar product $\scal{A}{B}\equiv\tr(AB)$. The three Pauli matrices themselves span a subspace of Hermitian matrices with vanishing trace.
\end{illustration}

Another important class of operators that builds upon the concept of Hermitian conjugation is that of \emph{unitary} operators. An operator $\hat U$ is called unitary if it satisfies
\begin{equation}
\scal{\hat U\vec u}{\hat U\vec v}=\scal{\vec u}{\vec v}
\end{equation}
for all vectors $\vec u,\vec v\in V$. Unitary operators generalize the geometric notion of transformations that preserve the length of vectors and the angles between different vectors. From the definition of Hermitian conjugation~\eqref{app:hermconj} it follows that
\begin{equation}
\hat U\he{\hat U}=\he{\hat U}\hat U=\hat\un\;,
\end{equation}
which is often used as an alternative definition of unitarity. Following the same steps as for Hermitian operators, one can show that the eigenvalues and eigenvectors of unitary operators have the properties:
\begin{enumerate}
\item[(U.1)] All eigenvalues of a unitary operator have absolute value 1.
\item[(U.2)] Eigenvectors of a unitary operator corresponding to different eigenvalues are mutually orthogonal.
\item[(U.3)] In the vector space on which a unitary operator acts, there is an orthonormal basis composed of its eigenvectors.
\end{enumerate}
The latter two properties are actually the same as for Hermitian operators. It is only the first property addressing the spectrum where Hermitian and unitary operators differ. Both classes of operators play a key role in quantum mechanics: Hermitian operators represent physical observables while unitary operators define symmetry transformations of the space of states of the quantum system.

\begin{illustration}%
It is easy to see that the product of two unitary matrices, that is matrices for which $U\he U=\he UU=\un$, is again unitary. Moreover, unitary matrices are by construction invertible with $U^{-1}=\he U$. These properties ensure that the set of all $n\times n$ unitary matrices furnishes a mathematical structure called \emph{Lie group}, denoted as $\gr{U}(n)$. Remarkably, for any Hermitian matrix $A$, the exponential $\exp(\I A)$ is unitary. See for yourself: $\he{\exp(\I A)}=\exp(-\I\he A)=\exp(-\I A)=\exp(\I A)^{-1}$. It turns out that, in fact, any unitary matrix $U$ can be expressed as $U=\exp(\I A)$ with some Hermitian $A$. If you wonder why, remember how to define functions of diagonalizable matrices and then set $A=-\I\log U$. This hints at an intimate relationship between the group of unitary matrices and the space of Hermitian matrices. Without going into further details, we remark that the group structure of unitary matrices defined by matrix multiplication is reflected in an algebraic structure, defined by the commutator of Hermitian matrices. This makes the space of Hermitian matrices into a \emph{Lie algebra}. The subspace of traceless Hermitian matrices gives rise to the subgroup of unitary matrices with unit determinant, called $\gr{SU}(n)$. This correspondence follows from our beloved identity~\eqref{app:exptr}.
\end{illustration}

%%%%%%%%%%%%%%%%%%%%%%%%%%%%%%%%%%%%%%%%%%%%%%%%%%%%%%%%%%%%

\subsection{Real Symmetric and Orthogonal Operators}

Much of what was said above about Hermitian and unitary operators can be translated to real vector spaces. A scalar product on a real vector space $V$ is defined just like before, except that it maps $V\times V$ to $\R$ and the symmetry property is now simply $\scal{\vec u}{\vec v}=\scal{\vec v}{\vec u}$. Similarly, linearity with respect to multiplication by a scalar reduces to $\scal{a\vec u}{\vec v}=\scal{\vec u}{a\vec v}=a\scal{\vec u}{\vec v}$. Finally, the scalar product of two vectors $\vec u,\vec v\in V$ can be expressed in terms of their components in an orthonormal basis $\vec e_i$ by the familiar formula $\scal{\vec u}{\vec v}=\sum_{i=1}^nu_iv_i$, which is a special case of~\eqref{app:scaluv} for real vectors.

On real vector spaces, we define the \emph{transpose}, $\smash{\hat A^T}$, of a given operator $\hat A$ by a formula analogous to \eqref{app:hermconj},
\begin{equation}
\scal{\vec u}{\hat A^T\vec v}=\scal{\hat A\vec u}{\vec v}\;.
\end{equation}
This is just a formalization of the concept of a transpose of a matrix. Indeed, following the same steps as in~\eqref{app:heAmatel}, one finds that the matrix elements of $\hat A^T$ in an orthonormal basis are $(\hat A^T)_{ij}=A_{ji}$. The transpose of operators has the expected properties, copying~\eqref{app:hermconj2},
\begin{equation}
(\hat A^T)^T=\hat A\;,\qquad
(\hat A\hat B)^T=\hat B^T\hat A^T\;.
\end{equation}
An operator $\hat A$ on a real vector space is called \emph{symmetric} if it is equal to its transpose, $\hat A^T=\hat A$. Symmetric operators on real spaces satisfy the same properties (H.1)--(H.3) as Hermitian operators on complex spaces. The real-space analogy of unitary operators are \emph{orthogonal} operators $\hat R$, which are defined by
\begin{equation}
\scal{\hat R\vec u}{\hat R\vec v}=\scal{\vec u}{\vec v}\quad\text{or}\quad
\hat R\hat R^T=\hat R^T\hat R=\hat\un\;.
\label{app:orthogonal}
\end{equation}

\begin{watchout}%
In any orthonormal basis $\vec e_i$ on the vector space, an orthogonal operator $\hat R$ is represented by an orthogonal matrix $R$, satisfying $RR^T=R^TR=\un$. In addition, the defining property $\scal{\hat R\vec u}{\hat R\vec v}=\scal{\vec u}{\vec v}$ guarantees that $\vec e_i'\equiv\hat R\vec e_i=R_{1i}\vec e_1+\dotsb+R_{ni}\vec e_n$ is still an orthonormal basis. As a consequence, any orthogonal operator in an $n$-dimensional real vector space can be thought of as a rotation of a Cartesian coordinate frame in $\R^n$. Combined with the fact that $\det R^T=\det R$, the condition $RR^T=R^TR=\un$ finally implies that $\det R=\pm1$. Rotations for which $\det R=1$ are called \emph{proper}. These represent motions of a rigid body around a fixed point, and can be thought of in terms of a continuous change of orientation of the coordinate frame. On the other hand, rotations with $\det R=-1$ are \emph{improper}. These can always be composed of a proper rotation and a reflection of one of the coordinate axes.
\end{watchout}

At the level of matrix elements, an orthogonal matrix is a special case of a unitary matrix that is real. The properties (U.1)--(U.3) therefore automatically apply to orthogonal \emph{matrices} as well. However, we have to face the fact that both the eigenvalues and the eigenvectors of an orthogonal matrix may be complex. This is the price for solving the characteristic equation in the complex domain, which guarantees a number of solutions for the eigenvalues equal to the dimension of the vector space. A complex eigenvector of the matrix $R$ may not map to any vector in the real vector space on which $\hat R$ acts. In other words, orthogonal operators are, up to rare exceptions, not diagonalizable. This underlines the importance of keeping in mind the distinction between an operator and its matrix representation.

\begin{illustration}%
Recall the matrix~\eqref{app:rotmatrix}, representing a rotation by angle $\vp$ in the Euclidean plane. The eigenvalues of this matrix are $\E^{\pm\I\vp}$, and the corresponding eigenvectors read, up to a choice of normalization,
\begin{equation}
\vec v_+=\begin{pmatrix}1\\-\I\end{pmatrix}\;,\qquad
\vec v_-=\begin{pmatrix}1\\\I\end{pmatrix}\;.
\end{equation}
Note that the eigenvectors match those of the second Pauli matrix; cf.~\eqref{app:Paulieigen}. This follows from the fact that our rotation matrix can be written as $\un\cos\vp-\I\tau_2\sin\vp$. The matrix has a unit determinant and as such represents a proper rotation. Let us use this opportunity to make a comment about rotations in \emph{three}-dimensional Euclidean space, relevant for the kinematics of rigid bodies. The determinant of a proper rotation must equal one, and at the same time all eigenvalues of the rotation matrix must have absolute value one. Moreover, complex eigenvalues can only appear in complex-conjugate pairs, being solutions to an algebraic equation with real coefficients. It follows that in three dimensions, the eigenvalues of a proper rotation are $\{1,\E^{\I\vp},\E^{-\I\vp}\}$ with some angle $\vp$. There is guaranteed to be one real eigenvector with eigenvalue one. This explains \emph{Euler's rotation theorem}, by which any displacement of a rigid body around a fixed point is equivalent to a single rotation $\hat R$ around some axis running through the fixed point. The direction of the axis is given by the eigenvector of $\hat R$ with eigenvalue one. The phase $\vp$ of the complex eigenvalues of $\hat R$ determines the angle of rotation around this axis.
\end{illustration}

This is a good place to point out a useful aspect that is common to Hermitian and real symmetric operators. Let us pick an arbitrary orthonormal basis $\vec e_i$ in the vector space. In this basis, a Hermitian operator $\hat A$ is represented by a Hermitian matrix $A$, and a real symmetric operator $\hat A$ is likewise represented by a real symmetric matrix $A$. In both cases, it is possible to switch to a new orthonormal basis $\vec e'_i$, consisting entirely of eigenvectors of $A$. The two bases are related by~\eqref{app:basischange}; the columns of the matrix $P$ collect the components of the eigenvectors of $\hat A$ in the original basis $\vec e_i$. The fact that the basis of eigenvectors is orthonormal means that the matrix $P$ is unitary ($\he PP=\un$) in case of Hermitian $\hat A$, or real orthogonal ($P^TP=\un$) in case of real symmetric $\hat A$. This ultimately implies that the transformation $A\to A'=P^{-1}AP$ that diagonalizes a Hermitian matrix can be realized by a unitary $P$, and one that diagonalizes a real symmetric matrix can be realized by a real orthogonal $P$.

%%%%%%%%%%%%%%%%%%%%%%%%%%%%%%%%%%%%%%%%%%%%%%%%%%%%%%%%%%%%

\section{Bilinear and Quadratic Forms}

Having covered the necessary basics of linear algebra, we conclude the \appendixname{} with an application that emerges in the study of small oscillations of mechanical systems, and also plays a role in describing the inertia of rotating rigid bodies. Let $V$ be a real vector space. A \emph{bilinear form} is a map $\o:V\times V\to\R$ that is linear in both of its arguments. This is a generalization of the concept of scalar product on real vector spaces where we do not impose the symmetry and positivity properties. The bilinear form is completely determined by its action on any chosen (not necessarily orthonormal) basis of $V$, $\vec e_1,\dotsc,\vec e_n$. Indeed, in terms of the coefficients $\O_{ij}\equiv\o(\vec e_i,\vec e_j)$ one finds, for any two vectors $\vec u,\vec v\in V$,
\begin{equation}
\omega(\vec u,\vec v)=\sum_{i,j=1}^nu_iv_j\o(\vec e_i,\vec e_j)=\sum_{i,j=1}^n\O_{ij}u_iv_j=
\begin{pmatrix}
u_1,\dotsc,u_n
\end{pmatrix}
\raisebox{-3ex}{$\begin{pmatrix}
\O_{11} & \dotsc & \O_{1n}\\
\vdots & \ddots & \vdots\\
\O_{n1} & \cdots & \O_{nn}
\end{pmatrix}$}
\raisebox{-3ex}{$\begin{pmatrix}
v_1\\
\vdots\\
v_n
\end{pmatrix}$}\;.
\label{app:bilinaux}
\end{equation}

In a given basis, both operators on $V$ and bilinear forms on $V$ are represented by a square matrix. It is therefore worthwhile to highlight the difference between the two types of objects. On the abstract level this is obvious: an operator is a map $V\to V$, whereas a bilinear form is a map $V\times V\to\R$. However, the difference can also be detected at the level of matrix elements by inspecting how the matrix representations of operators and bilinear forms transform under a change of basis in $V$. We know from before that when the basis is changed by a matrix $P$, the components of any vector in $V$ transform by~\eqref{app:basis_transfo}. Accordingly, by \eqref{app:optransfo}, the matrix representation of an operator $\hat A$ changes as $A\to A'=P^{-1}AP$. The matrix representation of bilinear forms transforms differently. Using~\eqref{app:bilinaux}, it follows that in the new basis, $\o(\vec u,\vec v)$ amounts to
\begin{equation}
\begin{pmatrix}
u_1',\dotsc,u_n'
\end{pmatrix}
\raisebox{-3ex}{$\begin{pmatrix}
\O_{11}' & \dotsc & \O_{1n}'\\
\vdots & \ddots & \vdots\\
\O_{n1}' & \cdots & \O_{nn}'
\end{pmatrix}$}
\raisebox{-3ex}{$\begin{pmatrix}
v_1'\\
\vdots\\
v_n'
\end{pmatrix}$}=
\begin{pmatrix}
u_1,\dotsc,u_n
\end{pmatrix}
P^{-1T}
\raisebox{-3ex}{$\begin{pmatrix}
\O_{11}' & \dotsc & \O_{1n}'\\
\vdots & \ddots & \vdots\\
\O_{n1}' & \cdots & \O_{nn}'
\end{pmatrix}$}
P^{-1}
\raisebox{-3ex}{$\begin{pmatrix}
v_1\\
\vdots\\
v_n
\end{pmatrix}$}\;.
\end{equation}
For $\o(\vec u,\vec v)$ to have a well-defined meaning independent of the choice of basis, the matrix representation $\O$ of the bilinear form must transform under the change of basis to $\O'=P^T\O P$.

A bilinear form $\o$ is called \emph{symmetric} if it satisfies the condition $\o(\vec u,\vec v)=\o(\vec v,\vec u)$ for any pair of vectors $\vec u,\vec v\in V$. Any symmetric bilinear form $\o$ gives naturally rise to a \emph{quadratic form} $K:V\to\R$, defined by
\begin{equation}
K(\vec v)\equiv\o(\vec v,\vec v)=\begin{pmatrix}
v_1,\dotsc,v_n
\end{pmatrix}
\raisebox{-3ex}{$\begin{pmatrix}
\O_{11} & \dotsc & \O_{1n}\\
\vdots & \ddots & \vdots\\
\O_{n1} & \cdots & \O_{nn}
\end{pmatrix}$}
\raisebox{-3ex}{$\begin{pmatrix}
v_1\\
\vdots\\
v_n
\end{pmatrix}$}\;.
\end{equation}
The set of matrix elements, $\O_{ij}$, of a symmetric bilinear form is a real symmetric matrix. We know from before that such a matrix can be diagonalized via $\O\to\O'=P^{-1}\O P$ with a real orthogonal $P$. From the orthogonality property of $P$ we get $P^{-1}=P^T$, hence $\O'=P^T\O P$, which exactly corresponds to a transformation of matrix elements of the bilinear form under a change of basis of $V$. As a consequence, any quadratic form can by a suitable change of basis be expressed as
\begin{equation}
K(\vec v)=\sum_{i=1}^n\lambda_iv_i^2\;,
\end{equation}
where $\lambda_i$ are the eigenvalues of $\O$. In fact, we can go even further and perform another change of basis where the matrix $P$ is taken to be diagonal,
\begin{equation}
P=\begin{pmatrix}
P_1 & 0 & 0 & \dotsb & 0\\
0 & P_2 & 0 & \dotsb & 0\\
0 & 0 & P_3 & \dotsb & 0\\[2pt]
\smash\vdots & \smash\vdots & \smash\vdots & \smash\ddots & \smash\vdots\\
0 & 0 & 0 & \dotsb & P_n
\end{pmatrix}\;.
\end{equation}
This corresponds to a mere rescaling of the basis vectors, which changes the eigenvalues $\lambda_i$ to $\lambda_iP_i^2$. The moral is that unlike for operators, the eigenvalues of (the matrix representation $\O$ of) a symmetric bilinear form are not a basis-independent concept. In fact, by a judicious choice of $P_i$, we can ensure that $\lambda_iP_i^2\in\{-1,0,+1\}$. Up to a change of basis, any quadratic form $K$ is therefore uniquely determined by the number of eigenvalues that are respectively negative, zero and positive. This data is called the \emph{signature} of $K$. The quadratic form is called \emph{positive-definite} if all its eigenvalues are positive.

\begin{watchout}%
We have used the concepts of a symmetric bilinear form and a quadratic form somewhat interchangeably, although they do seem distinct. A bilinear form is a function of two arguments, whereas a quadratic form is a function of a single argument. However, they do carry the same information. Above, we introduced a quadratic form as a restriction of a bilinear form to two equal arguments. Conversely, the bilinear form can be uniquely reconstructed from the descendant quadratic form using the identity
\begin{equation}
\o(\vec u,\vec v)=\frac14[K(\vec u+\vec v)-K(\vec u-\vec v)]\;.
\end{equation}
\end{watchout}

We will wrap up with one remarkable property of quadratic forms that plays a key role in the decomposition of small oscillations of mechanical systems into normal modes. In quantum mechanics, it is well-known that two Hermitian operators can be simultaneously diagonalized by a change of basis if and only if they commute. One might expect a similar property to hold for quadratic forms, but here the difference between operators and bilinear forms has a surprise for us: \emph{any} two quadratic forms $K$ and $L$ such that (at least) one of them is positive-definite can be simultaneously diagonalized. To see how this comes about, assume that it is $K$ that is positive-definite. We know from above that in that case, we can choose a basis in which $K$ is represented by the unit matrix. We would now like to perform an additional change of basis that diagonalizes $L$ while keeping $K$ diagonal. This can be accomplished with an orthogonal matrix $P$. Namely, on the one hand, $L$ is real symmetric and thus can be diagonalized by an orthogonal transformation. On the other hand, such an orthogonal transformation does not change $\O=\un$ since $\O\to\O'=P^T\O P=P^TP=\un$. At the end of the day, we find that there is a basis in which $K$ is represented by the unit matrix whereas $L$ is diagonal but with eigenvalues generally different from one.

\begin{illustration}%
For two quadratic forms $K$ and $L$ to be simultaneously diagonalizable, the assumption that one of them is positive-definite is essential. To see that the claim may not hold when this assumption is violated, it is sufficient to restrict to $V=\R^2$. Denoting the components of vectors in $\R^2$ simply as $x,y$, we define the two quadratic forms as $K(x,y)=x^2-y^2$ and $L(x,y)=xy$. It is easy to check by an explicit calculation that there is no invertible matrix $P$ that would keep $K$ diagonal and simultaneously make $L$ diagonal as well.
\end{illustration}
\bibliographystyle{spphys}
\bibliography{refs}

%%%%%%%%%%%%%%%%%%%%%%%%%%%%%%%%%%%%%%%%%%%%%%%%%%%%%%%%%%%%

\end{document}